\documentclass[14pt,fleqn]{extreport}

\usepackage[utf8]{inputenc}
\usepackage[T2A]{fontenc}
\usepackage[english,russian]{babel}

\usepackage[colorlinks=true,linkcolor=blue!75!black,pagecolor=red!75!black,citecolor=green!45!black]{hyperref}
\usepackage{amsmath}
\usepackage{amssymb}
\usepackage{amsthm}
\usepackage{array}
\usepackage{bm}
\usepackage{cite}
\usepackage{color}
\usepackage{dsfont}
\usepackage{enumerate}
\usepackage{enumitem}
\usepackage{epsf}
\usepackage{euscript}
\usepackage{graphicx} 
\usepackage{indentfirst}
\usepackage{latexsym}
\usepackage{makeidx}
\usepackage{mathrsfs}
\usepackage{mathtools}
\usepackage{ulem}  
\newcommand*\starletfill[1][]{\tikz[#1]{ 
  \draw [line width=0.05em,fill] 
       (18:0.34em) -- (54:0.1292em) -- (90:0.34em)
    -- (126:0.1292em) -- (162:0.34em) -- (198:0.1292em)
    -- (234:0.34em) -- (270:0.1292em) -- (306:0.34em)
    -- (342:0.1292em) -- cycle;
  \draw [line width=0.05em,fill] 
       (18:0.34em) -- (54:0.1292em) -- (90:0.34em)
    -- (126:0.1292em) -- (162:0.34em) -- (198:0.1292em)
    -- (234:0.34em) -- (270:0.1292em) -- (306:0.34em)
    -- (342:0.1292em) -- cycle;
}} 
\usepackage{wasysym}
\usepackage{wrapfig}
\usepackage{yfonts} 
\usepackage{tikz}
\usepackage{pgf}

\usetikzlibrary{calc}
\usetikzlibrary{intersections}
\usetikzlibrary{fadings}

\usepackage[left=2cm,right=2cm, top=2.5cm,bottom=2.5cm,bindingoffset=0cm]{geometry}

\SetEnumitemKey{midsep}{itemsep=0pt}
\setlist{midsep}

\newtheorem{rtheorem}{\indent\bf Теорема}

\newtheorem{theorem}{\indent\bf Теорема}[section]
\newtheorem{corollary}[theorem]{\indent\bf Следствие}
\newtheorem{lemma}[theorem]{\indent\bf Лемма}
\newtheorem{sublemma}[theorem]{\indent\bf Подлемма}

\newtheorem{proposition}[theorem]{\indent\bf Предложение}
\newtheorem{remark}[theorem]{\indent\bf Замечание}
\newtheorem{problem}{\indent\bf Проблема}

\newtheorem{hypothesis}{\indent\bf Гипотеза}

\definecolor{cg4}{gray}{0.99}              
\definecolor{cg6}{gray}{0.30}          
\definecolor{cg1}{gray}{0.60}           
\definecolor{cg2}{gray}{0.90}            
\definecolor{cg3}{gray}{0.75}             
\definecolor{cg5}{gray}{0.45}          

\colorlet{c1}{cg1}
\colorlet{c2}{cg2}
\colorlet{c3}{cg3}
\colorlet{c4}{cg4}
\colorlet{c5}{cg5}
\colorlet{c6}{cg6}

\newcounter{\theequation}[section]
\renewcommand{\theequation}{\thesection.\arabic{equation}}

\makeindex
\usepackage{cmap}
\usepackage{etoolbox}
\AtBeginEnvironment{thebibliography}{%
  \interlinepenalty=10000  
  \clubpenalty=10000       
  \widowpenalty=10000      
}

\newlength{\templength}
\newlength{\templengtha}
\newlength{\templengthb}
\newlength{\templengthc}
\newlength{\templengthd}
\newlength{\templengths}
\newlength{\templengthw}

\definecolor{mathcolor}{RGB}{0,75,105} 
\definecolor{argcolor} {RGB}{0,105,50} 
\definecolor{speccolor}{RGB}{0,40,125} 
\newcommand{\dismath} [1] {{\color{mathcolor}#1}}
\newcommand{\distext} [1] {{\color{speccolor}#1}}
\newcommand{\argument}[1] {{\color{argcolor!50!black}{#1}}}

\newcommand{\ulined}  [1] {\settowidth{\templengthw}{\mbox{#1}}
                           \hspace{0.25\templengthw}\underline{\hspace{-0.25\templengthw}#1\hspace{-0.3\templengthw}}\hspace{0.3\templengthw}}

\newcommand{\arrayitem}{\distext{\!\!\!\!\!\!\!\!\bullet}}
\newcommand{\arrayitemskip}     {\medskip}

\newcommand{\arraytab}{\hspace{1.69ex}} 

\newcommand{\defnotion}       [1]       {\distext{\textit{#1\/}}}

\newcommand{\remember}        [1]       {}
  \newcommand{\Rem}           [1]       {\remember{#1}}

\newcommand{\prop}{{\lang{P}}} 
\newcommand{\uclosure}        [1]       {\dismath{\bar{\forall}}#1} 

\newcommand{\bwedge}   {\boldsymbol{\wedge}}
\newcommand{\bvee}     {\boldsymbol{\vee}}
\newcommand{\bto}      {\boldsymbol{\to}}
\newcommand{\bbot}     {\boldsymbol{\bot}}
\newcommand{\bneg}     {\boldsymbol{\neg}}
\newcommand{\btop}     {\boldsymbol{\top}}

\newcommand{\bBox}     {\boldsymbol{\Box}}

\newcommand{\boxm}    [1] {[\argument{#1}]}
\newcommand{\diam}    [1] {\langle\argument{#1}\rangle}
\newcommand{\comp}        {\mathop{\dismath{;}}}
\newcommand{\choice}      {\mathop{\dismath{\cup}}}
\newcommand{\inter}       {\mathop{\dismath{\cap}}}

\newcommand{\Next}        {{\dismath{\ocircle}}}
\newcommand{\Until}       {{\dismath{\mathcal{U}}}}

\newcommand{\set}             [1]       {\dismath{\left\{ \argument{#1} \right\}}}
\newcommand{\setc}            [2]       {\set{ \argument{#1} \mathrel{\dismath{:}} \argument{#2} }}
\newcommand{\Power}           [1]       {\dismath{\mathscr{P}(\argument{#1})}}

\newcommand{\tuple}           [1]       {\dismath{\left( \argument{#1} \right)}}
\newcommand{\pair}            [2]       {\tuple{{#1},{#2}}}

\newcommand{\otuple}          [1]       {\dismath{\left\langle\argument{#1}\right\rangle}}
\newcommand{\opair}           [2]       {\otuple{{#1},{#2}}}

\newcommand{\function}        [3]       {\dismath{\argument{#1}\colon \argument{#2}\to \argument{#3}}}
\newcommand{\func}            [3]       {\function{#1}{#2}{#3}}

\newcommand{\intmodels}       [0]   {\mathrel{\dismath{|\hspace{-1.25pt}|\hspace{-2.8pt}{-}}}}
\newcommand{\clmodels}        [0]   
                                    {\mathrel{\dismath{|\hspace{-1.25pt}{\models}}}}
  
  \newcommand{\imodels}   [0]{\intmodels}
  \newcommand{\cmodels}   [0]{\clmodels}

\newcommand{\nameKFrame}      [1]       {\dismath{\mathfrak{#1}}}       
\newcommand{\nameKModel}      [1]       {\dismath{\mathfrak{#1}}}       
\newcommand{\nameClModel}     [1]       {\dismath{{#1}}}        
\newcommand{\kframe}          [1]       {\nameKFrame{#1}}       
\newcommand{\kmodel}          [1]       {\nameKModel{#1}}       
\newcommand{\cmodel}          [1]       {\nameClModel{#1}}      

\newcommand{\kFrame}          [1]       {\dismath{\bfrak{#1}}}       
\newcommand{\kModel}          [1]       {\dismath{\bfrak{#1}}}       
\newcommand{\cModel}          [1]       {\dismath{\bm{#1}}}      

\newcommand{\formula}         [1]       {\dismath{\bm{#1}}}             
\newcommand{\logic}           [1]       {\dismath{\mathbf{#1}}}         
\newcommand{\lang}            [1]       {\dismath{\mathcal{#1}}}        
\newcommand{\cclass}          [1]       {\dismath{\mathrm{#1}}}         
  \newcommand{\ccls}[1]{\cclass{#1}}
\newcommand{\sclass}          [1]       {\dismath{\mathscr{#1}}}        
  \newcommand{\scls}[1]{\sclass{#1}}
\newcommand{\Sclass}          [1]       { \settowidth{\templengths}{\mbox{$\mathscr{#1}$}}
\dismath{\mathscr{#1}\hspace{-0.975\templengths}\mathscr{#1}\hspace{-0.975\templengths}\mathscr{#1}}}        
  \newcommand{\Scls}[1]{\Sclass{#1}}

\newcommand{\numbers}         [1]       {\dismath{\mathds{#1}}}
\newcommand{\numN}                      {\numbers{N}}
\newcommand{\numNp}                     {\numbers{N}^+}
\newcommand{\numQ}                      {\numbers{Q}}
\newcommand{\numR}                      {\numbers{R}}

\newcommand{\Pow}             [1]       {\dismath{\mathscr{P}(\argument{#1})}}  

\newcommand{\otupleIs}        [2]       {\ensuremath \argument{#1} = \otuple{#2}}
\newcommand{\kframeIs}        [2]       {\otupleIs{\nameKFrame{#1}}{#2}}

\newcommand{\kfmodelIs}       [3]       {\otupleIs{\nameKModel{#1}}{\nameKFrame{#2},{#3}}}
\newcommand{\kmodelIs}        [2]       {\otupleIs{\nameKModel{#1}}{#2}}

\newcommand{\kframes}         [1]       {\dismath{\textnormal{\texttt{KF}}(\argument{#1})}}   
\newcommand{\kFrames}         [1]       {\dismath{\textnormal{\texttt{KFD}}(\argument{#1})}}  
\newcommand{\mlogic}          [1]       {\dismath{\textnormal{\texttt{ML}}(\argument{#1})}}   
\newcommand{\mPlogic}         [1]       {\dismath{\textnormal{\texttt{QML}}(\argument{#1})}}  
\newcommand{\mPlogicC}        [1]       {\dismath{\textnormal{\texttt{QML}}_{\mathrm{c}}(\argument{#1})}}  
\newcommand{\mPlogicx}        [2]       {\dismath{\textnormal{\texttt{QML}}^{#2}(\argument{#1})}}  
\newcommand{\mPlogicCx}       [2]       {\dismath{\textnormal{\texttt{QML}}_{\mathrm{c}}^{#2}(\argument{#1})}}  
\newcommand{\ilogic}          [1]       {\dismath{\textnormal{\texttt{IL}}(\argument{#1})}}   
\newcommand{\iPlogic}         [1]       {\dismath{\textnormal{\texttt{QIL}}(\argument{#1})}}  
\newcommand{\iPlogicC}        [1]       {\dismath{\textnormal{\texttt{QIL}}_{\mathrm{c}}(\argument{#1})}}  
\newcommand{\iPlogicx}        [2]       {\dismath{\textnormal{\texttt{QIL}}^{#2}(\argument{#1})}}  
\newcommand{\iPlogicCx}       [2]       {\dismath{\textnormal{\texttt{QIL}}_{\mathrm{c}}^{#2}(\argument{#1})}}  
\newcommand{\malg}            [1]       {\dismath{\textnormal{\texttt{MV}}(\argument{#1})}}   

\newcommand{\kflogic}         [1]       {\dismath{\logic{Q}^{\mathrm{K}}\argument{#1}}}  
\newcommand{\wfin}            [1]       {\dismath{\argument{#1}_{\mathit{wfin}}}}

\providecommand{\hash}{\#}
\providecommand{\QCLFs}{\logic{QCl}_{\mathit{fin}}^{\mathit{bin}\mbox{-}3}}
\providecommand{\langQCLFs}{\lang{QL}^{\mathit{bin}\mbox{-}3}}
\providecommand{\pQCLFs}{\logic{QCl}_{\mathit{fin}}^{\mathit{bin}\mbox{-}3{+}}}
\providecommand{\plangQCLFs}{\lang{QL}^{\mathit{bin}\mbox{-}3{+}}}

\newcommand{\boldfrak}        [1]       {\boldsymbol{\frak{#1}}}

\newcommand{\bfrak}           [1]       {\boldfrak{#1}}

\newcommand{\implication}               {\to}
  \newcommand{\imp} {\implication}
\renewcommand{\conjunction}             {\wedge}   
  \newcommand{\con} {\conjunction}
\newcommand{\disjunction}               {\vee}
  \newcommand{\dis} {\disjunction}

\newcommand{\equivalence}               {\leftrightarrow}
  \newcommand{\lra} {\equivalence}
\newcommand{\fusion}                    {\mathop{\dismath{\ast}}}

\newcommand{\leftsquare}[1]{%
\begin{tikzpicture}[scale=#1]
\draw (0,0)--(0,1)--(1,1)--(1,0)--cycle;
\draw [fill, color=black!25] (0,0)--(0.5,0.5)--(0,1);
\draw (0,0)--(0.5,0.5)--(0,1);
\draw (0,1)--(0,0);
\end{tikzpicture}
}
\newcommand{\rightsquare}[1]{%
\begin{tikzpicture}[scale=#1]
\draw (0,0)--(0,1)--(1,1)--(1,0)--cycle;
\draw [fill, color=black!25] (1,0)--(0.5,0.5)--(1,1);
\draw (1,0)--(0.5,0.5)--(1,1);
\draw (1,1)--(1,0);
\end{tikzpicture}
}

\newcommand{\upsquare}[1]{%
\begin{tikzpicture}[scale=#1]
\draw (0,0)--(0,1)--(1,1)--(1,0)--cycle;
\draw [fill, color=black!25] (0,1)--(0.5,0.5)--(1,1);
\draw (0,1)--(0.5,0.5)--(1,1);
\draw (1,1)--(0,1);
\end{tikzpicture}
}

\newcommand{\downsquare}[1]{%
\begin{tikzpicture}[scale=#1]
\draw (0,0)--(0,1)--(1,1)--(1,0)--cycle;
\draw [fill, color=black!25] (0,0)--(0.5,0.5)--(1,0);
\draw (0,0)--(0.5,0.5)--(1,0);
\draw (1,0)--(0,0);
\end{tikzpicture}
}

\newcommand{\leftsq} {\mathop{\leftsquare {0.3}}}
\newcommand{\rightsq}{\mathop{\rightsquare{0.3}}}
\newcommand{\upsq}   {\mathop{\upsquare   {0.3}}}
\newcommand{\downsq} {\mathop{\downsquare {0.3}}}

\renewcommand{\iff}                     {\mathrel{\dismath{\Longleftrightarrow}}} 
\newcommand{\imply}                     {\mathrel{\dismath{\Longrightarrow}}}     
\newcommand{\bydef}                     {\mathrel{\dismath{\leftrightharpoons}}}  

\newcommand{\logicK}                    {\logic{K}}
\newcommand{\logicT}                    {\logic{T}}
\newcommand{\logicKfour}                {\logic{K4}}
\newcommand{\logicSfour}                {\logic{S4}}
\newcommand{\logicSfive}                {\logic{S5}}
\newcommand{\logicKB}                   {\logic{KB}}
\newcommand{\logicKTB}                  {\logic{KTB}}

\newcommand{\logicKfive}                {\logic{K5}}
\newcommand{\logicGL}                   {\logic{GL}}
\newcommand{\logicGrz}                  {\logic{Grz}}
\newcommand{\logicwGrz}                 {\logic{wGrz}}

\newcommand{\MP}                        {\textrm{modus ponens\/}}
\newcommand{\mdepth}                    {\mathop{\mathit{md}}}

\makeatletter
\newcommand{\mref}{\@ifnextchar({\mref@i}{\mref@i({\Box},{p})}}
\def\mref@i(#1,#2){#1#2 \implication #2}
\makeatother
    \makeatletter
    \newcommand{\mrefp}{\@ifnextchar({\mrefp@i}{\mrefp@i({p})}}
    \def\mrefp@i(#1){\mref(\Box,#1)}
    \makeatother
\makeatletter
\newcommand{\FOref}{\@ifnextchar({\FOref@i}{\FOref@i({x},{P})}}
\def\FOref@i(#1,#2){\forall #1\,#2(#1,#1)}
\makeatother
    \makeatletter
    \newcommand{\FOrefp}{\@ifnextchar({\FOrefp@i}{\FOrefp@i({P})}}
    \def\FOrefp@i(#1){\FOref(x,#1)}
    \makeatother
\makeatletter
\newcommand{\FOrefi}{\@ifnextchar({\FOrefi@i}{\FOrefi@i({x},{P})}}
\def\FOrefi@i(#1,#2){\forall #1\,#1#2#1}
\makeatother
    \makeatletter
    \newcommand{\FOrefip}{\@ifnextchar({\FOrefip@i}{\FOrefip@i({P})}}
    \def\FOrefip@i(#1){\FOrefi(x,#1)}
    \makeatother

\makeatletter
\newcommand{\mtra}{\@ifnextchar({\mtra@i}{\mtra@i({\Box},{p})}}
\def\mtra@i(#1,#2){#1#2 \implication #1#1#2}
\makeatother
    \makeatletter
    \newcommand{\mtrap}{\@ifnextchar({\mtrap@i}{\mtrap@i({p})}}
    \def\mtrap@i(#1){\mtra(\Box,#1)}
    \makeatother
\makeatletter
\newcommand{\FOtra}{\@ifnextchar({\FOtra@i}{\FOtra@i({x},{y},{z},{P})}}
\def\FOtra@i(#1,#2,#3,#4){\forall #1\forall #2\forall #3\,(#4(#1,#2)\conjunction #4(#2,#3) \implication #4(#1,#3))}
\makeatother
    \makeatletter
    \newcommand{\FOtrap}{\@ifnextchar({\FOtrap@i}{\FOtrap@i({P})}}
    \def\FOtrap@i(#1){\FOtra(x,y,z,#1)}
    \makeatother
\makeatletter
\newcommand{\FOtrai}{\@ifnextchar({\FOtrai@i}{\FOtrai@i({x},{y},{z},{P})}}
\def\FOtrai@i(#1,#2,#3,#4){\forall #1\forall #2\forall #3\,(#1#4#2\conjunction #2#4#3 \implication #1#4#3)}
\makeatother
    \makeatletter
    \newcommand{\FOtraip}{\@ifnextchar({\FOtraip@i}{\FOtraip@i({P})}}
    \def\FOtraip@i(#1){\FOtrai(x,y,z,#1)}
    \makeatother

\makeatletter
\newcommand{\msym}{\@ifnextchar({\msym@i}{\msym@i({\Box},{\Diamond},{p})}}
\def\msym@i(#1,#2,#3){#3 \implication #1#2#3}
\makeatother
    \makeatletter
    \newcommand{\msymp}{\@ifnextchar({\msymp@i}{\msymp@i({p})}}
    \def\msymp@i(#1){\msym(\Box,\Diamond,#1)}
    \makeatother
\makeatletter
\newcommand{\FOsym}{\@ifnextchar({\FOsym@i}{\FOsym@i({x},{y},{P})}}
\def\FOsym@i(#1,#2,#3){\forall #1\forall #2\,(#3(#1,#2)\implication #3(#2,#1))}
\makeatother
    \makeatletter
    \newcommand{\FOsymp}{\@ifnextchar({\FOsymp@i}{\FOsymp@i({P})}}
    \def\FOsymp@i(#1){\FOsym(x,y,#1)}
    \makeatother
\makeatletter
\newcommand{\FOsymi}{\@ifnextchar({\FOsymi@i}{\FOsymi@i({x},{y},{P})}}
\def\FOsymi@i(#1,#2,#3){\forall #1\forall #2\,(#1#3#2 \implication #2#3#1)}
\makeatother
    \makeatletter
    \newcommand{\FOsymip}{\@ifnextchar({\FOsymip@i}{\FOsymip@i({P})}}
    \def\FOsymip@i(#1){\FOsymi(x,y,#1)}
    \makeatother

\makeatletter
\newcommand{\meuc}{\@ifnextchar({\meuc@i}{\meuc@i({\Box},{\Diamond},{p})}}
\def\meuc@i(#1,#2,#3){#2#3 \implication #1#2#3}
\makeatother
    \makeatletter
    \newcommand{\meucp}{\@ifnextchar({\meucp@i}{\meucp@i({p})}}
    \def\meucp@i(#1){\meuc(\Box,\Diamond,#1)}
    \makeatother
\makeatletter
\newcommand{\FOeuc}{\@ifnextchar({\FOeuc@i}{\FOeuc@i({x},{y},{z},{P})}}
\def\FOeuc@i(#1,#2,#3,#4){\forall #1\forall #2\forall #3\,(#4(#1,#2)\con #4(#1,#3)\implication #4(#2,#3))}
\makeatother
    \makeatletter
    \newcommand{\FOeucp}{\@ifnextchar({\FOeucp@i}{\FOeucp@i({P})}}
    \def\FOeucp@i(#1){\FOeuc(x,y,z,#1)}
    \makeatother
\makeatletter
\newcommand{\FOeuci}{\@ifnextchar({\FOeuci@i}{\FOeuci@i({x},{y},{z},{P})}}
\def\FOeuci@i(#1,#2,#3,#4){\forall #1\forall #2\forall #3\,(#1#4#2 \con #1#4#3 \implication #2#4#3)}
\makeatother
    \makeatletter
    \newcommand{\FOeucip}{\@ifnextchar({\FOeucip@i}{\FOeucip@i({P})}}
    \def\FOeucip@i(#1){\FOeuci(x,y,z,#1)}
    \makeatother

\makeatletter
\newcommand{\mser}{\@ifnextchar({\mser@i}{\mser@i({\Box},{\Diamond},{p})}}
\def\mser@i(#1,#2,#3){#1#3 \implication #2#3}
\makeatother
    \makeatletter
    \newcommand{\mserp}{\@ifnextchar({\mserp@i}{\mserp@i({p})}}
    \def\mserp@i(#1){\mser(\Box,\Diamond,#1)}
    \makeatother
\makeatletter
\newcommand{\FOser}{\@ifnextchar({\FOser@i}{\FOser@i({x},{y},{P})}}
\def\FOser@i(#1,#2,#3){\forall #1\exists #2\,#3(#1,#2)}
\makeatother
    \makeatletter
    \newcommand{\FOserp}{\@ifnextchar({\FOserp@i}{\FOserp@i({P})}}
    \def\FOserp@i(#1){\FOser(x,y,#1)}
    \makeatother
\makeatletter
\newcommand{\FOseri}{\@ifnextchar({\FOseri@i}{\FOseri@i({x},{y},{P})}}
\def\FOseri@i(#1,#2,#3){\forall #1\exists #2\,#1#3#2}
\makeatother
    \makeatletter
    \newcommand{\FOserip}{\@ifnextchar({\FOserip@i}{\FOserip@i({P})}}
    \def\FOserip@i(#1){\FOseri(x,y,#1)}
    \makeatother

\makeatletter
\newcommand{\mla}{\@ifnextchar({\mla@i}{\mla@i({\Box},{p})}}
\def\mla@i(#1,#2){#1(#1#2 \implication #2) \implication #1#2}
\makeatother
    \makeatletter
    \newcommand{\mlap}{\@ifnextchar({\mlap@i}{\mlap@i({p})}}
    \def\mlap@i(#1){\mla(\Box,#1)}
    \makeatother

\makeatletter
\newcommand{\mgrz}{\@ifnextchar({\mgrz@i}{\mgrz@i({\Box},{p})}}
\def\mgrz@i(#1,#2){#1(#1(#2 \implication #1#2) \implication #2) \implication #2}
\makeatother
    \makeatletter
    \newcommand{\mgrzp}{\@ifnextchar({\mgrzp@i}{\mgrzp@i({p})}}
    \def\mgrzp@i(#1){\mgrz(\Box,#1)}
    \makeatother

\makeatletter
\newcommand{\mwgrz}{\@ifnextchar({\mwgrz@i}{\mwgrz@i({\Box},{p})}}
\def\mwgrz@i(#1,#2){#1^+(#1(#2 \implication #1#2) \implication #2) \implication #2}
\makeatother
    \makeatletter
    \newcommand{\mwgrzp}{\@ifnextchar({\mwgrzp@i}{\mwgrzp@i({p})}}
    \def\mwgrzp@i(#1){\mwgrz(\Box,#1)}
    \makeatother

\makeatletter
\newcommand{\ExtDiamond}{\@ifnextchar({\ExtDiamond@i}{\ExtDiamond@i({p})}}
\def\ExtDiamond@i(#1){{\Diamond\!\!\!\!\Diamond}_{#1}}
\makeatother
\newcommand{\pDiamond}{\ExtDiamond(p)}
\newcommand{\PDiamond}{\ExtDiamond({\forall x\,P(x)})}

\usetikzlibrary{decorations.pathmorphing}
\usepackage{xcolor}
\usepackage{eso-pic} 

\renewcommand{\dismath} [1] {{#1}}
\renewcommand{\distext} [1] {{#1}}
\renewcommand{\argument}[1] {{#1}}

\newcounter{savefootnote}

\makeindex

\begin{document}


\thispagestyle{empty}

\begin{center}
{\sc
Федеральное государственное автономное\\ образовательное учреждение высшего образования\\ <<Московский физико-технический институт\\ (национальный исследовательский университет)>>
\\[5mm]
Высшая школа современной математики\\[7mm]
}
\end{center}

~\hfill ~
\\[25mm]

\begin{center}
{\large Рыбаков Михаил Николаевич}
\end{center}

\begin{center}

{\LARGE\bf Моделирование логических систем \\
средствами их фрагментов}
\\[12mm]

\centerline{\Large\sc ДИССЕРТАЦИЯ}

на соискание учёной степени доктора\\ физико-математических наук

\vfill

\textit{Версия с исправлениями и дополнениями}

\vfill

\vfill

Москва -- 2025
\end{center}


\clearpage
\thispagestyle{empty}
\addtocounter{page}{-1}

\definecolor{pageyellow}{RGB}{255,250,220}
\definecolor{textgreen}{RGB}{0,32,0}
\definecolor{frame}{RGB}{0,60,0}

\AddToShipoutPictureBG*{%
  \AtPageLowerLeft{%
    \begin{tikzpicture}[remember picture, overlay]
      \fill[pageyellow!60] (0,0) rectangle (\paperwidth,\paperheight);
      
      \draw[frame, line width=0.4pt, 
 				decorate,
            decoration={snake, amplitude=0.7mm, segment length=3mm, pre=lineto, pre length=.1mm, post=lineto, post length=.1mm}] 
            (15mm,15mm) rectangle (\paperwidth-15mm,\paperheight-15mm);
    \end{tikzpicture}%
  }%
}

\newgeometry{left=21mm, right=21mm, top=20mm, bottom=20mm} 
\color{textgreen}

~\\[-45pt]

\begin{center}
{\bf\Large Предисловие}
\end{center}


Диссертация была подготовлена в период работы автора в Тверском государственном университете, Научно-исследовательском институте <<Центрпрограммсистем>>, Институте проблем передачи информации имени А.\,А.~Харкевича Российской академии наук, Высшей школе экономики и Высшей школе современной математики Московского физико-технического института.

Многие результаты диссертации были получены в рамках научных проектов, поддержанных грантами Министерства образования Российской Федерации (сейчас это Министерство науки и высшего образования Российской Федерации), Российского фонда фундаментальных исследований, Российского гуманитарного научного фонда, Российского научного фонда, а также в рамках совместных проектов Российского фонда фундаментальных исследований и французского Национального центра научных исследований (Centre national de la recherche scientifique).

Исследования по теме диссертации были начаты автором под руководством Александра Васильевича Чагрова и совместно с ним. Часть результатов этих исследований вошла в кандидатскую диссертацию автора, научным руководителем которой был Александр Васильевич, а затем и в докторскую.

Защита диссертации состоялась 20 ноября 2025 года в диссертационном совете Национального исследовательского университета <<Высшая школа экономики>> по математике. 


Автор выражает глубокую благодарность членам комитета по защите диссертации и диссертационному совету за внимание к диссертационной работе и ценные замечания, способствовавшие улучшению текста.

Приводимый ниже текст содержит исправления неаккуратностей и неточностей, отмеченных членами 
комитета,
а внесённые в текст дополнения касаются лишь библиографических ссылок и исторических замечаний; математические утверждения, их доказательства, а также нумерация утверждений сохранены, и соответствуют официальной версии диссертации.

\vfill

{\it
\hfill 28.12.2025\,г.\medskip

\hfill Михаил Рыбаков

\vspace{7.5pt}
}

\restoregeometry

\clearpage

\tableofcontents

\chapter*{Введение}
\addcontentsline{toc}{chapter}{Введение}

\section*{Описание области исследований}
\addcontentsline{toc}{section}{Описание области исследований}

Проведённое автором исследование связано с выразительностью языков, логик и теорий, и прежде всего с алгоритмической выразительностью (в том числе вычислительной сложностью) определённых их фрагментов.

Многие естественные логические системы либо алгоритмически неразрешимы (причём иногда сильно неразрешимы), либо, будучи разрешимыми, имеют высокую сложность проблемы разрешения. Известно, что определённые ограничения, накладываемые на средства языка или используемую семантику, приводят к изменению алгоритмической сложности тех или иных задач. В~то же время иногда это не так: например, в неклассических логиках как неразрешимость, так и высокая сложность проблемы разрешения в случае разрешимости могут получаться при очень сильных ограничениях на средства языка.

Представляется актуальным не только нахождение границ, в рамках которых подобные проблемы оказываются алгоритмически простыми или наоборот остаются алгоритмически сложными, но и разработка общих методов, позволяющих получать оценки алгоритмической сложности фрагментов как отдельных логических систем, так и всех систем тех или иных бесконечных классов. Вместе с методами хотелось бы иметь общие признаки или критерии, позволяющие сразу говорить об алгоритмической сложности тех или иных фрагментов интересующей нас системы или хотя бы о потенциальной возможности или невозможности применения этих методов.

Подобные исследования проводились как в области пропозициональных систем, так и в области предикатных логик и теорий. В~диссертации предложены подходы и методы, позволяющие, в частности, ответить на многие вопросы об алгоритмической сложности фрагментов различных систем, оставшиеся открытыми в работах исследователей, занимавшихся их изучением.

Проблема разрешения многих <<стандартных>> модальных пропозициональных логик PSPACE-полна~\cite{Ladner77,ChZ,ZakharyaschevWolterChagrov-2001}; то же относится к интуиционистской пропозициональной логике и близким к ней~\cite{Statman-1979-1,Chagrov-1985-1-rus}. Кроме того, для некоторых модальных логик было установлено, что PSPACE-полным являются даже их фрагменты от одной переменной~\cite{BS93,Spaan-1993-1,Halpern95,Sve03}.

Вычислительная сложность проблемы разрешения таких логик оказалась тесно связана с вопросами сложности их аппроксимации шкалами Крипке. Вопросы о сложности аппроксимации неклассических логик были поставлены А.\,В.~Кузнецовым~\cite{Kuznetsov-1979-1,Kuznetsov75} в контексте интуиционистской логики.
При этом сам А.\,В.~Кузнецов предполагал, что интуиционистская логика полиномиально аппроксимируема, а следовательно, полиномиально разрешима, и даже предложил конструкцию, позволяющую полиномиально погрузить интуиционистскую логику в классическую в предположении справедливости этой гипотезы~\cite{Kuznetsov-1979-1}.
Чуть позже М.\,В.~Захарьящевым и С.\,В.~Поповым было показано, что интуиционистская логика не является полиномиально аппроксимируемой~\cite{ZakharyaschevPopov-1980-1-rus}; то же справедливо и для модальных логик.

Тем не менее, фрагмент интуиционистской логики от одной переменной является полиномиально разрешимым, что следует из конструкции Ригера--Ниши\-муры~\cite{Rieger52,Nishimura60}. Сложность фрагментов от большего числа переменных была неизвестна~\cite[проблема~18.4]{ChZ}, и А.\,В.~Чагровым высказывалась гипотеза о том, что каждый из этих фрагментов является полиномиально аппроксимируемым, а значит, полиномиально разрешимым.

Автор диссертации начал исследование этих вопросов под руководством А.\,В.~Чагрова примерно в 2001 году. В ходе исследований были получены результаты, касающиеся сложности проблемы разрешения для бесконечных семейств неклассических логик; кроме того, как для логик в полном языке, так и для их фрагментов в языке с конечным числом переменных, были получены результаты, касающиеся сложности их аппроксимации шкалами Крипке, а также связи сложности аппроксимации с вычислительной сложностью проблемы разрешения. Возникшие в ходе исследований методы были затем перенесены на полимодальные пропозициональные логики~\cite{MR:2007:JANCL,MR:2007:Tver,
MR:2018:SAICSIT,MR:2018:IGPL,MR:2021:JLC:1,MR:2021:arXiv,MR:2022:TCS,MR:2022:JLC} и предикатные логики~\cite{MR:2002:LI:2,MR:2015:LI, MR:2016:Tver, MR:2017:LI, MR:2019:SL, MR:2020:AiML, MR:2020:JLC:2, MR:2021:LI, MR:2021:JLC:3, MR:2021:JLC:2, MR:2023:SCAN:1, MR:2023:IGPL}.

Классическая логика предикатов $\logic{QCl}$ алгоритмически неразрешима~\cite{Church36,Turing36} (более точно, $\Sigma^0_1$-полна~\cite{Harel86}). Область исследования вопросов разрешимости её фрагментов известна как <<классическая проблема разрешения>>~\cite{BGG97}: имеется огромное количество результатов, дающих как разрешимые, так и неразрешимые фрагменты логики~$\logic{QCl}$, классических теорий первого порядка, а также других формальных предикатных систем~\cite{Suranyi43, Motohashi-1990-1, Gradel99, GKV97, Mortimer75, HWZ00, HWZ01, KKZ05, WZ01, Kripke62, Kremer97, GSh93}. Так, чтобы доказать неразрешимость логики $\logic{QCl}$, достаточно, чтобы её язык содержал одну бинарную предикатную букву и три предметные переменные~\cite{TG87}.
В то же время, следующие фрагменты логики $\logic{QCl}$ разрешимы:
\begin{itemize}
\item
монадический фрагмент, даже обогащённый равенством~\cite{Lowenheim15,BBJ07};
\item
различные охраняемые фрагменты~\cite{Motohashi-1990-1,ANB:1998,Gradel99};
\item
фрагмент с двумя предметными переменными~\cite{GKV97,Mortimer75} (см.\ также работы \cite{Seg73} и \cite{Shehtman:2012:rus}).
\end{itemize}
Неразрешимы также и многие первопорядковые теории, в частности, теории в языке лишь с одной бинарной предикатной буквой: например, теория симметричного иррефлексивного бинарного отношения (это извлекается из конструкций, приведённых в~\cite{NerodeShore80,Kremer97}) и, как следствие, теория симметричного рефлексивного бинарного отношения. Что касается теорий конечных моделей, они могут даже не быть рекурсивно перечислимыми~\cite{Trakhtenbrot50,Trakhtenbrot53} (похожие результаты для других языков можно найти, например, в~\cite{BM15,Hajek98}).

Алгоритмическая неразрешимость модальных предикатных логик мгновенно следует из неразрешимости классической логики предикатов. Для доказательства неразрешимости классической логики предикатов требуются предикатные буквы валентности не менее чем два, а также как минимум три предметные переменные. Поэтому возникает вопрос об алгоритмической сложности монадических фрагментов модальных предикатных логик, а также их фрагментов от не более чем двух предметных переменных. Фрагменты от одной предметной переменной могут быть разрешимы: как было замечено Д.\,М.~Габбаем и В.\,Б.~Шехтманом~\cite{GSh93}, из~\cite{Seg73} следует разрешимость фрагмента логики~$\logic{QS5}$ от одной предметной переменной, а аналогичный результат для $\logic{QS4}$ следует из~\cite{FS77,WZ01}; кроме того, в~\cite{AD90} доказана разрешимость фрагмента от одной предметной переменной логики~$\logic{QGL}$. Поэтому особый интерес представляет вопрос о разрешимости фрагментов с двумя предметными переменными, в частности, монадических.

Известно, что, в отличие от монадического фрагмента классической логики предикатов, монадические фрагменты модальных предикатных логик во многих <<естественных>> случаях алгоритмически неразрешимы, причём для соответствующего доказательства достаточно, чтобы их язык содержал две унарные предикатные буквы и бесконечно много предметных переменных~\cite{Kripke62} или две предметные переменные и бесконечно много унарных предикатных букв~\cite{GSh93,KKZ05}.

Из неразрешимости классической логики предикатов $\logic{QCl}$ следует и
неразрешимость интуиционистской предикатной логики $\logic{QInt}$, поскольку $\logic{QCl}$ погружается в~$\logic{QInt}$. Известно, что монадический фрагмент $\logic{QInt}$ неразрешим, и даже при одной унарной букве в языке~\cite{MMO-1965-1,Gabbay81}, а также известно, что неразрешим фрагмент $\logic{QInt}$ с двумя предметными переменными~\cite{GSh93,KKZ05}. Разрешимость фрагмента $\logic{QInt}$ с одной предметной переменной следует из~\cite{Fischer-Servi:1978-GISTFM}, а также из разрешимости аналогичного фрагмента модальной логики~$\logic{QS4}$.

В диссертации показано, что эти результаты о неразрешимости можно усилить и распространить на большой класс логик, включающий как модальные и суперинтуиционистские предикатные логики, так и расширения предикатных вариантов базисной и формальной логик Виссера. При этом предложены общие методы, позволяющие моделировать бинарные предикатные буквы унарными (в~духе метода Крипке~\cite{Kripke62} для модальных логик), а также унарные предикатные буквы формулами от одной унарной предикатной буквы.

Кроме того, в диссертации довольно большое внимание уделено вопросам алгоритмической сложности логик элементарно неопределимых классов шкал Крипке.
Известно, что если логика (любая, в том числе модальная, суперинтуиционистская, пропозициональная, предикатная) полна относительно элементарно определимого класса шкал Крипке, то такая логика погружается в классическую логику предикатов~\cite{Benthem85,MR:2000:IFRAS,GShS}, а следовательно, является рекурсивно перечислимой. Если же логика полна по Крипке, но при этом не полна относительно элементарно определимого класса шкал Крипке, то такая логика довольно часто не является рекурсивно перечислимой.

Имеются элементарно неопределимые классы шкал Крипке, содержащие в том числе бесконечные шкалы, модальные предикатные логики которых представляют определённый интерес. Например, к такого рода <<естественным>> классам относятся классы конечных шкал Крипке той или иной логики, различные классы деревьев, а также классы шкал таких логик как $\logic{QGL}$, $\logic{QGrz}$, $\logic{QwGrz}$ и других.

В диссертации рассмотрены логики, определяемые такого рода классами шкал Крипке, и продемонстрированы методы, позволяющие доказывать, что эти логики сильно неразрешимы даже при очень бедных средствах языка. Это позволило извлечь некоторые следствия, в том числе касающиеся полноты по Крипке различных исчислений и их фрагментов.


\section*{Цель работы}
\addcontentsline{toc}{section}{Цель работы}

Основная цель работы состоит в том, чтобы развить общие методы моделирования алгоритмически сложных проблем внутри логик и теорий минимальными средствами языка. В~частности, в работе предложены методы моделирования полных языков их ограниченными средствами, позволяющие получить погружения логик и теорий в соответствующие их фрагменты; в большинстве случаев эти погружения оказываются полиномиально вычислимыми.

Сопутствующая цель работы состоит в том, чтобы показать, что при дальнейшем ограничении средств языков подобные моделирования становятся невозможными. Таким образом, вместе с разработкой методов получается и описание своего рода границ возможности их применения.

К средствам языка, которые минимизируются, в первую очередь относятся следующие: число пропозициональных переменных в пропозициональных языках, число предметных переменных, а также число и валентность предикатных букв в языках первого порядка. Рассматриваются и некоторые ограничения на использование логических связок и кванторов.

Отметим, что попутно удаётся решить и другие задачи, не попадающие в рамки описанных целей работы. Так, в качестве одного из следствий получены результаты о неполноте по Крипке бесконечных классов модальных исчислений первого порядка, причём даже при сильных ограничениях на средства языка.


\section*{Основные результаты}
\addcontentsline{toc}{section}{Основные результаты}

В диссертации представлены результаты об алгоритмической сложности фрагментов как пропозициональных~\cite{MR:2019:IGPL, MR:2008:JANCL, MR:2007:JANCL, MR:2018:IGPL, MR:2022:JLC, MR:2023:JLC}, так и предикатных~\cite{MR:2022:DAN:eng, MR:2022:IGPL, MR:2017:LI, MR:2019:SL, MR:2022:SL, MR:2020:JLC:2, MR:2021:JLC:2, MR:2021:JLC:3, MR:2023:IGPL, MR:2020:JLC:1} логик.
Ниже используются стандартные обозначения для логик и формул~\cite{ChZ,GShS}; знак <<$+$>> понимается как объединение множеств формул с последующим замыканием по правилам modus ponens, обобщения и подстановки, а знак <<$\oplus$>>~--- как объединение множеств формул с последующим замыканием по правилам modus ponens, обобщения, подстановки и усиления.

\begin{rtheorem}[результат представлен в~{\cite{MR:2008:JANCL}}]
\label{th:Int:complexity}
Пусть $\logic{Int}\subseteq L\subseteq \logic{KC}$. Тогда позитивный фрагмент от двух переменных логики $L$ является\/ $\cclass{PSPACE}$-трудным.
\end{rtheorem}

\begin{rtheorem}[результат представлен в~{\cite{MR:2022:JLC}}]
Произведения и полупроизведения логик, первый сомножитель которых равен\/ $\logic{K}$, $\logic{T}$, $\logic{KB}$ или\/ $\logic{KTB}$, полиномиально погружаются в свой фрагмент от одной переменной.
\end{rtheorem}

Это даёт возможность получить ряд результатов об алгоритмической сложности фрагментов от конечного числа переменных для больших классов логик, учитывая результаты о сложности произведений логик в полном языке~\cite{Marx99,HHK02}.

\begin{rtheorem}[результат представлен в~{\cite{MR:2022:JLC}}]
Пусть логика $L$ такова, что\/ $\logic{K} \times \logic{K} \times \logic{K} \subseteq L \subseteq \logic{KTB} \times \logic{S5} \times \logic{S5}$.
Тогда фрагмент логики $L$ в языке с одной пропозициональной переменной неразрешим.
\end{rtheorem}

\begin{rtheorem}[результат представлен в~{\cite{MR:2022:JLC}}]
Пусть\/ $\logic{K} \times \logic{K} \subseteq L \subseteq \logic{KTB} \times \logic{S5}$.
Тогда фрагмент логики $L$ в языке с одной пропозициональной переменной является\/ $\cclass{coNEXPTIME}$-трудным.
\end{rtheorem}

Последнее утверждение имеет следующее уточнение, касающееся нескольких конкретных логик.

\begin{rtheorem}[результат представлен в~{\cite{MR:2022:JLC}}]
Фрагменты от одной переменной логик\/ $\logic{K} \times \logic{K}$, $\logic{K} \times \logic{K4}$ и\/ $\logic{K} \times \logic{S5}_2$ неэлементарны.
\end{rtheorem}

Для динамических логик получены следующие результаты.

\begin{rtheorem}[результат представлен в~{\cite{MR:2018:IGPL}}]
Константный фрагмент логики\/ $\logic{PDL}$ является\/ $\cclass{EXPTIME}$-полным, причём в языке с одной элементарной программой и итерацией.
\end{rtheorem}

\begin{rtheorem}[результат представлен в~{\cite{MR:2018:IGPL}}]
Константный фрагмент логики\/ $\logic{IPDL}$ является\/ $\cclass{2EXPTIME}$-полным, причём в языке с одной элементарной программой.
\end{rtheorem}

\begin{rtheorem}[результат представлен в~{\cite{MR:2018:IGPL}}]
Константный фрагмент логики\/ $\logic{PRSPDL}$ является неразрешимым.
\end{rtheorem}

Использованная техника моделирования переменных формулами от их фиксированного конечного числа позволяет получить экспоненциальную нижнюю оценку функции сложности для соответствующих фрагментов $\cclass{PSPACE}$-трудных мономодальных логик.

Тем не менее, прямой связи между сложностью аппроксимации логики и вычислительной сложностью проблемы её разрешения нет.

\begin{rtheorem}[результат представлен в~{\cite{MR:2023:JLC}}]
\label{th:complexity:function:K4:0}
Для любой степени неразрешимости существует нормальное линейно аппроксимируемое расширение логики\/~$\logic{K4}$, проблема принадлежности формул и проблема принадлежности константных формул которому имеют эту степень неразрешимости.
\end{rtheorem}

\begin{rtheorem}[результат представлен в~{\cite{MR:2023:JLC}}]
\label{th:complexity:function:KTB-GL-Grz:1}
Пусть $L\in\{\logic{KTB}, \logic{GL}, \logic{Grz}\}$.
Для любой степени неразрешимости существует нормальное линейно аппроксимируемое расширение логики~$L$, проблема принадлежности формул и проблема принадлежности формул от одной переменной которому имеют эту степень неразрешимости.
\end{rtheorem}

\begin{rtheorem}[результат представлен в~{\cite{MR:2023:JLC}}]
\label{th:complexity:function:Int:2}
Для любой степени неразрешимости существует линейно аппроксимируемое расширение логики\/~$\logic{Int}$, проблема принадлежности формул и проблема принадлежности формул от двух переменных которому имеют эту степень неразрешимости.
\end{rtheorem}

Теперь опишем основные результаты, полученные для модальных и суперинтуиционистских предикатных логик.

\begin{rtheorem}[результат представлен в~{\cite{MR:2017:LI}}]
\label{th:QS5:LI}
Пусть $L$~--- модальная предикатная логика, содержащая\/ $\logic{QCl}$ и содержащаяся в\/~$\logic{QS5}$, $\logic{QGL}\oplus\bm{bd}_2\oplus\bm{bf}$, $\logic{QGrz}\oplus\bm{bd}_2\oplus\bm{bf}$, $\logic{QGL.3}\oplus\bm{bf}$ или\/ $\logic{QGrz.3}\oplus\bm{bf}$. Тогда фрагмент логики $L$ в языке с одной унарной предикатной буквой алгоритмически неразрешим.
\end{rtheorem}

Этот результат позже был усилен для очень многих логик. Именно, была получена следующая теорема.

\begin{rtheorem}[результат представлен в~\cite{MR:2019:SL}]
Пусть $L$~--- модальная предикатная логика, содержащая\/ $\logic{QCl}$ и содержащаяся в\/~$\logic{QKTB}$, $\logic{QGL}\oplus\bm{bf}$ или\/ $\logic{QGrz}\oplus\bm{bf}$. Тогда фрагмент логики $L$ в языке с одной унарной предикатной буквой и двумя предметными переменными алгоритмически неразрешим.
\end{rtheorem}

В доказательстве было существенным, что у логики имеются сколь угодно большие (и~даже бесконечные) шкалы. Следующие наблюдения показывают, что это условие действительно является важным.

\begin{rtheorem}[результат представлен в~{\cite{MR:2017:LI}}]
\label{th:QS5:LI:fin}
Пусть $L$~--- модальная предикатная логика, определяемая одной конечной шкалой Крипке. Тогда монадический фрагмент логики $L$ алгоритмически разрешим.
\end{rtheorem}

\begin{rtheorem}[результат представлен в~{\cite{MR:2023:unary}}]
\label{th:QS5:LI:QAlt-n}
Для любого $n\in\numN$ монадические фрагменты логик\/ $\logic{QAlt}_n$, $\logic{QTAlt}_n$, $\logic{QAlt}_n\oplus\bm{bf}$ и\/ $\logic{QTAlt}_n\oplus\bm{bf}$ алгоритмически разрешимы.
\end{rtheorem}

Аналогичные результаты получены для монадических фрагментов модальных предикатных логик с равенством.

В целом, в случае логик бесконечных классов конечных шкал Крипке ситуация с разрешимостью иная.

\begin{rtheorem}[результат представлен в~{\cite{MR:2020:JLC:2}}]
Пусть $L$~--- модальная предикатная логика, содержащая\/ $\logic{QCl}$ и содержащаяся в\/ $\logic{QS5}$, $\logic{QGL.3}$ или\/ $\logic{QGrz.3}$. Тогда фрагменты логик $L_{\mathit{wfin}}$ и $L_{\mathit{wfin}}\oplus\bm{bf}$ в языке с тремя предметными переменными являются одновременно\/ $\Sigma^0_1$-трудным и\/ $\Pi^0_1$-трудным.
\end{rtheorem}

Отметим, что при этом дополнительно можно потребовать, чтобы язык логики~$L$ не содержал предикатных букв, кроме некоторой бинарной буквы.

Для монадических фрагментов модальных предикатных логик классов конечных шкал был получен следующий результат.

\begin{rtheorem}[результат представлен в~{\cite{MR:2020:JLC:2}}]
Пусть $L$~--- модальная предикатная логика, содержащая\/ $\logic{QCl}$ и содержащаяся в\/ $\logic{QKTB}$, $\logic{QGL}$ или\/ $\logic{QGrz}$. Тогда фрагменты логик $L_{\mathit{wfin}}$ и $L_{\mathit{wfin}}\oplus\bm{bf}$ в языке с одной унарной предикатной буквой и тремя предметными переменными являются\/ $\Pi^0_1$-трудным.
\end{rtheorem}

Близкие результаты были получены и для суперинтуиционистских предикатных логик, а также предикатных вариантов базисной и формальной логик Виссера~\cite{Visser81}, которые обозначим~$\logic{QBL}$ и $\logic{QFL}$ соответственно.



\begin{rtheorem}[результат представлен в~\cite{MR:2019:SL}]
\label{th:QBL:QKC:QFL}
Пусть\/ $\logic{QBL}\subseteq L\subseteq\logic{QKC}$ или\/ $\logic{QBL}\subseteq L\subseteq\logic{QFL}$. Тогда позитивный фрагмент логики~$L$ алгоритмически неразрешим в языке с одной унарной предикатной буквой и двумя предметными переменными.
\end{rtheorem}

Для суперинтуиционистских логик классов конечных шкал Крипке получаются результаты, сходные с модальным случаем.

\begin{rtheorem}[результат представлен в~\cite{MR:2021:JLC:2}]
Пусть $L$~--- суперинтуиционистская предикатная логика, содержащаяся в\/ $\logic{QLC}$. Тогда позитивные фрагменты логик $L_{\mathit{wfin}}$ и $L_{\mathit{wfin}}+\bm{cd}$ в языке с тремя предметными переменными являются одновременно\/ $\Sigma^0_1$-трудными и\/ $\Pi^0_1$-трудными.
\end{rtheorem}

\begin{rtheorem}[результат представлен в~\cite{MR:2021:JLC:2}]
Пусть $L$~--- суперинтуиционистская предикатная логика, содержащаяся в\/ $\logic{QKC}$. Тогда позитивные фрагменты логик $L_{\mathit{wfin}}$ и $L_{\mathit{wfin}}+\bm{cd}$ в языке с одной унарной предикатной буквой и тремя предметными переменными является\/ $\Pi^0_1$-трудным.
\end{rtheorem}

Используемые методы могут быть распространены и на логики, задаваемые другими классами шкал, не определимыми элементарно. В~диссертации представлены результаты исследования алгоритмических свойств модальных предикатных логик некоторых линейных порядков, а также нётеровых порядков.

\begin{rtheorem}[результат представлен в~\cite{MR:2021:JLC:3}]
Пусть $\alpha = \omega\cdot m + k$~--- ординал, где\/ $1\leqslant m < \omega$ и $k < \omega$.
Пусть $R$~--- бинарное отношение на $\alpha$, лежащее между отношениями $<$~и~$\leqslant$, где отношение $<$~--- естественный строгий порядок на $\alpha$, а отношение $\leqslant$~--- его рефлексивное замыкание. Тогда модальная предикатная логика шкалы Крипке\/ $\otuple{\alpha,R}$ является\/ $\Pi^1_1$-трудной в языке с одной унарной предикатной буквой, одной пропозициональной буквой и двумя предметными переменными.
\end{rtheorem}

Отметим, что использованное моделирование позволяет попутно получить результаты, касающиеся алгоритмической сложности монадических фрагментов логик шкал Крипке $\otuple{\numQ,<}$, $\otuple{\numQ,\leqslant}$, $\otuple{\numR,<}$, $\otuple{\numR,\leqslant}$ и некоторых других. Именно, логики таких шкал $\Sigma^0_1$-трудны в языке с одной унарной буквой, одной пропозициональной буквой и двумя предметными переменными. Известно~\cite{Corsi93}, что логикой шкалы Крипке $\otuple{\numQ,\leqslant}$ является аксиоматическая система $\logic{QS4.3}$, а логикой шкалы Крипке $\otuple{\numQ,<}$~--- система $\logic{QK4.3.D.X}$, и мы можем заключить, что указанные фрагменты этих логик $\Sigma^0_1$-полны.

Теперь скажем о результатах, полученных для логик нётеровых порядков. Для модальной предикатной логики $L$ определим модальную предикатную логику $L^\ast$ как логику класса шкал Крипке логики~$L$.

\begin{rtheorem}[результат представлен в~\cite{MR:2023:IGPL}]
Пусть модальная предикатная логика $L$ такова, что\/ $\logic{QwGrz}^\ast \subseteq L$, а также $L \subseteq \logic{QGL.3}^\ast\oplus\bm{bf}$ или $L \subseteq \logic{QGrz.3}^\ast\oplus\bm{bf}$. Тогда фрагмент логики $L$ в языке с одной унарной предикатной буквой, одной пропозициональной буквой и тремя предметными переменными является\/ $\Pi^1_1$-трудным.
\end{rtheorem}

\begin{rtheorem}[результат представлен в~\cite{MR:2023:IGPL}]
Пусть модальная предикатная логика $L$ такова, что\/ $\logic{QwGrz}^\ast \subseteq L$, а также $L \subseteq \logic{QGL.3}^\ast\oplus\bm{bf}$ или $L \subseteq \logic{QGrz.3}^\ast\oplus\bm{bf}$. Тогда фрагмент логики $L$ в языке с двумя унарными предикатными буквами, одной пропозициональной буквой и двумя предметными переменными является\/ $\Pi^1_1$-трудным.
\end{rtheorem}

\begin{rtheorem}[результат представлен в~\cite{MR:2023:IGPL}]
Пусть модальная предикатная логика $L$ такова, что\/ $\logic{QwGrz}^\ast \subseteq L$, а также $L \subseteq \logic{QGL}^\ast\oplus\bm{bf}$ или $L \subseteq \logic{QGrz}^\ast\oplus\bm{bf}$. Тогда фрагмент логики $L$ в языке с одной унарной предикатной буквой, одной пропозициональной буквой и двумя предметными переменными является\/ $\Pi^1_1$-трудным.
\end{rtheorem}

Как следствие получаем неполноту по Крипке всех рекурсивно перечислимых логик между $\logic{QwGrz}$ и $\logic{QGL.3}\oplus\bm{bf}$, а также между $\logic{QwGrz}$ и $\logic{QGrz.3}\oplus\bm{bf}$ (неполнота по Крипке некоторых из этих логик получена ранее в~\cite{Montagna84}, но другими методами), причём даже их соответствующих фрагментов.

С учётом этих результатов возникает вопрос о существовании рекурсивно перечислимых полных по Крипке модальных предикатных логик, не задаваемых элементарно определимыми классами шкал Крипке. Примеры таких логик были построены.

\begin{rtheorem}[результат представлен в~\cite{MR:2020:JLC:1}]
Существуют полные по Крипке рекурсивно перечислимые модальные предикатные логики, которые неполны относительно элементарно определимых классов шкал.
\end{rtheorem}


\section*{Описание методов}
\addcontentsline{toc}{section}{Описание методов}

Были использованы как общие методы, так и специальные. К общим методам относятся различные общематематические методы (например, часто использовавшийся метод математической индукции), а также синтаксический и семантический методы (например, активно использовалась семантика Крипке), методы теории алгоритмов, методы теории вычислительной сложности, методы теории решёток, методы теории замощений. Мы не будем давать подробных пояснений к этим методам, и остановимся подробнее на специальных методах.
\begin{itemize}
\item
Был использован метод, предложенный Дж.~Халперном~\cite{Halpern95}, позволяющий моделировать пропозициональные переменные языка формулами от одной пропозициональной переменной. Этот метод был сначала обобщён в классе пропозициональных логик~\cite{MR:2019:IGPL, MR:2007:JANCL, MR:2018:IGPL, MR:2022:JLC} (в~частности, в некоторых случаях подобное моделирование удалось получить с помощью константных формул), а затем модифицирован и применён в классе модальных предикатных логик~\cite{MR:2019:SL, MR:2020:JLC:2, MR:2021:LI, MR:2023:SCAN:1, MR:2023:IGPL}.
\item
Был использован метод моделирования переменных языка формулами от одной переменной, предложенный П.~Блэкбёрном и Э.~Спаан~\cite{BS93}. Модификации этого метода хорошо показали себя в исследованиях, связанных с модальными предикатными логиками, определяемыми линейными шкалами Крипке~\cite{MR:2021:JLC:3}.
\item
Был предложен метод моделирования пропозициональных переменных интуиционистского языка формулами от двух переменных~\cite{MR:2008:JANCL}. Затем этот метод был перенесён на суперинтуиционистские предикатные логики~\cite{MR:2020:JLC:2,MR:2021:JLC:2}.
\item
Был использован и модифицирован метод С.~Крипке~\cite{Kripke62}, позволяющий моделировать бинарные предикатные буквы в классических предикатных формулах с помощью модальности и унарных предикатных букв~\cite{MR:2017:LI}. Этот метод был перенесён на различные классы модальных и суперинтуиционистских предикатных логик, а также на предикатные варианты базисной и формальной логик Виссера~\cite{MR:2020:JLC:2,MR:2021:JLC:2}.
\item
Были использованы методы моделирования укладки домино~\cite{Berger66,Harel86}, описанные в работах Ф.~Волтера, М.~Захарьящева, Р.~Кончакова и А.~Куруш~\cite{WZ01,KKZ05}, с помощью модификаций которых были получены результаты о неразрешимости и сильной неразрешимости модальных и суперинтуиционистских предикатных логик~\cite{MR:2021:JLC:3,MR:2023:IGPL}.
\item
Был использован стандартный перевод модальных предикатных формул в формулы языка классической логики предикатов~\cite{Benthem85,MR:2000:IFRAS,GShS}, позволивший получить один из результатов работы~\cite{MR:2020:JLC:1}.
\item
Был модифицирован метод доказательства разрешимости монадического фрагмента классической логики предикатов~\cite{BBJ07}, что позволило получить верхнюю границу алгоритмической сложности неклассических предикатных логик, определяемых классами конечных шкал Крипке~\cite{MR:2017:LI}.
\end{itemize}

\section*{Направления возможного применения методов}
\addcontentsline{toc}{section}{Направления возможного применения методов}

Результаты и методы исследований могут быть перенесены на другие классы логик. Приведём несколько примеров.

Начнём с пропозициональных логик. Касающиеся их результаты содержатся в работах~\cite{MR:2019:IGPL, MR:2008:JANCL, MR:2007:JANCL, MR:2018:IGPL, MR:2022:JLC, MR:2023:JLC}, и вполне разумно говорить о дальнейшем применении соответствующих методов.
\begin{itemize}
\item
Было бы интересно распространить описанные методы на релевантные~\cite{RM73,Urquhart84,Urquhart07} и интуиционистские модальные логики~\cite{FS77,Prior57}. Частично это сделано, и уже получены некоторые результаты.
\item
Было бы интересно выявить такие ситуации и такие логики, когда методы моделирования переменных языка формулами от фиксированного конечного числа переменных не работают, а также понять, какова в каждом из случаев сложность проблемы разрешения.
\end{itemize}
При этом некоторые исследования уже проведены. Упомянем некоторые из полученных результатов.
\begin{itemize}
\item
Удалось распространить результат о $\cclass{PSPACE}$-трудности константных фрагментов модальных логик~\cite{MR:2003:AiML} на класс всех логик, лежащих между $\logic{K}$ и $\logic{wGrz}$~\cite{MR:2022:arXiv}.
\item
Сколь угодно сложные линейно аппроксимируемые логики найдены в классах ненормальных и квазинормальных логик~\cite{MR:2018:Tver:2}, а также в классах расширений базисной и формальной логик Виссера.
\item
Построены полиномиальные погружения модальных интуиционистских логик в их фрагменты от одной переменной~\cite{MR:2023:SCAN:3,MR:2025:JLC}.
\item
Построены погружения логики $\logic{HC}$ (известной как логика совместных задач и высказываний~\cite{Mel:I,Mel:II}) в её фрагменты от одной переменной; доказано, что эти фрагменты $\ccls{PSPACE}$-полны~\cite{MR:2024:Nsk:1}.
\end{itemize}

Теперь скажем о предикатных логиках. Здесь видится довольно много вопросов, оставшихся за рамками вошедших в диссертацию научных работ.
\begin{itemize}
\item
В классе неклассических предикатных логик остались нерешёнными вопросы об алгоритмической сложности <<стандартных>> логик в языке с одной предметной переменной, логик классов конечных шкал Крипке в языке с двумя предметными переменными, а также многих других логик, определяемых <<естественными>> элементарно неопределимыми классами шкал Крипке. Результаты работы показывают, что здесь можно ожидать разное: такие логики могут быть как рекурсивно перечислимыми в полном языке, так и сильно неразрешимыми в довольно бедном фрагменте языка.
\item
Доказано (но публикаций на момент написания этого текста ещё нет), что множество общезначимых формул первого порядка в языке с бинарной предикатной буквой и тремя предметными переменными рекурсивно неотделимо от множества формул в том же языке, опровергающихся в классе конечных моделей, где бинарная предикатная буква интерпретируется симметричным иррефлексивным бинарным отношением.
Это даёт возможность существенно расширить класс модальных и суперинтуиционистских предикатных логик, для которых известно, что их фрагмент от одной унарной предикатной буквы и трёх предметных переменных неразрешим.
\item
В.\,Б.~Шехтманом был поставлен вопрос о рекурсивной отделимости монадических фрагментов модальных предикатных логик и дополнений логик, определяемых классами конечных шкал Крипке исходных логик. Развитые в диссертации методы позволяют исследовать этот вопрос, и первые результаты в этом направлении уже имеются~\cite{MR:2024:MR,MR:2025:RecInsep}.
\item
Идеи доказательства разрешимости монадических фрагментов логик, определяемых одной конечной шкалой Крипке, были перенесены на логики с равенством. В результате удалось доказать разрешимость их монадических фрагментов с равенством при различных пониманиях равенства; аналогичные результаты были получены для полимодальных и суперинтуиционистских логик~\cite{MR:2023:unary,MR:2023:SCAN:1}.
\item
Кроме того, были получены обобщения конструкции Крипке~\cite{Kripke62}, позволяющей моделировать бинарные предикатные буквы языка с помощью модальности и унарных предикатных букв~\cite{MR:2023:Kripke}.
\item
Была доказана неразрешимость фрагмента предикатного варианта логики Гёделя--Дамметта $\logic{QLC}$ в языке с двумя предметными переменными (этот вопрос долго был открыт; см., например,~\cite{CMRT:2022}), причём не только для логики $\logic{QLC}$, но и для бесконечного класса её расширений; для доказательства использовался язык с бинарной предикатной буквой и бесконечным множеством унарных~\cite{MR:2024:CMCR,MR:2025:QLC-2var} или же две бинарные предикатные буквы~\cite{MR:2024:MR:2}.
\end{itemize}


\section*{Апробация и публикации}
\addcontentsline{toc}{section}{Апробация и публикации}

На основе результатов исследований были сделаны доклады на научных семинарах, а также российских и международных конференциях~\cite{%
MR:1999,
MR:2000:IFRAS,
MR:2000:Piter,
MR:2001:IFRAS,
MR:2001:Samara,
MR:2002:Piter,
MR:2002:Toulouse:AiML,
MR:2002:Math300,
MR:2003:AiML,
MR:2003:Kolmogorov,
MR:2003:SmirnovReadings,
MR:2004:Piter,
MR:2005:Poncelet,
MR:2006:AiML,
MR:2006:Piter,
MR:2007:SmirnovReadings,
MR:2007:FM,
MR:2008:Piter,
MR:2009:SmirnovReadings,
MR:2010:Piter,
MR:2012:ORFIQ,
MR:2015:SmirnovReadings:1,
MR:2015:SmirnovReadings:2,
MR:2017:SmirnovReadings:1,
MR:2017:SmirnovReadings:2,
MR:2017:SmirnovReadings:3,
MR:2018:SAICSIT,
MR:2018:LN,
MR:2019:AiML,
MR:2019:SAICSIT,
MR:2019:SmirnovReadings,
MR:2020:AiML,
MR:2020:SAICSIT,
MR:2021:Tver:conf,
MR:2021:SmirnovReadings,
MR:2021:LogicalPerspectives,
MR:2022:CMCR,
MR:2023:SmirnovReadings,
MR:2023:SCAN:1,
MR:2023:SCAN:2,
MR:2023:SCAN:3,
MR:2024:AiML,
MR:2024:MR,
MR:2024:MR:2,
MR:2024:CMCR,
MR:2024:Nsk:1}.
Научные центры, где были сделаны выступления в рамках научных семинаров: \mbox{ТвГУ}, ИФ~РАН, МГУ имени М.\,В.\,Ломоносова, МИАН, ИМ~СО~РАН, НИУ~ВШЭ, МФТИ, IRIT (Тулуза, Франция), WITS (Йоханнесбург, ЮАР).
Кроме того, некоторые из ранних результатов вошли в кандидатскую диссертацию автора~\cite{MR:2005:Diss} и некоторые более поздние~--- в PhD thesis~\cite{MR:2019:PhD}.


По теме диссертации опубликованы научные работы~\cite{
MR:2000:Tver,
MR:2001:LI,
MR:2002:LI:1,
MR:2002:LI:2,
MR:2003:LI,
MR:2004:LI,
MR:2007:JANCL,
MR:2007:Tver,
MR:2008:JANCL,
MR:2013:LI,
MR:2014:Tver,
MR:2015:LI,
MR:2016:CPS,
MR:2016:Tver,
MR:2017:LI,
MR:2018:IGPL,
MR:2018:Tver:1,
MR:2018:Tver:2,
MR:2019:IGPL,
MR:2019:SL,
MR:2020:JLC:1,
MR:2020:JLC:2,
MR:2021:arXiv,
MR:2021:Tver,
MR:2021:LI,
MR:2021:JLC:1,
MR:2021:JLC:2,
MR:2021:JLC:3,
MR:2022:arXiv,
MR:2022:DAN:rus,
MR:2022:DAN:eng,
MR:2022:TCS,
MR:2022:SL,
MR:2022:JLC,
MR:2022:IzVUZ:rus,
MR:2022:IzVUZ:eng,
MR:2022:IGPL,
MR:2023:IGPL,
MR:2023:LI,
MR:2023:Tver,
MR:2023:JLC,
MR:2025:JLC}.
Исследования продолжаются, и имеются работы, направленные в печать.


\pagebreak[3]

\setcounter{savefootnote}{\value{footnote}}

\part{Пропозициональные логики}
\chapter{Предварительные сведения}
\setcounter{footnote}{\value{savefootnote}}
  \section{Необходимые понятия}
    \subsection{Пропозициональные языки}

Как правило, для определения пропозициональных формул используются языки, содержащие бесконечное множество исходных символов (обычно требуется наличие бесконечного множества пропозициональных переменных). В~то же время алгоритмы работают со словами, записанными с помощью символов конечных алфавитов. Таким образом, чтобы формулы формально могли быть поданы на вход алгоритмам, нужно, чтобы они строились в языке с конечным числом исходных символов.

Для этих целей зафиксируем \defnotion{алфавит}\index{алфавит} 
$\mathcal{PA}$:
\label{alph_PA}
\label{alph:PA}
$$
\begin{array}{lcl}
\cal{PA} & = & \{p, l, m, |, (, )\}.
\end{array}
$$

Пусть $\numbers{N}=\{0,1,2,\dots\}$~--- множество натуральных чисел, $\numbers{N}^+$~--- множество положительных натуральных чисел. Для всякого $n\in\numbers{N}$ обозначим
$$
\begin{array}{lcl}
\bar{n} & = & \underbrace{||\dots|}_{n+1},
\end{array}
$$
т.\,е. $\bar{0} = |$, $\bar{1} = ||$, $\bar{2} = |||$ и т.\,д.

Слово вида $p(\bar{n})$ будем называть \defnotion{пропозициональной переменной}\index{переменная!пропозициональная} и вместо записи <<$p(\bar{n})$>> будем использовать более привычную~---~<<$p_n$>>. Аналогично тому, как использовалась буква $p$ для определения пропозициональных переменных, будем использовать букву $l$ для определения \defnotion{логических связок}.\index{связка!логическая} Именно, в качестве логических связок будем использовать слова вида $l(\bar{k})(\bar{n})$, записывая их более коротко как~$l^k_n$, где $k$~--- \defnotion{валентность}\index{бяа@валентность} (\defnotion{арность}\index{арность}) связки $l^k_n$, а $n$~--- её номер. Букву $m$ будем использовать для определения \defnotion{модальностей}\index{модальность}, считая их выражениями вида $m(\bar{k})(\bar{n})$ и записывая более коротко как~$m^k_n$, где $k$~--- валентность (арность) модальности $m^k_n$, а $n$~--- её номер.

В результате логические связки $\bot$, $\wedge$, $\vee$, $\to$ могут быть определены, например, так: ${\bot}=l^0_0$, ${\wedge}=l^2_0$, ${\vee}=l^2_1$, ${\to}=l^2_2$. Одноместная модальность $\Box_n$ может быть определена как~$m^1_n$. При необходимости в качестве индексов модальностей можно использовать не только коды натуральных чисел, но и иные выражения. Так, например, в динамических логиках используются одноместные модальности вида $[\alpha]$, где $\alpha$~--- некоторое выражение, называемое программой. В этом случае модальность $[\alpha]$ можно определить формально как выражение~$m(\bar{1})(\alpha)$, при необходимости расширяя исходный алфавит, чтобы ввести определение программы.

Далее мы не будем уделять большого внимания подобным техническим подробностям, полагая, что приведённых здесь пояснений достаточно для того, чтобы было понятно, как дать соответствующие формальные определения; важным является лишь то, что все рассматриваемые здесь формулы и языки могут быть определены с использованием подходящего конечного алфавита.

    \subsection{Разрешимость и классы сложности}

Формально под \defnotion{алгоритмами}\index{алгоритм} мы будем понимать машины Тьюринга или иную эквивалентную формализацию понятия алгоритма, см.,~например, \cite{ErshovPaliutin-1987-1-rus,Malcev-1965-1-rus,Mendelson-1976-1-rus,BJ-1994-1-rus}. Фактически мы будем пользоваться интуитивным понятием алгоритма, принимая при этом тезис Чёрча--Тьюринга, состоящий в том, что всякая функция, вычислимая алгоритмически в интуитивном смысле, вычислима с помощью некоторой машины Тьюринга.

Далее речь пойдёт только о \defnotion{задачах принадлежности} множествам слов.\index{еяд@задача!принадлежности} Содержательно, под задачей принадлежности слов в алфавите~$\Sigma$ множеству~$A$ будем понимать задачу, состоящую в следующем: по произвольному слову~$x$ выяснить, принадлежит ли~$x$ множеству~$A$. Формально задачу принадлежности слов множеству~$A$ будем отождествлять с самим множеством~$A$, и далее будем использовать как понятие <<множество слов>>, так и <<задача>>.

Множество слов $A$ в некотором алфавите $\Sigma$ называем \defnotion{разрешимым},\index{множество!разрешимое} или \defnotion{рекурсивным},\index{множество!рекурсивное} если существует такая алгоритмически вычислимая функция\footnote{Под функцией понимаем всюду определённую функцию.} $\function{f}{\Sigma^\ast}{\set{0,1}}$, что для каждого слова $x$ в алфавите~$\Sigma$
$$
\begin{array}{lcl}
f(x) & = & \left\{
           \begin{array}{ll}
           1, & \mbox{если $x\in A$}, \\
           0, & \mbox{если $x\not\in A$},
           \end{array}
           \right.
\end{array}
$$
т.\,е. если существует алгоритм, выясняющий по каждому слову в~$\Sigma$, принадлежит ли оно множеству~$A$; в противном случае множество $A$ называем \defnotion{неразрешимым}.\index{множество!неразрешимое}

Множество слов $A$ в некотором алфавите $\Sigma$ называем \defnotion{рекурсивно перечислимым},\index{множество!рекурсивно перечислимое} если $A=\varnothing$ или существует такая алгоритмически вычислимая функция $\function{f}{\numbers{N}}{\Sigma^\ast}$, что
$$
\begin{array}{lcl}
A & = & \{f(n) : n\in\numbers{N}\},
\end{array}
$$
т.\,е. если существует алгоритм, перечисляющий элементы множества~$A$.

Множество $A$ слов в алфавите $\mathcal{A}$ называется \defnotion{разрешимым относительно}\index{множество!разрешимое!относительно множества} множества $B$ слов в алфавите $\mathcal{B}$, если существует алгоритм с \defnotion{оракулом}\index{алгоритм!с оракулом}\index{оракул} во множестве $B$, решающий задачу принадлежности множеству~$A$; говорят также, что $A$ \defnotion{рекурсивно относительно}\index{множество!рекурсивное!относительно множества}~$B$. Наличие оракула во множестве $B$ означает, что алгоритм в процессе своей работы может использовать характеристическую функцию множества~$B$. Если $A$ разрешимо относительно $B$, то пишем $A\leqslant_{\mathrm{R}}B$.

Множество $A$ слов в алфавите $\mathcal{A}$ называем \defnotion{рекурсивно эквивалентным}\index{множество!рекурсивно эквивалентное множеству} множеству $B$ слов в алфавите $\mathcal{B}$, если $A\leqslant_{\mathrm{R}}B$ и $B\leqslant_{\mathrm{R}}A$; в этом случае пишем $A\sim_{\mathrm{R}}B$. Класс $[A]_{\mathrm{R}}=\set{B : B\sim_{\mathrm{R}}A}$ называем\footnote{\label{footnote:UnsolvabilityDegree}При введении дополнительных ограничений~--- например, требуя, что слова всех рассматриваемых множеств определяются над одним и тем же конечным алфавитом~--- можно добиться того, чтобы возникающий класс оказался множеством. В~литературе в подобных ситуациях часто рассматривают подмножества множества~$\numN$. Для нас это будет несущественно, т.к. мы будем рассматривать весьма ограниченный класс задач (связанных с множествами формул в тех или иных языках), и он является множеством; кроме того, нам будет интересен не столько этот класс сам по себе, сколько отдельные задачи.} \defnotion{степенью неразрешимости}.\index{степень!неразрешимости}

Содержательно, степень неразрешимости состоит из таких множеств, что если у нас появится способ выяснения принадлежности элементов одному из них, то это даст возможность получить такой способ и для любого другого множества из этой степени неразрешимости. Степени неразрешимости называются также \defnotion{степенями Тьюринга},\index{степень!Тьюринга} или \defnotion{T-степенями}\index{степень!t@$T$-степень}, а возникающее отношение $\leqslant_{\mathrm{R}}$~--- \defnotion{тьюринговой сводимостью},\index{сводимость!тьюрингова} или \defnotion{T-сводимостью}\index{сводимость!t@$T$-сводимость}, и обозначют также~$\leqslant_{\mathrm{T}}$.

Несмотря на то, что T-сводимость довольно полно отражает интуитивное понятие сводимости разрешимости одной задачи к другой, имеются другие, более сильные, сводимости.

Задача $A$ в алфавите $\mathcal{A}$ называется \defnotion{рекурсивно сводимой}\index{еяд@задача!сводимая!рекурсивно} к задаче $B$ в алфавите $\mathcal{B}$, если существует такая вычислимая функция $\function{f}{\mathcal{A}^\ast}{\mathcal{B}^\ast}$, что для всякого $x\in\mathcal{A}^\ast$
$$
\begin{array}{lcl}
x\in A & \iff & f(x)\in B.
\end{array}
$$
Рекурсивная сводимость известна также как \defnotion{m\nobreakdash-сводимость}\index{сводимость!m@m-сводимость}\footnote{Буква <<m>> здесь от <<many-one reducible>>.}. Если задача $A$ m-сводима к задаче $B$, то пишем $A\leqslant_{\mathrm{m}}B$. Для множества $A$ можно определить\footnote{См.~сноску~\ref{footnote:UnsolvabilityDegree}.} \defnotion{m-степень}\index{степень!m@m-степень} $[A]_{\mathrm{m}}=\set{B : \mbox{$A\leqslant_{\mathrm{m}}B$ и $B\leqslant_{\mathrm{m}}A$}}$.

Заметим, что если $A\leqslant_{\mathrm{m}}B$, то $A\leqslant_{\mathrm{R}}B$; обратное, вообще говоря, неверно: если $\mathcal{A}$~--- алфавит, то $\varnothing\leqslant_{\mathrm{R}}\mathcal{A}^\ast$, но $\varnothing\not\leqslant_{\mathrm{m}}\mathcal{A}^\ast$.

Пусть $C$~--- некоторый класс неразрешимых задач. Задача $A$ называется \defnotion{$C$\nobreakdash-трудной},\index{еяд@задача!c@$C$-трудная} если к ней m-сводима любая задача из класса~$C$. Задача $A$ называется \defnotion{$C$\nobreakdash-полной},\index{еяд@задача!c@$C$-полная} если она принадлежит классу~$C$ и при этом является $C$\nobreakdash-трудной.

Примерами классов неразрешимых задач являются \defnotion{арифметическая иерархия}\index{иерархия!арифметическая} и \defnotion{аналитическая иерархия},\index{иерархия!аналитическая} см.~\cite{Rogers}. Обычно и ту, и другую определяют не через алгоритмы, а через арифметические формулы, задействуя язык арифметики первого порядка в случае арифметической иерархии и язык арифметики второго порядка в случае аналитической иерархии. В арифметической иерархии нас будут интересовать в основном класс $\Pi^0_1$ и класс $\Sigma^0_1$, а точнее, $\Pi^0_1$-трудные и $\Sigma^0_1$-трудные задачи. Сами задачи при этом мы будем рассматривать не для множеств натуральных чисел, а для множеств формул; чтобы было возможно так делать, формально вместо формул можно, например, брать их \defnotion{гёделевы номера}\index{номер!гёделев}~\cite{Mendelson-1976-1-rus}, отождествляя формулы с ними. В~приводимых ниже построениях и доказательствах мы не будем уделять внимания обсуждению подобных технических деталей, т.е. не будем обращаться к гёделевой нумерации явно. В аналитической иерархии нас будут интересовать в основном $\Pi^1_1$-трудные и $\Sigma^1_1$-трудные задачи.

Нетрудно понять, что одна из T-степеней состоит в точности из разрешимых задач; в отношении таких задач нас будет интересовать \defnotion{сложность}\index{сложность} проблемы их разрешения.

Определим основные классы сложности, с которыми будем иметь дело.

Сложность разрешимой задачи в некотором алфавите $\Sigma$ будем оценивать через сложность алгоритмов, решающих эту задачу. Мы ограничимся лишь временн\'{о}й и ёмкостной сложностью. Под временн\'{о}й сложностью алгоритма $M$ понимаем функцию $\function{t_M}{\Sigma^\ast}{\numbers{N}}$, которая слову~$x$ сопоставляет число шагов вычисления алгоритма~$M$ на слове~$x$, а под ёмкостной сложностью алгоритма~$M$~--- функцию $\function{s_M}{\Sigma^\ast}{\numbers{N}}$, которая слову~$x$ сопоставляет число ячеек памяти, требуемых дополнительно (т.\,е. не считая памяти для хранения~$x$) для вычисления алгоритма~$M$ на слове~$x$. Под временн\'{о}й и ёмкостной сложностью в наихудшем случае понимаем, соответственно, функции $\function{t_M}{\numbers{N}}{\numbers{N}}$ и $\function{s_M}{\numbers{N}}{\numbers{N}}$, определённые следующим образом:
$$
\begin{array}{lcl}
t_M(n) & = & \max\{t_M(x) : x\in \Sigma^\ast ~\&~ |x|\leqslant n\}; \\
s_M(n) & = & \max\{s_M(x) : x\in \Sigma^\ast ~\&~ |x|\leqslant n\}, \\
\end{array}
$$
где $|x|$~--- длина слова $x$, т.\,е. суммарное число вхождений символов в~$x$.
Далее под временн\'{о}й и ёмкостной сложностью алгоритмов мы будем понимать временн\'{у}ю и ёмкостную сложность в наихудшем случае.

Пусть
\begin{itemize}
\item
$\ccls{DTIME}(f(n))$~--- класс задач, решаемых детерминированными машинами Тьюринга, временн\'{а}я сложность которых находится в классе~$\mathcal{O}(f(n))$;
\item
$\ccls{NTIME}(f(n))$~--- класс задач, решаемых недетерминированными машинами Тьюринга, временн\'{а}я сложность которых находится в классе~$\mathcal{O}(f(n))$;
\item
$\ccls{DSPACE}(f(n))$~--- класс задач, решаемых детерминированными машинами Тьюринга, ёмкостная сложность которых находится в классе~$\mathcal{O}(f(n))$;
\item
$\ccls{NSPACE}(f(n))$~--- класс задач, решаемых недетерминированными машинами Тьюринга, ёмкостная сложность которых находится в классе~$\mathcal{O}(f(n))$.
\end{itemize}
Следующие классы сложности определим стандартно\footnote{Двойное определение класса $\ccls{PSPACE}$ возможно благодаря теореме Савича~\cite{Savitch-1970-1}.}:
\[
\begin{array}{clcl}
\arrayitem
  & \ccls{P}
  & =
  & \displaystyle\bigcup\limits_{\mathclap{k=0}}^\infty \ccls{DTIME}(n^k);
  \arrayitemskip\\
\arrayitem
  & \ccls{NP}
  & =
  & \displaystyle\bigcup\limits_{\mathclap{k=0}}^\infty \ccls{NTIME}(n^k);
  \arrayitemskip\\
\arrayitem
  & \ccls{PSPACE}
  & =
  & \displaystyle\bigcup\limits_{\mathclap{k=0}}^\infty \ccls{DSPACE}(n^k)
    \arraytab
    {=}
    \arraytab
    \displaystyle\bigcup\limits_{\mathclap{k=0}}^\infty \ccls{NSPACE}(n^k);
  \arrayitemskip\\
\arrayitem
  & \ccls{EXPTIME}
  & =
  & \displaystyle\bigcup\limits_{\mathclap{k=0}}^\infty \ccls{DTIME}(2^{n^k});
  \arrayitemskip\\
\arrayitem
  & \ccls{2EXPTIME}
  & =
  & \displaystyle\bigcup\limits_{\mathclap{k=0}}^\infty \ccls{DTIME}(2^{2^{n^k}}).
\end{array}
\]
Для класса задач $C$ определим класс $\ccls{co}C$ как класс задач, дополнительных к задачам из~$C$: \[
\begin{array}{clcl}
\arrayitem
 & \ccls{co}C
 & =
 & \set{A : \bar{A}\in C},
\end{array}
\]
где дополнение множества берётся во множестве всех слов соответствующего алфавита.
Несложно понять, что имеют место равенства
$$
\begin{array}{lcl}
\ccls{coDTIME}(f(n)) = \ccls{DTIME}(f(n))
  & \mbox{и}
  & \ccls{coDSPACE}(f(n)) = \ccls{DSPACE}(f(n)),
\end{array}
$$
а следовательно, $\ccls{coP} = \ccls{P}$, $\ccls{coPSPACE} = \ccls{PSPACE}$, $\ccls{coEXPTIME} = \ccls{EXPTIME}$, $\ccls{co2EXPTIME} = \ccls{2EXPTIME}$. Совпадают ли классы $\ccls{NP}$ и $\ccls{coNP}$, автору (на момент написания этого текста) неизвестно.

Задача $A$ в алфавите $\mathcal{A}$ называется \defnotion{полиномиально сводимой}\index{еяд@задача!сводимая!полиномиально} к задаче $B$ в алфавите $\mathcal{B}$, если существует такая полиномиально вычислимая\footnote{Т.\,е. если существуют такие полином $p$ и детерминированный алгоритм $M$, что для каждого $x\in\mathcal{A}^\ast$ выполнено равенство $M(x)=f(x)$, причём время вычисления $M$ на входе $x$ ограничено сверху значением $p(|x|)$, где $|x|$~--- длина слова $x$.} функция $\function{f}{\mathcal{A}^\ast}{\mathcal{B}^\ast}$, что для всякого $x\in\mathcal{A}^\ast$
$$
\begin{array}{lcl}
x\in A & \iff & f(x)\in B.
\end{array}
$$
Если задача $A$ полиномиально сводима к задаче~$B$, то пишем $A\leqslant_{\mathrm{p}} B$.

Пусть $C$~--- некоторый класс разрешимых задач. Задача $A$ называется \defnotion{$C$\nobreakdash-трудной},\index{еяд@задача!c@$C$-трудная} если к ней полиномиально сводима любая задача из класса~$C$. Задача $A$ называется \defnotion{$C$\nobreakdash-полной},\index{еяд@задача!c@$C$-полная} если она принадлежит классу~$C$ и при этом является $C$\nobreakdash-трудной.

Более детально с теорией вычислительной сложности и с классами сложности можно ознакомиться в~\cite{LP98,Papadimitriou}, а с m\nobreakdash-сводимостью и T-степенями~--- в~\cite{Shoenfield1971}; см.~также работу~\cite{Post1944}, задавшую направление развития теории вычислимости.

    \subsection{Понятие логики}

Понятие логики можно определять по-разному, причём определения могут оказаться не эквивалентными между собой. Так, довольно часто под логикой понимают некоторый формальный язык вместе с некоторой фиксированной семантикой для этого языка.
Или же под логикой понимают отношение следования, оператор добавления следствий или решётку теорий (см., например,~\cite{Gorbunov:2018:LI}) в некотором формальном языке.
Мы будем смотреть на логику как на множество схем правильных (с~некоторой точки зрения) рассуждений, и формально под логикой будем понимать множество формул языка, замкнутое по правилу подстановки. В~пропозициональном случае это правило означает следующее: если в некоторой формуле, принадлежащей логике, заменить все вхождения любой переменной на вхождения одной и той же формулы (т.\,е. выполнить подстановку формулы вместо переменной), то получившаяся формула также будет принадлежать логике.

В этом смысле системы типа $\logic{PAL}$ (Public Announcement Logic, см., например,~\cite{Plaza89,DHK08}) не являются логиками по причине отсутствия замкнутости по правилу подстановки. В~то же время множества \defnotion{тавтологий}\index{тавтология} классической, интуиционистской и других пропозициональных логик~--- примеры того, что мы здесь и будем называть логиками.

Кроме того, мы будем предполагать, что логика замкнута также по правилу modus ponens (примером логики, для которой это не так, является, скажем, множество формул интуиционистского языка, истинных в шкале, состоящей из одного иррефлексивного мира, см.~раздел~\ref{ssec:int:Visser:logics}).

Отметим также, что многие результаты, приведённые в работе, будут справедливы и для теорий, не являющихся логиками.
В~подобных случаях будут сделаны соответствующие замечания.

  \section{Классическая логика высказываний}
  \label{sec:Cl}
    \subsection{Синтаксис и семантика}

Опишем язык $\lang{L}$\index{уян@язык!l@$\lang{L}$} классической логики высказываний.
Будем считать, что исходными символами языка $\lang{L}$ являются следующие: счётное множество пропозициональных переменных~$\prop$,
символы логических связок $\wedge$ (конъюнкция\index{конъюнкция}), $\vee$ (дизъюнкция\index{дизъюнкция}), $\to$ (импликация\index{импликация}), логическая константа $\bot$ (ложь\index{ложь}), а также скобки.

\defnotion{Формулы языка $\lang{L}$}, или \defnotion{$\lang{L}$-формулы}\index{уяа@формула!l@$\lang{L}$-формула} определяются рекурсивно:
\begin{itemize}[midsep]
\item если $p\in \prop$, то $p$~--- $\lang{L}$-формула;
\item $\bot$~--- $\lang{L}$-формула;
\item если $\varphi$ и $\psi$~--- $\lang{L}$-формулы, то $(\varphi \wedge \psi)$, $(\varphi \vee \psi)$, $(\varphi \to \psi)$~--- тоже $\lang{L}$-формулы.
\end{itemize}
Далее $\lang{L}$-формулы часто будем называть просто формулами.

Введём привычные связки $\neg$ (отрицание\index{отрицание}), $\leftrightarrow$ (эквивалентность\index{уял@эквивалентность}), а также константу $\top$ (истина\index{истина}) как следующие сокращения:
\[
\begin{array}{clcl}
\arrayitem
  & \neg\varphi
  & =
  & (\varphi \to \bot);
  \arrayitemskip\\
\arrayitem
  & \varphi \leftrightarrow \psi
  & =
  & (\varphi \to \psi) \wedge (\psi \to \varphi);
  \arrayitemskip\\
\arrayitem
  & \top
  & =
  & \neg \bot.
\end{array}
\]

При записи формул будем опускать внешние скобки, а также некоторые скобки внутри формул, считая, что $\neg$ связывает формулы сильнее, чем $\wedge$ и~$\vee$, которые, в свою очередь, сильнее, чем $\leftrightarrow$~и~$\to$; кроме того, мы будем записывать кратные конъюнкции и дизъюнкции формул, считая (для определённости), что конъюнктивные или, соответственно, дизъюнктивные члены формул записываются в порядке возрастания индексов, а скобки ассоциированы влево.

Под языком $\lang{L}$ будем понимать множество всех $\lang{L}$-формул.
Пропозициональные переменные, а также константу $\bot$ будем называть \defnotion{атомарными формулами}\index{уяа@формула!атомарная}, или \defnotion{элементарными формулами}\index{уяа@формула!уял@элементарная} языка~$\lang{L}$.

Общепринятая семантика для языка $\lang{L}$ предполагает, что каждой элементарной формуле приписывается значение $1$ (<<истина>>) или $0$ (<<ложь>>), причём константе $\bot$ всегда приписывается именно $0$, а остальным формулам значения приписываются следующим образом:
\begin{itemize}
\item формуле $\varphi \wedge \psi$ приписывается наименьшее из значений, приписанных формулам $\varphi$~и~$\psi$;
\item формуле $\varphi \vee \psi$ приписывается наибольшее из значений, приписанных формулам $\varphi$~и~$\psi$;
\item формуле $\varphi \to \psi$ приписывается $1$, если значение, приписанное формуле $\varphi$, не превосходит значения, приписанного формуле~$\psi$; в противном случае ей приписывается значение~$0$.
\end{itemize}
Формально мы определим другую семантику классического пропозиционального языка, но она будет тесно связана с только что описанной.

\defnotion{Классической моделью}\index{модель!классическая} языка $\lang{L}$ будем называть любое подмножество $\cmodel{M}$ множества пропозициональных переменных~$\prop$. Определим отношение $\cmodels$ между моделью и $\lang{L}$-формулами. Отношение $\cmodels$ называется \defnotion{отношением истинности}\index{отношение!истинности} формул в модели; если формула $\varphi$ истинна в модели $\cmodel{M}$, то пишем $\cmodel{M}\cmodels \varphi$. Определяется это отношение рекурсивно:
\[
\begin{array}{clcl}
\arrayitem & \cmodel{M}\cmodels p_i
           & \leftrightharpoons
           & p_i\in \cmodel{M};
           \arrayitemskip\\
\arrayitem & \cmodel{M}\not\cmodels \bot;
           \arrayitemskip\\
\arrayitem & \cmodel{M}\cmodels \varphi' \wedge \varphi''
           & \leftrightharpoons
           & \mbox{$\cmodel{M} \cmodels \varphi'$ и $\cmodel{M}\cmodels \varphi''$;}
           \arrayitemskip\\
\arrayitem & \cmodel{M}\cmodels \varphi' \vee \varphi''
           & \leftrightharpoons
           & \mbox{$\cmodel{M} \cmodels \varphi'$ или $\cmodel{M}\cmodels \varphi''$;}
           \arrayitemskip\\
\arrayitem & \cmodel{M}\cmodels \varphi' \to \varphi''
           & \leftrightharpoons
           & \mbox{$\cmodel{M} \not\cmodels \varphi'$ или $\cmodel{M}\cmodels \varphi''$.}
\end{array}
\]

В контексте описанной семантики $\lang{L}$-формулу $\varphi$ называют:
\begin{itemize}
\item \defnotion{выполнимой},\index{уяа@формула!бяа@выполнимая} если существует такая классическая модель $\cmodel{M}$, что $\cmodel{M}\cmodels \varphi$;
\item \defnotion{опровержимой},\index{уяа@формула!опровержимая} если существует такая модель $\cmodel{M}$, что $\cmodel{M}\not\cmodels \varphi$;
\item \defnotion{тождественно истинной},\index{уяа@формула!тождественно истинная} если для любой классической модели $\cmodel{M}$ верно, что $\cmodel{M}\cmodels \varphi$;
\item \defnotion{тождественно ложной},\index{уяа@формула!тождественно ложная} если для любой классической модели $\cmodel{M}$ верно, что $\cmodel{M}\not\cmodels \varphi$.
\end{itemize}

Под \defnotion{классической логикой высказываний}\index{логика!классическая!бяа@высказываний} $\logic{Cl}$ мы будем понимать множество тождественно истинных $\lang{L}$-формул.
Отметим, что классическая логика высказываний может быть задана в виде исчисления, см., например,~\cite[глава~1]{ChZ}.

Классическую логику высказываний называют также \defnotion{пропозициональной классической логикой}\index{логика!классическая!пропозициональная}, язык $\lang{L}$~--- \defnotion{классическим пропозициональным языком},\index{уян@язык!пропозициональный!классический} а $\lang{L}$-формулы~--- \defnotion{классическими пропозициональными формулами},\index{уяа@формула!пропозициональная!классическая} но за рамками раздела~\ref{sec:Cl} мы будем избегать такого названия как для~$\lang{L}$ так и для $\lang{L}$-формул, поскольку язык~$\lang{L}$ будет нам интересен как язык для нескольких классов неклассических логик.

    \subsection{Сложность некоторых задач}

Известно, что проблема выполнимости для классических пропозициональных формул является $\ccls{NP}$-полной, а проблема тождественной истинности~--- $\ccls{coNP}$-полной~\cite{Cook-1971-1,Levin-1973-1}, но ситуация меняется для некоторых фрагментов классического пропозиционального языка. Приведём некоторые примеры.

\defnotion{Литералом}\index{литерал} будем называть формулу, являющуюся переменной или отрицанием переменной. Формулу вида $l_1\wedge\ldots\wedge l_n$, где $l_1,\ldots, l_n$~--- литералы, будем называть \defnotion{элементарной конъюнкцией}.\index{конъюнкция!уял@элементарная} Если $n=0$, то получаем пустую элементарную конъюнкцию, которую считаем равной формуле~$\top$.
Говорим, что классическая пропозициональная формула находится в \defnotion{дизъюнктивной нормальной форме},\index{уяа@форма!нормальная!дизъюнктивная} если она имеет вид $\varphi_1\vee\ldots\vee\varphi_m$, где $\varphi_1,\ldots,\varphi_m$~--- элементарные конъюнкции; при $m=0$ считаем, что формула равна~$\bot$. Формула находится в \defnotion{совершенной дизъюнктивной нормальной форме},\index{уяа@форма!нормальная!совершенная дизъюнктивная} если она находится в дизъюнктивной нормальной форме, где дизъюнктивные члены не повторяются, а множество переменных, входящих в какой-либо её дизъюнктивный член, совпадает со множеством переменных, входящих в саму формулу. Для краткости для понятий дизъюнктивной нормальной формы и совершенной дизъюнктивной нормальной формы используют аббревиатуры ДНФ и СДНФ. Формулу вида $l_1\vee\ldots\vee l_n$, где $l_1,\ldots, l_n$~--- литералы, будем называть \defnotion{элементарной дизъюнкцией}.\index{дизъюнкция!уял@элементарная} Если $n=0$, то получаем пустую элементарную дизъюнкцию, которую считаем равной формуле $\bot$.
Говорим, что классическая пропозициональная формула находится в \defnotion{конъюнктивной нормальной форме},\index{уяа@форма!нормальная!конъюнктивная} если она имеет вид $\varphi_1\wedge\ldots\wedge\varphi_m$, где $\varphi_1,\ldots,\varphi_m$~--- элементарные дизъюнкции; при $m=0$ считаем, что формула равна~$\top$. Формула находится в \defnotion{совершенной конъюнктивной нормальной форме},\index{уяа@форма!нормальная!совершенная конъюнктивная} если она находится в конъюнктивной нормальной форме, где конъюнктивные члены не повторяются, а множество переменных, входящих в какой-либо её конъюнктивный член, совпадает со множеством переменных, входящих в саму формулу. Для краткости для понятий конъюнктивной нормальной формы и совершенной конъюнктивной нормальной формы используют аббревиатуры КНФ и СКНФ.

Несложно показать, что для СДНФ и СКНФ как проблема выполнимости, так и проблема тождественной истинности решаются полиномиально. То же можно сказать о проблеме выполнимости для ДНФ и проблеме тождественной истинности для КНФ. При этом проблема выполнимости для КНФ остаётся $\ccls{NP}$-полной, а проблема тождественной истинности для ДНФ~--- $\ccls{coNP}$-полной (и даже с некоторыми дополнительными ограничениями, см., например, конструкцию в~\cite{Tseitin:1968}).

Поставим вопрос: что происходит со сложностью задач выполнимости и тождественной истинности при ограничении числа переменных, используемых в формулах?

Для ответа на него рассмотрим следующий алгоритм проверки выполнимости классических пропозициональных формул: для формулы $\varphi$ нужно последовательно перебрать все наборы значений её переменных и на каждом из них вычислить значение $\varphi$ на этом наборе; если хотя бы один раз получим значение $1$, то $\varphi$ выполнима, если нет, то не выполнима.

Этот алгоритм вычисляет значение формулы не более чем $2^k$ раз, где $k$~--- число переменных формулы, при этом одно вычисление требует выполнить не более чем $m$ операций, где $m$~--- число вхождений связок в формулу. Таким образом, необходимые затраты времени на выполнение этого алгоритма выражаются функцией из класса $\mathcal{O}(n\cdot 2^n)$, где $n$~--- длина тестируемой формулы.
Заметим, что ограничение числа переменных тестируемых формул некоторым числом $k_0$ приведёт к тому, что время выполнения этого же алгоритма будет ограничено сверху линейной функцией, т.е. функцией из $\mathcal{O}(n)$, поскольку в этом случае экспоненциальный сомножитель $2^n$ превратится в константу~$2^{k_0}$.

Аналогична ситуация и с проблемой тождественной истинности классических пропозициональных формул: схожий алгоритм является экспоненциальным для случая, когда формула может содержать любое число переменных, и становится линейным, если множество используемых переменных является конечным.

%

Ниже мы покажем, что для больших классов неклассических логик ситуация в корне отличается от только что описанной.

\setcounter{savefootnote}{\value{footnote}}
\chapter{Модальные логики}
\setcounter{footnote}{\value{savefootnote}}
\label{chapter:ML}
  \section{Основные определения и факты}
    \subsection{Синтаксис}

Будем считать, что модальные пропозициональные языки являются расширениями классического пропозиционального языка с помощью особых средств языка~--- \defnotion{модальностей}.\index{модальность} Здесь мы рассмотрим случай \defnotion{мономодальных}\index{логика!мономодальная} логик, т.\,е. логик, язык которых получается из классического добавлением всего лишь одной модальности, которую мы обозначим $\Box$.
Мономодальные логики будем называть просто \defnotion{модальными}.\index{логика!модальная}

             \defnotion{Формулы модального пропозиционального языка}, или \defnotion{модальные пропозициональные формулы}\index{уяа@формула!пропозициональная!модальная} определяются рекурсивно:
             \begin{itemize}
                \item если $p\in \prop$, то $p$~--- формула;
                \item $\bot$~--- формула;
                \item если $\varphi$ и $\psi$~--- формулы, то $(\varphi \wedge \psi)$, $(\varphi \vee \psi)$, $(\varphi \to \psi)$~--- тоже формулы;
                \item если $\varphi$~--- формула, то $\Box\varphi$~--- тоже формула.
             \end{itemize}
             Формулу $\Box\varphi$ читаем как <<необходимо~$\varphi$>>, а новую (по отношению к классическому пропозициональному языку) связку $\Box$ называем \defnotion{необходимостью},\index{необходимость} или \defnotion{модальностью необходимости}.

             Связки $\neg$, $\leftrightarrow$, константу $\top$, а также модальность $\Diamond$ введём как следующие сокращения:
             \begin{itemize}
                \item $\neg\varphi = (\varphi \to \bot)$;
                \item $\varphi \leftrightarrow \psi = (\varphi \to \psi) \wedge (\psi \to \varphi)$;
                \item $\top = \neg \bot$;
                \item $\Diamond\varphi = \neg\Box\neg\varphi$.
             \end{itemize}
             Модальность $\Diamond$ будем называть \defnotion{возможностью},\index{бяб@возможность} а формулу $\Diamond\varphi$ читать как <<возможно~$\varphi$>>. При записи формул будем опускать некоторые скобки, считая, что модальности $\Box$ и $\Diamond$ связывают формулы сильнее, чем остальные связки.

             Множество всех модальных пропозициональных формул обозначим посредством~$\lang{ML}$ и иногда такие формулы будем называть $\lang{ML}$-формулами.\index{уяа@формула!ml@$\lang{ML}$-формула} Само множество $\lang{ML}$ будем называть \defnotion{модальным пропозициональным языком}.\index{уян@язык!пропозициональный!модальный} Заметим, что $\lang{L}\subseteq\lang{ML}$.\index{уян@язык!ml@$\lang{ML}$}

    \subsection{Семантика Крипке}

            Опишем реляционную семантику Крипке для модального пропозиционального языка, при этом мы будем придерживаться обозначений~\cite{ChZ}.

            \defnotion{Шкалой Крипке}\index{уяи@шкала!Крипке} будем называть пару $\kframe{F} = \langle
            W,R\rangle$, где $W$~--- некоторое непустое множество, а $R$~---
            бинарное отношение на этом множестве. Элементы множества $W$ будем
            называть \defnotion{мирами},\index{мир} а отношение $R$~--- \defnotion{отношением
            достижимости}\index{отношение!достижимости} на множестве~$W$. Если для некоторых $w,w'\in W$
            выполнено отношение $wRw'$, то говорим, что мир $w'$ достижим из
            мира~$w$.

            Для каждого мира $w$ шкалы Крипке $\kframe{F} = \langle W,R\rangle$ определим множество~$R(w)$, положив
            $$
            \begin{array}{lcl}
            R(w) & = & \set{w'\in W : wRw'}.
            \end{array}
            $$
            Множество $R(w)$ представляет собой множество всех миров шкалы $\kframe{F}$, достижимых из мира~$w$.

            В случае модального языка 
            \defnotion{моделью Крипке}\index{модель!Крипке}
            называется набор $\kmodel{M} = \langle\kframe{F},v\rangle$, где
            $\kframe{F} = \langle W,R\rangle$~--- шкала Крипке,
            а $v$~--- оценка пропозициональных переменных в $W$, т.\,е. функция,
            сопоставляющая каждой пропозициональной переменной некоторое
            подмножество множества~$W$.


            Пусть $\kmodel{M} = \langle\kframe{F},v\rangle$~--- модель Крипке,
            определённая на шкале $\kframe{F} = \langle W,R\rangle$, и пусть
            $\varphi$~--- модальная пропозициональная формула. Определим
            отношение~$(\kmodel{M},w)\models\varphi$:
            \[
            \begin{array}{clcl}
            \arrayitem &
            (\kmodel{M},w)\not\models\bot;\!\! &  &
            \arrayitemskip
            \\
            \arrayitem &
            (\kmodel{M},w)\models p_n
            & \leftrightharpoons &
            \parbox[t]{225pt}{$w\in v(p_n)$;}
            \arrayitemskip
            \\
            \arrayitem &
            (\kmodel{M},w)\models\varphi\wedge\psi & \leftrightharpoons &
            \parbox[t]{225pt}{$(\kmodel{M},w)\models\varphi$ \hphantom{л}и\hphantom{и}
            $(\kmodel{M},w)\models\psi$;}
            \arrayitemskip
            \\
            \arrayitem &
            (\kmodel{M},w)\models\varphi\vee\psi & \leftrightharpoons &
            \parbox[t]{225pt}{$(\kmodel{M},w)\models\varphi$ или
            $(\kmodel{M},w)\models\psi$;}
            \arrayitemskip
            \\
            \arrayitem &
            (\kmodel{M},w)\models\varphi\to\psi & \leftrightharpoons &
            \parbox[t]{225pt}{$(\kmodel{M},w)\not\models\varphi$ или
            $(\kmodel{M},w)\models\psi$;}
            \arrayitemskip
            \\
            \arrayitem &
            (\kmodel{M},w)\models\Box\varphi
            & \leftrightharpoons &
            \parbox[t]{225pt}{$(\kmodel{M},w')\models\varphi$ для всякого $w'\in R(w)$.}
            \end{array}
            \]

Если $(\kmodel{M},w)\models\varphi$, то говорим, что формула $\varphi$ \defnotion{истинна в мире} $w$ модели $\kmodel{M}$; в противном случае говорим, что $\varphi$ \defnotion{опровергается в мире} $w$ модели~$\kmodel{M}$. Формулу $\varphi$ считаем \defnotion{истинной в модели} $\kmodel{M}$, если для всякого $w\in W$ выполнено отношение $(\kmodel{M},w)\models\varphi$; в этом случае пишем $\kmodel{M}\models\varphi$. Формулу $\varphi$ считаем \defnotion{истинной в шкале} $\kframe{F}$, если $\varphi$ истинна в любой модели, определённой на шкале~$\kframe{F}$; в этом случае пишем $\kframe{F}\models\varphi$. Формулу $\varphi$ считаем \defnotion{истинной в мире шкалы} $\kframe{F}$, если для любой модели $\kmodel{M}$, определённой на шкале~$\kframe{F}$, выполнено отношение $(\kmodel{M},w)\models\varphi$; в этом случае пишем $(\kframe{F},w)\models\varphi$. Формулу $\varphi$ считаем \defnotion{истинной в классе шкал} $\sclass{C}$, если $\varphi$ истинна в каждой шкале из этого класса; в этом случае пишем $\sclass{C}\models\varphi$.

            Заметим, что для всякой модели $\kmodel{M} = \langle\kframe{F},v\rangle$, определённой на шкале $\kframe{F}=\langle W,R\rangle$ и всякого мира $w\in W$
            \[
            \begin{array}{clcl}
            \arrayitem &
            (\kmodel{M},w)\models\neg\varphi & \iff &
            \parbox[t]{225pt}{$(\kmodel{M},w)\not\models\varphi$;}
            \arrayitemskip
            \\
            \arrayitem &
            (\kmodel{M},w)\models\top;\!\! &  &
            \arrayitemskip
            \\
            \arrayitem &
            (\kmodel{M},w)\models\varphi\leftrightarrow\psi & \iff &
            \parbox[t]{280pt}{\mbox{$(\kmodel{M},w)\models\varphi$}, если и только если
            $(\kmodel{M},w)\models\psi$;}
            \arrayitemskip
            \\
            \arrayitem &
            (\kmodel{M},w)\models\Diamond\varphi
            & \iff &
            \parbox[t]{280pt}{$(\kmodel{M},w')\models\varphi$ для некоторого $w'\in R(w)$.}
            \end{array}
            \]

            Иногда вместо обозначения $(\kmodel{M},w)\models\varphi$ удобно использовать альтернативное:
            $$
            \begin{array}{lcl}
            (\kframe{F},w)\models^v\varphi & \leftrightharpoons & (\otuple{\kframe{F},v},w)\models\varphi;
            \end{array}
            $$
            в этом случае говорим, что в мире $w$ шкалы $\kframe{F}$ при оценке~$v$ истинна формула~$\varphi$.

            Кроме того, когда ясно, о какой именно модели или шкале идёт речь, саму модель опускают и вместо $(\kmodel{M},w)\models\varphi$ или $(\kmodel{F},w)\models^v \varphi$ пишут $w\models\varphi$ или $w\models^v \varphi$.

            Если имеется множество модальных формул $\varGamma$ и какая-либо структура $\mathfrak{S}$~--- шкала, модель или мир,~--- то будем говорить, что в~$\mathfrak{S}$ истинно~$\varGamma$, если в~$\mathfrak{S}$ истинна каждая формула из~$\varGamma$; в этом случае пишем $\mathfrak{S}\models\varGamma$.

            Пусть $\kframe{F}=\otuple{W,R}$ и $\kframe{F}'=\otuple{W',R'}$~--- шкалы Крипке, $X\subseteq W$. Тогда говорим, что
            \begin{itemize}
            \item
            $\kframe{F}'$ является \defnotion{подшкалой}\index{подшкала} шкалы $\kframe{F}$, если $W'\subseteq W$ и $R'=R\upharpoonright W'$; в этом случае пишем $\kframe{F}'\subseteq\kframe{F}$;
            \item
            $\kframe{F}'$ является \defnotion{порождённой подшкалой}\index{подшкала!порождённая} шкалы $\kframe{F}$, если $\kframe{F}'$ является подшкалой шкалы $\kframe{F}$ и $R(W')\subseteq W'$;
            \item
            $X$ \defnotion{порождает} подшкалу $\kframe{F}'$ шкалы $\kframe{F}$, если $R^\ast(X)=W'$, где $R^\ast$~--- рефлексивно-транзитивное замыкание отношения~$R$;
            \item
            $\kframe{F}'$ является \defnotion{корневой подшкалой}\index{подшкала!корневая} шкалы $\kframe{F}$, если $\kframe{F}'$ является подшкалой шкалы $\kframe{F}$, порождённой множеством $\{x\}$, где $x$~--- некоторый мир из~$W$; в этом случае $x$ называют \defnotion{корнем}\index{корень шкалы Крипке} шкалы~$\kframe{F}'$.
            \end{itemize}

    \subsection{Логики}

            Под \defnotion{модальной пропозициональной логикой}\index{логика!пропозициональная!модальная} понимаем множество модальных пропозициональных формул, замкнутое по \defnotion{правилу подстановки}\index{правило!подстановки}. Чтобы сформулировать это правило точно, введём понятие подстановки. \defnotion{Подстановкой} в модальном языке называем любую функцию $\func{s}{\prop}{\lang{ML}}$. Расширим эту функцию до функции $\func{\bm{s}}{\lang{ML}}{\lang{ML}}$:
            \[
            \begin{array}{clcl}
                \arrayitem & \bm{s}(p_i)               & = & s(p_i);
                           \arrayitemskip\\
                \arrayitem & \bm{s}(\bot)              & = & \bot;
                           \arrayitemskip\\
                \arrayitem & \bm{s}(\varphi\wedge\psi) & = &
                             \bm{s}(\varphi)\wedge\bm{s}(\psi);
                           \arrayitemskip\\
                \arrayitem & \bm{s}(\varphi\vee\psi)   & = &
                             \bm{s}(\varphi)\vee\bm{s}(\psi);
                           \arrayitemskip\\
                \arrayitem & \bm{s}(\varphi\to\psi)    & = &
                             \bm{s}(\varphi)\to\bm{s}(\psi);
                           \arrayitemskip\\
                \arrayitem & \bm{s}(\Box\varphi)       & = & \Box\bm{s}(\varphi).
            \end{array}
            \]
            Формулу $\bm{s}(\varphi)$ называем \defnotion{подстановочным примером} формулы~$\varphi$.
            Теперь \defnotion{правило подстановки} можно сформулировать следующим образом:
            \begin{itemize}
                \item
                по формуле $\varphi$ получается некоторый подстановочный пример формулы~$\varphi$.
            \end{itemize}

            Мы будем рассматривать не все модальные логики, а лишь логики из некоторых специальных классов.

            Определим следующие правила вывода. Правило \defnotion{modus ponens}:\index{правило!modus ponens}
            \begin{itemize}
                \item
                по формулам $\varphi$ и $\varphi\to\psi$ получается формула~$\psi$.
            \end{itemize}
            Следующее правило известно как \defnotion{правило Гёделя},\index{правило!Гёделя} или \defnotion{правило необходимости}:\index{правило!необходимости}
            \begin{itemize}
                \item
                по формуле $\varphi$ получается формула~$\Box\varphi$.
            \end{itemize}

            Будем считать, что все рассматриваемые модальные логики замкнуты по правилу modus ponens.

            Модальную пропозициональную логику $L$ будем называть \defnotion{нормальной},\index{логика!пропозициональная!нормальная модальная} если она
            \begin{itemize}
                \item содержит ${\bf Cl}$;
                \item содержит формулу $\Box(p\to q)\to(\Box p\to \Box q)$, где $p,q\in\prop$, $p\ne q$;
                \item замкнута по правилу подстановки, правилу modus ponens и по правилу Гёделя.
            \end{itemize}

            Минимальную нормальную модальную пропозициональную логику принято обозначать $\logic{K}$. Известно, что логика $\logic{K}$ совпадает со множеством всех $\lang{ML}$\nobreakdash-формул, истинных в классе всех шкал Крипке, см.~\cite[теорема~3.53]{ChZ}.

            Для множеств формул ${\Sigma}$ и ${\Sigma}'$ обозначим через
            ${\Sigma}+{\Sigma}'$ логику, получающуюся
            замыканием множества ${\Sigma}\cup{\Sigma}'$ по {\MP} и правилу подстановки, а через
            ${\Sigma}\oplus{\Sigma}'$~--- логику,
            получающуюся замыканием множества ${\Sigma}\cup{\Sigma}'$
            по \MP, правилу Гёделя и правилу подстановки.

            Следующие логики определим стандартно~\cite{ChZ} (см. также~\cite{Litak2007} в отношении~$\logicwGrz$):
            $$
            \begin{array}{lclcl}
            \logic{T}   & = & \logic{K}  & \!\!\oplus\!\! & \mref;                             \\
            \logic{D}   & = & \logic{K}  & \!\!\oplus\!\! & \Diamond \top;                     \\
            \logic{KB}  & = & \logic{K}  & \!\!\oplus\!\! & \msym;                             \\
            \logic{KTB} & = & \logic{T}  & \!\!\oplus\!\! & \logic{KB};                        \\
            \logic{K4}  & = & \logic{K}  & \!\!\oplus\!\! & \mtra;                             \\
            \logic{K5}  & = & \logic{K}  & \!\!\oplus\!\! & \meuc;                             \\
            \logic{S4}  & = & \logic{T}  & \!\!\oplus\!\! & \logic{K4};                        \\
            \logic{S5}  & = & \logic{S4} & \!\!\oplus\!\! & \msym;                             \\
            \logic{GL}  & = & \logic{K4} & \!\!\oplus\!\! & \mla;                              \\
            \logic{Grz} & = & \logic{S4} & \!\!\oplus\!\! & \mgrz;                             \\
            \logic{wGrz}& = & \logic{K4} & \!\!\oplus\!\! & \mwgrz.                            \\
            \end{array}
            $$
            Логика $\logic{GL}$ известна как логика Гёделя--Лёба, логика $\logic{Grz}$~--- как логика Гжегорчика, логика $\logic{wGrz}$~--- как слабая логика Гжегорчика.

    \subsection{Необходимые факты}

            Для дальнейшего изложения нам понадобятся факты о связи между нормальными модальными пропозициональными логиками и семантикой Крипке.

            Для логики $L$ определим $\kframes{L}$ как класс шкал Крипке, в которых истинны все формулы, принадлежащие логике~$L$. Если $\kframe{F}$~--- шкала Крипке и $\kframe{F}\models L$, то $\kframe{F}$ называем \defnotion{шкалой логики~$L$},\index{уяи@шкала!логики} или \defnotion{$L$-шкалой}.\index{уяи@шкала!l@$L$-шкала}

            Для класса шкал Крипке $\scls{C}$ определим $\mlogic{\scls{C}}$ как множество модальных пропозициональных формул, истинных в каждой шкале из~$\scls{C}$. Отметим, что $\mlogic{\scls{C}}$ является нормальной модальной логикой. Логику $\mlogic{\scls{C}}$ называем \defnotion{логикой класса~$\scls{C}$}.\index{логика!класса шкал}

            Логика $L$ называется \defnotion{корректной}\index{логика!корректная} относительно класса шкал~$\scls{C}$, если $\scls{C}\subseteq\kframes{L}$. Логика $L$ называется \defnotion{полной} относительно класса~$\scls{C}$ шкал Крипке, если $L\subseteq\mlogic{\scls{C}}$. Логика $L$ называется \defnotion{адекватной}\index{логика!адекватная} относительно класса $\scls{C}$, если $L=\mlogic{\scls{C}}$. Довольно часто используется терминология, когда логику называют полной относительно класса~$\scls{C}$, если $L=\mlogic{\scls{C}}$, т.\,е. если $L$ адекватна относительно $\scls{C}$; далее мы будем использовать понятие полноты именно в этом смысле. Будем говорить, что логика $L$ \defnotion{полна по Крипке},\index{логика!полная по Крипке} если существует такой класс $\scls{C}$ шкал Крипке, что $L=\mlogic{\scls{C}}$. Логика $L$ называется \defnotion{финитно аппроксимируемой},\index{логика!уяа@финитно аппроксимируемая} если существует такой класс $\scls{C}$ конечных шкал Крипке, что~$L=\mlogic{\scls{C}}$.

            Известны следующие факты~\cite{ChZ} (в отношении~$\logicwGrz$ см.~\cite{Litak2007}):
            \begin{itemize}
            \item
            $\kframes{\logicK}$~--- класс всех шкал Крипке;
            \item
            $\kframes{\logicT}$~--- класс всех рефлексивных шкал Крипке, т.\,е. в которых отношение достижимости $R$ удовлетворяет условию $\FOrefip(R)$;
            \item
            $\kframes{\logicKfour}$~--- класс всех транзитивных шкал Крипке, т.\,е. в которых отношение достижимости $R$ удовлетворяет условию $\FOtraip(R)$;
            \item
            $\kframes{\logicKfive}$~--- класс всех евклидовых шкал Крипке, т.\,е. в которых отношение достижимости $R$ удовлетворяет условию $\FOeucip(R)$;
            \item
            $\kframes{\logicSfour}$~--- класс всех рефлексивно-транзитивных шкал Крипке;
            \item
            $\kframes{\logic{D}}$~--- класс всех серийных шкал Крипке, т.\,е. в которых отношение достижимости $R$ удовлетворяет условию $\FOserip(R)$;
            \item
            $\kframes{\logicKB}$~--- класс всех симметричных шкал Крипке, т.\,е. в которых отношение достижимости $R$ удовлетворяет условию $\FOsymip(R)$;
            \item
            $\kframes{\logicKTB}$~--- класс всех рефлексивно-симметричных шкал Крипке;
            \item
            $\kframes{\logicSfive}$~--- класс всех рефлексивно-симметрично-транзитивных шкал Крипке;\-
            \item
            $\kframes{\logicGL}$~--- класс всех транзитивных иррефлексивных шкал Крипке, не имеющих бесконечно возрастающих цепей, т.е. класс всех строгих нётеровых порядков;
            \item
            $\kframes{\logicGrz}$~--- класс всех шкал Крипке, отношение достижимости в которых является рефлексивным замыканием отношения достижимости в шкалах Крипке из класса~$\kframes{\logicGL}$, т.е. класс всех нестрогих нётеровых порядков;
            \item
            $\kframes{\logicwGrz}$~--- класс всех шкал Крипке, рефлексивное замыкание отношения достижимости в которых даёт шкалы из класса~$\kframes{\logicGrz}$.
            \end{itemize}
            Кроме того, логики $\logicK$, $\logicT$, $\logicKfour$, $\logicSfour$, $\logic{D}$, $\logicKB$, $\logicKTB$, $\logicSfive$, $\logicGL$, $\logicGrz$, $\logicwGrz$ полны по Крипке и финитно аппроксимируемы.

            Логику $L$ называем \defnotion{сабфреймовой},\index{логика!сабфреймовая} если любая подшкала любой $L$-шкалы тоже является $L$-шкалой. Нетрудно понять, что логики $\logicK$, $\logicT$, $\logicKfour$, $\logicSfour$, $\logicKB$, $\logicKTB$, $\logicSfive$, $\logicGL$, $\logicGrz$ являются сабфреймовыми, а, скажем, логика $\logic{D}$ сабфреймовой не является.

            Логику $L$ называем \defnotion{табличной},\index{логика!табличная} если она полна относительно некоторой конечной шкалы Крипке.
            Логику $L$ называем \defnotion{локально табличной},\index{логика!локально табличная} если любой её фрагмент от конечного числа переменных является полным относительно некоторой конечной шкалы Крипке. Локально табличной является, например, логика $\logic{K5}$, а также все её нормальные расширения~\cite{NTh75}, в частности, логика~$\logic{S5}$.

    \subsection{Удобные обозначения}

В этом разделе мы опишем некоторые обозначения и понятия. Несмотря на то, что некоторые из них формально будут определены для модального пропозиционального языка, мы будем позже использовать их и для модального предикатного языка.

\defnotion{Длиной}\index{длина формулы} формулы $\varphi$ считаем\footnote{Возможно и другое определение. Например, иногда длиной формулы считают число вхождений связок в эту формулу. Отметим, что для дальнейшего изложения можно выбрать в качестве определения длины формулы и такое.} число вхождений символов в $\varphi$ как в слово; длину формулы $\varphi$ обозначаем~$|\varphi|$.

\defnotion{Множество подформул} формулы $\varphi$ будем обозначать $\mathop{\mathit{sub}}\varphi$. Отметим, что число подформул формулы $\varphi$ не превосходит~$|\varphi|$.

Для модальной формулы $\varphi$ определим её \defnotion{модальную глубину}\index{бяб@глубина!модальная} $\mdepth \varphi$ как наибольшее число вложенных модальностей в формуле~$\varphi$:
$$
\begin{array}{lcl}
  \mdepth p & = & 0, ~\hfill \mbox{если $p \in \prop$;} \\
  \mdepth \bot & = & 0;  \\
  \mdepth (\varphi_1 \con \varphi_2)
    & = & \max \{\mdepth \varphi_1, \mdepth \varphi_2 \};  \\
  \mdepth (\varphi_1 \dis \varphi_2)
    & = & \max \{\mdepth \varphi_1, \mdepth \varphi_2 \};  \\
  \mdepth (\varphi_1 \imp \varphi_2)
    & = & \max \{\mdepth \varphi_1, \mdepth \varphi_2 \};  \\
  \mdepth \Box \varphi_1 & = & \mdepth \varphi_1 + 1.
\end{array}
$$

Определим следующие модальности:
$$
\begin{array}{lclclcl}
  \Box^+ \varphi
    & =
    & \varphi\wedge\Box\varphi;
    & ~~~ &
  \Diamond^+ \varphi
    & =
    & \varphi\vee\Diamond\varphi;
    \\
  \Box^{0} \varphi
    & =
    & \varphi;
    & &
  \Diamond^0 \varphi
    & =
    & \varphi;
    \\
  \Box^{k+1} \varphi
    & =
    & \Box\Box^k\varphi;
    & &
  \Diamond^{k+1} \varphi
    & =
    & \Diamond\Diamond^k\varphi;
    \\
  \Box^{\leqslant 0} \varphi
    & =
    & \varphi;
    & &
  \Diamond^{\leqslant 0} \varphi
    & =
    & \varphi;
    \\
  \Box^{\leqslant k+1} \varphi
    & =
    & \Box^{\leqslant k}\varphi \wedge \Box^{k+1}\varphi;
    & &
  \Diamond^{\leqslant k+1} \varphi
    & =
    & \Diamond^{\leqslant k}\varphi \vee \Diamond^{k+1}\varphi;
    \\
  \Box^{=k} \varphi
    & =
    & \Box^{k+1}\varphi \to \Box^{k}\varphi;
    & &
  \Diamond^{=k} \varphi
    & =
    & \Diamond^{k}\varphi \wedge \neg\Diamond^{k+1}\varphi;
    \\
  \Box^{\simeq k} \varphi
    & =
    & \Box^{k-1}\varphi \to \Box^{k}\varphi;
    & &
  \Diamond^{\simeq k} \varphi
    & =
    & \Diamond^{k}\varphi \wedge \neg\Diamond^{k-1}\varphi.
\end{array}
$$

  \section{Сложность проблемы разрешения}
    \subsection{Сложность логик в полном языке}

        Начнём с того, что приведём краткое обоснование $\cclass{PSPACE}$-трудности проблемы разрешения логик (и вообще любых множеств формул), заключённых между $\logic{K}$ и $\logic{GL}$, между $\logic{K}$ и $\logic{Grz}$, а также между $\logic{K}$ и $\logic{KTB}$; нам важны будут возникающие модальные формулы. Для этого покажем, как, используя конструкцию Р.\,Ладнера~\cite{Ladner77}, полиномиально свести проблему истинности булевых формул с кванторами к произвольной логике указанного класса.

        Считаем, что булевы формулы с кванторами строятся из пропозициональных переменных множества $\prop$ и константы~$\bot$ с помощью связок $\wedge$, $\vee$, $\to$ и кванторов по пропозициональным переменным $\forall p$ и $\exists p$, где $p\in\prop$. Пусть $\lang{QBF}$~--- множество всех булевых формул с кванторами.

        В формулах вида $\forall p\,\varphi$ и $\exists p\,\varphi$ формула $\varphi$ называется областью действия квантора $\forall p$ или $\exists p$, соответственно. Вхождение переменной $p$ называется замкнутым, если оно находится в области действия квантора по этой переменной; иначе оно называется свободным. Свободными переменными формулы называются переменные, имеющие свободное вхождение в эту формулу. Если формула не имеет свободных переменных, то она называется замкнутой.

        Универсальным замыканием булевой формулы с кванторами $\varphi$ называем формулу $\forall q_1\ldots\forall q_n\,\varphi$, где $q_1,\ldots,q_n$~--- все свободные переменные формулы~$\varphi$. Универсальное замыкание формулы $\varphi$ определено однозначно с точностью до порядка следования добавляемых перед $\varphi$ кванторов всеобщности; будем считать, что этот порядок соответствует порядку индексов свободных переменных и
        для универсального замыкания формулы $\varphi$ будем использовать обозначение~$\uclosure{\varphi}$.

        Истинность булевых формул с кванторами определяется обычно. Именно, пусть $\cmodel{M}\subseteq\prop$ и $\varphi$~--- булева формула с кванторами; множество $\cmodel{M}$ называем моделью. Отношение $\cmodel{M}\cmodels\varphi$ задаётся согласно следующему рекурсивному определению:
        \[
          \begin{array}{clcl}
          \arrayitem & \mbox{$\cmodel{M}\cmodels p_i$}
                     & \leftrightharpoons
                     & p_i\in \cmodel{M};
                     \arrayitemskip\\
          \arrayitem & \mbox{$\cmodel{M}\not\cmodels \bot$;}
                     \arrayitemskip\\
          \arrayitem & \mbox{$\cmodel{M}\cmodels \varphi' \wedge \varphi''$}
                     & \leftrightharpoons
                     & \mbox{$\cmodel{M} \cmodels \varphi'$ $\phantom{\mbox{л}}$и$\phantom{\mbox{и}}$ $\cmodel{M}\cmodels \varphi''$;}
                     \arrayitemskip\\
          \arrayitem & \mbox{$\cmodel{M}\cmodels \varphi' \vee \varphi''$}
                     & \leftrightharpoons
                     & \mbox{$\cmodel{M} \cmodels \varphi'$ или $\cmodel{M}\cmodels \varphi''$;}
                     \arrayitemskip\\
          \arrayitem & \mbox{$\cmodel{M}\cmodels \varphi' \to \varphi''$}
                     & \leftrightharpoons
                     & \mbox{$\cmodel{M} \not\cmodels \varphi'$ или $\cmodel{M}\cmodels \varphi''$;}
                     \arrayitemskip\\
          \arrayitem & \mbox{$\cmodel{M}\cmodels \forall p_i\,\varphi'$}
                     & \leftrightharpoons
                     & \mbox{$\cmodel{M}\cup\{p_i\} \cmodels \varphi'$ $\phantom{\mbox{л}}$и$\phantom{\mbox{и}}$ $\cmodel{M}\setminus\{p_i\}\cmodels \varphi'$;}
                     \arrayitemskip\\
          \arrayitem & \mbox{$\cmodel{M}\cmodels \exists p_i\,\varphi'$}
                     & \leftrightharpoons
                     & \mbox{$\cmodel{M}\cup\{p_i\} \cmodels \varphi'$ или $\cmodel{M}\setminus\{p_i\}\cmodels \varphi'$.}
        \end{array}
        \]
        Булеву формулу с кванторами называем тождественно истинной, если она истинна в любой модели.

        Заметим, что тождественная истинность формулы $\varphi$ равносильна тому, что её универсальное замыкание истинно в некоторой модели, а также тому, что её универсальное замыкание истинно в любой модели. Пусть
        $$
        \begin{array}{lcl}
        \logic{TQBF} & = & \{\varphi\in\lang{QBF} : \varnothing\models\uclosure{\varphi}\},
        \end{array}
        $$
        т.\,е. $\logic{TQBF}$~--- множество всех тождественно истинных булевых формул с кванторами.

        Будем говорить, что булева формула с кванторами $\varphi$ находится в {\defnotion{префиксной форме},\index{уяа@форма!префиксная} если $\varphi=Q_1q_1\ldots Q_nq_n\,\varphi'$, где $Q_1,\ldots,Q_n\in \{\forall,\exists\}$, $q_1,\ldots,q_n\in\prop$, а $\varphi'$~--- бескванторная формула.

        Известно, что задача принадлежности формул множеству $\logic{TQBF}$ является $\cclass{PSPACE}$\nobreakdash-полной~\cite{Stockmeyer:Meyer:1973}; причём достаточно ограничиться рассмотрением замкнутых формул в префиксной форме. Используя этот факт, Р.\,Ладнер доказал $\cclass{PSPACE}$\nobreakdash-трудность логик $\logic{K}$, $\logic{T}$ и $\logic{S4}$~\cite{Ladner77}, а позднее его конструкция была обобщена и на другие модальные логики, см., например,~\cite[глава~18]{ChZ}.

        Для наших целей понадобится незначительная модификация конструкции Р.\,Ладнера; опишем её.

        Пусть $\varphi = Q_1p_1\ldots Q_np_n\,\varphi'$~--- замкнутая булева формула в префиксной форме. Пусть $q_0,\ldots,q_{n+1}$~--- пропозициональные переменные, не входящие в $\varphi$; содержательный смысл переменной $q_i$~--- в том, что мы <<раскрыли>> не менее чем $i$~кванторов в формуле~$\varphi$. Пусть $\varphi^\ast$~---
        \label{varphi_ast}
        \label{varphi:ast}
        конъюнкция следующих формул:
          \[
          \begin{array}{lcl}
A & = 
            & \displaystyle q_0\wedge\bigwedge\limits_{i=1}^n \neg p_i; \smallskip\\
B & =
            & \displaystyle
            \Box^{\leqslant n} \bigwedge\limits_{i = 1}^n ( q_i \imp q_{i-1}); \smallskip\\
C & = 
            & \displaystyle
            \Box^{\leqslant n-1} \bigwedge\limits_{\mathclap{{Q}_i = \exists}} ( q_{i-1}\wedge\neg q_{i} \imp \Diamond (q_{i}\wedge\neg q_{i+1}); \smallskip\\
D & =
            & \displaystyle
            \Box^{\leqslant n-1} \bigwedge\limits_{\mathclap{{Q}_i = \forall}} (q_{i-1}\wedge\neg q_{i} \imp \Diamond (q_{i}\wedge\neg q_{i+1}
            \con p_{i}) \con \Diamond (q_{i}\wedge\neg q_{i+1} \con \neg p_{i})); \smallskip\\
E & =
            & \displaystyle
            \Box^{\leqslant n-1} \bigwedge\limits_{i = 1}^{\mathclap{n-1}} ( q_i \imp
            \bigwedge\limits_{j \leqslant i} (\phantom{\neg}p_j \imp \Box (q_{i+1}\wedge\neg p_{n+1} \imp \phantom{\neg}p_j)) \con {} \\
            && \displaystyle \phantom{\Box^{\leqslant n-1} \bigwedge\limits_{i = 1}^{n-1} ( q_i \imp\,}
            \bigwedge\limits_{j \leqslant i} (\neg p_j \imp \Box
            (q_{i+1}\wedge\neg p_{n+1} \imp \neg p_j))); \smallskip\\
F & =
            & \displaystyle \Box^{n} (q_n\wedge\neg q_{n+1} \imp \varphi').
          \end{array}
          \]

            Формула $\varphi^\ast$ <<объясняет>>, как последовательно <<раскрываются>> кванторы формулы~$\varphi$, а также требует, чтобы после <<раскрытия>> кванторов формула~$\varphi'$ была истинной. Более точно, выполняется следующее утверждение.

            \begin{proposition}
            \label{prop:modal:complexity}
            Пусть $\logic{K}\subseteq L\subseteq\logic{GL}$
            или   $\logic{K}\subseteq L\subseteq\logic{Grz}$
            или   $\logic{K}\subseteq L\subseteq\logic{KTB}$.
            Тогда справедлива следующая эквивалентность:
            $$
            \begin{array}{lcl}
            \neg\varphi\in\logic{TQBF} & \iff & \neg\varphi^\ast\in L.
            \end{array}
            $$
            \end{proposition}

            \begin{proof}
            Покажем, что
            \begin{itemize}
                \item
                если $\neg\varphi\not\in\logic{TQBF}$, то $\neg\varphi^\ast\not\in\logic{GL}$, $\neg\varphi^\ast\not\in\logic{Grz}$ и $\neg\varphi^\ast\not\in\logic{KTB}$;
                \item
                если $\neg\varphi^\ast\not\in\logic{K}$, то $\neg\varphi\not\in\logic{TQBF}$,
            \end{itemize}
            откуда и будет следовать указанная эквивалентность.

            Пусть $\neg\varphi\not\in\logic{TQBF}$. Тогда $\varphi$~--- истинная формула. Это означает, что мы можем построить кванторное дерево для~$\varphi$; опишем его построение. В корне этого дерева находится вершина, являющаяся классической моделью $\cmodel{M}_\varphi^0=\varnothing$. Пусть уже построена вершина $\cmodel{M}^s_\psi$ для формулы $\psi=Q_kp_k\ldots Q_np_n\,\varphi'$, где $k\in\{1,\ldots,n\}$.
            Если $Q_k=\forall$, то добавляем вершины $\cmodel{M}^0_{\psi'} = \cmodel{M}^s_\psi$ и $\cmodel{M}^1_{\psi'} = \cmodel{M}^s_\psi \cup \{p_k\}$, где $\psi'$~--- формула, полученная из~$\psi$ удалением квантора $Q_kp_k$.
            Если же $Q_k=\exists$, то поступаем следующим образом: добавляем какую-нибудь из вершин $\cmodel{M}^0_{\psi'}$ и~$\cmodel{M}^1_{\psi'}$, в которой истинна формула~$\psi'$.
            В каждую из добавленных вершин $\cmodel{M}^0_{\psi'}$ и~$\cmodel{M}^1_{\psi'}$ проводим дугу из вершины~$\cmodel{M}^s_\psi$. Построение кванторного дерева заканчивается, когда не остаётся висячих вершин, помеченных формулами, содержащими хотя бы один квантор.

            Заметим, что если вершина $\cmodel{M}_\psi^s$ оказалась в кванторном дереве для~$\varphi$, то $\cmodel{M}_\psi\models\psi$.

            Определим шкалу $\otupleIs{\kframe{F}}{W,R}$, взяв в качестве $W$ множество всех вершин кванторного дерева для~$\varphi$, а в качестве~$R$~--- множество всех дуг в этом дереве. Определим оценку~$v$:
            $$
            \begin{array}{lcl}
            v(p_k) & = & \{w\in W : p_k\in w\}.
            \end{array}
            $$
            Пусть теперь $R^{t}$, $R^{rt}$ и $R^{rs}$~--- транзитивное, рефлексивно-транзитивное и рефлексивно-симметричное замыкания отношения~$R$, соответственно. Несложно понять, что в этом случае шкалы
            $\otupleIs{\kframe{F}^{t}}{W,R^{t}}$, $\otupleIs{\kframe{F}^{rt}}{W,R^{rt}}$ и $\otupleIs{\kframe{F}^{rs}}{W,R^{rs}}$ являются шкалами логик $\logic{GL}$, $\logic{Grz}$ и $\logic{KTB}$, соответственно, причём в корне~$\cmodel{M}_\varphi^0$ каждой из этих шкал при оценке~$v$ истинна формула~$\varphi^\ast$. Следовательно, $\neg\varphi^\ast\not\in\logic{GL}$, $\neg\varphi^\ast\not\in\logic{Grz}$ и $\neg\varphi^\ast\not\in\logic{KTB}$.

            Пусть $\neg\varphi^\ast\not\in\logic{K}$. Тогда существует модель Крипке, в некотором мире которой истинна формула~$\varphi^\ast$. Но истинность формулы~$\varphi^\ast$ позволяет <<шаг за шагом>> извлечь из этой модели кванторное дерево для формулы~$\varphi$, подтверждающее её истинность, а значит, $\neg\varphi\not\in\logic{TQBF}$.
            \end{proof}

            \label{observation:modal:complexity:heredity}
            Заметим, что в предложении~\ref{prop:modal:complexity} оценка $v$, определённая на шкалах $\otupleIs{\kframe{F}^{t}}{W,R^{t}}$, $\otupleIs{\kframe{F}^{rt}}{W,R^{rt}}$, удовлетворяет следующему условию, которое назовём условием наследственности <<вверх>>:
            $$
            \begin{array}{lcl}
            \mbox{$w\in v(p_k)$ и $wRw'$} & \imply & w'\in v(p_k).
            \end{array}
            $$
            Это наблюдение станет важным, когда мы будем устанавливать нижние границы сложности для фрагментов логик, полных относительно классов шкал, в которых одним из необходимых свойств отношения достижимости является транзитивность.

    \subsection{Интервал $[\logic{K},\logic{wGrz}]$}
    \label{subsection:K-K4}

            Пусть $\varphi$~--- булева формула с кванторами, $p_1,\dots,p_n$~---
            все переменные, входящие в~$\varphi$. Пусть $\varphi^\ast$~---
            формула, построенная по $\varphi$ так, как это описано на
            стр.\,\pageref{varphi:ast}.  Напомним, что переменными формулы
            $\varphi^\ast$ являются $p_1,\dots,p_{n},q_0,\ldots,q_{n+1}$. Для единообразия обозначений будем считать, что $q_0=p_{n+1},\ldots,q_{n+1}=p_{2n+2}$, т.\,е. переменными формулы $\varphi^\ast$ являются $p_1,\dots,p_{2n+2}$.

            Для всякого $k\in\numNp$ положим
            $$
            \begin{array}{lcl}
            \alpha_k
            & = &
            \Box(\Diamond^k\Box\bot \wedge \neg
            \Diamond^{k+1}\Box\bot \to \Box(\Diamond \top \to \Diamond\Box\bot))
            \end{array}
            $$
            и обозначим через $\varphi_\alpha^\ast$ формулу, получающуюся из
            $\varphi^\ast$ подстановкой $\alpha_1,\dots,\alpha_{2n+2}$ вместо
            $p_1,\dots,p_{2n+2}$ соответственно.

            \begin{lemma}
            \label{lem_K4(0)form_equiv}
            \label{lem:K4(0)form:equiv}
            Формула $\varphi_\alpha^\ast$ строится по $\varphi^\ast$
            некоторым алгоритмом за полиномиальное время от $|\varphi^\ast|$, при
            этом
            $$
            \begin{array}{rcl}
            \mbox{$\varphi_\alpha^\ast$ $\logic{wGrz}$-выполнима}
            & \iff &
            \mbox{$\varphi^\ast$ $\logic{wGrz}$-выполнима.}
            \end{array}
            $$
            \end{lemma}

            \begin{proof}
            Заметим, что формулы  $\alpha_1,\dots,\alpha_{2n+2}$ строятся по
            $\varphi^\ast$ полиномиально, поскольку для
            некоторой константы~$c$
            $$
            |\alpha_m|
              ~\leqslant~ c\cdot m
              ~\leqslant~ c\cdot(2n + 2)
              ~\leqslant~ c\cdot|\varphi^\ast|.
            $$
            Следовательно,
            $$
            |\varphi_\alpha^\ast|
              ~\leqslant~ \max\{|\alpha_1|,\dots,|\alpha_{2n+2}|\}\cdot|\varphi^\ast|
              ~\leqslant~ c\cdot|\varphi^\ast|^2,
            $$
            из чего несложно заключить, что для построения формулы
            $\varphi_\alpha^\ast$ по $\varphi^\ast$ достаточно
            полиномиального времени от $|\varphi^\ast|$.

            Пусть $\varphi^\ast$ не является $\logic{wGrz}$-выполнимой. Тогда
            $\neg\varphi^\ast\in\logic{wGrz}$. Поскольку $\neg\varphi^\ast_\alpha$
            получена подстановкой из $\neg\varphi^\ast$, то
            $\neg\varphi^\ast_\alpha\in\logic{wGrz}$, и следовательно,
            $\varphi^\ast_\alpha$ не является $\logic{K4}$\nobreakdash-выполнимой.

            Пусть теперь $\varphi^\ast$~--- $\logic{wGrz}$-выполнимая формула. Это
            означает, что булева формула с кванторами $\varphi$ истинна.
            Следовательно, $\varphi^\ast$ истинна в некотором мире $w_0$
            модели $\kfmodelIs{M}{F}{v}$, определённой на
            $\logic{Grz}$-шкале $\kframeIs{F}{W,R}$ высоты $n\,{+}\,1$ (см.~доказательство
            предложения~\ref{prop:modal:complexity}). Несложно заметить, что оценка
            $v$ является наследственной: для всякой переменной $p_i$, где $0\leqslant
            i\leqslant 2n\,{+}\,2$, и для всяких $w',w''\in W$ таких, что
            $w'Rw''$ и $(\kmodel{M}, w') \models p_i$, выполнено
            $(\kmodel{M}, w'') \models p_i$.

            По модели $\kmodel{M}$ построим $\logic{wGrz}$-шкалу $\kframe{F}'=\opair{W'}{R'}$ и модель $\kmodel{M}' = \opair{\mathfrak{F}'}{v'}$, определённую на этой шкале, в некотором мире которой истинна формула~$\varphi^\ast_\alpha$.

\begin{figure}
  \centering
  \begin{tikzpicture}[scale=1.50]

    \coordinate (a30)    at ( 0.0, 3);
    \coordinate (a31)    at ( 0.0, 4);
    \coordinate (a32)    at ( 0.0, 5);
    \coordinate (a34)    at ( 0.0, 6);
    \coordinate (a35)    at ( 0.0, 7);
    \coordinate (b30)    at (-1.0, 4);
    \coordinate (c30)    at ( 0.0, 2);

    \coordinate (ld)     at (-3.2, 2.5);
    \coordinate (lu)     at (-3.2, 7.5);
    \coordinate (ru)     at ( 1.0, 7.5);
    \coordinate (rd)     at ( 1.0, 2.5);

    \draw [fill]     (a30)   circle [radius=2.0pt] ;
    \draw [fill]     (a31)   circle [radius=2.0pt] ;
    \draw [fill]     (a32)   circle [radius=2.0pt] ;
    \draw [fill]     (a35)   circle [radius=2.0pt] ;
    \draw []     (b30)   circle [radius=2.0pt] ;
    \draw []     (c30)   circle [radius=2.0pt] ;

    \begin{scope}[>=latex]

      \draw [->, shorten >=  2.75pt, shorten <= 2.75pt] (c30) -- (a30) ;
      \draw [->, shorten >=  2.75pt, shorten <= 2.75pt] (a30) -- (a31) ;
      \draw [->, shorten >=  2.75pt, shorten <= 2.75pt] (a31) -- (a32) ;
      \draw [->, shorten >= 11.75pt, shorten <= 2.75pt] (a32) -- (a34) ;
      \draw [->, shorten >=  2.75pt, shorten <= 7.75pt] (a34) -- (a35) ;
      \draw [->, shorten >=  2.75pt, shorten <= 2.75pt] (a30) -- (b30) ;

    \end{scope}

      \node []         at (a34)  {$\vdots$};
      \node [left=2pt] at (a35)  {$\Box\bot$};

      \node [right=2pt] at (a30)  {$a^m_0$};
      \node [right=2pt] at (a31)  {$a^m_1$};
      \node [right=2pt] at (a32)  {$a^m_2$};
      \node [right=2pt] at (a35)  {$a^m_{m}$};

      \node [left=2pt]  at (b30)  {$\Diamond\top\wedge\Box\Diamond\top$};
      \node [above=2pt] at (b30)  {$~~b^m$};
      \node [left=2pt]  at (c30)  {$\phantom{c^m}\neg\alpha_m$};
      \node [right=2pt] at (c30)  {$c^m$};

      \draw [dashed, color=black!50!] (ld) -- (lu) -- (ru) -- (rd) -- cycle ;

    \end{tikzpicture}
    \caption{Шкала $\kframe{F}_m$}
    \label{fig1}
  \end{figure}

            Сначала заметим, что для того, чтобы формула $\alpha_m$ опровергалась
            в некотором мире (транзитивной) модели, достаточно, чтобы
            из этого мира была достижима шкала, изображённая на рис.~\ref{fig1}
            (чёрные кружк\'{и} соответствуют иррефлексивным мирам, светлые~---
            рефлексивным, отношение достижимости транзитивно). Обозначим
            обведённую на рис.~\ref{fig1} шкалу через $\kframe{F}_m^{\phantom{l}}$, а всю
            шкалу, изображённую на рис.~\ref{fig1},~--- через $\kframe{F}^+_m$.
            Формальное описание шкал $\kframe{F}_m^{\phantom{l}}$ и $\kframe{F}^+_m$ таково:
            $\kframe{F}_m^{\phantom{l}}=\opair{W_m^{\phantom{l}}}{R_m^{\phantom{l}}}$, $\kframe{F}^+_m=\opair{W^+_m}{R^+_m}$, где
            $$
            \begin{array}{rcl}
            W_m & = & \set{b^m,a^m_0,a^m_1,\dots,a^m_m}; \\
            W^+_m & = & W_m\cup\set{c^m},
            \end{array}
            $$
            а $R_m^{\phantom{l}}$ и $R^+_m$~--- транзитивные замыкания отношений
            $$
            \set{\opair{a^m_0}{b^m}, \opair{b^m}{b^m}} \cup
            \setc{\opair{a^m_i}{a^m_{i+1}}}{0\leqslant i\leqslant m-1}
            $$
            и
            $$
            R_m\cup \set{\opair{c^m}{c^m}, \opair{c^m}{a^m_0}}
            $$
            соответственно.

            Ясно, что если $\alpha_m$ истинна в некотором мире транзитивной
            модели, то она также истинна во всех мирах, достижимых из данного
            (поскольку главной связкой формулы $\alpha_m$ является $\Box$). Заметим
            также, что если $k \ne m$, то $\kframe{F}^+_m\models\alpha_k$.
            Используя эти наблюдения, мы построим требуемую модель $\kmodel{M}'$.
            Именно, чтобы получить $\kmodel{M}'$, расширим шкалу $\kframe{F}$, сделав
            достижимой копию шкалы $\kframe{F}_m$ из каждого мира множества $W$, в
            котором в $\kmodel{M}$ опровергается переменная $p_m$.

            Опишем модель $\kmodel{M}'$ формально.

            Для каждого $w\in W$ обозначим через $\kframe{F}_m^w$ копию шкалы
            $\kframe{F}_m$, помеченную миром $w$: положим $\kframe{F}_m^w=
            \langle W_m^w, R_m^w\rangle$, где $W^w_m= W_m\times\{w\}$ и
            для всяких $x,y\in W_m$
            $$
            \begin{array}{rcl}
            \otuple{x,w} R_m^w \otuple{y,w}
            & \bydef &
            xR_my.
            \end{array}
            $$
            Положим
            $$
            \begin{array}{rcl}
            W' & = & \displaystyle W \cup \bigcup\setc{W_m^w}{\mbox{$1\leqslant
            m\leqslant 2n + 2$, $w\in W$, $(\kmodel{M},w) \not\models p_m$}}.
            \end{array}
            $$
            На множестве $W'$ определим отношение $\tilde R$:
            $$
            \begin{array}{rcl}
            w\tilde{R}w'
              & \bydef
              & \mbox{либо $w,w'\in W$ и $wRw'$,}
            \\
              &
              & \mbox{либо $w,w'\in W_m^u$ и $wR_m^uw'$,}
            \\
              &
              & \mbox{либо $w\in W$, $(\kmodel{M},w) \not\models p_k$
                 и $w'= \otuple{a_0^m,w}$}.
            \end{array}
            $$
            Обозначим через $R'$ транзитивное замыкание отношения $\tilde R$.
            Пусть $\kframe{F}'=\otuple{W',R}$ и пусть $v'$~---
            произвольная оценка пропозициональных переменных в мирах из множества
            $W'$. Положим $\kmodel{M}'=\otuple{\kframe{F}',v'}$.

            Заметим, что для каждых $w\in W'$ и $m\in\set{1,\ldots,n+1}$
            $$
            \begin{array}{rcl}
            (\kmodel{M}', w) \not\models \alpha_m
            & \iff &
            \mbox{$w\in W$ и $(\kmodel{M}, w) \not\models p_m$.}
            \end{array}
            \eqno{\mbox{$(\ast)$}}
            $$
            Действительно, для опровержения $\alpha_m$ в $w$ миру $w$ требуется видеть мир, из которого можно попасть в слепой мир за $m$ шагов, но не больше (а значит, он иррефлексивен), а также можно видеть бесконечную цепь миров (например, рефлексивный мир); это возможно, только если мы находимся в корне копии шкалы $\kframe{F}_m$, а значит, эта копия $\tilde R$\nobreakdash-достижима из $w$ или мира, $R$\nobreakdash-достижимого из~$w$, причём находящегося в~$W$. Учитывая условие наследственности <<вверх>> для оценки в $\kmodel{M}$, получаем, что $(\kmodel{M}, w) \not\models p_i$ по построению модели~$\kmodel{M}'$. Если же $w\in W$ и $(\kmodel{M}, w) \not\models p_i$, то $(\kmodel{M}, w) \not\models \alpha_i$ по построению~$\kmodel{M}'$.

            Для всякой формулы $\psi$ от переменных
            $p_1,\dots,p_{2n+2}$ обозначим через $\psi_\alpha$ формулу,
            полученную из $\psi$ подстановкой $\alpha_1, \dots , \alpha_{n+2}$
            вместо $p_1, \dots , p_{n+2}$ соответственно. Тогда для всякой
            подформулы $\psi$ бескванторной булевой формулы $\varphi'$
            и всякого мира $w \in W$ уровня $n$ имеет место следующая
            эквивалентность:
            $$
            \begin{array}{rcl}
            (\kmodel{M}', w) \models \psi_\alpha
            & \iff &
            (\kmodel{M}, w) \models \psi.
            \end{array}
            $$
            Справедливость этого утверждения обосновывается индукцией по построению~$\psi$:
            случай, когда $\psi = \bot$, тривиален; если $\psi = p_m$,
            то $\psi_\alpha = \alpha_m$, и требуемое получаем по~\mbox{$(\ast)$};
            индукционный шаг также тривиален (напомним, что $\psi$~--- безмодальная
            пропозициональная формула).

            Таким образом, если $w$~--- мир модели $\kmodel{M}$ уровня $n$, то
            $(\kmodel{M}',w) \models \varphi'_\alpha$. Заметим, что, согласно~\mbox{$(\ast)$}, формула
            $\alpha_{2n+1} \wedge\neg \alpha_{2n+2}$ истинна в некотором мире
            $w$ модели $\kmodel{M}'$ только если $w$ является миром уровня $n$ в
            $\kmodel{M}$, поскольку миры модели $\kmodel{M}$, в которых истинна
            формула $q_n \wedge\neg q_{n+1}$,~--- это в точности миры
            уровня~$n$. Следовательно, $(\kmodel{M}',w_0) \models
            \Box^n(\alpha_{2n+1}\wedge\neg\alpha_{2n+2}\to\varphi'_\alpha)$, т.е.\ $(\kmodel{M}',w_0) \models F_\alpha$.

            Осталось проверить, что
            $$
            (\kmodel{M}', w_0) \models 
            A_\alpha \wedge
            B_\alpha \wedge
            C_\alpha \wedge
            D_\alpha \wedge
            E_\alpha.
            $$

            Так как $(\kmodel{M},w_0)\models q_0$, то, согласно~\mbox{$(\ast)$}, получаем $(\kmodel{M}',w_0)\models\alpha_{n+1}$ (напомним, что
            $p_{n+1}= q_0$).
            Поскольку $(\kmodel{M},w_0) \not\models q_i$ при $i>0$, то по~\mbox{$(\ast)$} получаем, что
            $(\kmodel{M}',w_0)\models\neg\alpha_{n+1+i}$. Значит, $(\kmodel{M}',w_0)\models A_\alpha$.

            Предположим, что $(\kmodel{M}',w_0)\not\models B_\alpha$.
            Тогда для некоторого $w\in W'$, достижимого из $w_0$, имеет место
            отношение 
            $$
            \begin{array}{c}
            \displaystyle
            (\kmodel{M}',w)\not\models
            \bigwedge\limits_{i=n+2}^{2n+2}(\alpha_i\to\alpha_{i-1}).
            \end{array}
            $$
            Заметим, что $w\in W$. Действительно, во всяком мире шкалы
            $\kframe{F}_k$ истинны все формулы $\alpha_i$, поэтому истинна и
            конъюнкция их импликаций друг к другу. Итак, $w\in W$, причём для
            некоторого $i\in\set{n+2,\dots,2n+2}$ имеет место
            отношение $(\kmodel{M}',w)\not\models\alpha_i\to\alpha_{i-1}$.
            Но тогда $(\kmodel{M}',w)\models\alpha_i$ и
            $(\kmodel{M}',w)\not\models\alpha_{i-1}$, а следовательно,
            $(\kmodel{M},w)\models p_i$ и
            $(\kmodel{M},w)\not\models p_{i-1}$, из чего несложно заключить,
            что 
            $(\kmodel{M},w_0)\not\models B$. Получили противоречие,
            следовательно, $(\kmodel{M}',w_0)\models B_\alpha$.

            Предположим, что $(\kmodel{M}',w_0)\not\models E_\alpha$.
            Тогда для некоторого $w\in W'$, достижимого из $w_0$, 
            %
            %
            и некоторого $i\in\set{1,\dots,n}$
            $$
            \begin{array}{l}
            (\kmodel{M}',w)\models\alpha_{n+1+i},
            \end{array}
            $$
            при этом 
            $$
            \begin{array}{l}
            (\kmodel{M}',w)\not\models
            \big(\alpha_i\to\Box(\alpha_{n+1+i}\wedge\neg
            \alpha_{2n+2}\to
            \alpha_i)\big)
            \end{array}
            $$
            или
            $$
            \begin{array}{l}
            (\kmodel{M}',w)\not\models
            \big(\neg \alpha_i\to\Box(\alpha_{n+1+i}\wedge\neg
            \alpha_{2n+2}\to\neg \alpha_i)\big).
            \end{array}
            $$

            Пусть $(\kmodel{M}',w) \not\models \alpha_i\wedge\neg
            \alpha_{2n+2} \to
            \Box(\alpha_{n+1+i} \to \alpha_i)$. Тогда
            $(\kmodel{M}',w)\models\alpha_i$ и существует $w'\in W'$ такой, что
            $wR'w'$, $(\kmodel{M}',w') \models \alpha_{n+1+i}\wedge\neg
            \alpha_{2n+2}$,
            $(\kmodel{M}',w') \not\models \alpha_i$. Так как
            $(\kmodel{M}',w') \not\models \alpha_i$, то $w'\in W$, а
            значит, и $w\in W$, откуда следует, что
            $$
            \begin{array}{rcl}
            (\kmodel{M},w) \not\models
            q_i\to\big(p_i\to\Box(q_i\wedge\neg
            q_{n+1}\to
            p_i)\big)\wedge\big(\neg p_i\to\Box(q_i\wedge\neg
            q_{n+1}\to\neg p_i)\big),
            \end{array}
            $$
            а значит, $(\kmodel{M},w_0) \not\models E$. Получили
            противоречие.

            Пусть $(\kmodel{M}',w) \not\models
            \neg \alpha_i\to\Box(\alpha_{n+1+i}\wedge\neg
            \alpha_{2n+2}\to\neg \alpha_i)$. Тогда
            $(\kmodel{M}',w) \not\models \alpha_i$ и существует $w'\in W'$
            такой, что $wR'w'$, $(\kmodel{M}',w')\models\alpha_{n+1+i}\wedge\neg
            \alpha_{2n+2}$,
            $(\kmodel{M}',w')\models\alpha_i$. Так как
            $(\kmodel{M}',w)\not\models \alpha_i$, то $w\in W$, а так как
            $(\kmodel{M}',w') \not\models \alpha_{2n+2}$, то $w'\in W$.
            Следовательно, $(\kmodel{M},w)\not\models p_i$,
            $(\kmodel{M},w)\models q_i$,
            $(\kmodel{M},w')\models q_i$,
            \linebreak
            $(\kmodel{M},w')\not\models q_{n+1}$,
            $(\kmodel{M},w')\models p_i$. Но тогда
            $$
            \begin{array}{rcl}
            (\kmodel{M},w) \not\models
            q_i\to\big(p_i\to\Box(q_i\wedge\neg
            q_{n+1}\to
            p_i)\big)\wedge\big(\neg p_i\to\Box(q_i\wedge\neg
            q_{n+1}\to\neg p_i)\big),
            \end{array}
            $$
            а значит, $(\kmodel{M},w_0) \not\models E$. Снова получили
            противоречие.

            Значит, $(\kmodel{M}',w_0)\models E_\alpha$.

            Предположим, что $(\kmodel{M}',w_0) \not\models D_\alpha$.
            В этом случае для некоторого
            $w\in W'$,
            достижимого из $w_0$, 
            и некоторого $i\in\set{0,\dots,n-1}$ такого, что $Q_{i+1}=
            \forall$, 
            $$
            (\kmodel{M}',w)\models \alpha_{n+1+i}\wedge\neg\alpha_{n+2+i},
            $$
            при этом
            $$
            (\kmodel{M}',w)\not\models
            \Diamond(\alpha_{n+2+i} \wedge \neg\alpha_{n+3+i} \wedge
            \alpha_{i+1})
            $$
            или
            $$
            (\kmodel{M}',w)\not\models
            \Diamond(\alpha_{n+2+i} \wedge \neg\alpha_{n+3+i} \wedge
            \neg \alpha_{i+1}).
            $$
            Рассмотрим первый случай; второй рассматривается аналогично.
            Так как $(\kmodel{M}',w)\not\models\alpha_{n+2+i}$, то $w\in
            W$, а так как $(\kmodel{M}',w)\models
            \alpha_{n+1+i}\wedge\neg\alpha_{n+2+i}$, то
            $(\kmodel{M},w)\models
            q_{i}\wedge\neg q_{i+1}$. Поскольку
            $(\kmodel{M},w_0)\models D$, получаем, что
            $(\kmodel{M},w)\models
            q_{i}\wedge\neg q_{i+1} \to \Diamond(q_{i+1}\wedge\neg
            q_{i+2}\wedge p_i)$, а следовательно, существует
            мир $w'\in W$ такой,
            что $wRw'$ и $(\kmodel{M},w')\models
            q_{i+1}\wedge\neg q_{i+2}\wedge p_i$.
            По
            построению
            модели
            $\kmodel{M}'$ в этом случае имеет место отношение
            $(\kmodel{M}',w')\models
            \alpha_{n+2+i} \wedge \neg\alpha_{n+3+i} \wedge
            \alpha_{i+1}$. Так как $wRw'$, то $wR'w'$, а значит,
            $(\kmodel{M}',w)\models
            \Diamond(\alpha_{n+2+i} \wedge \neg\alpha_{n+3+i} \wedge
            \alpha_{i+1})$. Получили противоречие, следовательно,
            $(\kmodel{M}',w_0)\models D_\alpha$.

            Истинность в мире $w_0$ модели $\kmodel{M}'$ формулы $C_\alpha$ 
            доказывается аналогично тому, как это было сделано в
            случае формулы~$D_\alpha$.

            В результате получаем, что $(\kmodel{M}', w_0) \models
            \varphi^\ast_\alpha$. Осталось заметить, что $\kframe{F}'$~--- шкала
            логики $\logic{wGrz}$, и значит,
            $\varphi^\ast_\alpha$ является $\logic{wGrz}$\nobreakdash-выполнимой.
            \end{proof}

            В качестве следствия сразу же получаем следующее утверждение.

            \begin{theorem}
            \label{th_PSPACE_K4}
            \label{th:PSPACE:K4}
            Проблема разрешения константного фрагмента логики $L\in [\logic{K},\logic{wGrz}]$ является\/ $\cclass{PSPACE}$\nobreakdash-трудной.
            \end{theorem}

            \begin{proof}
            Достаточно показать, что $\logic{TQBF}$ полиномиально сводится к проблеме $L$\nobreakdash-выполнимости константных формул. Без ограничений общности можем рассматривать только формулы вида $\neg Q_1p_1\ldots Q_np_n\varphi'$, где $\varphi'$~--- бескванторная формула от переменных $p_1,\ldots,p_n$. Действительно, булеву формулу с кванторами можно привести к префиксной нормальной форме, а затем взять её двойное отрицание, оставить внешнее отрицание, а внутреннее пронести через кванторную приставку.

            Пусть $\varphi = Q_1p_1\ldots Q_np_n\varphi'$. Покажем, что
            $$
            \begin{array}{rcl}
            \neg\varphi\in\logic{TQBF}
              & \iff
              & \neg\varphi^\ast_\alpha\in{L}.
            \end{array}
            $$

            Пусть $\neg\varphi\in\logic{TQBF}$. Тогда, согласно предложению~\ref{prop:modal:complexity}, $\neg\varphi^\ast\in\logic{K}$. Логика $\logic{K}$ замкнута по правилу подстановки, поэтому $\neg\varphi^\ast_\alpha\in\logic{K}$. Поскольку $\logic{K}\subseteq L$, получаем, что $\neg\varphi^\ast_\alpha\in L$.

            Пусть $\neg\varphi\not\in\logic{TQBF}$. Тогда, согласно предложению~\ref{prop:modal:complexity}, $\neg\varphi^\ast\not\in\logic{wGrz}$. В этом случае, согласно лемме~\ref{lem_K4(0)form_equiv}, $\neg\varphi^\ast_\alpha\not\in\logic{wGrz}$. Поскольку $L\subseteq\logic{wGrz}$, получаем, что $\neg\varphi^\ast_\alpha\not\in L$.

            Осталось заметить, что функция, сопоставляющая формуле $\varphi$ формулу $\varphi^\ast_\alpha$, является полиномиально вычислимой.
            \end{proof}

    \subsection{Анализ доказательства: важные моменты}

            Обратим внимание на важные моменты в доказательстве теоремы~\ref{th:PSPACE:K4}.

            Прежде всего, когда мы делали подстановку константных формул вместо переменных, нам достаточно было рассматривать только те формулы, которые были построены для моделирования некоторой $\cclass{PSPACE}$-трудной проблемы (в~нашем случае, $\logic{TQBF}$).

            В лемме~\ref{lem:K4(0)form:equiv} импликация $(\Rightarrow)$ очевидна в силу замкнутости логики по правилу подстановки; более того, в доказательстве теоремы~\ref{th:PSPACE:K4} соответствующую аргументацию всё равно пришлось повторить. Таким образом, в лемме~\ref{lem:K4(0)form:equiv} фактически важна лишь импликация~$(\Leftarrow)$.

            Необходимым условием, чтобы можно было доказать высокую сложность константного фрагмента, является наличие бесконечного множества попарно неэквивалентных в логике константных формул. Как мы увидим дальше, это условие не является достаточным. В случае, когда константный фрагмент логики алгоритмически прост, для получения аналогичного результата имеет смысл перейти к рассмотрению фрагмента от одной переменной или большего их числа.\-

            Наконец, условие наследственности <<вверх>>. Это очень важный момент в моделировании переменных формулами в случае требования транзитивности для шкал Крипке. С одной стороны, условие наследственности <<вверх>> можно бы было снять, например, для $\logic{K}$ или $\logic{T}$ (и мы этим ещё воспользуемся). С другой стороны, тот факт, что теорема~\ref{th:PSPACE:K4} доказана при этом условии, позволяет сделать предположение, что её аналог возможен и в интуиционистском случае (и мы покажем, что это действительно так).

            В целом, очень многие построения и рассуждения ниже будут основаны на идеях, изложенных в доказательстве теоремы~\ref{th:PSPACE:K4}, поэтому какие-то из них будут описаны более лаконично, без излишних деталей.

    \subsection{Интервалы $[\logic{K},\logic{GL}]$ и $[\logic{K},\logic{Grz}]$}
    \label{ssec:modprop:emb:GL:Grz}

            Константный фрагмент логики $\logic{GL}$ совпадает с константным фрагментом логики $\logic{GLLin}$ (в других обозначениях, $\logic{GL.3}$), а потому, хоть и содержит бесконечное множество попарно неэквивалентных формул (например, формул вида $\Box^n\bot$, где $n\in\numbers{N}$), является полиномиально разрешимым, что несложно получить, используя линейную аппроксимируемость $\logic{GL.3}$, см.~\cite[Теорема~18.25]{ChZ}. Мы вернёмся к подобным конструкциям позже, а пока наша цель~--- получить аналог теоремы~\ref{th:PSPACE:K4} для логики $\logic{GL}$ и её подлогик в языке с одной переменной.

            Константный фрагмент логики $\logic{Grz}$ устроен ещё проще. Дело в том, что в расширениях логики $\logic{D}$ любая константная формула эквивалентна либо константе~$\bot$, либо константе~$\top$, поскольку логике~$\logic{D}$ принадлежат формулы
            $$
            \begin{array}{cccc}
            \Box\bot\lra\bot,
              & \Diamond\bot\lra\bot,
              & \Box\top\lra\top,
              & \Diamond\top\lra\top,
            \end{array}
            $$
            а логика $\logic{Grz}$ является расширением логики~$\logic{D}$. Для фрагмента $\logic{Grz}$ от одной переменной аналог теоремы~\ref{th:PSPACE:K4}, тем не менее, справедлив.

            Мы предложим единый метод моделирования $\cclass{PSPACE}$-полной проблемы, чтобы получить доказательство $\cclass{PSPACE}$-трудности логик из интервалов $[\logic{K},\logic{GL}]$ и $[\logic{K},\logic{Grz}]$ в языке с одной переменной.

            Изменим $\alpha_i$, для чего положим
$$
\begin{array}{lcl}
\label{page:GL:delta}
\delta_1
  & =
  & \Diamond\Box^+ p;
  \\
\delta_{m+1}
  & =
  & \Diamond(p \wedge \Diamond(\neg p \wedge \delta_m)); \\
\alpha'_m
  & =
  & \neg p\to
    \Box^+(p \to \Box(\neg p \wedge \delta_m \wedge \neg\delta_{m+1} \to \Box\Diamond^+ p)).
\end{array}
$$

Пусть $\varphi^\ast_{\alpha'}$~--- формула, получающаяся из $\varphi^\ast$ подстановкой вместо каждой переменной $p_i$ формулы $\alpha'_i$.

\begin{figure}
  \centering
  \begin{tikzpicture}[scale=1.50]

    \coordinate (a30)    at ( 0.0, 3);
    \coordinate (a31)    at ( 0.0, 4);
    \coordinate (a32)    at ( 0.0, 5);
    \coordinate (a33)    at ( 0.0, 6);
    \coordinate (a34)    at ( 0.0, 7);
    \coordinate (a35)    at ( 0.0, 8);
    \coordinate (b30)    at (-1.0, 4);
    \coordinate (c30)    at ( 0.0, 2);

    \coordinate (ld)     at (-2.0, 2.5);
    \coordinate (lu)     at (-2.0, 8.5);
    \coordinate (ru)     at ( 1.0, 8.5);
    \coordinate (rd)     at ( 1.0, 2.5);

    \draw [fill]     (a30)   circle [radius=2.0pt] ;
    \draw [fill]     (a31)   circle [radius=2.0pt] ;
    \draw [fill]     (a32)   circle [radius=2.0pt] ;
    \draw [fill]     (a33)   circle [radius=2.0pt] ;
    \draw [fill]     (a35)   circle [radius=2.0pt] ;
    \draw [fill]     (b30)   circle [radius=2.0pt] ;
    \draw [fill]     (c30)   circle [radius=2.0pt] ;

    \begin{scope}[>=latex]

      \draw [->, shorten >=  2.75pt, shorten <= 2.75pt] (c30) -- (a30) ;
      \draw [->, shorten >=  2.75pt, shorten <= 2.75pt] (a30) -- (a31) ;
      \draw [->, shorten >=  2.75pt, shorten <= 2.75pt] (a31) -- (a32) ;
      \draw [->, shorten >=  2.75pt, shorten <= 2.75pt] (a32) -- (a33) ;
      \draw [->, shorten >= 11.75pt, shorten <= 2.75pt] (a33) -- (a34) ;
      \draw [->, shorten >=  2.75pt, shorten <= 7.75pt] (a34) -- (a35) ;
      \draw [->, shorten >=  2.75pt, shorten <= 2.75pt] (a30) -- (b30) ;

    \end{scope}

      \node [left=2pt] at (a30)  {$\phantom{a^{2m}_0}p$};
      \node [left=2pt] at (a31)  {$\phantom{a^{2m}_0}\neg p$};
      \node [left=2pt] at (a32)  {$\phantom{a^{2m}_0}p$};
      \node [left=2pt] at (a33)  {$\phantom{a^{2m}_0}\neg p$};
      \node []         at (a34)  {$\vdots$};
      \node [left=2pt] at (a35)  {$\phantom{a^{2m}_0}p$};

      \node [right=2pt] at (a30)  {$a^{2m}_0$};
      \node [right=2pt] at (a31)  {$a^{2m}_1$};
      \node [right=2pt] at (a32)  {$a^{2m}_2$};
      \node [right=2pt] at (a33)  {$a^{2m}_3$};
      \node [right=2pt] at (a35)  {$a^{2m}_{2m}$};

      \node [left=2pt]  at (b30)  {$\phantom{a^{2m}_0}\neg p$};
      \node [above=2pt] at (b30)  {$~~b^{2m}$};
      \node [left=2pt]  at (c30)  {$\phantom{c^{2m}}\neg\alpha_m$};
      \node [right=2pt] at (c30)  {$c^{2m}$};

      \draw [dashed, color=black!50!] (ld) -- (lu) -- (ru) -- (rd) -- cycle ;

  \end{tikzpicture}
~~~~
  \begin{tikzpicture}[scale=1.50]

    \coordinate (a30)    at ( 0.0, 3);
    \coordinate (a31)    at ( 0.0, 4);
    \coordinate (a32)    at ( 0.0, 5);
    \coordinate (a33)    at ( 0.0, 6);
    \coordinate (a34)    at ( 0.0, 7);
    \coordinate (a35)    at ( 0.0, 8);
    \coordinate (b30)    at (-1.0, 4);
    \coordinate (c30)    at ( 0.0, 2);

    \coordinate (ld)     at (-2.0, 2.5);
    \coordinate (lu)     at (-2.0, 8.5);
    \coordinate (ru)     at ( 1.0, 8.5);
    \coordinate (rd)     at ( 1.0, 2.5);

    \draw []     (a30)   circle [radius=2.0pt] ;
    \draw []     (a31)   circle [radius=2.0pt] ;
    \draw []     (a32)   circle [radius=2.0pt] ;
    \draw []     (a33)   circle [radius=2.0pt] ;
    \draw []     (a35)   circle [radius=2.0pt] ;
    \draw []     (b30)   circle [radius=2.0pt] ;
    \draw []     (c30)   circle [radius=2.0pt] ;

    \begin{scope}[>=latex]

      \draw [->, shorten >=  2.75pt, shorten <= 2.75pt] (c30) -- (a30) ;
      \draw [->, shorten >=  2.75pt, shorten <= 2.75pt] (a30) -- (a31) ;
      \draw [->, shorten >=  2.75pt, shorten <= 2.75pt] (a31) -- (a32) ;
      \draw [->, shorten >=  2.75pt, shorten <= 2.75pt] (a32) -- (a33) ;
      \draw [->, shorten >= 11.75pt, shorten <= 2.75pt] (a33) -- (a34) ;
      \draw [->, shorten >=  2.75pt, shorten <= 7.75pt] (a34) -- (a35) ;
      \draw [->, shorten >=  2.75pt, shorten <= 2.75pt] (a30) -- (b30) ;

    \end{scope}

      \node [left=2pt] at (a30)  {$\phantom{a^{2m}_0}p$};
      \node [left=2pt] at (a31)  {$\phantom{a^{2m}_0}\neg p$};
      \node [left=2pt] at (a32)  {$\phantom{a^{2m}_0}p$};
      \node [left=2pt] at (a33)  {$\phantom{a^{2m}_0}\neg p$};
      \node []         at (a34)  {$\vdots$};
      \node [left=2pt] at (a35)  {$\phantom{a^{2m}_0}p$};

      \node [right=2pt] at (a30)  {$a^{2m}_0$};
      \node [right=2pt] at (a31)  {$a^{2m}_1$};
      \node [right=2pt] at (a32)  {$a^{2m}_2$};
      \node [right=2pt] at (a33)  {$a^{2m}_3$};
      \node [right=2pt] at (a35)  {$a^{2m}_{2m}$};

      \node [left=2pt]  at (b30)  {$\phantom{a^{2m}_0}\neg p$};
      \node [above=2pt] at (b30)  {$~~b^{2m}$};
      \node [left=2pt]  at (c30)  {$\phantom{c^{2m}}\neg\alpha_m$};
      \node [right=2pt] at (c30)  {$c^{2m}$};

      \draw [dashed, color=black!50!] (ld) -- (lu) -- (ru) -- (rd) -- cycle ;

  \end{tikzpicture}
  \caption{Модели $\kmodel{M}_m$ и $\kmodel{M}'_m$}
  \label{fig1:2}
\end{figure}

\begin{lemma}
\label{lem:GL(1)form:equiv}
Если $\varphi\in\logic{TQBF}$, то $\varphi^\ast_{\alpha'}$ является $\logic{GL}$\nobreakdash-выполнимой и $\logic{Grz}$\nobreakdash-выполнимой.
\end{lemma}

\begin{proof}
Нужно повторить схему доказательства леммы~\ref{lem:K4(0)form:equiv}, сделав незначительные технические изменения в описанные построения.

Пусть $\varphi\in\logic{TQBF}$. Тогда $\varphi$ истинна в корне конечного иррефлексивного транзитивного дерева.

Переопределим шкалы $\kframe{F}_m = \otuple{W_m,R_m}$ и $\kframe{F}^+_m = \otuple{W^+_m,R^+_m}$, определённые в доказательстве леммы~\ref{lem_K4(0)form_equiv}, убрав из отношения $R_m$ пару $\opair{b^m}{b^m}$, а из отношения $R^+_m$~--- ещё и пару $\opair{c^m}{c^m}$; другими словами, сделаем миры этих шкал иррефлексивными. Далее из переопределённых таким образом шкал нас будут интересовать только шкалы с чётными индексами.

На полученных шкалах $\kframe{F}_{2m}$ и $\kframe{F}^+_{2m}$ рассмотрим модели $\kmodel{M}_m = \otuple{\kframe{F}_{2m},v_{m}}$ и $\kmodel{M}^+_m = \otuple{\kframe{F}^+_{2m},v^+_{m}}$, для которых выполнено условие
$$
\begin{array}{lclcl}
v_m(p) & = & v^+_m(p) & = & \setc{a^{2m}_{2k}}{0\leqslant k\leqslant m}.
\end{array}
$$
Тогда нетрудно видеть, что
$$
\begin{array}{lcl}
(\kmodel{M}^+_m,w)\not\models\alpha'_k
  & \iff
  & \mbox{$m=k$ и $w=c^{2m}$}.
\end{array}
$$
Теперь в конструкции, описанной в лемме~\ref{lem:K4(0)form:equiv}, копии шкалы $\kframe{F}_{m}$ заменим копиями шкалы $\kframe{F}_{2m}$, а оценку на этих копиях зададим в соответствии с оценкой~$v_m$, т.е. вместо произвольной оценки $v'$ нужно рассмотреть такую $v'$, при которой переменная $p$ истинна в мирах вида $\otuple{a^{2m}_{2k},w}$. В итоге мы получим $\logic{GL}$-модель, в корне которой истинна формула~$\varphi^\ast_{\alpha'}$.

Чтобы получить обоснование $\logic{Grz}$\nobreakdash-выполнимости формулы~$\varphi^\ast_{\alpha'}$, достаточно взять вместо конечного иррефлексивного дерева, где истинна $\varphi$, конечное рефлексивное дерево, а в качестве шкал $\kframe{F}_{2m}$ и $\kframe{F}^+_{2m}$ их рефлексивные замыкания $\kframe{F}^\ast_{2m}$ и $\kframe{F}^{+\ast}_{2m}$, на которых определить модели $\kmodel{M}'_m = \otuple{\kframe{F}^\ast_{2m},v_{m}}$ и $\kmodel{M}'^+_m = \otuple{\kframe{F}^{+\ast}_{2m},v^+_{m}}$.
\end{proof}

\begin{theorem}
\label{th_PSPACE_GL_Grz}
\label{th:PSPACE:GL:Grz}
Проблема разрешения фрагмента от одной переменной логики $L\in [\logic{K},\logic{GL}]\cup[\logic{K},\logic{Grz}]$ является\/ $\cclass{PSPACE}$\nobreakdash-трудной.
\end{theorem}

\begin{proof}
            Аналогично доказательству теоремы~\ref{th:PSPACE:K4}.

            Достаточно показать, что $\logic{TQBF}$ полиномиально сводится к проблеме $L$\nobreakdash-выполнимости формул от одной переменной. Без ограничений общности можем рассматривать только формулы вида $\neg Q_1p_1\ldots Q_np_n\varphi'$, где $\varphi'$~--- бескванторная формула от переменных $p_1,\ldots,p_n$.

            Пусть $\varphi = Q_1p_1\ldots Q_np_n\varphi'$. Покажем, что
            $$
            \begin{array}{rcl}
            \neg\varphi\in\logic{TQBF}
              & \iff
              & \neg\varphi^\ast_{\alpha'}\in{L}.
            \end{array}
            $$

            Пусть $\neg\varphi\in\logic{TQBF}$. Тогда, согласно предложению~\ref{prop:modal:complexity}, $\neg\varphi^\ast\in\logic{K}$. Логика $\logic{K}$ замкнута по правилу подстановки, поэтому $\neg\varphi^\ast_{\alpha'}\in\logic{K}$. Поскольку $\logic{K}\subseteq L$, получаем, что $\neg\varphi^\ast_{\alpha'}\in L$.

            Пусть $\neg\varphi\not\in\logic{TQBF}$. Тогда, согласно предложению~\ref{prop:modal:complexity}, $\neg\varphi^\ast\not\in\logic{wGrz}$. В~этом случае, согласно лемме~\ref{lem:GL(1)form:equiv}, $\neg\varphi^\ast_{\alpha'}\not\in\logic{GL}$ и $\neg\varphi^\ast_{\alpha'}\not\in\logic{Grz}$. Поскольку $L\subseteq\logic{GL}$ или $L\subseteq\logic{Grz}$, получаем, что $\neg\varphi^\ast_{\alpha'}\not\in L$.

            Осталось заметить, что функция, сопоставляющая формуле $\varphi$ формулу $\varphi^\ast_{\alpha'}$, является полиномиально вычислимой.
\end{proof}

    \subsection{Интервал $[\logic{K},\logic{KTB}]$}
    \label{ssec:modprop:emb:KTB}

Похожую технику можно применить и для логик симметричных шкал: нужно лишь подходящим образом изменить формулы $\alpha_i$ (или $\alpha'_i$), а также шкалы $\kframe{F}_i$ и модели $\kmodel{M}_i$ (или~$\kmodel{M}'_i$), использованные выше. Мы продемонстрируем немного другой подход. Прежде заметим, что доказательства теорем~\ref{th:PSPACE:K4} и~\ref{th:PSPACE:GL:Grz} основаны на том, что мы осуществляли подстановку константных формул или, соответственно, формул от одной переменной в специальные формулы, именно, в формулы вида $\varphi^\ast$ (напомним, что $\varphi^\ast$ строится определённым способом по замкнутой булевой формуле с кванторами $\varphi$, находящейся в префиксной форме). Так вот, мы покажем, что за счёт подобных подстановок в некоторых случаях можно построить полиномиальное погружение всей логики в свой фрагмент от конечного числа переменных. Сразу отметим, что так сделать получается не всегда, но более подробно мы обсудим это в разделе~\ref{ssec:modprop:emb:discuss}.

Пусть $\varphi$~--- некоторая $\lang{ML}$-формула от переменных $p_1,\ldots,p_n$.

Определим следующие формулы, зависящие от переменной~$p$:
$$
\begin{array}{lcl}
  \alpha_k & = & \Box^{k}\, \neg p \con
                               \Diamond^{k+1}\, (p \con \neg \Diamond \Box^{+} p) \con
                               \Diamond^{k+2}\, \Box^+ p;
\\
  \beta_k  & = & \neg p \con \Diamond^{k+2}\, \alpha_{k} \con
                 \neg \Diamond^{k+1}\, \alpha_{k}.
\end{array}
$$

Формулы $\beta_1, \ldots, \beta_n$ мы будем использовать для моделирования переменных
$p_1, \ldots, p_n$ в формулах (от $n$ переменных), при этом формула $\beta_{n+1}$ будет иметь особый статус, поэтому положим
$$
\begin{array}{lclcl}
  B_\varphi  & = & \beta_{n+1}
             & = & \displaystyle
                   \neg p \con \Diamond^{n+3}\, \alpha_{n+1} \con
                   \neg \Diamond^{n+2}\, \alpha_{n+1}.
\end{array}
$$

Определим рекурсивно перевод $\tau_\varphi$:
$$
\begin{array}{lcl}
  \tau_\varphi(p_k)            & = & \beta_k, \qquad \mbox{где $k \in \{1, \ldots, n \}$;} \\
  \tau_\varphi(\bot)           & = & \bot;  \\
  \tau_\varphi(\psi \con \chi) & = & \tau_\varphi(\psi) \con \tau_\varphi(\chi);  \\
  \tau_\varphi(\psi \dis \chi) & = & \tau_\varphi(\psi) \dis \tau_\varphi(\chi);  \\
  \tau_\varphi(\psi \imp \chi) & = & \tau_\varphi(\psi) \imp \tau_\varphi(\chi);  \\
  \tau_\varphi(\Box  \psi)     & = & \Box (B_\varphi \imp \tau_\varphi(\psi)).
\end{array}
$$

  \begin{figure}
  \label{fig:KTB}
  \centering
  \begin{tikzpicture}[scale=1.50]

    \coordinate (bkp1)    at ( 1, 1);   \node [below=2pt] at (bkp1)  {$b^k_{k+1}$} ;
    \coordinate (bk)      at ( 2, 1);   \node [below=2pt] at (bk)    {$b^k_k$}    ;
    \coordinate (b1)      at ( 4, 1);   \node [below=2pt] at (b1)    {$b^k_1$}    ;
    \coordinate (ak)      at ( 5, 1);   \node [below=2pt] at (ak)    {$a^k_{\phantom{k}}$}      ;
    \coordinate (c1)      at ( 6, 1);   \node [below=2pt] at (c1)    {$c^k_1$}    ;
    \coordinate (ck)      at ( 8, 1);   \node [below=2pt] at (ck)    {$c^k_k$}    ;
    \coordinate (ckp1)    at ( 9, 1);   \node [below=2pt] at (ckp1)   {$c^k_{k+1}$}    ;
    \coordinate (ckp2)    at (10, 1);   \node [below=2pt] at (ckp2)   {$c^k_{k+2}$}    ;

    \draw [] (bkp1)     circle [radius=2.0pt] ;
    \draw [] (bk)       circle [radius=2.0pt] ;
    \draw [] (b1)       circle [radius=2.0pt] ;
    \draw [] (ak)       circle [radius=2.0pt] ;
    \draw [] (c1)       circle [radius=2.0pt] ;
    \draw [] (ckp1)     circle [radius=2.0pt] ;
    \draw [] (ck)       circle [radius=2.0pt] ;
    \draw [] (c1)       circle [radius=2.0pt] ;
    \draw [] (ckp2)     circle [radius=2.0pt] ;

    \begin{scope}[>=latex]

      \draw [<->, shorten >= 2.75pt, shorten <= 2.75pt] (bkp1)   -- (bk) ;
      \draw [<->, shorten >= 2.75pt, shorten <= 2.75pt] (b1) -- (ak);
      \draw [<->, shorten >= 2.75pt, shorten <= 2.75pt] (ak)   -- (c1);
      \draw [<->, shorten >= 2.75pt, shorten <= 2.75pt] (ckp1)   -- (ck) ;
      \draw [<->, shorten >= 2.75pt, shorten <= 2.75pt] (ckp1) -- (ckp2);
      \draw [<->, shorten >= 2.75pt, shorten <= 2.75pt, dashed] (b1)   -- (bk) ;
      \draw [<->, shorten >= 2.75pt, shorten <= 2.75pt, dashed] (c1) -- (ck);

    \end{scope}

      \draw (1.75,1.15) -- (1.75, 1.25) -- (8.25, 1.25) -- (8.25, 1.15)      ;

      \node [above=10pt] at (ak)  {$\neg p$};

      \node [above=10pt] at (bkp1)   {$p$};
      \node [above=10pt] at (ckp1)   {$p$};
      \node [above=10pt] at (ckp2)   {$p$};

    \end{tikzpicture}

    \caption{Модель $\frak{M}_k^{\mathit{rs}}$}
  \end{figure}

\begin{lemma}
\label{lem:KTB(1)}
Имеет место следующая эквивалентность:
$$
\begin{array}{lcl}
  \varphi\in\logic{KTB} & \iff & B_\varphi\to\tau_\varphi(\varphi)\in \logic{KTB}.
\end{array}
$$
\end{lemma}

\begin{proof}
%
$(\Rightarrow)$
Пусть $B_\varphi\to\tau_\varphi(\varphi)\not\in\logic{KTB}$. Тогда существуют рефлексивная симметричная шкала $\kframe{F}=\otuple{W,R}$, модель $\kmodel{M}=\otuple{\kframe{F},v}$ и мир $x\in W$, такие, что $(\kmodel{M},x)\not\models B_\varphi\to\tau_\varphi(\varphi)$. Пусть
$$
\begin{array}{lcl}
W'          & = & \setc{y\in W}{(\kmodel{M},y)\models B_\varphi}; \\
R'          & = & R\upharpoonright W'; \\
\kframe{F}' & = & \otuple{W',R'}.
\end{array}
$$
Заметим, что $W'\ne\varnothing$, поскольку $x\in W'$.
Определим оценку $v'$, положив для каждой переменной $p_k$
$$
\begin{array}{lcl}
v'(p_k) & = & v(\beta_k)\cap W'. \\
\end{array}
$$
Наконец, положим
$$
\begin{array}{lcl}
\kmodel{M}' & = & \otuple{\kframe{F}',v'}.
\end{array}
$$

Теперь заметим, что для каждого $y\in W'$ и любой подформулы $\psi$ формулы~$\varphi$ имеет место следующая эквивалентность:
$$
\begin{array}{lcl}
(\kmodel{M}',y)\models\psi & \iff & (\kmodel{M},y)\models\tau_\varphi(\psi).
\end{array}
\eqno{\mbox{$({\ast})$}}
$$
Эквивалентность \mbox{$({\ast})$} обосновывается индукцией по построению~$\psi$; соответствующее доказательство является рутинной проверкой, и здесь опускается.

Но тогда из \mbox{$({\ast})$} получаем, что $(\kmodel{M}',x)\not\models \varphi$. Осталось заметить, что шкала $\kframe{F}'$ рефлексивна и симметрична, а значит, является шкалой логики $\logic{KTB}$, и следовательно $\varphi\not\in\logic{KTB}$.

$(\Leftarrow)$
Пусть $\varphi\not\in\logic{KTB}$. Тогда существуют рефлексивная симметричная шкала $\kframe{F}=\otuple{W,R}$, модель $\kmodel{M}=\otuple{\kframe{F},v}$ и мир $x\in W$, такие, что $(\kmodel{M},x)\not\models \varphi$. Построим модель, в которой опровергается формула $B_\varphi\to\tau_\varphi(\varphi)$.

Для каждого $k\in\numNp$ определим шкалу $\kframe{F}^{\mathit{rs}}_k=\otuple{W_k,R_k}$, положив
$$
\begin{array}{lcl}
W_k & = & \{ a^k\} \cup
          \{ b^k_{i} : 1 \leqslant i \leqslant k+1\}
          \cup \{ c^k_{i} : 1 \leqslant i \leqslant k+2\}
\end{array}
$$
и взяв в качестве $R_k$ рефлексивно-симметричное замыкание отношения
$$
  \begin{array}{l}
    \{ \langle \mathstrut b^{\mathstrut k}_i, b^{\mathstrut k}_{i+1}
    \rangle : 1 \leqslant i \leqslant k \}  \cup
    \{ \langle \mathstrut c^{\mathstrut k}_i, c^{\mathstrut k}_{i+1}
    \rangle : 1 \leqslant i \leqslant k+1 \} \cup
    \{ \langle \mathstrut a^{\mathstrut k}_{\phantom{i}}, b^{\mathstrut k}_{1}
    \rangle, \langle \mathstrut a^{\mathstrut k}_{\phantom{i}}, c^{\mathstrut k}_{1}
    \rangle \}.
  \end{array}
$$
На шкале $\kframe{F}^{\mathit{rs}}_k$ рассмотрим оценку $v_k$, удовлетворяющую условию
$$
\begin{array}{lcl}
v_k(p) & = & \{b^k_{k+1},c^k_{k+1},c^k_{k+2}\},
\end{array}
$$
и положим
$$
\begin{array}{lcl}
\kmodel{M}^{\mathit{rs}}_k & = & \otuple{\kframe{F}_k^{\mathit{rs}},v_k},
\end{array}
$$
см.~рис.~\ref{fig:KTB}.

Тогда если мы хотим сделать в каком-то новом (по отношению к $W_k$) мире $y$ истинной формулу $\beta_k$, достаточно, чтобы из этого мира был достижим мир $b^k_{k+1}$ модели $\kmodel{M}^{\mathit{rs}}_k$. Мы воспользуемся этим наблюдением.

Пусть
$$
\begin{array}{lcl}
W' & = & \displaystyle W\cup \bigcup\limits_{k=1}^{n+1} (W_k\times W) \\
\end{array}
$$
и пусть $R'$~--- рефлексивно-симметричное замыкание отношения
$$
\begin{array}{lcl}
R & \cup
  & \displaystyle
    \bigcup\limits_{k=1}^{n+1}\bigcup\limits_{w\in W}\set{\otuple{\otuple{y,w},\otuple{z,w}} : \mbox{$y,z\in W_k$ и $yR_kz$}} ~\cup {}
  \smallskip\\
  & \cup
  & \displaystyle
    \bigcup\limits_{k=1}^{n+1}\bigcup\limits_{w\in W}\set{\otuple{w,\otuple{b^k_{k+1},w}} : \mbox{$k=n+1$ или $(\kmodel{M},w)\models p_k$}}.
\end{array}
$$
Пусть $\kmodel{F}'=\otuple{W',R'}$. Определим на шкале $\kframe{F}'$ оценку $v'$ так, чтобы выполнялось следующее условие:
$$
\begin{array}{lcl}
v'(p)
  & =
  & \displaystyle
    \bigcup\limits_{k=1}^{n+1}\bigcup\limits_{w\in W}\set{\otuple{b^k_{k+1},w},\otuple{c^k_{k+1},w},\otuple{c^k_{k+2},w}},
\end{array}
$$
т.е. $v'$ в определённом смысле согласована с оценками $v_1,\ldots,v_{n+1}$.
Пусть $\kmodel{M}'=\otuple{\kframe{F}',v'}$.

Несложно понять, что для всякого $y\in W'$
$$
\begin{array}{lcl}
(\kmodel{M}',y)\models B_\varphi & \iff & y\in W;
\end{array}
\eqno{\mbox{$({\ast}{\ast})$}}
$$
кроме того, для всякого $y\in W$ и $k\in\set{1,\ldots,n}$
$$
\begin{array}{lcl}
(\kmodel{M},y)\models p_k & \iff & (\kmodel{M}',y)\models \beta_k.
\end{array}
\eqno{\mbox{$({\ast}{\ast}{\ast})$}}
$$

Теперь заметим, что для каждого $y\in W$ и любой подформулы $\psi$ формулы~$\varphi$ имеет место следующая эквивалентность:
$$
\begin{array}{lcl}
(\kmodel{M},y)\models\psi & \iff & (\kmodel{M}',y)\models\tau_\varphi(\psi).
\end{array}
\eqno{\mbox{$({\ast}{\ast}{\ast}{\ast})$}}
$$
Эквивалентность \mbox{$({\ast}{\ast}{\ast}{\ast})$} обосновывается индукцией по построению~$\psi$: условие \mbox{$({\ast}{\ast}{\ast})$} обеспечивает базис, а условие \mbox{$({\ast}{\ast})$}~--- обоснование индукционного шага для случая, когда $\psi=\Box\psi'$.

Из условий \mbox{$({\ast}{\ast})$} и \mbox{$({\ast}{\ast}{\ast}{\ast})$} получаем, что $(\kmodel{M}',x)\not\models B_\varphi\to\tau_\varphi(\varphi)$, а значит, $B_\varphi\to\tau_\varphi(\varphi)\not\in\logic{KTB}$.
\end{proof}


\begin{lemma}
\label{lem:KTB(1)K}
Имеет место следующая эквивалентность:
$$
\begin{array}{lcl}
  \varphi\in\logic{K} & \iff & B_\varphi\to\tau_\varphi(\varphi)\in \logic{K}.
\end{array}
$$
\end{lemma}

\begin{proof}
Достаточно повторить аргументацию доказательства леммы~\ref{lem:KTB(1)}, с той разницей, что в части $(\Leftarrow)$ при построении шкалы $\kframe{F}'$ вместо рефлексивно-симметричного замыкания отношения, на основе которого определяется~$R'$, нужно взять само это отношение.
\end{proof}

\begin{theorem}
\label{th:PSPACE:KTB}
Проблема разрешения фрагмента от одной переменной логики $L\in [\logic{K},\logic{KTB}]$ является\/ $\cclass{PSPACE}$\nobreakdash-трудной.
\end{theorem}

\begin{proof}
            Аналогично доказательству теоремы~\ref{th:PSPACE:K4}.

            Достаточно показать, что $\logic{TQBF}$ полиномиально сводится к проблеме $L$\nobreakdash-выполнимости формул от одной переменной. Снова можем ограничиться рассмотрением формул вида $\neg Q_1p_1\ldots Q_np_n\varphi'$, где $\varphi'$~--- бескванторная формула от переменных $p_1,\ldots,p_n$.

            Пусть $\varphi = Q_1p_1\ldots Q_np_n\varphi'$. Покажем, что
            $$
            \begin{array}{rcl}
            \neg\varphi\in\logic{TQBF}
              & \iff
              & B_{\neg\varphi^\ast} \to \tau_{\neg\varphi^\ast}(\neg\varphi^\ast)\in L.
            \end{array}
            $$

            Пусть $\neg\varphi\in\logic{TQBF}$. Тогда, согласно предложению~\ref{prop:modal:complexity}, $\neg\varphi^\ast\in\logic{K}$. Тогда, согласно лемме~\ref{lem:KTB(1)K}, $B_{\neg\varphi^\ast} \to \tau_{\neg\varphi^\ast}(\neg\varphi^\ast)\in \logic{K}$, а значит, $B_{\neg\varphi^\ast} \to \tau_{\neg\varphi^\ast}(\neg\varphi^\ast)\in L$, т.к. $\logic{K}\subseteq L$.

            Пусть $\neg\varphi\not\in\logic{TQBF}$. Тогда, согласно предложению~\ref{prop:modal:complexity}, $\neg\varphi^\ast\not\in\logic{KTB}$. В~этом случае, согласно лемме~\ref{lem:KTB(1)}, $B_{\neg\varphi^\ast} \to \tau_{\neg\varphi^\ast}(\neg\varphi^\ast)\not\in \logic{KTB}$, а значит, $B_{\neg\varphi^\ast} \to \tau_{\neg\varphi^\ast}(\neg\varphi^\ast)\not\in L$, т.к. $L\subseteq \logic{KTB}$.

            Осталось заметить, что функция, сопоставляющая формуле $\varphi$ формулу $B_{\neg\varphi^\ast} \to \tau_{\neg\varphi^\ast}(\neg\varphi^\ast)$, является полиномиально вычислимой.
\end{proof}

    \subsection{Обсуждение конструкций}
    \label{ssec:modprop:emb:discuss}

Обратим внимание на то, что изменилось в доказательстве теоремы~\ref{th:PSPACE:KTB} по отношению к доказательству теорем \ref{th:PSPACE:K4} и~\ref{th:PSPACE:GL:Grz}.

Во-первых, в доказательстве теорем \ref{th:PSPACE:K4} и~\ref{th:PSPACE:GL:Grz} моделирование всех переменных с помощью константных формул или, соответственно, формул от одной переменной происходило только в специальных формулах, а в теореме~\ref{th:PSPACE:KTB}~--- в произвольных. Это дало возможность полиномиально погрузить логики $\logic{K}$ и $\logic{KTB}$ в их собственные фрагменты от одной переменной (леммы~\ref{lem:KTB(1)K} и~\ref{lem:KTB(1)}, соответственно). Отметим, что для произвольной логики $L\in [\logic{K},\logic{KTB}]$ мы не можем гарантировать существование такого погружения, и поэтому для доказательства $\cclass{PSPACE}$-трудности таких логик всё-таки приходится использовать формулы специального вида. Тем не менее, если бы мы интересовались только конкретными логиками, то формулы специального вида могли бы и не понадобиться; но зато важна полнота по Крипке рассматриваемой логики и её сабфреймовость.
Фактически важна ещё и транзитивность, точнее, отсутствие требования транзитивности (и даже более слабых её вариантов). Применив описанный подход к конкретным модальным логикам, несложно убедиться, что с помощью описанной техники логика $\logic{K}$ полиномиально погружается в свой константный фрагмент, а $\logic{T}$, $\logic{KB}$, $\logic{KTB}$~--- в свои фрагменты от одной переменной.

Во-вторых, отметим, что в ряде случаев представленные выше результаты о сложности логик можно усилить. Так, например, можно показать~\cite{Ladner77,Shapirovsky2018,Shapirovsky2022}, что многие стандартные модальные пропозициональные логики $\cclass{PSPACE}$-полны в полном языке, поэтому результаты о $\cclass{PSPACE}$-трудности их фрагментов дают $\cclass{PSPACE}$-полноту этих фрагментов. В~частности, получаем следующее утверждение.

\begin{corollary}
Константные фрагменты логик\/ $\logic{K}$, $\logic{K4}$, $\logic{wGrz}$, а также фрагменты от одной переменной логик $\logic{T}$, $\logic{KD}$, $\logic{KB}$, $\logic{KTB}$, $\logic{GL}$, $\logic{Grz}$ являются\/ $\cclass{PSPACE}$-полными.
\end{corollary}

Из этого наблюдения, в частности, следует, что $\cclass{PSPACE}$-полные логики, шкалы которых транзитивны, тоже полиномиально погружаются в свои $\cclass{PSPACE}$-полные фрагменты от конечного числа переменных, но как построить в явном виде такое погружение, скажем, для $\logic{K4}$, $\logic{S4}$, $\logic{wGrz}$, $\logic{Grz}$ или $\logic{GL}$, автору неясно: хотелось бы, чтобы это погружение сохраняло структуру исходной формулы, а описанный метод в случае этих логик работает не для всех модальных формул.

В-третьих, мы использовали приём, который будем называть
\label{page:relativization}
\defnotion{релятивизацией}:\index{релятивизация} фактически, формула $B_\varphi$ позволила нам рассматривать не все миры модели Крипке, а только те, где она оказалась истинной. Мы увидим, что этот приём весьма эффективен.

В-четвёртых, что пока было не очень важно, но может оказаться технически удобно в дальнейшем, вместо условия наследственности <<вверх>> мы можем рассматривать условие наследственности <<вниз>>: если переменная $q$ истинна в мире $y$ и при этом $y$ достижим из $x$, то $q$ истинна и в~$x$. В явном виде мы это условие не использовали, но нетрудно понять, что, заменяя переменные их отрицаниями, можно перейти как от условия наследственности <<вверх>> к условию наследственности <<вниз>>, так и обратно.


    \subsection{Логики линейных шкал}
    \label{ssec:modprop:emb:NP}

До этого мы имели дело с логиками, проблема разрешения которых $\cclass{PSPACE}$-трудна. Отметим, что семантически рассмотренные нами логики устроены так, что их корневые шкалы Крипке не имеют ограничений ни на длину цепей, ни на ширину ветвления (или размер антицепей). Что происходит с логиками, где эти условия не выполняются? Оказывается, сложность фрагментов от конечного числа переменных может быть разной; пока мы рассмотрим несколько простых примеров, но позже (раздел~\ref{ssec:poly:approx}) ещё вернёмся к этому вопросу.

Нетрудно понять, что если логика $L$ локально таблична, то любой её фрагмент от конечного числа переменных является полиномиально разрешимым. Это означает, в частности, что фрагменты от любого конечного числа переменных любого расширения логики $\logic{K5}$ полиномиально разрешимы. Покажем, что имеются $\cclass{coNP}$-полные логики, не являющиеся локально табличными, которые имеют $\cclass{coNP}$-полные фрагменты от конечного числа переменных. Речь идёт о таких логиках, как $\logic{GLLin}$ (в других обозначениях $\logic{GL.3}$), $\logic{S4.3}$, $\logic{Grz.3}$. Нетрудно показать~\cite[теоремы~9,~11]{MR:2003:AiML}, что
\begin{itemize}
\item
константный фрагмент логики $\logic{GLLin}$ полиномиально разрешим;
\item
фрагменты от одной переменной логик $\logic{S4.3}$ и $\logic{Grz.3}$ полиномиально разрешимы.
\end{itemize}
Тем не менее, если мы увеличим число переменных в каждом случае хотя бы на одну, то получим $\cclass{coNP}$-полные фрагменты.

\begin{theorem}
Проблема выполнимости фрагмента от одной переменной логики\/ $\logic{GLLin}$ является\/ $\cclass{NP}$-полной.
\end{theorem}

\begin{proof}
Для доказательства достаточно свести к проблеме $\logic{GLLin}$-выполнимости проблему выполнимости булевых формул. Этого можно достичь, заменяя в булевых формулах каждую переменную $p_k$ формулой $\Diamond(\Box^{k+1}\bot\wedge \Diamond^k\top\wedge p)$.
\end{proof}

\begin{theorem}
Проблема выполнимости фрагмента от двух переменных каждой из логик\/ $\logic{S4.3}$ и\/ $\logic{Grz.3}$ является\/ $\cclass{NP}$-полной.
\end{theorem}

\begin{proof}
Аналогично, только в булевых формулах каждую переменную $p_k$ нужно заменить формулой $\Diamond(\delta_k\wedge\neg\delta_{k+1}\wedge q)$, где $\delta_k$ и $\delta_{k+1}$ определяются так, как на стр.~\pageref{page:GL:delta}.
\end{proof}

  \section{Ненормальные и квазинормальные логики}
    \subsection{Понятия ненормальной и квазинормальной логики}

Мы считаем\footnote{Возможно другое определение, когда вместо замкнутости по правилам вывода говорят о том, какие правила вывода постулированы. Такой подход является более тонким, но для наших целей достаточно тех определений, которые приводятся.} модальную пропозициональную логику
\begin{itemize}
\item
  \defnotion{нормальной},\index{логика!модальная!нормальная} если она является расширением минимальной нормальной модальной логики $\logic{K}$ и замкнута по правилу подстановки, правилу \MP\ и правилу Гёделя;
\item
  \defnotion{квазинормальной},\index{логика!модальная!квазинормальная} если она является расширением логики $\logic{K}$ и замкнута по правилу подстановки и правилу~\MP;
\item
  \defnotion{ненормальной},\index{логика!модальная!ненормальная} если она не является расширением логики $\logic{K}$ и замкнута по правилу подстановки и правилу~\MP.
\end{itemize}

Чтобы определить интересующие нас логики, нам понадобится внести некоторые изменения в определение семантики Крипке.

    \subsection{Семантика}

Переопределим некоторые понятия семантики Крипке для языка~$\lang{ML}$.

Набор $\kframeIs{F}{W,N,D,R}$ называем шкалой Крипке, если $W$~--- непустое множество, $N\subseteq W$, $D\subseteq W$, $D\ne\varnothing$, $R\subseteq N\times W$; при этом, как и раньше элементы множества $W$ называем мирами, элементы множества $N$~--- \defnotion{нормальными}\index{мир!нормальный} мирами, элементы множества $W\setminus N$~--- \defnotion{взрывающимися}\index{мир!взрывающийся} мирами, элементы множества $D$~--- \defnotion{выделенными}\index{мир!выделенный} мирами, $R$~--- отношением достижимости в~$\kframe{F}$.


Набор $\kmodelIs{M}{\nameKFrame{F},v}$ называем моделью Крипке, если $\nameKFrame{F}$~--- шкала Крипке, а $v$~--- оценка пропозициональных переменных в этой шкале (как и раньше, $v(p)\subseteq W$ для $p\in\prop$).

Определим отношение истинности $\lang{ML}$-формул в мирах модели $\kmodelIs{M}{\nameKFrame{F},v}$, определённой на шкале $\kframeIs{F}{W,N,D,R}$; если $\varphi$ истинна в мире $w$ модели $\nameKModel{M}$, то, как и раньше, пишем $(\nameKModel{M},w)\models\varphi$; отношение определим рекурсивно:
            \[
            \begin{array}{clcl}
            \arrayitem &
            (\kmodel{M},w)\not\models\bot;\!\! &  &
            \arrayitemskip
            \\
            \arrayitem &
            (\kmodel{M},w)\models p_n
            & \leftrightharpoons &
            \parbox[t]{225pt}{$w\in v(p_n)$;}
            \arrayitemskip
            \\
            \arrayitem &
            (\kmodel{M},w)\models\varphi\wedge\psi & \leftrightharpoons &
            \parbox[t]{225pt}{$(\kmodel{M},w)\models\varphi$ \hphantom{л}и\hphantom{и}
            $(\kmodel{M},w)\models\psi$;}
            \arrayitemskip
            \\
            \arrayitem &
            (\kmodel{M},w)\models\varphi\vee\psi & \leftrightharpoons &
            \parbox[t]{225pt}{$(\kmodel{M},w)\models\varphi$ или
            $(\kmodel{M},w)\models\psi$;}
            \arrayitemskip
            \\
            \arrayitem &
            (\kmodel{M},w)\models\varphi\to\psi & \leftrightharpoons &
            \parbox[t]{225pt}{$(\kmodel{M},w)\not\models\varphi$ или
            $(\kmodel{M},w)\models\psi$;}
            \arrayitemskip
            \\
            \arrayitem &
            (\kmodel{M},w)\models\Box\varphi
            & \leftrightharpoons &
            \mbox{$w\in N$ и $(\kmodel{M},w')\models\varphi$ для всякого $w'\in R(w)$.}
            \end{array}
            \]
Формулу $\varphi$ считаем истинной в модели $\kmodel{M}$, если $(\nameKModel{M},w)\models\varphi$ для каждого $w\in D$; в этом случае пишем $\nameKModel{M}\models\varphi$.
Формулу $\varphi$ считаем истинной в шкале $\nameKFrame{F}$, если $\varphi$ истинна в любой модели, определённой на~$\nameKFrame{F}$; в этом случае пишем $\nameKFrame{F}\models\varphi$.
Формулу $\varphi$ считаем истинной в классе шкал $\sclass{C}$, если $\varphi$ истинна в каждой шкале из этого класса; в этом случае пишем $\sclass{C}\models\varphi$.

Отметим, что обычную шкалу Крипке $\kframe{F}=\otuple{W,R}$ можно понимать как шкалу $\otuple{W,W,W,R}$ в смысле нового определения, т.к. в этом случае множество формул, истинных в $\kframe{F}$, совпадает со множеством формул, истинных~в~$\otuple{W,W,W,R}$.

Если $\kframeIs{F}{W,N,D,R}$ и $D=\set{w_0}$, то пишут также $\kframeIs{F}{W,N,w_0,R}$. Учитывая определение истинности формул в шкалах, нетрудно понять, что достаточно рассматривать только корневые шкалы вида $\kframeIs{F}{W,N,w_0,R}$, где $w_0$~--- корень. 

    \subsection{Логики}

Ниже приведены обозначения и определения некоторых модальных логик, рассматриваемых в литературе:
\begin{itemize}
\item
$\logic{E3}$~--- множество формул, истинных в шкалах, где отношение достижимости рефлексивно на множестве нормальных миров и транзитивно на множестве всех миров;
\item
$\logic{S2}$~--- множество формул, истинных в шкалах, где отношение достижимости рефлексивно на множестве нормальных миров и выделенные миры являются нормальными;
\item
$\logic{S3}$~--- множество формул, истинных в шкалах, где отношение достижимости рефлексивно на множестве нормальных миров и транзитивно на множестве всех миров, а выделенные миры являются нормальными;
\item
$\logic{S6}$~--- множество формул, истинных в шкалах, где отношение достижимости рефлексивно на множестве нормальных миров и для каждого нормального мира существует достижимый из него взрывающийся мир, а выделенные миры являются нормальными.
\end{itemize}

Для технического удобства введём в рассмотрение логику $\logic{K0}$, которую определим как множество формул, истинных в классе всех шкал Крипке (в смысле нового определения).\footnote{Автор не знаком с работами, где бы эта логика рассматривалась; предложенное обозначение взято лишь по аналогии с логикой~$\logic{K}$, и возможно, является не самым удачным.}

    \subsection{Перенос конструкции}

Результаты, полученные выше для нормальных модальных логик, автоматически переносятся на классы ненормальных и квазинормальных модальных логик. Действительно, в~случае квазинормальных модальных логик достаточно заметить, что теоремы~\ref{th:PSPACE:K4}, \ref{th:PSPACE:GL:Grz} и~\ref{th:PSPACE:KTB} справедливы не только для логик, попавших в указанные в их формулировках интервалы, но и для любых множеств формул, заключённых между логиками, являющимися границами этих интервалов. Что касается ненормальных логик, то достаточно заметить, что формула $\varphi^\ast$ (стр.~\pageref{varphi:ast}) содержит конъюнктивные члены вида $\Box^{n}\psi$, гарантирующие, что интересующие нас миры являются нормальными, т.е. доказательство в случае большого класса ненормальных логик также не изменится.


В итоге получаем следующие теоремы.

\begin{theorem}
\label{th:PSPACE:K0const}
Проблема разрешения константного фрагмента логики $L\in [\logic{K0},\logic{wGrz}]$ является\/ $\cclass{PSPACE}$\nobreakdash-трудной.
\end{theorem} 

\begin{theorem}
\label{th:PSPACE:K0:GL:Grz}
Проблема разрешения фрагмента от одной переменной логики $L\in [\logic{K0},\logic{GL}]\cup[\logic{K0},\logic{Grz}]\cup[\logic{K0},\logic{KTB}]$ является\/ $\cclass{PSPACE}$\nobreakdash-трудной.
\end{theorem}

Что касается таких логик, как, например, $\logic{S6}$, то достаточно применить релятивизацию, взяв в качестве ограничивающей формулы~$\Box\top$, т.е. заменив в используемых формулах (в~$\varphi^\ast$ и формулах, получающихся из~$\varphi^\ast$ путём описанных выше преобразований) подформулы вида $\Box\psi$ на $\Box(\Box\top\to\psi')$, где $\psi'$ получается из $\psi$ аналогичной заменой.

\begin{theorem}
\label{th:PSPACE:K0:S6}
Проблема разрешения фрагмента от одной переменной логики $L\in [\logic{K0},\logic{S6}]$ является\/ $\cclass{PSPACE}$\nobreakdash-трудной.
\end{theorem}

    \subsection{Некоторые замечания}

Описанная здесь техника моделирования переменных в модальных формулах переносится на другие языки и логики. Пока основная цель состояла в том, чтобы, с одной стороны, изложить эту технику в достаточно простом виде, а с другой стороны, показать возможности её модификаций. Так, несложно понять, что эту технику можно перенести не только на шкалы с выделенными или взрывающимися мирами, но и на обобщённые шкалы, захватив тем самым в рассмотрение, например, логику Соловея~$\logic{S}=\logic{GL}+\mref$ (см.~\cite[с.\,95]{ChZ}), для адекватного описания которой обычных шкал Крипке недостаточно, а также на системы с иными дополнительными требованиями к семантике Крипке, к каким относится, например, логика $\logic{S1}$~\cite{Cresswell-1972}; мы опускаем здесь описание каких бы то ни было деталей такого переноса, поскольку они, во-первых, не содержат нетривиальных решений, а во-вторых, не понадобятся для дальнейшего изложения.

Ниже будут описаны сходные по духу конструкции, позволяющие провести аналогичное моделирование в полимодальном и интуиционистском случаях, а также будет предложена их модификация для предикатных неклассических логик, позволяющая доказывать неразрешимость (и~даже высокую степень неразрешимости) в языках с сильными ограничениями, накладываемыми на используемые средства (прежде всего, число и арность предикатных букв, а также число предметных переменных). Но прежде чем мы перейдём к описанию соответствующих построений, скажем о некоторых попутных наблюдениях, связанных с семантикой.

  \section{Сложность аппроксимации}
    \subsection{Функция сложности}

Пусть $L$~--- полная по Крипке финитно аппроксимируемая модальная пропозициональная логика.
\defnotion{Функция сложности}\index{уяа@функция!сложности} $\function{f_L}{\numN}{\numN}$ для логики $L$ определяется следующим образом (см.~\cite[раздел~18.1]{ChZ}):
$$
\begin{array}{lcl}
f_L(n)
  & =
  & \displaystyle
    \max\big\{\min\{|\kframe{F}| : \mbox{$\kframe{F}\models L$, $\kframe{F}\not\models\varphi$}\}
         : \mbox{$|\varphi| \leqslant n$, $\varphi \not\in L$}\big\},
%
%
\end{array}
$$
где $|\kframe{F}|$~--- число миров в шкале $\kframe{F}$, а $|\varphi|$~---
длина формулы $\varphi$.

Отметим, что при определении функции сложности вместо логики в полном языке можно рассматривать её фрагмент, в частности, фрагмент в языке с конечным числом переменных; в этом случае можно говорить о функции сложности соответствующего фрагмента.

Мы оценивали сложность логик и их фрагментов, выясняя принадлежность проблемы разрешения (или выполнимости) тем или иным классам сложности. Из определения функции сложности видно, что она тоже некоторым образом описывает сложность проблемы разрешения логики~$L$. Более того, наблюдается связь между сложностью проблемы разрешения логики в смысле полноты этой проблемы в некотором классе сложности и сложностью проблемы разрешения той же логики в смысле функции сложности: так, функция сложности для многих логик с $\cclass{NP}$-полной проблемой выполнимости (например, $\logic{Cl}$, $\logic{S5}$, $\logic{GL.3}$, $\logic{S4.3}$, $\logic{Grz.3}$) ограничена некоторым полиномом, а функция сложности для многих логик с $\cclass{PSPACE}$-полной проблемой разрешения экспоненциальна~\cite[глава~18]{ChZ}.


Функция сложности интересна тем, что будь она полиномиальной для таких $\cclass{PSPACE}$-полных логик как $\logic{K}$, $\logic{T}$, $\logic{K4}$, $\logic{S4}$, $\logic{GL}$ и др., мы сразу получили бы равенство $\cclass{NP}=\cclass{PSPACE}$, которое представляется сомнительным. То же самое относится и к $\cclass{PSPACE}$-полным фрагментам таких логик, рассмотренным выше. Поэтому стоит ожидать, что функция сложности для этих фрагментов ограничена снизу экспонентой, т.е. функцией из класса $\varTheta(2^{n^c})$ для некоторого~$c\in\numR^+$.

Описанная выше техника действительно позволяет получить экспоненциальную нижнюю оценку функции сложности для рассмотренных выше $\cclass{PSPACE}$-трудных константных фрагментов или фрагментов от одной переменной.

\begin{theorem}
\label{th:complexity:function:K}
Функции сложности для константных фрагментов логик из интервала\/ $[\logic{K},\logic{wGrz}]$, а также для фрагментов от одной переменной логик из интервалов\/ $[\logic{K},\logic{GL}]$, $[\logic{K},\logic{Grz}]$, $[\logic{K},\logic{KTB}]$ ограничены снизу экспонентой.
\end{theorem}

\begin{proof}
Достаточно взять в описанной выше конструкции $\varphi=\forall p_1\ldots\forall p_n\,\top$. При увеличении $n$ длина модальных формул, которые строятся по $\varphi$, т.е. формул $\varphi^\ast$, $\varphi^\ast_\alpha$, $\varphi^\ast_{\alpha'}$ и $B_{\varphi^\ast}\wedge\tau_{\varphi^\ast}(\varphi^\ast)$, растёт полиномиально, а наименьший размер шкал Крипке, в которых истинны эти формулы, растёт экспоненциально.
\end{proof}

\begin{remark}
\label{remark:th:complexity:function:K}
В формулировке теоремы можно заменить\/ $\logic{K}$ на\/~$\logic{K0}$.
\end{remark}

    \subsection{Полиномиально аппроксимируемые логики}
    \label{ssec:poly:approx}
\subsubsection{Конструкция А.\,В.~Кузнецова}

Интерес к функции сложности во многом был мотивирован предположением, что её характер связан со сложностью разрешающих алгоритмов для логики. Так, А.В.\,Кузнецов предполагал, что функция сложности для интуиционистской логики полиномиальна, и если бы это действительно было так, то можно было бы построить полиномиальное погружение интуиционистской пропозициональной логики в классическую~\cite[проблемы~8 и~9]{Kuznetsov-1979-1}; то же относится и к модальным логикам. Почти в то же время было показано~\cite{ZakharyaschevPopov-1980-1-rus}, что функция сложности для интуиционистской логики экспоненциальна~(см. также~\cite[теорема~18.5]{ChZ}), но конструкция А.\,В.\,Кузнецова, тем не менее, работает для многих полиномиально аппроксимируемых логик (т.е. функция сложности которых ограничена сверху некоторым полиномом). Покажем это на примере логики~$\logic{S5}$.

Заметим, что функция сложности логики $\logic{S5}$ линейна. Действительно, если формула $\varphi$ не принадлежит $\logic{S5}$, то она опровергается в некотором мире $w$ шкалы $\kframe{F} = \otuple{W,R}$, где $R$~--- отношение эквивалентности. Для каждой подформулы $\psi$ формулы $\varphi$ выберем в $W$ некоторый мир, достижимый из $w$, в котором опровергается~$\psi$ (если он есть); для самой формулы $\varphi$ в качестве такого мира возьмём~$w$. Из выбранных миров построим модель, сохранив оценку и отношение достижимости на них. Тогда несложно понять, что в новой модели в $w$ будет опровергаться~$\varphi$, а размер этой модели не превосходит~$|\varphi|$.

Пусть $\varphi$~--- модальная формула и $m = f_{\logic{S5}}(|\varphi|)$; будем считать, что миры модели, определённой на $\logic{S5}$-шкале размера $m$, всегда занумерованы числами от $1$ до~$m$.

Для каждой подформулы $\psi$ формулы $\varphi$ и каждого
$i \in \set{1, \ldots, m}$ введём в рассмотрение переменную $q^{\psi}_i$, содержательный смысл которой <<формула~$\psi$ истинна в мире $i$ (при некоторой оценке)>>, а также для каждых $i, j \in \set{1, \ldots, m}$~--- переменную $r_{ij}$, содержательный смысл которой <<мир~$j$ достижим из мира~$i$>>.

Следующая формула выражает условия истинности подформул формулы $\varphi$ (используем обозначение $\mathop{\mathit{sub}} \varphi$ для множества подформул формулы $\varphi$) в моделях, определённых на $\logic{S5}$-шкалах с $m$ мирами:
$$
\begin{array}{lcl}

  \tau^\varphi & = &   \displaystyle \bigwedge\limits_{\mathclap{i=1}}^m
                 \bigg(
                   (q^\bot_i  \leftrightarrow  \bot)
                   \wedge

  \bigwedge\limits_{\mathclap{\psi\to\chi \in \mathop{\mathit{sub}} \varphi}}
                   \big(
                   q^{\psi\to\chi}_i  \leftrightarrow  (q^{\psi}_i\to q^{\chi}_i)
                   \big)

             \wedge  \bigwedge\limits_{\mathclap{\Box \psi \in \mathop{\mathit{sub}} \varphi}}
                   \big(
                   q^{\Box\psi}_i  \leftrightarrow  \displaystyle   \bigwedge\limits_{\mathclap{j=1}}^m  (r_{ij}\to q^{\psi}_j)
                   \big)
                 \bigg).
  \smallskip\\
\end{array}
$$

Опишем свойства отношения достижимости в $\logic{S5}$-шкалах (рефлексивность, симметричность и транзитивность):
$$
\begin{array}{lcl}
  \rho^\varphi & = & \displaystyle
                 \bigwedge\limits_{\mathclap{i=1}}^m r_{ii} \wedge
                 \bigwedge\limits_{\mathclap{i,j=1}}^m  (r_{ij}^{\phantom{i}}\to r_{ji}) \wedge
                 \bigwedge\limits_{\mathclap{i,j,k=1}}^m (r_{ij}\wedge r_{jk}\to r_{ik}).
\end{array}
$$

Наконец, определим формулу, утверждающую, что $\varphi$ истинна в каждом мире модели:
$$
\begin{array}{lcl}
  \theta^\varphi_{\phantom{i}} & = &  q^\varphi_1 \wedge \ldots \wedge q^\varphi_m.
\end{array}
$$

Несложно понять, что в этом случае
$$
\begin{array}{lcl}
\varphi \in \logic{S5}
  & \iff
  & \tau^\varphi \wedge \rho^\varphi \imp \theta^\varphi \in \logic{Cl}.
\end{array}
$$
Осталось заметить, что формула $\tau^\varphi \wedge \rho^\varphi \imp \theta^\varphi$ вычислима за полиномиальное время от $|\varphi|$, а значит, логика $\logic{S5}$ полиномиально погружается в логику~$\logic{Cl}$.

Нетрудно понять, что аналогичный трюк можно использовать, чтобы погрузить в $\logic{Cl}$ логики $\logic{K}$, $\logic{T}$, $\logic{KB}$, $\logic{KTB}$, $\logic{K4}$, $\logic{S4}$, $\logic{GL}$, $\logic{Grz}$, $\logic{wGrz}$ и др., но, поскольку функция сложности для каждой из них экспоненциальна, число $m$, возникающее в аналогах формул $\tau^\varphi$, $\rho^\varphi$ и $\theta^\varphi$, является экспонентой от~$|\varphi|$.

Описанные построения возможны, если у нас имеется не только ясное ограничение на размер опровергающих моделей (т.е. ясное ограничение функции сложности), но и возможность описать свойства отношения достижимости в интересующих нас моделях. Если же с описанием таких свойств имеются сложности, то построение таких погружений оказывается под сомнением.

Контрпримеры известны: так, Э.\,Спаан\footnote{Теперь Э.\,Хемаспаандра.}~\cite[теорема~2.1.1]{Spaan-1993-1}, развивая идеи А.\,Уркварта\footnote{А.\,Уркварт~\cite{Urquart81} доказал, используя мощностную аргументацию, существование неразрешимых финитно аппроксимируемых логик. Существование таких логик следует также из того, что К.\,Файн~\cite{Fine74} построил континуум модальных логик высоты три в расширениях~$\logic{K4}$, поскольку расширения~$\logic{K4}$ конечной высоты локально табличны, а следовательно, финитно аппроксимируемы.}, показала (с помощью мощностной аргументации), что полиномиально аппроксимируемые логики могут быть неразрешимы; более точно, она обосновала существование нормальных модальных расширений логики $\logic{K}$, которые линейно аппроксимируемы шкалами Крипке и при этом имеют неразрешимые фрагменты от одной переменной.

Используя формулы, сходные с теми, с помощью которых мы моделировали переменные, мы можем усилить этот результат сразу в нескольких направлениях. Именно, мы покажем, что в расширениях логик $\logic{KTB}$, $\logic{GL}$, $\logic{Grz}$ для каждой степени неразрешимости $C$ в ней существуют линейно аппроксимируемые логики, причём в случае расширений $\logic{KTB}$, $\logic{GL}$ и $\logic{Grz}$ уже их фрагменты от одной переменной находятся в~$C$, а в случае расширений $\logic{wGrz}$ уже их константные фрагменты находятся в~$C$.

\subsubsection{Неразрешимые расширения логики $\logic{K}$}

\providecommand{\vp}{\varphi}
\providecommand{\notin}{\not\in}
\providecommand{\nat}{\numN}

Мы начнём с класса расширений логики $\logic{K}$, поскольку в этом случае все построения будут очень простыми, и уже потом покажем модификации, позволяющие получить аналогичные результаты в классах расширений других логик.

Интересующие нас логики будем определять семантически.
Пусть $\numN^{++} = \numN\setminus\set{0,1}$ и пусть
для каждого $n \in \numN^{++}$
\settowidth{\templength}{\mbox{$W_n$}}
\begin{itemize}
\item $\parbox{\templength}{$W_n$}=\{w_0,\ldots,w_n, w^\ast\}$;
\item
  $\parbox{\templength}{$R_n$}=\{\langle w_k,w_{k+1} \rangle : 0\leqslant k < n\} \cup
  \{\langle w_0, w^\ast \rangle \}$;
\item $\parbox{\templength}{$\kframe{F}_n$} = \langle W_n, R_n \rangle$,
\end{itemize}
см.~рис.~\ref{fig}.

\begin{figure}
\centering
\begin{tikzpicture}[scale=1.5]

\coordinate (w1)   at (+0.0000,+0.0000);
\coordinate (w2)   at (+1.0000,+0.0000);
\coordinate (w3)   at (+2.0000,+0.0000);
\coordinate (w4ph) at (+3.0000,+0.0000);
\coordinate (dts)  at (+3.5000,+0.0000);
\coordinate (w5ph) at (+4.0000,+0.0000);
\coordinate (wn-1) at (+5.0000,+0.0000);
\coordinate (wn)   at (+6.0000,+0.0000);
\coordinate (w*)   at (+0.0000,+1.0000);

\begin{scope}[>=latex]
\draw [->,  shorten >= 1.5pt, shorten <= 1.5pt]
(w1) -- (w2);
\draw [->,  shorten >= 1.5pt, shorten <= 1.5pt]
(w2) -- (w3);
\draw [->,  shorten >= 1.5pt, shorten <= 1.5pt]
(w3) -- (w4ph);
\draw [->,  shorten >= 1.5pt, shorten <= 1.5pt]
(w5ph) -- (wn-1);
\draw [->,  shorten >= 1.5pt, shorten <= 1.5pt]
(wn-1) -- (wn);
\draw [->,  shorten >= 1.5pt, shorten <= 1.5pt]
(w1) -- (w*);
\end{scope}

\node [below      ] at (w1)   {${w_0}$}     ;
\node [below      ] at (w2)   {${w_1}$}     ;
\node [below      ] at (w3)   {${w_2}$}     ;
\node [           ] at (dts)  {${\cdots}$}  ;
\node [above      ] at (w*)   {${w^\ast}$}  ;
\node [below      ] at (wn-1) {${w_{n-1}}$} ;
\node [below      ] at (wn)   {${w_n}$}     ;


\filldraw [] (w1)   circle [radius=1.5pt]   ;
\filldraw [] (w2)   circle [radius=1.5pt]   ;
\filldraw [] (w3)   circle [radius=1.5pt]   ;
\filldraw [] (w*)   circle [radius=1.5pt]   ;
\filldraw [] (wn-1) circle [radius=1.5pt]   ;
\filldraw [] (wn)   circle [radius=1.5pt]   ;

\end{tikzpicture}

\caption{Шкала $\kframe{F}_n$}
\label{fig}
\end{figure}

Пусть также для каждого $n \in \numN^{++}$
$$
\begin{array}{lcl}
\alpha_n & = & \Diamond \Box\bot \con \Diamond^n\Box\bot.
\end{array}
$$

\begin{lemma}
\label{lem:alphas}
Для любых $n,k \in \numN^{++}$ имеет место следующая эквивалентность:
$$
\begin{array}{lcl}
  (\kframe{F}_n, x) \models \alpha_k & \iff & \mbox{$k = n$ и $x = w_0$}.
\end{array}
$$
\end{lemma}

\begin{proof}
  Формула $\alpha_k$ утверждает, что последний иррефлексивный мир виден за один шаг и за $k$ шагов; это возможно в точности при описанных в формулировке условиях.
\end{proof}

Для подмножества $A$ множества $\numN^{++}$ положим
$\tilde{A} = \numN^{++}\setminus\set{2n : n\in A}$,
$\mathscr{C}_{A} = \{\kframe{F}_n : n\in \tilde{A}\}$ и
$\mathbf{L}_{A} = \mlogic{\mathscr{C}_A}$.

\begin{lemma}
  \label{lem:alphas-I}
  Для каждого $n \in \numN^{++}$ имеют место следующие эквивалентности:
  $$
  \begin{array}{lclcl}
  \neg \alpha_{2n} \in \mathbf{L}_{A}
    & \iff
    & \kframe{F}_{2n} \notin \mathscr{C}_A
    & \iff
    & n \in A.
  \end{array}
  $$
\end{lemma}

\begin{proof}
  Следует непосредственно из леммы~\ref{lem:alphas}, а также определений класса $\mathscr{C}_A$ и логики $\mathbf{L}_{A}$.
\end{proof}

\begin{lemma}
  \label{lem:lfp}
  Для любого подмножества $A$ множества $\numN^{++}$ логика $\mathbf{L}_{A}$ является линейно аппроксимируемой.
\end{lemma}

\begin{proof}
Покажем, что если $\vp\notin\mathbf{L}_{A}$ то $\vp$ опровергается в некоторой $\mathbf{L}_{A}$-шкале размера не более чем $2 \mdepth \varphi + 5$.

Пусть $\vp \notin \mathbf{L}_{A}$.  Тогда $(\kframe{F}_n, x) \not\models^v \vp$ для некоторой шкалы $\kframe{F}_n$ из класса $\mathscr{C}_A$, мира $x$ из $\kframe{F}_n$ и некоторой оценки $v$ в~$\kframe{F}_n$. Рассмотрим три возможных случая.

Случай, когда $x = w_0$.

Если $n \leqslant \mdepth  \vp$, то $\vp$ опровергается в шкале из $\mathscr{C}_A$ размера не более $\mdepth  \varphi + 2$.  Пусть $n > \mdepth \vp$.  Рассмотрим оценку $v'$ на $\kframe{F}_{2\mdepth \vp+3}$, такую, что для каждой переменной $p$ и каждого $w\in\{w^\ast,w_0,\ldots,w_{\mdepth \vp}\}$,
  $$
  \begin{array}{lcl}
  (\kframe{F}_{2\mdepth \vp+3},w)\models^{v'}p
    & \iff
    & (\kframe{F}_{n},w)\models^{v}p.
  \end{array}
  $$

Поскольку истинность формулы $\psi$ в мире $w$ не зависит от миров, достижимых из $w$ за более чем   $\mdepth \psi$ шагов, получаем, что $(\kframe{F}_{2\mdepth \vp+3},w_0)\not\models^{v'}\vp$.
Поскольку число $2\mdepth \vp+3$ нечётно, оно находится в $\tilde{A}$; значит, $\kframe{F}_{2\mdepth \vp+3} \models \mathbf{L}_{A}$.

Случай, когда $x = w_k$ для некоторого $k \in \{1, \ldots, n\}$.

Пусть $n-k\leqslant\mdepth \vp$. Рассмотрим оценку $v'$ на $\kframe{F}_{2\mdepth \vp+3}$ такую, что для каждой переменной $p$ и каждого $i\in\set{0,\ldots,n-k}$
  $$
  \begin{array}{lcl}
  (\kframe{F}_{2\mdepth \vp+3},w_{2\mdepth \vp+3-n+k+i})\models^{v'}p
    & \iff
    & (\kframe{F}_{n},w_{k+i})\models^{v}p.
  \end{array}
  $$
Тогда $(\kframe{F}_{2\mdepth \vp+3},w_{2 \mdepth \vp+3-n+k})\not\models^{v'}\vp$, и мы получаем требуемое.

Пусть $n-k > \mdepth \vp$. Рассмотрим оценку $v'$ на $\kframe{F}_{2\mdepth \vp+3}$, такую, что для каждой переменной $p$ и каждого $i\in\{0,\ldots,\mdepth \vp\}$,
  $$
  \begin{array}{lcl}
  (\kframe{F}_{2\mdepth \vp+3},w_{1+i})\models^{v'}p
    & \iff
    & (\kframe{F}_{n},w_{k+i})\models^{v}p.
  \end{array}
  $$
Тогда $(\kframe{F}_{2\mdepth \vp+3},w_{1})\not\models^{v'}\vp$, и мы снова получаем требуемое.

Случай, когда $x = w^\ast$.

В этом случае достаточно заметить, что $(\kframe{F}_{3}, w^\ast) \not\models \vp$, т.к. $w^\ast$ в обеих шкалах является последним иррефлексивным миром.
\end{proof}


\begin{theorem}
  \label{thr:ext-K}
Для любой степени неразрешимости $C$ существует линейно аппроксимируемое нормальное расширение логики\/ $\logic{K}$, находящееся в $C$, константный фрагмент которого тоже находится в~$C$.
\end{theorem}

\begin{proof}
Пусть $A$~--- подмножество множества $\numN^{++}$, находящееся ~$C$.  Покажем, что $\mathbf{L}_A$~--- требуемая логика.

По лемме~\ref{lem:lfp}, логика $\mathbf{L}_A$ является линейно аппроксимируемой. Поскольку каждая формула $\alpha_n$ не содержит переменных, константный фрагмент логики $\mathbf{L}_A$ находится в~$C$, согласно лемме~\ref{lem:alphas-I}.

Покажем, что $\mathbf{L}_A$ находится в $C$. Заметим, что существует алгоритм, строящий по формуле $\vp$ множество
  $$
  \begin{array}{lcl}
    X
    & =
    & \{ n \in \nat : \mbox{$2 \leqslant n \leqslant 2\mdepth
      \varphi + 3$, $\kframe{F}_n \not\models \vp$} \}.
  \end{array}
  $$
Действительно, достаточно перебрать последовательно шкалы с номерами от $2$ до $2\mdepth \varphi + 3$ и оценки переменных формулы $\vp$ в каждой из них.

Теперь, чтобы определить, верно ли, что $\vp \notin \mathbf{L}_A$, нужно лишь выяснить, содержит ли $X$ нечётное число и существует ли такое $n \in \nat$, что $2n \in X$ и $n \notin A$. Следовательно, проблема непринадлежности формул множеству $\mathbf{L}_A$ находится в дополнении к~$C$. Значит, проблема принадлежности формул множеству $\mathbf{L}_A$ находится~в~$C$.
\end{proof}

\subsubsection{Неразрешимые расширения логики $\logic{KTB}$}

Усилим описанную выше конструкцию, распространив её на класс нормальных расширений логики~$\mathbf{KTB}$.

\begin{figure}
\centering
\begin{tikzpicture}[scale=1.5]

\coordinate (w1bar)at (+0.0000,+0.0000);
\coordinate (w1)   at (+1.0000,+0.0000);
\coordinate (w2bar)at (+2.0000,+0.0000);
\coordinate (w2)   at (+3.0000,+0.0000);
\coordinate (wnbar)at (+4.0000,+0.0000);
\coordinate (wn)   at (+5.0000,+0.0000);
\coordinate (dts)  at (+3.5000,+0.0000);
\coordinate (a)    at (-1.0000,+0.0000);
\coordinate (abar) at (+6.0000,+0.0000);
\coordinate (b1)   at (-1.8500,+0.5000);
\coordinate (b2)   at (-1.0000,+1.0000);
\coordinate (b1bar)at (-1.8500,-0.5000);
\coordinate (b2bar)at (-1.0000,-1.0000);
\coordinate (c1)   at (+6.8500,+0.5000);
\coordinate (c2)   at (+6.0000,+1.0000);
\coordinate (c1bar)at (+6.8500,-0.5000);
\coordinate (c2bar)at (+6.0000,-1.0000);

\begin{scope}[>=latex]
\draw [<->,  shorten >= 2pt, shorten <= 2pt]
(w1bar) -- (w1);
\draw [<->,  shorten >= 2pt, shorten <= 2pt]
(w1) -- (w2bar);
\draw [<->,  shorten >= 2pt, shorten <= 2pt]
(w2bar) -- (w2);
\draw [<->,  shorten >= 2pt, shorten <= 2pt]
(wnbar) -- (wn);
\draw [<->,  shorten >= 2pt, shorten <= 2pt]
(a) -- (w1bar);
\draw [<->,  shorten >= 2pt, shorten <= 2pt]
(wn) -- (abar);
\draw [<->,  shorten >= 2pt, shorten <= 2pt]
(a) -- (b1);
\draw [<->,  shorten >= 2pt, shorten <= 2pt]
(b1) -- (b2);
\draw [<->,  shorten >= 2pt, shorten <= 2pt]
(a) -- (b1bar);
\draw [<->,  shorten >= 2pt, shorten <= 2pt]
(b1bar) -- (b2bar);
\draw [<->,  shorten >= 2pt, shorten <= 2pt]
(b1) -- (b1bar);
\draw [<->,  shorten >= 2pt, shorten <= 2pt]
(abar) -- (c1);
\draw [<->,  shorten >= 2pt, shorten <= 2pt]
(c1) -- (c2);
\draw [<->,  shorten >= 2pt, shorten <= 2pt]
(abar) -- (c1bar);
\draw [<->,  shorten >= 2pt, shorten <= 2pt]
(c1bar) -- (c2bar);
\draw [<->,  shorten >= 2pt, shorten <= 2pt]
(c1) -- (c1bar);

\end{scope}

\node [above      ] at (a)       {${a}$}            ;
\node [above      ] at (w1)      {${w_0}$}          ;
\node [above      ] at (w2)      {${w_1}$}          ;
\node [above      ] at (wn)      {${w_n}$}          ;
\node [above      ] at (abar)    {${\bar a}$}       ;
\node [above      ] at (w1bar)   {${\bar w_0}$}     ;
\node [above      ] at (w2bar)   {${\bar w_1}$}     ;
\node [above      ] at (wnbar)   {${\bar w_n}$}     ;
\node [           ] at (dts)     {${\cdots}$}       ;
\node [      left ] at (b1)      {${b_1}$}          ;
\node [      right] at (b2)      {${b_2}$}          ;
\node [      left ] at (b1bar)   {${\bar b_1}$}     ;
\node [      right] at (b2bar)   {${\bar b_2}$}     ;
\node [      right] at (c1)      {${c_1}$}          ;
\node [      right] at (c1bar)   {${\bar c_1}$}     ;
\node [      left ] at (c2)      {${c_2}$}          ;
\node [      left ] at (c2bar)   {${\bar c_2}$}     ;


\draw [] (w1)    circle [radius=1.5pt]   ;
\draw [] (w2)    circle [radius=1.5pt]   ;
\draw [] (wn)    circle [radius=1.5pt]   ;
\draw [] (w1bar) circle [radius=1.5pt]   ;
\draw [] (w2bar) circle [radius=1.5pt]   ;
\draw [] (wnbar) circle [radius=1.5pt]   ;
\draw [] (a)     circle [radius=1.5pt]   ;
\draw [] (b1)    circle [radius=1.5pt]   ;
\draw [] (b2)    circle [radius=1.5pt]   ;
\draw [] (c1)    circle [radius=1.5pt]   ;
\draw [] (c2)    circle [radius=1.5pt]   ;
\draw [] (abar)  circle [radius=1.5pt]   ;
\draw [] (b1bar) circle [radius=1.5pt]   ;
\draw [] (b2bar) circle [radius=1.5pt]   ;
\draw [] (c1bar) circle [radius=1.5pt]   ;
\draw [] (c2bar) circle [radius=1.5pt]   ;

\end{tikzpicture}
\caption{Шкала $\kframe{F}^{rs}_n$}
\label{fig-F-rs}
\end{figure}

Пусть для каждого $n \in \numNp$
\settowidth{\templength}{$W^{rs}_n$}
\begin{itemize}
\item
  $\parbox{\templength}{$W^{rs}_n$} = \{ w_0, \ldots, w_n \} \cup \{
  \bar{w}_0, \ldots, \bar{w}_n \} \cup \{ a, \bar{a},
  b_1, \bar{b}_1, b_2, \bar{b}_2, c_1, \bar{c}_1, c_2,
  \bar{c}_2\}$;
\item $R^{rs}_n$~--- рефлексивно-симметричное замыкание отношения, являющегося объединением следующих множеств:
$$
\begin{array}{l}
  \{ \langle \bar{w}_k, w_k \rangle : 0 \leqslant k \leqslant n
  \},
  \\
  \{ \langle w_k, \bar{w}_{k+1} \rangle : 0 \leqslant k < n
  \},
  \\
  \{ \langle a, \bar{w}_0 \rangle, \langle \bar{a},
  {w}_n \rangle \},
  \\
  \{ \langle a, {b}_1 \rangle, \langle a,
  \bar{b}_1 \rangle, \langle {b}_1, \bar{b}_1
  \rangle, \langle b_1, {b}_2 \rangle, \langle
  \bar{b}_1, \bar{b}_2 \rangle \},
  \\
  \{ \langle \bar{a}, {c}_1 \rangle, \langle
  \bar{a}, \bar{c}_1 \rangle, \langle {c}_1,
  \bar{c}_1 \rangle, \langle c_1, {c}_2 \rangle,
  \langle \bar{c}_1, \bar{c}_2 \rangle \};
\end{array}
$$
\item $\parbox{\templength}{$\kframe{F}^{rs}_n$} = \langle W^{rs}_n, R^{rs}_n \rangle$,
\end{itemize}
см. рис.~\ref{fig-F-rs}.

До конца этого раздела будем использовать модальность $\Diamond^{\simeq 2}$, которую определяем как следующее сокращение: $\Diamond^{\simeq 2}\psi = \Diamond^2\psi\wedge \neg\Diamond\psi$.

Определим рекурсивно последовательность формул
$$
\begin{array}{lcl}
  \zeta_0 & = & \neg p \con \Diamond^{\simeq 2} \Box p \con \Diamond^{\simeq 2} \Box \neg p; \\
  \zeta_{k+1} & = & \neg p \con \Diamond (p \con \Diamond \zeta_k).
\end{array}
$$

Для каждого $n \in \numNp$ положим
$$
\begin{array}{lcl}
  \beta_n
  & =
  & \displaystyle
    p \con \Diamond^{\simeq 2} \Box p \con \Diamond^{\simeq 2} \Box \neg p
    \con \Diamond \zeta_n \con \bigwedge\limits_{\mathclap{k = 0}}^{\mathclap{n-1}} \neg \Diamond
    \zeta_k.
\end{array}
$$

\begin{lemma}
  \label{lem:betas}
  Для любых $n,k \in\numNp$ имеет место следующая эквивалентность:
  $$
  \begin{array}{lcl}
    (\kframe{F}^{rs}_n, x) \not\models \neg \beta_k
      & \iff
      & \mbox{$k = n$ и $x \in \{a, \bar{a} \}$.}
  \end{array}
  $$
\end{lemma}

\begin{proof}
Для доказательства достаточно заметить, что имеется только два способа сделать формулу $\beta_k$ истинной в мирах шкалы $\kframe{F}^{rs}_n$ для некоторого $n \geqslant 1$. Оба требуют взять шкалу $\kframe{F}^{rs}_k$ и такую оценку $v$, что
  $$
  \begin{array}{lclclcl}
    x \in v(p) &\iff& \bar{x} \notin v(p); && b_1 \in v(p)
    &\iff&
           b_2 \in v(p); \\
    a \in v(p) &\iff& w_0 \in v(p); && c_1 \in v(p)
    &\iff& c_2 \in v(p); \\
  \end{array}
  $$
при этом формула $\beta_k$ может быть истинной только либо в $a$ (если $a\in v(p)$), либо в~$\bar{a}$ (если $\bar{a}\in v(p)$).
\end{proof}

Для каждого подмножества $A$ множества $\numN^{++}$ положим
$\mathscr{C}^{rs}_{A} = \{ \kframe{F}^{rs}_n : n \in \tilde{A} \}$ и
$\mathbf{L}^{rs}_{A} = \mlogic{\mathscr{C}^{rs}_A}$.

\begin{lemma}
  \label{lem:betas-S}
Для любого $n \in \numN^{++}$ имеют место следующие эквивалентности:
  $$
  \begin{array}{lclcl}
  \neg \beta_{2n} \in \mathbf{L}^{rs}_A
    & \iff
    & \kframe{F}^{rs}_{2n} \notin \mathscr{C}^{rs}_{A}
    & \iff
    & n \in A.
  \end{array}
  $$
\end{lemma}

\begin{proof}
Следует непосредственно из леммы~\ref{lem:betas} и определений класса $\mathscr{C}^{rs}_{A}$ и логики~$\mathbf{L}^{rs}_A$.
\end{proof}

\begin{lemma}
  \label{lem:lfp-rs}
Для каждого подмножества $A$ множества $\numN^{++}$ логика $\mathbf{L}^{rs}_{A}$ является линейно аппроксимируемой.
\end{lemma}

\begin{proof}
Аналогично доказательству леммы~\ref{lem:lfp}.
\end{proof}

\begin{theorem}
  \label{thr:ext-KTB}
Для любой степени неразрешимости $C$, существует линейно аппроксимируемое нормальное расширение логики\/ $\logic{KTB}$, находящееся в $C$, фрагмент от одной переменной которого тоже находится в~$C$.
\end{theorem}

\begin{proof}
Аналогично доказательству теоремы~\ref{thr:ext-K}, но надо использовать леммы~\ref{lem:lfp-rs} и~\ref{lem:betas-S}.
\end{proof}

\subsubsection{Неразрешимые расширения логик $\logic{wGrz}$ и $\logic{GL}$}

В случае расширений логики $\logic{K4}$ не получится воспользоваться модальной глубиной формулы, как мы сделали это выше. Тем не менее, мы можем использовать некоторую модификацию описанной выше конструкции для логики~$\logic{K}$.

\begin{figure}
\centering
\begin{tikzpicture}[scale=1.5]

\coordinate (w1)   at (+0.0000,+0.0000);
\coordinate (w2)   at (+1.0000,+0.0000);
\coordinate (w3)   at (+2.0000,+0.0000);
\coordinate (w4ph) at (+3.0000,+0.0000);
\coordinate (dts)  at (+3.5000,+0.0000);
\coordinate (w5ph) at (+4.0000,+0.0000);
\coordinate (wn-1) at (+5.0000,+0.0000);
\coordinate (wn)   at (+6.0000,+0.0000);
\coordinate (w*)   at (+0.0000,+1.0000);

\begin{scope}[>=latex]
\draw [->,  shorten >= 1.5pt, shorten <= 1.5pt]
(w1) -- (w2);
\draw [->,  shorten >= 1.5pt, shorten <= 1.5pt]
(w2) -- (w3);
\draw [->,  shorten >= 1.5pt, shorten <= 1.5pt]
(w3) -- (w4ph);
\draw [->,  shorten >= 1.5pt, shorten <= 1.5pt]
(w5ph) -- (wn-1);
\draw [->,  shorten >= 1.5pt, shorten <= 1.5pt]
(wn-1) -- (wn);
\draw [->,  shorten >= 1.5pt, shorten <= 1.5pt]
(w1) -- (w*);
\end{scope}

\node [below      ] at (w1)   {${w_0}$}     ;
\node [below      ] at (w2)   {${w_1}$}     ;
\node [below      ] at (w3)   {${w_2}$}     ;
\node [           ] at (dts)  {${\cdots}$}  ;
\node [above      ] at (w*)   {${w^\ast}$}  ;
\node [below      ] at (wn-1) {${w_{n-1}}$} ;
\node [below      ] at (wn)   {${w_n}$}     ;


\filldraw [] (w1)   circle [radius=1.5pt]   ;
\filldraw [] (w2)   circle [radius=1.5pt]   ;
\filldraw [] (w3)   circle [radius=1.5pt]   ;
\draw [] (w*)   circle [radius=1.5pt]   ;
\filldraw [] (wn-1) circle [radius=1.5pt]   ;
\filldraw [] (wn)   circle [radius=1.5pt]   ;
\end{tikzpicture}

\caption{Шкала $\kframe{F}^t_n$}
\label{fig-4}
\end{figure}

Пусть для каждого $n \in \numNp$
\settowidth{\templength}{\mbox{$W^t_n$}}
\begin{itemize}
\item $\parbox{\templength}{$W^t_n$}=\{w_0, \ldots, w_n, w^\ast\}$;
\item $R^t_n$~--- транзитивное замыкание отношения
  $$
  \{\langle w_k,w_{k+1} \rangle : 0\leqslant k < n\} \cup \{\langle
  w_0, w^\ast \rangle, \langle w^\ast, w^\ast \rangle \};
  $$
\item $\parbox{\templength}{$\kframe{F}^t_n$} = \langle W^t_n, R^t_n \rangle$,
\end{itemize}
см. рис.~\ref{fig-4}; стрелки, которые можно восстановить по транзитивности, опущены; чёрные кружк\'{и} соответствуют иррефлексивным мирам, белый~--- рефлексивному миру.

Для каждого $n \in \numNp$ положим
$$
\begin{array}{lcl}
\gamma_n & = & \Diamond (\Diamond \top \wedge \Box \Diamond \top) \con
\Diamond^n \Box \bot \con \neg \Diamond^{n+1} \Box \bot.
\end{array}
$$

\begin{lemma}
\label{lem:gammas}
Для любых $n,k \in \numNp$ имеет место следующая эквивалентность:
$$
\begin{array}{lcl}
  (\kframe{F}^t_n, x) \models \gamma_k
  & \iff
  & \mbox{$k = n$ and $x = w_0$.}
\end{array}
$$
\end{lemma}

\begin{proof}
Следует непосредственно из определения $\kframe{F}^t_n$ и~$\gamma_k$.
\end{proof}

Для каждого подмножества $A$ множества $\numN^{++}$ положим
$\mathscr{C}^t_{A} = \{ \kframe{F}^t_n : n\in \tilde{A} \}$ и
$\mathbf{L}^t_{A} = \mlogic{\mathscr{C}^t_A}$.

\begin{lemma}
  \label{lem:gammas-S}
Для любого $n \in \numN^{++}$ имеет место следующая эквивалентность:
  $$
  \begin{array}{lclcl}
  \neg \gamma_{2n} \in \mathbf{L}^t_{A}
    & \iff
    & \kframe{F}_{2n} \notin \mathscr{C}^t_A
    & \iff
    & n \in A.
  \end{array}
  $$
\end{lemma}

\begin{proof}
Следует из леммы~\ref{lem:gammas} и определений класса $\mathscr{C}^t_A$ и логики~$\mathbf{L}^t_{A}$.
\end{proof}

\begin{lemma}
  \label{lem:lfp-K4}
Для каждого подмножества $A$ множества $\numN^{++}$ логика
  $\mathbf{L}^t_{A}$ является линейно аппроксимируемой.
\end{lemma}

\begin{proof}
Покажем, что если $\vp\notin \mathbf{L}^t_{A}$, то $\vp$ опровергается в некоторой $\mathbf{L}^t_{A}$-шкале размера не более чем~$|\vp|+6$.

Пусть $\vp \notin \mathbf{L}^t_{A}$. Тогда $(\kframe{F}^t_n, x) \not\models^v \vp$ для некоторой шкалы $\kframe{F}^t_n$ из класса $\mathscr{C}^t_A$, некоторого мира $x$ из $\kframe{F}^t_n$ и некоторой оценки $v$ в~$\kframe{F}^t_n$. Рассмотрим только случай, когда $n > |\vp|+4$ (если $n \leqslant |\vp|+4$, то $|\kframe{F}^t_n| \leqslant |\vp|+6$, и доказывать нечего).

Пусть $\Box \psi_1, \ldots, \Box \psi_m$~--- список всех формул вида $\Box \psi$ из $\mathop{\mathit{sub}} \vp$. Для каждого $k \in \{1, \ldots, m\}$ положим $U_k = \{ w_j\in W_n^t : (\kframe{F}^t_n, w_j) \not\models \psi_k, 0 \leqslant j \leqslant n \}$ и
  $$
    \begin{array}{lcl}
      U'_k
      & =
      & \left\{
        \begin{array}{rl}
          \{ w_s \}, & \mbox{если $U_k \ne \varnothing$ и $s = \max\{j : w_j\in U_k\}$;}\\
          \varnothing, & \mbox{если не так.}\\
        \end{array}
      \right.
    \end{array}
    $$
    Пусть
    \settowidth{\templength}{\mbox{$W'$}}
    \begin{itemize}
    \item
      $\parbox{\templength}{$W'$} = \displaystyle\bigcup\limits_{\mathclap{k = 1}}^{m} U'_k
      \cup \{ w_0,w_1,w_2,w^\ast,x \}$;
    \item
      $\parbox{\templength}{$l$} = |W' - \{w^\ast,w_0\}| = |W'|-2$;
      $w'$~--- некоторый мир из $W_n^t - W'$ (такой мир существует, поскольку, по предположению, $n > |\vp|+ 4$, а значит, $|W_n^t| > |\vp|+ 6$, но
      $m \leqslant |\mathop{\mathit{sub}} \vp| \leqslant |\vp|$, поэтому
      $|W'| \leqslant |\vp| + 5$);
    $$
    \begin{array}{lcl}
      W''
      & =
      & \left\{
        \begin{array}{rl}
          W',             & \mbox{если $l \in \tilde{A}$;}\\
          W' \cup \{w'\}, & \mbox{если $l \not\in \tilde{A}$;}\\
        \end{array}
      \right.
    \end{array}
    $$
  \item $\parbox{\templength}{$R'$} = R_n^t \upharpoonright W''$;
  \item $\parbox{\templength}{$\kframe{F}'$} = \langle W'', R' \rangle$;
  \item $v'$~--- оценка в $\kframe{F}'$, определённая условием
    $v'(p) = v(p) \cap W''$ для каждой переменной~$p$.
  \end{itemize}

  Индукцией по построению $\psi$ несложно показать, что для любого мира $w \in W''$ и любой формулы $\psi \in \mathop{\mathit{sub}} \vp$
  $$
  \begin{array}{lcl}
  (\kframe{F}^t_n, w) \models^v \psi
    & \iff
    & (\kframe{F}', w) \models^{v'} \psi.
  \end{array}
  $$
  Следовательно, $(\kframe{F}', x) \not\models^{v'} \vp$.

Теперь покажем, что $\kframe{F}'$ изоморфна некоторой шкале из класса~$\mathscr{C}^t_A$. Если $l \in \tilde{A}$, то $\kframe{F}'$ изоморфна шкале $\kframe{F}^t_l$ и $\kframe{F}^t_l \in \mathscr{C}^t_A$. Если $l \not\in \tilde{A}$, то $\kframe{F}'$ изоморфна шкале $\kframe{F}^t_{l+1}$; более того, в этом случае число $l$ чётно (иначе $l$ было бы в~$\tilde{A}$), следовательно, $l+1$ нечётно, а значит, $\kframe{F}^t_{l+1} \in \mathscr{C}^t_A$.

Таким образом, $\vp$ опровергается в $\mathbf{L}^t_A$-шкале размера не более~$|\vp| + 6$.
\end{proof}


\begin{theorem}
  \label{thr:ext-K4}
Для любой степени неразрешимости $C$ существует линейно аппроксимируемое нормальное расширение логики\/ $\logic{wGrz}$, находящееся в $C$, константный фрагмент которого тоже находится в~$C$.
\end{theorem}

\begin{proof}
Аналогично доказательству теоремы~\ref{thr:ext-K} с применением лемм~\ref{lem:lfp-K4} и~\ref{lem:gammas-S}.
%
%
\end{proof}

\begin{figure}
\centering
\begin{tikzpicture}[scale=1.5]

\coordinate (w1)   at (+0.0000,+0.0000);
\coordinate (w2)   at (+1.0000,+0.0000);
\coordinate (w3)   at (+2.0000,+0.0000);
\coordinate (w4ph) at (+3.0000,+0.0000);
\coordinate (dts)  at (+3.5000,+0.0000);
\coordinate (w5ph) at (+4.0000,+0.0000);
\coordinate (wn-1) at (+5.0000,+0.0000);
\coordinate (wn)   at (+6.0000,+0.0000);
\coordinate (w*)   at (+0.0000,+1.0000);

\begin{scope}[>=latex]
\draw [->,  shorten >= 1.5pt, shorten <= 1.5pt]
(w1) -- (w2);
\draw [->,  shorten >= 1.5pt, shorten <= 1.5pt]
(w2) -- (w3);
\draw [->,  shorten >= 1.5pt, shorten <= 1.5pt]
(w3) -- (w4ph);
\draw [->,  shorten >= 1.5pt, shorten <= 1.5pt]
(w5ph) -- (wn-1);
\draw [->,  shorten >= 1.5pt, shorten <= 1.5pt]
(wn-1) -- (wn);
\draw [->,  shorten >= 1.5pt, shorten <= 1.5pt]
(w1) -- (w*);
\end{scope}

\node [below      ] at (w1)   {${w_0}$}     ;
\node [below      ] at (w2)   {${w_1}$}     ;
\node [below      ] at (w3)   {${w_2}$}     ;
\node [           ] at (dts)  {${\cdots}$}  ;
\node [above      ] at (w*)   {${w^\ast}$}  ;
\node [below      ] at (wn-1) {${w_{n+1}}$} ;
\node [below      ] at (wn)   {${w_{n+2}}$}     ;


\filldraw [] (w1)   circle [radius=1.5pt]   ;
\filldraw [] (w2)   circle [radius=1.5pt]   ;
\filldraw [] (w3)   circle [radius=1.5pt]   ;
\filldraw [] (w*)   circle [radius=1.5pt]   ;
\filldraw [] (wn-1) circle [radius=1.5pt]   ;
\filldraw [] (wn)   circle [radius=1.5pt]   ;
\end{tikzpicture}

\caption{Шкала $\kframe{F}^{N}_n$}
\label{fig-gl}
\end{figure}

Для случая расширений логики $\logic{GL}$ построения аналогичны.

Для каждого $n \in\numNp$ пусть
\settowidth{\templength}{\mbox{$W^t_n$}}
\begin{itemize}
\item
  $\parbox{\templength}{$W^{N}_n$}=\{w_0, \ldots, w_{n+2}, w^\ast\}$;
\item $R^{N}_n$~--- транзитивное замыкание отношения
  $$
  \{\langle w_k,w_{k+1} \rangle : 0\leqslant k < n+2\} \cup \{\langle
  w_0, w^\ast \rangle \};
  $$
\item $\parbox{\templength}{$\kframe{F}^{N}_n$} = \langle W^{N}_n, R^{N}_n \rangle$,
\end{itemize}
см. рис.~\ref{fig-gl}; стрелки, которые можно восстановить по транзитивности, опущены.

Для каждого $n \in\numNp$ положим
$$
\begin{array}{lcl}
  \delta_n & = & \Diamond (\neg p \wedge \Box \bot) \con \Box^2 p \con
                 \Diamond^{n+2} \Box \bot \con \neg \Diamond^{n+3} \Box \bot.
\end{array}
$$

\begin{lemma}
  \label{lem:deltas}
Для любых $n,k \in\numNp$ имеют место следующие эквивалентности:
$$
\begin{array}{lcl}
  (\kframe{F}^{N}_n, x) \not\models \neg \delta_k
  & \iff
  & \mbox{$k = n$ и $x = w_0$.}
\end{array}
$$
\end{lemma}

\begin{proof}
Как и раньше, простая проверка.
\end{proof}

Для каждого подмножества $A$ множества $\numN^{++}$ положим $\mathscr{C}^{N}_{A} = \{ \kframe{F}^{N}_n : n\in \tilde{A} \}$ и
$\mathbf{L}^{N}_{A} = \mathsf{Log}(\mathscr{C}^{N}_A)$.

\begin{lemma}
  \label{lem:deltas-S}
Для любого $n \in \numN^{++}$ имеют место следующие эквивалентности:
  $$
  \begin{array}{lclcl}
    \neg \delta_{2n} \in \mathbf{L}^{N}_{A}
    & \iff
    & \kframe{F}_{2n} \notin \mathscr{C}^{N}_A
    & \iff
    & n \in A.
  \end{array}
  $$
\end{lemma}

\begin{proof}
Следует из леммы~\ref{lem:deltas} и определений класса $\mathscr{C}^{N}_A$ и логики~$\mathbf{L}^{N}_{A}$.
\end{proof}

\begin{lemma}
  \label{lem:lfp-GL}
Для каждого подмножества $A$ множества $\numN^{++}$ логика
  $\mathbf{L}^{N}_{A}$ является линейно аппроксимируемой.
\end{lemma}

\begin{proof}
Аналогично доказательству леммы~\ref{lem:lfp-K4}.
\end{proof}


\begin{theorem}
  \label{thr:ext-GL}
Для любой степени неразрешимости $C$ существует линейно аппроксимируемое нормальное расширение логики\/ $\logic{GL}$, находящееся в $C$, фрагмент от одной переменной которого тоже находится в~$C$.
\end{theorem}

\begin{proof}
Аналогично доказательству теоремы~\ref{thr:ext-K4} с использованием лемм~\ref{lem:deltas-S} и~\ref{lem:lfp-GL}.
\end{proof}

\subsubsection{Неразрешимые расширения логики $\logic{Grz}$}

Аналогичные результаты справедливы и для класса расширений логики $\logic{Grz}$; они следуют из построений, которые мы чуть позже (раздел~\ref{ssec:poly:approx:int}) предложим для класса расширений интуиционистской логики. Поэтому пока мы приводим лишь формулировку, а доказательство будет дано ниже (см.~предложение~\ref{prop:grz}).

\begin{theorem}
  \label{thr:ext-Grz:pre}
Для любой степени неразрешимости $C$ существует линейно аппроксимируемое нормальное расширение логики\/ $\logic{Grz}$, находящееся в $C$, фрагмент от одной переменной которого тоже находится в~$C$.
\end{theorem}

\subsubsection{Сложно разрешимые расширения логик $\logic{wGrz}$, $\logic{Grz}$, $\logic{GL}$, $\logic{KTB}$}

Приведённую выше аргументацию, увы, не получается повторить полностью для случая, когда $C$~--- класс сложности, а не степень неразрешимости. Дело в том, что в случае рассмотрения вопросов сложности важно, чтобы натуральные числа кодировались записями в некоторой позиционной системе счисления (или каким-то сходным образом). Но тогда функция, сопоставляющая натуральному числу~$n$ формулу~$\alpha_n$ (или $\beta_n$, или $\gamma_n$, или $\delta_n$), будет не полиномиально вычислимой, а экспоненциально вычислимой. Тем не менее, рассмотрение очень сложных задач (например, когда временная сложность описывается башней экспонент), приводит нас и к очень сложно разрешимым линейно аппроксимируемым логикам в расширениях логик $\logic{wGrz}$, $\logic{Grz}$, $\logic{GL}$, $\logic{KTB}$, причём сложными в этих логиках будут уже их фрагменты от одной переменной, а в некоторых случаях и константные фрагменты. Но мы не можем гарантировать $C$-полноту этих фрагментов. Поэтому можно сформулировать следующие вопросы.

\begin{problem}
Пусть $C$~--- класс разрешимых задач, замкнутый по полиномиальной сводимости задач и содержащий $C$-полную задачу. Существует ли в расширениях логики\/ $\logic{KD}$, или\/ $\logic{T}$, или\/ $\logic{KB}$, или\/ $\logic{Grz}$, или\/ $\logic{GL}$, или\/ $\logic{KTB}$ полиномиально аппроксимируемая $C$-полная логика с $C$-трудным фрагментом от одной переменной?
\end{problem}

\begin{problem}
Пусть $C$~--- класс разрешимых задач, замкнутый по полиномиальной сводимости задач и содержащий $C$-полную задачу. Существует ли в расширениях логики\/ $\logic{K}$, или\/ $\logic{K4}$, или\/ $\logic{wGrz}$ полиномиально аппроксимируемая $C$-полная логика с $C$-трудным константным фрагментом?
\end{problem}

    \subsection{Модальные алгебры}

Используя описанные конструкции, можно получить некоторые результаты, касающиеся модальных алгебр.

Алгебра $\bm{A}=\langle A,\bwedge,\bvee,\bneg,\bbot,\btop\rangle$ называется \defnotion{булевой},\index{алгебра!булева} если для всяких $x,y,z\in A$
\[
\begin{array}{ll}
\arrayitem &
\mbox{$x\bwedge y = y\bwedge x$, $x\bvee y = y\bvee x$;} \\
\arrayitem &
\mbox{$x\bwedge (y\bwedge z) = (x\bwedge y)\bwedge z$, $x\bvee (y\bvee z) = (x\bvee y)\bvee z$;} \\
\arrayitem &
\mbox{$(x\bwedge y)\bvee y = y$, $(x\bvee y)\bwedge y = y$;} \\
\arrayitem &
\mbox{$x\bwedge (y\bvee z) = (x\bwedge y)\bvee (x\bwedge z)$, $x\bvee (y\bwedge z) = (x\bvee y)\bwedge (x\bvee z)$;} \\
\arrayitem &
\mbox{$x\bwedge \bneg x = \bbot$, $x\bvee \bneg x = \btop$.} \\
\end{array}
\]

Алгебра $\bm{A}=\langle A,\bwedge,\bvee,\bneg,\bbot,\btop,\bBox\rangle$ называется \defnotion{модальной},\index{алгебра!модальная} если $\langle A,\bwedge,\bvee,\bneg,\bbot,\btop\rangle$~--- булева алгебра и для всяких $x,y\in A$
\[
\begin{array}{llcl}
\arrayitem &
\bBox \btop & = & \btop; \\
\arrayitem &
\bBox(x\bwedge y) & = & \bBox x\bwedge \bBox y.
\end{array}
\]

В контексте алгебр $\lang{ML}$-формулы будем называть также \defnotion{$\lang{ML}$-термами}, или просто \defnotion{термами};\index{терм!ml@$\lang{ML}$-терм} переменные в термах будем обозначать буквами $x$, $y$, $z$ (иногда с индексами). Нам будет интересна проблема равенства $\lang{ML}$-термов в модальных алгебрах и классах модальных алгебр. Чтобы описать эту проблему, введём понятия оценки и значения терма при оценке в модальной алгебре.

Функцию $v$, которая каждой переменной $x$ сопоставляет элемент $v(x)\in A$, называем \defnotion{оценкой}\index{оценка} в алгебре~$\bm{A}$. Расширим $v$ на множество всех $\lang{ML}$\nobreakdash-термов:
\[
\begin{array}{llcl}
\arrayitem &
v(\bot) & = & \bbot; \\
\arrayitem &
v(t\wedge s) & = & v(t)\bwedge v(s); \\
\arrayitem &
v(t\vee s) & = & v(t)\bvee v(s); \\
\arrayitem &
v(t\to s) & = & \bneg v(t)\bvee v(s); \\
\arrayitem &
v(\Box t) & = & \bBox v(t).
\end{array}
\]
Элемент $v(t)$ называем \defnotion{значением терма} $t$ в алгебре $\bm{A}$ при оценке~$v$.

Говорим, что в $\bm{A}$ справедливо равенство $t=s$ термов $t$ и $s$ при оценке $v$, если $v(t)=v(s)$; этот факт обозначаем как $\bm{A}\models^v t=s$. Говорим, что в $\bm{A}$ справедливо равенство $t=s$, если $\bm{A}\models^v t=s$ для любой оценки $v$ в $\bm{A}$; этот факт обозначаем как $\bm{A}\models t=s$. Говорим, что равенство $t=s$ справедливо в классе алгебр, если оно справедливо в каждой алгебре этого класса. Если равенство не является справедливым (при оценке, в алгебре, в классе алгебр), то говорим, что оно опровергается.

Под проблемой равенства термов в классе алгебр $\scls{C}$ понимаем множество равенств термов, которые справедливы в~$\scls{C}$.

Пусть $L$~--- нормальная модальная логика. Алгебру $\bm{A}$ называем \defnotion{$L$-алгеброй}, если для каждой формулы $\varphi\in L$ в $\bm{A}$ справедливо равенство $\varphi=\top$. Класс всех $L$-алгебр обозначим\footnote{Этот класс является \defnotion{многообразием}\index{многообразие}, см.~\cite[раздел~7.6]{ChZ}; в качестве обозначения взяты первые буквы слов <<modal>> и <<variaty>>.}~$\malg{L}$.

Каждой шкале Крипке $\kframe{F}=\otuple{W,R}$ сопоставим модальную алгебру $\kframe{F}^+ = \otuple{\Pow{W},\cap,\cup,\overline{\,\cdot\,},\varnothing,W,\bBox}$, где
$$
\begin{array}{lcl}
\Pow{W} & = & \set{X : X\subseteq W}; \\
\bBox X & = & \set{w\in W : R(w)\subseteq X}.
\end{array}
$$

Пусть $L$~--- одна из логик $\logic{K}$, $\logic{wGrz}$, $\logic{GL}$, $\logic{Grz}$, $\logic{KTB}$; для нас важно следующее: $\varphi\in L$ тогда и только тогда, когда равенство $\varphi=\top$ справедливо в любой модальной алгебре, построенной описанным способом по некоторой $L$-шкале, см.~\cite[раздел~7.5]{ChZ}.

Пусть $\bm{A}=\langle A,\bwedge,\bvee,\bneg,\bbot,\btop,\bBox\rangle$~--- модальная алгебра, $X\subseteq A$. Наименьшая модальная подалгебра алгебры $\bm{A}$, содержащая все элементы множества $X$, называется \defnotion{подалгеброй} алгебры $\bm{A}$, \defnotion{порождённой множеством~$X$};\index{подалгебра!порождённая} в этом случае $X$ называется множеством, \defnotion{порождающим} эту подалгебру. Алгебра $\bm{A}$ называется \defnotion{$n$\nobreakdash-порождённой},\index{подалгебра!n@$n$-порождённая} где $n\in\mathds{N}$, если существует множество $X$ из $n$ элементов, порождающее алгебру~$\bm{A}$; в частности, если $X=\varnothing$, то $\bm{A}$ называется $0$\nobreakdash-порождённой алгеброй.


\begin{theorem}
Пусть $L\in [\logic{K},\logic{wGrz}]$. Тогда
\begin{itemize}
\item проблема равенства константных\/ $\lang{ML}$-термов в классе\/ $\malg{L}$ является\/ $\cclass{PSPACE}$-трудной;
\item проблема равенства произвольных\/ $\lang{ML}$-термов в классе\/ $0$-порождённых $L$-алгебр является\/ $\cclass{PSPACE}$-трудной.
\end{itemize}
\end{theorem}

\begin{proof}
Следствие теоремы~\ref{th:PSPACE:K4}.
\end{proof}

\pagebreak[3]

\begin{theorem}
Пусть $L\in [\logic{K},\logic{GL}] \cup [\logic{K},\logic{Grz}] \cup [\logic{K},\logic{KTB}]$. Тогда
\begin{itemize}
\item проблема равенства\/ $\lang{ML}$-термов от одной переменной в классе\/ $\malg{L}$ является\/ $\cclass{PSPACE}$-трудной;
\item проблема равенства произвольных\/ $\lang{ML}$-термов в классе\/ $1$-порождённых $L$-алгебр является\/ $\cclass{PSPACE}$-трудной.
\end{itemize}
\end{theorem}

\begin{proof}
Следствие теорем~\ref{th:PSPACE:GL:Grz} и~\ref{th:PSPACE:KTB}.
\end{proof}

  \section{Замечания}

Техника, использованная в этой главе, основана на идеях Дж.\,Халперна~\cite{Halpern95}, показавшего, что логики $\logic{K}$, $\logic{T}$ и $\logic{S4}$ являются $\cclass{PSPACE}$-полными в языке с одной переменной. Помимо работы Дж.\,Халперна, имеются и другие, где рассматривается сложность модальных логик в языках с конечным числом переменных~\cite{Sve03,BS93,Spaan-1993-1}. Несколько позже мы уделим внимание моделированию, предложенному в работе П.\,Блэкбёрна и Э.\,Спаан~\cite{BS93}, поскольку там описан метод моделирования переменных, не требующий увеличения ширины шкал, но сделаем это уже для предикатных логик, шкалы Крипке которых линейны. 

Стоит отметить, что мы сосредоточились в этой главе на логиках, которые являются $\cclass{PSPACE}$-полными: $\logic{K}$, $\logic{T}$, $\logic{K4}$, $\logic{S4}$, $\logic{GL}$ и др. Рассмотрение в первую очередь именно таких логик связано не столько с особенностями описанного выше метода моделирования, сколько со свойствами этих логик. Как мы увидим позже, метод моделирования всех переменных формулы формулами, построенными из переменных некоторого фиксированного конечного множества, несложно распространяется и на логики, проблема разрешения которых лежит в классах, содержащих более сложные задачи (скажем, на логики с $\cclass{EXPTIME}$-полной или $\cclass{2EXPTIME}$-полной проблемой разрешения), а также на неразрешимые и даже не являющиеся рекурсивно перечислимыми.

\setcounter{savefootnote}{\value{footnote}}
\chapter{Логики Виссера и суперинтуиционистские логики}
\setcounter{footnote}{\value{savefootnote}}
\label{chapter:IL}
  \section{Основные определения и факты}
    \subsection{Синтаксис}

Здесь мы будем рассматривать логики в языке $\lang{L}$, т.е. в том же языке, в котором определили классическую логику~$\logic{Cl}$. В контексте исследования логик, которые определим ниже, $\lang{L}$\nobreakdash-формулы будем называть не классическими, а \defnotion{интуиционистскими}.\index{уяа@формула!пропозициональная!интуиционистская}
\index{уяа@формула!l@$\lang{L}$-формула}

Интуиционистскую формулу $\varphi$ называем \defnotion{позитивной},\index{уяа@формула!позитивная} если она не содержит вхождений~$\bot$. Отметим, что поскольку $\neg\psi=\psi\to\bot$, позитивные формулы не содержат отрицания.

Для множества интуиционистских формул $X$ его подмножество, состоящее из позитивных формул, обозначим~$X^+$ и будем называть \defnotion{позитивным фрагментом}\index{уяа@фрагмент!позитивный} множества~$X$. 

    \subsection{Семантика Крипке}

Как и в модальном случае, для оценки истинности формул мы будем использовать семантику Крипке.

Пусть $\kframe{F}=\otuple{W,R}$~--- шкала с транзитивным и антисимметричным\footnote{В контексте рассмотрения интуиционистской логики добавляется ещё условие рефлексивности; мы хотим включить в рассмотрение и иные логики, поэтому пока требования рефлексивности нет.}
отношением достижимости~$R$. Оценку $v$ в $\kframe{F}$ называем \defnotion{наследственной},\index{оценка!наследственная} или \defnotion{интуиционистской},\index{оценка!интуиционистская} если она удовлетворяет следующему условию: для всякой переменной $p$ и всяких миров $w,w'\in W$
$$
\begin{array}{lcl}
\mbox{$wRw'$ и $w\in v(p)$} & \imply & w'\in v(p).
\end{array}
$$

Модель Крипке $\kmodel{M}=\otuple{\kframe{F},v}$ на транзитивной антисимметричной шкале $\kframe{F}$ называем \defnotion{интуиционистской},\index{модель!интуиционистская} если оценка $v$ является наследственной.

Пусть $\kmodel{M}=\otuple{\kframe{F},v}$~--- интуиционистская модель Крипке, определённая на транзитивной антисимметричной шкале $\kframe{F}=\otuple{W,R}$. Для всякой интуиционистской формулы $\varphi$ и всякого $w\in W$ определим отношение~$(\kmodel{M},w)\imodels\varphi$:
\[
\begin{array}{clcl}
\arrayitem &
(\kmodel{M},w) \not\imodels \bot; &  &
\arrayitemskip
\\
\arrayitem &
(\kmodel{M},w)\imodels p_n
& \bydef &
\parbox[t]{225pt}{$w\in v(p_n)$;}
\arrayitemskip
\\
\arrayitem &
(\kmodel{M},w)\imodels\varphi\wedge\psi & \bydef &
\parbox[t]{225pt}{$(\kmodel{M},w)\imodels\varphi$ и
$(\kmodel{M},w)\imodels\psi$;}
\arrayitemskip
\\
\arrayitem &
(\kmodel{M},w)\imodels\varphi\vee\psi & \bydef &
\parbox[t]{225pt}{$(\kmodel{M},w)\imodels\varphi$ или
$(\kmodel{M},w)\imodels\psi$;}
\arrayitemskip
\\
\arrayitem &
(\kmodel{M},w)\imodels\varphi\to\psi & \bydef &
\parbox[t]{225pt}{для всякого $w'\in R(w)$
справедливо $(\kmodel{M},w') \not\imodels \varphi$ или
$(\kmodel{M},w')\imodels\psi$.}
\end{array}
\]

Истинность интуиционистских формул в модели Крипке, в шкале Крипке и в классе шкал Крипке определяется аналогично модальному случаю.
Если $(\kmodel{M},w)\imodels\varphi$, то говорим, что формула $\varphi$ \defnotion{истинна в мире} $w$ модели $\kmodel{M}$; в противном случае говорим, что $\varphi$ \defnotion{опровергается в мире} $w$ модели~$\kmodel{M}$. Формулу $\varphi$ считаем \defnotion{истинной в модели} $\kmodel{M}$, если для всякого $w\in W$ выполнено отношение $(\kmodel{M},w)\imodels\varphi$; в этом случае пишем $\kmodel{M}\imodels\varphi$. Формулу $\varphi$ считаем \defnotion{истинной в шкале} $\kframe{F}$, если $\varphi$ истинна в любой модели, определённой на шкале~$\kframe{F}$; в этом случае пишем $\kframe{F}\imodels\varphi$. Формулу $\varphi$ считаем \defnotion{истинной в мире шкалы} $\kframe{F}$, если для любой модели $\kmodel{M}$, определённой на шкале~$\kframe{F}$, выполнено отношение $(\kmodel{M},w)\imodels\varphi$; в этом случае пишем $(\kframe{F},w)\imodels\varphi$. Формулу $\varphi$ считаем \defnotion{истинной в классе шкал} $\sclass{C}$, если $\varphi$ истинна в каждой шкале из этого класса; в этом случае пишем $\sclass{C}\imodels\varphi$.

Если имеется множество интуиционистских формул $\varGamma$ и какая-либо структура $\mathfrak{S}$~--- шкала, модель или мир,~--- то будем говорить, что в~$\mathfrak{S}$ истинно~$\varGamma$, если в~$\mathfrak{S}$ истинна каждая формула из~$\varGamma$; в этом случае пишем $\mathfrak{S}\imodels\varGamma$.

    \subsection{Логики}
    \label{ssec:int:Visser:logics}

Под логикой в языке $\lang{L}$ понимаем множество $\lang{L}$-формул, замкнутое по правилу подстановки.

Определим следующие логики стандартно~\cite{ChZ,Visser81}:
\begin{itemize}
\item
$\logic{Int}$~--- множество $\lang{L}$-формул, истинных в классе всех рефлексивных транзитивных антисимметричных шкал Крипке;
\item
$\logic{BPL}$~--- множество $\lang{L}$-формул, истинных в классе всех транзитивных антисимметричных шкал Крипке;
\item
$\logic{FPL}$~--- множество $\lang{L}$-формул, истинных в классе всех иррефлексивных транзитивных антисимметричных шкал Крипке, не имеющих бесконечно возрастающих цепей, т.е. в классе всех строгих нётеровых порядков.
\end{itemize}

Логика $\logic{Int}$ называется \defnotion{интуиционистской пропозициональной логикой},\index{логика!пропозициональная!интуиционистская} иногда обозначается также~$\logic{H}$. Логика $\logic{BPL}$ называется \defnotion{базисной пропозициональной логикой Виссера}\index{логика!пропозициональная!базисная Виссера}, а логика $\logic{FPL}$~--- \defnotion{формальной пропозициональной логикой Виссера}.\index{логика!пропозициональная!формальная Виссера} Все эти три логики замкнуты по правилу \MP.

Расширения интуиционистской логики называются \defnotion{суперинтуиционистскими}\index{логика!суперинтуиционистская} логиками.

Следующие суперинтуиционистские пропозициональные логики определим стандартно~\cite{ChZ}.
$$
\begin{array}{lclcl}
\logic{KC} & = & \logic{Int} & \!\!+\!\! & \neg p\vee\neg\neg p    \\
\logic{LC} & = & \logic{Int} & \!\!+\!\! & (p\to q)\vee(q\to p)    \\
\logic{KP} & = & \logic{Int} & \!\!+\!\! & (\neg p\to q\vee r)\to (\neg p\to q)\vee(\neg p\to r) \\
\logic{Cl} & = & \logic{Int} & \!\!+\!\! & p\vee\neg p    \\
\end{array}
$$
Логика $\logic{KC}$, обозначаемая также $\logic{HJ}$, известна как логика слабого закона исключённого третьего и как логика Янкова, $\logic{LC}$~--- как линейная логика, $\logic{KP}$~--- как логика Крайзеля--Патнэма; $\logic{Cl}$~--- классическая пропозициональная логика.

Отметим, что все непротиворечивые суперинтуиционистские логики находятся между $\logic{Int}$ и $\logic{Cl}$, а потому называются также \defnotion{промежуточными};\index{логика!промежуточная} все они замкнуты по правилу \MP. Расширения же базисной и формальной логик Виссера не обязательно замкнуты по \MP. В качестве примера логики, не замкнутой по \MP, можно взять логику шкалы, состоящей из одного иррефлексивного мира: ей принадлежат формулы $\top$ и $\top\to\bot$, но не принадлежит формула~$\bot$.

    \subsection{Необходимые факты}

            Пусть, как и в модальном случае, $\kframes{L}$~--- класс шкал Крипке, в которых истинны все формулы, принадлежащие логике~$L$; если $\kframe{F}$~--- шкала Крипке и $\kframe{F}\imodels L$, то $\kframe{F}$ называем шкалой логики~$L$, или $L$-шкалой.

            Для класса шкал Крипке $\scls{C}$ определим $\ilogic{\scls{C}}$ как множество интуиционистских пропозициональных формул, истинных в~$\scls{C}$. Отметим, что $\ilogic{\scls{C}}$ является нормальной логикой; если $\scls{C}$ не содержит шкал с иррефлексивными мирами, то $\ilogic{\scls{C}}$~--- суперинтуиционистская логика.

            Понятия  корректности, полноты, адекватности, финитной аппроксимируемости для расширений базисной логики Виссера определяются аналогично тому, как это было сделано в модальном случае.


            Известны следующие факты~\cite{ChZ}:
            \begin{itemize}
            \item
            $\kframes{\logic{KC}}$~--- класс всех $\logic{Int}$-шкал, в которых отношение достижимости $R$ удовлетворяет условию $\forall x\forall y\forall z\,(xRy\wedge xRz \to \exists u\,(yRu \wedge zRu))$, известному как условие Чёрча--Россера (шкалы Крипке, удовлетворяющие этому условию, называют также конвергентными);
            \item
            $\kframes{\logic{LC}}$~--- класс всех $\logic{Int}$-шкал, в которых отношение достижимости $R$ удовлетворяет условию $\forall x\forall y\forall z(xRy\wedge xRz \to yRz\vee zRy)$;
            \item
            $\kframes{\logic{KP}}$~--- класс всех $\logic{Int}$-шкал, в которых отношение достижимости $R$ удовлетворяет условию
            $$
            \begin{array}{l}
            \forall x,y,z\, (xRy\wedge xRz\wedge \neg yRz\wedge\neg zRy\to {}
            \\
            ~~~\to \exists u\, (xRu\wedge uRy\wedge uRz\wedge \forall v\, (uRv\to \exists w\, (vRw\wedge (yRw\vee zRw)))));
            \end{array}
            $$
            \item
            $\kframes{\logic{Cl}}$~--- класс всех $\logic{Int}$-шкал, в которых отношение достижимости $R$ удовлетворяет условию $\forall x\forall y\,(xRy\to x=y)$.
            \end{itemize}
            Кроме того, логики $\logic{BPL}$, $\logic{FPL}$, $\logic{Int}$, $\logic{KC}$, $\logic{LC}$, $\logic{KP}$, $\logic{Cl}$ полны по Крипке и финитно аппроксимируемы.

Расширения логик $\logic{BPL}$, $\logic{FPL}$, $\logic{Int}$ связаны с модальными логиками.
Рассмотрим \defnotion{перевод Гёделя}\index{перевод!Гёделя}~$\mathsf{T}$, определённый следующим образом:
$$
\begin{array}{rcl}
\mathsf{T}(\bot) & = & \Box\bot;
\\
\mathsf{T}(p) & = & \Box p, ~\hfill \mbox{где $p\in\prop$;}
\\
\mathsf{T}(\varphi\wedge\psi) & = & \mathsf{T}(\varphi)\wedge\mathsf{T}(\psi);
\\
\mathsf{T}(\varphi\vee\psi) & = & \mathsf{T}(\varphi)\vee\mathsf{T}(\psi);
\\
\mathsf{T}(\varphi\to\psi) & = & \Box(\mathsf{T}(\varphi)\to\mathsf{T}(\psi)).
\end{array}
$$
Несложно понять, что для всякой формулы $\varphi$
$$
\begin{array}{lclcl}
\varphi\in\logic{Int}
  & \iff
  & \mathsf{T}(\varphi)\in\logic{S4}
  & \iff
  & \mathsf{T}(\varphi)\in\logic{Grz};
  \\
\varphi\in\logic{BPL}
  & \iff
  & \mathsf{T}(\varphi)\in\logic{K4}
  & \iff
  & \mathsf{T}(\varphi)\in\logic{wGrz};
  \\
\varphi\in\logic{FPL}
  & \iff
  & \mathsf{T}(\varphi)\in\logic{GL}.
\end{array}
$$
Для расширения $L$ логики $\logic{S4}$ множество формул $\mathsf{T}^{-1}(L)$ является суперинтуиционистской логикой, которую называют \defnotion{суперинтуиционистским фрагментом}\index{уяа@фрагмент!суперинтуиционистский} логики~$L$; по аналогии, для расширения $L$ логики $\logic{K4}$ множество формул $\mathsf{T}^{-1}(L)$ тоже будем называть суперинтуиционистским фрагментом логики~$L$. Модальную логику $L$, суперинтуиционистский фрагмент которой совпадает с расширением $L'$ логики $\logic{BPL}$, будем называть \defnotion{модальным напарником}\index{напарник!модальный} логики~$L'$.

Известно~\cite{Jankov-1968-1-rus}, что $\logic{KC}$ является наибольшей из суперинтуиционистских логик, позитивный фрагмент которых совпадает с позитивным фрагментом логики $\logic{Int}$, т.е. для произвольной суперинтуиционистской логики $L$ выполняется следующая эквивалентность:
$$
\begin{array}{lcl}
  L^+ = \logic{Int}^+
    & \iff
    & L\subseteq \logic{KC}.
\end{array}
$$

Также известно~\cite{Statman-1979-1}, что логика $\logic{Int}$ является $\cclass{PSPACE}$-полной, причём $\cclass{PSPACE}$-трудным является уже фрагмент~$\logic{Int}^+$. С учётом эквивалентности выше это означает, что все логики из интервала $[\logic{Int},\logic{KC}]$ являются $\cclass{PSPACE}$-трудными, а их позитивные фрагменты~--- $\cclass{PSPACE}$-полными. Ситуация с базисной и формальной логиками Виссера аналогична~\cite{Chagrov-1985-1-rus}: их позитивные фрагменты $\cclass{PSPACE}$-полны. Как следствие, $\cclass{PSPACE}$-трудны все логики из класса $[\logic{BPL},\logic{KC}]\cup[\logic{BPL},\logic{FPL}]$, причём таковыми являются уже их позитивные\footnote{Из~\cite{Statman-1979-1} и~\cite{Chagrov-1985-1-rus} следует, что $\cclass{PSPACE}$-трудны даже импликативные фрагменты этих логик.} фрагменты.

    \subsection{Удобные обозначения}

Как и в модальном случае, длиной формулы $\varphi$ считаем число вхождений символов в $\varphi$ как в слово, а длину формулы $\varphi$ обозначаем~$|\varphi|$. Также $\mathop{\mathit{sub}}\varphi$ используем как обозначение для множества подформул формулы~$\varphi$.

Введём $\Box$ как следующее сокращение: $\Box\varphi = \top\to\varphi$. В мирах интуиционистских моделей формула $\Box\varphi$ оценивается сходным образом с модальным случаем: если $\kframe{F}$~--- $\logic{BPL}$-шкала, $w$~--- мир, а $v$~--- оценка в~$\kframe{F}$, то
$$
\begin{array}{lcl}
(\kframe{F},w)\imodels^v \Box\varphi 
 & \iff 
 & \mbox{$(\kframe{F},w')\imodels^v\varphi$ для любого $w'\in R(w)$.}
\end{array}
$$
Отметим, что в случае $\logic{Int}$-шкал формулы $\Box\varphi$ и $\varphi$ истинны или опровергаются в мирах моделей, определённых на этих шкалах, одновременно, и связка $\Box$ будет удобной для построений, связанных с расширениями логик $\logic{BPL}$ и~$\logic{FPL}$. Как и в модальном случае, считаем, что $\Box^0\varphi = \varphi$ и $\Box^{n+1}\varphi = \Box\Box^n\varphi$.

  \section{Сложность проблемы разрешения}
    \subsection{Логика ${\logic{BPL}}$}
    \label{ssec:BPL}

Для всякого $n\in\numNp$ определим формулу $\alpha_n$ (напомним, что $\Box\psi=\top\to\psi$):
$$
\begin{array}{lcl}
\label{alpha_n_BPL}
\alpha_n
  & =
  & (\Box^{n+2}\bot\to\Box^{n+1}\bot) \to (\Box^{n+1}\bot\to\Box^n\bot)\vee\Box^{n+2}\bot.
\end{array}
$$
Сразу заметим, что длина формулы $\alpha_n$ ограничена сверху некоторой линейной функцией от~$n$.

Пусть $\varphi$~--- некоторая позитивная формула от переменных $p_1,\dots,p_n$. Через $\varphi_\alpha$
\label{phi_alpha_BPL}
обозначим формулу, получающуюся из $\varphi$ подстановкой формул $\alpha_1,\dots,\alpha_n$ вместо переменных $p_1,\dots,p_n$ соответственно.

\begin{lemma}
\label{lem_BPL_alpha}
Для всякой позитивной формулы $\varphi$ имеет место следующая эквивалентность:
$$
\begin{array}{rcl}
\varphi\in\logic{BPL}
& \iff &
\varphi_\alpha\in\logic{BPL}.
\end{array}
$$
\end{lemma}

\begin{proof}
Пусть $\varphi$~--- позитивная формула от переменных $p_1,\dots,p_n$.
Если $\varphi\in\logic{BPL}$, то $\varphi_\alpha\in\logic{BPL}$, поскольку логика $\logic{BPL}$ замкнута относительно правила подстановки, а $\varphi_\alpha$~--- подстановочный пример~$\varphi$.

Пусть $\varphi\not\in \logic{BPL}$. Покажем, что в этом случае $\varphi_\alpha\not\in \logic{BPL}$. Так как $\varphi\not\in \logic{BPL}$, то существует модель $\kmodel{M}=\langle\kframe{F},v\rangle$, построенная на $\logic{BPL}$-шкале $\kframe{F}=\otuple{W,R}$, такая, что для некоторого мира $w_0\in W$ имеет место отношение $(\kmodel{M},w_0)\not\imodels \varphi$. По модели $\kmodel{M}$ построим $\logic{BPL}$-шкалу $\kframe{F}^\ast$, такую, что $\kframe{F}^\ast\not\imodels \varphi_\alpha$.

Пусть $\kframe{F}_k=\otuple{W_k,R_k}$~--- шкала Крипке, где
$$
\begin{array}{rcl}
W_k & = & \{a_1^k,\dots,a_{k+2}^k,b_k\};
\smallskip \\
wR_kw' & \iff & \mbox{либо $w= a_s^k$, $w'= a_t^k$ и $s\,{>}\,t$,}
\\
& & \mbox{либо $w= a_{k+2}^k$ и $w'= b_k$,}
\\
& & \mbox{либо $w= w'= b_k$.}
\end{array}
$$

\begin{figure}
  \centering
  \begin{tikzpicture}[scale=1.50]

    \coordinate (a30)    at ( 0.0, 3);
    \coordinate (a31)    at ( 0.0, 4);
    \coordinate (a32)    at ( 0.0, 5);
    \coordinate (a34)    at ( 0.0, 6);
    \coordinate (a35)    at ( 0.0, 7);
    \coordinate (b30)    at (-1.0, 4);

    \draw [fill]     (a30)   circle [radius=2.0pt] ;
    \draw [fill]     (a31)   circle [radius=2.0pt] ;
    \draw [fill]     (a32)   circle [radius=2.0pt] ;
    \draw [fill]     (a35)   circle [radius=2.0pt] ;
    \draw []     (b30)   circle [radius=2.0pt] ;

    \begin{scope}[>=latex]

      \draw [->, shorten >=  2.75pt, shorten <= 2.75pt] (a30) -- (a31) ;
      \draw [->, shorten >=  2.75pt, shorten <= 2.75pt] (a31) -- (a32) ;
      \draw [->, shorten >= 11.75pt, shorten <= 2.75pt] (a32) -- (a34) ;
      \draw [->, shorten >=  2.75pt, shorten <= 7.75pt] (a34) -- (a35) ;
      \draw [->, shorten >=  2.75pt, shorten <= 2.75pt] (a30) -- (b30) ;

    \end{scope}

      \node []         at (a34)  {$\vdots$};
      \node [left=2pt] at (a35)  {$\Box\bot$};

      \node [right=2pt] at (a30)  {$a^k_{k+2}$};
      \node [right=2pt] at (a31)  {$a^k_{k+1}$};
      \node [right=2pt] at (a32)  {$a^k_{k}$};
      \node [right=2pt] at (a35)  {$a^k_{1}$};

      \node [left=2pt] at (b30)  {$b_k$};

    \end{tikzpicture}
    \caption{Шкала $\kframe{F}_k$}
    \label{fig_BPLcase}
  \end{figure}

Шкала $\kframe{F}_k$ изображена на рис.~\ref{fig_BPLcase}; чёрными кружк\'{а}ми изображены иррефлексивные миры, светлым~--- рефлексивный мир, отношениям достижимости между мирами соответствуют стрелки, при этом те стрелки, которые восстанавливаются по транзитивности
отношения $R_k$, опущены.

Для каждого $w\in W$ обозначим через $\kframe{F}_k^w$ копию шкалы $\kframe{F}_k$, помеченную миром $w$: положим $\kframe{F}_k^w=\otuple{W_k^w, R_k^w}$, где $W^w_k= W_k\times\{w\}$ и
для всяких $x,y\in W_k$
$$
\begin{array}{rcl}
\otuple{x,w} R_k^w \otuple{y,w}
& \iff &
xR_ky.
\end{array}
$$
Положим
$$
\begin{array}{rcl}
W^\ast
  & =
  & \displaystyle
    W\cup\bigcup\{W_k^w : \mbox{$1\leqslant k\leqslant n$, $w\in W$, $(\kmodel{M},w)\not\imodels p_k$}\} \cup\bigcup\{W_{n+1}^w : w\in W\}.
\end{array}
$$
На множестве $W^\ast$ определим отношение $R'$:
$$
\begin{array}{rcl}
wR'w' & \iff &
\mbox{либо $w,w'\in W$ и $wRw'$,}
\\
& &
\mbox{либо $w,w'\in W_k^u$
и $wR_k^uw'$,}
\\
& &
\mbox{либо $w\in W$, $(\kmodel{M},w)\not\imodels p_k$ и $w'=
\langle{a_{k+2}^k,w}\rangle$,}
\\
& &
\mbox{либо $w\in W$ и $w'= \langle{a_{n+3}^{n+1},w}\rangle$.}
\end{array}
$$
Пусть $R^\ast$~--- транзитивное замыкание отношения $R'$. Положим $\kframe{F}^\ast=\otuple{W^\ast,R^\ast}$. Ясно, что $\kframe{F}^\ast$~--- шкала логики $\logic{BPL}$.

Покажем, что $(\kframe{F}^\ast,w_0)\not\imodels \varphi_\alpha$. Заметим, что $\varphi_\alpha$ не содержит переменных, поэтому истинность её подформул в мирах любой шкалы не зависит от оценки в этой шкале.

Для всякой подформулы $\psi$ формулы $\varphi$ через $\psi_\alpha$ обозначим формулу, получающуюся из $\psi$ с помощью подстановки формул $\alpha_1,\dots,\alpha_n$ вместо переменных $p_1,\dots,p_n$.

Покажем, что для всякой подформулы $\psi$ формулы $\varphi$ и для всякого $w\in W$ имеет место следующая эквивалентность:
$$
\begin{array}{rcl}
(\kframe{F}^\ast,w)\imodels\psi_{\alpha}
& \iff &
(\kmodel{M},w)\imodels\psi.
\end{array}
\eqno{\mbox{$({\ast})$}}
$$
Доказательство проведём индукцией по построению~$\psi$.

Базис: $\psi= p_k$.

Пусть $(\kmodel{M},w)\not\imodels p_k$. Тогда в шкале $\kframe{F}^\ast$ из мира $w$ достижима копия шкалы~$\kframe{F}_k$. Нетрудно видеть, что
\[
\begin{array}{l}
(\kframe{F}^\ast,\langle a^k_{k+2},w\rangle)\imodels \Box^{k+2}\bot\to\Box^{k+1}\bot, \\
(\kframe{F}^\ast,\langle a^k_{k+2},w\rangle)\not\imodels (\Box^{k+1}\bot\to\Box^k\bot)\vee\Box^{k+2}\bot,
\end{array}
\]
а поскольку $wR^\ast\langle a^k_{k+2},w\rangle$, то $(\kframe{F}^\ast,w)\not\imodels \alpha_k$, т.\,е. $(\kframe{F}^\ast,w)\not\imodels \psi_\alpha$.

Пусть $(\kmodel{M},w)\imodels p_k$. Нам нужно показать, что в этом случае $(\kframe{F}^\ast,w)\imodels \alpha_k$. Предположим, что $(\kframe{F}^\ast,w)\not\imodels \alpha_k$. Тогда для некоторого мира $w'\in W^\ast$, такого, что $wR^\ast w'$ имеют место отношения
\[
\begin{array}{l}
(\kframe{F}^\ast,w')\imodels \Box^{k+2}\bot\to\Box^{k+1}\bot, \\
(\kframe{F}^\ast,w')\not\imodels (\Box^{k+1}\bot\to\Box^k\bot)\vee\Box^{k+2}\bot.
\end{array}
\]
Заметим, что $w'\not\in W$. Действительно, пусть $w'\in W$. По определению шкалы $\kframe{F}^\ast$, из всякого мира множества $W$ достижима копия шкалы $\kframe{F}_{n+1}$; нетрудно видеть, что
\[
\begin{array}{l}
(\kframe{F}^\ast,\langle a^{n+1}_{k+3},w'\rangle)\not\imodels \Box^{k+2}\bot\to\Box^{k+1}\bot,
\end{array}
\]
а следовательно,
\[
\begin{array}{l}
(\kframe{F}^\ast,w')\not\imodels \Box^{k+2}\bot\to\Box^{k+1}\bot,
\end{array}
\]
что даёт противоречие с полученным выше. Итак, $w'\not\in W$, т.\,е. $w'\in W_m^u$ для некоторого $m\in\set{1,\dots,n\,{+}\,1}$ и некоторого $u\in W$, причём $u= w$ или $wRu$.

Из шкал $\kframe{F}_1,\dots,\kframe{F}_{n+1}$ имеется ровно одна, в которой существует мир~$x$, такой, что
\[
\begin{array}{l}
(\kframe{F}^\ast,\langle x,u\rangle)\imodels \Box^{k+2}\bot\to\Box^{k+1}\bot, \\
(\kframe{F}^\ast,\langle x,u\rangle)\not\imodels (\Box^{k+1}\bot\to\Box^k\bot)\vee\Box^{k+2}\bot,
\end{array}
\]
а именно, это шкала $\kframe{F}_k$, а в качестве такого мира $x$ нужно взять мир~$a_{k+2}^k$. Таким образом, $w'= \langle a_{k+2}^k,u \rangle$. Но последнее равенство означает, что из мира $u$ достижима копия шкалы $\kframe{F}_k$, что, в свою очередь, означает, что $(\kmodel{M},u)\not\imodels p_k$. Поскольку $w= u$ или $wRu$, по определению оценки получаем, что $(\kmodel{M},w)\not\imodels p_k$. Получили противоречие, следовательно, $(\kframe{F}^\ast,w)\imodels \alpha_k$.

Пусть формулы $\psi'$ и $\psi''$ таковы, что для всякого мира $w\in W$
$$
\begin{array}{rcl}
(\kframe{F}^\ast,w)\imodels\psi'_{\alpha}
& \iff &
(\kmodel{M},w)\imodels\psi',
\smallskip \\
(\kframe{F}^\ast,w)\imodels\psi''_{\alpha}
& \iff &
(\kmodel{M},w)\imodels\psi''.
\end{array}
$$

Если $\psi= \psi'\wedge\psi''$, то для всякого $w\in W$ справедливы следующие эквивалентности:
$$
\begin{array}{rcl}
(\kframe{F}^\ast,w)\imodels\psi_{\alpha}
& \iff &
(\kframe{F}^\ast,w)\imodels\psi'_{\alpha}
\mbox{ и }
(\kframe{F}^\ast,w)\imodels\psi''_{\alpha}
\smallskip \\
& &
(\kmodel{M},w)\imodels\psi'
\mbox{ и }
(\kmodel{M},w)\imodels\psi''
\smallskip \\
& &
(\kmodel{M},w)\imodels\psi.
\end{array}
$$

Если $\psi= \psi'\vee\psi''$, то для всякого $w\in W$ справедливы следующие эквивалентности:
$$
\begin{array}{rcl}
(\kframe{F}^\ast,w)\imodels\psi_{\alpha}
& \iff &
(\kframe{F}^\ast,w)\imodels\psi'_{\alpha}
\mbox{ или }
(\kframe{F}^\ast,w)\imodels\psi''_{\alpha}
\smallskip \\
& &
(\kmodel{M},w)\imodels\psi'
\mbox{ или }
(\kmodel{M},w)\imodels\psi''
\smallskip \\
& &
(\kmodel{M},w)\imodels\psi.
\end{array}
$$

Пусть теперь $\psi=\psi'\to\psi''$. Если $(\kmodel{M},w)\not\imodels \psi$, то существует мир $w'$ такой, что $wRw'$, $(\kmodel{M},w')\imodels\psi'$ и $(\kmodel{M},w')\not\imodels \psi''$. По индукционному предположению получаем, что $(\kframe{F}^\ast,w')\imodels\psi'_\alpha$ и $(\kframe{F}^\ast,w')\not\imodels \psi''_\alpha$, а следовательно, $(\kframe{F}^\ast,w)\not\imodels \psi_\alpha$.

Пусть $(\kframe{F}^\ast,w)\not\imodels \psi_\alpha$. Тогда существует мир $w'\in W^\ast$ такой, что $wR^\ast w'$, $(\kframe{F}^\ast,w')\imodels\psi'_\alpha$ и $(\kframe{F}^\ast,w')\not\imodels \psi''_\alpha$.

Заметим, что для всяких $k,m\in\numNp$ имеет место отношение $\kframe{F}_k\imodels\alpha_m$. Действительно, чтобы в некотором мире шкалы $\kframe{F}_k$ опровергалась формула $\alpha_m$, нужно, чтобы из этого мира был достижим мир $a^m_{m+2}$ шкалы $\kframe{F}_m$, что невозможно. Ясно, что в этом случае в шкале $\kframe{F}_k$ будут истинными все формулы, построенные из формул $\alpha_i$ с помощью конъюнкции, дизъюнкции и импликации (т.\,е. без использования~$\bot$). В частности, во всех мирах шкал $\kframe{F}_1,\dots,\kframe{F}_{n+1}$ будет истинна формула $\psi''_\alpha$. Поскольку $(\kframe{F}^\ast,w')\not\imodels \psi''_\alpha$, то можем сделать вывод, что $w'\in W$. Используя индукционное предположение, получаем, что $(\kmodel{M},w')\imodels\psi'$ и $(\kmodel{M},w')\not\imodels \psi''$, а следовательно, $(\kmodel{M},w)\not\imodels \psi$.

Эквивалентность \mbox{$({\ast})$} доказана.

Так как $(\kmodel{M},w_0)\not\imodels \varphi$, получаем, что $(\kframe{F}^\ast,w_0)\not\imodels \varphi_\alpha$.
\end{proof}

Утверждение следующей леммы должно быть очевидным.

\begin{lemma}
Существует алгоритм, который по всякой позитивной формуле $\varphi$ строит формулу $\varphi_\alpha$, при этом время работы алгоритма ограничено некоторым полиномом от длины входной формулы.
\end{lemma}

В качестве следствия из предыдущих двух лемм получаем следующую теорему.

\begin{theorem}
\label{th_BPL(0)}
Константный фрагмент логики\/ $\logic{BPL}$ является\/ $\cclass{PSPACE}$-полным.
\end{theorem}

Заметим, что поскольку логики $\logic{K4}$ и $\logic{wGrz}$ являются модальными напарниками логики $\logic{BPL}$, из теоремы~\ref{th_BPL(0)} следует $\cclass{PSPACE}$-трудность константных фрагментов логик из интервала $[\logic{K4},\logic{wGrz}]$, что является частным случаем теоремы~\ref{th:PSPACE:K4}; теорема~\ref{th_BPL(0)} из теоремы~\ref{th:PSPACE:K4} напрямую, конечно, не следует.

    \subsection{Интервал $[\logic{BPL},\logic{FPL}]$}

Константный фрагмент логики~$\logic{FPL}$ полиномиально разрешим, поскольку $\logic{FPL}$ является суперинтуиционистским фрагментом логики~$\logic{GL}$. В~целом, расширение базисной логики Виссера не может быть сложнее своего модального напарника; это касается и фрагментов от конечного числа переменных. Но мы видели, что фрагмент логики~$\logic{GL}$ от одной переменной является $\cclass{PSPACE}$-полным; покажем, что то же самое можно сказать и о логике~$\logic{FPL}$.

Нам будет важно, что для доказательства $\cclass{PSPACE}$-трудности позитивных фрагментов логик из интервала $[\logic{BPL},\logic{FPL}]$ используются одни и те же формулы: ситуация аналогична той, которую мы подробно рассмотрели в модальном случае, и можно полиномиально свести проблему $\logic{TQBF}$ к логикам из этого интервала, используя одну и ту же сводящую функцию. Мы не будем здесь приводить детали соответствующего полиномиального сведения $\logic{TQBF}$ к логикам из этого интервала; их можно найти, например, в~\cite{Chagrov-1985-1-rus,MR:2003:LI}.

Пусть $p$~--- некоторая пропозициональная переменная. Для всякого $n\in\numNp$ положим (напомним, что $\Box\psi=\top\to\psi$)
$$
\begin{array}{lcl}
\label{alpha_i_FPLcase}
\alpha_n & = & \Box^{2}p\to(\Box^{n+1}\bot\to\Box^n\bot\vee p).
\end{array}
$$
Сразу заметим, что длина формулы $\alpha_n$ ограничена сверху некоторой линейной функцией от~$n$.

Пусть $\varphi$~--- некоторая позитивная формула от переменных $p_1,\dots,p_n$. Через $\varphi_\alpha$
\label{varphi_alpha_FPLcase}
обозначим формулу, получающуюся из $\varphi$ подстановкой формул $\alpha_1,\dots,\alpha_n$ вместо переменных $p_1,\dots,p_n$ соответственно.

\begin{lemma}
\label{lem_base_for_FPL}
Для всякой позитивной формулы $\varphi$ имеет место следующая эквивалентность:
$$
\begin{array}{rcl}
\varphi\in\logic{FPL}
& \iff &
\varphi_\alpha\in\logic{FPL}.
\end{array}
$$
\end{lemma}

\begin{proof}
Пусть $\varphi$~--- позитивная формула от переменных $p_1,\dots,p_n$. Если $\varphi\in\logic{FPL}$, то $\varphi_\alpha\in\logic{FPL}$, поскольку логика $\logic{FPL}$ замкнута относительно правила подстановки, а $\varphi_\alpha$~--- подстановочный пример $\varphi$.

Пусть $\varphi\not\in \logic{FPL}$. Покажем, что в этом случае $\varphi_\alpha\not\in \logic{FPL}$. Так как $\varphi\not\in \logic{FPL}$, то существует модель $\kmodel{M}=\langle\kframe{F},v\rangle$, определённая на шкале $\kframe{F}=\otuple{W,R}$ логики $\logic{FPL}$, такая, что $(\kmodel{M},w_0)\not\imodels \varphi$ для некоторого мира $w_0\in W$. По модели $\kmodel{M}$
построим шкалу $\kframe{F}^\ast$ логики $\logic{FPL}$ такую, что $\kframe{F}^\ast\not\imodels \varphi_\alpha$.

Пусть $\kframe{F}_k=\otuple{W_k,R_k}$~--- шкала Крипке, где
$$
\begin{array}{rcl}
W_k & = & \{a_1^k,\dots,a_{k+2}^k\};
\medskip \\
a_s^kR_ka_t^k & \iff & s\,{>}\,t.
\end{array}
$$

Для каждого $w\in W$ обозначим через $\kframe{F}_k^w$ копию шкалы $\kframe{F}$, помеченную миром~$w$: положим $\kframe{F}_k^w=\langle W_k^w, R_k^w\rangle$, где $W^w_k= W_k\times\{w\}$ и для всяких $x,y\in W$
$$
\begin{array}{rcl}
\langle x,w \rangle R_k^w \langle y,w \rangle
& \iff &
xR_ky.
\end{array}
$$
Положим
$$
\begin{array}{rcl}
W^\ast 
  & = 
  & \displaystyle
    W \cup \bigcup\{\mbox{$W_k^w : 1\leqslant k\leqslant n$, $w\in W$, $(\kmodel{M},w)\not\imodels p_k$}\} \cup \bigcup\{W_{n+1}^w : w\in W\}.
\end{array}
$$
На множестве $W^\ast$ определим отношение $R'$:
$$
\begin{array}{rcl}
wR'w' & \iff &
\mbox{либо $w,w'\in W$ и $wRw'$,}
\\
& &
\mbox{либо $w,w'\in W_k^u$
и $wR_k^uw'$,}
\\
& &
\mbox{либо $w\in W$, $(\kmodel{M},w)\not\imodels p_k$ и $w'=
\langle a_{k+2}^k,w\rangle$,}
\\
& &
\mbox{либо $w\in W$ и $w'= \langle a_{n+3}^{n+1},w\rangle$.}
\end{array}
$$
Пусть $R^\ast$~--- транзитивное замыкание отношения~$R'$. По\-ло\-жим
\linebreak[3]
$\kframe{F}^\ast=\langle W^\ast,R^\ast\rangle$. Тогда $\kframe{F}^\ast$ является конечной иррефлексивной шкалой, а значит, шкалой логики $\logic{FPL}$.

На шкале $\kframe{F}^\ast$ определим модель $\kmodel{M}^\ast=\langle\kframe{F}^\ast,v^\ast\rangle$, положив
$$
\begin{array}{rcl}
(\kmodel{M}^\ast,w)\imodels p
& \iff &
\mbox{$w=\langle a^k_m,u \rangle$ и $k\geqslant m$.}
\end{array}
$$
Покажем, что $(\kmodel{M},w_0)\not\imodels \varphi_\alpha$.

Для всякой подформулы $\psi$ формулы $\varphi$ через $\psi_\alpha$ обозначим формулу, получающуюся из $\psi$ с помощью подстановки формул $\alpha_1,\dots,\alpha_n$ вместо переменных $p_1,\dots,p_n$.

Покажем, что для всякой подформулы $\psi$ формулы $\varphi$ и для всякого $w\in W$ имеет место следующая эквивалентность:
$$
\begin{array}{rcl}
(\kmodel{M}^\ast,w)\imodels\psi_{\alpha}
& \iff &
(\kmodel{M},w)\imodels\psi.
\end{array}
\eqno{\mbox{$({\ast})$}}
$$
Доказательство проведём индукцией по построению~$\psi$.

Базис: $\psi= p_k$.

Пусть $(\kmodel{M},w)\not\imodels p_k$. Тогда в шкале $\kframe{F}^\ast$ из мира $w$ достижима копия шкалы~$\kframe{F}_k$. Нетрудно видеть, что
\[
\begin{array}{l}
(\kmodel{M}^\ast,\langle a^k_{k+2},w\rangle)\imodels \Box^{2}p, \\
(\kmodel{M}^\ast,\langle a^k_{k+2},w\rangle)\not\imodels \Box^{k+1}\bot\to\Box^k\bot\vee p,
\end{array}
\]
а поскольку $wR^\ast\langle a^k_{k+2},w\rangle$, то $(\kmodel{M}^\ast,w)\not\imodels \alpha_k$, т.\,е. $(\kmodel{M}^\ast,w)\not\imodels \psi_\alpha$.

Пусть $(\kmodel{M},w)\imodels p_k$. Нам нужно показать, что в этом случае $(\kmodel{M}^\ast,w)\imodels \alpha_k$. Предположим, что $(\kmodel{M}^\ast,w)\not\imodels \alpha_k$. Тогда для некоторого мира $w'\in W^\ast$ такого, что $wR^\ast w'$ имеют место отношения
\[
\begin{array}{l}
(\kmodel{M}^\ast,w')\imodels \Box^{2}p, \\
(\kmodel{M}^\ast,w')\not\imodels \Box^{k+1}\bot\to\Box^k\bot\vee p.
\end{array}
\]
Заметим, что $w'\not\in W$. Действительно, пусть $w'\in W$. По определению шкалы $\kframe{F}^\ast$, из всякого мира множества $W$ достижима копия шкалы $\kframe{F}_{n+1}$; нетрудно видеть, что
\[
\begin{array}{l}
(\kmodel{M}^\ast,\langle a^{n+1}_{n+3},w'\rangle)\not\imodels \Box^{2}p,
\end{array}
\]
а следовательно,
\[
\begin{array}{l}
(\kmodel{M}^\ast,w')\not\imodels \Box^{2}p,
\end{array}
\]
что противоречит полученному выше. Итак, $w'\not\in W$, т.\,е. $w'\in W_m^u$ для некоторого $m\in\{1,\dots,n + 1\}$ и некоторого $u\in W$, причём $u= w$ или~$wRu$.

Из шкал $\kframe{F}_1,\dots,\kframe{F}_{n+1}$ имеется ровно одна, в которой существует мир $x$ такой, что
\[
\begin{array}{l}
(\kmodel{M}^\ast,\langle x,u\rangle)\imodels \Box^{2}p, \\
(\kmodel{M}^\ast,\langle x,u\rangle)\not\imodels \Box^{k+1}\bot\to\Box^k\bot\vee p,
\end{array}
\]
а именно, это шкала $\kframe{F}_k$, а в качестве такого мира $x$ нужно взять мир $a_{k+2}^k$. Таким образом, $w'= \langle a_{k+2}^k,u \rangle$. Но последнее равенство означает, что из мира $u$ достижима копия шкалы $\kframe{F}_k$, что, в свою очередь, означает, что $(\kmodel{M},u)\not\imodels p_k$. Поскольку $w= u$ или $wRu$, по определению оценки получаем, что $(\kmodel{M},w)\not\imodels p_k$. Получили противоречие, следовательно, $(\kmodel{M}^\ast,w)\imodels \alpha_k$.

Пусть формулы $\psi'$ и $\psi''$ таковы, что для всякого мира $w\in W$
$$
\begin{array}{rcl}
(\kmodel{M}^\ast,w)\imodels\psi'_{\alpha}
& \iff &
(\kmodel{M},w)\imodels\psi',
\smallskip \\
(\kmodel{M}^\ast,w)\imodels\psi''_{\alpha}
& \iff &
(\kmodel{M},w)\imodels\psi''.
\end{array}
$$

Если $\psi= \psi'\wedge\psi''$, то для всякого $w\in W$ справедливы следующие эквивалентности:
$$
\begin{array}{rcl}
(\kmodel{M}^\ast,w)\imodels\psi_{\alpha}
& \iff &
(\kmodel{M}^\ast,w)\imodels\psi'_{\alpha}
\mbox{ и }
(\kmodel{M}^\ast,w)\imodels\psi''_{\alpha}
\smallskip \\
& &
(\kmodel{M},w)\imodels\psi'
\mbox{ и }
(\kmodel{M},w)\imodels\psi''
\smallskip \\
& &
(\kmodel{M},w)\imodels\psi.
\end{array}
$$

Если $\psi= \psi'\vee\psi''$, то для всякого $w\in W$ справедливы следующие эквивалентности:
$$
\begin{array}{rcl}
(\kmodel{M}^\ast,w)\imodels\psi_{\alpha}
& \iff &
(\kmodel{M}^\ast,w)\imodels\psi'_{\alpha}
\mbox{ или }
(\kmodel{M}^\ast,w)\imodels\psi''_{\alpha}
\smallskip \\
& &
(\kmodel{M},w)\imodels\psi'
\mbox{ или }
(\kmodel{M},w)\imodels\psi''
\smallskip \\
& &
(\kmodel{M},w)\imodels\psi.
\end{array}
$$

Пусть теперь $\psi=\psi'\to\psi''$. Если $(\kmodel{M},w)\not\imodels \psi$, то существует мир $w'$ такой, что $wRw'$, $(\kmodel{M},w')\imodels\psi'$ и $(\kmodel{M},w')\not\imodels \psi''$. По индукционному предположению получаем, что $(\kmodel{M}^\ast,w')\imodels\psi'_\alpha$ и $(\kmodel{M}^\ast,w')\not\imodels \psi''_\alpha$, а следовательно, $(\kmodel{M}^\ast,w)\not\imodels \psi_\alpha$.

Пусть $(\kmodel{M}^\ast,w)\not\imodels \psi_\alpha$. Тогда существует мир $w'\in W^\ast$ такой, что $wR^\ast w'$, $(\kmodel{M}^\ast,w')\imodels\psi'_\alpha$ и $(\kmodel{M}^\ast,w')\not\imodels \psi''_\alpha$.

Заметим, что для всякого $u\in W^\ast\setminus W$ и всякого $m\in\numNp$ имеет место отношение $\kmodel{M}_k\imodels\alpha_m$. Действительно, чтобы в некотором мире $u=\langle a^k_s,u' \rangle$ модели $\kmodel{M}^\ast$ опровергалась формула $\alpha_m$, нужно, чтобы из этого мира был достижим мир $\langle a^m_{m+2},u'\rangle$, что невозможно. Ясно, что в этом случае в каждом мире $u\in W^\ast \setminus W$ будут истинными все формулы, построенные из формул $\alpha_i$ с помощью конъюнкции, дизъюнкции и импликации. В частности, во всех мирах множества $W^\ast \setminus W$ будет истинна формула $\psi''_\alpha$. Поскольку $(\kmodel{M}^\ast,w')\not\imodels \psi''_\alpha$, то $w'\in W$. Используя индукционное предположение, получаем, что $(\kmodel{M},w')\imodels\psi'$ и
$(\kmodel{M},w')\not\imodels \psi''$, а следовательно, $(\kmodel{M},w)\not\imodels \psi$.

Эквивалентность \mbox{$({\ast})$} доказана.

Так как $(\kmodel{M},w_0)\not\imodels \varphi$, то получаем, что $(\kmodel{M}^\ast,w_0)\not\imodels \varphi_\alpha$, а значит, $\varphi_\alpha\not\in \logic{FPL}$.
\end{proof}

Утверждение следующей леммы очевидно.

\begin{lemma}
Существует алгоритм, который по всякой позитивной формуле $\varphi$ строит формулу $\varphi_\alpha$, при этом время работы алгоритма ограничено некоторым полиномом от длины входной формулы.
\end{lemma}

В качестве следствия из предыдущих двух лемм получаем следующие теоремы.

\begin{theorem}
\label{th_FPL(1)}
Фрагмент логики\/ $\logic{FPL}$ от одной переменной является\/ $\cclass{PSPACE}$-полным.
\end{theorem}


\begin{theorem}
\label{th_BPL_FPL(1)}
Пусть $L\in[\logic{BPL},\logic{FPL}]$. Тогда фрагмент $L$ от одной переменной является \/ $\cclass{PSPACE}$-трудным.
\end{theorem}

\begin{proof}
Следует из того, что логики из интервала $[\logic{BPL},\logic{FPL}]$ содержат один и тот же $\cclass{PSPACE}$-полный фрагмент, состоящий из позитивных формул.~\cite{Chagrov-1985-1-rus,MR:2003:LI}.
\end{proof}

Заметим, что поскольку $\logic{K4}$ и $\logic{GL}$ являются модальными напарниками логик $\logic{BPL}$ и $\logic{FPL}$, из теоремы~\ref{th_BPL_FPL(1)} следует $\cclass{PSPACE}$-трудность фрагментов от одной переменной логик из интервала $[\logic{K4},\logic{GL}]$, что является частным случаем теоремы~\ref{th:PSPACE:GL:Grz}; теорема~\ref{th_BPL_FPL(1)} из теоремы~\ref{th:PSPACE:GL:Grz} напрямую не следует.

    \subsection{Логика ${\logic{Int}}$}
    \label{ssec:Int-positive}

Обратим внимание на то, что доказательства $\cclass{PSPACE}$-трудности фрагментов от одной переменной логик из интервала $[\logic{BPL},\logic{FPL}]$ и их модальных напарников из интервала $[\logic{K4},\logic{GL}]$ очень близки; то же относится и к доказательствам $\cclass{PSPACE}$-полноты константных фрагментов логик $\logic{BPL}$, $\logic{K4}$ и~$\logic{wGrz}$. Что касается суперинтуиционистских логик, то, при условии, что $\cclass{P}\ne\cclass{PSPACE}$, так не получится. Так, минимальная суперинтуиционистская логика $\logic{Int}$ является суперинтуиционистским фрагментом каждой модальной логики $L\in[\logic{S4},\logic{Grz}]$, и, как следствие, фрагмент от одной переменной логики $\logic{Int}$ является суперинтуиционистским фрагментом фрагмента от одной переменной логики~$L$. Но фрагмент от одной переменной такой логики $L$ является $\cclass{PSPACE}$-полным, а фрагмент от одной переменной логики $\logic{Int}$ полиномиально разрешим, что несложно получается с помощью решётки Ригера--Нишимуры~\cite{Rieger52,Nishimura60}.

Ввиду сказанного мы обратимся к вопросу о сложности фрагмента логики $\logic{Int}$ от двух переменных. Покажем, что уже позитивный фрагмент $\logic{Int}$ от двух переменных является $\cclass{PSPACE}$-трудным.

Пусть $p$ и $q$~--- две пропозициональные переменные. Определим следующие формулы:
$$
\begin{array}{rcl}
\label{p-q-formulas_for_KC}
D_1 & = & p\to q; \phantom{A^1_1} \smallskip \\
D_2 & = & q\to p; \phantom{A^1_1} \smallskip \\
D_3 & = & D_1\wedge D_2 \to p\vee q;\phantom{A^1_1} \smallskip \\
A_1^0 & = & D_2\to D_1 \vee D_3; \smallskip \\
A_2^0 & = & D_3\to D_1 \vee D_2; \smallskip \\
B_1^0 & = & D_1\to D_2 \vee D_3; \smallskip \\
B_2^0 & = & A_1^0\wedge A_2^0\wedge B_1^0 \to
            D_1\vee D_2 \vee D_3;
\end{array}
~~~
\begin{array}{rcl}
A_1^1 & = & A_1^0\wedge A_2^0 \to B_1^0\vee B_2^0;
\smallskip \\
A_2^1 & = & A_1^0\wedge B_1^0 \to A_2^0\vee B_2^0;
\smallskip \\
A_3^1 & = & A_1^0\wedge B_2^0 \to A_2^0\vee B_1^0;
\smallskip \\
B_1^1 & = & A_2^0\wedge B_1^0 \to A_1^0\vee B_2^0;
\smallskip \\
B_2^1 & = & A_2^0\wedge B_2^0 \to A_1^0\vee B_1^0;
\smallskip \\
B_3^1 & = & B_1^0\wedge B_2^0 \to A_1^0\vee A_2^0.
\smallskip \\
\phantom{B_3^1}
\end{array}
$$
Пусть определены формулы $A_1^k,\dots,A_m^k$ и $B_1^k,\dots,B_m^k$, где $k\in \numNp$. Определим формулы $A_1^{k+1},\dots,A_n^{k+1}$ и $B_1^{k+1},\dots,B_n^{k+1}$, где $n = (m-1)^2$. Для этого упорядочим лексико-графически пары индексов $\pair{i}{j}$, перенумеруем их числами от $1$ до $n$, и если пара $\pair{i}{j}$ имеет в этой нумерации номер~$m$, положим
$$
\begin{array}{rcl}
A_m^{k+1} & = & A_1^k\,{\to}\,B_1^k\,{\vee}\,A_i^k\,{\vee}\,B_j^k; \\
B_m^{k+1} & = & B_1^k\,{\to}\,A_1^k\,{\vee}\,A_i^k\,{\vee}\,B_j^k;
\end{array}
$$
Формулы $A_i^k$ и $B_i^k$ будем называть формулами уровня $k$. Обозначим через $N_k$ число формул $A^k_i$ уровня~$k$ (ясно, что $N_k$ будет также и числом формул $B^k_i$ уровня~$k$). Нетрудно видеть, что
$$
\begin{array}{ccc}
N_0 = 2, & N_1 = 3, & N_{k+2} = (N_{k+1}-1)^2.
\end{array}
$$
Пусть
$$
\begin{array}{rcl}
l_0 & = & \displaystyle\max_{\mathclap{1\leqslant i\leqslant 2}}\{|A_i^0|+|B_i^0|\}.
\end{array}
$$
Нетрудно видеть, что $|A_i^k|\,{<}\,5^k\cdot l_0$ и $|B_i^k|\,{<}\,5^k\cdot l_0$.

Заметим, что существует такое $k_0$, что $5^{k}\cdot l_0 < N_k$ для всякого $k\geqslant k_0$, поскольку
$$
\begin{array}{rcl}
\displaystyle\lim\limits_{k\to\infty}\frac{5^k}{N_k} & = & 0.
\end{array}
$$
Фактически, можно взять $k_0 = 6$.

Теперь построим модель, в которой будут опровергаться все определённые выше формулы, при этом для каждой формулы будет существовать наибольший мир, в котором она опровергается. Пусть
$$
\begin{array}{rcl}
W & = & \{c_0, d_1, d_2, d_3\} \cup
        \{a_i^k,b_i^k : \mbox{$k\geqslant 0$, $1\leqslant i\leqslant{N_k}$}\}.
\end{array}
$$
Элементы $a_i^k$ и $b_j^k$ будем называть мирами уровня~$k$. На множестве $W$ определим отношение достижимости~$R$. Положим
$$
\begin{array}{rcl}
R_0   & = & \{\langle d_1, c_0\rangle, \hfill
            \langle d_2, c_0\rangle,   \hfill
            \langle d_3, c_0\rangle,   \hfill
            \langle b_2^0, d_1\rangle, \hfill
            \langle b_2^0, d_2\rangle, \hfill
            \langle b_2^0, d_3\rangle, \hfill~
            \smallskip \\
      & &    
             ~ \hfill
             \langle a_1^0, d_1\rangle, \hfill
             \langle a_1^0, d_3\rangle, \hfill
             \langle a_2^0, d_1\rangle, \hfill
             \langle a_2^0, d_2\rangle, \hfill
             \langle b_1^0, d_2\rangle, \hfill
             \langle b_1^0, d_3\rangle\};
            \medskip \\
R_1^a & = & \{\langle a_1^1,b_1^0\rangle,
            \langle a_1^1,b_2^0\rangle,
            \langle a_2^1,a_2^0\rangle,
            \langle a_2^1,b_2^0\rangle,
            \langle a_3^1,a_2^0\rangle,
            \langle a_3^1,b_1^0\rangle\};
            \smallskip \\
R_1^b & = & \{\langle b_1^1,a_1^0\rangle, \hfill
            \langle b_1^1,b_2^0\rangle, \hfill
            \langle b_2^1,a_1^0\rangle, \hfill
            \langle b_2^1,b_1^0\rangle, \hfill
            \langle b_3^1,a_1^0\rangle, \hfill
            \langle b_3^1,a_2^0\rangle\};
            \smallskip \\
R_1   & = & R_1^a\cup R_1^b.
\end{array}
$$
Пусть для всякого $k\geqslant 1$
$$
\begin{array}{rcl}
R_{k+1}^a & = & \{\langle a^{k+1}_m,b^k_1\rangle,
                  \langle a^{k+1}_m,a^k_i\rangle,
                  \langle a^{k+1}_m,b^k_j\rangle :
                  A^{k+1}_m = A^k_1\to B^k_1\vee A^k_i\vee B^k_j\};
                  \smallskip \\
R_{k+1}^b & = & \{\langle b^{k+1}_m,a^k_1\rangle, \hfill
                  \langle b^{k+1}_m,a^k_i\rangle, \hfill
                  \langle b^{k+1}_m,b^k_j\rangle  \hfill: \hfill
                  B^{k+1}_m = B^k_1\to A^k_1\vee A^k_i\vee B^k_j\};
                  \smallskip \\
R_{k+1} & = & R_{k+1}^a\cup R_{k+1}^b.
\end{array}
$$
Обозначим
$$
\begin{array}{rcl}
R' & = & \displaystyle\bigcup\limits_{k=0}^{\infty}R_{k}
\end{array}
$$
и в качестве отношения достижимости $R$ возьмём реф\-лек\-сив\-но-тран\-зи\-тив\-ное замыкание отношения $R'$. Положим
$\kframe{F}=\langle W,R\rangle$, $\kmodel{M}=\langle\kframe{F},v\rangle$, где оценка $v$ определена следующим образом:
$$
\begin{array}{rcl}
(\kmodel{M},w)\imodels p & \iff &
\mbox{$w= c_0$ или $w= d_1$;}
\smallskip \\
(\kmodel{M},w)\imodels q & \iff &
\mbox{$w= c_0$ или $w= d_2$,}
\end{array}
$$
см. рис.~\ref{fig:F-2}.

\begin{figure}
  \centering
  \begin{tikzpicture}[scale=1.5]

\coordinate (c0)  at (3, 8);
\coordinate (d1)  at (2, 7);
\coordinate (d3)  at (3, 7);
\coordinate (d2)  at (4, 7);
\coordinate (a01) at (2, 6);
\coordinate (a02) at (3, 6);
\coordinate (b01) at (4, 6);
\coordinate (b02) at (5, 6);
\coordinate (b13) at (1, 5);
\coordinate (b12) at (2, 5);
\coordinate (b11) at (3, 5);
\coordinate (a13) at (4, 5);
\coordinate (a12) at (5, 5);
\coordinate (a11) at (6, 5);

\coordinate (dots1) at (3.5, 4.25);

\coordinate (ak1) at (1, 3.5);
\coordinate (aki) at (2, 3.5);
\coordinate (bkj) at (5, 3.5);
\coordinate (bk1) at (6, 3.5);

\coordinate (dots1ak) at (1.5, 3.5);
\coordinate (dots2ak) at (2.5, 3.5);
\coordinate (dots1bk) at (5.5, 3.5);
\coordinate (dots2bk) at (4.5, 3.5);

\coordinate (akp1m) at (5, 2.5);
\coordinate (bkp1m) at (2, 2.5);

\coordinate (dots1akp1m) at (4.5, 2.5);
\coordinate (dots2akp1m) at (5.5, 2.5);
\coordinate (dots1bkp1m) at (1.5, 2.5);
\coordinate (dots2bkp1m) at (2.5, 2.5);

\coordinate (dots2) at (3.5, 1.75);

\draw [] (c0)  circle [radius=2.0pt] ;
\draw [] (d1)  circle [radius=2.0pt] ;
\draw [] (d3)  circle [radius=2.0pt] ;
\draw [] (d2)  circle [radius=2.0pt] ;
\draw [] (a01) circle [radius=2.0pt] ;
\draw [] (a02) circle [radius=2.0pt] ;
\draw [] (b01) circle [radius=2.0pt] ;
\draw [] (b02) circle [radius=2.0pt] ;
\draw [] (b13) circle [radius=2.0pt] ;
\draw [] (b12) circle [radius=2.0pt] ;
\draw [] (b11) circle [radius=2.0pt] ;
\draw [] (a13) circle [radius=2.0pt] ;
\draw [] (a12) circle [radius=2.0pt] ;
\draw [] (a11) circle [radius=2.0pt] ;

\draw [] (dots1) node {$\ldots$};

\draw [] (ak1) circle [radius=2.0pt] ;
\draw [] (aki) circle [radius=2.0pt] ;
\draw [] (bkj) circle [radius=2.0pt] ;
\draw [] (bk1) circle [radius=2.0pt] ;

\draw [] (dots1ak) node {$\ldots$};
\draw [] (dots2ak) node {$\ldots$};
\draw [] (dots1bk) node {$\ldots$};
\draw [] (dots2bk) node {$\ldots$};

\draw [] (akp1m) circle [radius=2.0pt] ;
\draw [] (bkp1m) circle [radius=2.0pt] ;

\draw [] (dots1akp1m) node {$\ldots$};
\draw [] (dots2akp1m) node {$\ldots$};
\draw [] (dots1bkp1m) node {$\ldots$};
\draw [] (dots2bkp1m) node {$\ldots$};

\draw [] (dots2) node {$\ldots$};

\begin{scope}[>=latex]
\draw [->, shorten >= 2.75pt, shorten <= 2.75pt] (d1) -- (c0);
\draw [->, shorten >= 2.75pt, shorten <= 2.75pt] (d2) -- (c0);
\draw [->, shorten >= 2.75pt, shorten <= 2.75pt] (d3) -- (c0);
\draw [->, shorten >= 2.75pt, shorten <= 2.75pt] (a01) -- (d1);
\draw [->, shorten >= 2.75pt, shorten <= 2.75pt] (a01) -- (d3);
\draw [->, shorten >= 2.75pt, shorten <= 2.75pt] (a02) -- (d1);
\draw [->, shorten >= 2.75pt, shorten <= 2.75pt] (a02) -- (d2);
\draw [->, shorten >= 2.75pt, shorten <= 2.75pt] (b01) -- (d3);
\draw [->, shorten >= 2.75pt, shorten <= 2.75pt] (b01) -- (d2);
\draw [->, shorten >= 2.75pt, shorten <= 2.75pt] (b02) -- (d1);
\draw [->, shorten >= 2.75pt, shorten <= 2.75pt] (b02) -- (d2);
\draw [->, shorten >= 2.75pt, shorten <= 2.75pt] (b02) -- (d3);

\draw [->, shorten >= 2.75pt, shorten <= 2.75pt] (b13) -- (a01);
\draw [->, shorten >= 2.75pt, shorten <= 2.75pt] (b13) -- (a02);
\draw [->, shorten >= 2.75pt, shorten <= 2.75pt] (b12) -- (a01);
\draw [->, shorten >= 2.75pt, shorten <= 2.75pt] (b12) -- (b01);
\draw [->, shorten >= 2.75pt, shorten <= 2.75pt] (b11) -- (a01);
\draw [->, shorten >= 2.75pt, shorten <= 2.75pt] (b11) -- (b02);
\draw [->, shorten >= 2.75pt, shorten <= 2.75pt] (a13) -- (a02);
\draw [->, shorten >= 2.75pt, shorten <= 2.75pt] (a13) -- (b01);
\draw [->, shorten >= 2.75pt, shorten <= 2.75pt] (a12) -- (a02);
\draw [->, shorten >= 2.75pt, shorten <= 2.75pt] (a12) -- (b02);
\draw [->, shorten >= 2.75pt, shorten <= 2.75pt] (a11) -- (b01);
\draw [->, shorten >= 2.75pt, shorten <= 2.75pt] (a11) -- (b02);

\draw [->, shorten >= 2.75pt, shorten <= 2.75pt] (bkp1m) -- (ak1);
\draw [->, shorten >= 2.75pt, shorten <= 2.75pt] (bkp1m) -- (aki);
\draw [->, shorten >= 2.75pt, shorten <= 2.75pt] (bkp1m) -- (bkj);
\draw [->, shorten >= 2.75pt, shorten <= 2.75pt] (akp1m) -- (bk1);
\draw [->, shorten >= 2.75pt, shorten <= 2.75pt] (akp1m) -- (aki);
\draw [->, shorten >= 2.75pt, shorten <= 2.75pt] (akp1m) -- (bkj);

\end{scope}

\node [      left =2pt] at (c0)  {$c_0$}    ;
\node [      left =2pt] at (d1)  {$d_1$}    ;
\node [      right=2pt] at (d2)  {$d_2$}    ;
\node [      left =2pt] at (d3)  {$d_3$}    ;
\node [      left =2pt] at (a01) {$a^0_1$}  ;
\node [      left =2pt] at (a02) {$a^0_2$}  ;
\node [      right=2pt] at (b01) {$b^0_1$}  ;
\node [      right=2pt] at (b02) {$b^0_2$}  ;

\node [below right] at (a11) {$a^1_1$} ;
\node [below right] at (a12) {$a^1_2$} ;
\node [below right] at (a13) {$a^1_3$} ;
\node [below left ] at (b11) {$b^1_1$} ;
\node [below left ] at (b12) {$b^1_2$} ;
\node [below left ] at (b13) {$b^1_3$} ;

\node [above left ] at (ak1) {$a^k_1$} ;
\node [above left ] at (aki) {$a^k_i$} ;
\node [above right] at (bk1) {$b^k_1$} ;
\node [above right] at (bkj) {$b^k_j$} ;

\node [below right] at (akp1m) {$a^{k+1}_m$} ;
\node [below left ] at (bkp1m) {$b^{k+1}_m$} ;

\node [above] at (d1)  {$p$}    ;
\node [above] at (d2)  {$q$}    ;
\node [above right] at (c0)  {$p,q$}  ;

\end{tikzpicture}

\caption{Модель $\kmodel{M}$}
  \label{fig:F-2}
\end{figure}

\begin{lemma}
\label{lem_akbk}
Пусть $w$~--- мир модели $\kmodel{M}$. Тогда для всякого $i\in\{1,2,3\}$
$$
\begin{array}{rcl}
(\kmodel{M},w)\not\imodels D_i & \iff & wRd_i.
\end{array}
$$
\end{lemma}

\begin{proof}
Поскольку $d_1$~--- единственный мир, в котором истинно $p$ и не истинно $q$, а $d_2$~--- единственный мир, в котором истинно $q$ и не истинно $p$, то
$$
\begin{array}{rcl}
(\kmodel{M},w)\not\imodels D_1 & \iff & wRd_1;
\smallskip \\
(\kmodel{M},w)\not\imodels D_2 & \iff & wRd_2.
\end{array}
$$
Точно так же, $d_3$~--- единственный мир, в котором опровергаются и $p$, и $q$, из которого при этом не достижимы миры $d_1$ и $d_2$, а значит,
$$
\begin{array}{rcl}
(\kmodel{M},w)\not\imodels D_3 & \iff & wRd_3.
\end{array}
$$
\end{proof}

\begin{lemma}
Пусть $w$~--- мир модели $\kmodel{M}$. Тогда
$$
\begin{array}{rcl}
(\kmodel{M},w)\not\imodels A^k_m & \iff & wRa^k_m;
\smallskip \\
(\kmodel{M},w)\not\imodels B^k_m & \iff & wRb^k_m.
\end{array}
$$
\end{lemma}

\begin{proof}
Индукция по $k$. Пусть $k= 0$.

Имеется всего четыре формулы нулевого уровня: $A_1^0$, $A_2^0$, $B_1^0$ и~$B_2^0$.

Поскольку $(\kmodel{M},a_1^0)\imodels D_2$ и $(\kmodel{M},a_1^0)\not\imodels D_1\vee D_3$, то $(\kmodel{M},a_1^0)\not\imodels A_1^0$, а следовательно, для всякого мира $w$ такого, что $wRa_1^0$, имеет место отношение $(\kmodel{M},w)\not\imodels A_1^0$. Пусть теперь для некоторого мира $w\in W$ выполнено отношение $(\kmodel{M},w)\not\imodels A_1^0$. Тогда существует мир $w'$, достижимый из $w$, такой, что $(\kmodel{M},w')\imodels D_2$ и $(\kmodel{M},w')\not\imodels D_1\vee D_3$. Но тогда из $w'$ должны быть достижимы миры $d_1$ и $d_3$ и должен быть недостижим мир~$d_2$. Заметим, что $w'$ не может быть миром уровня $1$ или выше, поскольку из всякого такого мира достижим мир~$d_2$; но среди оставшихся миров модели $\kmodel{M}$ указанными свойствами обладает только мир~$a_1^0$.
Следовательно, $w'= a_1^0$, и значит,~$wRa_1^0$.

Поскольку $(\kmodel{M},a_2^0)\imodels D_3$ и $(\kmodel{M},a_2^0)\not\imodels D_1\vee D_2$, то $(\kmodel{M},a_2^0)\not\imodels A_2^0$, а следовательно, для всякого мира $w$ такого, что $wRa_2^0$, имеет место отношение $(\kmodel{M},w)\not\imodels A_2^0$. Пусть теперь для некоторого
мира $w\in W$ выполнено отношение $(\kmodel{M},w)\not\imodels A_2^0$. Тогда существует мир $w'$,
достижимый из $w$, такой, что $(\kmodel{M},w')\imodels D_3$ и $(\kmodel{M},w')\not\imodels D_1\vee D_2$. Но тогда из $w'$ должны быть достижимы миры $d_1$ и $d_2$ и должен быть недостижим мир~$d_3$. Заметим, что $w'$ не может быть миром уровня $1$ или выше, поскольку из всякого такого мира достижим мир $d_2$; но среди оставшихся миров модели $\kmodel{M}$ указанными свойствами обладает только мир~$a_2^0$. Следовательно, $w'= a_2^0$, и значит,~$wRa_2^0$.

Поскольку $(\kmodel{M},b_1^0)\imodels D_1$ и $(\kmodel{M},b_1^0)\not\imodels D_2\vee D_3$, то $(\kmodel{M},b_1^0)\not\imodels B_1^0$, а следовательно, для всякого мира $w$ такого, что $wRb_1^0$, имеет место отношение $(\kmodel{M},w)\not\imodels B_1^0$. Пусть теперь для некоторого мира $w\in W$ выполнено отношение $(\kmodel{M},w)\not\imodels B_1^0$. Тогда существует мир $w'$, достижимый из $w$, такой, что $(\kmodel{M},w')\imodels D_1$ и $(\kmodel{M},w')\not\imodels D_2\vee D_3$. Но тогда из $w'$ должны быть достижимы миры $d_2$ и $d_3$ и должен быть недостижим мир~$d_1$. Заметим, что $w'$ не может быть миром уровня $1$ или выше, поскольку из всякого такого мира достижим мир~$d_2$; но среди оставшихся миров модели $\kmodel{M}$ указанными свойствами обладает только мир~$b_1^0$.
Следовательно, $w'= b_1^0$, и значит,~$wRb_1^0$.

Поскольку из $b_2^0$ не достижимы миры $a_1^0$, $a_2^0$, $b_1^0$, то из доказанного выше следует, что $(\kmodel{M},b_2^0)\imodels A_1^0\wedge A_2^0 \wedge B_1^0$, а поскольку из $b_2^0$ достижимы миры $d_1$, $d_2$, $d_3$, то $(\kmodel{M},b_2^0)\not\imodels D_1\vee D_2\vee D_3$. Но тогда $(\kmodel{M},b_2^0)\not\imodels B_2^0$, а следовательно, для всякого мира $w$ такого, что $wRb_2^0$, имеет место отношение $(\kmodel{M},w)\not\imodels B_2^0$. Пусть теперь для некоторого мира $w\in W$ выполнено отношение $(\kmodel{M},w)\not\imodels B_2^0$. Тогда существует мир $w'$, достижимый из $w$, такой, что $(\kmodel{M},w')\imodels A_1^0\wedge A_2^0 \wedge B_1^0$ и $(\kmodel{M},w')\not\imodels D_1\vee D_2\vee D_3$. Но тогда из $w'$ должны быть достижимы миры $d_1$, $d_2$, $d_3$ и должны быть недостижимы миры $a_1^0$,~$a_2^0$,~$b_1^0$. Заметим, что $w'$ не может быть миром уровня $1$ или выше, поскольку из всякого такого мира достижим хотя бы один из миров $a_1^0$,~$a_2^0$,~$b_1^0$. Среди оставшихся миров модели $\kmodel{M}$ указанными свойствами обладает только мир~$b_2^0$.  Следовательно, $w'= b_2^0$, и значит,~$wRb_2^0$.

Пусть для всякого мира $w$ модели $\kmodel{M}$ и всякого $m\leqslant N_k$ имеют место следующие эквивалентности:
$$
\begin{array}{rcl}
(\kmodel{M},w)\not\imodels A^k_m & \iff & wRa^k_m;
\smallskip \\
(\kmodel{M},w)\not\imodels B^k_m & \iff & wRb^k_m.
\end{array}
$$
Пусть
$$
\begin{array}{cc}
A_m^{k+1} = A_1^k\to B_1^k\vee A_i^k\vee B_j^k,
&
B_m^{k+1} = B_1^k\to A_1^k\vee A_i^k\vee B_j^k,
\end{array}
$$
где $i,j\in\{2,\dots,N_k\}$. Покажем, что
$$
\begin{array}{rcl}
(\kmodel{M},w)\not\imodels A^{k+1}_m
& \iff &
wRa^{k+1}_m;
\smallskip \\
(\kmodel{M},w)\not\imodels B^{k+1}_m
& \iff &
wRb^{k+1}_m.
\end{array}
$$

Поскольку $a_m^{k+1}Rb_1^k$, $a_m^{k+1}Ra_i^k$, $a_m^{k+1}Rb_j^k$, то по индукционному предположению $(\kmodel{M},a_m^{k+1})\not\imodels B_1^k\vee A_i^k\vee B_j^k$, а поскольку неверно, что $a_m^{k+1}Ra_1^k$, то по индукционному предположению $(\kmodel{M},a_m^{k+1})\imodels A_1^k$.  Следовательно, для всякого $w$ такого, что $wRa_m^{k+1}$, имеет место отношение $(\kmodel{M},w)\not\imodels A^{k+1}_m$.

Аналогично обосновывается, что для всякого $w$ такого, что $wRb_m^{k+1}$, имеет место отношение $(\kmodel{M},w)\not\imodels B^{k+1}_m$.

Пусть $(\kmodel{M},w)\not\imodels A^{k+1}_m$ для некоторого мира $w$ модели $\kmodel{M}$. Тогда существует достижимый из $w$ мир $w'$ такой, что $(\kmodel{M},w')\imodels A_1^k$ и $(\kmodel{M},w')\not\imodels B_1^k\vee A_i^k\vee B_j^k$. По индукционному предположению, из $w'$ должны быть достижимы миры $b_1^k$, $a_i^k$, $b_j^k$ и должен быть недостижим мир~$a_1^k$. Поскольку различные миры одного и того же уровня друг из друга не достижимы, а миры более высокого уровня не достижимы из миров более низкого уровня, то $w'$ является миром уровня не ниже, чем $k+1$. С другой стороны, для всякого мира $w''$ более высокого уровня, чем $k + 1$, имеет место отношение $w''Ra_1^k$, и значит, $(\kmodel{M},w'')\not\imodels A_1^k$. Следовательно, $w'$ является миром уровня не выше, чем~$k + 1$. Итак, $w'$~--- мир уровня $k + 1$, из которого достижимы миры
$b_1^k$, $a_i^k$, $b_j^k$ и недостижим мир~$a_1^k$. По определению шкалы $\kframe{F}$ это возможно только в том случае, когда $w'= a_s^{k+1}$, где $A_s^{k+1} = A_1^k\to B_1^k\vee A_i^k\vee B_j^k$. Ясно, что в нашем случае $s= m$, т.\,е. $w'= a_m^{k+1}$. Следовательно,~$wRa_m^{k+1}$.

Аналогично доказывается, что если $(\kmodel{M},w)\not\imodels B^{k+1}_m$, то $wRb_m^{k+1}$.
\end{proof}

Пусть $\varphi$~--- позитивная формула, $p_1,\dots,p_n$~--- все её переменные. Обозначим через $k$
наименьшее натуральное число, удовлетворяющее отношению
$|\varphi|\,{<}\,5^{k}\cdot l_0$.
Заметим, что
$$
N_{k+k_0} ~>~ 5^{k+k_0}\cdot l_0 ~>~ 5^{k_0}\cdot |\varphi| ~>~
|\varphi| ~>~ n,
$$
поэтому следующее определение корректно: для всякого $i\in\{1,\dots,n\}$ определим формулу $\alpha_i$, положив
$$
\begin{array}{lcl}
\alpha_i & = & A_i^{k+k_0}\vee B_i^{k+k_0}.
\end{array}
\label{def_alpha_IntKC}
$$
Обозначим через $\varphi_\alpha$
\label{def_varphi_alpha_IntKC}
формулу, которая получается из $\varphi$ подстановкой формул $\alpha_1,\dots,\alpha_n$ вместо переменных $p_1,\dots,p_n$ соответственно.

\begin{lemma}
Для некоторого натурального числа $m$ имеет место отношение $|\varphi_\alpha| < m\cdot |\varphi|^2$.
\end{lemma}

\begin{proof}
Так как $k$~--- наименьшее натуральное число, удовлетворяющее отношению
$|\varphi| < 5^{k}\cdot l_0$, то
$5^{k-1}\cdot l_0 \leqslant |\varphi|$,
и следовательно,
$$
5^{k+k_0}\cdot l_0 ~ \leqslant ~ 5^{k_0+1}\cdot|\varphi|.
$$
Поскольку $|A_i^{k+k_0}| < 5^{k+k_0}\cdot l_0$ и $|B_i^{k+k_0}| < 5^{k+k_0}\cdot l_0$, то
$$
|\alpha_i| ~<~ 2\cdot 5^{k+k_0}\cdot l_0 ~\leqslant~ 2\cdot 5^{k_0+1}\cdot |\varphi|.
$$
Следовательно,
$$
|\varphi_\alpha| ~\leqslant~ |\varphi|\cdot \max_{\mathclap{1\leqslant i\leqslant n}}|\alpha_i|
~<~ 2\cdot 5^{k_0+1}\cdot |\varphi|^2,
$$
т.\,е. в качестве $m$ можно взять число $2\cdot 5^{k_0+1}$.
\end{proof}

\begin{lemma}
\label{lem_mainforInt}
Для позитивной формулы $\varphi$ имеет место следующая эквивалентность:
$$
\begin{array}{rcl}
\varphi\in\logic{Int}
& \iff &
\varphi_\alpha\in\logic{Int}.
\end{array}
$$
\end{lemma}

\begin{proof}
Если $\varphi\in\logic{Int}$, то $\varphi_\alpha\in\logic{Int}$, поскольку $\varphi_\alpha$ является подстановочным примером формулы~$\varphi$.

Пусть $\varphi$~--- позитивная интуиционистская формула и пусть $\varphi\not\in \logic{Int}$. Тогда существует интуиционистская модель $\kmodel{M}_\varphi=\langle\kframe{F}_\varphi,v_\varphi\rangle$, определённая на $\logic{Int}$-шкале $\kframe{F}_\varphi=\langle W_\varphi,R_\varphi\rangle$, такая, что $(\kmodel{M}_\varphi,w_0)\not\imodels \varphi$ для некоторого $w_0\in W_\varphi$.

Построим интуиционистскую модель, в некотором мире которой будет опровергаться формула~$\varphi_\alpha$. Без ограничений общности можем считать, что $W\cap W_\varphi=\varnothing$. Положим $W^\ast= W\cup W_\varphi$. На множестве $W^\ast$ рассмотрим следующее отношение~$R'$:
$$
\begin{array}{rcl}
R' & = & \{\langle w,a_i^{k+k_0}\rangle,\langle w,b_i^{k+k_0}\rangle :
\mbox{$w\in W_\varphi$, $(\kmodel{M}_\varphi,w)\not\imodels p_i$, $1\leqslant i\leqslant n$}\}\cup {}
\smallskip \\
& &
{}\cup\{\langle w,a_{n+1}^{k+k_0}\rangle,
\langle w,b_{n+1}^{k+k_0}\rangle : w\in W_\varphi\}.
\end{array}
$$
Пусть $R^\ast$~--- рефлексивно-транзитивное замыкание отношения $R\cup R_\varphi\cup R'$ и пусть $\kframe{F}^\ast=\langle W^\ast,R^\ast\rangle$. На шкале $\kframe{F}^\ast$ определим модель $\kmodel{M}^\ast=\langle\kframe{F}^\ast,v^\ast\rangle$,
\label{KC-model}
положив для всякого $w\in W^\ast$
$$
\begin{array}{rcl}
(\kmodel{M}^\ast,w)\imodels p & \iff &
\mbox{$w= c_0$ или $w= d_1$;}
\smallskip \\
(\kmodel{M}^\ast,w)\imodels q & \iff &
\mbox{$w= c_0$ или $w= d_2$.}
\end{array}
$$

Для всякой подформулы $\psi$ формулы $\varphi$ через $\psi_\alpha$ обозначим формулу, получающуюся из $\psi$ подстановкой формул $\alpha_1,\dots,\alpha_n$ вместо переменных $p_1,\dots,p_n$ соответственно. Индукцией по построению $\psi$ докажем, что для всякого $w\in W_\varphi$
$$
\begin{array}{rcl}
(\kmodel{M}^\ast,w)\imodels\psi_\alpha
& \iff &
(\kmodel{M}_\varphi,w)\imodels\psi.
\end{array}
$$

Пусть $\psi= p_m$. Если $(\kmodel{M}_\varphi,w)\not\imodels p_m$, то в модели $\kmodel{M}^\ast$ из мира $w$ достижимы миры $a_m^{k+k_0}$ и $b_m^{k+k_0}$. Но $(\kmodel{M},a_m^{k+k_0})\not\imodels A_m^{k+k_0}$, $(\kmodel{M},b_m^{k+k_0})\not\imodels B_m^{k+k_0}$. Понятно, что в этом случае также имеют место отношения $(\kmodel{M}^\ast,a_m^{k+k_0})\not\imodels A_m^{k+k_0}$ и $(\kmodel{M}^\ast,b_m^{k+k_0})\not\imodels B_m^{k+k_0}$, а следовательно, $(\kmodel{M}^\ast,w)\not\imodels \alpha_m$, т.\,е. $(\kmodel{M}^\ast,w)\not\imodels \psi_\alpha$.

Пусть теперь $(\kmodel{M}^\ast,w)\not\imodels \alpha_m$ для некоторого мира $w\in W_\varphi$; пусть при этом
$$
\begin{array}{lcl}
A_m^{k+k_0} & = & A_1^{k+k_0-1}\to B_1^{k+k_0-1}\vee A_i^{k+k_0-1}\vee B_{j}^{k+k_0-1}; \\
B_m^{k+k_0} & = & B_1^{k+k_0-1}\to A_1^{k+k_0-1}\vee A_i^{k+k_0-1}\vee B_{j}^{k+k_0-1}.
\end{array}
$$
Тогда в модели
$\kmodel{M}^\ast$ существуют миры $w'$ и $w''$, такие, что $wR^\ast w'$, $wR^\ast w''$, и при этом
$$
\begin{array}{rrr}
(\kmodel{M}^\ast,w')\imodels A_1^{k+k_0-1}, &
(\kmodel{M}^\ast,w')\not\imodels  B_1^{k+k_0-1}, &
(\kmodel{M}^\ast,w')\not\imodels  A_i^{k+k_0-1}\vee B_j^{k+k_0-1};
\smallskip \\
(\kmodel{M}^\ast,w'')\not\imodels  A_1^{k+k_0-1}, &
(\kmodel{M}^\ast,w'')\imodels B_1^{k+k_0-1}, &
(\kmodel{M}^\ast,w'')\not\imodels  A_i^{k+k_0-1}\vee B_j^{k+k_0-1}.
\end{array}
$$
Заметим, что $w',w''\not\in W_\varphi$. Действительно, для всякого мира $u\in W_\varphi$ имеют место отношения $uR^\ast a_{n+1}^{k+k_0}$ и $uR^\ast b_{n+1}^{k+k_0}$; но $(\kmodel{M}^\ast,a_{n+1}^{k+k_0})\not\imodels B_{1}^{k+k_0-1}$, а $(\kmodel{M}^\ast,b_{n+1}^{k+k_0})\not\imodels A_{1}^{k+k_0-1}$, поэтому $(\kmodel{M}^\ast,u)\not\imodels A_{1}^{k+k_0-1}$ и $(\kmodel{M}^\ast,u)\not\imodels B_{1}^{k+k_0-1}$, и поскольку $(\kmodel{M}^\ast,w')\imodels A_1^{k+k_0-1}$, а $(\kmodel{M}^\ast,w'')\imodels B_1^{k+k_0-1}$, получаем, что $w',w''\not\in W_\varphi$.

Так как миры $w'$ и $w''$ достижимы по $R^\ast$ из некоторого мира множества $W_\varphi$, то они не могут быть мирами более высокого уровня, чем $k + k_0$. С другой стороны, $(\kmodel{M}^\ast,w')\not\imodels  A_m^{k+k_0}$, а $(\kmodel{M}^\ast,w'')\not\imodels  B_m^{k+k_0}$, поэтому $w'$ и $w''$ не могут быть мирами уровня ниже, чем $k + k_0$. Таким образом, $w'$ и $w''$~--- миры уровня $k + k_0$. Но среди миров уровня $k + k_0$ формула $A_m^{k+k_0}$ опровергается только в мире $a_m^{k+k_0}$, а формула $B_m^{k+k_0}$~--- только в мире $b_m^{k+k_0}$, следовательно $w'= a_m^{k+k_0}$, а $w''= b_m^{k+k_0}$.

Итак, $wR^\ast a_m^{k+k_0}$. Это означает, что при построении шкалы $\kframe{F}^\ast$ мы положили $uR^\ast a_m^{k+k_0}$ для некоторого $u\in W_\varphi$ такого, что $wR_\varphi u$. Но если при построении шкалы $\kframe{F}^\ast$ мы положили $uR^\ast a_m^{k+k_0}$, то мы также должны были положить $uR^\ast b_m^{k+k_0}$, и сделать это мы могли только в том случае, когда $(\kmodel{M}_\varphi,u)\not\imodels p_m$. Поскольку $wR_\varphi u$, то получаем, что $(\kmodel{M}_\varphi,w)\not\imodels p_m$, и тем самым базис индукции обоснован.

Пусть теперь $\psi'$ и $\psi''$~--- подформулы формулы $\varphi$, такие, что для всякого мира $w\in W_\varphi$
$$
\begin{array}{rcl}
(\kmodel{M}^\ast,w)\imodels\psi'_\alpha
& \iff &
(\kmodel{M}_\varphi,w)\imodels\psi';
\\
(\kmodel{M}^\ast,w)\imodels\psi''_\alpha
& \iff &
(\kmodel{M}_\varphi,w)\imodels\psi''.
\end{array}
$$

Пусть $\psi=\psi'\wedge\psi''$. Тогда для всякого $w\in W_\varphi$
$$
\begin{array}{rcl}
(\kmodel{M}^\ast,w)\imodels\psi_\alpha
& \iff &
(\kmodel{M}^\ast,w)\imodels\psi'_\alpha
\mbox{ и }
(\kmodel{M}^\ast,w)\imodels\psi''_\alpha
\smallskip \\
& \iff &
(\kmodel{M}_\varphi,w)\imodels\psi'
\mbox{ и }
(\kmodel{M}_\varphi,w)\imodels\psi''
\smallskip \\
& \iff &
(\kmodel{M}_\varphi,w)\imodels\psi.
\end{array}
$$

Пусть $\psi=\psi'\vee\psi''$. Тогда для всякого $w\in W_\varphi$
$$
\begin{array}{rcl}
(\kmodel{M}^\ast,w)\imodels\psi_\alpha
& \iff &
(\kmodel{M}^\ast,w)\imodels\psi'_\alpha
\mbox{ или }
(\kmodel{M}^\ast,w)\imodels\psi''_\alpha
\smallskip \\
& \iff &
(\kmodel{M}_\varphi,w)\imodels\psi'
\mbox{ или }
(\kmodel{M}_\varphi,w)\imodels\psi''
\smallskip \\
& \iff &
(\kmodel{M}_\varphi,w)\imodels\psi.
\end{array}
$$

Пусть $\psi=\psi'\to\psi''$. Если $(\kmodel{M}_\varphi,w)\not\imodels \psi$, то в модели $\kmodel{M}_\varphi$ существует мир $w'$, достижимый из $w$, такой, что $(\kmodel{M}_\varphi,w')\imodels\psi'$ и $(\kmodel{M}_\varphi,w')\not\imodels \psi''$; но тогда $(\kmodel{M}^\ast,w')\imodels\psi'_\alpha$ и $(\kmodel{M}^\ast,w')\not\imodels \psi''_\alpha$, а следовательно, $(\kmodel{M}^\ast,w)\not\imodels \psi_\alpha$.

Пусть $(\kmodel{M}^\ast,w)\not\imodels \psi_\alpha$ для некоторого $w\in W_\varphi$. Тогда существует $w'\in W^\ast$ такой, что $wR^\ast w'$, $(\kmodel{M}^\ast,w')\imodels\psi'_\alpha$ и $(\kmodel{M}^\ast,w')\not\imodels \psi''_\alpha$. Заметим, что во всяком мире $u\in W$ уровня не выше $k + k_0$ формулы $\alpha_1,\dots,\alpha_n$ истинны. Следовательно, во всяком таком мире будут истинны и все формулы, построенные из $\alpha_1,\dots,\alpha_n$ с помощью конъюнкции, дизъюнкции и импликации, в частности, во всяком таком мире должна быть истинной формула~$\psi''_\alpha$. Поскольку $(\kmodel{M}^\ast,w')\not\imodels \psi''_\alpha$, то $w'$ не может быть миром из множества $W$, и следовательно, $w'\in W_\varphi$. Тогда, применяя индукционное предположение, получаем, что $(\kmodel{M}_\varphi,w')\imodels\psi'$ и $(\kmodel{M}_\varphi,w')\not\imodels \psi''$, и значит, $(\kmodel{M}_\varphi,w)\not\imodels \psi$.
\end{proof}

\begin{remark}
\label{rem:BPL:fa}
Заметим, что модель $\kmodel{M}^\ast$, построенная в доказательстве, является бесконечной, чего можно избежать: если исходная модель $\kmodel{M}_\varphi$ конечна (а этого можно добиться ввиду финитной аппроксимируемости логики~$\logic{Int}$), то в описанной выше конструкции нужно <<присоединять>> к ней не всю модель $\kmodel{M}$, а только ту её подмодель, миры которой достижимы в модели $\kmodel{M}^\ast$ из миров модели~$\kmodel{M}_\varphi$.
\end{remark}

В качестве следствия получаем следующую теорему.

\begin{theorem}
\label{th_PSPACE_Int+(2)}
Фрагмент от двух переменных логики\/ $\logic{Int}^+$ является $\cclass{PSPACE}$-полным.
\end{theorem}

    \subsection{Интервал $[\logic{BPL},\logic{KC}]$}
    \label{ssec:BPL-KC}

Напомним, что логика слабого закона исключённого третьего $\logic{KC}$ определяется
синтаксически добавлением к $\logic{Int}$ слабого закона исключённого
третьего:
$$
\begin{array}{rcl}
\logic{KC} & = & \logic{Int} + \neg p\vee\neg\neg p.
\end{array}
$$
Семантически $\logic{KC}$ описывается корневыми конвергентными шкалами Крипке, т.\,е. такими, в которых для любых двух миров существует мир, достижимый из каждого из них, см.~\cite{ChZ}.

Известно~\cite{Jankov-1968-1-rus}, что логика $\logic{KC}$ является наибольшей логикой, имеющей тот же позитивный фрагмент, что и~$\logic{Int}$, поэтому из теоремы~\ref{th_PSPACE_Int+(2)} получаем следующее утверждение.

\begin{corollary}
\label{cor_Int-KC}
Пусть $L\in[\logic{Int},\logic{KC}]$. Тогда позитивный фрагмент от двух переменных логики $L$ является\/ $\cclass{PSPACE}$-трудным.
\end{corollary}

Поскольку все логики из интервала $[\logic{BPL},\logic{KC}]$ содержат некоторый общий $\cclass{PSPACE}$-полный позитивный фрагмент~\cite{Chagrov-1985-1-rus,MR:2003:LI}, получаем более сильное утверждение.

\begin{theorem}
\label{th_BPL-KC}
Пусть $L\in[\logic{BPL},\logic{KC}]$. Тогда позитивный фрагмент от двух переменных логики $L$ является\/ $\cclass{PSPACE}$-трудным.
\end{theorem}

В случае самой логики $\logic{KC}$ формулировку последних утверждений можно уточнить.

\begin{theorem}
\label{th_KC}
Позитивный фрагмент от двух переменных логики\/ $\logic{KC}$ является $\cclass{PSPACE}$-полным.
\end{theorem}

Отметим, что следствие~\ref{cor_Int-KC} можно доказать и напрямую, т.\,е. не используя тот факт, что $\logic{Int}^+ = \logic{KC}^+$. Для этого достаточно заметить, что шкала $\kframe{F}^\ast$, возникающая в доказательстве леммы~\ref{lem_mainforInt}, является конвергентной.

Приведём примеры логик, которые удовлетворяют условию следствия~\ref{cor_Int-KC}.

Прежде всего, это логики, аксиоматизируемые над $\logic{Int}$ формулой от одной переменной. В~\cite{Nishimura60} описаны все такие логики, и из~\cite{Nishimura60} следует, что все непротиворечивые суперинтуиционистские логики, аксиоматизируемые над $\logic{Int}$ формулой от одной переменной и отличные от $\logic{Cl}$, содержатся в $\logic{KC}$. В результате получаем следующее утверждение.

\begin{corollary}
Пусть $\varphi$~--- интуиционистская формула от одной переменной и пусть логика $L + \varphi$ непротиворечива и отлична от\/ $\logic{Cl}$. Тогда фрагмент от двух переменных логики $L$ является $\cclass{PSPACE}$-трудным.
\end{corollary}

Ещё один класс логик, содержащихся в $\logic{KC}$,~--- это класс логик, аксиоматизируемых безымпликативными формулами. Формулу $\varphi$ называем \defnotion{безымпликативной},\index{уяа@формула!безымпликативная} если она построена из переменных с помощью $\wedge$, $\vee$ и~$\neg$. Все безымликативные формулы (с~точностью до дедуктивной эквивалентности) описаны в~\cite{Carpinska-1981-1}. Именно, для всяких $k,i\in\numNp$, таких, что $i\leqslant k$, положим
$$
\begin{array}{rcl}
\delta_k^i & = &
\neg(p_1\wedge\dots\wedge p_{i-1}\wedge \neg p_i
\wedge p_{i+1}\wedge\dots\wedge p_k).
\end{array}
$$
Теперь положим
$$
\begin{array}{rclcrcl}
\tau_{-1} & = & p_1, & &
\tau_0    & = & p_1 \vee \neg p_1; \\
\end{array}
$$
и для всякого $k\in\numNp$ положим
$$
\begin{array}{rcl}
\tau_k    & = & \delta_k^1 \vee \dots \vee \delta_k^k.
\end{array}
$$
Множество $\{\tau_{-1},\tau_0,\tau_1,\tau_2,\dots\}$ обладает следующими свойствами: формулы этого множества являются безымпликативными, они попарно дедуктивно неэквивалентны над $\logic{Int}$, любая безымпликативная формула дедуктивно эквивалентна над $\logic{Int}$ одной из формул этого множества.

Ясно, что $\logic{Int}+\tau_{-1}$~--- противоречивая логика, а $\logic{Int}+\tau_{0}=\logic{Cl}$.  Несложно понять, что если $k\in\numNp$, то $\logic{Int} + \tau_{k}\subseteq \logic{KC}$, поэтому мы получаем ещё одно следствие.

\begin{corollary}
Пусть $L = \logic{Int} + \tau_k$, где $k\in\numNp$. Тогда позитивный фрагмент от двух переменных логики $L$ является $\cclass{PSPACE}$-трудным.
\end{corollary}

Теперь рассмотрим ещё две логики~--- логику Крайзеля--Патнэма $\logic{KP}$ и логику Медведева~$\logic{ML}$. Напомним, что
$$
\begin{array}{rcl}
\logic{KP} & = & \logic{Int} +
(\neg p\to q\vee r)\to(\neg p\to q)\vee(\neg p\to r).
\end{array}
$$

Логика конечных задач Медведева, или просто логика Медведева $\logic{ML}$ посредством шкал Крипке определяется так. Для всякого $n\in\numNp$ обозначим через $W_n$ множество непустых подмножеств множества $\{1,\dots,n\}$, а через $R_n$~--- отношение на $W_n$, заданное следующим образом: для всяких $w,w'\in W_n$
$$
\begin{array}{rcl}
wR_nw' & \bydef & w'\subseteq w.
\end{array}
$$
Пусть $\kframe{F}_n = \otuple{W_n,R_n}$. Ясно, что отношение достижимости в $\kframe{F}_n$ рефлексивно, транзитивно и антисимметрично. Обозначим
$\scls{C}_{M} = \{\kframe{F}_n : n\in\numNp\}$.
Тогда
$$
\begin{array}{rcl}
\logic{ML} & = & \set{\varphi : \scls{C}_M\imodels\varphi}.
\end{array}
$$

Несложно понять, что множество интуиционистских формул, не принадлежащих $\logic{ML}$, рекурсивно перечислимо. При этом неизвестно, является ли логика $\logic{ML}$ разрешимой.

Логика $\logic{KP}$ разрешима~\cite{Gabbay-1970-1}. При этом независимо от того, разрешима $\logic{ML}$ или нет, можно утверждать, что проблема разрешения для обеих логик достаточно сложна: во-первых, неизвестно, принадлежит ли проблема разрешения логики $\logic{KP}$ (а тем более, $\logic{ML}$) классу $\cclass{PSPACE}$, а во-вторых, для аппроксимации логик из интервала $[\logic{KP},\logic{ML}]$ требуются шкалы Крипке с нижней оценкой числа миров порядка $2^{2^n}$, где $n$~--- длина тестируемой формулы, см.~\cite{ChZ}.

Из~\cite{Szatkowski-1981-1} следует, что $\logic{ML}^+ = \logic{Int}^+$, и в результате, учитывая~\cite{Jankov-1968-1-rus}, получаем, что $\logic{ML}\subseteq\logic{KC}$. Поскольку $\logic{KP}\subseteq\logic{ML}$, мы получаем ещё одно следствие.

\begin{corollary}
Позитивные фрагменты от двух переменных логик\/ $\logic{KP}$ и\/ $\logic{ML}$ являются\/ $\cclass{PSPACE}$-трудными.
\end{corollary}

    \subsection{Модальные напарники логики $\logic{KC}$}

Теперь обратимся к модальным напарникам рассмотренных суперинтуиционистских логик, и прежде всего к модальным напарникам~$\logic{KC}$, поскольку это наибольшая логика среди тех, для которых мы обосновали $\cclass{PSPACE}$-трудность их фрагментов от двух переменных. Модальными напарниками логики $\logic{KC}$ являются логики из интервала $[\logic{S4.2},\logic{Grz.2}]$, где
$$
\begin{array}{lclcl}
\logic{S4.2}  & = & \logic{S4}  & \!\!\oplus\!\! & \Diamond\Box p \to \Box\Diamond p; \\
\logic{Grz.2} & = & \logic{Grz} & \!\!\oplus\!\! & \Diamond\Box p \to \Box\Diamond p. \\
\end{array}
$$

Из теоремы~\ref{th_KC} автоматически получаем, что фрагменты от двух переменных логик из интервала $[\logic{S4},\logic{Grz.2}]$ являются $\cclass{PSPACE}$-трудными.

\begin{theorem}
\label{S4.2}
Пусть $L\in[\logic{K},\logic{Grz.2}]$. Тогда фрагмент от двух переменных логики $L$ является\/ $\cclass{PSPACE}$-трудным.
\end{theorem}

Аналогичные результаты получаются и для расширений логик $\logic{GL}$ и $\logic{wGrz}$. Чтобы сформулировать их, определим логики $\logic{GL.2}^\ast$ и $\logic{wGrz.2}$, расширяющие логику~$\logic{K4.2}$:
$$
\begin{array}{lclcl}
\logic{K4.2}   & = & \logic{K4}   & \!\!\oplus\!\! & \Diamond(p\wedge\Box q) \to \Box(p\vee\Diamond q); \\
\logic{GL.2}^\ast   & = & \logic{GL.2} & \!\!\oplus\!\! & \Diamond(p\wedge\Box^+ q) \to \Box(p\vee\Diamond^+ q); \\
\logic{wGrz.2} & = & \logic{K4.2} & \!\!\oplus\!\! & \logic{wGrz}. \\
\end{array}
$$
Тогда справедлива следующая теорема.

\begin{theorem}
\label{GL.2}
Пусть $L\in[\logic{K},\logic{GL.2}^\ast]\cup [\logic{K},\logic{wGrz.2}]$. Тогда фрагмент от двух переменных логики $L$ является\/ $\cclass{PSPACE}$-трудным.
\end{theorem}


Конечно, можно расширить указанные результаты и на ненормальные логики: именно, в формулировках всех теорем этого раздела можно заменить логику $\logic{K}$ на~$\logic{K0}$.

  \section{Сложность аппроксимации}
    \subsection{Функция сложности}

Функция сложности $\function{f_L}{\numN}{\numN}$ логики $L$, являющейся расширением базисной логики Виссера, определяется аналогично тому, как это было сделано в модальном случае (см.~\cite[раздел~18.1]{ChZ}). Именно, если $L$~--- полная по Крипке финитно аппроксимируемая логика, то
$$
\begin{array}{lcl}
f_L(n)
  & =
  & \displaystyle
    \max\big\{\min\{|\kframe{F}| : \mbox{$\kframe{F}\imodels L$, $\kframe{F}\not\imodels\varphi$}\}
         : \mbox{$|\varphi| \leqslant n$, $\varphi \not\in L$}\big\},
\end{array}
$$
где $|\kframe{F}|$~--- число миров в шкале $\kframe{F}$, а $|\varphi|$~--- длина формулы $\varphi$.

\begin{theorem}
\label{th:complexity:function:BPL}
Функции сложности для константного фрагмента логики $\logic{BPL}$, фрагментов от одной переменной логик из интервала\/ $[\logic{BPL},\logic{FPL}]$, а также для фрагментов от двух переменных логик из интервала\/ $[\logic{BPL},\logic{KC}]$ ограничены снизу экспонентой.
\end{theorem}

\begin{proof}
Достаточно взять в описанной выше конструкции формулы, которые строятся в~\cite{Statman-1979-1}, \cite{Chagrov-1985-1-rus} или~\cite{MR:2003:LI} по булевым формулам с кванторами вида $\forall p_1\ldots\forall p_n\,\top$. При увеличении $n$ длина соответствующих интуиционистских формул растёт полиномиально, а наименьший размер шкал Крипке, в которых опровергаются эти формулы, растёт экспоненциально (кванторы всеобщности моделируются формулами, требующими бинарного ветвления в мире модели), см.~также замечание~\ref{rem:BPL:fa}. Доказательство также получается, если взять формулы М.\,В.\,Захарьящева~\cite{ZakharyaschevPopov-1980-1-rus} (см.~также~\cite[раздел~18.2]{ChZ}), предварительно промоделировав в них константу $\bot$ с помощью новой пропозициональной переменной (см.~предложение~\ref{prop:Int:positivization}).
\end{proof}

    \subsection{Полиномиально аппроксимируемые логики}
    \label{ssec:poly:approx:int}

Как и в случае модальных логик, идеи А.\,В.\,Кузнецова позволяют строить полиномиальные погружения полиномиально аппроксимируемых суперинтуиционистских логик (а также других расширений базисной логики Виссера) в классическую пропозициональную логику; пример можно найти в~\cite[раздел~18.1]{ChZ}.

Покажем, что существуют линейно аппроксимируемые суперинтуиционистские логики, фрагмент от двух переменных которых может быть сколь угодно сложным. Для этого мы будем использовать шкалу Крипке, известную как \defnotion{лестница Ригера--Нишимуры},\index{лестница Ригера--Нишимуры} где особым образом определена оценка, а также формулы Ригера--Нишимуры, соответствующие мирам получающейся модели~\cite{Rieger52,Nishimura60}.

\begin{figure}
\centering
\begin{tikzpicture}[scale=1.5]

\coordinate (w0)   at (+0.0000,+0.0000);
\coordinate (w1)   at (+1.0000,+0.0000);
\coordinate (w2)   at (+0.0000,-1.0000);
\coordinate (w3)   at (+1.0000,-1.0000);
\coordinate (w4)   at (+0.0000,-2.0000);
\coordinate (w5)   at (+1.0000,-2.0000);
\coordinate (w6)   at (+0.0000,-3.0000);
\coordinate (w7)   at (+1.0000,-3.0000);
\coordinate (w8)   at (+0.0000,-4.0000);
\coordinate (w9)   at (+1.0000,-4.0000);
\coordinate (w10*) at (+0.5000,-4.0000);
\coordinate (dots) at (+0.5000,-4.5000);

\begin{scope}[>=latex]

\draw [->,  shorten >= 1.5pt, shorten <= 1.5pt] (w2)  -- (w0);
\draw [->,  shorten >= 1.5pt, shorten <= 1.5pt] (w4)  -- (w2);
\draw [->,  shorten >= 1.5pt, shorten <= 1.5pt] (w6)  -- (w4);
\draw [->,  shorten >= 1.5pt, shorten <= 1.5pt] (w8)  -- (w6);

\draw [->,  shorten >= 1.5pt, shorten <= 1.5pt] (w3)  -- (w1);
\draw [->,  shorten >= 1.5pt, shorten <= 1.5pt] (w5)  -- (w3);
\draw [->,  shorten >= 1.5pt, shorten <= 1.5pt] (w7)  -- (w5);
\draw [->,  shorten >= 1.5pt, shorten <= 1.5pt] (w9)  -- (w7);

\draw [->,  shorten >= 1.5pt, shorten <= 1.5pt] (w3)  -- (w0);
\draw [->,  shorten >= 1.5pt, shorten <= 1.5pt] (w5)  -- (w2);
\draw [->,  shorten >= 1.5pt, shorten <= 1.5pt] (w7)  -- (w4);
\draw [->,  shorten >= 1.5pt, shorten <= 1.5pt] (w9)  -- (w6);

\draw [->,  shorten >= 1.5pt, shorten <= 1.5pt] (w4)  -- (w1);
\draw [->,  shorten >= 1.5pt, shorten <= 1.5pt] (w6)  -- (w3);
\draw [->,  shorten >= 1.5pt, shorten <= 1.5pt] (w8)  -- (w5);
\draw [->,  shorten >= 1.5pt, shorten <= 1.5pt] (w10*)-- (w7);

\end{scope}

\node [      right] at (w1)   {${w_1}$}     ;
\node [      right] at (w3)   {${w_3}$}     ;
\node [      right] at (w5)   {${w_5}$}     ;
\node [      right] at (w7)   {${w_7}$}     ;
\node [      right] at (w9)   {${w_9}$}     ;

\node [      left ] at (w0)   {${w_0}$}     ;
\node [      left ] at (w2)   {${w_2}$}     ;
\node [      left ] at (w4)   {${w_4}$}     ;
\node [      left ] at (w6)   {${w_6}$}     ;
\node [      left ] at (w8)   {${w_8}$}     ;

\node [above      ] at (w0)   {${p}$}       ;
\node               at (dots) {$\cdots$}    ;


\draw [] (w0)   circle [radius=1.5pt]   ;
\draw [] (w1)   circle [radius=1.5pt]   ;
\draw [] (w2)   circle [radius=1.5pt]   ;
\draw [] (w3)   circle [radius=1.5pt]   ;
\draw [] (w4)   circle [radius=1.5pt]   ;
\draw [] (w5)   circle [radius=1.5pt]   ;
\draw [] (w6)   circle [radius=1.5pt]   ;
\draw [] (w7)   circle [radius=1.5pt]   ;
\draw [] (w8)   circle [radius=1.5pt]   ;
\draw [] (w9)   circle [radius=1.5pt]   ;

\end{tikzpicture}

\caption{Лестница Ригера--Нишимуры}
\label{fig:Rieger_Nishimura}
\end{figure}

Лестница Ригера--Нишимуры~--- это интуиционистская шкала $\kframe{L} = \langle V, S \rangle$ (см.~рис.~\ref{fig:Rieger_Nishimura}), где
\begin{itemize}
\item $V = \{ w_n : n \in \numN \}$;
\item $S$~--- рефлексивно-транзитивное замыкание отношения
  $$
  \begin{array}{l}
  \{ \langle w_2, w_0 \rangle, \langle w_3, w_0 \rangle, \langle w_3,
  w_1 \rangle \} \cup {}\\
    {}\cup\{ \langle w_{2n}, w_{2n - 2} \rangle, \langle
  w_{2n}, w_{2n - 3} \rangle : n \geqslant 2 \}
    \cup {}\\
    {} \cup \{ \langle w_{2n+1}, w_{2n - 1} \rangle, \langle
  w_{2n+1}, w_{2n - 2} \rangle : n \geqslant 2 \}.
  \end{array}
  $$
\end{itemize}

Формулы Ригера--Нишимуры определяются рекурсивно с использованием единственной пропозициональной переменной~$p$:
$$
\begin{array}{lcll}
  \nu_0 & = & p; \\
  \nu_1 & = & \neg p; \\
  \nu_2 & = & \neg \neg p; \\
  \nu_3 & = & \neg \neg p \imp p; \\
  \nu_{n+4} & = & \nu_{n+3} \imp \nu_n \vee \nu_{n+1},
                 & \mbox{где $n \in \numN$.}
\end{array}
$$

Нам будет интересна модель Ригера--Нишимуры $\kmodel{N} = \langle \kframe{L}, \bar{v} \rangle$, где оценка $\bar{v}$ определена следующим образом:
\begin{itemize}
\item $\bar{v} (p) = \{w_0\}$.
\end{itemize}

Следующий факт является хорошо известным~\cite{Nishimura60}:
  \label{prop:nishimura}
для каждого $n \in \numN$
  $$
  \begin{array}{lcl}
  (\kmodel{N},u) \imodels \nu_n & \iff & u \in S(w_n).
  \end{array}
  \eqno{\mbox{$({\ast})$}}
  $$

\begin{figure}
\centering

\begin{tikzpicture}[scale=1.5]
\coordinate (u0)   at (+0.0000,+0.0000);
\begin{scope}[>=latex]
\end{scope}
\node [ left ] at (u0)   {${u_0}$}     ;
\draw [] (u0)   circle [radius=1.5pt]   ;
\node [ right] at (u0)   {$~p$}         ;
\end{tikzpicture}
~~~~~~~~~
\begin{tikzpicture}[scale=1.5]
\coordinate (u0)   at (+0.0000,+0.0000);
\begin{scope}[>=latex]
\end{scope}
\node [ left ] at (u0)   {${u_0}$}     ;
\draw [] (u0)   circle [radius=1.5pt]   ;
\end{tikzpicture}
~~~~~~~~~
\begin{tikzpicture}[scale=1.5]
\coordinate (u0)   at (+0.0000,+0.0000);
\coordinate (u1)   at (+0.0000,-1.0000);
\begin{scope}[>=latex]
\draw [->,  shorten >= 1.5pt, shorten <= 1.5pt] (u1) -- (u0);
\end{scope}
\node [ left ] at (u0)   {${u_0}$}     ;
\node [ left ] at (u1)   {${u_1}$}     ;
\node [ right] at (u0)   {$p$}         ;
\draw [] (u0)   circle [radius=1.5pt]   ;
\draw [] (u1)   circle [radius=1.5pt]   ;
\end{tikzpicture}
~~~~~~~~~
\begin{tikzpicture}[scale=1.5]
\coordinate (u0)   at (+0.0000,+0.0000);
\coordinate (u1)   at (+1.0000,+0.0000);
\coordinate (u2)   at (+0.5000,-1.0000);
\begin{scope}[>=latex]
\draw [->,  shorten >= 1.5pt, shorten <= 1.5pt] (u2) -- (u0);
\draw [->,  shorten >= 1.5pt, shorten <= 1.5pt] (u2) -- (u1);
\end{scope}
\node [ left ] at (u0)   {${u_0}$}     ;
\node [ left ] at (u1)   {${u_1}$}     ;
\node [ left ] at (u2)   {${u_2}$}     ;
\node [ right] at (u1)   {$p$}         ;
\draw [] (u0)   circle [radius=1.5pt]   ;
\draw [] (u1)   circle [radius=1.5pt]   ;
\draw [] (u2)   circle [radius=1.5pt]   ;
\end{tikzpicture}

\caption{Модели  $\kmodel{L}_0$, $\kmodel{L}_1$, $\kmodel{L}_2$, $\kmodel{L}_3$}
\label{fig:models}
\end{figure}

Определим четыре модели (см. рис.~\ref{fig:models}), которые понадобятся нам для технических целей:
\begin{itemize}
\item $\kmodel{L}_0 = \langle U_0, Q_0, v_0 \rangle$, где
  \[
  \begin{array}{lcl}
    U_0     & = & \{ u_0 \}, \\
    Q_0     & = & \{ \langle u_0, u_0 \rangle \}, \\
    v_0 (p) & = & \{u_0\}; \\
  \end{array}
  \]
\item $\kmodel{L}_1 = \langle U_1, Q_1, v_1 \rangle$, где
  \[
  \begin{array}{lcl}
    U_1 & = & \{ u_0 \}, \\
    Q_1 & = & \{ \langle u_0, u_0 \rangle \}, \\
    v_1 (p) & = & \varnothing; \\
  \end{array}
  \]
\item $\kmodel{L}_2 = \langle U_2, Q_2, v_2 \rangle$, где
  \[
  \begin{array}{lcl}
    U_2 & = & \{ u_0, u_1 \}, \\
    Q_2 & = & \{ \langle u_1, u_1\rangle, \langle u_1, u_0\rangle, \langle u_0, u_0 \rangle \}, \\
    v_2 (p) & = & \{u_0\}; \\
  \end{array}
  \]
\item $\kmodel{L}_3 = \langle U_3, Q_3, v_3 \rangle$, где
  \[
  \begin{array}{lcl}
    U_3 & = & \{ u_0, u_1, u_2 \}, \\
    Q_3 & = & \{ \langle u_2, u_2\rangle, \langle u_2, u_1\rangle, \langle u_2, u_0\rangle, \langle u_1, u_1\rangle, \langle u_0, u_0 \rangle \}, \\
    v_3 (p) & = & \{u_1\}. \\
  \end{array}
  \]
\end{itemize}

\begin{lemma}
  \label{lem:models}
Пусть $k\in\{0,1,2,3\}$ и $n\geqslant k+2$. Тогда $\kmodel{L}_{k} \imodels \nu_n$.
\end{lemma}

\begin{proof}
  Рассмотрим случай для каждого из возможных значений $k$ отдельно.

  Пусть $k=0$.
  Заметим, что $\kmodel{L}_0 \imodels \{\nu_0, \nu_2, \nu_3\}$. Если
  $n \geqslant 4$, то заключение формулы $\nu_n$ содержит или
  $\nu_0$, или $\nu_2$, или $\nu_k$ для некоторого $k \geqslant 3$.

  Пусть $k=1$.
  Заметим, что $\kmodel{L}_1 \imodels \{\nu_1, \nu_3\}$.  Если
  $n \geqslant 4$, то заключение формулы $\nu_n$ содержит или
  $\nu_1$, или $\nu_k$ для некоторого $k \geqslant 3$.

  Пусть $k=2$.
  Заметим, что $\kmodel{L}_2 \imodels \{\nu_2, \nu_4 \}$.  Если
  $n \geqslant 5$, то заключение формулы $\nu_n$ содержит или
  $\nu_2$, или $\nu_k$ для некоторого $k \geqslant 4$.

  Пусть $k=3$.
  Заметим, что $\kmodel{L}_3 \imodels \{\nu_3, \nu_5 \}$.  Если
  $n \geqslant 6$, то заключение формулы $\nu_n$ содержит или
  $\nu_3$, или $\nu_k$ для некоторого $k \geqslant 5$.
\end{proof}

Для дальнейшего изложения нам будет удобным воспользоваться понятием \defnotion{псевдоморфизма},\index{морфизм!p@псевдоморфизм} или, коротко, p\nobreakdash-морфизма. Пусть $\kmodel{M} = \langle W, R, v \rangle$ и $\frak{M}' = \langle W', R', v' \rangle$~--- модели на $\logic{Int}$-шкалах.  Отображение $\function{f}{W}{W'}$ называется \defnotion{p\nobreakdash-морфизмом}\index{морфизм!p@p-морфизм} модели $\kmodel{M}$ на модель $\kmodel{M}'$, если оно сюръективно и для любых $w, u \in W$ и $q \in \prop$
\[
\begin{array}{llrl}
\arrayitem & w R u & \Longrightarrow & f(w) R' f(u); \\
\arrayitem & f(w) R' f(u) & \Longrightarrow & \mbox{существует такой $y \in R(w)$, что $f(y) = f(u)$;} \\
\arrayitem & w \in v(q) & \iff & f(w) \in v'(q).
\end{array}
\]
Следующее утверждение является хорошо известным~\cite[теорема~2.15]{ChZ} и легко проверяется:
если $\function{f}{W}{W'}$~--- p\nobreakdash-морфизм модели $\kmodel{M}$ на модель $\kmodel{M}'$, то для каждого $w \in W$ и каждой $\lang{L}$\nobreakdash-формулы~$\varphi$
$$
  \begin{array}{lcl}
  (\kmodel{M}, w)  \imodels \varphi & \iff & (\kmodel{M}', f(w))  \imodels \varphi.
  \end{array}
  \eqno{\mbox{$({\ast}{\ast})$}}
$$

\begin{lemma}
  \label{lem:ladder-p-morph}
  Пусть $v$~--- оценка в шкале $\kframe{L}$, такая, что $v(p) \ne \bar{v} (p)$. Тогда существует p-морфизм модели $\langle \kframe{L}, v \rangle$ на одну из моделей $\kmodel{L}_0$, $\kmodel{L}_1$, $\kmodel{L}_2$, $\kmodel{L}_3$.
\end{lemma}

\begin{proof}
  Если $v(p) = \varnothing$, то отображение $f\colon V \to U_1$, определённое условием
  $$
  \begin{array}{lcll}
  f(w) & = & u_0 & \mbox{для каждого $w \in V$,}
  \end{array}
  $$
  является p-морфизмом модели $\langle \kframe{L}, v \rangle$ на модель~$\kmodel{L}_1$.

  Если $v(p) = V$, то отображение $f\colon V \to U_0$, определённое условием
  $$
  \begin{array}{lcll}
  f(w) & = & u_0 & \mbox{для каждого $w \in V$,}
  \end{array}
  $$
  является p-морфизмом модели $\langle \kframe{L}, v \rangle$ на модель~$\kmodel{L}_0$.

  Если $v(p) = \{w_1\}$, то отображение $f\colon V \to U_3$, определённое условием
  $$
  \begin{array}{lcl}
    f(w)
      & =
      & \left\{
        \begin{array}{rl}
          u_0, & \mbox{если $w=w_0$ или $w=w_2$;}\\
          u_1, & \mbox{если $w=w_1$;}\\
          u_2, & \mbox{если $w \in V \setminus \{w_0, w_1, w_2\}$,}
        \end{array}
        \right.
  \end{array}
  $$
  является p-морфизмом модели $\langle \kframe{L}, v \rangle$ на модель~$\kmodel{L}_3$.

  Если $v(p) = \{w_0, w_1\}$, то отображение $f\colon V \to U_2$, определённое условием
  $$
  \begin{array}{lcl}
    f(w)
      & =
      & \left\{
        \begin{array}{rl}
          u_0, & \mbox{если $w=w_0$ или $w=w_1$;}\\
          u_1, & \mbox{если $w \in V \setminus \{w_0, w_1\},\phantom{,w_2}$}
        \end{array}
        \right.
  \end{array}
  $$
  является p-морфизмом модели $\langle \kframe{L}, v \rangle$ на модель~$\kmodel{L}_2$.

  Если $v(p) = \{w_0, w_2\}$, то отображение $f\colon V \to U_3$, определённое условием
  $$
  \begin{array}{lcl}
    f(w)
      & =
      & \left\{
        \begin{array}{rl}
          u_0, & \mbox{если $w=w_1$;}\\
          u_1, & \mbox{если $w=w_0$ или $w=w_2$;}\\
          u_2, & \mbox{если $w \in V \setminus \{w_0, w_1, w_2\}$,}
        \end{array}
        \right.
  \end{array}
  $$
  является p-морфизмом модели $\langle \kframe{L}, v \rangle$ на модель~$\kmodel{L}_3$.

  Во всех остальных случаях модель $\langle \kframe{L}, v \rangle$ p-морфно отображается на модель~$\kmodel{L}_2$.
\end{proof}

\begin{lemma}
  \label{lem:failure-nish}
  Пусть $n \geqslant 5$. Тогда
  $$
  \begin{array}{lcl}
  (\langle \kframe{L}, v \rangle, x) \not\imodels \nu_n
    & \iff
    & \mbox{$v(p) = \bar{v}(p)$ и $x \notin S(w_n)$.}
  \end{array}
  $$
\end{lemma}

\begin{proof}
  Следует из~лемм~\ref{lem:models} и~\ref{lem:ladder-p-morph} с учётом \mbox{$({\ast})$} и~\mbox{$({\ast}{\ast})$}.
\end{proof}

\begin{figure}
\centering

\begin{tikzpicture}[scale=1.5]

\coordinate (w0)   at (+0.0000,+0.0000);
\coordinate (w1)   at (+1.0000,+0.0000);
\coordinate (w2)   at (+0.0000,-1.0000);
\coordinate (w3)   at (+1.0000,-1.0000);
\coordinate (w4)   at (+0.0000,-2.0000);
\coordinate (w5)   at (+1.0000,-2.0000);
\coordinate (w6)   at (+0.0000,-3.0000);
\coordinate (w7)   at (+1.0000,-3.0000);
\coordinate (w9)   at (+1.0000,-4.0000);
\coordinate (w*)   at (+2.0000,-3.0000);
\coordinate (dots) at (+0.5000,-4.5000);

\begin{scope}[>=latex]

\draw [->,  shorten >= 1.5pt, shorten <= 1.5pt] (w2)  -- (w0);
\draw [->,  shorten >= 1.5pt, shorten <= 1.5pt] (w4)  -- (w2);
\draw [->,  shorten >= 1.5pt, shorten <= 1.5pt] (w6)  -- (w4);

\draw [->,  shorten >= 1.5pt, shorten <= 1.5pt] (w3)  -- (w1);
\draw [->,  shorten >= 1.5pt, shorten <= 1.5pt] (w5)  -- (w3);
\draw [->,  shorten >= 1.5pt, shorten <= 1.5pt] (w7)  -- (w5);
\draw [->,  shorten >= 1.5pt, shorten <= 1.5pt] (w9)  -- (w7);

\draw [->,  shorten >= 1.5pt, shorten <= 1.5pt] (w3)  -- (w0);
\draw [->,  shorten >= 1.5pt, shorten <= 1.5pt] (w5)  -- (w2);
\draw [->,  shorten >= 1.5pt, shorten <= 1.5pt] (w7)  -- (w4);
\draw [->,  shorten >= 1.5pt, shorten <= 1.5pt] (w9)  -- (w6);

\draw [->,  shorten >= 1.5pt, shorten <= 1.5pt] (w4)  -- (w1);
\draw [->,  shorten >= 1.5pt, shorten <= 1.5pt] (w6)  -- (w3);
\draw [->,  shorten >= 1.5pt, shorten <= 1.5pt] (w9)  -- (w*);

\end{scope}

\node [      right] at (w1)   {${w_1}$}     ;
\node [      right] at (w3)   {${w_3}$}     ;
\node [      right] at (w5)   {${w_5}$}     ;
\node [      right] at (w7)   {${w_7}$}     ;
\node [      right] at (w9)   {${w_9}$}     ;

\node [      left ] at (w0)   {${w_0}$}     ;
\node [      left ] at (w2)   {${w_2}$}     ;
\node [      left ] at (w4)   {${w_4}$}     ;
\node [      left ] at (w6)   {${w_6}$}     ;
\node [      right] at (w*)   {${w^\ast_{\phantom{i}}}$}  ;

\node [above      ] at (w0)   {${p}$}       ;
\node [above      ] at (w*)   {${q}$}      ;
\node               at (dots) {$~~~~~~$}    ;


\draw [] (w0)   circle [radius=1.5pt]   ;
\draw [] (w1)   circle [radius=1.5pt]   ;
\draw [] (w2)   circle [radius=1.5pt]   ;
\draw [] (w3)   circle [radius=1.5pt]   ;
\draw [] (w4)   circle [radius=1.5pt]   ;
\draw [] (w5)   circle [radius=1.5pt]   ;
\draw [] (w6)   circle [radius=1.5pt]   ;
\draw [] (w7)   circle [radius=1.5pt]   ;
\draw [] (w9)   circle [radius=1.5pt]   ;
\draw [] (w*)   circle [radius=1.5pt]   ;

\end{tikzpicture}

\caption{Модель на шкале~$\kframe{G}_9$}
\label{fig:Rieger_Nishimura:center}
\end{figure}

Для каждого $n \geqslant 9$ определим шкалу $\langle V_n, S_n \rangle$ как подшкалу шкалы $\kframe{L}$, порождённую множеством~$\{w_n\}$, и положим
\[
\begin{array}{llcl}
\arrayitem
  & V^\ast_n & = & V_n \cup \{ w^\ast \}; \\
\arrayitem
  & S^\ast_n & = & S_n \cup \{ \langle w_n, w^\ast \rangle, \langle w^\ast, w^\ast \rangle \}; \\
\arrayitem & \kframe{G}_n & = & \langle V^\ast_n, S^\ast_n \rangle,
\end{array}
\]
см. рис.~\ref{fig:Rieger_Nishimura:center}, где показан пример такой шкалы для~$n = 9$.

Пусть $q$~--- фиксированная пропозициональная переменная, отличная от~$p$. Для каждого $n \geqslant 5$ положим
$$
\begin{array}{lcl}
\eta_n & = & \nu_{n} \wedge (\nu_{n-1} \imp \neg q ) \imp \nu_{n-3} \vee
\nu_{n-2} \vee \neg q.
\end{array}
$$

\begin{lemma}
  \label{lem:etas}
  Пусть $n \geqslant 9$ и $k \geqslant 9$. Тогда
  $$
  \begin{array}{lcl}
  (\kframe{G}_n, x) \not\imodels \eta_k & \iff & \mbox{$k = n$ и $x = w_n$.}
  \end{array}
  $$
\end{lemma}

\begin{proof}
  Из леммы~\ref{lem:failure-nish} следует, что $\eta_k$ опровергается в мире $x$ шкалы~$\kframe{G}_n$ при оценке~$v$, только если $\kframe{G}_n=\kframe{G}_k$, $x=w_k$, а оценка $v$ удовлетворяет условиям $v(p) = \{w_0\}$ и $v(q) = \{w^\ast\}$.
\end{proof}

Пусть $\function{h}{\numN}{\numN}$~--- отображение, заданное правилом $h(n) = 2n + 9$.
для каждого подмножества~$A$ множества~$\numN$ положим
\[
\begin{array}{llcl}
\arrayitem & \mathscr{C}^{si}_A & = & \{ \kframe{G}_{h(2n)} : n \notin {A} \}
                                      \cup \set{\kframe{G}_{h(2n+1)} : n\in\numN}; \\
\arrayitem & \mathbf{L}^{si}_A & = & \ilogic{\mathscr{C}^{si}_A}; \\
\arrayitem & \mathbf{L}^{m}_A & = & \mlogic{\mathscr{C}^{si}_A}. \\
\end{array}
\]

\begin{lemma}
  \label{lem:etas-S}
  Для каждого $n \in \numN$ справедливы следующие эквивалентности:
  $$
  \begin{array}{lclcl}
    \eta_{h(2n)} \in \mathbf{L}^{si}_{A}
    & \iff
    & \kframe{G}_{h(2n)} \not\in \mathscr{C}^{si}_A
    & \iff
    & n \in A.
  \end{array}
  $$
\end{lemma}

\begin{proof}
  Следует из леммы~\ref{lem:etas}, определения класса $\mathscr{C}^{si}_A$ и определения логики~$\mathbf{L}^{si}_{A}$.
\end{proof}


\begin{lemma}
  \label{lem:g-etas-S}
  Для каждого $n \in \numN$ справедлива эквивалентность
  $$
  \begin{array}{lcl}
    \mathsf{T}(\eta_{h(2n)}) \in \mathbf{L}^{m}_{A}
    & \iff
    & n \in A.
  \end{array}
  $$
\end{lemma}

\begin{proof}
Из определения логик $\mathbf{L}^{si}_{A}$ и $\mathbf{L}^{m}_{A}$ следует, что логика~$\mathbf{L}^{m}_{A}$ является модальным напарником логики~$\mathbf{L}^{si}_{A}$. Остальное получаем из леммы~\ref{lem:etas-S}.
\end{proof}

\begin{lemma}
  \label{lem:lfp-grz}
  Для каждого подмножества $A$ множества $\numN$ логика $\mathbf{L}^{m}_{A}$ является линейно аппроксимируемой.
\end{lemma}

\begin{proof}
Покажем, что существует такая константа $c$, что каждая модальная формула $\varphi$, не принадлежащая логике $\mathbf{L}^{m}_{A}$, опровергается в шкале из класса~$\mathscr{C}^{si}_A$, число миров в которой не превосходит~$c \cdot |\varphi|$.

Доказательство будет похоже на доказательство леммы~\ref{lem:lfp-K4}.

Пусть $\varphi \notin \mathbf{L}^{m}_{A}$.  Тогда
  $(\kframe{G}_l, w) \not\models^v \varphi$ для некоторой шкалы $\kframe{G}_l \in \mathscr{C}^{si}_A$, некоторого мира
  $w \in V^\ast_l$ и некоторой оценки $v$ в~$\kframe{G}_l$.

Если $\varphi$ не содержит вхождений модальности $\Box$ или если $w = w^\ast$,
  то $\varphi$ опровергается в любой шкале класса $\mathscr{C}^{si}_A$, например, в~$\kframe{G}_{11}$.  Поэтому будем считать, что $\varphi$ содержит вхождения модальности $\Box$ и что~$w \ne w^\ast$.  Пусть
  $\Box \psi_1, \ldots, \Box \psi_m$~--- список всех подформул вида
  $\Box \psi$ из $\mathop{\mathit{sub}} \varphi$.

  Введём вспомогательную терминологию и обозначения.

  Пусть $n \geqslant 9$ и пусть $\kmodel{L}$~--- некоторая модель, определённая на шкале~$\kframe{G}_n$.

  Для каждого $i \in \{1, \ldots, m\}$ определим множество $Y_i$, положив
  $$
  \begin{array}{lcl}
    w\in Y_i
    & \bydef
    & \left\{
      \begin{array}{l}
        w \in V^\ast_n \setminus \{w^\ast\}; \\
        (\kmodel{L}, w\phantom{'}) \not\models \psi_i; \\
        \mbox{$(\kmodel{L}, w') \models \psi_i$~ для каждого $w' \in S^\ast_n(w) \setminus \{ w \}$.}
      \end{array}
      \right.
  \end{array}
  $$
  Будем использовать обозначение $Y^{\kmodel{L}}_i$ вместо $Y_i$ в случаях, когда требуется уточнить, по какой модели это множество построено.

  Подмножество $Z$ множества $V^\ast_n$ называем \defnotion{уровнем шкалы}~$\kframe{G}_n$, если $Z = \{w_{2t},w_{2t+1}\} \cap V^\ast_n$ для некоторого~$t \in \numN$; \defnotion{уровнем мира} $w \in V^\ast_n$ называем уровень шкалы~$\kframe{G}_n$, содержащей~$w$ (заметим, что у каждого мира может быть не более одного уровня). Заметим, что для каждого $i \in \{1, \ldots, m\}$ множество~$Y_i$ содержит миры из не более чем двух уровней.

  Для каждого $i \in \{1, \ldots, m\}$ определим множество $X_i$ как объединение уровней тех миров, которые оказались в~$Y_i$; пусть также $X_0 = \{ w_0, w_1, w_2, w_3 \}$ и $X_{m+1} = \{ w_n \}$. Нетрудно видеть, что $|X_i| \leqslant 4$ для каждого $i \in \{0, \ldots, m+1\}$;
  более того, если $w_s,w_t\in X_i$, то $|s - t| \leqslant 3$.
  Мы будем использовать обозначение $X^{\kmodel{L}}_i$ в место $X_i$ в случаях, когда требуется уточнить, по какой модели это множество построено.

  Пусть $i, j \in \{0, \ldots, m+1\}$; говорим, что мир $w \in V^\ast_n$ \defnotion{лежит строго между $X_i$ и~$X_j$}, если
  \begin{itemize}
  \item $w \notin X_i \cup X_j$;
  \item существуют такие $x \in X_i$ и $y\in X_j$, что $x S^\ast_n w S^\ast_n y$ или $y S^\ast_n w S^\ast_n x$;
  \end{itemize}
  говорим, что уровень $Z$ \defnotion{лежит строго между $X_i$ и~$X_j$}, если каждый мир из $Z$ лежит строго между $X_i$ и~$X_j$; наконец, говорим, что множества $X_i$ и $X_j$ являются \defnotion{соседними}, если
  \begin{itemize}
  \item $X_i\ne\varnothing$ и $X_j\ne\varnothing$;
  \item $X_i \cap X_j = \varnothing$;
  \item не существует таких $s \in \{0, \ldots, m+1\}$ и $w \in X_s$, что $w$ лежит строго между $X_i$ и~$X_j$.
  \end{itemize}

  Пусть теперь $\varphi$ опровергается в модели $\kmodel{L}'$, определённой на шкале
  $\kframe{G}_n$, где для некоторых $i, j \in \{0, \ldots, m+1\}$ число уровней, лежащих строго между соседними множествами $X^{\kmodel{L}'}_i$ и $X^{\kmodel{L}'}_j$, больше одного. Покажем, что в этом случае $\varphi$ опровергается в некоторой модели $\kmodel{L}''$, определённой на шкале
  $\kframe{G}_{n-2}$, в которой между соседними множествами $X^{\kmodel{L}''}_i$ и $X^{\kmodel{L}''}_j$ имеется на один уровень меньше, причём для каждого множества $X^{\kmodel{L}'}_k$ существует множество $X^{\kmodel{L}''}_k$, состоящее из такого же количества миров. Ясно, что в этом случае, уменьшая раз за разом количество уровней на один, мы получим модель, где опровергается~$\varphi$, причём число миров в этой модели достаточно невелико.

  Пусть $(\kmodel{L}', w) \not\models \varphi$ для некоторой модели
  $\kmodel{L}'$ на шкале $\kframe{G}_n$ и некоторого мира $w \in V^{\ast}_n$; пусть $\kmodel{L}'$ содержит два уровня, лежащих между соседними множествами $X_i$ и $X_j$; пусть $\{w_{2t},w_{2t+1}\}$~--- один из этих уровней. Определим модель $\kmodel{L}''$ на
  $\kframe{G}_{n-2}$ так, что для каждой пропозициональной переменной~$q$
  $$
  \begin{array}{lclcl}
    (\kmodel{L}'', w_r) \models q
    & \iff
    & (\kmodel{L}', w_r) \models q
    & & \mbox{при $r < 2t$;}
    \\
    (\kmodel{L}'', w_r) \models q
    & \iff
    &  (\kmodel{L}', w_{r+2}) \models q
    & & \mbox{при $r > 2t + 1$};
    \\
    (\kmodel{L}'', w^\ast) \models q
    & \iff
    & (\kmodel{L}', w^\ast) \models q.
  \end{array}
  $$

  Тогда для каждой формулы
  $\psi \in \mathop{\mathit{sub}} \varphi$
  $$
  \begin{array}{lcl}
    (\kmodel{L}'', w^\ast) \models \psi
    & \iff
    & (\kmodel{L}', w^\ast) \models \psi
  \end{array}
  $$
  а также для каждой $\psi \in \mathop{\mathit{sub}} \varphi$ и каждого
  $w_r\in X^{\kmodel{L}'}_0\cup\ldots\cup X^{\kmodel{L}'}_{m+1}$
  $$
  \begin{array}{lclcl}
    (\kmodel{L}'', w_r) \models \psi
    & \iff
    & (\kmodel{L}', w_r) \models \psi
    && \mbox{при $r < 2t$;}
    \\
    (\kmodel{L}'', w_r) \models \psi
    & \iff
    & (\kmodel{L}', w_{r+2}) \models \psi
    && \mbox{при $r > 2t + 1$,}
  \end{array}
  $$
  что обосновывается индукцией по построению~$\psi$.

  Как следствие получаем доказательство леммы. Действительно, если $l$ достаточно велико, то нужно применить описанную конструкцию сначала к $\otuple{\kframe{G}_l,v}$, затем (при необходимости) к получившейся модели на шкале $\kframe{G}_{l-2}$, затем (при необходимости) к получившейся модели на шкале $\kframe{G}_{l-4}$ и так далее; рано или поздно мы получим модель на шкале с небольшим индексом вида~$h(2n+1)$.
\end{proof}

\begin{lemma}
  \label{lem:lfp-si}
  Для каждого подмножества $A$ множества\/ $\numN$ логика
  $\mathbf{L}^{si}_{A}$ является линейно аппроксимируемой.
\end{lemma}

\begin{proof}
  Следует из леммы~\ref{lem:lfp-grz} с учётом того, что $\mathbf{L}^{si}_{A}$ является суперинтуиционистским фрагментом логики~$\mathbf{L}^{m}_{A}$.
\end{proof}

\begin{theorem}
  \label{thr:si}
Для любой степени неразрешимости $C$ существует линейно аппроксимируемая суперинтуиционистская логика, находящаяся в $C$, фрагмент от двух переменных которой тоже находится в~$C$.
\end{theorem}

\begin{proof}
  Аналогично доказательству теоремы~\ref{thr:ext-K4}, но с использованием лемм~\ref{lem:etas-S} и~\ref{lem:lfp-si}.
\end{proof}

Ясно, что из лемм~\ref{lem:g-etas-S} и~\ref{lem:lfp-grz} следует существование линейно аппроксимируемых логик в классе расширений логики~$\mathbf{Grz}$, имеющих $C$-полные фрагменты от двух переменных, где $C$~--- произвольная степень неразрешимости. Но выше мы сформулировали теорему~\ref{thr:ext-Grz:pre}, утверждающую, что аналогичный результат справедлив для фрагментов от одной переменной. Покажем это.

\begin{figure}
\centering
\begin{tikzpicture}[scale=1.5]

\coordinate (w0)   at (+0.0000,+0.0000);
\coordinate (w1)   at (+1.0000,+0.0000);
\coordinate (w2)   at (+0.0000,-1.0000);
\coordinate (w3)   at (+1.0000,-1.0000);
\coordinate (w4)   at (+0.0000,-2.0000);
\coordinate (w5)   at (+1.0000,-2.0000);
\coordinate (w6)   at (+0.0000,-3.0000);
\coordinate (w7)   at (+1.0000,-3.0000);
\coordinate (w9)   at (+1.0000,-4.0000);
\coordinate (w*)   at (+2.0000,-3.0000);
\coordinate (w**)  at (+2.0000,-2.0000);
\coordinate (dots) at (+0.5000,-4.5000);

\begin{scope}[>=latex]

\draw [->,  shorten >= 1.5pt, shorten <= 1.5pt] (w2)  -- (w0);
\draw [->,  shorten >= 1.5pt, shorten <= 1.5pt] (w4)  -- (w2);
\draw [->,  shorten >= 1.5pt, shorten <= 1.5pt] (w6)  -- (w4);

\draw [->,  shorten >= 1.5pt, shorten <= 1.5pt] (w3)  -- (w1);
\draw [->,  shorten >= 1.5pt, shorten <= 1.5pt] (w5)  -- (w3);
\draw [->,  shorten >= 1.5pt, shorten <= 1.5pt] (w7)  -- (w5);
\draw [->,  shorten >= 1.5pt, shorten <= 1.5pt] (w9)  -- (w7);

\draw [->,  shorten >= 1.5pt, shorten <= 1.5pt] (w3)  -- (w0);
\draw [->,  shorten >= 1.5pt, shorten <= 1.5pt] (w5)  -- (w2);
\draw [->,  shorten >= 1.5pt, shorten <= 1.5pt] (w7)  -- (w4);
\draw [->,  shorten >= 1.5pt, shorten <= 1.5pt] (w9)  -- (w6);

\draw [->,  shorten >= 1.5pt, shorten <= 1.5pt] (w4)  -- (w1);
\draw [->,  shorten >= 1.5pt, shorten <= 1.5pt] (w6)  -- (w3);
\draw [->,  shorten >= 1.5pt, shorten <= 1.5pt] (w9)  -- (w*);
\draw [->,  shorten >= 1.5pt, shorten <= 1.5pt] (w*)  -- (w**);

\end{scope}

\node [      right] at (w1)   {${w_1}$}     ;
\node [      right] at (w3)   {${w_3}$}     ;
\node [      right] at (w5)   {${w_5}$}     ;
\node [      right] at (w7)   {${w_7}$}     ;
\node [      right] at (w9)   {${w_9}$}     ;

\node [      left ] at (w0)   {${w_0}$}     ;
\node [      left ] at (w2)   {${w_2}$}     ;
\node [      left ] at (w4)   {${w_4}$}     ;
\node [      left ] at (w6)   {${w_6}$}     ;
\node [      right] at (w*)   {${w^\ast_{\phantom{i}}}$}  ;
\node [      right] at (w**)  {${w^{{\ast}{\ast}}_{\phantom{i}}}$}  ;

\node [above      ] at (w0)   {${p}$}       ;
\node [left       ] at (w*)   {${p}$}      ;
\node               at (dots) {$~~~~~~$}    ;


\draw [] (w0)   circle [radius=1.5pt]   ;
\draw [] (w1)   circle [radius=1.5pt]   ;
\draw [] (w2)   circle [radius=1.5pt]   ;
\draw [] (w3)   circle [radius=1.5pt]   ;
\draw [] (w4)   circle [radius=1.5pt]   ;
\draw [] (w5)   circle [radius=1.5pt]   ;
\draw [] (w6)   circle [radius=1.5pt]   ;
\draw [] (w7)   circle [radius=1.5pt]   ;
\draw [] (w9)   circle [radius=1.5pt]   ;
\draw [] (w*)   circle [radius=1.5pt]   ;
\draw [] (w**)  circle [radius=1.5pt]   ;

\end{tikzpicture}

\caption{Модель на шкале~$\kframe{G}^+_9$}
\label{fig:Rieger_Nishimura:right}
\end{figure}

Для каждого $n \geqslant 9$ положим
$$
\begin{array}{lcl}
  \eta^+_n
    & =
    & \Diamond(p\wedge\Diamond\neg p\wedge \Box(\neg p\to \Box\neg p))
      \to (g(\nu_{n})\imp g(\nu_{n-3} \vee \nu_{n-2})).
\end{array}
$$
\pagebreak[3]

Для каждого $n \geqslant 9$ определим также шкалу $\kframe{G}^+_n$ (см.~рис.~\ref{fig:Rieger_Nishimura:right}, где показан пример шкалы $\kframe{G}^+_n$ для~$n = 9$); для этого положим
\[
\begin{array}{llcl}
\arrayitem
  & V^+_n
  & =
  & V^\ast_n \cup \{ w^{\ast \ast} \}; \\
\arrayitem
  & S^+_n
  & =
  & S^\ast_n \cup \{ \langle w^\ast, w^{\ast \ast} \rangle, \langle w^{\ast \ast}, w^{\ast \ast} \rangle, \langle w_n, w^{\ast \ast} \rangle \}; \\
\arrayitem
  & \kframe{G}^+_n
  & =
  & \langle V^+_n, S^+_n \rangle.
\end{array}
\]

\begin{lemma}
  \label{lem:etas-plus}
  Пусть $n \geqslant 9$ и $k \geqslant 9$. Тогда
  $$
  \begin{array}{lcl}
    \kframe{G}^+_n, x \not\imodels \eta^+_k & \iff & \mbox{$k = n$ и $x = w_n$.}
  \end{array}
  $$
\end{lemma}

\begin{proof}
  Аналогично доказательству леммы~\ref{lem:etas}.
\end{proof}

Для каждого подмножества $A$ множества $\numN$ положим
\[
\begin{array}{llcl}
\arrayitem & \mathscr{C}^{+}_A & = & \{ \kframe{G}^+_{h(2n)} : n \not\in A \}
                                     \cup\{\kframe{G}^+_{h(2n+1)} : n\in\numN\}; \\
\arrayitem & \mathbf{L}^{+}_A  & = & \mlogic{\mathscr{C}^{+}_A}.
\end{array}
\]

\begin{lemma}
  \label{lem:g-etas-S-plus}
  Для каждого $n \in \numN$ имеет место следующая эквивалентность:
  $$
  \begin{array}{lcl}
    \eta^+_{h(2n)} \in \mathbf{L}^{+}_{A}
    & \iff
    & n \in A.
  \end{array}
  $$
\end{lemma}

\begin{proof}
  Следует из леммы~\ref{lem:etas-plus}, определения класса
  $\mathscr{C}^{+}_A$ и определения логики~$\mathbf{L}^{+}_{A}$.
\end{proof}

\begin{lemma}
  \label{lem:lfp-grz-plus}
  Для каждого подмножества $A$ множества\/ $\numN$ логика\/ $\mathbf{L}^{+}_{A}$ является линейно аппроксимируемой.
\end{lemma}

\begin{proof}
  Аналогично доказательству леммы~\ref{lem:lfp-grz}.
\end{proof}

Теперь мы можем доказать теорему~\ref{thr:ext-Grz:pre}, которую сформулируем ещё раз в виде следующего предложения.

\begin{proposition}
  \label{prop:grz}
Для любой степени неразрешимости $C$ существует линейно аппроксимируемое нормальное расширение логики\/ $\logic{Grz}$, находящееся в $C$, фрагмент от одной переменной которого тоже находится в~$C$.
\end{proposition}

\begin{proof}
  Аналогично доказательству теоремы~\ref{thr:ext-K4}, но с использованием
  лемм~\ref{lem:g-etas-S-plus} и~\ref{lem:lfp-grz-plus}.
\end{proof}

Отметим, что приведённые здесь результаты несложно распространить и на расширения логик $\logic{BPL}$ и $\logic{FPL}$.
В случае $\logic{BPL}$ достаточно рассмотреть формулы, которые мы уже использовали выше (стр.~\pageref{alpha_n_BPL}):
$$
\begin{array}{lcl}
\alpha_n
  & =
  & (\Box^{n+2}\bot\to\Box^{n+1}\bot) \to (\Box^{n+1}\bot\to\Box^n\bot)\vee\Box^{n+2}\bot,
\end{array}
$$
где $\Box\psi=\top\to\psi$.
В случае $\logic{FPL}$ можно взять модификацию формул, определённых на стр.~\pageref{alpha_i_FPLcase}:
$$
\begin{array}{lcl}
\alpha'_n & = & \Box^{2}p\to(\Box^{n+1}\bot\to\Box^n\bot\vee p)\vee(\Box\bot\to p).
\end{array}
$$
Мы не будем приводить технические детали всех построений, лишь сформулируем результаты, аналогичные теореме~\ref{thr:si}.

\begin{theorem}
  \label{thr:BPL}
Для любой степени неразрешимости $C$ существует линейно аппроксимируемое расширение логики\/ $\logic{BPL}$, находящееся в $C$, константный фрагмент которого тоже находится в~$C$.
\end{theorem}

\begin{theorem}
  \label{thr:FPL}
Для любой степени неразрешимости $C$ существует линейно аппроксимируемое расширение логики\/ $\logic{FPL}$, находящееся в $C$, фрагмент от одной переменной которого тоже находится в~$C$.
\end{theorem}

    \subsection{Гейтинговы алгебры}

Пусть $\bm{A}=\langle A,\bwedge,\bvee,\bto,\bbot\rangle$~--- алгебра с бинарными операциями $\bwedge$, $\bvee$, $\bto$ и выделенным элементом~$\bbot$. Введём отношение $\leqslant$ на $A$, положив для всяких $x,y\in A$
$$
\begin{array}{lcl}
x\leqslant y & \leftrightharpoons & x\bwedge y = x.
\end{array}
$$
Алгебра $\bm{A}=\langle A,\bwedge,\bvee,\bto,\bbot\rangle$ называется \defnotion{гейтинговой},\index{алгебра!бяб@гейтингова} или \defnotion{псевдобулевой},\index{алгебра!псевдобулева} если для любых $x,y,z\in A$
\[
\begin{array}{ll}
\arrayitem &
\mbox{$x\bwedge y = y\bwedge x$, $x\bvee y = y\bvee x$;} \\
\arrayitem &
\mbox{$x\bwedge (y\bwedge z) = (x\bwedge y)\bwedge z$, $x\bvee (y\bvee z) = (x\bvee y)\bvee z$;} \\
\arrayitem &
\mbox{$(x\bwedge y)\bvee y = y$, $(x\bvee y)\bwedge y = y$;} \\
\arrayitem &
\mbox{$z\bwedge x\leqslant y$ тогда и только тогда, когда $z\leqslant x\bto y$;} \\
\arrayitem &
\mbox{$\bbot\leqslant x$.} \\
\end{array}
\]

В контексте гейтинговых алгебр $\lang{L}$\nobreakdash-формулы будем называть также \defnotion{$\lang{L}$\nobreakdash-термами}.\index{терм!l@$\lang{L}$-терм} Как и в случае с модальными алгебрами, нам будет интересна проблема равенства $\lang{L}$\nobreakdash-термов в гейтинговых алгебрах и классах гейтинговых алгебр.

Пусть $v$~--- оценка переменных в гейтинговой алгебре~$\bm{A}$. Расширим $v$ на множество всех $\lang{L}$\nobreakdash-термов:
\[
\begin{array}{llcl}
\arrayitem &
v(\bot) & = & \bbot; \\
\arrayitem &
v(t\wedge s) & = & v(t)\bwedge v(s); \\
\arrayitem &
v(t\vee s) & = & v(t)\bvee v(s); \\
\arrayitem &
v(t\to s) & = & v(t)\bto v(s).
\end{array}
\]




Из доказанных выше утверждений (см.~следствие~\ref{cor_Int-KC}) получаем следующие теоремы.

\begin{theorem}
\label{MRybakov:th:Int-one}
Проблема равенства позитивных\/ $\lang{L}$\nobreakdash-термов от двух переменных в классе всех гейтинговых алгебр является\/ $\cclass{PSPACE}$\nobreakdash-полной.
\end{theorem}

\begin{theorem}
\label{MRybakov:th:Int-two}
Проблема равенства позитивных\/ $\lang{L}$\nobreakdash-термов в классе всех\/ $2$\nobreakdash-порождённых гейтинговых алгебр является\/ $\cclass{PSPACE}$\nobreakdash-полной.
\end{theorem}

\begin{theorem}
\label{MRybakov:th:Int-three}
Если равенство позитивных\/ $\lang{L}$\nobreakdash-термов $t=s$ опровергается в некоторой гейтинговой алгебре, то оно опровергается и в некоторой\/ $2$\nobreakdash-порождённой гейтинговой алгебре.
\end{theorem}

Теорема~\ref{MRybakov:th:Int-three} следует также из~\cite{Mardaev-1987-1-rus}; сложностные вопросы в~\cite{Mardaev-1987-1-rus} не рассматривались.

\pagebreak[3]

\begin{theorem}
\label{MRybakov:prop:Hext}
Пусть $\scls{C}$~--- произвольный класс гейтинговых алгебр, содержащий все\/ $\logic{KC}$\nobreakdash-алгебры. Тогда
\begin{itemize}
\item
проблема равенства\/ $\lang{L}$-термов от двух переменных в классе $\scls{C}$ является\/ $\cclass{PSPACE}$-трудной;
\item
проблема равенства произвольных\/ $\lang{L}$-термов в классе всех\/ $2$\nobreakdash-порож\-дён\-ных гейтинговых алгебр из класса $\scls{C}$ является\/ $\cclass{PSPACE}$\nobreakdash-трудной.
\end{itemize}
\end{theorem}

  \section{Замечания}
  \label{sec:notes:int}

Некоторые представленные в этой главе результаты (разделы~\ref{ssec:BPL}--\ref{ssec:BPL-KC}) являются частью кандидатской диссертации автора~\cite{MR:2005:Diss} и изложены с очень незначительными изменениями и дополнениями. В~дальнейшем нам будет важна конструкция, использованная для моделирования всех переменных позитивных $\lang{L}$\nobreakdash-формул позитивными формулами от двух переменных (раздел~\ref{ssec:Int-positive}): ниже (глава~\ref{ch:QInt}) мы покажем, как её можно модифицировать для получения сходных результатов в предикатном случае.

Отметим, что вместо погружения позитивного фрагмента интуиционистской логики (и близких к ней) в её фрагмент от двух переменных можно было погружать в этот фрагмент всю логику. Для этого предварительно можно промоделировать константу $\bot$ в формулах с помощью новой пропозициональной переменной.

Действительно, пусть $\varphi$~--- интуиционистская формула, $p_1,\ldots,p_n$~--- список всех входящих в неё переменных. Возьмём новую переменную~$q$. Заменим каждое вхождение формулы $\bot$ в формулу $\varphi$ вхождением~$q$, получившуюся формулу обозначим~$\varphi_q$. Пусть
$$
\begin{array}{lcl}
\varphi^q
  & =
  & \displaystyle
    \bigwedge\limits_{\mathclap{k=1}}^{n}(q\to p_k)\to \varphi_q.
\end{array}
$$
Заметим, что формула $\varphi^q$ является позитивной и строится по формуле $\varphi$ полиномиально.

\begin{proposition}
\label{prop:Int:positivization}
Имеет место следующая эквивалентность:
$$
\begin{array}{lcl}
\varphi\in\logic{Int}
  & \iff
  & \varphi^q \in\logic{Int}.
\end{array}
$$
\end{proposition}

\begin{proof}
Пусть $\varphi^\bot$~--- формула, получающаяся из $\varphi^q$ подстановкой $\bot$ вместо~$q$.

Если $\varphi^q\in\logic{Int}$, то $\varphi^\bot\in\logic{Int}$. Но посылка формулы~$\varphi^\bot$ принадлежит $\logic{Int}$, поскольку является конъюнкцией формул вида $\bot\to p_k$, а заключение совпадает с~$\varphi$, откуда получаем, что $\varphi\in\logic{Int}$.

Пусть теперь $\varphi^q\not\in\logic{Int}$. Тогда существуют такие интуиционистская модель $\kmodel{M}=\langle W,R,v\rangle$ и мир $w\in W$, что $(\kmodel{M},w)\not\imodels \varphi^q$. Пусть $\kmodel{M}'=\langle W',R',v'\rangle$~--- подмодель модели $\kmodel{M}$, в которой $W'=\{u\in W : (\kmodel{M},u)\not\imodels q\}$. Заметим, что $w\in W'$; несложно понять, что $\kmodel{M}'\imodels q\leftrightarrow\bot$ и, как следствие, $\kmodel{M}',w\not\imodels \varphi^\bot$. Поскольку $\varphi^\bot=\varphi$, получаем, что $\varphi\not\in\logic{Int}$.
\end{proof}

В этом разделе мы обошлись без указанного моделирования, поскольку $\cclass{PSPACE}$-полнота интуиционистской логики (и близких к ней) была изначально доказана для позитивного фрагмента языка; но это моделирование (точнее, его аналог) окажется удобным в предикатном случае, где нам понадобятся позитивные формулы, а результаты, на которые мы хотели бы опираться в своих построениях, получены для формул, содержащих отрицание (константу~$\bot$).

Теперь обратим внимание на связки, использованные в позитивных формулах. Исходно $\cclass{PSPACE}$-полнота интуиционистской логики (и близких к ней) была изначально доказана не просто для позитивного фрагмента языка, а для его импликативного фрагмента~\cite{Chagrov-1985-1-rus,Statman-1979-1}. Мы же использовали моделирование, в котором задействованы связки $\wedge$, $\vee$ и~$\to$, и поэтому естественно спросить следующее: можно ли получить похожее моделирование (в языке с двумя или с несколько большим числом переменных) при меньшем наборе связок, например, для импликативного фрагмента?

Несложно понять, что в формулах $A^k_m$ и $B^k_m$ можно избавиться от конъюнкции, поскольку в интуиционистской логике формулы $p\wedge q\to r$ и $p\to (q\to r)$ эквивалентны (в логиках $\logic{BPL}$ и $\logic{FPL}$ эквивалентность этих формул отсутствует, но мы и так не использовали конъюнкцию в моделировании переменных как константными формулами, так и формулами от одной переменной). Импликация нужна, поскольку без неё мы получим фрагмент, не отличающийся от соответствующего фрагмента классической логики. Дизъюнкция, как это ни странно, тоже нужна: в силу теоремы Диего~\cite{Diego} (см.~также~\cite[раздел~5.4]{ChZ}), интуиционистская логика в языке без дизъюнкции является локально табличной, а значит, её бездизъюнктивные фрагменты от любого фиксированного числа переменных полиномиально разрешимы; конечно же, то же относится к фрагментам с ограничением на глубину вложенности дизъюнкции (в предложенном выше моделировании такого ограничения в случае интуиционистской логики нет). Таким образом, существенно усилить описанные выше результаты не получится (при условии, что $\cclass{P}\ne\cclass{PSPACE}$).

Наконец, отметим, что представленные результаты переносятся и на другие логики, близкие к интуиционистской. К таким логикам относятся, например, \defnotion{совместная логика задач и высказываний} $\logic{HC}$ и близкая к ней логика $\logic{H4}$ \mbox{\cite{Mel:I,Mel:II}}, а также интуиционистские эпистемические логики $\logic{IEL}^-$, $\logic{IEL}$, $\logic{IEL}^+$~\cite{ArtemovProtopopescu} (логика $\logic{IEL}^+$ совпадает с~$\logic{H4}$).

\setcounter{savefootnote}{\value{footnote}}
\chapter{Полимодальные логики}
\setcounter{footnote}{\value{savefootnote}}
\label{chapter:PML}
  \section{Общие определения}
    \subsection{Синтаксис}
    \label{ssec:syntax-polymodal}

Будем считать, что \defnotion{полимодальные}\index{уян@язык!пропозициональный!полимодальный} языки являются расширениями классического пропозиционального языка с помощью модальностей; пока будем считать, что модальности берутся из некоторого множества $\setc{\Box_k}{k\in\mathbb{I}}$, где $\mathbb{I}$~--- некоторое (индексное) множество, и являются одноместными\footnote{Позже рассмотрим также языки, где имеются двухместные модальности.}.

При определении формулы полимодального языка с модальностями из множества $\setc{\Box_k}{k\in\mathbb{I}}$ добавляется следующий пункт:
             \begin{itemize}
                \item если $\varphi$~--- формула и $k\in\mathbb{I}$, то $\Box_k\varphi$~--- тоже формула.
             \end{itemize}

Для модальности $\Box_k$ определим двойственную к ней модальность $\Diamond_k$ как следующее сокращение:
             \begin{itemize}
                \item $\Diamond_k\varphi = \neg\Box_k\neg\varphi$.
             \end{itemize}

Полимодальные языки будем обозначать по-разному, не предлагая единой системы обозначений; обозначения будут варьироваться в зависимости от особенностей языков и логик, которые мы будем рассматривать. Формулы полимодальных языков будем называть также по-разному, используя для них общий термин \defnotion{модальные формулы}; то же относится и к полимодальным языкам.

Особо выделим языки вида $\lang{ML}_n$\index{уян@язык!mln@$\lang{ML}_n$} для $n\in\numN$, определяя $\lang{ML}_n$ как модальный язык с множеством модальностей $\setc{\Box_k}{1\leqslant k\leqslant n}$; в этом случае $\lang{ML}=\lang{ML}_1$ и $\lang{L}=\lang{ML}_0$.

    \subsection{Семантика Крипке}
    \label{ssec:sem-polymodal}

Семантика Крипке для полимодальных пропозициональных языков сходна с семантикой Крипке для мономодального пропозиционального языка: вместо одного отношения достижимости обычно нужно рассматривать некоторое множество отношений достижимости, содержащее для каждой модальности своё отношение. Тем не менее, в некоторых случаях возникают нюансы, поэтому мы предпочитаем не давать здесь общего определения семантики Крипке для полимодальных языков, а делать нужные уточнения, требующиеся для языков, которые будем рассматривать, по ходу изложения.

Здесь мы определим семантику Крипке для полимодальных пропозициональных языков вида $\lang{ML}_n$, где $n\in\numNp$.

            \defnotion{Шкалой Крипке}\index{уяи@шкала!Крипке} для полимодального языка $\lang{ML}_n$ будем называть набор $\kframe{F} = \langle
            W,R_1,\ldots,R_n\rangle$, где $W$~--- некоторое непустое множество, а $R_1,\ldots,R_n$~---
            бинарные отношения на этом множестве. Элементы множества $W$, как и раньше, будем
            называть \defnotion{мирами},\index{мир} а отношения $R_1,\ldots,R_n$~--- \defnotion{отношениями достижимости}\index{отношение!достижимости} на множестве~$W$.

            Для каждого мира $w$ шкалы Крипке $\kframe{F} = \langle W,R_1,\ldots,R_n\rangle$ определим множество~$R_k(w)$, положив
            $$
            \begin{array}{lcl}
            R_k(w) & = & \set{w'\in W : wR_kw'}.
            \end{array}
            $$

            \defnotion{Моделью Крипке}\index{модель!Крипке} для языка $\lang{ML}_n$
            называется набор $\kmodel{M} = \langle\kframe{F},v\rangle$, где
            $\kframe{F} = \langle W,R_1,\ldots,R_n\rangle$~--- шкала Крипке,
            а $v$~--- оценка пропозициональных переменных в~$W$.


            Пусть $\kmodel{M} = \langle\kframe{F},v\rangle$~--- модель,
            определённая на шкале $\kframe{F} = \langle W,R_1,\ldots,R_n\rangle$, и пусть
            $\varphi$~--- модальная пропозициональная $\lang{ML}_n$-формула. Определим
            отношение~$(\kmodel{M},w)\models\varphi$:
            \[
            \begin{array}{clcl}
            \arrayitem &
            (\kmodel{M},w)\not\models\bot;\!\! &  &
            \arrayitemskip
            \\
            \arrayitem &
            (\kmodel{M},w)\models p_i
            & \leftrightharpoons &
            \parbox[t]{225pt}{$w\in v(p_i)$;}
            \arrayitemskip
            \\
            \arrayitem &
            (\kmodel{M},w)\models\varphi\wedge\psi & \leftrightharpoons &
            \parbox[t]{225pt}{$(\kmodel{M},w)\models\varphi$ \hphantom{л}и\hphantom{и}
            $(\kmodel{M},w)\models\psi$;}
            \arrayitemskip
            \\
            \arrayitem &
            (\kmodel{M},w)\models\varphi\vee\psi & \leftrightharpoons &
            \parbox[t]{225pt}{$(\kmodel{M},w)\models\varphi$ или
            $(\kmodel{M},w)\models\psi$;}
            \arrayitemskip
            \\
            \arrayitem &
            (\kmodel{M},w)\models\varphi\to\psi & \leftrightharpoons &
            \parbox[t]{225pt}{$(\kmodel{M},w)\not\models\varphi$ или
            $(\kmodel{M},w)\models\psi$;}
            \arrayitemskip
            \\
            \arrayitem &
            (\kmodel{M},w)\models\Box_k\varphi
            & \leftrightharpoons &
            \parbox[t]{235pt}{$(\kmodel{M},w')\models\varphi$ для всякого $w'\in R_k(w)$.}
            \end{array}
            \]

Если $(\kmodel{M},w)\models\varphi$, то говорим, что формула $\varphi$ \defnotion{истинна в мире} $w$ модели $\kmodel{M}$; в противном случае говорим, что $\varphi$ \defnotion{опровергается в мире} $w$ модели~$\kmodel{M}$. Формулу $\varphi$ считаем \defnotion{истинной в модели} $\kmodel{M}$, если для всякого $w\in W$ выполнено отношение $(\kmodel{M},w)\models\varphi$; в этом случае пишем $\kmodel{M}\models\varphi$. Формулу $\varphi$ считаем \defnotion{истинной в шкале} $\kframe{F}$, если $\varphi$ истинна в любой модели, определённой на шкале~$\kframe{F}$; в этом случае пишем $\kframe{F}\models\varphi$. Формулу $\varphi$ считаем \defnotion{истинной в мире шкалы} $\kframe{F}$, если для любой модели $\kmodel{M}$, определённой на шкале~$\kframe{F}$, выполнено отношение $(\kmodel{M},w)\models\varphi$; в этом случае пишем $(\kframe{F},w)\models\varphi$. Формулу $\varphi$ считаем \defnotion{истинной в классе шкал} $\sclass{C}$, если $\varphi$ истинна в каждой шкале из этого класса; в этом случае пишем $\sclass{C}\models\varphi$.

            Как и в мономодальном случае, иногда вместо обозначения $(\kmodel{M},w)\models\varphi$ удобно использовать альтернативное:
            $$
            \begin{array}{lcl}
            (\kframe{F},w)\models^v\varphi & \leftrightharpoons & (\otuple{\kframe{F},v},w)\models\varphi;
            \end{array}
            $$
            в этом случае говорим, что в мире $w$ шкалы $\kframe{F}$ при оценке~$v$ истинна формула~$\varphi$.

Другие обозначения, введённые для мономодального языка, также сохраняются.



Понятия, связанные с подшкалами, требуют уточнений.

            Пусть $\kframe{F}=\otuple{W,R_1,\ldots,R_n}$ и $\kframe{F}'=\otuple{W',R'_1,\ldots,R'_n}$~--- шкалы Крипке, $X\subseteq W$. Тогда говорим, что
            \begin{itemize}
            \item
            $\kframe{F}'$ является \defnotion{подшкалой}\index{подшкала} шкалы $\kframe{F}$, если $W'\subseteq W$ и $R'_k=R_k\upharpoonright W'$ для каждого $k\in\set{1,\ldots,n}$;
            \item
            $\kframe{F}'$ является \defnotion{порождённой подшкалой}\index{подшкала!порождённая} шкалы $\kframe{F}$, если $\kframe{F}'$ является подшкалой шкалы $\kframe{F}$ и $R(W')\subseteq W'$, где $R=R_1\cup\ldots\cup R_n$;
            \item
            $X$ \defnotion{порождает} подшкалу $\kframe{F}'$ шкалы $\kframe{F}$, если $R^\ast(X)=W'$, где $R^\ast$~--- рефлексивно-транзитивное замыкание отношения~$R=R_1\cup\ldots\cup R_n$;
            \item
            $\kframe{F}'$ является \defnotion{корневой подшкалой}\index{подшкала!корневая} шкалы $\kframe{F}$, если $\kframe{F}'$ является подшкалой шкалы $\kframe{F}$, порождённой множеством $\{x\}$, где $x$~--- некоторый мир из~$W$; в этом случае $x$ называют \defnotion{корнем}\index{корень шкалы Крипке} шкалы~$\kframe{F}'$.
            \end{itemize}

  \section{Слияния логик и их обогащения}
    \subsection{Слияния логик}
    \label{ssec:fusions}

Логика $L$ в языке $\lang{ML}_n$ называется \defnotion{нормальной},\index{логика!пропозициональная!нормальная модальная}\index{логика!модальная!нормальная} если она содержит $\logic{Cl}$, формулу $\Box_k(p\to q)\to(\Box_k p\to \Box_k q)$ для каждого $k\in\set{1,\ldots,n}$, а также замкнута по \MP, правилу подстановки и правилу Гёделя для каждой модальности из множества $\set{\Box_1,\ldots,\Box_n}$.

Пусть $L_1,\ldots,L_n$~--- нормальные мономодальные логики в языке~$\lang{ML}$. Переобозначим модальности их языка так, чтобы модальность $\Box$ логики $L_k$ теперь обозначалась~$\Box_k$; полученные логики обозначим $L'_1,\ldots,L'_n$. Логику $L$ в языке $\lang{ML}_n$ будем называть \defnotion{слиянием}\index{слияние}\footnote{В английском используется термин fusion.} логик $L_1,\ldots,L_n$, если $L$~--- наименьшая нормальная модальная логика, содержащая $L'_1,\ldots,L'_n$. Слияние двух логик $L_1$ и $L_2$ обозначаем $L_1\fusion L_2$, а слияние $n$ логик $L_1,\ldots,L_n$ обозначаем $L_1\fusion\ldots\fusion L_n$.

В литературе часто рассматривают слияния одинаковых логик; в этом случае используют индекс $n$ рядом с именем логики для указания на то, слияние какого количества соответствующих логик получено. Так, логики $\logic{K}_n$, $\logic{T}_n$, $\logic{S4}_n$~--- это слияния, соответственно, $n$ логик $\logic{K}$, $n$ логик $\logic{T}$, $n$ логик $\logic{S4}$.
Для технического удобства будем считать, что индекс $0$ означает, что мы берём безмодальный фрагмент логики; так, $\logic{K}_0 = \logic{T}_0 = \logic{S4}_0 = \logic{Cl}$.

Нетрудно понять, что если $L$~--- слияние логик $L_1,\ldots,L_n$, одна из которых имеет $\cclass{PSPACE}$-трудный фрагмент от $k$ переменных (выше мы видели, что обычно достаточно, чтобы $k$ не превосходило двух), то фрагмент $L$ от $k$ переменных тоже является $\cclass{PSPACE}$-трудным. Но слияние логик, фрагменты от $k$ переменных которых разрешимы полиномиально, может тоже иметь $\cclass{PSPACE}$-трудный фрагмент от $k$ переменных. Так, например, в логике $\logic{S5}_2$ модальность $\Box = \Box_1\Box_2$ рефлексивна, но не является ни транзитивной, ни симметричной, что позволяет повторить для неё конструкцию, которую можно было применить к логике $\logic{T}$, см.~теоремы~\ref{th:PSPACE:GL:Grz} и~\ref{th:PSPACE:KTB} (логика $\logic{T}$ удовлетворяет условиям обеих теорем). В~итоге получаем, что фрагмент от одной переменной логики $\logic{S5}_2$ является $\cclass{PSPACE}$-трудным\footnote{Несложно показать, что этот фрагмент является и $\cclass{PSPACE}$-полным.}, хотя фрагменты логики $\logic{S5}$ от любого конечного числа переменных полиномиально разрешимы в силу её локальной табличности\footnote{Логика $L$ \defnotion{локально таблична},\index{логика!локально табличная} если любой фрагмент от конечного числа переменных логики $L$ является \defnotion{табличным},\index{логика!табличная} т.е. полным относительно одной фиксированной конечной модели (например, шкалы Крипке). Это, в частности, означает, что при конечном числе переменных в языке логики $L$ имеется лишь конечное множество классов эквивалентности, состоящих из эквивалентных в $L$ формул. Подробнее см.~\cite{ChZ}.}.

Обобщая это наблюдение, получаем следующую теорему.

\begin{theorem}
\label{th:PSPACE:S5-2}
Пусть $n\geqslant 2$. Тогда проблема разрешения фрагмента от одной переменной логики $L\in [\logic{K}_n,\logic{S5}_n]$ является\/ $\cclass{PSPACE}$\nobreakdash-трудной.
\end{theorem}

Ниже мы покажем, что описанная выше техника распространяется не только на слияния модальных логик, но и более сложно устроенные полимодальные логики, позволяя доказывать, что их фрагменты от небольшого числа переменных имеют ту же сложность, что и логика в языке с бесконечным числом переменных.

    \subsection{Логика $\logic{K}^\ast$}
    \label{ssec:K-ast}

Язык логики $\logic{K}^\ast$ получается обогащением языка $\lang{ML}$ логики $\logic{K}$ модальностью~$\Box^\ast$; при этом определяем двойственную к $\Box^\ast$ модальность $\Diamond^\ast$ как сокращение: $\Diamond^\ast\varphi = \neg\Box^\ast\neg\varphi$. Обозначим получившийся язык~$\lang{ML}^{\ast}$.

Шкалами и моделями Крипке для $\lang{ML}^{\ast}$-формул являются шкалы и модели Крипке для $\lang{ML}$-формул. При определении истинности $\lang{ML}^{\ast}$-формул в мире $w$ модели $\kmodel{M}=\otuple{\kframe{F},v}$, где $\kframe{F}=\otuple{W,R}$, нужно добавить следующее:
            \[
            \begin{array}{clcl}
            \arrayitem &
            (\kmodel{M},w)\models\Box^\ast\varphi
            & \bydef &
            \mbox{$(\kmodel{M},w')\models\varphi$ для каждого $w'\in R^\ast(w)$,}
            \end{array}
            \]
где $R^\ast$~--- рефлексивно-транзитивное замыкание отношения~$R$.

Логика $\logic{K}^\ast$~--- это множество $\lang{ML}^{\ast}$-формул, истинных в классе всех шкал Крипке.

\begin{theorem}
\label{maintheorem:K-ast}
Константный фрагмент логики $\logic{K}^\ast$ является $\cclass{EXPTIME}$-полным.
\end{theorem}

\begin{proof}
Тот факт, что константный фрагмент логики $\logic{K}^\ast$ принадлежит классу $\cclass{EXPTIME}$, следует из того, что проблема разрешения для логики $\logic{K}^\ast$ является $\cclass{EXPTIME}$-полной~\cite{FL79} (см.~также~\cite{BlackburndeRijkeVenema-2001-1}). Покажем, что константный фрагмент логики $\logic{K}^\ast$ является $\cclass{EXPTIME}$-трудным. Для этого сначала сведём проблему принадлежности логике $\logic{K}^\ast$ к проблеме принадлежности логике $\logic{K}^\ast$ формул специального вида.

Пусть $\varphi$~--- произвольная $\lang{ML}^\ast$-формула. Пусть $p_1,\ldots,p_n$~--- список пропозициональных переменных, входящих в~$\varphi$. Пусть $p_{n+1}$~--- новая переменная, т.е. не входящая в~$\varphi$. Используя~$p_{n+1}$, для каждой подформулы~$\psi$ формулы~$\varphi$ определим формулу~$\psi^\ast$ следующим образом:
$$
\begin{array}{lcl}
\bot^\ast & = & \bot;
\\

p_i^\ast  & = & p_i,~ \qquad \mbox{где $i\in\{1,\ldots,n\}$;}
\\

(\psi_1\wedge\psi_2)^\ast & = & \psi_1^\ast\wedge\psi_2^\ast;
\\

(\psi_1\vee\psi_2)^\ast & = & \psi_1^\ast\vee\psi_2^\ast;
\\

(\psi_1\to\psi_2)^\ast & = & \psi_1^\ast\to\psi_2^\ast;
\\

(\Box\psi_1)^\ast & = & \Box(p_{n+1}\to\psi_1^\ast);
\\

(\Box^\ast\psi_1)^\ast & = & \Box^\ast(p_{n+1}\to\psi_1^\ast).
\end{array}
$$

Определим формулу $\varphi^\#$, положив
$$
\begin{array}{rcl}
\varphi^\# & = & (p_{n+1}\wedge\Box^\ast(\neg
p_{n+1}\to\Box^\ast\neg p_{n+1}))\to\varphi^\ast.
\end{array}
$$

Формула $\varphi^\#$ отличается от $\varphi$, по сути, лишь тем, что все вхождения модальностей в $\varphi$ заменяются вхождениями соответствующих ограниченных модальностей, где в роли ограничения выступает новая переменная~$p_{n+1}$; выше (см.~стр.~\pageref{page:relativization}) мы назвали такой приём релятивизацией, только вместо переменной~$p_{n+1}$ использовалась формула~$B_\varphi$. Снимая это ограничение, мы получаем формулу, эквивалентную в $\logic{K}^\ast$ исходной формуле~$\varphi$. Более точно, пусть $\varphi^\#_\top$~--- формула, получающаяся из~$\varphi^\#$ подстановкой формулы~$\top$ вместо переменной~$p_{n+1}$.

\begin{lemma}
\label{lem1a}
Справедливо следующее:
$\varphi^\#_\top\leftrightarrow\varphi\in\logic{K}^\ast$.
\end{lemma}

\begin{proof}
Достаточно заметить, что $\top\wedge\Box^\ast(\neg
\top\to\Box^\ast\neg \top)\in\logic{K}^\ast$ и $(\top\to q)\leftrightarrow q \in\logic{K}^\ast$.
\end{proof}

%

\begin{lemma}
\label{lem1}
Имеет место следующая эквивалентность:
$$
\begin{array}{rcl}
\varphi\in\logic{K}^\ast
  & \iff
  & \varphi^\#\in\logic{K}^\ast.
\end{array}
$$
\end{lemma}

\begin{proof}
Если $\varphi^\#\in\logic{K}^\ast$, то $\varphi^\#_\top\in\logic{K}^\ast$, откуда, согласно лемме~\ref{lem1a}, получаем, что $\varphi\in\logic{K}^\ast$.

Пусть теперь $\varphi\in\logic{K}^\ast$. Предположим, что $\varphi^\#\not\in\logic{K}^\ast$. Тогда существует шкала $\kframe{F}=\langle W,R\rangle$ и модель $\kmodel{M} = \langle\kframe{F},v\rangle$, определённая на этой шкале, такая, что $(\kmodel{M},x_0)\not\models\varphi^\#$ для некоторого $x_0\in W$.
Пусть
$$
\begin{array}{rcl}
W' & = & \{x\in W : \mbox{$x_0R^\ast x$ и $(\kmodel{M},x)\models p_{n+1}$}\}.
\end{array}
$$
Заметим, что $W'\ne\varnothing$, т.\,к. $x_0\in W'$. На множестве $W'$ определим отношение $R'$, положив для всяких $x,y\in W'$
$$
\begin{array}{rcl}
xR'y & \leftrightharpoons & xRy.
\end{array}
$$
Пусть $\kframe{F}' = \langle W',R'\rangle$. В шкале $\kframe{F}'$ определим оценку $v'$, положив для всякой переменной $p$ и всякого $x\in W'$
$$
\begin{array}{rcl}
x\in v'(p) & \leftrightharpoons & x\in v(p).
\end{array}
$$
Пусть $\kmodel{M}'=\langle\kframe{F}',v'\rangle$.

\pagebreak[3]

Индукцией по построению подформулы $\psi$ формулы $\varphi$ покажем, что для всякого $x\in
W'$ имеет место эквивалентность
$$
\begin{array}{rcl}
(\kmodel{M}',x)\models\psi
  & \iff
  & (\kmodel{M},x)\models\psi^\ast.
\end{array}
$$

Случай, когда $\psi=\bot$ тривиален. Если $\psi=p_i$, то указанная эквивалентность выполняется в силу определения оценки~$v'$.

Пусть формулы $\psi_1$ и $\psi_2$ таковы, что для всякого $x\in W'$ имеют место эквивалентности
$$
\begin{array}{rcl}
(\kmodel{M}',x)\models\psi_1 & \iff &
(\kmodel{M},x)\models\psi_1^\ast;
\medskip \\
(\kmodel{M}',x)\models\psi_2 & \iff &
(\kmodel{M},x)\models\psi_2^\ast.
\end{array}
$$

Если $\psi = \psi_1\wedge \psi_2$, то для всякого $x\in W'$ имеем
следующую цепочку эквивалентностей:
$$
\begin{array}{rcl}
(\kmodel{M}',x)\models\psi & \iff &
(\kmodel{M}',x)\models\psi_1 \mbox{ и } (\kmodel{M}',x)\models\psi_2
\medskip \\
& \iff & (\kmodel{M},x)\models\psi_1^\ast
\mbox{ и } (\kmodel{M},x)\models\psi_2^\ast
\medskip \\
& \iff & (\kmodel{M},x)\models\psi^\ast.
\end{array}
$$

Случаи $\psi = \psi_1\vee \psi_2$ и $\psi = \psi_1\to \psi_2$ рассматриваются аналогично.

Пусть $\psi = \Box\psi_1$ и пусть $x\in W'$.

Если $(\kmodel{M}',x)\not\models\psi$, то существует мир $y\in W'$, такой, что $xR'y$ и $(\kmodel{M}',y)\not\models\psi_1$, и, согласно индукционному предположению, $(\kmodel{M},y)\not\models\psi_1^\ast$. Но $y\in W'$, поэтому $(\kmodel{M},y)\models p_{n+1}$, а следовательно, $(\kmodel{M},y)\not\models p_{n+1}\to\psi_1^\ast$, а т.\,к. $xRy$, получаем, что
$(\kmodel{M},x)\not\models\Box(p_{n+1}\to\psi_1^\ast)$, т.\,е. $(\kmodel{M},x)\not\models\psi^\ast$.

Пусть теперь $(\kmodel{M},x)\not\models\psi^\ast$. Тогда существует мир $y\in W$, такой, что $xRy$, $(\kmodel{M},y)\models p_{n+1}$ и $(\kmodel{M},y)\not\models\psi_1^\ast$. Поскольку $(\kmodel{M},y)\models p_{n+1}$, получаем, что $y\in W'$, и, по индукционному предположению, $(\kmodel{M}',y)\not\models\psi_1$, а т.\,к. $xR'y$, получаем, что $(\kmodel{M}',x)\not\models\psi$.

Пусть $\psi = \Box^\ast\psi_1$ и пусть $x\in W'$.

Если $(\kmodel{M}',x)\not\models\psi$, то аналогично предыдущему случаю (просто заменяя в доказательстве $R$ на $R^\ast$) получаем, что $(\kmodel{M},x)\not\models\psi^\ast$.

Пусть теперь $(\kmodel{M},x)\not\models\psi^\ast$. Тогда существует мир $y\in W$ такой, что $xR^\ast y$, $(\kmodel{M},y)\models p_{n+1}$ и $(\kmodel{M},y)\not\models\psi_1^\ast$. Поскольку $(\kmodel{M},y)\models p_{n+1}$, получаем, что $y\in W'$. Покажем, что $xR'^\ast y$. Утверждение $xR^\ast y$ означает, что выполнено хотя бы одно из следующих условий:
\begin{itemize}
\item $x=y$;
\item $xRy$;
\item существуют $x_1,\ldots,x_k\in W$, такие, что $xRx_1R\ldots Rx_kRy$.
\end{itemize}
Ясно, что в первом и втором случаях $xR'^\ast y$. Рассмотрим третий случай. Покажем, что $x_1,\ldots,x_k\in W'$. Предположим, что это не так, т.\,е. $x_i\not\in W'$ для некоторого $i\in\{1,\ldots,k\}$. Но тогда $(\kmodel{M},x_i)\not\models p_{n+1}$. Поскольку $(\kmodel{M},x_0)\not\models\varphi^\#$, получаем, что $(\kmodel{M},x_0)\models p_{n+1}\wedge\Box^\ast(\neg p_{n+1}\to\Box^\ast\neg p_{n+1})$. Последнее, в частности, означает, что $(\kmodel{M},x_i)\models\neg p_{n+1}\to\Box^\ast\neg p_{n+1}$, и из того, что $(\kmodel{M},x_i)\models\neg p_{n+1}$, получаем, что $(\kmodel{M},x_i)\models\Box^\ast\neg p_{n+1}$, а следовательно, $(\kmodel{M},y)\models\neg p_{n+1}$, что невозможно, т.\,к. $y\in W'$. Таким образом, $x_1,\ldots,x_k\in W'$, а значит, и в третьем случае $xR'^\ast y$. Осталось заметить, что, по индукционному предположению, $(\kmodel{M}',y)\not\models\psi_1$, а значит, $(\kmodel{M}',x)\not\models\psi$.

Из доказанного следует, что $(\kmodel{M},x_0)\not\models\varphi$, что невозможно, т.\,к. $\varphi\in\logic{K}^\ast$. Следовательно, $\varphi^\#\in\logic{K}^\ast$.
\end{proof}

Отметим ещё одно свойство формулы $\varphi^\#$.

\begin{lemma}
\label{lem2}
Если $\varphi^\#\not\in\logic{K}^\ast$, то существует модель $\kmodel{M}=\langle\kframe{F},v\rangle$, определённая на некоторой шкале $\kframe{F}=\langle W,R\rangle$, такая, что $\kmodel{M}\not\models\varphi^\#$ и $v(p_{n+1})=W$.
\end{lemma}

\begin{proof}
Если $\varphi^\#\not\in\logic{K}^\ast$, то, согласно лемме~\ref{lem1}, $\varphi\not\in\logic{K}^\ast$, и, согласно лемме~\ref{lem1a}, $\varphi^\#_\top\not\in\logic{K}^\ast$, откуда следует справедливость доказываемого утверждения.
\end{proof}

\begin{figure}
  \centering
  \begin{tikzpicture}[scale=1.50]

    \coordinate (a30)    at ( 0.0, 3);
    \coordinate (a31)    at ( 0.0, 4);
    \coordinate (a32)    at ( 0.0, 5);
    \coordinate (a34)    at ( 0.0, 6);
    \coordinate (a35)    at ( 0.0, 7);
    \coordinate (b30)    at (-1.0, 4);

    \draw [fill]     (a30)   circle [radius=2.0pt] ;
    \draw [fill]     (a31)   circle [radius=2.0pt] ;
    \draw [fill]     (a32)   circle [radius=2.0pt] ;
    \draw [fill]     (a35)   circle [radius=2.0pt] ;
    \draw []     (b30)   circle [radius=2.0pt] ;

    \begin{scope}[>=latex]

      \draw [->, shorten >=  2.75pt, shorten <= 2.75pt] (a30) -- (a31) ;
      \draw [->, shorten >=  2.75pt, shorten <= 2.75pt] (a31) -- (a32) ;
      \draw [->, shorten >= 11.75pt, shorten <= 2.75pt] (a32) -- (a34) ;
      \draw [->, shorten >=  2.75pt, shorten <= 7.75pt] (a34) -- (a35) ;
      \draw [->, shorten >=  2.75pt, shorten <= 2.75pt] (a30) -- (b30) ;

    \end{scope}

      \node []         at (a34)  {$\vdots$};

      \node [right=2pt] at (a30)  {$a^0_{m}$};
      \node [right=2pt] at (a31)  {$a^1_{m}$};
      \node [right=2pt] at (a32)  {$a^2_{m}$};
      \node [right=2pt] at (a35)  {$a^m_{m}$};

      \node [left=2pt] at (b30)  {$b_m$};

    \end{tikzpicture}
    \caption{Шкала $\kframe{F}_m$}
    \label{fig:K-ast}
  \end{figure}

Для каждого $m\in\numNp$ определим формулы $\alpha_m$ и $\beta_m$ следующим образом:
$$
\begin{array}{rcl}
\alpha_m & = & \Diamond^m\Box\bot \wedge
\neg\Diamond^{m+1}\Box\bot \wedge \Diamond(\Diamond\top \wedge
\Box\Diamond\top);
\\

\beta_m & = & \Diamond\alpha_m.
\end{array}
$$
Кроме того, для каждого $m\in\numNp$ определим шкалу $\kframe{F}_m =
\langle W_m,R_m \rangle$, положив
$$
\begin{array}{rcl}
W_m & = & \{a_m^0,a_m^1,\ldots,a_m^m\}\cup\{b_m\};
\medskip \\
xR_my & \leftrightharpoons & \mbox{либо $x=a_m^i$, $y=a_m^j$ и
$i<j$,}
\\
& & \mbox{либо $x=a_m^0$, $y=b_m$,}
\\
& & \mbox{либо $x=y=b_m$.}
\end{array}
$$
Шкала $\kframe{F}_m$ изображена на рис.~\ref{fig:K-ast}, и уже использовалась выше (единственное, её миры были обозначены иначе); чёрные кружк\'{и} соответствуют иррефлексивным мирам $a_m^0,\ldots,a_m^m$, светлый~--- рефлексивному миру $b_m$, отношение достижимости
транзитивно.

\begin{lemma}
\label{lem3}
Для всяких $k,m\in\numNp$ имеет место следующая эквивалентность:
$$
\begin{array}{rcl}
(\kframe{F}_m,x)\models\alpha_k & \iff &
\mbox{$k=m$ и $x=a_m^0$.}
\end{array}
$$
\end{lemma}

\begin{proof}
Прямая проверка.
\end{proof}

Определим $\varphi^\#_\beta$ как формулу, получающуюся подстановкой формул $\beta_1,\ldots,\beta_{n+1}$ в формулу $\varphi^\#$ вместо переменных $p_1,\ldots,p_{n+1}$ соответственно.

\begin{lemma}
Имеет место следующая эквивалентность:
$$
\begin{array}{rcl}
\varphi\in\logic{K}^\ast & \iff &
\varphi^\#_\beta\in\logic{K}^\ast.
\end{array}
$$
\end{lemma}

\begin{proof}
Пусть $\varphi\in\logic{K}^\ast$. Тогда, по лемме~\ref{lem1}, $\varphi^\#\in\logic{K}^\ast$, а поскольку формула $\varphi^\#_\beta$ является подстановочным примером формулы $\varphi^\#$, получаем, что $\varphi^\#_\beta\in\logic{K}^\ast$.

Пусть $\varphi\not\in\logic{K}^\ast$. Тогда, согласно лемме~\ref{lem1}, $\varphi^\#\not\in\logic{K}^\ast$. Следовательно, существует модель $\kmodel{M}=\langle\kframe{F},v\rangle$, определённая на некоторой шкале $\kframe{F}=\langle W,R\rangle$, такая, что формула $\varphi^\#$ опровергается в некотором мире $x_0$ этой модели. Согласно лемме~\ref{lem2}, можем считать, что $\kmodel{M}\models p_{n+1}$. Построим шкалу, в некотором мире которой будет опровергаться формула~$\varphi^\#_\beta$.

Без ограничений общности можем (и будем) считать, что множества миров шкал Крипке $\kframe{F},\kframe{F}_1,\ldots,\kframe{F}_{n+1}$ попарно не пересекаются. Пусть $\kframe{F}_\beta = \langle W_\beta,R_\beta\rangle$, где
$$
\begin{array}{rcl}
W_\beta & = & W\cup W_1\cup\ldots\cup W_{n+1};
\smallskip
\\
xR_\beta y & \leftrightharpoons & \mbox{либо $x,y\in W$ и $xRy$,}
\\
& & \mbox{либо $x,y\in W_m$ для некоторого $m\in\{1,\ldots,n+1\}$ и
$xR_my$,}
\\
& & \mbox{либо $x\in W$, $(\kmodel{M},x)\models p_k$ и $y=a_k^0$.}
\end{array}
$$

Для всякой подформулы $\psi$ формулы $\varphi$ обозначим через $\psi_\beta$ формулу, получающуюся из $\psi^\ast$ подстановкой формул $\beta_1,\ldots,\beta_{n+1}$ вместо $p_1,\ldots,p_{n+1}$ соответственно. Индукцией по построению подформулы $\psi$ формулы $\varphi$ докажем, что для всякого $x\in W$
$$
\begin{array}{rcl}
(\kframe{F}_\beta,x)\models\psi_\beta & \iff
& (\kmodel{M},x)\models\psi^\ast.
\end{array}
\eqno{\mbox{$({\ast})$}}
$$

Если $\psi=\bot$, то $\psi^\ast=\bot$ и $\psi_\beta=\bot$, поэтому указанная эквивалентность выполняется тривиально.

Пусть $\psi=p_k$, где $k\in\{1,\ldots,n\}$. В этом случае $\psi^\ast=p_k$, $\psi_\beta=\beta_k$.

Пусть $x\in W$ и $(\kmodel{M},x)\models p_k$. В этом случае $xR_\beta a_k^0$. Тогда, согласно лемме~\ref{lem3}, $(\kframe{F}_k,a_k^0)\models\alpha_k$; ясно, что также $(\kframe{F}_\beta,a_k^0)\models\alpha_k$, а поэтому $(\kframe{F}_\beta,x)\models\Diamond\alpha_k$, т.\,е. $(\kframe{F}_\beta,x)\models\beta_k$.

Пусть $x\in W$ и $(\kframe{F}_\beta,x)\models\beta_k$. Тогда существует мир $y\in W_\beta$, такой, что $xR_\beta y$ и $(\kframe{F}_\beta,y)\models\alpha_k$. Заметим, что $y\not\in W$. Действительно, предположим, что $y\in W$. Тогда $yR_\beta a_{n+1}^0$ (поскольку $\kmodel{M}\models p_{n+1}$), поэтому $(\kframe{F}_\beta,y)\not\models\neg\Diamond^{k+1}\Box\bot$, и следовательно, $(\kframe{F}_\beta,y)\not\models\alpha_k$. Получаем противоречие, а значит, $y\not\in W$. Получаем, что $y\in W_m$ для некоторого $m\in\{1,\ldots,n+1\}$, и в силу леммы~\ref{lem3}, $y=a_k^0$. Осталось заметить, что, по определению отношения $R_\beta$, $xR_\beta a_k^0$ только в том случае, когда $(\kmodel{M},x)\models p_k$.

Пусть подформулы $\chi$ и $\xi$ формулы $\varphi$ таковы, что для всякого $z\in W$
$$
\begin{array}{rcl}
(\kframe{F}_\beta,z)\models\chi_\beta & \iff
& (\kmodel{M},z)\models\chi^\ast;
\medskip \\
(\kframe{F}_\beta,z)\models\xi_\beta & \iff &
(\kmodel{M},z)\models\xi^\ast.
\end{array}
$$

Пусть $\psi=\chi\wedge\xi$, $x\in W$. Тогда имеют место следующие эквивалентности:
$$
\begin{array}{rcl}
(\kframe{F}_\beta,x)\models\psi_\beta & \iff
& (\kframe{F}_\beta,x)\models\chi_\beta \mbox{ и }
(\kframe{F}_\beta,x)\models\xi_\beta
\medskip \\
& \iff & (\kmodel{M},x)\models\chi^\ast
\mbox{ и } (\kmodel{M},x)\models\xi^\ast
\medskip \\
& \iff & (\kmodel{M},x)\models\psi^\ast.
\end{array}
$$

Случаи, когда $\psi=\chi\vee\xi$ и $\psi=\chi\to\xi$, рассматриваются аналогично.

Пусть $\psi=\Box\chi$, $x\in W$. Если $(\kmodel{M},x)\not\models(\Box\chi)^\ast$, то существует мир $y\in W$, такой, что $xRy$, $(\kmodel{M},y)\models p_{n+1}$ и $(\kmodel{M},y)\not\models\chi^\ast$. По индукционному предположению, $(\kframe{F}_\beta,y)\not\models\chi_\beta$. С другой стороны, т.\,к. $(\kmodel{M},y)\models p_{n+1}$, получаем, что $yR_\beta a_{n+1}^0$, откуда следует, что
$(\kframe{F}_\beta,y)\models\beta_{n+1}$. Значит, $(\kframe{F}_\beta,y)\not\models\beta_{n+1}\to\chi_\beta$, и следовательно, $(\kframe{F}_\beta,x)\not\models\Box(\beta_{n+1}\to\chi_\beta)$, т.\,е. $(\kframe{F}_\beta,x)\not\models\psi_\beta$.

Пусть теперь $(\kframe{F}_\beta,x)\not\models\psi_\beta$. Тогда существует мир $y\in W_\beta$, такой, что $xR_\beta y$, $(\kframe{F}_\beta,y)\models\beta_{n+1}$ и $(\kframe{F}_\beta,y)\not\models\chi_\beta$. Нетрудно убедиться, что $\beta_{n+1}$ опровергается в каждом мире каждой из шкал $\kframe{F}_1,\ldots,\kframe{F}_{n+1}$, откуда несложно заключить, что $y\in W$. Следовательно, по индукционному предположению, $(\kmodel{M},y)\not\models\chi^\ast$. Учитывая, что $\kmodel{M}\models p_{n+1}$, получаем, что $(\kmodel{M},y)\not\models p_{n+1}\to\chi^\ast$, а значит, $(\kmodel{M},y)\not\models \Box(p_{n+1}\to\chi^\ast)$, т.\,е. $(\kmodel{M},y)\not\models \psi^\ast$.

Случай, когда $\psi=\Box^\ast\chi$, рассматривается аналогично.

На этом обоснование эквивалентности~\mbox{$({\ast})$} завершено.

Теперь вернёмся к формуле $\varphi^\#$. Из того, что $(\kmodel{M},x_0)\not\models\varphi^\#$, следует, что $(\kmodel{M},x_0)\models p_{n+1}\wedge\Box^\ast(\neg p_{n+1}\to\Box^\ast\neg p_{n+1})$ и $(\kmodel{M},x_0)\not\models\varphi^\ast$. Последнее, с учётом~\mbox{$({\ast})$}, даёт, что $(\kframe{F}_\beta,x_0)\not\models\varphi_\beta$. Кроме того, нетрудно видеть, что $(\kframe{F}_\beta,x_0)\models \beta_{n+1}\wedge\Box^\ast(\neg \beta_{n+1}\to\Box^\ast\neg \beta_{n+1})$, а значит, $(\kframe{F}_\beta,x_0)\not\models\varphi_\beta^\#$. Следовательно, $\varphi^\#_\beta\not\in\logic{K}^\ast$.
\end{proof}

Для завершения доказательства теоремы~\ref{maintheorem:K-ast} осталось заметить, что формула $\varphi^\#_\beta$ строится по формуле~$\varphi$ полиномиально.
\end{proof}

Теорема~\ref{maintheorem:K-ast} и метод её доказательства позволяют получить следствия, связанные со сложностью фрагментов от небольшого числа переменных различных полимодальных логик, язык которых обогащён модальностями, близкими к~$\Box^\ast$. Некоторые из таких логик содержат $\logic{K}^\ast$ в качестве своего фрагмента (с~точностью до обозначения модальностей), а для некоторых можно провести сходное моделирование, но с помощью формул от одной переменной.

    \subsection{Логики знания}

Логики знания интерпретируют модальность $\Box_k$ языка $\lang{ML}_n$ как модальность знания \defnotion{агента}~$k$.\index{агент} Предполагается, что имеются агенты, имена которых для удобства обозначим натуральными числами $1,\ldots,n$, и каждый из них может знать какие-то факты, в том числе факты о знаниях других агентов. Вместо $\Box_1,\ldots,\Box_n$ используются обозначения $K_1,\ldots,K_n$, и $\lang{ML}_n$-формула $K_i\varphi$ читается как <<агент~$i$ знает, что~$\varphi$>>. В зависимости от того, какие свойства приписываются модальностям $K_1,\ldots,K_n$ (аксиоматически или семантически), мы получаем те или иные \defnotion{логики знания};\index{логика!еяд@знания} обычно считают, что агенты однотипны, и мы получаем в качестве логик знания такие логики, как $\logic{T}_n$, $\logic{KD45}_n$, $\logic{S4}_n$, $\logic{S5}_n$ и другие~\cite{FHMV95}; при этом, как правило, считают, что $n\geqslant 2$.

В языке $\lang{ML}_n$ можно формулировать некоторые утверждения о знаниях, отличных от утверждений о знаниях отдельного агента. Например, можно выразить модальности $S$ и $E$, где $S\varphi$ и $E\varphi$ понимаются как <<кто-то из агентов знает, что~$\varphi$>> и <<каждый из агентов знает, что~$\varphi$>>:
$$
\begin{array}{lcl}
S\varphi & = & K_1\varphi\vee\ldots\vee K_n\varphi; \\
E\varphi & = & K_1\varphi\wedge\ldots\wedge K_n\varphi. \\
\end{array}
$$
Обычным образом введём сокращения для $S^k$ и $E^k$, где $k\in\numNp$:
$$
\begin{array}{lclclcll}
S^1\varphi & = & S\varphi; & & E^1\varphi & = & E\varphi; \\
S^{m+1}\varphi & = & SS^m\varphi; & & E^{m+1}\varphi & = & EE^m\varphi, & \mbox{где $m\in\numN$.} \\
\end{array}
$$
Отметим, что итерирование модальностей $S$ и $E$ приводит к экспоненциальному росту длины формул, если эти модальности вводить как указанные сокращения, но этого не произойдёт, если их добавить к языку $\lang{ML}_n$ в качестве самостоятельных символов языка.

Очень часто языки логик знания обогащают другими модальностями, которые описывают иные виды знания, не выразимые $\lang{ML}_n$-формулами. Мы рассмотрим в качестве примера две такие модальности: модальность $D$ \defnotion{распределённого знания}\index{модальность!распределённого знания} и модальность $C$ \defnotion{общего знания}\index{модальность!общего знания} агентов. Языки, обогащённые такими модальностями будем обозначать $\lang{ML}_n^{D}$, $\lang{ML}_n^{C}$, $\lang{ML}_n^{DC}$, где в верхнем индексе присутствуют те модальности, которые были добавлены к языку~$\lang{ML}_n$. Эти модальности позволяют строить формулы вида $D\varphi$ и $C\varphi$, которые можно читать как <<в системе распределено знание о~$\varphi$>> и <<общеизвестно, что~$\varphi$>>.

Определим модальности $D$ и $C$ семантически. Пусть $\kframe{F}=\otuple{W,R_1,\ldots,R_n}$~--- шкала Крипке, $\kmodel{M}=\otuple{\kframe{F},v}$~--- модель Крипке на этой шкале, $w$~--- мир модели~$\kmodel{M}$. Пусть $R_D=R_1\cap\ldots\cap R_n$ и $R_C$~--- транзитивное замыкание отношения $R_1\cup\ldots\cup R_n$. Тогда полагаем, что
$$
\begin{array}{lcl}
(\kmodel{M},w)\models D\varphi
  & \bydef
  & \mbox{$(\kmodel{M},w')\models \varphi$ для любого $w'\in R_D(w)$;} \\
(\kmodel{M},w)\models C\varphi
  & \bydef
  & \mbox{$(\kmodel{M},w')\models \varphi$ для любого $w'\in R_C(w)$.} \\
\end{array}
$$

Приведём несколько примеров, демонстрирующих некоторые свойства модальностей $D$ и~$C$.

Модальность $D$. Если, например, агенты $1$ и $2$ знают, соответственно, что $\varphi\to\psi$ и $\varphi$ (т.е. в модели верны формулы $K_1(\varphi\to\psi)$ и $K_2\varphi$), то знание факта~$\psi$ является распределённым (т.е. в модели верна формула~$D\psi$), причём даже если ни один из агентов не знает, что~$\psi$ верно (т.е. формулы $K_1\psi$ и $K_2\psi$ опровергаются в модели). Кроме того, нетрудно понять, что формулы вида $K_i\varphi\to D\varphi$, где $i\in\set{1,\ldots,n}$, верны в любом мире любой модели Крипке, а также что модальность $D$ является нормальной: если в модели истинна формула $\varphi$, то в этой модели истинна и формула~$D\varphi$.

Модальность $C$. Заметим, что модальность $C$ можно определить, используя модальность~$E$: если $w$~--- мир некоторой модели~$\kmodel{M}$, то
$$
\begin{array}{lcl}
(\kmodel{M},w)\models C\varphi
  & \iff
  & \mbox{$(\kmodel{M},w)\models E^k\varphi$ для любого $k\in \numNp$,}
\end{array}
$$
из чего несложно заключить, что
\settowidth{\templength}{\mbox{$(\kmodel{M},w)\models E^k\varphi$ для любого $k\in \numNp$}}
$$
\begin{array}{lcl}
(\kmodel{M},w)\models C\varphi
  & \iff
  & \parbox[t]{\templength}{$(\kmodel{M},w)\models E(\varphi\wedge C\varphi)$.}
\end{array}
$$
Кроме того, нетрудно понять, что формулы вида $C\varphi\to K_i\varphi$, где $i\in\set{1,\ldots,n}$, верны в любом мире любой модели Крипке, а также что модальность $C$ тоже является нормальной: если в модели истинна формула $\varphi$, то в этой модели истинна и формула~$C\varphi$.

Для полной по Крипке нормальной модальной логики $L$ в языке $\lang{ML}_n$ определим логики $L^D$, $L^C$, $L^{DC}$ как множества формул, соответственно, в языках $\lang{ML}_n^{D}$, $\lang{ML}_n^{C}$, $\lang{ML}_n^{DC}$, истинных в классе всех шкал Крипке логики~$L$.

Известно~\cite{HalpernMoses-1992-1}, что добавление модальности $D$ к логикам знания не увеличивает вычислительной сложности проблемы разрешения. Поэтому мы можем получить аналог теоремы~\ref{th:PSPACE:S5-2}.

\begin{theorem}
\label{th:PSPACE:S5-2-D}
Пусть $n\geqslant 2$. Тогда проблема разрешения фрагмента от одной переменной логики $L\in [\logic{K}_n^D,\logic{S5}_n^D]$ является\/ $\cclass{PSPACE}$\nobreakdash-трудной.
\end{theorem}

Для части логик можно сделать уточнение.

\begin{theorem}
\label{th:PSPACE:wGrz-2-D}
Пусть $n\geqslant 2$. Тогда проблема разрешения константного фрагмента логики $L\in [\logic{K}_n^D,(\logic{wGrz}\fusion\logic{S5}_{n-1})^D]$ является\/ $\cclass{PSPACE}$\nobreakdash-трудной.
\end{theorem}

В случае наличия в языке модальности $C$ можно промоделировать в нём модальность $\Box^\ast$ языка логики $\logic{K}^\ast$ с помощью модальности $C^+$, где $C^+\varphi = \varphi\wedge C\varphi$. Хотя такое моделирование дублирует формулу, стоящую под модальностью $\Box^\ast$, и возникающий при этом перевод $\lang{ML}^\ast$ формул в $\lang{ML}_1^C$-формулы является экспоненциальным, реально интересующие нас формулы (т.е. позволяющие доказать $\cclass{EXPTIME}$-трудность логики~$\logic{K}^\ast$), не содержат итераций модальности~$\Box^\ast$~\cite{FL79}, а значит, такой перевод для них будет полиномиальным. Как следствие, получаем следующие теоремы.

\begin{theorem}
\label{th:PSPACE:S5-2-DC}
Пусть $n\geqslant 2$. Тогда проблема разрешения фрагмента от одной переменной логики $L\in [\logic{K}_n^C,\logic{S5}_n^C]\cup[\logic{K}_n^{DC},\logic{S5}_n^{DC}]$ является\/ $\cclass{EXPTIME}$\nobreakdash-трудной.
\end{theorem}

\begin{theorem}
\label{th:PSPACE:wGrz-2-DC}
Пусть $n\geqslant 2$. Тогда проблема разрешения константного фрагмента логики $L\in [\logic{K}_n^C,(\logic{wGrz}\fusion\logic{S5}_{n-1})^C] \cup [\logic{K}_n^{DC},(\logic{wGrz}\fusion\logic{S5}_{n-1})^{DC}]$ является\/ $\cclass{EXPTIME}$\nobreakdash-трудной.
\end{theorem}

    \subsection{Логики с универсальной модальностью}

Добавим к языку $\lang{ML}_n$, где $n\in\numNp$, \defnotion{универсальную модальность} $A$;\index{модальность!универсальная} получившийся язык обозначим $\lang{ML}_n^A$. Модальность $A$ позволяет строить формулы вида~$A\varphi$.

Определим модальность $A$ семантически. Пусть $\kframe{F}=\otuple{W,R_1,\ldots,R_n}$~--- шкала Крипке, $\kmodel{M}=\otuple{\kframe{F},v}$~--- модель Крипке на этой шкале, $w$~--- мир модели~$\kmodel{M}$. Тогда полагаем, что
$$
\begin{array}{lcl}
(\kmodel{M},w)\models A\varphi
  & \bydef
  & \mbox{$(\kmodel{M},w')\models \varphi$ для любого $w'\in W$.} \\
\end{array}
$$

Модальность $A$ ведёт себя как $\logic{S5}$-модальность $\Box$ в связной шкале. Доказательство $\cclass{EXPTIME}$-трудности логик с универсальной модальностью не отличается от доказательства $\cclass{EXPTIME}$-трудности логик с модальностью типа $\Box^\ast$ или~$C$. То же относится и к моделированию переменных $\lang{ML}_n^A$-формул константными формулами или формулами от одной переменной. Поэтому в качестве следствия описанной выше конструкции для~$\logic{K}^\ast$ получаем следующие теоремы.

\begin{theorem}
\label{th:PSPACE:S5-2-A}
Пусть $n\geqslant 2$. Тогда проблема разрешения фрагмента от одной переменной логики $L\in [\logic{K}_n^A,\logic{S5}_n^A]$ является\/ $\cclass{EXPTIME}$\nobreakdash-трудной.
\end{theorem}

\begin{theorem}
\label{th:PSPACE:wGrz-2-A}
Пусть $n\geqslant 2$. Тогда проблема разрешения константного фрагмента логики $L\in [\logic{K}_n^A,(\logic{wGrz}\fusion\logic{S5}_{n-1})^A]$ является\/ $\cclass{EXPTIME}$\nobreakdash-трудной.
\end{theorem}

\begin{theorem}
\label{th:PSPACE:T-1-A}
Пусть $n\geqslant 1$. Тогда проблема разрешения фрагмента от одной переменной логики $L\in [\logic{K}_n^A,(\logic{T}\fusion\logic{S5}_{n-1})^A]$ является\/ $\cclass{EXPTIME}$\nobreakdash-трудной.
\end{theorem}

\begin{theorem}
\label{th:PSPACE:K-1-A}
Проблема разрешения константного фрагмента логики $\logic{K}^A$ является\/ $\cclass{EXPTIME}$\nobreakdash-трудной.
\end{theorem}

На самом деле, эти теоремы можно расширить и на другие логики, например, заменяя в формулировках логики вида $\logic{S5}_{k+2}$ (при $k\in\numNp$) логикой $\logic{S5}_2\fusion\logic{Triv}_k$ или логикой $\logic{S5}_2\fusion\logic{Verum}_k$. Мы не будем <<вылавливать>> тонкости подобного рода, поскольку такие логики весьма редко оказываются в поле интереса исследований в области полимодальных логик.

    \subsection{Сложность аппроксимации логики $\logic{K}^\ast$}

Функция сложности для полимодальных логик определяется так же, как и для мономодальных или суперинтуиционистских: если $L$~--- полимодальная логика, то 
$$
\begin{array}{lcl}
f_L(n)
  & =
  & \displaystyle
    \max\big\{\min\{|\kframe{F}| : \mbox{$\kframe{F}\models L$, $\kframe{F}\not\models\varphi$}\}
         : \mbox{$|\varphi| \leqslant n$, $\varphi \not\in L$}\big\},
\end{array}
$$
где $|\kframe{F}|$~--- число миров в шкале $\kframe{F}$, а $|\varphi|$~---
длина формулы~$\varphi$.

Мы здесь коснёмся вопроса сложности аппроксимации шкалами Крипке логики $\logic{K}^\ast$ и её фрагментов, но приведённые ниже наблюдения легко переносятся и на логики знания с модальностью $C$, и на логики с универсальной модальностью.

Поскольку $\lang{ML}$-фрагмент логики $\logic{K}^\ast$ совпадает с логикой $\logic{K}$, функция сложности логики $\logic{K}^\ast$ ограничена снизу экспонентой. Тем не менее, сложность аппроксимации логики шкалами Крипке описывается не только числом миров в контрмоделях для формул, не принадлежащих этой логике. Дело в том, что функцию сложности для логики $\logic{K}^\ast$ можно ограничить экспонентой не только снизу, но и сверху (см., например,~\cite{HalpernMoses-1992-1}, где это показано для логик знания с модальностью~$C$). Тем не менее, даже при, казалось бы, одинаковых ограничениях на функцию сложности проблема разрешения для логики $\logic{K}$ (а также логик $\logic{T}$, $\logic{K4}$, $\logic{S4}$ и~др{.}) является $\cclass{PSPACE}$-полной, а для логики $\logic{K}^\ast$ (а также логик $\logic{K}_n^C$, $\logic{T}_n^C$ и др{.})~--- $\cclass{EXPTIME}$-полной. Такая разница в вычислительной сложности находит отражение и в устройстве контрмоделей для формул этих логик.

В~языке $\lang{ML}^\ast$ существуют формулы, требующие для своего опровержения модели, содержащие цепи миров, длина которых выражается в виде экспоненты от длины формулы; в случае языка $\lang{ML}$ таких формул нет, что как раз и позволяет получить разрешающий алгоритм для $\logic{K}$ (и~многих других логик), использующий объём памяти, ограниченный полиномом от длины тестируемой формулы~\cite{Ladner77}: память требуется для <<просмотра>> цепей в контрмодели для формулы, а длина таких цепей в случае $\logic{K}$ может быть ограничена сверху модальной глубиной формулы.

Чтобы показать это, рассмотрим следующие $\lang{ML}^\ast$-формулы от пропозициональных переменных $p_1,\dots,p_n$, см.~\cite{BlackburndeRijkeVenema-2001-1}:
$$
\begin{array}{rcl}
A_n^0 & = & \displaystyle \neg p_1\to \bigg(\Box
p_1\wedge\bigwedge\limits_{i=2}^n\big((p_i\to\Box p_i)\wedge(\neg
p_i\to\Box\neg p_i)\big)\bigg);
\medskip \\

A_n^k & = & \displaystyle \neg p_{k+1}\wedge\bigwedge\limits_{i=1}^k p_i \to 
{}
\smallskip \\
& & \displaystyle 
\to  \bigg(\Box\big(p_{k+1}\wedge\bigwedge\limits_{i=1}^k \neg
p_i\big)\wedge \bigwedge\limits_{\mathclap{i=k+2}}^n\big((p_i\to\Box p_i)\wedge(\neg
p_i\to\Box\neg p_i)\big)\bigg),
\end{array}
$$
где $k\in\{1,\dots,n-1\}$.

Поясним <<устройство>> формул $A_n^i$. Если смотреть на набор значений переменных $p_n,\dots,p_1$ как на двоичное число (где $p_i$ соответствует цифре~<<$1$>>, а $\neg p_i$~--- цифре~<<$0$>>), то каждую формулу~$A_n^i$ можно понимать как описание операции прибавления единицы к двоичному числу,
заканчивающемуся на $i$ единиц, перед которыми стоит ноль. Результат этой операции будет описываться набором значений переменных $p_n,\dots,p_1$ в мире, достижимом из мира, в котором истинна формула~$A_n^i$. Несложно убедиться, что в этом случае формула
$$
\begin{array}{rcl}
\varphi_n & = & \displaystyle \neg\bigg(\bigwedge\limits_{\mathclap{i=1}}^{n}\neg p_i 
\wedge  \Box^\ast\Diamond\top  \wedge 
\Box^\ast\bigwedge\limits_{\mathclap{i=1}}^{\mathclap{n-1}} A_n^i\bigg)
\end{array}
$$
требует для своего опровержения модель, число миров в которой не менее чем~$2^n$, в то время как длина формул $\varphi_n$ ограничена сверху полиномом (второй степени) от~$n$.

Аналогичная конструкция возможна и для константного фрагмента языка $\lang{ML}^\ast$: достаточно рассмотреть формулы~$(\varphi_n)^\#_\beta$ (см.~раздел~\ref{ssec:K-ast}). 

\begin{corollary}
Для $k\in\numN$ существует константная $\lang{ML}^\ast$-формула $\varphi_k$, не принадлежащая логике\/ $\logic{K}^\ast$, такая, что любая модель Крипке, опровергающая формулу $\varphi_k$, содержит цепь из не менее чем\/ $2^k$ попарно различных миров, при этом длина формул $\varphi_k$ ограничена сверху некоторым полиномом от~$k$.
\end{corollary}

Несложно понять, что формулы, подобные $\varphi_n$, можно построить и для других упомянутых выше $\cclass{EXPTIME}$-полных логик, а также получить для них аналогичные следствия.

  \section{Динамические логики}
    \subsection{Предварительные сведения}

Пропозициональная динамическая логика была предложена в~\cite{FL79}, и с тех пор используется как один из инструментов для описания поведения программ. Позже она была расширена различными способами для возможности описания работы более широкого спектра программ~\cite{Streett82, HS82, Vardi85, Peleg87, Goldblatt92}. Кроме того, были разработаны различные формализмы, тесно связанные с пропозициональной динамической логикой, используемые для приложений в областях, отличных от описаний программ; например, это представление знаний~\cite{GL94,DLNN97,Massacci01}, запросы к полуструктурированным данным~\cite{ADdeR03}, анализ данных~\cite{CO85} и лингвистика~\cite{Kracht95}.

Сложность проблемы выполнимости для различных вариантов пропозициональной динамической логики имеет решающее значение для их применения в этих областях. Как правило, для формул, содержащих произвольное число пропозициональных переменных, сложность проблемы выполнимости для таких логик довольно высока: эта проблема может быть от $\cclass{EXPTIME}$-полной~\cite{FL79} до неразрешимой~\cite{TinBal2014}.

Ниже будет показано, что при ограничениях на число переменных в языке (и даже при ограничениях на число атомарных программ в языке) сложность пропозициональных динамических логик существенно не меняется. При этом мы не сможем (и не ставим такой целью) рассмотреть все известные варианты и модификации динамических логик, и выберем лишь некоторые из них. Наша основная цель будет состоять в том, чтобы показать, что описанные выше методы моделирования переменных в формулах константными формулами применимы и к динамическим логикам.

Мы рассмотрим обычную пропозициональную динамическую логику $\logic{PDL}$, пропозициональную динамическую логику с пересечением $\logic{IPDL}$ и пропозициональную динамическую логику с параллельной композицией~$\logic{PRSPDL}$. Выбор этих логик из многих возможных связан с тем, что они имеют разную сложность: $\logic{PDL}$ является $\cclass{EXPTIME}$-полной, $\logic{IPDL}$~--- $\cclass{2EXPTIME}$-полной, а $\logic{PRSPDL}$~--- неразрешимой. Наша цель состоит в том, чтобы для каждой из этих логик построить её погружение в её константный фрагмент.

    \subsection{Синтаксис и семантика}

\providecommand{\mmodel}[1]{\mbox{$\kmodel{#1}$}}
\providecommand{\states}[1]{\dismath{#1}}
\providecommand{\rel}   [1]{\dismath{#1}}
\providecommand{\sat}   [3]{\mbox{$\dismath{(\mmodel{#1},#2)\models{#3}}$}}
\providecommand{\notsat}[3]{\mbox{$\dismath{(\mmodel{#1},#2)\not\models{#3}}$}}
\providecommand{\sameas}[0]{\mbox{$\bydef$}}

Язык $\lang{IPDL}$ логики $\logic{IPDL}$ является расширением языка классической пропозициональной логики модальностями вида $\boxm{\alpha}$, где $\alpha$ является \defnotion{программным термом},\index{терм!программный} или, короче, \defnotion{программой},\index{программа} который строится из \defnotion{атомарных программных термов}\index{терм!программный!атомарный} множества $\mathit{AP} = \{a_1, a_2, a_3,\ldots\}$, формул с использованием операций $?$~(тест), а также с помощью операций $\comp$~(композиция), $\choice$ (выбор), $\inter$~(пересечение) и $^\ast$~(итерация). Подразумеваемое значение формулы $\boxm{\alpha} \varphi$ следующее: каждый запуск программы  $\alpha$ в текущем состоянии вычислительной системы приводит к состоянию этой системы, где формула~$\varphi$ истинна. Формально понятие формулы и программы можно определить одновременной рекурсией с помощью следующих BNF-выражений, где $\varphi$~--- формула, а $\alpha$~--- программа:
\[
\begin{array}{llcl}
\arrayitem &
  \varphi & := & p \mid \bot \mid  (\varphi \con \varphi)\mid  (\varphi \dis \varphi)\mid  (\varphi \imp \varphi) \mid \boxm{\alpha} \varphi;
  \arrayitemskip\\
\arrayitem &
  \alpha  & := & a \mid \varphi? \mid (\alpha \comp \alpha) \mid (\alpha \choice \alpha)
                   \mid (\alpha \inter \alpha) \mid \alpha^\ast,
\end{array}
\]
где $p\in\prop$ и $a\in\mathit{AP}$.

\defnotion{Модель Крипке}\index{модель!Крипке}~--- это набор $\mmodel{M} = \langle\states{S}, \{\rel{R}_a\}_{a \in \mathit{AP}}, v\rangle$, где $\states{S}$~--- непустое множество \defnotion{состояний},\index{состояние} $\rel{R}_a$~--- бинарное \defnotion{отношение достижимости}\index{отношение!достижимости} на~$\states{S}$ и $v$~--- \defnotion{оценка},\index{оценка} т.е. функция $v: \prop \to \Power{\states{S}}$. Отношения достижимости для неатомарных программ и отношение истинности формул в мирах модели определяются одновременной рекурсией. Отношение, соответствующее программе $\alpha$, обозначаем $\rel{R}_\alpha$; для отношения $\rel{R}_\alpha$ его рефлексивно-транзитивное замыкание обозначим~$\rel{R}_\alpha^\ast$. Тогда
\[
\begin{array}{llcl}
\arrayitem
  & \otuple{s,t} \in \rel{R}_{\varphi?}
  & \bydef
  & \mbox{$s = t$ и $\sat{M}{s}{\varphi}$};
  \arrayitemskip\\
\arrayitem
  & \otuple{s,t} \in \rel{R}_{\alpha \comp \beta}
  & \bydef
  & \mbox{$\otuple{s,u} \in \rel{R}_\alpha$
    и $\otuple{u,t} \in \rel{R}_\beta$ для некоторого $u \in \states{S}$;}
  \arrayitemskip\\
\arrayitem
  & \otuple{s,t} \in \rel{R}_{\alpha \choice \beta}
  & \bydef
  & \mbox{$\otuple{s,t} \in \rel{R}_\alpha$ или $\otuple{s,t} \in \rel{R}_\beta$;}
  \arrayitemskip\\
\arrayitem
  & \otuple{s,t} \in \rel{R}_{\alpha \inter \beta}
  & \bydef
  & \mbox{$\otuple{s,t} \in \rel{R}_\alpha$ и $\otuple{s,t} \in \rel{R}_\beta$;}
  \arrayitemskip\\
\arrayitem
  & \otuple{s,t} \in \rel{R}_{\alpha^*}
  & \bydef
  & \mbox{$\otuple{s,t} \in \rel{R}^\ast_{\alpha}$;}
  \arrayitemskip\\
\arrayitem
  & \sat{M}{s}{p_i}
  & \bydef
  & s \in v(p_i);
  \arrayitemskip\\
\arrayitem
  & \notsat{M}{s}{\bot};
  \arrayitemskip\\
\arrayitem
  & \sat{M}{s}{\varphi \con \psi}
  & \bydef
  & \mbox{$\sat{M}{s}{\varphi}$ \phantom{л}и\phantom{и} $\sat{M}{s}{\psi}$;}
  \arrayitemskip\\
\arrayitem
  & \sat{M}{s}{\varphi \dis \psi}
  & \bydef
  & \mbox{$\sat{M}{s}{\varphi}$ или $\sat{M}{s}{\psi}$;}
  \arrayitemskip\\
\arrayitem
  & \sat{M}{s}{\varphi \imp \psi}
  & \bydef
  & \mbox{$\notsat{M}{s}{\varphi}$ или $\sat{M}{s}{\psi}$;}
  \arrayitemskip\\
\arrayitem
  & \sat{M}{s}{\boxm{\alpha} \varphi}
  & \bydef
  & \mbox{$\sat{M}{t}{\varphi}$ для любого $t\in\rel{R}_{\alpha}(s)$.}
\end{array}
\]

Формула языка $\lang{IPDL}$ \defnotion{выполнима}, если она истинна в некотором состоянии некоторой модели. Формула языка $\lang{IPDL}$ \defnotion{общезначима}, если она истинна в каждом состоянии каждой модели. Логика $\logic{IPDL}$~--- это множество всех общезначимых $\lang{IPDL}$-формул.

Язык $\lang{PDL}$ логики $\logic{PDL}$ отличается от языка $\lang{IPDL}$ тем, что не содержит операций $\inter$ и~$?$ (мы рассматриваем $\lang{PDL}$ без оператора~$?$, хотя часто этот оператор включается в язык $\lang{PDL}$); семантика языка $\lang{PDL}$ изменяется соответствующим образом. Логика $\logic{PDL}$~--- это множество общезначимых $\lang{PDL}$-формул.

Язык $\lang{PRSPDL}$ логики $\logic{PRSPDL}$ интерпретируется на моделях, состоящих из состояний, обладающих внутренней структурой: состояние $s$ представляет собой \defnotion{композицию} $x \ast y$ состояний $x$ и $y$, если $s$ можно разделить на компоненты $x$ и $y$; в общем случае нет требования, чтобы отношение $\ast$ было функцией на множестве состояний. Программы строятся из атомарных программ, а также четырёх специальных программных термов $r_1$, $r_2$ (\defnotion{восстановление}, соответственно, первого и второго $\ast$-компонентов состояния~$x \ast y$), $s_1$ и $s_2$ (\defnotion{сохранение} составного состояния~$x \ast y$ в виде, соответственно, первого и второго $\ast$-компонентов), используя операции $?$~(тест), ${}^\ast$ (итерацию) и $||$~(параллельную композицию). Отметим, что, в частности, язык $\lang{PRSPDL}$ не содержит ни операции выбора программ, ни операции пересечения программ.

Модель Крипке для языка $\lang{PRSPDL}$~--- это набор
$\mmodel{M} = \langle\states{S}, \{\rel{R}_a\}_{a \in \mathit{AP}}, \ast, v\rangle$, где
$\states{S}$, $\rel{R}_a$ и $v$ определяются так же, как в моделях Крипке для языка $\lang{IPDL}$, а $\ast$~--- композиция, т.е. функция $\function{\ast}{\states{S} \times \states{S}}{\Power{\states{S}}}$. Определение истинности для $||$, $r_1$, $r_2$, $s_1$ и $s_2$ следующее:
\[
\begin{array}{llcl}
\arrayitem
  & \otuple{s,t} \in \rel{R}_{\alpha\, ||\, \beta}
  & \bydef
  & \parbox[t]{320pt}{существуют такие  $x_1, y_1, x_2, y_2 \in \states{S}$, что $s \in x_1 \ast x_2$, $t \in y_1 \ast y_2$, $\otuple{x_1, y_1} \in \rel{R}_\alpha$ и $\otuple{x_2, y_2} \in \rel{R}_\beta$;}
  \medskip\\
\arrayitem
  & \otuple{s,t} \in \rel{R}_{r_1}
  & \bydef
  & \parbox[t]{320pt}{существует такое $u \in \states{S}$, что $s \in t \ast u$;}
  \medskip\\
\arrayitem
  & \otuple{s,t} \in \rel{R}_{r_2}
  & \bydef
  & \parbox[t]{320pt}{существует такое $u \in \states{S}$, что $s \in u \ast t$;}
  \medskip\\
\arrayitem
  & \otuple{s,t} \in \rel{R}_{s_1}
  & \bydef
  & \parbox[t]{320pt}{существует такое $u \in \states{S}$, что $t \in s \ast u$;}
  \medskip\\
\arrayitem
  & \otuple{s,t} \in \rel{R}_{s_2}
  & \bydef
  & \parbox[t]{320pt}{существует такое $u \in \states{S}$, что $t \in u \ast s$.}
\end{array}
\]

Модели для языка $\lang{PRSPDL}$, определённые описанным способом, названы в~\cite{TinBal2014} $\ast$-разделёнными\footnote{В оригинальном английском варианте они называются $\ast$-separated.}.
Отметим, что авторы~\cite{TinBal2014} рассматривают различные логики в языке $\lang{PRSPDL}$, отличающиеся условиями, накладываемыми на функцию $\ast$ при определении семантики, и хотя мы рассмотрим только одну из таких логик, результаты, приведённые ниже для неё, можно получить и для остальных.

Понятия выполнимости и общезначимости $\lang{PRSPDL}$-формул определяются так же, как для $\lang{PDL}$-формул и $\lang{IPDL}$-формул. Логика $\logic{PRSPDL}$~--- это множество общезначимых $\lang{PRSPDL}$-формул.

Под константным фрагментом для каждой из введённых логик понимаем множество формул в языке соответствующей логики, не содержащих пропозициональных переменных; при этом ограничений на использование программ, вообще говоря, нет.

Ниже для удобства результат подстановки формулы $\psi$ в формулу $\varphi$ вместо переменной $p$ будем обозначать~$[\psi/p]\varphi$. Кроме того, вместе с модальностями вида $\boxm{\alpha}$ будем использовать двойственные им модальности вида $\diam{\alpha}$, определяя их как сокращения: $\diam{\alpha}\varphi=\neg\boxm{\alpha}\neg\varphi$.

    \subsection{Сложность проблемы выполнимости}

Покажем, что константные фрагменты логик $\logic{PDL}$, $\logic{IPDL}$ и $\logic{PRSPDL}$ имеют такую же вычислительную сложность, как и логики в полных языках. Для этого мы построим полиномиально вычислимые погружения логик $\logic{PDL}$ и $\logic{IPDL}$ в их константные фрагменты, а также алгоритмически вычислимое погружение логики $\logic{PRSPDL}$ в её константный фрагмент.

Мы начнём с логики $\logic{IPDL}$.

Пусть $\varphi$~--- произвольная $\lang{IPDL}$-формула, $p_1,\ldots,p_n$~--- список входящих в неё пропозициональных переменных, $a_1, \ldots, a_l$~--- список входящих в неё атомарных программ. Пусть $\gamma = a_1 \choice \ldots \choice a_l$. Определим рекурсивно перевод $\cdot'$ программ и формул:
$$
\begin{array}{llll}
    {a_j}'                  & = & a_j & \mbox{для $j \in \{1, \ldots, l \}$;} \\
    (\alpha \comp \beta)'   & = & \alpha' \comp \beta'; & \\
    (\alpha \choice \beta)' & = & \alpha' \choice \beta'; & \\
    (\alpha \inter \beta)'  & = & \alpha' \inter \beta'; & \\
    (\alpha^*)'             & = & (\alpha')^*; & \\
    (\phi?)'                & = & (\phi')?; & \\
    {p_i}'                  & = & p_i & \mbox{для $i \in \{1, \ldots, n \}$;} \\
    {\bot}'                 & = & \bot; & \\
    (\phi \con \psi)'       & = & \phi' \con \psi'; & \\
    (\phi \dis \psi)'       & = & \phi' \dis \psi'; & \\
    (\phi \imp \psi)'       & = & \phi' \imp \psi'; & \\
    (\boxm{\alpha} \phi)'   & = & \boxm{\alpha'} (p_{n+1} \imp \phi'). &
\end{array}
$$
Положим
$$
\begin{array}{lcl}
\Theta            & = & p_{n+1} \con \boxm{\gamma^*} (\diam{\gamma} p_{n+1} \imp p_{n+1}); \\
\widehat{\varphi} & = & \Theta \con \varphi'.
\end{array}
$$
Такой подход~--- это уже знакомая нам релятивизация (см.~стр.~\pageref{page:relativization}).

\begin{lemma}
  \label{lem:pdl-varphi-truth}
  Имеет место следующая эквивалентность:
$$
\begin{array}{lcl}
\mbox{$\varphi$ $\logic{IPDL}$-выполнима}
  & \iff
  & \mbox{$\widehat{\varphi}$ $\logic{IPDL}$-выполнима.}
\end{array}
$$
\end{lemma}

\begin{proof}
  Пусть $\widehat{\varphi}$ не является выполнимой.  Тогда $\neg \widehat{\varphi} \in \logic{IPDL}$, а поскольку $\logic{IPDL}$ замкнута по правилу подстановки, получаем, что $[\top/p_{n+1}]\neg \widehat{\varphi} \in \logic{IPDL}$.  Теперь осталось заметить, что   $[\top/p_{n+1}]\widehat{\varphi}  \equivalence \varphi \in \logic{IPDL}$, а значит, $\neg \varphi \in \logic{IPDL}$; таким образом, $\varphi$ тоже не является выполнимой.

  Пусть $\widehat{\varphi}$ является выполнимой. Тогда $\sat{M}{s_0}{\widehat{\varphi}}$ для некоторой модели $\mmodel{M}$ и некоторого мира $s_0$ в~$\mmodel{M}$. Пусть $\mmodel{M}'$~--- наименьшая подмодель модели~$\mmodel{M}$, такая, что
  \begin{itemize}
  \item $s_0$ входит в $\mmodel{M'}$;
  \item если $x$ входит в $\mmodel{M'}$, $x \rel{R}_{\gamma} y$ и $\sat{M}{y}{p_{n+1}}$, то $y$ тоже входит в~$\mmodel{M'}$.
  \end{itemize}
  Заметим, что переменная $p_{n+1}$ истинна в $\mmodel{M}'$. Тогда несложно убедиться, что для каждой подформулы $\psi$ формулы $\varphi$ и каждого состояния $s$ модели $\mmodel{M}'$
$$
\begin{array}{lcl}
   \sat{M}{s}{\psi'} & \iff & \sat{M'}{s}{\psi}.
\end{array}
$$
  Поскольку $\sat{M}{s_0}{\varphi'}$, сразу получаем, что $\sat{M'}{s_0}{\varphi}$; следовательно, формула $\varphi$ тоже выполнима.
\end{proof}


Введём в рассмотрение уже знакомые нам модели. Пусть $b$~--- атомарная программа, не входящая в~$\varphi$. Для каждого $m \in \{1, \ldots, n + 1\}$, определим модель $\mmodel{M}_m = \langle\states{S}_m, \{\rel{R}_a\}_{a \in \mathit{AP}}, v_m\rangle$, положив
  \begin{itemize}
  \item $\states{S}_m = \{r_m, t^m, s_1^m, s_2^m, \ldots, s_{m}^m\}$;
  \item $\rel{R}_{b}$~--- транзитивное замыкание отношения
        $$
        \{\langle r_m, t^m \rangle, \langle t^m, t^m \rangle, \langle r_m, s_1^m \rangle \} \cup \{\langle s_i^m, s_{i+1}^m \rangle : 1 \leqslant i \leqslant m - 1 \};
        $$
  \item $\rel{R}_{a} = \varnothing$, если $a \neq b$;
  \item $v_m (p) = \varnothing$ для каждой переменной $p \in \prop$.
  \end{itemize}

  Модель $\mmodel{M}_m$ изображена на рис.~\ref{fig_Mm}, где стрелки соответствуют отношению $\rel{R}_{b}$, причём те стрелки, которые восстанавливаются по транзитивности, на рисунке отсутствуют; белый кружок соответствует $\rel{R}_{b}$-рефлексивному состоянию, чёрные~--- $\rel{R}_{b}$-иррефлексивным.

\begin{figure}
  \centering
  \begin{tikzpicture}[scale=1.50]

    \coordinate (a30)    at ( 0.0, 3);
    \coordinate (a31)    at ( 0.0, 4);
    \coordinate (a32)    at ( 0.0, 5);
    \coordinate (a34)    at ( 0.0, 6);
    \coordinate (a35)    at ( 0.0, 7);
    \coordinate (b30)    at (-1.0, 4);

    \draw [fill]     (a30)   circle [radius=2.0pt] ;
    \draw [fill]     (a31)   circle [radius=2.0pt] ;
    \draw [fill]     (a32)   circle [radius=2.0pt] ;
    \draw [fill]     (a35)   circle [radius=2.0pt] ;
    \draw []     (b30)   circle [radius=2.0pt] ;

    \begin{scope}[>=latex]

      \draw [->, shorten >=  2.75pt, shorten <= 2.75pt] (a30) -- (a31) ;
      \draw [->, shorten >=  2.75pt, shorten <= 2.75pt] (a31) -- (a32) ;
      \draw [->, shorten >= 11.75pt, shorten <= 2.75pt] (a32) -- (a34) ;
      \draw [->, shorten >=  2.75pt, shorten <= 7.75pt] (a34) -- (a35) ;
      \draw [->, shorten >=  2.75pt, shorten <= 2.75pt] (a30) -- (b30) ;

    \end{scope}

      \node []         at (a34)  {$\vdots$};

      \node [right=2pt] at (a30)  {$r_{m}\models A_m$};
      \node [right=2pt] at (a31)  {$s^m_{1}$};
      \node [right=2pt] at (a32)  {$s^m_{2}$};
      \node [right=2pt] at (a35)  {$s^m_{m}$};

      \node [left=2pt] at (b30)  {$t^m$};

    \end{tikzpicture}
    \caption{Модель $\mmodel{M}_m$}
    \label{fig_Mm}
  \end{figure}

Для каждого $j \in \numN$, определим рекурсивно формулу $\langle b \rangle^j \psi$:
$$
\begin{array}{lcl}
  \langle b \rangle^0 \psi & = & \psi; \\
  \langle b \rangle^{k+1} \psi & = & \diam{b} \diam{b}^k \psi.
\end{array}
$$
Для каждого $m \in \{1, \ldots, n + 1\}$ положим
$$
\begin{array}{lcl}
A_m & = & \langle b \rangle^m \boxm{b} \bot \con \neg
\langle b \rangle^{m+1} \boxm{b} \bot \con \diam{b}
(\diam{b} \top \con \boxm{b} \diam{b} \top).
\end{array}
$$

\begin{lemma}
  \label{lem:alphas-2}
  Пусть $k,m \in \{1, \ldots, n + 1\}$, $x$~--- мир модели $\mmodel{M}_k$. Тогда
$$
\begin{array}{lcl}
  \mmodel{M}_k, x \models A_m & \iff & \mbox{$k = m$ и $x = r_m$.}
\end{array}
$$
\end{lemma}

\begin{proof}
  Получается непосредственной проверкой.
\end{proof}

Пусть теперь
$$
\begin{array}{lcl}
B_m & = & \diam{b} A_m.
\end{array}
$$

Пусть $\sigma$~--- функция (подстановка), заменяющая в $\lang{IPDL}$-формуле $\psi$ каждое вхождение переменной $p_i$ вхождением формулы~$B_i$, где $1 \leqslant i \leqslant n + 1$. Положим
$$
\begin{array}{lcl}
\varphi^* & = & \sigma(\widehat{\varphi}).
\end{array}
$$

\begin{lemma}
  \label{lem:pdl-main_lemma}
  Имеет место следующая эквивалентность:
$$
\begin{array}{lcl}
\mbox{$\varphi$ $\logic{IPDL}$-выполнима}
  & \iff
  & \mbox{$\varphi^*$ $\logic{IPDL}$-выполнима.}
\end{array}
$$
\end{lemma}

\begin{proof}
  Пусть $\varphi$ не является выполнимой. Тогда, по лемме~\ref{lem:pdl-varphi-truth}, $\widehat{\varphi}$ тоже невыполнима, а значит, $\neg \widehat{\varphi} \in \logic{IPDL}$. Поскольку логика $\logic{IPDL}$ замкнута по правилу подстановки, получаем, что $\neg \varphi^* \in \logic{IPDL}$, а следовательно, $\varphi^*$ не является выполнимой.

  Пусть $\varphi$ выполнима. Тогда, с учётом доказательства леммы~\ref{lem:pdl-varphi-truth},   $\sat{M}{s_0}{\widehat{\varphi}}$ для некоторой модели $\mmodel{M}$, в которой истинна переменная $p_{n+1}$, и некоторого состояния $s_0$ модели~$\mmodel{M}$. Определим модель \mmodel{M'} следующим образом. Добавим к модели \mmodel{M} все модели $\mmodel{M}_1,\ldots,\mmodel{M}_{n+1}$; после этого для каждого состояния $x$ модели $\mmodel{M}$ положим $x \rel{R}_{b} r_m$ (где $r_m$~--- корень модели~$\mmodel{M}_m$) в точности тогда, когда $\sat{M}{x}{p_m}$.  Заметим, что состояние $r_{n+1}$ достижимо в \mmodel{M'} из любого состояния $x$ модели~$\mmodel{M}$.

  Для завершения доказательства достаточно показать, что \sat{M'}{s_0}{\varphi^*}. Несложно убедиться, что \sat{M'}{s_0}{\sigma(\Theta)}. Тогда остаётся показать, что   \sat{M'}{s_0}{\sigma(\varphi')}. Для этого достаточно доказать, что для каждой подформулы~$\psi$ формулы~$\varphi$ и каждого состояния $x$ модели $\mmodel{M}$
$$
\begin{array}{lcl}
  \sat{M}{x}{\psi'} & \iff & \sat{M'}{x}{\sigma (\psi')}.
\end{array}
$$
  Последняя эквивалентность может быть доказана индукцией по построению формулы~$\psi$; мы рассмотрим только базис индукции, поскольку обоснование индукционного шага не содержит каких либо трудностей.

  Пусть \sat{M'}{x}{B_i}. Тогда $x \rel{R}'_{b} y$ и \sat{M'}{y}{A_i} для некоторого состояния $y$ модели $\mmodel{M}'$. Это возможно только если $y$ не входит в~\mmodel{M}. Действительно, если это не так, то \sat{M'}{y}{p_{n+1}}, поэтому $y \rel{R}'_{b} r_{n+1}$; следовательно, \sat{M'}{y}{\langle b \rangle^{i+1} \boxm{b} \bot}, а значит, \notsat{M'}{y}{A_i}, что даёт противоречие. Итак, $y$ входит в модель $\mmodel{M}_m$ для некоторого $m \in \{1, \ldots, n + 1\}$. Тогда, по лемме~\ref{lem:alphas-2}, $y = r_i$; следовательно, по определению модели $\mmodel{M}'$, получаем, что \sat{M}{x}{p_i}.

  Справедливость обратной импликации мгновенно следует из определения модели~\mmodel{M'}.
\end{proof}

Как следствие получаем следующую теорему.

\begin{theorem}
  \label{thr:ipdl}
  Логика\/ $\logic{IPDL}$ полиномиально погружается в свой константный фрагмент.
\end{theorem}

Извлечём следствие из теоремы~\ref{thr:ipdl}, касающееся сложности фрагментов логики~$\logic{IPDL}$.  Известно~\cite{LL05}, что фрагмент логики~$\logic{IPDL}$ в языке с одной атомарной программой является $\cclass{2EXPTIME}$-полным. Это даёт нам следующую теорему.

\begin{theorem}
Проблема выполнимости константного фрагмента логики\/ $\logic{IPDL}$, содержащего одну атомарную программу, является $\cclass{2EXPTIME}$-полной.
\end{theorem}

Теперь покажем, как описанное моделирование для логики $\logic{IPDL}$ может быть использовано для получения сходных результатов для логик $\logic{PDL}$ и $\logic{PRSPDL}$.

Несложно убедиться, что конструкция, описанная выше для $\logic{IPDL}$, работает также и для $\logic{PDL}$: достаточно убрать из неё всё, что касается операции пересечения программ.
В результате получаем следующее.

\begin{theorem}
  Логика\/ $\logic{PDL}$ полиномиально погружается в свой константный фрагмент.
\end{theorem}

Поскольку проблема выполнимости для логики $\logic{PDL}$ в языке с одной атомарной программой является $\cclass{EXPTIME}$-полной~\cite{FL79}, получаем следующую теорему.

\begin{theorem}
  \label{cor:pdl}
Проблема выполнимости константного фрагмента логики\/ $\logic{PDL}$, содержащего одну атомарную программу, является $\cclass{EXPTIME}$-полной.
\end{theorem}

Отметим, что этот результат следует также из теоремы~\ref{maintheorem:K-ast} с учётом того, что модальности $\Box$ и $\Box^\ast$ языка $\lang{ML}^\ast$ могут быть промоделированы в языке $\lang{PDL}$, соответственно, модальностями $\boxm{a}$ и $\boxm{a^\ast}$, где $a$~--- некоторая атомарная программа.

Теперь покажем, какие модификации можно выполнить, чтобы получить аналогичные результаты для логики~$\logic{PRSPDL}$.
Прежде всего построим аналог формулы~$\widehat{\varphi}$. Для этого переопределим перевод~$\cdot'$ следующим образом:
$$
\begin{array}{lcll}
    {a_i}'                 & = & a_i & \mbox{для $i \in \{1, \ldots, l \}$;} \\
    {r_i}'                 & = & r_i & \mbox{для $i \in \{1, 2 \}$;} \\
    {s_i}'                 & = & s_i & \mbox{для $i \in \{1, 2 \}$;} \\
    (\alpha\, ||\, \beta)' & = & \alpha'\, ||\, \beta';  \\
    (\alpha^*)'            & = & (\alpha')^*;  \\
    (\phi?)'               & = & (\phi')?;  \\
    {p_i}'                 & = & p_i & \mbox{для $i \in \{1, \ldots, n \}$;} \\
    {\bot}'                & = & \bot;  \\
    (\phi \con \psi)'      & = & \phi' \con \psi';  \\
    (\phi \dis \psi)'      & = & \phi' \dis \psi';  \\
    (\phi \imp \psi)'      & = & \phi' \imp \psi';  \\
    (\boxm{\alpha} \phi)'  & = & \boxm{\alpha'} (p_{n+1} \imp \phi').
\end{array}
$$
Теперь мы можем переопределить формулу~$\Theta$. Поскольку язык логики $\logic{PRSPDL}$ не содержит операции выбора программы, мы поступим следующим образом. Пусть имеется список всех списков программ, соответствующих максимальным спискам вложенных модальностей в~$\varphi$:
$$
\langle\alpha^1_1, \dots, \alpha^1_{n_1}\rangle,
  \ldots,
\langle\alpha^k_1, \dots, \alpha^k_{n_k}\rangle.
$$
Тогда положим
$$
\begin{array}{lcl}
\Theta
  & =
  & \displaystyle
  p_{n+1} \con \bigwedge_{i=1}^k \bigwedge_{j=1}^{n_k - 1 } \boxm{\alpha^i_1} \dots \boxm{\alpha^i_j} (\diam{\alpha^i_{j+1}} p_{n+1} \imp p_{n+1})
\end{array}
$$
и переопределим формулу $\widehat{\varphi}$:
$$
\begin{array}{lcl}
\widehat{\varphi} & = & \Theta \con \varphi'.
\end{array}
$$
Теперь можно повторить аргументацию, приведённую для случая $\logic{IPDL}$; в итоге получим следующие теоремы.

\begin{theorem}
  Логика\/ $\logic{PRSPDL}$ рекурсивно погружается в свой константный фрагмент.
\end{theorem}

\begin{theorem}
  Константный фрагмент логики\/ $\logic{PRSPDL}$ алгоритмически неразрешим.
\end{theorem}

Напомним, что при определении логики $\logic{PRSPDL}$ мы использовали $\ast$\nobreakdash-раз\-де\-лён\-ные модели; аналогичные результаты для других логик в языке $\lang{PRSPDL}$, рассмотренных в~\cite{TinBal2014}, могут быть получены с использованием той же аргументации. Уточним, что неразрешимость здесь означает $\Pi^1_1$-полноту.


  \section{Темпоральные логики}
    \subsection{Предварительные сведения}

\providecommand{\CTL}    [0]{\mbox{$\logic{CTL}$}}
\providecommand{\ATL}    [0]{\mbox{$\logic{ATL}$}}
\providecommand{\CTLstar}[0]{\mbox{$\logic{CTL}^\ast$}}
\providecommand{\ATLstar}[0]{\mbox{$\logic{ATL}^\ast$}}
\providecommand{\LTL}    [0]{\mbox{$\logic{LTL}$}}

Пропозициональные логики ветвящегося времени\footnote{Английский вариант названия~--- computational tree logics; такие логики называют также временн\'{ы}ми, или темпоральными.} $\CTL$~\cite{CE81,DGL16} и $\CTLstar$~\cite{EH86,DGL16}, а также логика линейного времени $\LTL$~\cite{Pnueli77} долгое время использовались для формальной спецификации и верификации (параллельных) не завершающихся компьютерных программ~\cite{HR04,DGL16}, таких как компоненты операционных систем, а также в формальной спецификации и проверке аппаратного обеспечения. Недавно возникшие временные логики $\ATL$ и~$\ATLstar$~\cite{AHK02,DGL16} стали использоваться для спецификации и верификации мультиагентных систем~\cite{SLB08} и, в более широком смысле, так называемых открытых систем, т.е. систем, корректность которых зависит от действий внешних объектов и субъектов, таких как окружающая среда или другие агенты.

Логики $\LTL$, $\CTL$, $\CTLstar$, $\ATL$ и $\ATLstar$ находят приложения в проектировании компьютерных систем. Так, задача проверки соответствия системы некоторой спецификации может быть выполнена за счёт того, что формула, выражающая эту спецификацию, истинна в структуре, моделирующей систему (для проверки программы такая структура обычно моделирует пути выполнения программы). Эта задача соответствует задаче, известной как проблема \defnotion{model checking}~\cite{CGP00} для логики. Ещё одна задача~--- задача реализуемости спецификации в виде некоторой системы; эта задача соответствует проблеме выполнимости формул для логики. Возможность проверить реализуемость спецификации является довольно значимой, поскольку позволяет избежать напрасных усилий при попытке внедрить неудовлетворительные системы. Более того, как правило, алгоритм, проверяющий выполнимость формулы, выражающей спецификацию, строит, явно или неявно, модель для этой формулы, предоставляя тем самым формальную модель системы, соответствующей этой спецификации; такая модель впоследствии может быть использована на этапе реализации. Есть надежда, что в один прекрасный день такие модели можно будет использовать как часть процедуры <<нажатия одной кнопки>>, обеспечивающей гарантированно правильную реализацию на основе модели спецификации, полностью избегая необходимости последующей проверки. Так, алгоритмы проверки выполнимости формул, разработанные для рассматриваемых темпоральных логик~\cite{EH85,Reynolds09,GorSh09,David15}, основаны на построении семантических таблиц, и суть их как раз в том, что они неявно пытаются строить модель для формулы, выполнимость которой проверяется.

Хорошо известно, что для формул, которые могут содержать произвольное количество пропозициональных переменных, сложность проблемы выполнимости для всех этих логик довольно высока: проблема выполнимости $\cclass{PSPACE}$-полна для $\LTL$~\cite{SislaClarke85}, $\cclass{EXPTIME}$-полна для $\CTL$~\cite{FL79,EH85}, $\cclass{2EXPTIME}$-полна для $\CTLstar$~\cite{VardiStockmeyer85}, $\cclass{EXPTIME}$-полна для $\ATL$~\cite{GD06,WLWW06} и $\cclass{2EXPTIME}$-полна для $\ATLstar$~\cite{Schewe08}.

Однако было замечено (см., например, \cite{DS02}), что на практике формулы, выражающие формальные спецификации, обычно содержат лишь очень небольшое количество пропозициональных переменных (как правило, две или три). И~это несмотря на то, что сами формулы могут быть довольно длинными и даже содержать длинные цепочки вложенных временных операторов. Таким образом, закономерно возникает вопрос о том, может ли ограничение на число используемых пропозициональных переменных существенно понизить сложность проблемы выполнимости для $\LTL$, $\CTL$, $\CTLstar$, $\ATL$ и~$\ATLstar$.

Ответ на этот вопрос состоит в том, что сложность каждой из этих логик совпадает со сложностью её фрагмента от одной переменной, и ниже мы покажем, как это доказать, используя описанную выше технику. Единственным исключением будет логика $\LTL$, где требуется другой метод, основанный не на идеях, изложенных в работе Дж.\,Халперна~\cite{Halpern95}, которые мы в основном и развивали выше для модальных логик, а, скорее, на идеях, изложенных в работе П.\,Блэкбёрна и Э.\,Спаан~\cite{BS93}; сам метод, описанный в~\cite{BS93}, весьма эффективен и мы к нему ещё обратимся, а пока более подробное обсуждение сложности фрагментов логики $\LTL$ в языках с конечным числом переменных отложим до раздела~\ref{ssec:LTL}.

    \subsection{Сложность фрагментов логик $\logic{CTL}^\ast$ и $\logic{CTL}$}
    \label{s4-4-2}

\providecommand{\AX}[0]{\mbox{$\dismath\forall\Next$}}
\providecommand{\EX}[0]{\mbox{$\dismath\exists\Next$}}
\providecommand{\AU}[0]{\mbox{$\dismath\forall\Until$}}
\providecommand{\EU}[0]{\mbox{$\dismath\exists\Until$}}
\providecommand{\AG}[0]{\mbox{$\dismath\forall\Box$}}
\providecommand{\EF}[0]{\mbox{$\dismath\exists\Diamond$}}

\subsubsection{Синтаксис и семантика}

Язык $\lang{CTL}^\ast$ логики $\logic{CTL}^\ast$ является обогащением языка $\lang{L}$ классической пропозициональной логики \defnotion{квантором пути}\index{квантор!пути} $\forall$, а также \defnotion{темпоральными связками}\index{связка!темпоральная} $\Next$ и $\Until$, которые мы будем называть <<next>> и <<until>>, не предлагая перевода их названий на русский язык. Имеется два вида формул в языке $\lang{CTL}^\ast$: \defnotion{формулы состояния}\index{уяа@формула!состояния} и \defnotion{формулы пути},\index{уяа@формула!пути} которые называются так, поскольку их истинность оценивается в \defnotion{состояниях}\index{состояние} и \defnotion{путях}\index{путь} моделей Крипке для языка~$\lang{CTL}^\ast$.
Формулы состояния и формулы пути можно определить одновременной рекурсией с помощью BNF-выражений (ниже $\varphi$~--- формула состояния, $\vartheta$~--- формула пути):
\[
\begin{array}{llcl}
\arrayitem &
\varphi & := & p  \mid \bot \mid  (\varphi \con \varphi)\mid  (\varphi \dis \varphi)\mid  (\varphi \imp \varphi) \mid \forall \vartheta; \\
\arrayitem &
\vartheta & := & \varphi \mid (\vartheta \con \vartheta)\mid (\vartheta \dis \vartheta)\mid (\vartheta \imp \vartheta) \mid (\vartheta \Until \vartheta)
\mid \Next \vartheta,
\end{array}
\]
где $p\in\prop$. Помимо обычных сокращений для других булевых связок, мы также определим следующие:
$\Diamond \vartheta = (\top \Until \vartheta)$, $\Box \vartheta = \neg \Diamond \neg \vartheta$ и $\exists \vartheta = \neg \forall \neg \vartheta$.

Для оценки истинности $\lang{CTL}^\ast$-формул мы будем пользоваться шкалами и моделями Крипке. Шкалы Крипке определяются так же, как и в мономодальном случае, с той разницей, что требуется серийность отношения достижимости в них, т.е. для шкалы $\kframe{F}=\otuple{\states{S},R}$ дополнительно требуется, чтобы отношение $R$ удовлетворяло условию $\forall x\exists y\, xRy$; при этом $\states{S}$ называем множеством \defnotion{состояний}\index{состояние} шкалы~$\kframe{F}$.

Бесконечная последовательность $s_0, s_1, s_2, \ldots$ состояний шкалы $\kframe{F}=\otuple{S,R}$, такая, что $s_i R s_{i+1}$ для каждого $i \in\numN$, называется \defnotion{путём}\index{путь} в~$\kframe{F}$. Пусть $\pi$~--- путь в~$\kframe{F}$ и $i \in\numN$; $i$\nobreakdash-й элемент пути~$\pi$ обозначаем~$\pi[i]$, а часть пути~$\pi$, начинающуюся с элемента~$\pi[i]$, обозначаем $\pi[i,\infty]$. Элемент $\pi[0]$ называем \defnotion{началом}\index{начало пути} пути~$\pi$. Для состояния $s \in \states{S}$ множество всех путей, начинающихся в $s$ (т.е. началом которых является~$s$), обозначаем~$\Pi(s)$; поскольку отношение достижимости в шкале является серийным, множество~$\Pi(s)$ непусто.

Модель Крипке для языка~$\lang{CTL}^\ast$~--- это набор $\mmodel{M} = \langle\kframe{F}, v\rangle$, где $\kframe{F}=\otuple{S,R}$~--- шкала Крипке (с~серийным отношением достижимости), а $v$~--- оценка переменных в состояниях шкалы $\kframe{F}$, т.е. функция $\function{v}{\prop}{\Power{\states{S}}}$. Поскольку нам в основном будут нужны модели Крипке, мы будем писать $\mmodel{M} = \langle\states{S}, R, v\rangle$ вместо $\mmodel{M} = \langle\kframe{F}, v\rangle$.

Далее под путями в модели $\mmodel{M} = \langle\states{S}, R, v\rangle$ понимаем пути в шкале~$\otuple{S,R}$.

Отношение истинности формул состояния в состояниях модели $\mmodel{M}$ и формул пути в путях модели~$\mmodel{M}$ определим рекурсивно (ниже считаем, что $\varphi_1$ и $\varphi_2$~--- формулы состояния, $\vartheta_1$ и $\vartheta_2$~--- формулы пути, $s$~--- состояние модели~$\mmodel{M}$, а $\pi$~--- путь в модели~$\mmodel{M}$):
\[
\begin{array}{llcl}
\arrayitem
  & \sat{M}{s}{p_i}
  & \bydef
  & s \in v(p_i);
  \medskip\\
\arrayitem
  & \notsat{M}{s}{\bot};
  \medskip\\
\arrayitem
  & \sat{M}{s}{\varphi_1 \con \varphi_2}
  & \bydef
  & \mbox{$\sat{M}{s}{\varphi_1}$ \phantom{л}и\phantom{и} $\sat{M}{s}{\varphi_2}$;}
  \medskip\\
\arrayitem
  & \sat{M}{s}{\varphi_1 \dis \varphi_2}
  & \bydef
  & \mbox{$\sat{M}{s}{\varphi_1}$ или $\sat{M}{s}{\varphi_2}$;}
  \medskip\\
\arrayitem
  & \sat{M}{s}{\varphi_1 \imp \varphi_2}
  & \bydef
  & \mbox{$\notsat{M}{s}{\varphi_1}$ или $\sat{M}{s}{\varphi_2}$;}
  \medskip\\
\arrayitem
  & \sat{M}{s}{\forall \vartheta_1}
  & \bydef
  & \mbox{$\sat{M}{\pi}{\vartheta_1}$ для каждого пути $\pi \in \Pi(s)$.}
  \medskip\\
\arrayitem
  & \sat{M}{\pi}{\varphi_1}
  & \bydef
  & \sat{M}{\pi[0]}{\varphi_1};
  \medskip\\
\arrayitem
  & \sat{M}{\pi}{\vartheta_1 \con \vartheta_2}
  & \bydef
  & \mbox{$\sat{M}{\pi}{\vartheta_1}$ \phantom{л}и\phantom{и} $\sat{M}{\pi}{\vartheta_2}$;}
  \medskip\\
\arrayitem
  & \sat{M}{\pi}{\vartheta_1 \dis \vartheta_2}
  & \bydef
  & \mbox{$\sat{M}{\pi}{\vartheta_1}$ или $\sat{M}{\pi}{\vartheta_2}$;}
  \medskip\\
\arrayitem
  & \sat{M}{\pi}{\vartheta_1 \imp \vartheta_2}
  & \bydef
  & \mbox{$\notsat{M}{\pi}{\vartheta_1}$ или $\sat{M}{\pi}{\vartheta_2}$;}
  \medskip\\
\arrayitem
  & \sat{M}{\pi}{\Next\vartheta_1}
  & \bydef
  & \sat{M}{\pi[1, \infty]}{\vartheta_1};
  \medskip\\
\arrayitem
  & \sat{M}{\pi}{\vartheta_1 \Until \vartheta_2}
  & \bydef
  & \parbox[t]{280pt}{$\sat{M}{\pi[i,\infty]}{\vartheta_2}$ для некоторого $i \in\numN$
    и $\sat{M}{\pi[j,\infty]}{\vartheta_1}$ для каждого $j$, такого, что $0 \leqslant j < i$.}
\end{array}
\]
Формально под $\lang{CTL}^\ast$-формулами понимаем формулы состояния в языке~$\lang{CTL}^\ast$; $\lang{CTL}^\ast$-формулу называем \defnotion{выполнимой}, если она истинна в некотором состоянии некоторой модели; $\lang{CTL}^\ast$-формулу называем \defnotion{общезначимой}, если она истинна в каждом состоянии каждой модели.  Под логикой $\CTLstar$ понимаем множество всех общезначимых $\lang{CTL}^\ast$-формул.

Язык $\lang{CTL}$ логики $\CTL$ можно считать фрагментом языка $\lang{CTL}^\ast$, состоящим из формул, в которых перед каждой темпоральной связкой стоит квантор пути. Благодаря этому ограничению, в частности, формулы пути становятся невозможными, а возникающие композиции кванторов пути и темпоральных связок можно понимать как особые модальности. Такими модальностями являются $\AX$, $\EX$, $\AU$, $\EU$, $\exists \Diamond$ и~$\forall \Box$, позволяя строить формулы $\AX\varphi$, $\EX\varphi$, $\varphi\AU\psi$, $\varphi\EU\psi$, $\exists \Diamond \varphi$ и $\forall \Box \varphi$, где $\varphi\AU\psi=\forall(\varphi\Until\psi)$ и $\varphi\EU\psi=\exists(\varphi\Until\psi)$. В~формальных построениях мы будем считать, что $\lang{CTL}$ содержит модальности~$\AX$, $\EX$, $\AU$, $\EU$, а модальности $\exists \Diamond$ и~$\forall \Box$ являются сокращениями.

Истинность $\lang{CTL}$-формул в состояниях модели определяется как истинность этих формул, понимаемых как формулы языка $\lang{CTL}^\ast$; понятия выполнимости и общезначимости определяются так~же. Под логикой $\CTL$ понимаем множество общезначимых $\lang{CTL}$-формул.

Как и раньше, будем использовать обозначение $[\psi/p]\varphi$ для формулы, получающейся подстановкой формулы $\psi$ в формулу $\varphi$ вместо переменной~$p$.

\subsubsection{Константные фрагменты логик $\CTLstar$ и $\CTL$}
\label{sec:ctl-variable-free-fragment}

Прежде всего заметим, что проблема выполнимости константных формул как для $\logic{CTL}$, так и для $\logic{CTL}^\ast$ решается полиномиально по времени. Действительно, несложно убедиться, что любая константная формула в языке каждой из этих формул эквивалентна либо формуле~$\bot$, либо формуле~$\top$. Для этого достаточно заметить, что логике~$\CTLstar$ принадлежат следующие эквивалентности формул:
$$
\begin{array}{lclclcl}
  (\top \con \top)   & \lra & \top; & ~~~~ & (\top \con \bot)   & \lra & \bot; \\
  (\bot \con \top)   & \lra & \bot; & ~~~~ & (\bot \con \bot)   & \lra & \bot; \medskip\\
  (\top \dis \top)   & \lra & \top; & ~~~~ & (\top \dis \bot)   & \lra & \top; \\
  (\bot \dis \top)   & \lra & \top; & ~~~~ & (\bot \dis \bot)   & \lra & \bot; \medskip\\
  (\top \imp \top)   & \lra & \top; & ~~~~ & (\top \imp \bot)   & \lra & \bot; \\
  (\bot \imp \top)   & \lra & \top; & ~~~~ & (\bot \imp \bot)   & \lra & \top; \medskip\\
  \forall \top       & \lra & \top; & ~~~~ & \forall \bot       & \lra & \bot; \\
  \Next \top         & \lra & \top; & ~~~~ & \Next \bot         & \lra & \bot; \\
  (\top \Until \top) & \lra & \top; & ~~~~ & (\top \Until \bot) & \lra & \bot; \\
  (\bot \Until \top) & \lra & \top; & ~~~~ & (\bot \Until \bot) & \lra & \bot.
\end{array}
$$

Таким образом, чтобы проверить, выполнима ли константная формула~$\varphi$, достаточно рекурсивно заменить каждую подформулу формулы~$\varphi$ либо на~$\bot$, либо на~$\top$, что даёт нам алгоритм, квадратичный по времени от длины тестируемой формулы~$\varphi$. Поскольку логика $\logic{CTL}$ является $\cclass{EXPTIME}$-полной, логика $\CTLstar$~--- $\cclass{2EXPTIME}$-полной, а $\cclass{P}\ne\cclass{EXPTIME}\subseteq\cclass{2EXPTIME}$, получаем, что константные фрагменты этих логик существенно проще каждой из них в полном языке.

\subsubsection{Фрагменты логик $\CTLstar$ и $\CTL$ от одной переменной}
\label{sec:ctl-single-variable-fragment}

Покажем, что ситуация существенно меняется, если в языке имеется хотя бы одна пропозициональная переменная.

Пусть $\varphi$~--- произвольная $\lang{CTL}^\ast$-формула, $p_1, \ldots, p_n$~--- входящие в неё переменные, $p_{n+1}$~--- новая переменная. Определим рекурсивно перевод~$\cdot'$ следующим образом:
$$
\begin{array}{lcll}
  {p_i}'                 & = & p_i & \mbox{для $i \in \{1, \ldots, n \}$;} \\
  \bot'                  & = & \bot; & \\
  (\phi \con \psi)'      & = & \phi' \con \psi'; & \\
  (\phi \dis \psi)'      & = & \phi' \dis \psi'; & \\
  (\phi \imp \psi)'      & = & \phi' \imp \psi'; & \\
  (\forall \alpha)'      & = & \forall (\Box p_{n+1} \imp \alpha'); &\\
  (\Next \alpha)'        & = & \Next \alpha'; & \\
  (\alpha \Until \beta)' & = & \alpha' \Until \beta'.  &
\end{array}
$$

Теперь положим
$$
\begin{array}{lcl}
\Theta            & = & p_{n+1} \con \AG (\EX p_{n+1} \lra p_{n+1});~~ \\
\widehat{\varphi} & = & \Theta \con \varphi'.
\end{array}
$$

Мы снова используем релятивизацию (см.~стр.~\pageref{page:relativization}), но учитываем в $\Theta$ требование серийности к отношению достижимости. Заметим, что формула~$\varphi$ эквивалентна в логике~$\logic{CTL}^\ast$ формуле~$[\top/p_{n+1}]\widehat{\varphi}$.

\begin{lemma}
  \label{lem:atl-varphi-truth}
  Имеет место следующая эквивалентность:
$$
\begin{array}{lcl}
\mbox{$\varphi$ $\CTLstar$-выполнима}
  & \iff
  & \mbox{$\widehat{\varphi}$ $\CTLstar$-выполнима.}
\end{array}
$$
\end{lemma}

\begin{proof}
  Пусть формула $\widehat{\varphi}$ невыполнима.  Тогда $\neg \widehat{\varphi} \in \CTLstar$, и поскольку логика $\CTLstar$ замкнута по правилу подстановки, получаем, что $[\top/p_{n+1}] \neg \widehat{\varphi} \in \CTLstar$.  Учитывая, что формула $[\top/p_{n+1}] \neg \widehat{\varphi}$ эквивалентна в $\CTLstar$ формуле $\neg\varphi$, заключаем, что $\neg \varphi \in \CTLstar$, а значит, $\varphi$ тоже невыполнима.

  Пусть формула $\widehat{\varphi}$ выполнима. Тогда $\sat{M}{s_0}{\widehat{\varphi}}$ для некоторой модели~$\mmodel{M}$ и некоторого состояния~$s_0$ в~$\mmodel{M}$. Пусть $\mmodel{M}'$~--- наименьшая подмодель модели~$\mmodel{M}$, такая, что
  \begin{itemize}
  \item $s_0$ находится в $\mmodel{M'}$;
  \item если $x$ входит в $\mmodel{M'}$, $x R y$ и $\sat{M}{y}{p_{n+1}}$, то $y$ тоже входит в~$\mmodel{M'}$.
  \end{itemize}
  Из того, что $\sat{M}{s_0}{p_{n+1} \con \AG (\EX p_{n+1} \lra p_{n+1})}$, получаем, что отношение достижимости в модели $\mmodel{M'}$ является серийным, а также что переменная $p_{n+1}$ истинна в каждом состоянии модели~$\mmodel{M}'$.

  Покажем, что $\sat{M'}{s_0}{\varphi}$. Для этого достаточно показать, что для каждого состояния   $x$ в модели~$\mmodel{M'}$ и каждой подформулы состояния $\psi$ формулы $\varphi$
  $$
  \begin{array}{lcl}
  \sat{M}{x}{\psi'} & \iff & \sat{M'}{x}{\psi},
  \end{array}
  $$
  а также что для каждого пути~$\pi$ в модели~$\mmodel{M'}$ и каждой подформулы пути~$\alpha$ формулы~$\varphi$
  $$
  \begin{array}{lcl}
  \sat{M}{\pi}{\alpha'} & \iff & \sat{M'}{\pi}{\alpha}.
  \end{array}
  $$

  Обе эти эквивалентности обосновываются одновременной индукцией по построению подформул $\psi$ и $\alpha$ формулы~$\varphi$.
  Мы разберём только случай, когда $\psi = \forall \alpha$; в остальных случаях обоснование не вызывает затруднений.

  Пусть $\psi = \forall \alpha$. Тогда $\psi' = \forall (\Box p_{n+1} \imp \alpha')$.

  Пусть $\notsat{M}{x}{\forall (\Box p_{n+1} \imp \alpha')}$. Тогда $\notsat{M}{\pi}{\alpha'}$ для некоторого пути $\pi \in \Pi(x)$, такого, что $\sat{M}{\pi[i]}{p_{n+1}}$ для каждого $i \in \numN$.  Согласно построению модели~$\mmodel{M'}$, путь $\pi$ является также и путём в $\mmodel{M'}$; применяя индукционное предположение, получаем, что $\notsat{M'}{\pi}{\alpha}$.  Следовательно, $\notsat{M'}{x}{\forall \alpha}$, что и требовалось доказать.

  Пусть теперь $\notsat{M'}{x}{\forall \alpha}$. Тогда $\notsat{M'}{\pi}{\alpha}$ для некоторого пути $\pi \in \Pi(x)$. Ясно, что $\pi$ является и путём в~$\mmodel{M}$. Поскольку переменная~$p_{n+1}$ истинна в каждом состоянии модели~$\mmodel{M'}$, а значит, и в каждом состоянии пути~$\pi$, используя индукционное предположение, получаем, что $\notsat{M}{x}{\forall (\Box p_{n+1} \imp \alpha')}$.
\end{proof}



Для каждого $m \in \{1, \ldots, n+1\}$ определим модель $\mmodel{M}_m = (\states{S}_m, R, v_m)$ следующим образом:
\[
\begin{array}{llcl}
  \arrayitem
    & \states{S}_m
    & =
    & \{r_m, b^m, a_1^m, a_2^m, \ldots, a_{2m}^m\};
    \medskip\\
  \arrayitem
    & R
    & =
    & \{ \langle r_m, b^m \rangle, \langle r_m, a_1^m \rangle \} \cup {}
    \\
  \phantom{\arrayitem}
    &
    &
    & {} \cup \{ \langle a_i^m, a_{i+1}^m \rangle : 1 \leqslant i \leqslant 2m - 1 \}
      \cup
      \{ \langle s, s \rangle : s \in \states{S}_m \};
    \medskip\\
  \arrayitem
    & s \in v_m (p)
    & \bydef
    & \mbox{$s = r_m$ или $s = a_{2k}^m$ для некоторого $k\in\{1,\ldots,m\}$.}
\end{array}
\]

\begin{figure}
  \centering
  \begin{tikzpicture}[scale=1.50]

    \coordinate (a30)    at ( 0.0, 3);
    \coordinate (a31)    at ( 0.0, 4);
    \coordinate (a32)    at ( 0.0, 5);
    \coordinate (a34)    at ( 0.0, 6);
    \coordinate (a35)    at ( 0.0, 7);
    \coordinate (a36)    at ( 0.0, 8);
    \coordinate (b30)    at (-1.0, 4);

    \draw []     (a30)   circle [radius=2.0pt] ;
    \draw []     (a31)   circle [radius=2.0pt] ;
    \draw []     (a32)   circle [radius=2.0pt] ;
    \draw []     (a34)   circle [radius=2.0pt] ;
    \draw []     (a36)   circle [radius=2.0pt] ;
    \draw []     (b30)   circle [radius=2.0pt] ;

    \begin{scope}[>=latex]

      \draw [->, shorten >=  2.75pt, shorten <= 2.75pt] (a30) -- (a31) ;
      \draw [->, shorten >=  2.75pt, shorten <= 2.75pt] (a31) -- (a32) ;
      \draw [->, shorten >=  2.75pt, shorten <= 2.75pt] (a32) -- (a34) ;
      \draw [->, shorten >= 11.75pt, shorten <= 2.75pt] (a34) -- (a35) ;
      \draw [->, shorten >=  2.75pt, shorten <= 7.75pt] (a35) -- (a36) ;
      \draw [->, shorten >=  2.75pt, shorten <= 2.75pt] (a30) -- (b30) ;

    \end{scope}

      \node []         at (a35)  {$\vdots$};

      \node [right=2pt] at (a30)  {$r_{m}\models A_m$};
      \node [right=2pt] at (a31)  {$a^m_{1}$};
      \node [right=2pt] at (a32)  {$a^m_{2}$};
      \node [right=2pt] at (a34)  {$a^m_{3}$};
      \node [right=2pt] at (a36)  {$a^m_{2m}$};

      \node [above=2pt] at (b30)  {$~~b^m$};

      \node [left=2pt] at (a30)  {$\phantom{^{m}}p$};
      \node [left=2pt] at (a31)  {$\phantom{^{m}}\neg p$};
      \node [left=2pt] at (a32)  {$\phantom{^{m}}p$};
      \node [left=2pt] at (a34)  {$\phantom{^{m}}\neg p$};
      \node [left=2pt] at (a36)  {$\phantom{^{m}}p$};

      \node [left=2pt] at (b30)  {$\phantom{^{m}}\neg p$};

    \end{tikzpicture}
    \caption{Модель $\mmodel{M}_m$}
    \label{fig_Mm_ctl_star}
  \end{figure}

  Модель $\mmodel{M}_m$ изображена на рис.~\ref{fig_Mm_ctl_star}, где миры представлены белыми кружк\'{а}ми. Определим рекурсивно следующие формулы:
  $$
  \begin{array}{lcll}
    \chi_0     & = & \forall \Box p; \\
    \chi_{k+1} & = & p \con \EX ( \neg p \con \EX \chi_k) & \mbox{для $k\in\numN$}.
  \end{array}
  $$
Для каждого $m \in \{1, \ldots, n + 1\}$ положим
$$
\begin{array}{lcl}
A_m & = & \chi_m \con \EX \AG \neg p.
\end{array}
$$

\begin{lemma}
  \label{lem:roots}
  Пусть $m,k \in \{1, \ldots, n + 1\}$ и пусть $x$~--- состояние модели~$\mmodel{M}_k$. Тогда
  $$
  \begin{array}{lcl}
  \mmodel{M}_k, x \models A_m & \iff & \mbox{$k = m$ и $x = r_m$.}
  \end{array}
  $$
\end{lemma}

\begin{proof}
  Состоит в непосредственной проверке.
\end{proof}

Для каждого $m \in \{1, \ldots, n+1\}$ положим
$$
  \begin{array}{lcl}
  B_m & = & \EX A_m.
  \end{array}
$$
Путь $\sigma$~--- подстановка формул $B_1,\ldots,B_{n+1}$ вместо переменных $p_1,\ldots,p_{n+1}$ соответственно. Положим
$$
\begin{array}{lcl}
\varphi^\ast & = & \sigma (\widehat{\varphi}).
\end{array}
$$

\begin{lemma}
  \label{lem:main_lemma_ctl_star}
  Имеет место следующая эквивалентность:
$$
\begin{array}{lcl}
\mbox{$\varphi$ $\logic{CTL}^\ast$-выполнима}
  & \iff
  & \mbox{$\varphi^\ast$ $\logic{CTL}^\ast$-выполнима.}
\end{array}
$$
\end{lemma}

\begin{proof}
  Если $\varphi$ невыполнима, то, по лемме~\ref{lem:atl-varphi-truth}, $\widehat{\varphi}$ невыполнима; тогда $\neg \widehat{\varphi} \in \CTLstar$, и в силу замкнутости логики $\CTLstar$ по правилу подстановки получаем, что $\neg \varphi^\ast \in \CTLstar$, а значит, $\varphi^\ast$ тоже невыполнима.

  Пусть $\varphi$ выполнима.  Тогда, с учётом леммы~\ref{lem:atl-varphi-truth} и её доказательства, $\widehat{\varphi}$ истинна в некотором состоянии $s$ некоторой модели $\mmodel{M} = \langle\states{S}, R, v\rangle$, такой, что $v(p_{n+1})=\states{S}$. Без ограничений общности можем (и~будем) считать, что модель $\mmodel{M}$ порождается миром~$s$. Определим модель $\mmodel{M'} = \langle\states{S}', R', v'\rangle$ следующим образом. Добавим к $\mmodel{M}$ все модели $\mmodel{M}_1,\ldots,\mmodel{M}_{n+1}$ (т.е. возьмём их раздельное объединение) и для каждого $x \in \states{S}$ сделаем $r_m$ достижимым из $x$ в $\mmodel{M'}$ в точности тогда, когда $\sat{M}{x}{p_m}$. Оценку переменной $p$ определим так: для состояний модели $\mmodel{M}_m$ оценка такая же, как в $\mmodel{M}_m$, а для каждого $x \in \states{S}$ положим $x \notin v'(p)$.

  Покажем, что $\sat{M'}{s}{\varphi^\ast}$. Нетрудно видеть, что $\sat{M'}{s}{\sigma(\Theta)}$, поэтому фактически нужно доказать, что $\sat{M'}{s}{\sigma(\varphi')}$. Поскольку $\sat{M}{s}{\varphi'}$, достаточно показать, что для каждой подформулы состояния $\psi$ формулы $\varphi$ и каждого состояния $x$ модели~$\mmodel{M}$
  $$
  \begin{array}{lcl}
  \sat{M}{x}{\psi'} & \iff & \sat{M'}{x}{\sigma(\psi')},
  \end{array}
  $$
  а также что для каждой подформулы пути $\alpha$ формулы $\varphi$ и каждого пути~$\pi$ в модели~$\mmodel{M}$
  $$
  \begin{array}{lcl}
  \sat{M}{\pi}{\alpha'} & \iff & \sat{M'}{\pi}{\sigma(\alpha')}.
  \end{array}
  $$

  Обе эти эквивалентности обосновываются одновременной рекурсией по построению подформул $\psi$ и $\alpha$ формулы~$\varphi$; обоснование не сталкивается с трудностями, и мы рассмотрим только случаи, когда $\psi = p_i$ и когда $\psi = \forall \alpha$.

  Пусть $\psi = p_i$; тогда $\psi' = p_i$ и $\sigma(\psi') = B_i$. Пусть $\sat{M}{x}{p_i}$. Тогда согласно построению модели~$\mmodel{M'}$ получаем, что $\sat{M'}{x}{B_i}$. Пусть теперь $\sat{M'}{x}{B_i}$. Поскольку $\sat{M'}{x}{B_i}$ влечёт, что $\sat{M'}{x}{\EX p}$, и поскольку  $\notsat{M}{y}{p}$ для каждого $y \in \states{S}$, заключаем, что $x R' r_m$ для некоторого $m \in \{1, \ldots, n+1\}$. Поскольку $(\mmodel{M'},r_m) \models A_i$ только при $m = i$ (по лемме~\ref{lem:roots}), получаем по построению модели~$\mmodel{M'}$, что $\sat{M}{x}{p_i}$.

  Пусть $\psi = \forall \alpha$; тогда $\psi' = \forall (\Box p_{n+1} \imp \alpha')$ и  $\sigma(\psi') = \forall (\Box B_{n+1} \imp \sigma(\alpha'))$. Пусть $\notsat{M}{x}{\forall (\Box p_{n+1} \imp \alpha')}$. Тогда $\notsat{M}{\pi}{\alpha'}$ для некоторого пути $\pi \in \Pi(x)$, такого, что $\sat{M}{\pi[i]}{p_{n+1}}$ для любого $i \in\numN$. Ясно, что $\pi$ является также и путём в $\mmodel{M'}$, а значит, $\sat{M'}{\pi[i]}{B_{n+1}}$ для любого $i \in\numN$ и $\notsat{M'}{\pi}{\sigma(\alpha')}$. Следовательно, $\notsat{M'}{x}{\forall (\Box B_{n+1} \imp \sigma(\alpha'))}$, что и требовалось. Пусть теперь $\notsat{M'}{x}{\forall (\Box B_{n+1} \imp \sigma(\alpha'))}$.  Тогда существует путь $\pi \in \Pi(x)$, такой, что $\sat{M'}{\pi[i]}{B_{n+1}}$ для каждого $i \in\numN$ и $\notsat{M'}{\pi}{\sigma(\alpha')}$. Заметим, что, согласно построению модели~$\mmodel{M}'$, ни в одном состоянии из $\states{S}'\setminus\states{S}$ формула $B_{n+1}$ не является истинной, и поэтому $\pi$~--- это путь и в~$\mmodel{M}$. Значит, можем применить индукционное предположение, и тогда получим, что $\notsat{M}{x}{\forall (\Box p_{n+1} \imp \alpha')}$.
\end{proof}

Используя лемму~\ref{lem:main_lemma_ctl_star} и тот факт, что $\varphi^\ast$ строится по~$\varphi$ полиномиально, получаем следующие теоремы.

\begin{theorem}
  \label{thr:CTLstar}
  Логика\/ $\CTLstar$ полиномиально погружается в свой фрагмент от одной переменной.
\end{theorem}

\begin{theorem}
  \label{thr:ctlstar-complexity}
  Проблема\/ $\CTLstar$-выполнимости формул от одной переменной является\/ $\cclass{2EXPTIME}$-полной.
\end{theorem}

\begin{proof}
Полнота в классе $\cclass{2EXPTIME}$ следует из теоремы~\ref{thr:CTLstar} и $\cclass{2EXPTIME}$-полноты проблемы выполнимости для логики $\CTLstar$~\cite{VardiStockmeyer85}.  \end{proof}

Покажем, что похожая аргументация применима и в случае логики~$\CTL$. Доказательство $\cclass{EXPTIME}$-полноты фрагмента $\CTL$ от одной переменной можно получить, следуя доказательству теоремы~\ref{maintheorem:K-ast} и используя только модальности $\AX$ и $\AG$, которые ведут себя сходным образом с модальностями $\Box$ и $\Box^\ast$, соответственно; единственная разница состоит в требовании серийности к отношению достижимости, что легко учитывается с помощью замены иррефлексивных миров рефлексивными, а константных формул~--- формулами от одной переменной. Но мы получим более общий результат: мы хотим построить полиномиальное погружение логики $\CTL$ в её фрагмент от одной переменной. Для этого мы слегка подправим доказательство, приведённое выше для логики~$\CTLstar$.

Сначала переопределим перевод $\cdot'$ следующим образом:
$$
\begin{array}{lcll}
  {p_i}'            & = & p_i & \mbox{для $i \in \{1, \ldots, n \}$;} \\
  {\bot}'           & = & \bot; & \\
  (\phi \con \psi)' & = & \phi' \con \psi'; & \\
  (\phi \dis \psi)' & = & \phi' \dis \psi'; & \\
  (\phi \imp \psi)' & = & \phi' \imp \psi'; & \\
  (\AX \phi)'       & = & \AX (p_{n+1} \imp \phi'); & \\
  (\EX \phi)'       & = & \EX (p_{n+1} \con \phi'); & \\
  (\phi \AU \psi)'  & = & \phi' \AU (p_{n+1} \con \psi'); & \\
  (\phi \EU \psi)'  & = & \phi' \EU (p_{n+1} \con \psi'). &
\end{array}
$$

Затем положим
$$
\begin{array}{lcl}
\Theta            & = & p_{n+1} \con \AG (\EX p_{n+1} \lra p_{n+1}). \\
\widehat{\varphi} & = & \Theta \con \varphi'.
\end{array}
$$

Тогда можно доказать аналог леммы~\ref{lem:atl-varphi-truth}.

\begin{lemma}
  \label{lem:varphi-truth-ctl}
  Имеет место следующая эквивалентность:
$$
\begin{array}{lcl}
\mbox{$\varphi$ $\CTL$-выполнима}
  & \iff
  & \mbox{$\widehat{\varphi}$ $\CTL$-выполнима.}
\end{array}
$$
\end{lemma}

\begin{proof}
Аналогично доказательству леммы~\ref{lem:atl-varphi-truth}.
\end{proof}

Теперь заметим, что формулы $A_1,\ldots,A_{n+1}$ и $B_1,\ldots,B_{n+1}$ можно понимать как $\lang{CTL}$-формулы; определим формулу $\varphi^\ast$ аналогично тому, как выше: $\varphi^\ast = \sigma(\widehat{\varphi})$. Справедлив аналог леммы~\ref{lem:main_lemma_ctl_star}.

\begin{lemma}
  \label{lem:main_lemma_ctl}
  Имеет место следующая эквивалентность:
$$
\begin{array}{lcl}
\mbox{$\varphi$ $\logic{CTL}$-выполнима}
  & \iff
  & \mbox{$\varphi^\ast$ $\logic{CTL}$-выполнима.}
\end{array}
$$
\end{lemma}

\begin{proof}
  Аналогично доказательству леммы~\ref{lem:main_lemma_ctl_star}.
\end{proof}

В результате мы получаем следующие теоремы.

\begin{theorem}
  \label{thr:ctl}
  Логика\/ $\CTL$ полиномиально погружается в свой фрагмент от одной переменной.
\end{theorem}

\begin{theorem}
  \label{thr:ctl-complexity}
  Проблема\/ $\CTL$-выполнимости формул от одной переменной является\/ $\cclass{EXPTIME}$-полной.
\end{theorem}

\begin{proof}
  Следует из теоремы~\ref{thr:ctl}, $\cclass{EXPTIME}$-трудности проблемы $\CTL$-выполнимости~\cite{FL79} и принадлежности проблемы $\CTL$-выполнимости классу $\cclass{EXPTIME}$~\cite{EH85}.
\end{proof}

    \subsection{Сложность фрагментов логик $\logic{ATL}^\ast$ и $\logic{ATL}$}

\providecommand{\coalname}[1]{\mathbb{#1}}
\providecommand{\cname}[1]{\coalname{#1}}
\providecommand{\coal}[1]{\langle\!\langle\coalname{#1}\rangle\!\rangle}
\providecommand{\Coal}[1]{\langle\!\langle{#1}\rangle\!\rangle}

\subsubsection{Предварительные сведения}

Временн\'{ы}е логики\footnote{На английском они называются alternating-time temporal logics.} $\ATLstar$ и $\ATL$ можно понимать как обобщения логик $\CTLstar$ и~$\CTL$ соответственно. Их модели содержат переходы между состояниями, вызванные одновременными действиями \defnotion{агентов}\index{агент} (или \defnotion{игроков}\index{игрок}) в системе; соответственно, в языках этих логик можно строить утверждения о путях, возникающих в результате таких выборов, а не обо всех путях или некоторых путях, как в случае~$\CTLstar$ и~$\CTL$. Однако в языках этих логик в определённом смысле сохраняется возможность говорить как обо всех, так и о некоторых путях, из-за чего $\ATLstar$ и $\ATL$ и можно понимать как обобщения логик~$\CTLstar$ и~$\CTL$.

В литературе имеется несколько подходов к тому, как определять, что такое $\ATLstar$ и $\ATL$. Одно из отличий состоит в том, сколько агентов имеется в языке. Мы сначала рассмотрим случай, когда в языке имеется фиксированное конечное число агентов~\cite{DGL16} (системы для $n$ агентов обозначим $\logic{ATL}^\ast_n$ и $\ATL_n$), а затем~--- когда агентов бесконечно много~\cite{WLWW06} (такие системы обозначим $\logic{ATL}^\ast_\infty$ и $\ATL_\infty$). Другое отличие связано с тем, как понимать задачу выполнимости формул языков этих систем. Дело в том, что помимо обычного понимания проблемы выполнимости в логике (формула выполнима, если её отрицание не принадлежит логике), в случае $\ATLstar$ и $\ATL$ предлагается несколько задач, идущих от семантики; внешне их формулировки похожи на утверждения, эквивалентные проблеме выполнимости формул в полной по Крипке логике. Такие задачи представляют интерес с точки зрения того, как можно использовать $\ATLstar$ и $\ATL$, т.е. с точки зрения прагматики этих логик. Таким образом, $\ATLstar$ и $\ATL$ интересны не столько тем, какие логические системы они собой представляют как множества формул, а их языком и задачами для этого языка, возникающими в семантике. Конечно, подобное утверждение справедливо и в отношении многих других логик, используемых в приложениях, но отличающиеся друг от друга проблемы локальной\footnote{Т.е. когда формула истинна в некотором состоянии некоторой модели, а не в каждом её состоянии.} выполнимости в модели обсуждались для $\ATLstar$ и $\ATL$ специально~\cite{DGL16,WLWW06}, и мы уделим внимание этому моменту.

Отметим, что предложенные обозначения $\logic{ATL}^\ast_1, \logic{ATL}^\ast_2, \logic{ATL}^\ast_3,\ldots,\logic{ATL}^\ast_\infty$ и $\logic{ATL}_1, \logic{ATL}_2, \logic{ATL}_3,\ldots,\logic{ATL}_\infty$ введены здесь для удобства; в литературе используются обозначения $\ATLstar$ и $\ATL$ в каждом из случаев.


\subsubsection{Синтаксис и семантика $\logic{ATL}^\ast_n$ и $\ATL_n$}
\label{sec:ctlatl:syntax:semantics}


Пусть $n\in\numNp$. Язык $\lang{ATL}_n^\ast$ логики $\logic{ATL}_n^\ast$ является обогащением языка $\lang{L}$; он содержит дополнительно непустое конечное множество $\cname{AG}_n$ имён агентов, подмножества которого называются \defnotion{коалициями},\index{коалиция} кванторы вида $\coal{C}$ для каждой коалиции~$\cname{C}$, темпоральные связки $\Next$, $\Until$ и~$\Box$. В~этом языке имеется два вида формул: \defnotion{формулы состояния}\index{уяа@формула!состояния} и \defnotion{формулы пути}.\index{уяа@формула!пути} Их можно определить одновременной рекурсией с помощью следующих BNF-выражений (ниже $\varphi$ используется для обозначения формулы состояния, а $\vartheta$~--- для обозначения формулы пути):
\[
\begin{array}{llcl}
\arrayitem &
\varphi
  & :=
  & p \mid \bot \mid  (\varphi \con \varphi) \mid  (\varphi \dis \varphi)
      \mid (\varphi \imp \varphi) \mid \coal{C} \vartheta;
  \arrayitemskip\\
\arrayitem &
\vartheta
  & :=
  & \varphi \mid (\vartheta \con \vartheta) \mid (\vartheta \dis \vartheta)
    \mid (\vartheta \imp \vartheta) \mid (\vartheta \Until \vartheta)
    \mid \Next \vartheta \mid \Box \vartheta,
\end{array}
\]
где $\cname{C}$ может быть любой коалицией агентов из $\cname{AG}_n$, а $p$~--- любой переменной из~$\prop$.  Другие связки определяются аналогично тому, как в случае языка~$\lang{CTL}^\ast$.

Для оценки истинности $\lang{ATL}_n^\ast$-формул используются \defnotion{модели параллельных игр}\index{модель!параллельных игр}\footnote{Исходное название следующее: concurrent game models.}. Моделью параллельных игр для языка $\lang{ATL}_n^\ast$ называется набор $\mmodel{M} = \otuple{\states{S}, \mathit{Act}, \mathit{act}, \delta, v}$, где
\begin{itemize}
\item $\states{S}$~--- непустое конечное множество \defnotion{состояний}\index{состояние};
\item $\mathit{Act}$~--- непустое конечное множество \defnotion{действий}\index{действие}, содержащее \defnotion{фиктивное действие}\index{действие!уяа@фиктивное}~$\#$;
\item $\mathit{act}$~--- отображение $\function{act}{\cname{AG}_n \times \states{S}}{\Power{\mathit{Act}}}$, называемое \defnotion{функцией управления действиями}\index{уяа@функция!управления действиями}; содержательно, эта функция указывает, какие действия доступны каждому из агентов в каждом из состояний;
\item $\delta$~--- \defnotion{функция перехода}\index{уяа@функция!перехода}, сопоставляющая каждому состоянию $s \in \states{S}$ и каждому \defnotion{профилю действий} $\alpha = \{\alpha_a\}_{a\in\cname{AG}_n}$, где $\alpha_a \in \mathit{act}(a,s)$ для каждого $a\in\cname{AG}_n$, некоторое состояние~$\delta(s,\alpha)$;
\item $v$~--- оценка, т.е. функция $\function{v}{\prop}{\Power{\states{S}}}$.
\end{itemize}


\defnotion{Путём}\index{путь} называем бесконечную последовательность $s_0, s_1, s_2, \ldots$ состояний модели $\mmodel{M}$, такую, что для каждого $i \in\numN$ существует такой профиль действий $\alpha$, что $s_{i+1} \in \delta(s_i, \alpha)$. Множество всех путей в модели $\mmodel{M}$ обозначим $\Pi_{\kmodel{M}}$. Для пути $\pi$ будем использовать обозначения $\pi[i]$ и $\pi[i, \infty]$, понимая их так же, как в случае моделей для языка $\lang{CTL}^\ast$. \defnotion{Начальный отрезок}\index{начальный отрезок пути} $\pi[0,i]$ пути $\pi$~--- это список $\pi[0],\ldots,\pi[i]$, который называется также \defnotion{историей};\index{история} для истории $h$ её последний элемент будем обозначать $\mathit{last}(h)$. Множество всех историй в модели~$\kmodel{M}$ обозначим~$H_{\kmodel{M}}$.

%

Пусть имеются состояние $s \in \states{S}$ и коалиция $\cname{C} \subseteq \cname{AG}_n$. Тогда \defnotion{$\cname{C}$-действие}\index{действие!c@$\cname{C}$-действие} в состоянии $s$~--- это такой набор $\alpha_{\cname{C}}=\{\alpha_{\cname{C}}(a)\}_{a\in\cname{AG}_n}$, что $\alpha_{\cname{C}}(a) \in \mathit{act}(a,s)$ для каждого $a \in \cname{C}$ и $\alpha_{\cname{C}}(a)=\#$ для каждого $a \not\in \cname{C}$. Пусть $\mathit{act}(\cname{C},s)$~--- множество всех $\cname{C}$-действий в~$s$.
Говорим, что профиль действий $\alpha$ \defnotion{расширяет} $\cname{C}$-действие $\alpha_C$, если $\alpha(a) = \alpha_{\cname{C}}(a)$ для каждого $a \in \cname{C}$; в этом случае пишем $\alpha_{\cname{C}} \sqsubseteq \alpha$.  \defnotion{Множеством результатов $\cname{C}$-действия} $\alpha_{\cname{C}}$ в состоянии~$s$ называем множество состояний
%
%
$$
\begin{array}{lcl}
\mathit{out}(s,\alpha_{\cname{C}})
  & =
  & \{ \delta(s, \alpha ) : \mbox{$\alpha \in \mathit{act}(\cname{AG}_n, s)$ и $\alpha_{\cname{C}} \sqsubseteq \alpha$}\}.
\end{array}
$$

\defnotion{Стратегией}\index{стратегия} агента $a\in\cname{AG}_n$ называется функция $\function{\mathit{str}_{a}}{H_{\kmodel{M}}}{\mathit{Act}}$, такая, что $\mathit{str}_{a}(h)\in \mathit{act}(a,\mathit{last}(h))$ для каждой истории $h\in H_{\kmodel{M}}$; \defnotion{$\cname{C}$-стратегией}\index{стратегия!c@$\cname{C}$-стратегия} называется набор стратегий $\mathit{str}_{\cname{C}} = \{\mathit{str}_{a}\}_{a\in\cname{C}}$.
Множество всех возможных путей из состояниях $s$ для коалиции $\cname{C}$, следующей стратегии $\mathit{str}_{\cname{C}}$, обозначим $\Pi_{\kmodel{M}}(s,\mathit{str}_{\cname{C}})$ и определим следующим образом:
$$
\begin{array}{lcl}
\Pi_{\kmodel{M}}(s,\mathit{str}_{\cname{C}})
  & =
  & \{ \pi\in\Pi_{\kmodel{M}} : \mbox{$\pi[0]=s$ и $\forall j\in\numN~ \pi[j+1]\in\mathit{out}(\pi[j],\mathit{str}_{\cname{C}}(\pi[0,j]))$}\}.
\end{array}
$$

Отношение истинности формул состояния в состояниях модели $\mmodel{M}$ и формул пути в путях модели~$\mmodel{M}$ определим рекурсивно (ниже считаем, что $\varphi_1$ и $\varphi_2$~--- формулы состояния, $\vartheta_1$ и $\vartheta_2$~--- формулы пути, $s$~--- состояние модели~$\mmodel{M}$, а $\pi$~--- путь в модели~$\mmodel{M}$):
\[
\begin{array}{llcl}
\arrayitem
  & \sat{M}{s}{p_i}
  & \bydef
  & s \in v(p_i);
  \arrayitemskip\\
\arrayitem
  & \notsat{M}{s}{\bot};
  \arrayitemskip\\
\arrayitem
  & \sat{M}{s}{\varphi_1 \con \varphi_2}
  & \bydef
  & \mbox{$\sat{M}{s}{\varphi_1}$ \phantom{л}и\phantom{и} $\sat{M}{s}{\varphi_2}$;}
  \arrayitemskip\\
\arrayitem
  & \sat{M}{s}{\varphi_1 \dis \varphi_2}
  & \bydef
  & \mbox{$\sat{M}{s}{\varphi_1}$ или $\sat{M}{s}{\varphi_2}$;}
  \arrayitemskip\\
\arrayitem
  & \sat{M}{s}{\varphi_1 \imp \varphi_2}
  & \bydef
  & \mbox{$\notsat{M}{s}{\varphi_1}$ или $\sat{M}{s}{\varphi_2}$;}
  \arrayitemskip\\
\arrayitem
  & \sat{M}{s}{\coal{C} \vartheta_1}
  & \bydef
  & \parbox[t]{280pt}{существует $\cname{C}$-стратегия $\mathit{str}_{\cname{C}}$, такая, что $\sat{M}{\pi}{\vartheta_1}$ для каждого $\pi \in \Pi(s, \mathit{str}_{\cname{C}})$;}
  \arrayitemskip\\
\arrayitem
  & \sat{M}{\pi}{\varphi_1}
  & \bydef
  & \sat{M}{\pi[0]}{\varphi_1};
  \arrayitemskip\\
\arrayitem
  & \sat{M}{\pi}{\vartheta_1 \con \vartheta_2}
  & \bydef
  & \mbox{$\sat{M}{\pi}{\vartheta_1}$ \phantom{л}и\phantom{и} $\sat{M}{\pi}{\vartheta_2}$;}
  \arrayitemskip\\
\arrayitem
  & \sat{M}{\pi}{\vartheta_1 \dis \vartheta_2}
  & \bydef
  & \mbox{$\sat{M}{\pi}{\vartheta_1}$ или $\sat{M}{\pi}{\vartheta_2}$;}
  \arrayitemskip\\
\arrayitem
  & \sat{M}{\pi}{\vartheta_1 \imp \vartheta_2}
  & \bydef
  & \mbox{$\notsat{M}{\pi}{\vartheta_1}$ или $\sat{M}{\pi}{\vartheta_2}$;}
  \arrayitemskip\\
\arrayitem
  & \sat{M}{\pi}{\Next \vartheta_1}
  & \bydef
  & \sat{M}{\pi[1,\infty]}{\vartheta_1};
  \arrayitemskip\\
\arrayitem
  & \sat{M}{\pi}{\Box \vartheta_1}
  & \bydef
  & \mbox{$\sat{M}{\pi[i,\infty]}{\vartheta_1}$ для каждого $i\in\numN$;}
  \arrayitemskip\\
\arrayitem
  & \sat{M}{\pi}{\vartheta_1 \Until \vartheta_2}
  & \bydef
  & \parbox[t]{280pt}{$\sat{M}{\pi[i, \infty]}{\vartheta_2}$ для некоторого $i\in\numN$ и   $\sat{M}{\pi[j,\infty]}{\vartheta_1}$ для каждого $j$, такого, что  $0 \leqslant j < i$.}
\end{array}
\]

Считаем, что $\lang{ATL}_n^\ast$-формулы~--- это формулы состояния; $\lang{ATL}_n^\ast$-формулу считаем \defnotion{выполнимой}, если она истинна в некотором состоянии некоторой модели; $\lang{ATL}_n^\ast$-формулу считаем \defnotion{общезначимой}, если она истинна в каждом состоянии каждой модели. Под логикой $\logic{ATL}^\ast_n$ понимаем множество всех общезначимых $\lang{ATL}_n^\ast$-формул.

Язык $\lang{ATL}_n$ логики $\ATL_n$ можно понимать как фрагмент языка $\lang{ATL}_n^\ast$, содержащий только формулы, в которых кванторы по коалициям используются только совместно с темпоральными связками. Это, в частности, означает, что в языке $\lang{ATL}_n$ нет формул пути. Таким образом, язык $\lang{ATL}_n$ содержит модальности вида $\coal{C}\Next$, $\coal{C}\Box$ и~$\coal{C} \Until$, где $\varphi\coal{C}\Until\psi = \coal{C}(\varphi\Until\psi)$, а определение $\lang{ATL}_n$-формулы в виде BNF-выражения выглядит следующим образом:
\[
\begin{array}{llcl}
\arrayitem
  & \varphi & := & p \mid \bot \mid  (\varphi \con \varphi)\mid  (\varphi \dis \varphi)\mid  (\varphi \imp \varphi) \mid \coal{C}\Next\varphi \mid \coal{C}\Box \varphi \mid (\varphi \coal{C}\Until \varphi),
\end{array}
\]
где $\cname{C}$ может быть произвольным подмножеством множества $\cname{AG}_n$, а~$p$~--- произвольной переменной из~$\prop$.

Истинность $\lang{ATL}_n$-формул в состояниях модели определяется как истинность этих формул, понимаемых как формулы языка $\lang{ATL}_n^\ast$; понятия выполнимости и общезначимости определяются так~же. Под логикой $\ATL_n$ понимаем множество общезначимых $\lang{ATL}_n$-формул.


\subsubsection{Константные фрагменты логик $\logic{ATL}^\ast_n$ и $\ATL_n$}
\label{sec:atl-variable-free-fragment}

Проблема выполнимости константных формул и для $\logic{ATL}_n$, и для $\logic{ATL}_n^\ast$ решается полиномиально по времени. Действительно, любая константная формула в языке каждой из этих формул эквивалентна либо формуле~$\bot$, либо формуле~$\top$. Для этого достаточно заметить, что логике~$\logic{ATL}^\ast_n$ принадлежат следующие эквивалентности формул: во-первых, классические эквивалентности
$$
\begin{array}{lclclcl}
  (\top \con \top)   & \lra & \top; & ~~~~ & (\top \con \bot)   & \lra & \bot; \\
  (\bot \con \top)   & \lra & \bot; & ~~~~ & (\bot \con \bot)   & \lra & \bot; 
  \\
  (\top \dis \top)   & \lra & \top; & ~~~~ & (\top \dis \bot)   & \lra & \top; \\
  (\bot \dis \top)   & \lra & \top; & ~~~~ & (\bot \dis \bot)   & \lra & \bot; 
  \\
  (\top \imp \top)   & \lra & \top; & ~~~~ & (\top \imp \bot)   & \lra & \bot; \\
  (\bot \imp \top)   & \lra & \top; & ~~~~ & (\bot \imp \bot)   & \lra & \top, 
  \\
\end{array}
$$
а во-вторых, эквивалентности для кванторов и темпоральных связок
$$
\begin{array}{lclclcl}
  \forall \top       & \lra & \top; & ~~~~ & \forall \bot       & \lra & \bot; \\
  \Next \top         & \lra & \top; & ~~~~ & \Next \bot         & \lra & \bot; \\
  \Box \top          & \lra & \top; & ~~~~ & \Box \bot          & \lra & \bot; \\
  \coal{C} \top      & \lra & \top; & ~~~~ & \coal{C} \bot      & \lra & \bot; \\
  (\top \Until \top) & \lra & \top; & ~~~~ & (\top \Until \bot) & \lra & \bot; \\
  (\bot \Until \top) & \lra & \top; & ~~~~ & (\bot \Until \bot) & \lra & \bot.
\end{array}
$$

Таким образом, чтобы проверить, выполнима ли константная формула, достаточно рекурсивно заменить каждую её подформулу либо на~$\bot$, либо на~$\top$, что даёт нам алгоритм, квадратичный по времени от длины тестируемой константной формулы. Поскольку логика $\logic{ATL}_n$ является $\cclass{EXPTIME}$-полной, логика $\logic{ATL}^\ast_n$~--- $\cclass{2EXPTIME}$-полной, а $\cclass{P}\ne\cclass{EXPTIME}\subseteq\cclass{2EXPTIME}$, получаем, что константные фрагменты этих логик существенно проще каждой из них в полном языке.

\subsubsection{Фрагменты логик $\logic{ATL}^\ast_n$ и $\ATL_n$ от одной переменной}
\label{sec:atl-single-variable-fragment}

Заметим, что логики $\CTLstar$ и $\CTL$ погружаются, соответственно, в логики $\logic{ATL}^\ast_n$ и $\ATL_n$ с помощью перевода, заменяющего в формулах квантор~$\forall$ на $\coal{\varnothing}$, а квантор~$\exists$~--- на~$\Coal{\cname{AG}_n}$. Как следствие, мы получаем следующие теоремы.

\begin{theorem}
  \label{thr:atlstar-complexity}
  Проблема\/ $\logic{ATL}^\ast_n$-выполнимости формул от одной переменной является\/ $\cclass{2EXPTIME}$-полной.
\end{theorem}

\begin{proof}
Полнота в классе $\cclass{2EXPTIME}$ следует из теоремы~\ref{thr:ctlstar-complexity} и $\cclass{2EXPTIME}$-полноты проблемы выполнимости для логики $\logic{ATL}^\ast_n$~\cite{Schewe08}.  \end{proof}

\begin{theorem}
  \label{thr:atl-complexity}
  Проблема\/ $\ATL_n$-выполнимости формул от одной переменной является\/ $\cclass{EXPTIME}$-полной.
\end{theorem}

\begin{proof}
  Следует из теоремы~\ref{thr:ctl-complexity}, $\cclass{EXPTIME}$-трудности проблемы $\ATL_n$-выполнимости~\cite{FL79} и принадлежности проблемы $\ATL_n$-выпол\-ни\-мости классу $\cclass{EXPTIME}$~\cite{GD06,WvdH03}.
\end{proof}

Отметим, что мы не формулируем и не доказываем утверждения для $\logic{ATL}^\ast_n$ и $\ATL_n$, аналогичные теоремам~\ref{thr:CTLstar} и~\ref{thr:ctl}. Нетрудно понять, что аналогичные утверждения для $\logic{ATL}^\ast_n$ и $\ATL_n$, конечно же, справедливы, поскольку любые две проблемы, полные в классе $\cclass{2EXPTIME}$ или полные в классе $\cclass{EXPTIME}$, полиномиально сводятся друг к другу. Тем не менее, предложить такое полиномиальное сведение, сохраняющее структуру формул, как в случае логик $\CTLstar$ и $\CTL$, автор не может. Сложность состоит в том, что неясно, как устроить релятивизацию. Дело в том, что квантор $\coal{C}$ ведёт себя и не как квантор всеобщности, и не как квантор существования, а как их комбинация вида~<<$\exists\forall$>>: $\coal{C}\vartheta$ означает, что существует $\cname{C}$-стратегия, такая, что $\vartheta$ выполняется для любого ответа агентов из множества $\cname{AG}_n\setminus\cname{C}$.

\subsubsection{Логики $\logic{ATL}^\ast_\infty$ и $\ATL_\infty$}
\label{sec:atlstar-complexity:altsemantics}

Мы рассмотрели вариант $\ATL$ и $\ATLstar$, когда множество агентов в языке и моделях конечно и зафиксировано изначально; этот подход соответствует описанному в~\cite{DGL16}. Но в литературе обсуждались и другие подходы к определению $\logic{ATL}^\ast$-формул и их выполнимости.

Так, в~\cite{WLWW06} описывается вариант $\ATL$ и $\ATLstar$, когда множество агентов в языке бесконечно; обозначим такие языки $\lang{ATL}_\infty$ и $\lang{ATL}^\ast_\infty$. Язык $\lang{ATL}^\ast_\infty$ логики $\logic{ATL}^\ast_\infty$ является объединением языков вида $\lang{ATL}^\ast_n$, где $n\in\numNp$; он содержит счётное множество $\cname{AG}$ имён агентов (конечные подмножества которого по-прежнему называем \defnotion{коалициями}\index{коалиция}), кванторы вида $\coal{C}$ для каждой коалиции~$\cname{C}$, темпоральные связки $\Next$, $\Until$ и~$\Box$. В~этом языке можно строить как \defnotion{формулы состояния},\index{уяа@формула!состояния} так и \defnotion{формулы пути};\index{уяа@формула!пути} определим их одновременной рекурсией с помощью BNF-выражений (ниже $\varphi$ используется для обозначения формулы состояния, а $\vartheta$~--- для обозначения формулы пути):
\[
\begin{array}{llcl}
\arrayitem &
\varphi
  & :=
  & p \mid \bot \mid  (\varphi \con \varphi) \mid  (\varphi \dis \varphi)
      \mid (\varphi \imp \varphi) \mid \coal{C} \vartheta;
  \arrayitemskip\\
\arrayitem &
\vartheta
  & :=
  & \varphi \mid (\vartheta \con \vartheta) \mid (\vartheta \dis \vartheta)
    \mid (\vartheta \imp \vartheta) \mid (\vartheta \Until \vartheta)
    \mid \Next \vartheta \mid \Box \vartheta,
\end{array}
\]
где $\cname{C}$ может быть любой коалицией агентов из $\cname{AG}$, а $p$~--- любой переменной из~$\prop$. Считаем, что $\lang{ATL}^\ast_\infty$-формулы~--- это формулы состояния.

Язык $\lang{ATL}_\infty$ является фрагментом языка $\lang{ATL}^\ast_\infty$, определяемым тем же условием, что и раньше: за квантором сразу следует темпоральная связка.

Модели определяются не для всего языка $\lang{ATL}^\ast_\infty$, а только для его фрагментов от конечного множества агентов; на модели дополнительно накладывается требование конечности и всех других составляющих: множества состояний, действий, и т.п. Формула считается \defnotion{выполнимой}, если она истинна в некотором состоянии некоторой модели, содержащей агентов, присутствующих в формуле; формула считается \defnotion{общезначимой}, если она истинна во всех мирах всех таких моделей. Формально под логиками $\logic{ATL}^\ast_\infty$ и $\logic{ATL}_\infty$ понимаем, соответственно, множество общезначимых $\lang{ATL}^\ast_\infty$-формул и множество общезначимых $\lang{ATL}_\infty$-формул.

При таком определении мы получаем проблему выполнимости, отличающуюся от рассмотренной нами выше для $\lang{ATL}^\ast_n$ (и для $\lang{ATL}^\ast_n$) при $n\in\numNp$. Поскольку у нас в языке уже нет возможности образовать коалицию из всех агентов, мы не можем устроить погружение $\CTL$ и $\CTLstar$ в $\ATL$ и $\ATLstar$ так, как это было описано для логик $\logic{ATL}^\ast_n$ и $\ATL_n$.

В целом, авторы~\cite{WLWW06} выделяют три задачи, которые называют вариациями проблемы выполнимости $\ATLstar$-формул:
\begin{itemize}
\item[$(S1)$]
по формуле и множеству агентов выяснить, выполнима ли эта формула в классе моделей для этого множества агентов;
\item[$(S2)$]
по формуле выяснить, выполнима ли она в классе всех моделей;
\item[$(S3)$]
по формуле выяснить, выполнима ли она в классе моделей, множество агентов в которых совпадает со множеством агентов в формуле.
\end{itemize}

Вариант $(S2)$ соответствует задачам $\logic{ATL}^\ast_\infty$-выполнимости и $\ATL_\infty$-выпол\-ни\-мости. Две другие задачи близки по формулировке, но не являются проблемами выполнимости для так определённой логики; тем не менее, они представляют интерес с точки зрения приложений.

Отметим, что задачи $\logic{ATL}^\ast_\infty$-выполнимости и $\ATL_\infty$-выполнимости не соответствуют ни~$(S1)$, ни~$(S2)$, ни~$(S3)$. Тем не менее, из наших построений следуют результаты о сложности задач $(S1)$ и $(S3)$ для языков $\lang{ATL}^\ast_\infty$ и $\lang{ATL}_\infty$ с одной переменной. Действительно, для обеих задач достаточно заметить, что $\CTLstar$ обладает свойством конечных моделей~\cite{DGL16}, а описанные выше построения (возникающие при моделировании переменных формулами от одной переменной) позволяют из конечной модели получить конечную модель; тогда можно построить тот же перевод $\CTLstar$-формул в $\lang{ATL}^\ast_\infty$-формулы (а также перевод $\CTL$-формул в $\lang{ATL}_\infty$-формулы), взяв в качестве множества агентов двухэлементное множество (и сделав его в задаче $(S1)$ её вторым параметром). Это даст нижние оценки как для $(S1)$, так и для~$(S3)$. Верхние следуют из~\cite{WLWW06}.

В итоге мы получаем следующие результаты.

\begin{corollary}
Задачи\/ $(S1)$ и\/ $(S3)$ являются\/ $\cclass{2EXPTIME}$-полными для фрагмента языка\/ $\lang{ATL}^\ast_\infty$ с одной переменной.
\end{corollary}

\begin{corollary}
Задачи\/ $(S1)$ и\/ $(S3)$ являются\/ $\cclass{EXPTIME}$-полными для фрагмента языка\/ $\lang{ATL}_\infty$ с одной переменной.
\end{corollary}

    \subsection{Сложность фрагментов логики $\logic{LTL}$}
    \label{ssec:LTL}

\subsubsection{Синтаксис и семантика}

Язык $\lang{LTL}$ логики $\LTL$ можно считать фрагментом языка $\lang{CTL}^\ast$, состоящим из формул пути, т.е. не содержащих кванторов пути $\forall$ и~$\exists$. 

Истинность $\lang{LTL}$-формулы $\varphi$ на пути модели определяется как истинность этих формул, понимаемых как формулы языка $\lang{CTL}^\ast$; $\lang{LTL}$-формула $\varphi$ считается выполнимой, если выполнима $\lang{CTL}^\ast$-формула~$\exists\varphi$; $\lang{LTL}$-формула $\varphi$ считается общезначимой, если общезначима $\lang{CTL}^\ast$-формула~$\forall\varphi$. 

Под логикой $\LTL$ понимаем множество всех общезначимых $\lang{LTL}$-формул.

Отметим, что $\LTL$, как и $\CTL$, является фрагментом логики $\CTLstar$, при этом логики $\LTL$ и $\CTL$ не сравнимы друг с другом по включению.

\subsubsection{Константный фрагмент логики $\LTL$}
\label{sec:ltl-variable-free-fragment}

Константный фрагмент логики $\LTL$ является фрагментом константного фрагмента логики~$\CTLstar$, который, как мы видели, полиномиально разрешим. Следовательно, полиномиально разрешим и константный фрагмент логики~$\LTL$.

\subsubsection{Фрагмент логики $\LTL$ от одной переменной}
\label{sec:ltl-single-variable-fragment}

В случае логики $\LTL$ мы не можем применить моделирование переменных формулами от одной переменной, описанное и многократно применённое выше, по той причине, что оно предполагает ветвление в моделях, а оценка истинности $\lang{LTL}$-формул происходит в выбранном пути, который не меняется в процессе оценки. Тем не менее, можно применить другое моделирование, идею которого мы пока не затрагивали. Если коротко, то идея состоит в том, чтобы при моделировании сложных проблем формулами логики заменить возникающие при этом подформулы вида $\Next\vartheta$ подформулами вида $\Next^{n+1}\vartheta$. Такая замена даёт возможность вместо цепи состояний вида $xRy$, где $R$~--- отношение достижимости в некоторой модели, а $x$ и $y$~--- состояния в ней, рассматривать цепь вида $xRz_1R\ldots Rz_nRy$, используя новые состояния $z_1,\ldots,z_n$ для моделирования в $x$ значений переменных $p_1,\ldots,p_n$: если переменная $p_k$ была истинной в состоянии $x$ в исходной модели, то в новой модели полагаем переменную $p$ истинной в состоянии $z_k$, т.е.\ в новой модели в состоянии $x$ будет истинной формула~$\Next^k p$. Несмотря на простоту идеи, её техническая реализация требует внимания, поскольку нужно ещё суметь отделить <<старые>> состояния от <<новых>>, т.е. устроить релятивизацию. Это возможно как с помощью введения новой переменной, так и за счёт рассмотрения особых формул, которые моделируют некоторую сложную проблему и устройство которых мы можем контролировать так, что становится возможным учитывать движение по пути <<шагами>> длины $n+1$.

Как доказано в~\cite{SislaClarke85}, проблема $\LTL$-выполнимости $\cclass{PSPACE}$-полна. Применяя описанную выше идею к формулам и моделям, предложенным в~\cite{SislaClarke85}, можно получить доказательство $\cclass{PSPACE}$-полноты фрагмента логики $\LTL$ в языке с одной пропозициональной переменной~\cite{MR:2018:SAICSIT}. Этот результат был получен ранее (и~другими методами) в~\cite{DS02} среди прочих результатов, связанных со сложностью логик линейного времени, и мы не будем приводить здесь детали ни доказательства из~\cite{MR:2018:SAICSIT}, ни детали доказательства из~\cite{DS02}, а ограничимся лишь формулировкой.

\begin{theorem}
Проблема\/ $\LTL$-выполнимости формул от одной переменной является\/ $\cclass{PSPACE}$-полной.
\end{theorem}

Что касается описанной выше идеи моделирования всех переменных формулами от одной переменной, мы применим её в предикатном случае; там же будут даны подробные описания технических деталей.

  \section{Произведения и полупроизведения логик}	
    \subsection{Предварительные сведения}

\providecommand{\funcInd}{\mathop{\mathit{ind}}}
\providecommand{\indbr} [1] {\funcInd (#1)} 
\providecommand{\ind}   [1] {\funcInd #1} 

\providecommand{\pw}[1]{\bar{#1}}
\providecommand{\repr}[1]{\ensuremath{\tilde{#1}}}
\providecommand{\emptytuple}{\ensuremath{\curlywedge}}
\providecommand{\emptystring}{\ensuremath{\varepsilon}}

\providecommand{\bigcon}{\bigwedge}
\providecommand{\bigdis}{\bigvee}
\providecommand{\psneg}{\sim}

\providecommand{\nat}{\numN}
\providecommand{\sameas}{\bydef}
\providecommand{\thickdot}{\boldsymbol{.}}


\providecommand{\al}{\ensuremath{\alpha}}
\providecommand{\be}{\ensuremath{\beta}}
\providecommand{\Si}{\ensuremath{\Sigma}}
\providecommand{\si}{\ensuremath{\sigma}}
\providecommand{\D}{\ensuremath{\Delta}}
\providecommand{\ga}{\ensuremath{\gamma}}
\providecommand{\vp}{\ensuremath{\varphi}}
\providecommand{\La}{\ensuremath{\Lambda}}
\providecommand{\Th}{\ensuremath{\Theta}}

\providecommand{\var}{\mathop{\mathit{var}}}
\providecommand{\sub}{\mathop{\mathit{sub}}}
\providecommand{\md}{\mathop{\mathit{md}}}


Произведения модальных пропозициональных логик~\cite{GKWZ,Kurucz08} появились в \mbox{1970-х} годах в работах К.~Сегерберга~\cite{Seg73} и В.\,Б.~Шехтмана~\cite{Shehtman78rus}. Произведения логик нашли приложения в нескольких областях теоретической информатики, связанных с пространственно-временными рассуждениями~\cite{BCWZ02,KKZW07}, темпорально-эпистемическими рассуждениями~\cite{FHMV95, HV89, WvdH03,GorSh09b} и представлениями знаний~\cite{WZ99, WZ2000, LWZ08}. В~области математической логики интерес к произведениям модальных пропозициональных логик в значительной степени мотивирован их связью с модальными предикатными логиками~\cite{GSh98,GKWZ}.

Сложность произведений модальных пропозициональных логик обычно довольно высока: такие логики могут быть $\cclass{coNEXPTIME}$-полными (например, это так для произведения $\logic{K} \times \logic{S5}$ логики $\logic{K}$ с логикой $\logic{S5}$~\cite{Marx99}), неэлементарными, но разрешимыми (как, например, $\logic{K} \times \logic{K}$, $\logic{K} \times \logic{K4}$ и $\logic{K} \times \logic{S5}_2$~\cite{GJL10}), и даже неразрешимыми (как, например, произведения двух логик, одна из которых совпадает с $\logic{K4}$, $\logic{S4}$, $\logic{GL}$ или $\logic{Grz}$~\cite{RZ01,GKWZ05}, или произведения большего числа логик, см.~\cite{HHK02}).


Иногда понижение сложности возможно за счёт ограничений на средства языка; в пропозициональных модальных логиках наиболее распространенными способами понижения сложности являются ограничение модальной глубины и количества пропозициональных переменных в формулах. Для произведений логик вопрос ограничения модальной глубины формул был изучен в~\cite{MM01}, где было показано, что фрагмент логики $\logic {K} \times \logic {K}$, состоящий из формул глубины не более двух, является $\cclass{coNEXPTIME}$-полным; аналогичные результаты для логик $\logic{K} \times \logic{K4}$ и $\logic{K} \times\logic{S5}_2$ следуют из~\cite{GJL10}. Здесь будет показано, что происходит со сложностью произведений модальных пропозициональных логик при ограничении на число переменных.

Используемая здесь техника несложно распространяется и на полупроизведения логик\footnote{Английское название следующее: expanding relativized products. В случае двух логик используется также термин semiproducts, который и был выбран для перевода понятия на русский язык. Известно, что полупроизведения логик~\cite{GKZW06} иногда имеют меньшую сложность, чем соответствующие произведения.} \cite{GKWZ}, когда хотя бы один из сомножителей совпадает с $\logic{K}$, $\logic{KT}$, $\logic{KB}$ или $\logic{KTB}$. Полупроизведения логик связаны с модальными предикатными логиками, определяемыми классами шкал Крипке с расширяющимися предметными областями~\cite{Shehtman19, ShehtmanShkkatov20, ShSh:2023:rus}; похожим образом произведения логик связаны с модальными предикатными логиками, определяемыми шкалами Крипке с постоянными предметными областями~\cite{GSh93,GSh98}. Отметим, что описанная ниже техника естественным образом распространяется также на произведения и полупроизведения полимодальных логик, содержащих модальность, ведущую себя как модальность одной из логик $\logic{K}$, $\logic{KT}$, $\logic{KB}$ или~$\logic{KTB}$.

    \subsection{Синтаксис и семантика}

Мы будем рассматривать логики в языках $\lang{ML}_n$, где $n\in \numNp$ (см.~раздел~\ref{ssec:syntax-polymodal}); мы определяли слияния логик в этих языках (раздел~\ref{ssec:fusions}), но, в отличие от слияний логик, произведения логик определяются семантически.

Мы будем использовать стандартные сокращения для модальностей. Для каждого $i \in \{1, \ldots, n\}$ положим $\Diamond_i \vp = \neg \Box_i \neg \vp$, и пусть также для каждого $k \in \nat$
$$
\begin{array}{rclcrcl}
  \Box_i^0 \vp & = & \vp,
  & &
    \Box_i^{k+1} \vp & = & \Box_i^{\phantom{i}} \Box_i^k \vp,
  \smallskip \\
  \Diamond_i^k \vp & = & \neg \Box_i^k \neg \vp,
  & &
    \Box_i^{+} \vp & = & \vp \con \Box_i^{\phantom{i}} \vp,
  \smallskip\\
  \Box_i^{\leqslant 0} \vp & = & \vp,
  & &
    \Box_i^{\leqslant k + 1}\vp & = & \Box_i^{\leqslant k} \vp \con \Box_i^{k+1} \vp.
\end{array}
$$
Для формулы $\vp$ множество входящих в неё переменных обозначаем $\var\vp$, а множество её подформул~--- $\sub\vp$.
Модальную глубину формулы $\vp$ обозначаем $\md \varphi$ и определяем как наибольшее число вложенных в~$\vp$ модальностей:
$$
\begin{array}{lcll}
  \md  p                    & = & 0 & \mbox{для $p\in\prop$;} \\
  \md  \bot                 & = & 0;  \\
  \md  (\vp_1 \con \vp_2)   & = & \max \{\md \vp_1, \md \vp_2\};  \\
  \md  (\vp_1 \dis \vp_2)   & = & \max \{\md \vp_1, \md \vp_2\};  \\
  \md  (\vp_1 \imp \vp_2)   & = & \max \{\md \vp_1, \md \vp_2\};  \\
  \md  \Box_i \vp_1         & = & \md \vp_1 + 1 & \mbox{для $i \in \{1, \ldots, n \}$}.
\end{array}
$$



Для оценки истинности формул будем использовать шкалы и модели Крипке.
Понятия шкалы Крипке, модели Крипке и связанные с ними определяются обычно, см.~раздел~\ref{ssec:sem-polymodal}; то же самое относится и к определению истинности формул в мирах модели, в моделях, шкалах, классах шкал. Шкалу Крипке вида $\kframe{F}=\otuple{W,R_1,\ldots,R_n}$ называем также \defnotion{$n$\nobreakdash-шкалой}.\index{уяи@шкала!n@$n$-шкала}

Нам понадобятся стандартные обозначения для бинарных отношений на непустых множествах: если $W \ne \varnothing$, $W'$~--- непустое модмножество множества~$W$ и $R, S \subseteq W \times W$, то
\begin{itemize}
\item $R \circ S$~--- \defnotion{композиция}\index{композиция} отношений $R$ и $S$;
\item $R^1 = R$, $R^{k+1} = R \circ R^k$ для $k\in\numNp$;
\item $R^\ast$~--- \defnotion{рефлексивно-транзитивное замыкание}\index{еяд@замыкание!рефлексивно-транзитивное} отношения~$R$;
\item $R \upharpoonright W' = R \cap (W' \times W')$;
\item $R(W') = \set{y \in W : \mbox{$xRy$ для некоторого $x \in W'$}}$.
\end{itemize}
Будем писать $R(x)$ вместо $R(\{x\})$.

Пусть для каждого $i\in\{1\ldots,n\}$ имеется $1$\nobreakdash-шкала $\kframe{F}_i = \langle W_i, R_i \rangle$; \defnotion{произведением}\index{произведение!уяи@шкал} шкал $\kframe{F}_1, \ldots, \kframe{F}_n$ называется $n$\nobreakdash-шкала
$$
\begin{array}{lcl}
  \kframe{F}_1 \times \ldots \times \kframe{F}_n
  & = &
        \langle W_1 \times \ldots \times W_n, \bar{R}_1, \ldots, \bar{R}_n \rangle,
\end{array}
$$
где для каждого $i \in \{1, \ldots, n\}$
$$
\begin{array}{lcl}
  \langle x_1, \ldots, x_n \rangle \bar{R}_i \langle y_1, \ldots, y_n \rangle
  & \leftrightharpoons
  & \mbox{$x_i R_i y_i$ и $x_k = y_k$, где $1\leqslant k \leqslant n$ и $k\ne i$.}
\end{array}
$$

В дальнейшем для удобства будем использовать следующее обозначение: если $\pw{x} \in W_1 \times \ldots \times W_n$ и $s \in \{1, \ldots, n\}$, то $x_s$~--- это $s$\nobreakdash-й компонент набора~$\pw{x}$. Элементы произведений шкал будем называть \defnotion{точками}.\index{точка}

Отметим, что произведения шкал удовлетворяют следующему \defnotion{условию коммутативности}:\index{условие!коммутативности} для любых $i, j \in \{1, \ldots, n\}$
$$
\begin{array}{lcl}
\bar{R}_i \circ \bar{R}_j & \subseteq & \bar{R}_j \circ \bar{R}_i.
\end{array}
\eqno{(1)}
$$

Пусть $L_1, \ldots, L_n$~--- полные по Крипке мономодальные логики; \defnotion{произведением}\index{произведение!логик} логик $L_1, \ldots, L_n$ называется $n$-модальная логика класса всех произведений шкал Крипке логик $L_1, \ldots, L_n$:
$$
\begin{array}{lcl}
  L_1 \times \ldots \times L_n
    & =
    & \mlogic {\set{ \kframe{F}_1 \times \ldots \times \kframe{F}_n :
      \mbox{$\kframe{F}_i \models L_i$, где $1\leqslant i \leqslant n$}}}.
\end{array}
$$
Заметим, что если в произведении $n$ логик поменять местами сомножители, то мы получим логику, отличающуюся, разве что, переобозначением модальностей. Поэтому ниже при исследовании произведений логик, в которых какой-то из сомножителей равен конкретной логике~$L$, мы можем считать, что логике~$L$ равен первый сомножитель.

Пусть $d,n\in\numNp$ и $d \leqslant n$ и пусть для каждого $i\in\{1\ldots,n\}$ имеется $1$\nobreakdash-шкала $\kframe{F}_i = \langle W_i, R_i \rangle$; $n$\nobreakdash-шкала
$\kframe{G} = \langle \bar{V}, \bar{S}_1, \ldots, \bar{S}_n \rangle$ называется \defnotion{$d$-полупроизведением}\index{полупроизведение!d@$d$-полупроизведение!шкал}\footnote{Английский термин следующий: $d$-expanding relativized product; для случая, когда $n=2$ и $d=1$ используется также термин semiproduct, который и был взят для перевода на русский.} шкал $\kframe{F}_1, \ldots, \kframe{F}_n$, если
$\kframe{G} \subseteq \kframe{F}_1\times\ldots\times\kframe{F}_n$ и при этом для каждого $i \in \{1, \ldots, d\}$
$$
\begin{array}{lcl}
  \mbox{$\pw{x} \in \bar{V}$ и $y \in R_i(x_i)$}
  & \imply
  & \mbox{$\langle x_1, \ldots, x_{i-1}, y, x_{i+1}, \ldots, x_n \rangle \in \bar{V}$,}
\end{array}
\eqno{(2)}
$$
т.е. $\bar{R}_i (\bar{V}) \subseteq \bar{V}$. Индекс $i \in \{1, \ldots, n\}$ называем \defnotion{выделенным},\index{индекс!бяа@выделенный} если $i \leqslant d$.  Класс всех $d$-полупроизведений $1$\nobreakdash-шкал $\kframe{F}_1,\ldots,\kframe{F}_n$ обозначаем
$(\kframe{F}_1^{\phantom{i}} \times \ldots \times
\kframe{F}_n^{\phantom{i}})_d^{\mathsf{EX}}$.

Известно, что в общем случае полупроизведения шкал не удовлетворяют условию~$(1)$, но они удовлетворяют более слабому \defnotion{условию левой коммутативности}:\index{условие!левой коммутативности} для каждого $i \in \{1, \ldots, d\}$
и каждого $j \in \{1, \ldots, n\}$,
$$
\begin{array}{lcl}
\bar{R}_j \circ \bar{R}_i & \subseteq & \bar{R}_i \circ \bar{R}_j.
\end{array}
\eqno{(3)}
$$

Пусть $L_1,\ldots,L_n$~--- полные по Крипке мономодальные логики, $d\in\set{1,\ldots,n}$; тогда \defnotion{$d$-полупроизведением}\index{полупроизведение!d@$d$-полупроизведение!логик}\footnote{Английский термин тот же, что и для шкал: $d$-expanding relativized product; для случая, когда $n=2$ и $d=1$ используется также термин semiproduct.} логик $L_1,\ldots,L_n$ называется логика
$$
\begin{array}{lcl}
  (L_1^{\phantom{i}} \times \ldots \times L_n^{\phantom{i}})_d^{\mathsf{EX}}
    & =
    & \mlogic{\set{\kframe{F}\in (\kframe{F}_1^{\phantom{i}} \times \ldots \times \kframe{F}_n^{\phantom{i}})_d^{\mathsf{EX}} :
    \mbox{$\kframe{F}_i^{\phantom{i}} \models L_i^{\phantom{i}}$, где $1\leqslant i\leqslant n$}}}.
\end{array}
$$
Заметим, что, поскольку истинность формул в мирах шкалы сохраняется при взятии её порождённых подшкал,
$(L_1^{\phantom{i}} \times \ldots \times L_n^{\phantom{i}})_n^{\mathsf{EX}} = L_1^{\phantom{i}} \times \ldots
\times L_n^{\phantom{i}}$.  Таким образом, произведения логик являются частным случаем полупроизведений логик.

Нетрудно видеть, что если мы изменим порядок сомножителей в полупроизведении логик, имеющих выделенные индексы, то получим логику, отличающуюся, разве что, переименованием модальностей; то же произойдёт, если мы изменим порядок остальных сомножителей. Следовательно, если логика $L$ входит в полупроизведение и имеет в нём выделенный индекс, мы можем без ограничений общности считать, что она имеет индекс~$1$.

Ниже для конечного алфавита $\Sigma$ мы будем использовать следующие стандартные обозначения:
\begin{itemize}
\item $\Sigma^\ast$~--- множество всех \defnotion{слов} (т.е. конечных последовательностей) в~$\Sigma$;
\item $\emptystring$~--- \defnotion{пустое слово};
\item $\lambda \cdot \mu$~--- \defnotion{конкатенация} слов $\lambda$ и~$\mu$.
\end{itemize}

    \subsection{Сложность произведений модальных логик}
    \label{sec:fragments-products}

Покажем, что если один из сомножителей произведения логик равен $\logic{K}$,
$\logic{KT}$, $\logic{KB}$ или $\logic{KTB}$, то это произведение логик полиномиально погружается в свой фрагмент от одной переменной. В~качестве следствий мы получим результаты о неразрешимости и сложности фрагментов от одной переменной логик из интервалов, ограниченных с одной стороны некоторым произведением с сомножителем~$\logic{K}$, а с другой~--- произведением с сомножителем~$\logic{KTB}$.

Пусть $\vp$~--- $\lang{ML}_n$-формула, $\var \vp \subseteq \set{p_1,\ldots,p_m}$ и $p$~--- переменная, отличная от~$p_1, \ldots, p_m$.

Определим формулы, которые будем использовать в погружении (их смысл прояснит лемма~\ref{lem:gamma}; пока может быть полезен рис.~\ref{fig:Bk}:
формула $\alpha_k$ соответствует точке $a^{k}$ шкалы
$\kframe{A}_k$). Положим для каждого $k \in \{1, \ldots, m+1\}$
$$
\begin{array}{lcl}
  \alpha_k^{\phantom{i}} & = & \Box_1^{k}\, \neg p \con
                               \Diamond_1^{k+1}\, (p \con \neg \Diamond_1^{\phantom{i}} \Box_1^{+} p) \con
                               \Diamond_1^{k+2}\, \Box_1^+ p;
\smallskip\\
\beta_k^{\phantom{i}}  & = & \neg p \con \Diamond_1^{k+2}\, \alpha_{k}^{\phantom{i}} \con
                 \neg \Diamond_1^{k+1}\, \alpha_{k}^{\phantom{i}}.
\end{array}
$$
Формулы $\beta_1, \ldots, \beta_m$ будем использовать для моделирования переменных $p_1, \ldots, p_m$. Формула $\beta_{m+1}$ нужна для релятивизации; поэтому пусть
$$
\begin{array}{lclcl}
  B  & = & \beta_{m+1}^{\phantom{i}} & = &\displaystyle
                             \neg p \con \Diamond_1^{m+3}\, \alpha_{m+1}^{\phantom{i}} \con
                             \neg \Diamond_1^{m+2}\, \alpha_{m+1}^{\phantom{i}}.
\end{array}
$$

Определим рекурсивно перевод~$\tau$:
$$
\begin{array}{lcll}
  \tau(p_k) & = & \beta_k & \mbox{для $k \in \{1, \ldots, m \}$;} \\
  \tau(\bot) & = & \bot;  \\
  \tau(\psi \imp \chi) & = & \tau(\psi) \imp \tau(\chi);  \\
  \tau(\Box_i  \psi) & = & \Box_i (B \imp \tau(\psi)) & \mbox{для $i \in \{1, \ldots, n \}$}.
\end{array}
$$

Для непустой последовательности $\lambda = i_1 \cdot \ldots \cdot i_s$ символов из алфавита $\{1, \dots, n\}$ положим
$$
\begin{array}{lcl}
  \Box_\lambda \psi & = & \Box_{i_1} \ldots \Box_{i_s} \psi; \\
  \Diamond_\lambda \psi & = & \neg\Box_\lambda \neg\psi,
\end{array}
$$
а также
$$
\begin{array}{lclcl}
\Box_\emptystring \psi & = & \Diamond_\emptystring \psi & = & \psi.
\end{array}
$$

Для каждой $\theta \in \sub \vp$ определим множество
$\ind{\theta} \subseteq \{1, \dots, n\}^{\ast}$, состоящее из конечных последовательностей чисел, соответствующих конечным последовательностям переходов по отношениям достижимости, возникающим при оценке истинности формулы $\theta$ в точках моделей Крипке; определение рекурсивно:
$$
\begin{array}{lcll}
  \ind{q}
  & =
  & \{\emptystring\}
  & \mbox{для $q\in\var \vp$;}
  \\
  \ind{\bot}
  & =
  & \{\emptystring\};
  \\
  \indbr{\psi \con \chi}
  & =
  & \ind{\psi} \cup \ind{\chi};
  \\
  \indbr{\psi \dis \chi}
  & =
  & \ind{\psi} \cup \ind{\chi};
  \\
  \indbr{\psi \imp \chi}
  & =
  & \ind{\psi} \cup \ind{\chi};
  \\
  \ind{\Box_i \psi}
  & =
  & \{ \emptystring \}
    \cup \{ i \cdot \lambda :\,  \lambda \in \ind{\psi}\},
  & \mbox{где $i \in \{1, \ldots, n \}$}.
\end{array}
$$
Так, например,
$\ind{\Box_1 (\Box_2 q \imp \Box_3 \Box_2 \bot)} = \{ \emptystring,
\langle 1 \rangle, \langle 1, 2 \rangle, \langle 1, 3 \rangle, \langle
1, 3, 2 \rangle \}$.

Заметим, что $\ind{\theta} \ne \varnothing$ для любой формулы $\theta \in \sub  \vp$. Кроме того, множество $\ind{\vp}$ замкнуто по операции взятия префиксов последовательностей: для любых $\mu, \lambda \in \{1, \dots, n\}^{\ast}$
$$
\begin{array}{lcl}
  \mu \cdot \lambda \in \ind{\vp}
  & \imply
  & \mu \in \ind{\vp}.
\end{array}
\eqno{(4)}
$$
Отметим также, что число элементов во множестве $\ind{\vp}$ ограничено сверху длиной формулы~$\vp$.

Для каждой последовательности $\kappa \in \ind{\varphi}$ определим
$\widehat{\kappa} \in \{2, \dots, n\}^{\ast}$ как последовательность, полученную из~$\kappa$ удалением всех вхождений символа~$1$.  Определим модальности $\Box^\star$~и~$\Diamond^\star$:
$$
\begin{array}{lclclcl}
  \Box^\star\psi & = & {\displaystyle\bigwedge_{\mathclap{\kappa \in
                       \ind{\vp}}}\Box_{\widehat{\kappa}}\, \psi}; & &
  \Diamond^\star\psi & = & \neg \Box^\star \neg \psi.
  \\
\end{array}
$$
Теперь положим
$$
\begin{array}{lcl}
  A
  & =
  & \displaystyle
    B \con
    \Box_1^{\leqslant\md\varphi} (B \imp \Box^\star B)
    \con
    \Box_1^{\leqslant\md\varphi} (\Diamond^\star B
    \imp B).
\end{array}
$$
Пусть, наконец,
$$
\begin{array}{lcl}
e(\vp) & = & A \imp \tau(\vp).
\end{array}
$$

Поскольку число элементов в $\ind{\vp}$ ограничено длиной формулы~$\vp$, формула $e(\vp)$ вычисляется полиномиально по времени от длины~$\vp$.  Для нас будет важна следующая техническая лемма.

\begin{lemma}
  \label{lem:gamma}
  Пусть $L = L_1 \times L_2 \times \ldots \times L_n$, где
  $L_1 \in \{\logic{K}, \logic{KT}, \logic{KB}, \logic{KTB}\}$, а
  $L_2, \ldots, L_n$~--- полные по Крипке модальные логики. Тогда
  $$
  \begin{array}{lcl}
  \vp \in L & \iff & e(\vp) \in L.
  \end{array}
  $$
\end{lemma}

\begin{proof}
  $(\Leftarrow)$ Пусть $\vp \notin L$. Тогда
  $(\kframe{F},\pw{w}) \not\models^v \vp$ для некоторого произведения
  $\kframe{F} = \langle \bar{W}, \bar{R}_1, \ldots, \bar{R}_n \rangle$
  шкал $\kframe{F}_1, \ldots, \kframe{F}_n$, таких, что
  $\kframe{F}_i \models L_i$ для каждого
  $i \in \{1, \ldots, n\}$, некоторой точки $\pw{w} \in \bar{W}$ и некоторой оценки~$v$ в~$\kframe{F}$.  Мы построим $L$\nobreakdash-шкалу, опровергающую формулу~$A \imp \tau(\vp)$. Далее считаем, что $\kframe{F}_i = \langle W_i, R_i \rangle$  для каждого
  $i \in \{1, \ldots, n\}$.

  Для каждого $k \in \{1, \ldots, m + 1\}$ определим $1$\nobreakdash-шкалу $\frak{A}_{k} = \langle U_{k}, S_{k} \rangle$, положив
  $$
  \begin{array}{llcl}
    U_{k} & = & \{ a^k\} \cup \{ b^k_{i} : 1
                \leqslant i \leqslant k+1\} \cup \{ c^k_{i} : 1
                         \leqslant i \leqslant k+2\}
  \end{array}
  $$
  и взяв в качестве $S_{k}$ рефлексивно-симметричное замыкание отношения
  $$
  \begin{array}{l}
    \{ \langle \mathstrut b^{\mathstrut k}_i, b^{\mathstrut k}_{i+1}
    \rangle : 1 \leqslant i \leqslant k \}  \cup
    \{ \langle \mathstrut c^{\mathstrut k}_i, c^{\mathstrut k}_{i+1}
    \rangle : 1 \leqslant i \leqslant k+1 \} \cup
    \{ \langle \mathstrut a^{\mathstrut k}_{\phantom{i}}, b^{\mathstrut k}_{1}
    \rangle, \langle \mathstrut a^{\mathstrut k}_{\phantom{i}}, c^{\mathstrut k}_{1}
    \rangle \},
  \end{array}
  $$
см. рис.~\ref{fig:Bk}.

  \begin{figure}
  \centering
  \begin{tikzpicture}[scale=1.50]

    \coordinate (bkp1)    at ( 1, 1);   \node [below=2pt] at (bkp1)  {$b^k_{k+1}$} ;
    \coordinate (bk)      at ( 2, 1);   \node [below=2pt] at (bk)    {$b^k_k$}    ;
    \coordinate (b1)      at ( 4, 1);   \node [below=2pt] at (b1)    {$b^k_1$}    ;
    \coordinate (ak)      at ( 5, 1);   \node [below=2pt] at (ak)    {$a^k_{\phantom{k}}$}      ;
    \coordinate (c1)      at ( 6, 1);   \node [below=2pt] at (c1)    {$c^k_1$}    ;
    \coordinate (ck)      at ( 8, 1);   \node [below=2pt] at (ck)    {$c^k_k$}    ;
    \coordinate (ckp1)    at ( 9, 1);   \node [below=2pt] at (ckp1)   {$c^k_{k+1}$}    ;
    \coordinate (ckp2)    at (10, 1);   \node [below=2pt] at (ckp2)   {$c^k_{k+2}$}    ;

    \draw [] (bkp1)     circle [radius=2.0pt] ;
    \draw [] (bk)       circle [radius=2.0pt] ;
    \draw [] (b1)       circle [radius=2.0pt] ;
    \draw [] (ak)       circle [radius=2.0pt] ;
    \draw [] (c1)       circle [radius=2.0pt] ;
    \draw [] (ckp1)     circle [radius=2.0pt] ;
    \draw [] (ck)       circle [radius=2.0pt] ;
    \draw [] (c1)       circle [radius=2.0pt] ;
    \draw [] (ckp2)     circle [radius=2.0pt] ;

    \begin{scope}[>=latex]

      \draw [<->, shorten >= 2.75pt, shorten <= 2.75pt] (bkp1)   -- (bk) ;
      \draw [<->, shorten >= 2.75pt, shorten <= 2.75pt] (b1) -- (ak);
      \draw [<->, shorten >= 2.75pt, shorten <= 2.75pt] (ak)   -- (c1);
      \draw [<->, shorten >= 2.75pt, shorten <= 2.75pt] (ckp1)   -- (ck) ;
      \draw [<->, shorten >= 2.75pt, shorten <= 2.75pt] (ckp1) -- (ckp2);
      \draw [<->, shorten >= 2.75pt, shorten <= 2.75pt, dashed] (b1)   -- (bk) ;
      \draw [<->, shorten >= 2.75pt, shorten <= 2.75pt, dashed] (c1) -- (ck);

    \end{scope}

      \draw (1.75,1.15) -- (1.75, 1.25) -- (8.25, 1.25) -- (8.25, 1.15)      ;

      \node [above=10pt] at (ak)  {$\neg p$};

      \node [above=10pt] at (bkp1)   {$p$};
      \node [above=10pt] at (ckp1)   {$p$};
      \node [above=10pt] at (ckp2)   {$p$};

    \end{tikzpicture}

    \caption{Шкала~$\kframe{A}_k$ с проекцией оценки~$v'$}
    \label{fig:Bk}
  \end{figure}

  Для каждого $x \in W_1^{\phantom{i}}$ определим шкалу
  $\kframe{A}_{k}^x = \langle U_{k}^{\phantom{i}} \times \{x\}, S_{k}^x
  \rangle$ как изоморфную копию шкалы $\kframe{A}_{k}^{\phantom{i}}$ при отображении $f\colon u \mapsto \langle u, x \rangle$.  Пусть
  $$
  \begin{array}{lcl}
    W'_1
    & =
    & \displaystyle
      W_{1}^{\phantom{i}} \cup \bigcup\limits_{{k = 1}}^{{m+1}}
      (U_{k} \times W_{1}^{\phantom{i}}),
   \medskip\\
    R'_{1} & =
    & \displaystyle
      R_{1}^{\phantom{i}} \cup \bigcup\limits_{{x \in W_1}} \bigcup\limits_{\mathclap{k
      = 1}}^{\mathclap{m+1}} \big(S_{k}^x \cup \{\langle x,\langle b^k_{k+1}, x\rangle
      \rangle, \langle \langle b^k_{k+1}, x\rangle, x
      \rangle \}\big)\,  \\
  \end{array}
  $$
и $\kframe{F}'_1 = \langle W'_1, R'_1 \rangle$.  Таким образом, для любых $x \in W_1^{\phantom{i}}$ и $k \in \{1, \ldots, m+1\}$ корень
$\langle b^k_{k+1}, x \rangle$ шкалы $\kframe{A}^x_k$ является
$R'_1$-достижимым из~$x$ в~$\kframe{F}'_1$.  Наконец, положим\footnote{\label{footnote:product}Заметим, что в произведении шкал каждое отношение достижимости зависит не только от того, каким было отношение достижимости в соответствующем сомножителе, но также и от множества точек в получившейся шкале. Отношения достижимости $\bar{R}'_2, \ldots, \bar{R}'_n$ в шкале $\kframe{F}'$ отличаются от отношений достижимости $\bar{R}_2, \ldots, \bar{R}_n$ в шкале $\kframe{F}$, поскольку $\kframe{F}'$ и $\kframe{F}$ имеют разные множества точек.}
$$
 \begin{array}{lcl}
   \kframe{F}' & = & \kframe{F}'_1 \times \kframe{F}_2^{\phantom{i}} \times \ldots
   \times \kframe{F}_n^{\phantom{i}} = \langle \bar{W}', \bar{R}'_1,
   \ldots, \bar{R}'_n \rangle.
 \end{array}
$$
  В результате получаем, что для любых $\pw{x} \in \bar{W}$ и
  $k \in \{1, \ldots, m+1\}$ корень
  $\langle \langle b^k_{k+1}, x_1^{\phantom{i}} \rangle,
  x_2^{\phantom{i}}, \ldots, x_n^{\phantom{i}} \rangle$ изоморфной копии шкалы $\kframe{A}_{k}$ является
  $\bar{R}'_1$-достижимым из точки $\pw{x}$ в~$\kframe{F}'$.

  Заметим, что $\kframe{F}\subseteq\kframe{F}'$.  Кроме того, $\kframe{F}'_1 \models L_1^{\phantom{i}}$, а значит,
  $\kframe{F}' \models L$.

  Пусть $v'$~--- оценка в $\kframe{F}'$, такая, что $\bar{x} \in v'(p)$ в точности тогда, когда выполнено одно из следующих условий:
  \begin{itemize}
  \item
    существуют такие $k \in \{1, \ldots, m\}$, $\pw{z} \in \bar{W}$ и $u \in \{ b_{k+1}^{k}, c_{k+1}^{k}, c_{k+2}^{k} \}$, для которых
    $(\kframe{F},\pw{z}) \models^{v} p_k^{\phantom{i}}$
    и
    $\bar{x} = \langle \langle u,
    z_1^{\phantom{i}} \rangle, z_2^{\phantom{i}}, \ldots,
    z_n^{\phantom{i}} \rangle$;
  \item
    существуют такие $\pw{z} \in \bar{W}$ и $u \in \{ b_{m+2}^{m+1},
    c_{m+2}^{m+1}, c_{m+3}^{m+1} \}$, для которых
    $\bar{x} = \langle \langle u, z_1^{\phantom{i}} \rangle, z_2^{\phantom{i}}, \ldots, z_n^{\phantom{i}} \rangle$.
  \end{itemize}

  Проекция оценки $v'$ на изоморфную копию шкалы $\kframe{A}_{k}$, где $k$ таково, что либо
  $k = m + 1$, либо $1 \leqslant k \leqslant m$ и
  $(\kframe{F},\pw{z}) \models^{v} p_k$ для некоторого $\pw{z} \in \bar{W}$, из которого достижима эта изоморфная копия, показана на рис.~\ref{fig:Bk}; если же $1 \leqslant k \leqslant m$ и
  $(\kframe{F},\pw{z}) \not\models^{v} p_k$, то проекция оценки $v'$ на $\kframe{A}_{k}$, достижимую из $\pw{z}$, такова, что
  $p$ ложна в каждой точке шкалы~$\kframe{A}_{k}$.


  \begin{sublemma}
    \label{sublem:alpha_k}
    Пусть $\pw{x} \in \pw{W}'$ и $k \in \{1, \ldots, m\}$. Тогда
    \settowidth{\templength}{\mbox{$(\kframe{F}', \pw{x}) \models^{v'} \alpha_{m+1}$}}
    \settowidth{\templengtha}{\mbox{$\exists \pw{z}\in\bar{W}~ \big( (\kframe{F},\pw{z}) \models^{v} p_k ~\&~ \bar{x} = \langle \langle a^k, z_1 \rangle,
        z_2,\ldots,z_n\rangle \big).$}}
    $$
    \begin{array}{lcl}
      \parbox{\templength}{$(\kframe{F}',\pw{x}) \models^{v'} \alpha_k$}
      & \iff
      &
        \parbox{\templengtha}{$\exists \pw{z}\in\bar{W}~ \big( (\kframe{F},\pw{z}) \models^{v} p_k ~\&~ \bar{x} = \langle \langle a^k, z_1 \rangle,
        z_2,\ldots,z_n\rangle \big)$;}
        \\
      \parbox{\templength}{$(\kframe{F}',\pw{x}) \models^{v'} \alpha_{m+1}$}
      & \iff
      &
        \parbox{\templengtha}{$\exists \pw{z}\in\bar{W}
        ~ \bar{x} = \langle \langle a^{m+1}, z_1 \rangle,
        z_2,\ldots,z_n\rangle$.}
    \end{array}
    $$
  \end{sublemma}

  \begin{proof}
    Из определения оценки $v'$ получаем, что для любого
    $\pw{y} \in \pw{W}'$ отношение
    $(\kframe{F}',\pw{y}) \models^{v'} \Box_1^+ p$ справедливо тогда и только тогда, когда выполнено одно из следующих условий:
    $$
    \begin{array}{ll}
      \bullet & \exists i \in \{1, \ldots, m\}~ \exists \pw{z} \in \bar{W}~
                \big( (\kframe{F},\pw{z}) \models^{v} p_i^{\phantom{i}} ~\&~ \bar{y} = \langle
                \langle c^{i}_{i+2}, z_1^{\phantom{i}} \rangle, z_2^{\phantom{i}}, \ldots, z_n^{\phantom{i}}
                \rangle \big); \medskip\\
      \bullet &  \exists \pw{z} \in \bar{W}
                ~ \bar{y} = \langle \langle c^{m+1}_{m+3}, z_1^{\phantom{i}}
                \rangle, z_2^{\phantom{i}}, \ldots, z_n^{\phantom{i}} \rangle.
    \end{array}
    \eqno (5)
    $$

    Из определения оценки $v'$ и $(5)$ следует, что для любого
    $\pw{y} \in \pw{W}'$ отношение
    $(\kframe{F}',\pw{x}) \models^{v'} p \con \neg \Diamond_1 \Box^{+} p$
    справедливо тогда и только тогда, когда выполнено одно из следующих условий:
    $$
    \begin{array}{ll}
      \bullet & \exists i \in \{1, \ldots, m\}~ \exists \pw{z} \in \bar{W}~
                \big( (\kframe{F},\pw{z}) \models^{v} p_i^{\phantom{i}} ~\&~ \bar{y} = \langle
                \langle b^{i}_{i+1}, z_1^{\phantom{i}} \rangle, z_2^{\phantom{i}}, \ldots, z_n^{\phantom{i}}
                \rangle \big); \medskip\\
      \bullet & \exists \pw{z} \in \bar{W}
                ~ \bar{y} = \langle \langle b^{m+1}_{m+2}, z_1^{\phantom{i}}
                \rangle, z_2^{\phantom{i}}, \ldots, z_n^{\phantom{i}} \rangle.
    \end{array}
    \eqno (6)
    $$

    Пусть теперь $\pw{x} \in \bar{W}'$ и $k \in \{1, \ldots, m+1\}$.
    Пусть $(\kframe{F}',\pw{x}) \models^{v'} \alpha_k$.  Тогда $\pw{x}$
    видит за $k+2$ шага точку, удовлетворяющую условиям ($5$), и за $k+1$ шаг точку, удовлетворяющую условиям ($6$), но не видит за $k$ шагов точку, в которой истинна переменная~$p$. Это возможно, только если
    $\bar{x} = \langle \langle a^k, z_1 \rangle,
    z_2,\ldots,z_n\rangle$ для некоторого $\pw{z} \in \pw{W}$ (заметим, что если $\bar{x} \in \bar{W}$, то $x$ видит за одни шаг, а значит, и за $k$ шагов, рефлексивную точку
    $\langle \langle b^{m+1}_{m+2}, x_1^{\phantom{i}} \rangle,
    x_2^{\phantom{i}}, \ldots, x_n^{\phantom{i}} \rangle$, в которой истинна переменная~$p$). Справедливость обратной импликации следует из определения оценки~$v'$.
  \end{proof}

  \begin{sublemma}
    \label{sublem:B}
    Если $\pw{x} \in \pw{W}'$, то
    $$
    \begin{array}{lcl}
      (\kframe{F}',\pw{x})^{\phantom{i}} \models^{v'} B & \iff & \pw{x} \in \pw{W}.
    \end{array}
    $$
  \end{sublemma}

  \begin{proof}
    Пусть $\pw{x} \in \pw{W}'$. Пусть
    $(\kframe{F}',\pw{x}) \models^{v'} B$. Тогда, учитывая, что
    $(\kframe{F}',\pw{x}) \models^{v'} \Diamond_1^{m+3}\,
    \alpha_{m+1} \con \neg \Diamond_1^{m+2}\, \alpha_{m+1}$,
    из подлеммы~\ref{sublem:alpha_k} получаем, что $\pw{x} \in \pw{W}$ или
    $\pw{x} = \langle \langle c^{m+1}_{m+3}, z_1^{\phantom{i}} \rangle, z_2^{\phantom{i}}, \ldots,
    z_n^{\phantom{i}} \rangle$ для некоторого $\pw{z} \in \pw{W}$. Поскольку
    $(\kframe{F}',\pw{x}) \not\models^{v'} p$, получаем, что $\pw{x} \in \pw{W}$.

    Обратная импликация следует из определения оценки~$v'$.
  \end{proof}

  \begin{sublemma}
    \label{sublem:gamma_k}
    Если $\pw{x} \in \pw{W}$ и $k \in \{1, \ldots, m\}$, то
    $$
    \begin{array}{lcl}
    (\kframe{F}',\pw{x})^{\phantom{i}} \models^{v'} \beta_k & \iff &
    (\kframe{F}, \pw{x})^{\phantom{i}} \models^{v}p_k.
    \end{array}
    $$
  \end{sublemma}

  \begin{proof}
    Аналогично доказательству подлеммы~\ref{sublem:B}.
  \end{proof}

Теперь покажем, что для каждой формулы $\theta \in \sub \vp$ и каждой точки~$\pw{x} \in \bar{W}$
  $$
  \begin{array}{lcl}
  (\kframe{F},\pw{x}) \models^v \theta
  & \iff &
  (\kframe{F}',\pw{x}) \models^{v'} \tau(\theta).
  \end{array}
  \eqno (7)
  $$

  Доказательство проведём индукцией по построению формулы~$\theta$.

  Если $\theta = {\bot}$, то $\tau(\theta) = {\bot}$, и доказываемое утверждение очевидно.

  Если $\theta = p_k$ для некоторого $k \in \{1, \ldots, m\}$, то $\tau(\theta) = \beta_k$, и доказываемое утверждение следует из подлеммы~\ref{sublem:gamma_k}.

  Если $\theta = \psi \con \chi$, $\theta = \psi \dis \chi$ или $\theta = \psi \imp \chi$, то обоснование тривиально, и мы его опускаем.

  Пусть $\theta = \Box_i \psi$; тогда $\tau(\theta) = \Box_i (B \imp \tau(\psi))$. Пусть
  $\pw{x}\in \bar W$.

  Предположим, что $(\kframe{F},\pw{x}) \not\models^v \Box_i \psi$. Тогда
  $(\kframe{F},\pw{y}) \not\models^v \psi$ для некоторой точки
  $\pw{y} \in \bar{R}_i (\pw{x}) \subseteq \bar{W}$.  Поскольку
  $\pw{y} \in \bar{W}$, согласно подлемме~\ref{sublem:B} получаем, что
  $(\kframe{F}',\pw{y}) \models^{v'} B$, а значит, по предположению индукции,
  $(\kframe{F}',\pw{y}) \not\models^{v'} \tau(\psi)$. Поскольку
  $\kframe{F} \subseteq \kframe{F}'$, получаем, что
  $\pw{y} \in \bar{R}'_i(\pw{x})$. Следовательно,
  $(\kframe{F}',\pw{x}) \not\models^{v'} \Box_i (B \imp \tau(\psi))$.

  Теперь предположим, что
  $(\kframe{F}',\pw{x}) \not\models^{v'} \Box_i (B \imp \tau(\psi))$.
  Тогда $(\kframe{F}',\pw{y}) \models^{v'} B$ и
  $(\kframe{F}',\pw{y}) \not\models^{v'} \tau(\psi)$ для некоторой точки
  $\pw{y} \in \bar{R}'_i (\pw{x})$.  Согласно подлемме~\ref{sublem:B},
  $\pw{y} \in \bar{W}$. Значит, по предположению индукции,
  $(\kframe{F},\pw{y}) \not\models^v \psi$.  Из определения $\bar{R}'_i$
  следует, что $\bar{R}'_i \upharpoonright \bar{W} = \bar{R}_i$.
  Поскольку $\pw{y} \in \bar{R}'_i(\pw{x})$ и $\pw{x} \in \bar{W}$, получаем, что
  $\pw{y} \in \bar{R}_i(\pw{x})$.  Следовательно,
  $(\kframe{F},\pw{x}) \not\models^v \Box_i \psi$.

  Обоснование индукционного шага завершено, и эквивалентность~$(7)$ доказана. Поскольку   $(\kframe{F},\pw{w}) \not\models^{v} \vp$ и $\pw{w} \in \bar{W}$, из $(7)$ следует, что  $(\kframe{F}',\pw{w}) \not\models^{v'} \tau(\vp)$.

  Теперь покажем, что $(\kframe{F}',\pw{w}) \models^{v'} A$.  Во-первых, поскольку
  $\pw{w} \in \bar{W}$, согласно подлемме~\ref{sublem:B}, получаем, что
  $(\kframe{F}', \pw{w}) \models^{v'} B$.  Во-вторых, по определению шкалы
  $\kframe{F}'$, для любых $\bar{x}, \bar{y} \in \bar{W}'$
  $$
  \begin{array}{lcl}
    \bar{x}\, (\bar{R}'_2 \cup \ldots \cup \bar{R}'_n)^\ast\, \bar{y} ~\&~ \bar{x} \in \bar{W}
    & \Longrightarrow
    & \bar{y} \in \bar{W}.
  \end{array}
  \eqno(8)
  $$
  Согласно определению шкалы $\kframe{F}'$, единственным отношением достижимости, связывающим точки множества $\bar{W}$ с точками множества $\bar{W}' \setminus \bar{W}$, является~$\bar{R}'_1$.  Тогда, согласно подлемме~\ref{sublem:B} и $(8)$, для любой точки~$\bar{x} \in \bar{W}'$
  $$
  \begin{array}{lcl}
  (\kframe{F}',\pw{x}) \models^{v'} B \imp \Box^\star B
    & \mbox{и}
    & (\kframe{F}',\pw{x}) \models^{v'} \Diamond^\star B \imp B.
  \end{array}
  $$
  Следовательно, $(\kframe{F}',\pw{w}) \models^{v'} A$.

  Таким образом, $(\kframe{F}',\pw{w}) \not\models^{v'} A \imp \tau(\vp)$.  Поскольку, как мы видели, $\kframe{F}' \models L$, получаем, что $A \imp \tau(\vp) \notin L$.

  $(\Rightarrow)$ Пусть $A \imp \tau(\vp) \notin L$.  Тогда
  $(\kframe{F},\pw{w}) \not\models^v A \imp \tau(\vp)$ для некоторого произведения
  $\kframe{F} = \langle \bar{W}, \bar{R}_1, \ldots, \bar{R}_n \rangle$
  шкал $\kframe{F}_1, \ldots, \kframe{F}_n$, таких, что
  $\kframe{F}_i = \langle W_i, R_i \rangle \models L_i$ для каждого
  $i \in \{1, \ldots, n\}$, некоторой точки $\pw{w} \in \bar{W}$ некоторой оценки~$v$ в~$\kframe{F}$.  Мы построим $L$\nobreakdash-шкалу, опровергающую формулу~$\vp$.

  Для непустой последовательности $\lambda = i_1 \cdot \ldots \cdot i_s$ символов из алфавита $\{1, \dots, n\}$ положим
  $$
  \begin{array}{lcl}
    \bar{R}_{\lambda} & = & \bar{R}_{i_1} \circ \ldots \circ \bar{R}_{i_s},
  \end{array}
  $$
  а также положим
  $$
  \begin{array}{lcl}
    \bar{R}_{\emptystring} & = & \{ \langle \pw{x}, \pw{x} \rangle :\, \pw{x} \in \pw{W} \}.
  \end{array}
  $$
  Для каждой точки $\pw{x}\in\bar{W}$ определим множество~$\ind{\pw{x}}$:
  $$
  \begin{array}{lcl}
  \ind{\pw{x}}
    & =
    & \left\{\lambda \in \{1, \ldots, n\}^{\ast} :
      \exists \mu \in \{1, \ldots, n\}^{\ast} ~\big(\mu
      \cdot \lambda \in \ind{\varphi} ~\&~
      \pw{w}\bar{R}_{\mu}\pw{x}\big) \right\}.
  \end{array}
  $$
  Пусть
  $$
  \begin{array}{lcl}
    W'_1
    & =
    & \{ x_1^{\phantom{i}} \in W_1^{\phantom{i}} :\,
      \mbox{$(\kframe{F},\pw{x}) \models^v B$ и $\ind{\pw{x}}\ne \varnothing$} \}.
  \end{array}
  $$

  Поскольку $(\kframe{F},\pw{w}) \models A$, получаем, что $(\kframe{F},\pw{w}) \models B$. Нетрудно видеть, что $\emptystring \in \ind{\pw{w}}$, т.е. $\ind{\pw{w}} \ne \varnothing$, а значит,
  $w_1^{\phantom{i}} \in W'_1$, и $W'_1 \ne \varnothing$.
  Пусть\footnote{См.~сноску~\ref{footnote:product}.}
  \settowidth{\templength}{\mbox{$R'_1$}}
  \begin{itemize}
  \item ${\parbox{\templength}{$R'_1$}} = R_1^{\phantom{i}} \upharpoonright W'_1$;
  \item ${\parbox{\templength}{$\kframe{F}'_1$}} = \langle W'_1, R'_1 \rangle$;
  \item
    ${\parbox{\templength}{$\kframe{F}'$}} = \kframe{F}'_1 \times \kframe{F}_2^{\phantom{i}} \times
    \ldots \times \kframe{F}_n^{\phantom{i}} = \langle \bar{W}',
    \bar{R}'_1, \ldots, \bar{R}'_n \rangle$.
  \end{itemize}

  Тогда $\kframe{F}'_1 \subseteq \kframe{F}_1^{\phantom{i}}$. Кроме того, $L_1$ является сабфреймовой логикой, а значит, $\kframe{F}'_1 \models L_1^{\phantom{i}}$, откуда следует, что  $\kframe{F}' \models L$.  Заметим также, что $\kframe{F}' \subseteq \kframe{F}$.

  \begin{sublemma}
    \label{sublem:D}
    Если\/ $\pw{y}\in \bar{W}'$ и\/ $\ind{\pw{y}}\ne \varnothing$, то
    $(\kframe{F},\pw{y})\models^{v} B$.
  \end{sublemma}

  \begin{proof}
    Поскольку $\pw{y}\in \bar{W}'$, получаем, что существует точка $\pw{x}\in \bar{W}$, такая, что
    \begin{itemize}
    \item $(\kframe{F},\pw{x}) \models^v B$;
    \item $\ind{\pw{x}}\ne \varnothing$;
    \item $x_1 = y_1$.
    \end{itemize}

    Поскольку $\ind{\pw{x}}\ne \varnothing$, существуют $\mu, \lambda \in \{1, \ldots, n\}^{\ast}$, такие, что $\mu \cdot \lambda \in \ind{\varphi}$ и $\pw{w}\bar{R}_\mu\pw{x}$.

    Пусть $\mu' \in \{1, \ldots, n\}^{\ast}$ получена из $\mu$ сдвигом всех вхождений символа~$1$ влево, т.е. в самое начало последовательности.  Иначе говоря, $\mu' = 1^k \cdot \nu$ для некоторых $k \leqslant \md \varphi$ и $\nu \in \{2, \ldots, n\}^{\ast}$.


    Поскольку $\mu \cdot \lambda \in \ind{\varphi}$, используя $(4)$, получаем, что $\mu \in \ind{\varphi}$. Кроме того, $\nu = \widehat{\mu}$. Поскольку $(\kframe{F},\pw{w}) \models^v A$, получаем, что $(\kframe{F},\pw{w}) \models^v \Box_1^{\leqslant\md\varphi} (\Diamond_\nu B \imp B)$.  Поскольку $\kframe{F}$ является произведением шкал, получаем, согласно $(1)$\footnote{\label{footnote:about(3)insteadof(1)}Заметим, что тот же вывод можно сделать, применяя $(3)$ вместо~$(1)$; мы воспользуемся этим в разделе~\ref{sec:fragments-semiproducts}.}, что $\pw{w}\bar{R}_{\mu'}\pw{x}$.

    Пусть $\pw{z}$~--- точка из $\bar{W}$, такая, что
    \begin{itemize}
    \item $z_1 = x_1$ (и значит, $z_1 = y_1$);
    \item $z_s = w_s$ для каждого $s\in\{2,\ldots,n\}$.
    \end{itemize}

    Для такой точки $\pw{z}$ условие $\pw{w}\bar{R}_{\mu'}\pw{x}$ означает, что    $\pw{w}\bar{R}_1^k \pw{z} \bar{R}_{\nu}\pw{x}$.

    Поскольку $(\kframe{F},\pw{w}) \models^v \Box_1^{\leqslant\md\varphi} (\Diamond_\nu B \imp B)$ и $k \leqslant \md \varphi$, получаем, что $(\kframe{F},\pw{z}) \models^v \Diamond_\nu B \imp B$.  Поскольку $\pw{z} \bar{R}_{\nu}\pw{x}$ и $(\kframe{F},\pw{x}) \models^v B$, получаем, что $(\kframe{F},\pw{z}) \models^v B$.

    Если $\ind{\pw{y}}\ne \varnothing$ для некоторых $\zeta, \delta \in \{1, \ldots, n\}^{\ast}$, то $\zeta \cdot \delta \in \ind{\varphi}$ и $\pw{w}\bar{R}_\zeta\pw{y}$.

    Пусть $\zeta' \in \{1, \ldots, n\}^{\ast}$ получена из $\zeta$ сдвигом всех вхождений символа~$1$ влево, т.е. в самое начало последовательности. Иначе говоря, $\zeta' = 1^l \cdot \eta$ для некоторых $l \leqslant \md \vp$ и~$\eta \in \{2, \ldots, n\}^{\ast}$.

    Поскольку $\zeta \cdot \delta \in \ind{\varphi}$, получаем, согласно $(4)$, что $\zeta \in \ind{\varphi}$.  Кроме того, $\eta = \widehat{\zeta}$. Поскольку $(\kframe{F},\pw{w}) \models^v A$, получаем, что $(\kframe{F},\pw{w}) \models^v \Box_1^{\leqslant\md\varphi} (B \imp \Box_\eta B)$.

    Поскольку $\kframe{F}$ является произведением шкал, получаем, согласно~($1$)\footnote{\label{footnote:about(3)insteadof(1)2}См.~сноску~\ref{footnote:about(3)insteadof(1)}.}, что $\pw{w}\bar{R}_{\zeta'}\pw{y}$, т.е.
    $\pw{w}\bar{R}_1^l \pw{v} \bar{R}_{\eta}\pw{y}$ для некоторого $\bar{v} \in \bar{W}$. Из того, что $x_1 = y_1 = z_1$, следует, что  $\bar{v} = \bar{z}$; значит, $\pw{z} \bar{R}_{\eta}\pw{y}$.

    Поскольку $(\kframe{F},\pw{w}) \models^v \Box_1^{\leqslant\md\varphi} (B \imp \Box_\eta B)$ и $l \leqslant \md \varphi$, получаем, что $(\kframe{F},\pw{z}) \models^v B \imp \Box_\eta B$. Как мы видели, $(\kframe{F},\pw{z}) \models^v B$ и $\pw{z} \bar{R}_{\eta}\pw{y}$.
    Следовательно, $(\kframe{F},\pw{y}) \models^v B$.
  \end{proof}

  Пусть $v'$~--- оценка в $\kframe{F}'$, такая, что для каждого $k \in \{1, \ldots, m\}$
  $$
  \begin{array}{lcl}
  v'(p_k) & = & \{\pw{x}\in\bar{W}' :
  (\kframe{F},\pw{x})\models^{v} \beta_k\}.
  \end{array}
  $$

  Покажем, что для любой формулы $\theta \in \sub  \varphi$ и любой точки $\pw{x}\in\bar{W}'$, такой, что $\ind{\theta}\subseteq \ind{\pw{x}}$,
  $$
  \begin{array}{lcl}
  (\kframe{F}',\pw{x})\models^{v'} \theta
    & \iff
    & (\kframe{F},\pw{x})\models^{v}\tau(\theta).
  \end{array}
  \eqno{(9)}
  $$

  Обоснование проведём индукцией по построению формулы~$\theta$.

  Если $\theta = {\bot}$, то $\tau(\theta) = {\bot}$, и доказываемое утверждение очевидно.

  Если $\theta = p_k$, то $\tau(\theta) = \beta_k$, и тогда справедливость доказываемого утверждения следует из определения оценки~$v'$.

  Пусть $\theta = \psi \imp \chi$. Тогда $\tau(\theta) = \tau(\psi) \imp \tau(\chi)$. Пусть   $\bar{x} \in \bar{W}'$ и $\ind{\theta}\subseteq \ind{\pw{x}}$.
  Поскольку $\ind{\theta} = \ind{\psi} \cup \ind{\chi}$, получаем, что $\ind{\psi}\subseteq \ind{\pw{x}}$ и $\ind{\chi}\subseteq \ind{\pw{x}}$; но тогда справедливость доказываемого утверждения следует из индукционного предположения.

  Если $\theta = \psi \con \chi$ или $\theta = \psi \dis \chi$, то нужно рассуждать аналогично.

  Пусть $\theta = \Box_i \psi$. Тогда $\tau(\theta) = \Box_i(B \to\tau(\psi))$. Пусть   $\pw{x}\in\bar{W}'$ и $\ind{\theta}\subseteq \ind{\pw{x}}$.

  Предположим, что $(\kframe{F}',\pw{x})\not\models^{v'} \Box_i \psi$. Тогда   $(\kframe{F}',\pw{y})\not\models^{v'}\psi$ для некоторой точки $\bar{y}\in \bar{R}'_i(\bar{x})$.

  Поскольку $\kframe{F}' \subseteq \kframe{F}$, получаем, что $\bar{y}\in \bar{R}_i(\bar{x})$.  Покажем, что $\ind{\psi} \subseteq \ind{\pw{y}}$ (и~значит,~$\ind{\pw{y}} \ne \varnothing$).

  Пусть $\lambda \in \ind{\psi}$. Тогда $i \cdot \lambda \in \ind{\theta}$. По условию,
  $\ind{\theta}\subseteq \ind{\pw{x}}$; следовательно,
  $i \cdot \lambda \in \ind{\bar{x}}$. Таким образом, $\mu \cdot i \cdot \lambda \in \ind{\vp}$ и   $\pw{w} \bar{R}_{\mu} \bar{x}$ для некоторого $\mu \in \{1, \ldots, n\}^{\ast}$. Поскольку   $\bar{y}\in \bar{R}_i(\bar{x})$, получаем, что $\pw{w} \bar{R}_{\mu \cdot i} \bar{y}$, а это вместе с условием $\mu \cdot i \cdot \lambda \in \ind{\vp}$ влечёт, что $\lambda \in \ind{\pw{y}}$. Значит, $\ind{\psi} \subseteq \ind{\pw{y}}$.

  Применяя индукционное предположение, получаем, что $(\kframe{F},\pw{y})\not\models^{v} \tau(\psi)$.  Поскольку $\bar{y} \in \bar{W}'$ и, как мы видели, $\ind{\pw{y}} \ne \varnothing$, по подлемме~\ref{sublem:D} получаем, что $(\kframe{F},\pw{y})\models^{v} B$. Следовательно,   $(\kframe{F},\pw{x})\not\models^{v} \Box_i (B \imp \tau(\psi))$.

  Теперь предположим, что $(\kframe{F},\pw{x})\not\models^{v} \Box_i (B \imp \tau(\psi))$. Тогда $(\kframe{F},\pw{y}) \models^{v} B$ и $(\kframe{F},\pw{y})\not\models^{v}\tau(\psi)$ для некоторой точки $\bar{y}\in \bar{R}_i(\bar{x})$. Нетрудно видеть, что $\ind{\psi} \subseteq \ind{\pw{y}}$, а значит, $\ind{\pw{y}} \ne \varnothing$. Тогда $y_1^{\phantom{i}} \in W'_1$ согласно определению множества~$W'_1$; значит, $\bar{y} \in \bar{W}'$. Следовательно, по индукционному предположению, $(\kframe{F}',\pw{y})\not\models^{v'} \psi$. Поскольку по условию $\bar{x}\in \bar{W}'$, получаем, что $\bar{y}\in \bar{R}'_i(\bar{x})$. Значит, $(\kframe{F}',\pw{x})\not\models^{v'} \Box_i \psi$.

  Эквивалентность~$(9)$ доказана.

  Поскольку $\pw{w} \in \bar{W}'$, с использованием $(9)$ получаем, что $(\kframe{F}', \pw{w}) \not\models^{v'} \vp$.  Как мы видели, $\kframe{F}' \models L$; значит, $\vp \notin L$.
\end{proof}

\begin{theorem}
  \label{thr:products-KTB}
  Пусть $L = L_1 \times L_2 \times \ldots \times L_n$, где
  $L_1 \in \{\logic{K}, \logic{KT}, \logic{KB}, \logic{KTB}\}$ и
  $L_2, \ldots, L_n$~--- полные по Крипке мономодальные логики. Тогда $L$ полиномиально погружается в свой фрагмент от одной переменной.
\end{theorem}

\begin{proof}
  По лемме~\ref{lem:gamma}, $\vp \in L$ тогда и только тогда, когда $e(\vp) \in L$. Осталось заметить, что $e(\vp)$ вычисляется по $\vp$ полиномиально по времени.
\end{proof}


\begin{corollary}
  \label{cor:HHK}
  Пусть $L\in [\logic{K} \times \logic{K} \times \logic{K}, \logic{KTB} \times \logic{S5} \times \logic{S5}]$. Тогда фрагмент логики $L$ в языке с одной переменной неразрешим.
\end{corollary}

\begin{proof}
  Покажем, что можно свести к $L$ неразрешимую проблему представимости конечных простых реляционных алгебр~\cite{HH01}.

  Существует алгоритм, который по каждой конечной простой реляционной алгебре~$\kframe{A}$ строит~\cite[леммы~5~и~6]{HHK02} $\lang{ML}_3$-формулу~$\vp_{\kframe{A}}$, такую, что
  \settowidth{\templength}{\mbox{$a$}}
  \begin{itemize}
  \item[$(a)$] $\neg \vp_{\kframe{A}} \notin \logic{S5} \times \logic{S5} \times \logic{S5}$, если $\kframe{A}$ представима;
  \item[$(\hspace{0.25pt}b\hspace{0.25pt})$]
    $\neg \vp_{\kframe{A}} \in \logic{K} \times \logic{K} \times \logic{K}$, если
    $\kframe{A}$ непредставима.
  \end{itemize}

  Пусть $\kframe{A}$~--- конечная простая реляционная алгебра. Предположим, что~$\kframe{A}$ представима. Тогда, согласно~$(a)$, $\neg \vp_{\kframe{A}} \notin \logic{S5} \times \logic{S5} \times \logic{S5}$. Следовательно, $\neg \vp_{\kframe{A}} \notin \logic{KTB} \times \logic{S5} \times \logic{S5}$, а значит, по лемме~\ref{lem:gamma}, $e(\neg \vp_{\kframe{A}}) \notin \logic{KTB} \times \logic{S5} \times \logic{S5}$, т.е. $e(\neg \vp_{\kframe{A}}) \notin L$.
  Теперь предположим, что $\kframe{A}$ непредставима. Тогда, согласно~$(b)$, $\neg \vp_{\kframe{A}} \in \logic{K} \times \logic{K} \times \logic{K}$, а значит, по лемме~\ref{lem:gamma},
  $e(\neg \vp_{\kframe{A}}) \in \logic{K} \times \logic{K} \times
  \logic{K}$, т.е. $e(\neg \vp_{\kframe{A}}) \in L$.
\end{proof}

\begin{corollary}
  \label{cor:marx}
  Пусть $L\in [\logic{K} \times \logic{K}, \logic{KTB} \times \logic{S5}]$.
  Тогда фрагмент логики $L$ в языке с одной переменной является\/ $\cclass{coNEXPTIME}$-трудным.
\end{corollary}

\begin{proof}
  Следует из теоремы~\ref{thr:products-KTB} и~\cite[теорема~3.2]{Marx99}; рассуждение аналогично доказательству следствия~\ref{cor:HHK}.
\end{proof}


\begin{corollary}
  Фрагменты от одной переменной логик\/ $\logic{K} \times \logic{K}$,\/ $\logic{K} \times \logic{K4}$ и\/ $\logic{K} \times \logic{S5}_2$ не являются элементарными.
\end{corollary}

\begin{proof}
  Следует из теоремы~\ref{thr:products-KTB} и, соответственно, \cite[следствие~3.7]{GJL10}, \cite[теорема~4.2]{GJL10} и~\cite[теорема~4.5]{GJL10}.
\end{proof}

\begin{corollary}
  Фрагмент от одной переменной логики\/ $\logic{K} \times \logic{S5}$ является\/
  $\cclass{coNEXPTIME}$-полным.
\end{corollary}

\begin{proof}
  Следует из теоремы~\ref{thr:products-KTB} и~\cite[теорема~4.5]{Marx99}.
\end{proof}

    \subsection{Сложность полупроизведений модальных логик}
    \label{sec:fragments-semiproducts}

В этом разделе мы покажем, что теорему~\ref{thr:products-KTB} (а фактически и её доказательство) можно обобщить на полупроизведения мономодальных логик, в которых один из выделенных сомножителей совпадает с $\logic{K}$, $\logic{KT}$, $\logic{KB}$ или~$\logic{KTB}$.

Пусть $\vp$~--- $\lang{ML}_n$-формула, $\var \vp \subseteq \set{p_1,\ldots,p_m}$ и $p$~--- переменная, отличная от~$p_1, \ldots, p_m$. Пусть формулы $A$, $B$ и $\beta_k$ для каждого $k \in \{1, \ldots, m\}$, а также перевод~$\tau$ и функция~$e$ определены так же, как в разделе~\ref{sec:fragments-products}.

\begin{lemma}
  \label{lem:beta-sp}
  Пусть $L = (L_1^{\phantom{i}} \times L_2^{\phantom{i}} \times \ldots
  \times L_n^{\phantom{i}})_d^{\mathsf{EX}}$, где $d \in \{1, \ldots, n\}$,
  $L_1 \in \{\logic{K}, \logic{KT}, \logic{KB}, \logic{KTB}\}$, а
  $L_2, \ldots, L_n$~--- полные по Крипке модальные логики. Тогда
  $$
  \begin{array}{lcl}
  \vp \in L & \iff & e(\vp) \in L.
  \end{array}
  $$
\end{lemma}

\begin{proof}
  $(\Leftarrow)$ Пусть $\vp \notin L$. Тогда
  $(\kframe{G}, \pw{w}) \not\models^v \vp$ для некоторой шкалы~$\kframe{G} = \langle \bar{V}, \bar{S}_1, \ldots, \bar{S}_n
  \rangle$, такой, что
  $\kframe{G} \in (\kframe{F}_1^{\phantom{i}} \times \ldots \times
  \kframe{F}_n^{\phantom{i}})_d^{\mathsf{EX}}$, где
  $\kframe{F}_i = \langle W_i, R_i \rangle \models L_i$ для каждого
  $i \in \{1, \ldots, n\}$, некоторой точки $\pw{w} \in \bar{V}$ и некоторой оценки~$v$ в~$\kframe{G}$.

  Считаем, что
  $\kframe{F}_1 \times \ldots \times \kframe{F}_n = \langle \bar{W},
  \bar{R}_1, \ldots, \bar{R}_n \rangle$. Поскольку
  $\kframe{G} \in (\kframe{F}_1^{\phantom{i}} \times \ldots \times
  \kframe{F}_n^{\phantom{i}})_d^{\sf EX}$, согласно определению
  $d$-полупроизведений шкал, $\bar{R}_s (\bar{V}) \subseteq \bar{V}$ для каждого
  $s \in \{1, \ldots, d\}$. Мы построим $L$-шкалу, опровергающую формулу~$A \imp \tau(\vp)$.

  Для каждого $k \in \{1, \ldots, m+1\}$ определим $1$-шкалу $\kframe{A}_k = \langle U_k, S_k \rangle$ так, как это сделано в доказательстве леммы~\ref{lem:gamma}; см.~также рис.~\ref{fig:Bk}.

  Для каждого $x \in W_1^{\phantom{i}}$ определим
  $\kframe{A}_k^x = \langle U_k^{\phantom{i}} \times \{x\}, S_k^x
  \rangle$ как изоморфную копию шкалы $\kframe{A}_k$ при отображении
  $f\colon u \mapsto \langle u, x \rangle$. Пусть, как и в доказательстве леммы~\ref{lem:gamma},
  $$
  \begin{array}{lcl}
    W'_1  & =  & \displaystyle W_{1}^{\phantom{i}} \cup \bigcup\limits_{{k = 1}}^{{m+1}} (U_{k}^{\phantom{i}} \times W_{1}^{\phantom{i}}), \\
    R'_{1} & = & \displaystyle R_{1}^{\phantom{i}} \cup \bigcup\limits_{x
                 \in W_1} \bigcup\limits_{k = 1}^{m+1} \big( S_{k}^x \cup
                 \{\langle x,\langle b^k_{k+1}, x\rangle \rangle, \langle \langle
                 b^k_{k+1}, x\rangle, x \rangle \} \big) \\
  \end{array}
  $$
  и $\kframe{F}'_1 = \langle W'_1, R'_1 \rangle$. Положим\footnote{См.~сноску~\ref{footnote:product}.}
  $$
  \begin{array}{lcl}
  \kframe{F}'
    & =
    & \kframe{F}'_1 \times \kframe{F}_2^{\phantom{i}} \times
  \ldots \times \kframe{F}_n^{\phantom{i}} = \langle \bar{W}',
  \bar{R}'_1, \ldots, \bar{R}'_n \rangle.
  \end{array}
  $$

Пусть теперь
\[ 
  \begin{array}{llcl}
    \arrayitem & \bar{V}'
    & =
    & \bar{V} \cup \{ \langle \langle u, x_1 \rangle, x_2,
      \ldots, x_n \rangle \in \bar{W}' :
      \mbox{$u \in \bigcup\limits_{\mathclap{k = 1}}^{\mathclap{m+1}} U_{k}$, $\pw{x} \in \bar{V}$} \};
      \medskip\\
    \arrayitem & \bar{S}'_i
    & =
    & \mbox{$\bar{R}'_i \upharpoonright \bar{V}'$
      для каждого $i \in \{1, \ldots, n \}$;}
      \medskip\\
    \arrayitem & \kframe{G}'
    & =
    & \langle \bar{V}', \bar{S}'_1, \ldots, \bar{S}'_n \rangle.
    \end{array}
\]

    Нетрудно видеть, что $\kframe{G} \subseteq \kframe{G}'$.  Покажем, что
    $\kframe{G}' \in (\kframe{F}'_1 \times \kframe{F}_2^{\phantom{i}} \times
    \ldots \times \kframe{F}_n^{\phantom{i}})_d^{\mathsf{EX}}$.

    Согласно определению,
    $\kframe{G}' \subseteq \kframe{F}'_1 \times \kframe{F}_2^{\phantom{i}}
    \times \ldots \times \kframe{F}_n$.  Нам надо показать, что выполняется условие~$(2)$, т.е. что $\bar{R}'_s (\bar{V}') \subseteq \bar{V}'$ для каждого $s \in \{1, \ldots, d\}$. Пусть $\pw{y} \in \bar{V}'$, $s \in \{1, \ldots, d\}$ и $\pw{z} \in \bar{R}'_s (\pw{y})$.
    Рассмотрим два случая: $s=1$ и $s \in \{2, \ldots, d\}$.

    Пусть $s = 1$.

    Предположим, что $\pw{y} \in \bar{V}$. Тогда либо
    $\pw{z} = \langle \langle b^k_{k+1}, y_1^{\phantom{i}} \rangle,
    y_2^{\phantom{i}}, \ldots, y_n^{\phantom{i}} \rangle$ для некоторого
    $k \in \{1, \ldots, m + 1\}$, либо $\pw{z} \in \bar{W}$. Из определения множества $\bar{V}'$ следует, что $\langle \langle b^k_{k+1}, y_1^{\phantom{i}} \rangle,
    y_2^{\phantom{i}}, \ldots, y_n^{\phantom{i}} \rangle \in
    \bar{V}'$ для каждого $k \in \{1, \ldots, m + 1\}$. Пусть
    $\pw{z} \in \bar{W}$. Тогда $\pw{z} \in \bar{R}_1(\pw{y})$.
    Поскольку $\pw{y} \in \bar{V}$ и $\bar{R}_1 (\bar{V}) \subseteq \bar{V}$, получаем, что
    $\pw{z} \in \bar{V}$, а значит, $\pw{z} \in \bar{V}'$.

    Предположим теперь, что $\pw{y} = \langle \langle u, x_1 \rangle, x_2, \ldots, x_n
    \rangle$ для некоторых $u \in U_{k}$, где $k \in \{1, \ldots, m + 1\}$,
    и $\pw{x} \in \bar{V}$. Тогда либо $\pw{z} = \langle \langle u', x_1 \rangle, x_2, \ldots, x_n
    \rangle$ для некоторого $u' \in U_{k}$, либо $\pw{z} \in \pw{V}$ (в~случае, когда $u = b^k_{k+1}$). В~любом случае из определения множества~$\bar{V}'$ получаем, что~$\pw{z} \in \bar{V}'$.

    Пусть $s \in \{2, \ldots, d\}$.

    Предположим, что $\pw{y} \in \bar{V}$. По условию,
    $\bar{R}_s (\bar{V}) \subseteq \bar{V}$; значит,
    $\pw{z} \in \bar{V}$, и следовательно, $\pw{z} \in \bar{V}'$.

    Предположим теперь, что
    $\pw{y} = \langle \langle u, x_1 \rangle, \ldots, x_s, \ldots, x_n
    \rangle$ для некоторых $u \in U_{k}$, где $k \in \{1, \ldots, m + 1\}$,
    и $\pw{x} \in \bar{V}$. Тогда для некоторого $x' \in R_s(x_s)$
    $$
    \begin{array}{lcl}
    \pw{z}
      & =
      & \langle \langle u, x_1 \rangle, \ldots, x_{s-1}, x', x_{s+1}, \ldots, x_n \rangle.
    \end{array}
    $$
    Поскольку
    $\bar{R}_s (\bar{V}) \subseteq \bar{V}$, получаем, что
    $$
    \langle x_1 \ldots, x_{s-1}, x', x_{s+1}, \ldots, x_n \rangle \in
    \bar{V}.
    $$
    Значит, $\pw{z} \in \bar{V}'$ по определению множества~$\bar{V}'$.

    Таким образом, $\kframe{G}' \in (\kframe{F}'_1 \times \kframe{F}_2^{\phantom{i}} \times     \ldots \times \kframe{F}_n^{\phantom{i}})_d^{\mathsf{EX}}$. Заметим, что $\kframe{F}'_1 \models L_1^{\phantom{i}}$; значит, $\kframe{G}' \models L$.

  Пусть $v'$~--- оценка в~$\kframe{G}'$, такая, что $\bar{x} \in v'(p)$ в точности тогда, когда выполнено одно из следующих условий:
    \begin{itemize}
      \item
    существуют такие $k \in \{1, \ldots, m\}$, $\pw{z} \in \bar{V}$ и $u \in \{ b_{k+1}^{k}, c_{k+1}^{k}, c_{k+2}^{k} \}$, для которых
    $(\kframe{G}, \pw{z}) \models^{v} p_k^{\phantom{i}}$ и $\bar{x} = \langle \langle u,
    z_1^{\phantom{i}} \rangle, z_2^{\phantom{i}}, \ldots,
    z_n^{\phantom{i}} \rangle$;
  \item
    существуют такие $\pw{z} \in \bar{V}$ и $u \in \{ b_{m+2}^{m+1},c_{m+2}^{m+1},c_{m+3}^{m+1} \}$, для которых $\bar{x} = \langle \langle u, z_1^{\phantom{i}} \rangle, z_2^{\phantom{i}}, \ldots, z_n^{\phantom{i}} \rangle$.
  \end{itemize}

  Теперь можно доказать аналоги лемм~\ref{sublem:B} и~\ref{sublem:gamma_k}.

  \begin{sublemma}
    \label{sublem:B-sp}
    Если $\pw{x} \in \bar{V}'$, то
    $$
    \begin{array}[lcl]{lcl}
      (\kframe{G}', \pw{x}) \models^{v'} B  & \iff & \pw{x} \in \bar{V}.
    \end{array}
    $$
  \end{sublemma}

  \begin{proof}
    Аналогично доказательству подлеммы~\ref{sublem:B}.
  \end{proof}

  \begin{sublemma}
    \label{sublem:gamma_k-sp}
    Если $\pw{x} \in \pw{V}$ и $k \in \{1, \ldots, m\}$, то
    $$
    \begin{array}{lcl}
    (\kframe{G}', \pw{x}^{\phantom{i}}) \models^{v'} \beta_k & \iff &
    (\kframe{G}, \pw{x}^{\phantom{i}}) \models^{v} p_k.
    \end{array}
    $$
  \end{sublemma}

  \begin{proof}
    Аналогично доказательству подлеммы~\ref{sublem:gamma_k}.
  \end{proof}

  Используя подлеммы~\ref{sublem:B-sp} и~\ref{sublem:gamma_k-sp}, можно доказать аналог эквивалентности~$(7)$, а именно, что для любой формулы~$\theta \in \sub \vp$ и любой точки~$\pw{x} \in \bar{V}$
  $$
  \begin{array}{lcl}
  (\kframe{G}, \pw{x}) \models^v \theta
  & \iff &
  (\kframe{G}', \pw{x}) \models^{v'} \tau(\theta).
  \end{array}
  $$

  Поскольку $(\kframe{G}, \pw{w}) \not\models^{v} \vp$ и
  $\pw{w} \in \bar{V}$, получаем, что
  $(\kframe{G}', \pw{w}) \not\models^{v'} \tau(\vp)$.

  Кроме того, можно показать, как и в доказательстве леммы~\ref{lem:gamma}, что
  $(\kframe{G}', \pw{w}) \models^{v'} A$.  Следовательно,
  $(\kframe{G}', \pw{w}) \not\models^{v'} A \imp \tau(\vp)$.  Поскольку $\kframe{G}' \models L$, получаем, что  $A \imp \tau(\vp) \notin L$.

  $(\Rightarrow)$ Пусть $A \imp \tau(\vp) \notin L$. Тогда
  $(\kframe{G}, \pw{w}) \not\models^v A \imp \tau(\vp)$ для некоторой шкалы~$\kframe{G} = \langle \bar{V}, \bar{S}_1, \ldots, \bar{S}_n
  \rangle$, такой, что
  $\kframe{G} \in (\kframe{F}_1^{\phantom{i}} \times \ldots \times
  \kframe{F}_n^{\phantom{i}})_d^{\mathsf{EX}}$ и
  $\kframe{F}_i = \langle W_i, R_i \rangle \models L_i$ для каждого
  $i \in \{1, \ldots, n\}$, некоторой точки $\pw{w} \in \bar{V}$ и некоторой оценки~$v$ в~$\kframe{G}$.

  Считаем, что
  $\kframe{F}_1 \times \ldots \times \kframe{F}_n = \langle \bar{W},
  \bar{R}_1, \ldots, \bar{R}_n \rangle$.  Поскольку
  $\kframe{G} \in (\kframe{F}_1^{\phantom{i}} \times \ldots \times
  \kframe{F}_n^{\phantom{i}})_d^{\sf EX}$, по определению
  $d$-полупроизведения,
  $\bar{R}_s (\bar{V}) \subseteq \bar{V}$ для каждого
  $s \in \{1, \ldots, d\}$. Мы построим $L$-шкалу, опровергающую формулу~$\vp$.

  Для непустой последовательности $\lambda = i_1 \cdot \ldots \cdot i_s$ символов из алфавита $\{1, \dots, n\}$ положим
  $$
  \begin{array}{lcl}
    \bar{S}_{\lambda} & = & \bar{S}_{i_1} \circ \ldots \circ \bar{S}_{i_s};
  \end{array}
  $$
  пусть также
  $$
  \begin{array}{lcl}
    \bar{S}_{\emptystring} & = & \{ \langle \pw{x}, \pw{x} \rangle :\, \pw{x} \in \pw{V} \}.
  \end{array}
  $$
  Для каждой точки $\pw{x}\in\bar{V}$ определим множество~$\ind{\pw{x}}$:
  $$
  \begin{array}{lcl}
    \ind{\pw{x}}
    & =
    & \left\{\lambda \in \{1, \ldots, n\}^{\ast} :
      \exists \mu \in \{1, \ldots, n\}^{\ast} ~\big(\mu
      \cdot \lambda \in \ind{\varphi} ~\&~
                                                  \pw{w}\bar{S}_{\mu}\pw{x}\big) \right\}.
  \end{array}
  $$
  Пусть
  $$
  \begin{array}{lcl}
    W'_1 & = & \{ x_1 :
               \mbox{$\pw{x} \in \pw{V} ~\&~ 
               (\kframe{G}, \pw{x}) \models^v B ~\&~ 
               \ind{\pw{x}}\ne \varnothing$} \}.
  \end{array}
  $$

  Поскольку $(\kframe{G}, \pw{w}) \models A$, получаем, что
  $(\kframe{G}, \pw{w}) \models B$.  Поскольку
  $\emptystring \in \ind{\varphi}$, нетрудно видеть, что
  $\emptystring \in \ind{\pw{w}}$, а значит,
  $\ind{\pw{w}} \ne \varnothing$.  Следовательно,
  $w_1^{\phantom{i}} \in W'_1$, т.е. $W'_1 \ne \varnothing$.  Пусть\footnote{См.~сноску~\ref{footnote:product}.}
  \settowidth{\templength}{\mbox{$R'_1$}}
  \begin{itemize}
  \item ${\parbox{\templength}{$R'_1$}} = R_1^{\phantom{i}} \upharpoonright W'_1$;
  \item ${\parbox{\templength}{$\kframe{F}'_1$}} = \langle W'_1, R'_1 \rangle$;
  \item
    ${\parbox{\templength}{$\kframe{F}'$}} = \kframe{F}'_1 \times \kframe{F}_2^{\phantom{i}} \times
    \ldots \times \kframe{F}_n^{\phantom{i}} = \langle \bar{W}',
    \bar{R}'_1, \ldots, \bar{R}'_n \rangle$.
  \end{itemize}

  Ясно, что $\kframe{F}'_1 \subseteq \kframe{F}_1^{\phantom{i}}$; кроме того, логика~$L_1$ является сабфреймовой; значит, $\kframe{F}'_1 \models L_1^{\phantom{i}}$. Пусть
  \begin{itemize}
  \item ${\parbox{\templength}{$\bar{V}'$}} = \bar{V} \cap \bar{W}'$;
  \item ${\parbox{\templength}{$\bar{S}'_i$}} = \bar{R}'_i \upharpoonright \bar{V}'$ для каждого $i \in \{1, \ldots, n\}$;
  \item
    ${\parbox{\templength}{$\kframe{G}'$}} = \langle \bar{V}', \bar{S}'_1, \ldots, \bar{S}'_n
    \rangle$.
  \end{itemize}

  Покажем, что
  $\kframe{G}' \in (\kframe{F}'_1 \times \kframe{F}_2^{\phantom{i}} \times
  \ldots \times \kframe{F}_n^{\phantom{i}})_d^{\mathsf{EX}}$.  По определению,
  $\kframe{G}' \subseteq \kframe{F}'_1 \times \kframe{F}_2^{\phantom{i}}
  \times \ldots \times \kframe{F}_n^{\phantom{i}}$. Поскольку
  $\bar{R}_s (\bar{V}') \subseteq \bar{R}_s (\bar{V}) \subseteq
  \bar{V}$ для каждого $s \in \{1, \ldots, d\}$, получаем, что
  $$
  \begin{array}{lclclcl}
  \bar{R}'_s (\bar{V}') & =
                        & \bar{R}_s^{\phantom{i}} (\bar{V}') \cap \bar{W}'
                        & \subseteq
                        & \bar{V} \cap \bar{W}'
                        & =
                        & \bar{V}'.
  \end{array}
  $$
Значит, $\kframe{G}'$ удовлетворяет условию~$(2)$. Следовательно,
$\kframe{G}' \in (\kframe{F}'_1 \times \kframe{F}_2^{\phantom{i}} \times
\ldots \times \kframe{F}_n^{\phantom{i}})_d^{\mathsf{EX}}$. Поскольку $\kframe{F}'_1 \models L_1^{\phantom{i}}$, получаем, что~$\kframe{G}' \models L$.

  Докажем аналог подлеммы~\ref{sublem:D}.

  \begin{sublemma}
    \label{sublem:D-sp}
    Если $\pw{y}\in \bar{V}'$ и $\ind{\pw{y}}\ne \varnothing$, то
    $(\kframe{G},\pw{y})\models^{v} B$.
  \end{sublemma}

  \begin{proof}
    Заметим, что каждый раз, когда в доказательстве подлеммы~\ref{sublem:D} мы использовали условие~$(1)$, было достаточно использовать более слабое условие~$(3)$, справедливое для $d$-полупроизведений (см.~сноски~\ref{footnote:about(3)insteadof(1)} и~\ref{footnote:about(3)insteadof(1)2}). Используя это наблюдение, заключаем, что можно повторить доказательство подлеммы~\ref{sublem:D}.
  \end{proof}

  Пусть $v'$~--- оценка в~$\kframe{G}'$, такая, что для каждого $k \in \{1, \ldots, m\}$,
  $$
  \begin{array}{lcl}
  v'(p_k) & = & \{\pw{x}\in\bar{V}' : (\kframe{G},\pw{x})\models^{v} \beta_k\}.
  \end{array}
  $$

  Покажем, что для любой $\theta \in \sub \varphi$ и любой точки $\pw{x}\in\bar{V}'$, такой, что   $\ind{\theta}\subseteq \ind{\pw{x}}$,
  $$
  \begin{array}{lcl}
  (\kframe{G}',\pw{x})\models^{v'} \theta
    & \iff
    & (\kframe{G},\pw{x})\models^{v}\tau(\theta).
  \end{array}
  \eqno{(10)}
  $$

  Доказательство проведём индукцией по построению формулы~$\theta$.
  Рассмотрим только случай, когда $\theta = \Box_i \psi$; в~остальных случаях нужно поступать так же, как при обосновании эквивалентности~$(9)$.

  Пусть $\theta = \Box_i \psi$. Тогда $\tau(\theta) = \Box_i(B \to\tau(\psi))$. Пусть   $\pw{x}\in\bar{V}'$ и $\ind{\theta}\subseteq \ind{\pw{x}}$.

  Предположим, что $(\kframe{G}',\pw{x})\not\models^{v'} \Box_i \psi$. Тогда   $(\kframe{G}',\pw{y})\not\models^{v'}\psi$ для некоторой точки $\bar{y}\in \bar{S}'_i(\bar{x})$.  Поскольку $\kframe{G}' \subseteq \kframe{G}$, получаем, что $\bar{y}\in \bar{S}_i(\bar{x})$. Покажем, что $\ind{\psi} \subseteq \ind{\pw{y}}$ (и~значит,
  $\ind{\pw{y}} \ne \varnothing$). Пусть $\lambda \in \ind{\psi}$. Тогда
  $i \cdot \lambda \in \ind{\theta}$. По условию, $\ind{\theta}\subseteq \ind{\pw{x}}$; значит, $i \cdot \lambda \in \ind{\bar{x}}$. Значит, $\mu \cdot i \cdot \lambda \in \ind{\vp}$ и $\pw{w} \bar{S}_{\mu} \bar{x}$ для некоторого $\mu \in \{1, \ldots, n\}^{\ast}$. Поскольку $\bar{y}\in \bar{S}_i(\bar{x})$, получаем, что $\pw{w} \bar{S}_{\mu \cdot i} \bar{y}$, и учитывая, что $\mu \cdot i \cdot \lambda \in \ind{\vp}$, заключаем, что $\lambda \in \ind{\pw{y}}$. Значит,   $\ind{\psi} \subseteq \ind{\pw{y}}$. Тогда, согласно индукционному предположению,   $(\kframe{G},\pw{y})\not\models^{v} \tau(\psi)$. Поскольку $\bar{y} \in \bar{V}'$ и $\ind{\pw{y}} \ne \varnothing$, по подлемме~\ref{sublem:D-sp}, $(\kframe{G},\pw{y})\models^{v} B$.
  Значит, $(\kframe{G},\pw{x})\not\models^{v} \Box_i (B \imp \tau(\psi))$.

  Теперь предположим, что $(\kframe{G},\pw{x})\not\models^{v} \Box_i (B \imp \tau(\psi))$. Тогда   $(\kframe{G},\pw{y}) \models^{v} B$ и
  $(\kframe{G},\pw{y})\not\models^{v}\tau(\psi)$ для некоторой точки $\bar{y}\in \bar{S}_i(\bar{x})$.  Нетрудно видеть, что в этом случае $\ind{\psi} \subseteq \ind{\pw{y}}$, а значит,
  $\ind{\pw{y}} \ne \varnothing$. Следовательно, $y_1^{\phantom{i}} \in W'_1$ по определению множества~$W'_1$, а значит, $\bar{y} \in \bar{V}'$. Тогда, по индукционному предположению, $(\kframe{G}',\pw{y})\not\models^{v'} \psi$. Покажем, что $\bar{S}'_i = \bar{S}_i \upharpoonright \pw{V}'$. Действительно,
  $$
  \begin{array}{lclclclclcl}
  \bar{S}'_i
    & =
    & \bar{R}'_i \upharpoonright \pw{V}'
    & =
    & (\bar{R}_i \upharpoonright \pw{W}') \upharpoonright \pw{V}'
    & =
    & (\bar{R}_i \upharpoonright \pw{V}) \upharpoonright \pw{W}'
    & =
    & \bar{S}_i \upharpoonright \pw{W}'
    & =
    & \bar{S}_i \upharpoonright \pw{V}'.
  \end{array}
  $$
Поскольку $\bar{y}\in \bar{S}_i(\bar{x})$, $\bar{y} \in \bar{V}'$ и $\bar{x}\in \bar{V}'$, получаем, что $\bar{y}\in \bar{S}'_i(\bar{x})$. Значит, $(\kframe{G}',\pw{x})\not\models^{v'} \Box_i \psi$.

Таким образом, эквивалентность~$(10)$ обоснована. Поскольку $w_1^{\phantom{i}} \in W'_1$, получаем, что $\pw{w} \in \bar{V}'$. Тогда, согласно~$(10)$, $(\kframe{G}', \pw{w}) \not\models^{v'} \vp$. Учитывая, что $\kframe{G}' \models L$, получаем, что~$\vp \notin L$.
\end{proof}

\begin{theorem}
  \label{thr:semiproducts-KTB}
  Пусть $L = (L_1^{\phantom{i}} \times L_2^{\phantom{i}} \times \ldots
  \times L_n^{\phantom{i}})_d^{\mathsf{EX}}$, где $d\in\set{1,\ldots,n}$,
  $L_1 \in \{\logic{K}, \logic{KT}, \logic{KB}, \logic{KTB}\}$ и
  $L_2, \ldots, L_n$~--- полные по Крипке мономодальные логики. Тогда $L$ полиномиально погружается в свой фрагмент от одной переменной.
\end{theorem}

\begin{proof}
  Аналогично доказательству теоремы~\ref{thr:products-KTB}.
\end{proof}

  \section{Замечания}

Мы рассмотрели в этой главе далеко не все виды полимодальных пропозициональных логик, но такая цель и не ставилась: задача была лишь в том, чтобы показать, как описанные ранее методы модифицируются для более богатых систем. 

В случае логик $\CTLstar$, $\CTL$, $\LTL$, $\ATLstar$, $\ATL$ мы не стали рассматривать классы расширений этих логик, в том числе интервалы логик. Причина этого состоит не в том, что имеются какие-то сложности с перенесением методов на расширения указанных логик, а в том, что обычно такие расширения не рассматриваются. Что касается возможности перенесения методов (и результатов) на расширения этих логик, то это, конечно же, возможно, и более того, не вызывает технических трудностей.

В случае произведений логик мы, наоборот, рассматривали именно классы логик, и здесь ситуация иная. Автору неясно, можно ли распространить описанные выше методы на более широкие классы произведений логик, в частности, на случай, когда все сомножители произведения мономодальных логик являются расширениями логики~$\logic{K4}$.

\begin{problem}
Какова сложность фрагментов от конечного числа переменных произведений мономодальных логик, являющихся расширениями логики\/~$\logic{K4}$?
\end{problem}

Кроме того, для произведений логик автору не удалось получить нетривиальные оценки сложности их константных фрагментов. Ясно, что если одним из сомножителей в произведении является логика, лежащая между $\logic{K}$ и $\logic{wGrz}$, то её константный фрагмент, как минимум, $\cclass{PSPACE}$-труден. Но можно ли говорить о большей сложности константных фрагментов произведений логик?

\begin{problem}
Какова сложность константных фрагментов произведений мономодальных логик, среди сомножителей которых имеются логики, имеющие бесконечно много попарно неэквивалентных константных формул?
\end{problem}

Отметим также результаты, связанные с логикой Джапаридзе~$\logic{GLP}$~\cite{Japaridze:1986:diss}, расширяющей логику~$\logic{GL}$. Как уже было сказано в разделе~\ref{ssec:modprop:emb:GL:Grz}, константный фрагмент логики~$\logic{GL}$ полиномиально разрешим. Оказалось, то же справедливо для $\logic{GLP}_n$ при любом $n\in\numNp$ (логика $\logic{GLP}_1$ совпадает с~$\logic{GL}$), хотя и логика $\logic{GLP}$, и её константный фрагмент являются $\ccls{PSPACE}$-полными~\cite{Shapirovsky:2008,Pahomov:2014}.

\setcounter{savefootnote}{\value{footnote}}

\part{Предикатные логики}
\chapter{Классическая логика предикатов}
\setcounter{footnote}{\value{savefootnote}}
      \label{ch:6}
  \section{Синтаксис и семантика}
  \label{sec:QCl:syntax:semantics}

Будем считать, что исходными символами \defnotion{классического предикатного языка}\index{уян@язык!предикатный!классический} $\lang{QL}$\index{уян@язык!ql@$\lang{QL}$} являются следующие: счётное множество \defnotion{предметных переменных},\index{переменная!предметная} счётное множество \defnotion{предикатных букв}\index{буква!предикатная} любой \defnotion{валентности},\index{бяа@валентность} константа~$\bot$, бинарные логические связки $\wedge$, $\vee$, $\to$, а также \defnotion{кванторные символы}\index{символ!кванторный} $\forall$ и~$\exists$. Валентность предикатной буквы называется также её \defnotion{арностью},\index{арность} или \defnotion{местностью},\index{местность} и представляет собой натуральное число. Если $n$~--- валентность предикатной буквы~$P$, то букву $P$ называем \defnotion{$n$\nobreakdash-арной},\index{буква!предикатная!4n@$n$-арная} или \defnotion{$n$\nobreakdash-местной};\index{буква!предикатная!4n@$n$-местная} $0$\nobreakdash-арные предикатные буквы называем также \defnotion{пропозициональными буквами}\index{буква!пропозициональная} (фактически их можно не отличать от пропозициональных переменных языка~$\lang{L}$, что мы и планируем делать; мы вернёмся к этому чуть ниже). Предметные переменные называют также \defnotion{индивидными};\index{переменная!индивидная} классический предикатный язык~$\lang{QL}$ называют также \defnotion{классическим языком первого порядка}.\index{уян@язык!классический!первого порядка} Обогатим множество символов языка $\lang{QL}$ бинарным предикатным \defnotion{символом равенства}\index{символ!равенства}~$=$; получившийся язык будем обозначать $\lang{QL}^=$ и называть \defnotion{классическим предикатным языком с равенством}.\index{уян@язык!предикатный!классический с равенством}

\defnotion{Формулы} в языке $\lang{QL}^=$ определяются обычным образом; определение рекурсивно:
\begin{itemize}
\item
\defnotion{элементарными}\index{уяа@формула!элементарная} формулами считаем выражения вида $\bot$, $x=y$ и $P(x_1,\ldots,x_n)$, где $x$, $y$ и $x_1,\ldots,x_n$~--- предметные переменные, а $P$~--- $n$\nobreakdash-мест\-ная предикатная буква;
\item
если $\varphi$ и $\psi$~--- формулы, то $(\varphi\wedge\psi)$, $(\varphi\vee\psi)$, $(\varphi\to\psi)$~--- тоже формулы;
\item
если $\varphi$~--- формула, а $x$~--- предметная переменная, то $\forall x\,\varphi$ и $\exists x\,\varphi$~--- тоже формулы.
\end{itemize}

Будем использовать $\neg$, $\top$, $\leftrightarrow$ как обычные сокращения: $\neg\varphi = \varphi\to \bot$, $\top=\neg\bot$, $\varphi \leftrightarrow \psi = (\varphi\to \psi)\wedge (\psi\to\varphi)$.

При записи формул будем пользоваться некоторыми договорённостями. Иногда в целях лучшей читаемости формул вместо $x=y$ будем писать $(x=y)$; по этим же причинам для пропозициональной буквы $p$ элементарную формулу $p(\curlywedge)$, где $\curlywedge$~--- \defnotion{пустой список},\index{список!пустой} будем обозначать как~$p$ (и~тогда можно считать языки $\lang{QL}$ и~$\lang{QL}^=$ расширениями пропозиционального языка~$\lang{L}$). Как и в пропозициональном случае, будем опускать некоторые скобки в формулах, считая, что кванторы имеют такой же приоритет, как и отрицание.

Элементарные формулы называют также \defnotion{атомарными}\index{уяа@формула!атомарная}, или \defnotion{атомами}. Если $x$~--- предметная переменная, то выражения $\forall x$ и $\exists x$ называем, соответственно, \defnotion{квантором всеобщности}\index{квантор!бяа@всеобщности} и \defnotion{квантором существования}\index{квантор!существования} по переменной~$x$; такие кванторы называем \defnotion{кванторами первого порядка}\index{квантор!первого порядка}\footnote{Кванторы второго порядка имеют вид $\forall P$ и $\exists P$, где $P$~--- некоторая предикатная буква.}.
В формулах $\forall x\,\varphi$ и $\exists x\,\varphi$ формулу $\varphi$ называем \defnotion{областью действия},\index{область!действия квантора} соответственно, квантора всеобщности по~$x$ или квантора существования по~$x$.

\defnotion{Вхождение} переменной~$x$ в формулу~$\varphi$ называем \defnotion{связным},\index{бяа@вхождение!переменной!связное} если оно находится в области действия некоторого квантора по~$x$; при этом оно \defnotion{связывается} самым внутренним вхождением квантора по~$x$, в области действия которого находится. Вхождение переменной~$x$ в формулу~$\varphi$ называем \defnotion{свободным},\index{бяа@вхождение!переменной!свободное} если оно не является связным. Переменная~$x$ называется \defnotion{свободной переменной формулы}\index{переменная!свободная}~$\varphi$, если существует свободное вхождение~$x$ в~$\varphi$. Формула~$\varphi$ называется \defnotion{замкнутой}\index{уяа@формула!еяд@замкнтурая}, если она не содержит свободных переменных.

Формулы языка $\lang{QL}^=$ называем также $\lang{QL}^=$-формулами.\index{уяа@формула!ql@$\lang{QL}^=$-формула} Формально под языком $\lang{QL}^=$\index{уян@язык!ql@$\lang{QL}^=$} понимаем множество всех $\lang{QL}^=$-формул.

Язык $\lang{QL}$\index{уян@язык!ql@$\lang{QL}$} определяем как фрагмент языка $\lang{QL}^=$, состоящий из формул, не содержащих вхождений символа равенства. Формулы языка $\lang{QL}$ называем также $\lang{QL}$-формулами.\index{уяа@формула!ql@$\lang{QL}$-формула} Все понятия, введённые для $\lang{QL}^=$-формул, сохраняются и для $\lang{QL}$-формул.

В дальнейшем мы в основном будем интересоваться языком без равенства; тем не менее, в некоторых случаях нам будет удобно пользоваться языком с равенством, и он введён в рассмотрение именно для таких случаев.

\defnotion{Моделью}\index{модель!классического предикатного языка} классического предикатного языка называется набор $\cModel{M} = \langle D,I\rangle$, где $D$~--- непустое множество \defnotion{индивидов}\index{индивид} модели $\cModel{M}$, а $I$~--- \defnotion{интерпретация предикатных букв}\index{интерпретация!предикатных букв} в $D$, т.е. функция, сопоставляющая каждой $n$-местной предикатной букве $P$ некоторое $n$-местное отношение $I(P)$ на~$D$; иначе говоря, $I(P)\subseteq D^n$, в частности, если $p$~--- пропозициональная буква, то $I(P)\subseteq D^0 = \set{\otuple{\curlywedge}}$.
Индивиды модели $\cModel{M} = \langle D,I\rangle$ называются также \defnotion{предметами},\index{предмет} а множество $D$~--- \defnotion{индивидной},\index{область!индивидная} или \defnotion{предметной областью}\index{область!предметная} модели~$\cModel{M}$.

\defnotion{Приписыванием}\index{приписывание} в модели $\cModel{M} = \langle D,I\rangle$ называется функция $g$, сопоставляющая каждой предметной переменной $x$ некоторый индивид $g(x)\in D$. Приписывания называются также \defnotion{интерпретациями предметных переменных}\index{интерпретация!предметных переменных}, \defnotion{оценками}\index{оценка} и \defnotion{означиваниями}\index{означивание}. Будем использовать запись $g' \stackrel{x}{=} g$, если приписывание $g'$ отличается от приписывания $g$, разве что, значением на переменной~$x$.

Истинность формулы $\varphi$ в модели $\cModel{M} = \langle D,I\rangle$ при приписывании $g$ определяется рекурсивно:
\settowidth{\templength}{\mbox{$\cModel{M}\cmodels^g\varphi'$ или $\cModel{M}\cmodels^g\varphi''$;}}
\settowidth{\templengtha}{\mbox{$w$}}
\settowidth{\templengthb}{\mbox{$\cModel{M}\cmodels^{h}\varphi'$ для некоторого $h$, такого, что $h \stackrel{x}{=} g$}}
\settowidth{\templengthc}{\mbox{$\cModel{M}\cmodels^g P(x_1,\ldots,x_n)$}}
$$
\begin{array}{lcl}
\cModel{M}\cmodels^g P(x_1,\ldots,x_n)
  & \leftrightharpoons
  & \parbox{\templengthb}{$\langle g(x_1),\ldots,g(x_n)\rangle \in I(P)$;} \\
\cModel{M}\cmodels^g x=y
  & \leftrightharpoons
  & \parbox{\templengthb}{$\langle g(x),g(y)\rangle \in \mathit{id}_D$,} \\
\end{array}
$$
где $P$~--- $n$-местная предикатная буква, а $\mathit{id}_D=\{\otuple{a,a}:a\in D\}$;
\settowidth{\templength}{\mbox{$\cModel{M}\cmodels^g\varphi'$ или $\cModel{M}\cmodels^g\varphi''$;}}
\settowidth{\templengtha}{\mbox{$w$}}
\settowidth{\templengthb}{\mbox{$\cModel{M}\cmodels^{g}\varphi'\to\varphi''$}}
\settowidth{\templengthc}{\mbox{$\cModel{M}\cmodels^g P(x_1,\ldots,x_n)$}}
$$
\begin{array}{lcl}
\parbox{\templengthc}{{}\hfill\parbox{\templengthb}{$\cModel{M} \not\cmodels^g \bot;$}}
  \\
\parbox{\templengthc}{{}\hfill\parbox{\templengthb}{$\cModel{M}\cmodels^g\varphi' \wedge \varphi''$}}
  & \leftrightharpoons
  & \parbox[t]{\templength}{$\cModel{M}\cmodels^g\varphi'$\hfill и\hfill $\cModel{M}\cmodels^g\varphi''$;}
  \\
\parbox{\templengthc}{{}\hfill\parbox{\templengthb}{$\cModel{M}\cmodels^g\varphi' \vee \varphi''$}}
  & \leftrightharpoons
  & \parbox[t]{\templength}{$\cModel{M}\cmodels^g\varphi'$ или $\cModel{M}\cmodels^g\varphi''$;}
  \\
\parbox{\templengthc}{{}\hfill\parbox{\templengthb}{$\cModel{M}\cmodels^g\varphi' \to \varphi''$}}
  & \leftrightharpoons
  & \parbox[t]{\templength}{$\cModel{M}\not\cmodels^g\varphi'$ или $\cModel{M}\cmodels^g\varphi''$;}
  \\
\parbox{\templengthc}{{}\hfill\parbox{\templengthb}{$\cModel{M}\cmodels^g\forall x\,\varphi'$}}
  & \leftrightharpoons
  & \mbox{$\cModel{M}\cmodels^{h}\varphi'$ для каждого $h$, такого, что $h \stackrel{x}{=} g$;}
  \\
\parbox{\templengthc}{{}\hfill\parbox{\templengthb}{$\cModel{M}\cmodels^g\exists x\,\varphi'$}}
  & \leftrightharpoons
  & \mbox{$\cModel{M}\cmodels^{h}\varphi'$ для некоторого $h$, такого, что $h \stackrel{x}{=} g$.}
\end{array}
$$

Отметим, что такие модели, где равенство интерпретируется в предметной области $D$ \defnotion{диагональю}\index{диагональ множества} $\mathit{id}_D$ множества $D$, называются \defnotion{нормальными}.\index{модель!нормальная} Вообще говоря, равенство можно интерпретировать отношением конгруэнтности, причём не обязательно совпадающим с диагональю предметной области, но известно (см., например,~\cite{Mendelson-1976-1-rus}), что можно ограничиться рассмотрением нормальных моделей; как следует из данного выше определения, мы будем предполагать, что все модели классического языка являются нормальными.

Если $\varphi$~--- формула, то иногда будем использовать запись $\varphi(x_1,\ldots,x_n)$, чтобы указать, что в $\varphi$ нет свободных переменных, отличных от $x_1, \ldots, x_n$.

Для формулы $\varphi(x_1,\ldots,x_n)$ и предметов $a_1, \ldots, a_n$ модели $\cModel{M}$ будем писать $\cModel{M} \cmodels \varphi (a_1, \ldots, a_n)$, если $\cModel{M} \cmodels^g \varphi (x_1, \ldots, x_n)$ для некоторого приписывания~$g$, такого, что $g(x_1) = a_1, \ldots, g(x_n) = a_n$.  Такое обозначение не приводит к двусмысленности, поскольку истинность формулы $\varphi(x_1, \ldots, x_n)$ в $\cModel{M}$ не зависит от значений переменных, отличных от $x_1, \ldots, x_n$.

Для модели $\cModel{M}$, класса $\mathscr{C}$ моделей, формулы $\varphi$ и множества~$X$ формул положим
\settowidth{\templengtha}{\mbox{$\cModel{M}\cmodels^{h}\varphi'$ для некоторого $h$, такого, что $h \stackrel{x}{=} g$}}
\settowidth{\templengthd}{\mbox{$\cModel{M}$}}
\settowidth{\templengthb}{\mbox{$\cModel{M}\cmodels X$}}
$$
\begin{array}{lcl}
\parbox{\templengthc}{{}\hfill\parbox{\templengthb}{$\cModel{M}\cmodels \varphi$}}
  & \leftrightharpoons
  & \parbox[t]{\templengtha}{$\cModel{M}\cmodels^g\varphi$ для любого~$g$;}
  \\
\parbox{\templengthc}{{}\hfill\parbox{\templengthb}{$\cModel{M}\cmodels X$}}
  & \leftrightharpoons
  & \parbox[t]{\templengtha}{$\cModel{M}\cmodels\varphi$ для любой $\varphi\in X$;}
  \\
\parbox{\templengthc}{{}\hfill\parbox{\templengthb}{$\parbox{\templengthd}{\hfill$\mathscr{C}$}\cmodels \varphi$}}
  & \leftrightharpoons
  & \parbox[t]{\templengtha}{$\cModel{M}\cmodels\varphi$ для любой $\cModel{M}\in \mathscr{C}$.}
  \\
\end{array}
$$

Если $\cModel{M}\cmodels \varphi$, говорим, что $\varphi$ \defnotion{истинна} в~$\cModel{M}$; в противном случае говорим, что $\varphi$ \defnotion{опровергается} в~$\cModel{M}$. Формулу~$\varphi$ называем \defnotion{общезначимой},\index{уяа@формула!общезначимая} или \defnotion{тождественно истинной},\index{уяа@формула!тождественно истинная} если $\varphi$ истинна в классе всех моделей. Замкнутую формулу~$\varphi$ называем \defnotion{выполнимой},\index{уяа@формула!бяа@выполнимая} если $\varphi$ истинна в некоторой модели. Замкнутую формулу~$\varphi$ называем \defnotion{опровержимой},\index{уяа@формула!опровержимая} если $\varphi$ опровергается в некоторой модели.

Модель $\cModel{M} = \langle D,I\rangle$ называем \defnotion{конечной},\index{модель!конечная} если её носитель $D$ является конечным множеством.

Множество $\lang{QL}$-формул называем\footnote{Довольно часто требуют, чтобы теория состояла только из замкнутых формул; для наших целей такое ограничение несущественно.} \defnotion{теорией первого порядка},\index{теория!первого порядка} множество $\lang{QL}^=$-формул\footnote{Вообще говоря, такое множество может и не содержать формул, в которые входит равенство; считаем, тем не менее, что такого не случается (можно включить в теорию общезначимую формулу, содержащую равенство).} называем \defnotion{теорией первого порядка с равенством}.\index{теория!первого порядка!с равенством}
Модель $\cModel{M}$ называем \defnotion{моделью теории~$\Gamma$},\index{модель!теории} если $\cModel{M}\cmodels\Gamma$.
Замкнутую формулу $\varphi$ называем \defnotion{выполнимой в теории}\index{уяа@формула!бяа@выполнимая!в теории} $\Gamma$, если существует модель теории~$\Gamma$, в которой истинна~$\varphi$.

Для класса $\mathscr{C}$ моделей положим $\mathit{Th}(\mathscr{C}) = \{\varphi\in\lang{QL} : \mathscr{C}\cmodels\varphi\}$. Множество $\mathit{Th}(\mathscr{C})$ называем \defnotion{теорией первого порядка класса~$\mathscr{C}$}. Аналогично определяется \defnotion{теория первого порядка с равенством класса~$\mathscr{C}$}: $\mathit{Th}^=(\mathscr{C}) = \{\varphi\in\lang{QL}^= : \mathscr{C}\cmodels\varphi\}$. В~дальнейшем теории первого порядка будем называть просто теориями; то же относится к теориям первого порядка с равенством.

Определим логики $\logic{QCl}$ и $\logic{QCl}_{\mathit{fin}}$ как, соответственно, теорию класса всех моделей и теорию класса всех конечных моделей; определим логики $\logic{QCl}^=$ и $\logic{QCl}^=_{\mathit{fin}}$ как, соответственно, теорию с равенством класса всех моделей и теорию с равенством класса всех конечных моделей. Все эти логики замкнуты по правилу предикатной подстановки. Логика $\logic{QCl}$ называется \defnotion{классической логикой предикатов первого порядка} (или \defnotion{классической логикой предикатов}, или \defnotion{логикой первого порядка}\index{логика!классическая!первого порядка}), логика $\logic{QCl}_{\mathit{fin}}$~--- \defnotion{классической логикой конечных моделей},\index{логика!классическая!конечных моделей} или \defnotion{теорией конечных моделей};\index{теория!конечных моделей} названия для $\logic{QCl}^=$ и $\logic{QCl}^=_{\mathit{fin}}$ аналогичны. Для логик $\logic{QCl}$ и $\logic{QCl}^=$ известны различные аксиоматики (см.,~например,~\cite{Mendelson}), а логики $\logic{QCl}_{\mathit{fin}}$ и $\logic{QCl}^=_{\mathit{fin}}$ не являются рекурсивно перечислимыми~\cite{Trakhtenbrot50,Trakhtenbrot53}, поэтому не имеют ни конечных, ни даже рекурсивных аксиоматик.

    \section{Неразрешимость}
    \subsection{Предварительные сведения}
    \label{sec:s6-2-1}

Известно, что классическая предикатная логика $\logic{QCl}$ является алгоритмически неразрешимой~\cite{Church36,Turing36} (более точно, $\Sigma^0_1$-полной, см.~\cite{Harel86} и~\cite[приложение]{MR:2021:LI}).
Но как только удаётся получить подобные <<негативные>> результаты, возникает естественный вопрос о том, как можно, ограничивая использование средств языка, избежать неразрешимости. Сейчас исследования подобных вопросов образуют целую область, которую можно назвать <<классической проблемой разрешения>>~\cite{BGG97}: имеется огромное количество результатов, дающих как разрешимые, так и неразрешимые фрагменты логики~$\logic{QCl}$, классических теорий первого порядка, а также других формальных предикатных систем~\cite{Lowenheim15, Suranyi43, Motohashi-1990-1, Gradel99, GKV97, Mortimer75, HWZ00, HWZ01, KKZ05, WZ01, MR:2013:LI, Kripke62, Kremer97, MR:2016:Tver, MR:2021:LI, GSh93}. Так, чтобы доказать неразрешимость логики $\logic{QCl}$, достаточно, чтобы её язык содержал какие-либо из перечисленных ниже наборов средств:
\begin{itemize}
\item
одну бинарную предикатную букву и бесконечно много предметных переменных (см.~\cite[глава~21]{BBJ07} или~\cite[глава~25]{BJ-1994-1-rus});
\item
одну бинарную предикатную букву, бесконечно много унарных предикатных букв и три предметные переменные~\cite{Suranyi43};
\item
и даже одну бинарную предикатную букву и три предметные переменные~\cite{TG87}.
\end{itemize}
В то же время, следующие фрагменты логики $\logic{QCl}$ разрешимы:
\begin{itemize}
\item
\defnotion{монадический} фрагмент (т.е. содержащий лишь одноместные предикатные буквы), даже обогащённый равенством~\cite{Lowenheim15} (см.\ также~\cite[глава~21]{BBJ07} или~\cite[глава~25]{BJ-1994-1-rus});
\item
различные \defnotion{охраняемые} фрагменты\footnote{В английском языке используется термин guarded fragments.}~\cite{Motohashi-1990-1,ANB:1998,Gradel99};
\item
фрагмент с двумя предметными переменными~\cite{GKV97,Mortimer75} (см.\ также работы \cite{Seg73} и \cite{Shehtman:2012:rus}).
\end{itemize}
Неразрешимы также и многие первопорядковые теории (различные арифметические теории, теория групп, различные аксиоматические теории множеств и др.), в частности, теории в языке лишь с одной бинарной предикатной буквой: например, теория симметричного иррефлексивного бинарного отношения (см.~\cite{NerodeShore80} и \cite[приложение]{Kremer97}) и, как следствие, теория симметричного рефлексивного бинарного отношения. Что касается теорий конечных моделей, они могут даже не быть рекурсивно перечислимыми~\cite{Trakhtenbrot50,Trakhtenbrot53} (см. также~\cite{MR:2019:SAICSIT} и \cite[приложение]{MR:2021:JLC:2}; похожие результаты для других языков можно найти, например, в~\cite{BM15,Hajek98}).

Наша основная цель, связанная с классической логикой предикатов, будет состоять в том, чтобы дать ответ на следующий вопрос:
\begin{itemize}
\item
\textit{разрешима ли теория симметричного иррефлексивного предиката в языке с одной бинарной буквой и тремя предметными переменными?}
\end{itemize}

Автору не удалось найти ответ на этот вопрос в литературе, хотя, как было отмечено выше, \defnotion{диадический} фрагмент логики предикатов (т.е. логика бинарного отношения) в языке с тремя предметными переменными является неразрешимым.
Предлагаемые ниже доказательства основаны на свед\'{е}нии к логике $\logic{QCl}$ в языке с одной бинарной предикатной буквой и тремя предметными переменными некоторой неразрешимой проблемы \defnotion{укладки плиток домино}\footnote{В английском используется термин tiling problem.}. При этом, в сравнении с другими известными автору доказательствами неразрешимости диадического фрагмента логики предикатов, оно не использует вспомогательные фрагменты языка классической логики предикатов, т.е. соответствующее свед\'{е}ние будет построено напрямую. Кроме того, предложенный метод позволяет доказать неразрешимость различных теорий бинарного предиката в языке с тремя предметными переменными, определяемых такими дополнительными требованиями к бинарному отношению, как рефлексивность, иррефлексивность, симметричность, антисимметричность, связность, серийность, транзитивность, интранзитивность, даже при некоторых их комбинациях, что и даст ответ на поставленный вопрос.


Опишем неразрешимую проблему укладки домино, которую мы будем использовать.

Считаем, что все \defnotion{плитки домино}\index{плитка домино} имеют квадратную форму одного и того же фиксированного размера, причём для каждой такой плитки зафиксирована ориентация её сторон: указано, какая именно сторона является \defnotion{верхней}, \defnotion{нижней}, \defnotion{правой} и \defnotion{левой}. Каждая плитка домино имеет некоторый \defnotion{тип} $t$, определяемый \defnotion{цветами} $\leftsq t$ (цвет левой грани), $\rightsq t$ (цвет правой грани), $\upsq t$ (цвет верхней грани) и $\downsq t$ (цвет нижней грани). \defnotion{Задача домино}\index{еяд@задача!домино} определяется набором $T = \{t_0, \ldots, t_{n}\}$ типов плиток домино и состоит в том, что нужно выяснить, существует ли  \defnotion{$T$-укладка}\index{укладка плиток домино} плиток домино, типы которых представлены в~$T$, т.е. функция $f\colon \mathds{N} \times \mathds{N} \to T$, такая, что для любых $i, j \in \mathds{N}$,
\begin{itemize}
\item[]$(T_1)$\quad $\rightsq f(i,j) = \leftsq f(i+1,j)$;
\item[]$(T_2)$\quad $\upsq f(i,j) = \downsq f(i,j+1)$.
\end{itemize}

Отметим, что имеются и другие задачи домино; они отличаются тем, какие требования предъявляются к укладке. Пока нам важно, что проблема домино, состоящая из задач домино, для которых существует укладка с условиями $(T_1)$~и~$(T_2)$ является неразрешимой~\cite{Berger66}, а именно, $\Pi^0_1$\nobreakdash-полной.

    \subsection{Теорема Чёрча и теорема Трахтенброта}

\newcommand{\drawtileflat} [9]
{
\draw [white, opacity = 0, name path = diag 1] #1--#3;
\draw [white, opacity = 0, name path = diag 2] #2--#4;
\draw [name intersections = {of = diag 1 and diag 2, by = {tcenter}}];

\foreach \c in {0.84}
{
  \coordinate (ttl)     at ($(tcenter)+\c*#4-\c*(tcenter)$);
  \coordinate (tbr)     at ($(tcenter)+\c*#2-\c*(tcenter)$);
  \coordinate (tbl)     at ($(tcenter)+\c*#1-\c*(tcenter)$);
  \coordinate (ttr)     at ($(tcenter)+\c*#3-\c*(tcenter)$);
  \coordinate (sttl)    at ($(tcenter)+0.4*(ttl)-0.4*(tcenter)$);
  \coordinate (stbr)    at ($(tcenter)+0.4*(tbr)-0.4*(tcenter)$);
  \coordinate (stbl)    at ($(tcenter)+0.4*(tbl)-0.4*(tcenter)$);
  \coordinate (sttr)    at ($(tcenter)+0.4*(ttr)-0.4*(tcenter)$);
}

\fill [#5, fill opacity=0.75] (ttl)--(tbl)--(stbl)--(sttl)--cycle;
\fill [#6, fill opacity=0.75] (tbl)--(tbr)--(stbr)--(stbl)--cycle;
\fill [#7, fill opacity=0.75] (tbr)--(ttr)--(sttr)--(stbr)--cycle;
\fill [#8, fill opacity=0.75] (ttl)--(ttr)--(sttr)--(sttl)--cycle;


\draw (ttl)--(tbl)--(tbr)--(ttr)--cycle;
\draw (sttl)--(stbl)--(stbr)--(sttr)--cycle;
\draw (ttl)--(sttl);
\draw (ttr)--(sttr);
\draw (tbl)--(stbl);
\draw (tbr)--(stbr);

\node [] at (tcenter) {#9};
}

\newcommand{\drawtile} [9]
{
\draw [white, opacity = 0, name path = diag 1] #1--#3;
\draw [white, opacity = 0, name path = diag 2] #2--#4;
\draw [name intersections = {of = diag 1 and diag 2, by = {bcenter}}];

\foreach \c in {0.84}
{
  \coordinate (tcenter) at ($(bcenter)+(0,#9)$);
  \coordinate (btl)     at ($(bcenter)+\c*#4-\c*(bcenter)$);
  \coordinate (bbr)     at ($(bcenter)+\c*#2-\c*(bcenter)$);
  \coordinate (bbl)     at ($(bcenter)+\c*#1-\c*(bcenter)$);
  \coordinate (btr)     at ($(bcenter)+\c*#3-\c*(bcenter)$);
  \coordinate (ttl)     at ($(btl)+(0,#9)$);
  \coordinate (tbr)     at ($(bbr)+(0,#9)$);
  \coordinate (tbl)     at ($(bbl)+(0,#9)$);
  \coordinate (ttr)     at ($(btr)+(0,#9)$);
  \coordinate (sttl)    at ($(tcenter)+0.4*(ttl)-0.4*(tcenter)$);
  \coordinate (stbr)    at ($(tcenter)+0.4*(tbr)-0.4*(tcenter)$);
  \coordinate (stbl)    at ($(tcenter)+0.4*(tbl)-0.4*(tcenter)$);
  \coordinate (sttr)    at ($(tcenter)+0.4*(ttr)-0.4*(tcenter)$);
}

\fill [#5, fill opacity=0.75] (ttl)--(tbl)--(stbl)--(sttl)--cycle;
\fill [#6, fill opacity=0.75] (tbl)--(tbr)--(stbr)--(stbl)--cycle;
\fill [#7, fill opacity=0.75] (tbr)--(ttr)--(sttr)--(stbr)--cycle;
\fill [#8, fill opacity=0.75] (ttl)--(ttr)--(sttr)--(sttl)--cycle;

\fill [white, fill opacity=0.75] (ttl)--(btl)--(bbl)--(tbl)--cycle;
\fill [white, fill opacity=0.75] (tbl)--(bbl)--(bbr)--(tbr)--cycle;
\fill [white, fill opacity=0.75] (stbl)--(stbr)--(sttr)--(sttl)--cycle;

\draw (ttl)--(tbl)--(tbr)--(ttr)--cycle;
\draw (ttl)--(btl)--(bbl)--(bbr)--(tbr);
\draw (bbl)--(tbl);
\draw (sttl)--(stbl)--(stbr)--(sttr)--cycle;
\draw (ttl)--(sttl);
\draw (ttr)--(sttr);
\draw (tbl)--(stbl);
\draw (tbr)--(stbr);
}

\newcommand{\drawtilemodela} [4]
{
\draw [>=latex, ->, shorten >= 1.96pt, shorten <= 1.96pt] #1--($#1 + (0,#2)$);
\shade [ball color=black] ($#1 + (0,#2)$) circle [radius = #3];
}

\newcommand{\drawtilemodelb} [4]
{
\draw [>=latex, ->, shorten >= 1.96pt, shorten <= 1.96pt] #1--($#1 + (0,#2)$);
\draw [>=latex, ->, shorten >= 1.96pt, shorten <= 1.96pt] ($#1 + (0,#2)$)--($#1 + (0,#2) + (0,#4)$);
\shade [ball color=black] ($#1 + (0,#2)$) circle [radius = #3];
\shade [ball color=black] ($#1 + (0,#2) + (0,#4)$) circle [radius = #3];
}

\newcommand{\drawtilemodelc} [4]
{
\draw [>=latex, ->, shorten >= 1.96pt, shorten <= 1.96pt] #1--($#1 + (0,#2)$);
\draw [>=latex, ->, shorten >= 1.96pt, shorten <= 1.96pt] ($#1 + (0,#2)$)--($#1 + (0,#2) + (0,#4)$);
\draw [>=latex, ->, shorten >= 1.96pt, shorten <= 1.96pt] ($#1 + (0,#2) + (0,#4)$)--($#1 + (0,#2) + 2*(0,#4)$);
\shade [ball color=black] ($#1 + (0,#2)$) circle [radius = #3];
\shade [ball color=black] ($#1 + (0,#2) + (0,#4)$) circle [radius = #3];
\shade [ball color=black] ($#1 + (0,#2) + 2*(0,#4)$) circle [radius = #3];
}

\subsubsection{Моделирование сетки $\numN\times\numN$}
\label{sec:grid}

Чтобы промоделировать проблему укладки домино, нам нужна сетка $\numN \times \numN$; чтобы описать её в предметной области модели, введём некоторые сокращения.


Зафиксируем бинарную предикатную~$P$. Можно понимать $P$ как транзитивное отношение (мы потребуем транзитивность для $P$ ниже). Пусть
$$
\begin{array}{rcl}
  W(x)
    & =
    & P(x,x); \\
  B(x)
    & =
    & \neg W(x);
    \\
  x\simeq y
    & =
    & \forall z\,((P(x,z)\leftrightarrow P(y,z))\wedge(P(z,x)\leftrightarrow P(z,y)));
    \\
  x\approx y
    & =
    & W(x)\leftrightarrow W(y);
    \\
  G(x)
    & =
    & \exists y\,(W(y)\wedge P(x,y));
    \\
  x\prec y
    & =
    & P(x,y)\wedge x\not\simeq y \wedge \forall z\,(P(x,z)\wedge P(z,y)\to z\simeq x\vee z\simeq y);
    \\
  x\prec_H y
    & =
    & x\prec y \wedge x\approx y;
    \\
  x\prec_V y
    & =
    & x\prec y \wedge x\not\approx y,
\end{array}
$$
где $x\not\simeq y$ и $x\not\approx y$ означают $\neg(x\simeq y)$ и $\neg(x\approx y)$ соответственно.
Подразумеваемый смысл введённых сокращений следующий:
$$
\begin{array}{rcl}
W(x)       & \mbox{---} & \mbox{$x$ белый;} \\
B(x)       & \mbox{---} & \mbox{$x$ чёрный;} \\
x\simeq y  & \mbox{---} & \mbox{$x$ и $y$ неразличимы;} \\
x\approx y & \mbox{---} & \mbox{$x$ и $y$ одного цвета;} \\
G(x)       & \mbox{---} & \mbox{$x$ является элементом сетки;} \\
x\prec y   & \mbox{---} & \mbox{$y$ следует за $x$;} \\
x\prec_H y & \mbox{---} & \mbox{$y$ находится непосредственно справа от $x$;} \\
x\prec_V y & \mbox{---} & \mbox{$y$ находится непосредственно над $x$.} \\
\end{array}
$$

Обратим внимание, что для того, чтобы определить $x\prec z$, надо в определении для $x\prec y$ одновременно заменить $y$ на $z$ и $z$ на $y$: это позволит не вводить новую переменную, а использовать только $x$, $y$ и~$z$. Аналогично надо поступить и с определением $x\prec_H z$, $x\prec_V z$, $y\prec z$, и~т.д.

Теперь можно описать сетку $\numN\times\numN$:
$$
\begin{array}{rcl}
  G_0
    & =
    & \forall x\forall y\forall z\,(P(x,y)\wedge P(y,z)\to P(x,z));
    \\
  G_1
    & =
    & \forall x\,(G(x)\to \exists y\,(G(y)\wedge x\prec_H y));
    \\
  G_2
    & =
    & \forall x\,(G(x)\to \exists y\,(G(y)\wedge x\prec_V y));
    \\
  G_3
    & =
    & \forall x\forall y\forall z\,(G(x)\wedge G(y)\wedge G(z)\to (x\prec_H y\wedge x\prec_H z \to y\simeq z));
    \\
  G_4
    & =
    & \forall x\forall y\forall z\,(G(x)\wedge G(y)\wedge G(z)\to (x\prec_V y\wedge x\prec_V z \to y\simeq z));
    \\
  G_5
    & =
    & \forall x\forall y\,(G(x)\wedge G(y)\to {}
    \\
    &
    & \hfill
    {} \to (\exists z\,(x\prec_Hz\wedge z\prec_Vy)\leftrightarrow \exists z\,(x\prec_Vz\wedge z\prec_Hy)));
    \\
  G_6
    & =
    & \exists x\,(B(x)\wedge G(x));
    \\
  \mathit{Grid}
    & =
    & G_0\wedge G_1\wedge G_2\wedge G_3\wedge G_4\wedge G_5\wedge G_6.
\end{array}
$$

Нетрудно видеть, что если формула $\mathit{Grid}$ истинна в некоторой модели $\cModel{M}=\langle D,I\rangle$, то $\cModel{M}$ содержит сетку $\numN\times\numN$, элементы первой строки которой состоят из <<чёрных>> предметов, второй~--- из <<белых>>, третьей~--- из <<чёрных>>, и т.д.; см.~рис.~\ref{MRybakov:fig:Grid1}, слева. Элементами этой сетки являются \defnotion{$\simeq$-кластеры}, определяемые следующим образом:
$$
\begin{array}{lcl}
\mathit{cl}_{\cModel{M}}(a) & = & \set{b\in D : \cModel{M}\cmodels b\simeq a}.
\end{array}
$$

\newcommand{\drawslantedgrid} [0]
{
\coordinate (dx)    at (1,0.5);
\coordinate (dy)    at (-0.83,+0.83);
\coordinate (g00)   at (0,0.0);
\coordinate (g10)   at ($(g00) + 1*(dx)$);
\coordinate (g20)   at ($(g00) + 2*(dx)$);
\coordinate (g30)   at ($(g00) + 3*(dx)$);
\coordinate (g40)   at ($(g00) + 4*(dx)$);
\coordinate (g50)   at ($(g00) + 5*(dx)$);
\coordinate (g01)   at ($(g00) + 1*(dy) + 0*(dx)$);
\coordinate (g11)   at ($(g00) + 1*(dy) + 1*(dx)$);
\coordinate (g21)   at ($(g00) + 1*(dy) + 2*(dx)$);
\coordinate (g31)   at ($(g00) + 1*(dy) + 3*(dx)$);
\coordinate (g41)   at ($(g00) + 1*(dy) + 4*(dx)$);
\coordinate (g51)   at ($(g00) + 1*(dy) + 5*(dx)$);
\coordinate (g02)   at ($(g00) + 2*(dy) + 0*(dx)$);
\coordinate (g12)   at ($(g00) + 2*(dy) + 1*(dx)$);
\coordinate (g22)   at ($(g00) + 2*(dy) + 2*(dx)$);
\coordinate (g32)   at ($(g00) + 2*(dy) + 3*(dx)$);
\coordinate (g42)   at ($(g00) + 2*(dy) + 4*(dx)$);
\coordinate (g52)   at ($(g00) + 2*(dy) + 5*(dx)$);
\coordinate (g03)   at ($(g00) + 3*(dy) + 0*(dx)$);
\coordinate (g13)   at ($(g00) + 3*(dy) + 1*(dx)$);
\coordinate (g23)   at ($(g00) + 3*(dy) + 2*(dx)$);
\coordinate (g33)   at ($(g00) + 3*(dy) + 3*(dx)$);
\coordinate (g43)   at ($(g00) + 3*(dy) + 4*(dx)$);
\coordinate (g53)   at ($(g00) + 3*(dy) + 5*(dx)$);
\coordinate (g04)   at ($(g00) + 4*(dy) + 0*(dx)$);
\coordinate (g14)   at ($(g00) + 4*(dy) + 1*(dx)$);
\coordinate (g24)   at ($(g00) + 4*(dy) + 2*(dx)$);
\coordinate (g34)   at ($(g00) + 4*(dy) + 3*(dx)$);
\coordinate (g44)   at ($(g00) + 4*(dy) + 4*(dx)$);
\coordinate (g54)   at ($(g00) + 4*(dy) + 5*(dx)$);
\coordinate (g05)   at ($(g00) + 5*(dy) + 0*(dx)$);
\coordinate (g15)   at ($(g00) + 5*(dy) + 1*(dx)$);
\coordinate (g25)   at ($(g00) + 5*(dy) + 2*(dx)$);
\coordinate (g35)   at ($(g00) + 5*(dy) + 3*(dx)$);
\coordinate (g45)   at ($(g00) + 5*(dy) + 4*(dx)$);
\coordinate (g55)   at ($(g00) + 5*(dy) + 5*(dx)$);

\begin{scope}[>=latex, ->, shorten >= 1.5pt, shorten <= 1.5pt]
\draw [] (g00) -- (g10);
\draw [] (g10) -- (g20);
\draw [] (g20) -- (g30);
\draw [] (g30) -- (g40);
\draw [shorten >= 14.5pt] (g40) -- (g50);
\draw [] (g01) -- (g11);
\draw [] (g11) -- (g21);
\draw [] (g21) -- (g31);
\draw [] (g31) -- (g41);
\draw [shorten >= 14.5pt] (g41) -- (g51);
\draw [] (g02) -- (g12);
\draw [] (g12) -- (g22);
\draw [] (g22) -- (g32);
\draw [] (g32) -- (g42);
\draw [shorten >= 14.5pt] (g42) -- (g52);
\draw [] (g03) -- (g13);
\draw [] (g13) -- (g23);
\draw [] (g23) -- (g33);
\draw [] (g33) -- (g43);
\draw [shorten >= 14.5pt] (g43) -- (g53);
\draw [] (g04) -- (g14);
\draw [] (g14) -- (g24);
\draw [] (g24) -- (g34);
\draw [] (g34) -- (g44);
\draw [shorten >= 14.5pt] (g44) -- (g54);
\draw [] (g00) -- (g01);
\draw [] (g01) -- (g02);
\draw [] (g02) -- (g03);
\draw [] (g03) -- (g04);
\draw [shorten >= 14.5pt] (g04) -- (g05);
\draw [] (g10) -- (g11);
\draw [] (g11) -- (g12);
\draw [] (g12) -- (g13);
\draw [] (g13) -- (g14);
\draw [shorten >= 14.5pt] (g14) -- (g15);
\draw [] (g20) -- (g21);
\draw [] (g21) -- (g22);
\draw [] (g22) -- (g23);
\draw [] (g23) -- (g24);
\draw [shorten >= 14.5pt] (g24) -- (g25);
\draw [] (g30) -- (g31);
\draw [] (g31) -- (g32);
\draw [] (g32) -- (g33);
\draw [] (g33) -- (g34);
\draw [shorten >= 14.5pt] (g34) -- (g35);
\draw [] (g40) -- (g41);
\draw [] (g41) -- (g42);
\draw [] (g42) -- (g43);
\draw [] (g43) -- (g44);
\draw [shorten >= 14.5pt] (g44) -- (g45);
\end{scope}

\node [] at (g50)   {$\cdots$}     ;
\node [] at (g51)   {$\cdots$}     ;
\node [] at (g52)   {$\cdots$}     ;
\node [] at (g53)   {$\cdots$}     ;
\node [] at (g54)   {$\cdots$}     ;

\node [] at (g05)   {$\vdots$}     ;
\node [] at (g15)   {$\vdots$}     ;
\node [] at (g25)   {$\vdots$}     ;
\node [] at (g35)   {$\vdots$}     ;
\node [] at (g45)   {$\vdots$}     ;


\filldraw [] (g00) circle [radius=1.5pt]   ;
\filldraw [] (g10) circle [radius=1.5pt]   ;
\filldraw [] (g20) circle [radius=1.5pt]   ;
\filldraw [] (g30) circle [radius=1.5pt]   ;
\filldraw [] (g40) circle [radius=1.5pt]   ;

\draw [] (g01) circle [radius=1.5pt]   ;
\draw [] (g11) circle [radius=1.5pt]   ;
\draw [] (g21) circle [radius=1.5pt]   ;
\draw [] (g31) circle [radius=1.5pt]   ;
\draw [] (g41) circle [radius=1.5pt]   ;

\filldraw [] (g02) circle [radius=1.5pt]   ;
\filldraw [] (g12) circle [radius=1.5pt]   ;
\filldraw [] (g22) circle [radius=1.5pt]   ;
\filldraw [] (g32) circle [radius=1.5pt]   ;
\filldraw [] (g42) circle [radius=1.5pt]   ;

\draw [] (g03) circle [radius=1.5pt]   ;
\draw [] (g13) circle [radius=1.5pt]   ;
\draw [] (g23) circle [radius=1.5pt]   ;
\draw [] (g33) circle [radius=1.5pt]   ;
\draw [] (g43) circle [radius=1.5pt]   ;

\filldraw [] (g04) circle [radius=1.5pt]   ;
\filldraw [] (g14) circle [radius=1.5pt]   ;
\filldraw [] (g24) circle [radius=1.5pt]   ;
\filldraw [] (g34) circle [radius=1.5pt]   ;
\filldraw [] (g44) circle [radius=1.5pt]   ;

}

\Rem{





}

\begin{figure}
\centering
\begin{tikzpicture}[scale=1.32]

\coordinate (g00)   at (0,0);
\coordinate (g10)   at (1,0);
\coordinate (g20)   at (2,0);
\coordinate (g30)   at (3,0);
\coordinate (g40)   at (4,0);
\coordinate (g50)   at (5,0);
\coordinate (g01)   at (0,1);
\coordinate (g11)   at (1,1);
\coordinate (g21)   at (2,1);
\coordinate (g31)   at (3,1);
\coordinate (g41)   at (4,1);
\coordinate (g51)   at (5,1);
\coordinate (g02)   at (0,2);
\coordinate (g12)   at (1,2);
\coordinate (g22)   at (2,2);
\coordinate (g32)   at (3,2);
\coordinate (g42)   at (4,2);
\coordinate (g52)   at (5,2);
\coordinate (g03)   at (0,3);
\coordinate (g13)   at (1,3);
\coordinate (g23)   at (2,3);
\coordinate (g33)   at (3,3);
\coordinate (g43)   at (4,3);
\coordinate (g53)   at (5,3);
\coordinate (g04)   at (0,4);
\coordinate (g14)   at (1,4);
\coordinate (g24)   at (2,4);
\coordinate (g34)   at (3,4);
\coordinate (g44)   at (4,4);
\coordinate (g54)   at (5,4);
\coordinate (g05)   at (0,5);
\coordinate (g15)   at (1,5);
\coordinate (g25)   at (2,5);
\coordinate (g35)   at (3,5);
\coordinate (g45)   at (4,5);
\coordinate (g55)   at (5,5);

\begin{scope}[>=latex, ->, shorten >= 1.5pt, shorten <= 1.5pt]
\draw [] (g00) -- (g10);
\draw [] (g10) -- (g20);
\draw [] (g20) -- (g30);
\draw [] (g30) -- (g40);
\draw [shorten >= 14.5pt] (g40) -- (g50);
\draw [] (g01) -- (g11);
\draw [] (g11) -- (g21);
\draw [] (g21) -- (g31);
\draw [] (g31) -- (g41);
\draw [shorten >= 14.5pt] (g41) -- (g51);
\draw [] (g02) -- (g12);
\draw [] (g12) -- (g22);
\draw [] (g22) -- (g32);
\draw [] (g32) -- (g42);
\draw [shorten >= 14.5pt] (g42) -- (g52);
\draw [] (g03) -- (g13);
\draw [] (g13) -- (g23);
\draw [] (g23) -- (g33);
\draw [] (g33) -- (g43);
\draw [shorten >= 14.5pt] (g43) -- (g53);
\draw [] (g04) -- (g14);
\draw [] (g14) -- (g24);
\draw [] (g24) -- (g34);
\draw [] (g34) -- (g44);
\draw [shorten >= 14.5pt] (g44) -- (g54);
\draw [] (g00) -- (g01);
\draw [] (g01) -- (g02);
\draw [] (g02) -- (g03);
\draw [] (g03) -- (g04);
\draw [shorten >= 14.5pt] (g04) -- (g05);
\draw [] (g10) -- (g11);
\draw [] (g11) -- (g12);
\draw [] (g12) -- (g13);
\draw [] (g13) -- (g14);
\draw [shorten >= 14.5pt] (g14) -- (g15);
\draw [] (g20) -- (g21);
\draw [] (g21) -- (g22);
\draw [] (g22) -- (g23);
\draw [] (g23) -- (g24);
\draw [shorten >= 14.5pt] (g24) -- (g25);
\draw [] (g30) -- (g31);
\draw [] (g31) -- (g32);
\draw [] (g32) -- (g33);
\draw [] (g33) -- (g34);
\draw [shorten >= 14.5pt] (g34) -- (g35);
\draw [] (g40) -- (g41);
\draw [] (g41) -- (g42);
\draw [] (g42) -- (g43);
\draw [] (g43) -- (g44);
\draw [shorten >= 14.5pt] (g44) -- (g45);
\end{scope}

\node [] at (g50)   {$\cdots$}     ;
\node [] at (g51)   {$\cdots$}     ;
\node [] at (g52)   {$\cdots$}     ;
\node [] at (g53)   {$\cdots$}     ;
\node [] at (g54)   {$\cdots$}     ;

\node [] at (g05)   {$\vdots$}     ;
\node [] at (g15)   {$\vdots$}     ;
\node [] at (g25)   {$\vdots$}     ;
\node [] at (g35)   {$\vdots$}     ;
\node [] at (g45)   {$\vdots$}     ;


\filldraw [] (g00) circle [radius=1.5pt]   ;
\filldraw [] (g10) circle [radius=1.5pt]   ;
\filldraw [] (g20) circle [radius=1.5pt]   ;
\filldraw [] (g30) circle [radius=1.5pt]   ;
\filldraw [] (g40) circle [radius=1.5pt]   ;

\draw [] (g01) circle [radius=1.5pt]   ;
\draw [] (g11) circle [radius=1.5pt]   ;
\draw [] (g21) circle [radius=1.5pt]   ;
\draw [] (g31) circle [radius=1.5pt]   ;
\draw [] (g41) circle [radius=1.5pt]   ;

\filldraw [] (g02) circle [radius=1.5pt]   ;
\filldraw [] (g12) circle [radius=1.5pt]   ;
\filldraw [] (g22) circle [radius=1.5pt]   ;
\filldraw [] (g32) circle [radius=1.5pt]   ;
\filldraw [] (g42) circle [radius=1.5pt]   ;

\draw [] (g03) circle [radius=1.5pt]   ;
\draw [] (g13) circle [radius=1.5pt]   ;
\draw [] (g23) circle [radius=1.5pt]   ;
\draw [] (g33) circle [radius=1.5pt]   ;
\draw [] (g43) circle [radius=1.5pt]   ;

\filldraw [] (g04) circle [radius=1.5pt]   ;
\filldraw [] (g14) circle [radius=1.5pt]   ;
\filldraw [] (g24) circle [radius=1.5pt]   ;
\filldraw [] (g34) circle [radius=1.5pt]   ;
\filldraw [] (g44) circle [radius=1.5pt]   ;


\end{tikzpicture}
\begin{tikzpicture}[scale=1.32]

\coordinate (g00)   at (0,0);
\coordinate (g10)   at (1,0);
\coordinate (g20)   at (2,0);
\coordinate (g30)   at (3,0);
\coordinate (g40)   at (4,0);
\coordinate (g50)   at (5,0);
\coordinate (g01)   at (0,1);
\coordinate (g11)   at (1,1);
\coordinate (g21)   at (2,1);
\coordinate (g31)   at (3,1);
\coordinate (g41)   at (4,1);
\coordinate (g51)   at (5,1);
\coordinate (g02)   at (0,2);
\coordinate (g12)   at (1,2);
\coordinate (g22)   at (2,2);
\coordinate (g32)   at (3,2);
\coordinate (g42)   at (4,2);
\coordinate (g52)   at (5,2);
\coordinate (g03)   at (0,3);
\coordinate (g13)   at (1,3);
\coordinate (g23)   at (2,3);
\coordinate (g33)   at (3,3);
\coordinate (g43)   at (4,3);
\coordinate (g53)   at (5,3);
\coordinate (g04)   at (0,4);
\coordinate (g14)   at (1,4);
\coordinate (g24)   at (2,4);
\coordinate (g34)   at (3,4);
\coordinate (g44)   at (4,4);
\coordinate (g54)   at (5,4);
\coordinate (g05)   at (0,5);
\coordinate (g15)   at (1,5);
\coordinate (g25)   at (2,5);
\coordinate (g35)   at (3,5);
\coordinate (g45)   at (4,5);
\coordinate (g55)   at (5,5);

\begin{scope}[>=latex, ->, shorten >= 1.96pt, shorten <= 1.96pt]
\draw [] (g00) -- (g10);
\draw [] (g10) -- (g20);
\draw [] (g20) -- (g30);
\draw [] (g30) -- (g40);
\draw [shorten >= 14.5pt] (g40) -- (g50);
\draw [] (g01) -- (g11);
\draw [] (g11) -- (g21);
\draw [] (g21) -- (g31);
\draw [] (g31) -- (g41);
\draw [shorten >= 14.5pt] (g41) -- (g51);
\draw [] (g02) -- (g12);
\draw [] (g12) -- (g22);
\draw [] (g22) -- (g32);
\draw [] (g32) -- (g42);
\draw [shorten >= 14.5pt] (g42) -- (g52);
\draw [] (g03) -- (g13);
\draw [] (g13) -- (g23);
\draw [] (g23) -- (g33);
\draw [] (g33) -- (g43);
\draw [shorten >= 14.5pt] (g43) -- (g53);
\draw [] (g04) -- (g14);
\draw [] (g14) -- (g24);
\draw [] (g24) -- (g34);
\draw [] (g34) -- (g44);
\draw [shorten >= 14.5pt] (g44) -- (g54);
\draw [] (g00) -- (g01);
\draw [] (g01) -- (g02);
\draw [] (g02) -- (g03);
\draw [] (g03) -- (g04);
\draw [shorten >= 14.5pt] (g04) -- (g05);
\draw [] (g10) -- (g11);
\draw [] (g11) -- (g12);
\draw [] (g12) -- (g13);
\draw [] (g13) -- (g14);
\draw [shorten >= 14.5pt] (g14) -- (g15);
\draw [] (g20) -- (g21);
\draw [] (g21) -- (g22);
\draw [] (g22) -- (g23);
\draw [] (g23) -- (g24);
\draw [shorten >= 14.5pt] (g24) -- (g25);
\draw [] (g30) -- (g31);
\draw [] (g31) -- (g32);
\draw [] (g32) -- (g33);
\draw [] (g33) -- (g34);
\draw [shorten >= 14.5pt] (g34) -- (g35);
\draw [] (g40) -- (g41);
\draw [] (g41) -- (g42);
\draw [] (g42) -- (g43);
\draw [] (g43) -- (g44);
\draw [shorten >= 14.5pt] (g44) -- (g45);
\end{scope}

\node [] at (g50)   {$\cdots$}     ;
\node [] at (g51)   {$\cdots$}     ;
\node [] at (g52)   {$\cdots$}     ;
\node [] at (g53)   {$\cdots$}     ;
\node [] at (g54)   {$\cdots$}     ;

\node [] at (g05)   {$\vdots$}     ;
\node [] at (g15)   {$\vdots$}     ;
\node [] at (g25)   {$\vdots$}     ;
\node [] at (g35)   {$\vdots$}     ;
\node [] at (g45)   {$\vdots$}     ;


\filldraw [] (g00) circle [radius=1.5pt]   ;
\filldraw [] (g10) circle [radius=1.5pt]   ;
\filldraw [] (g20) circle [radius=1.5pt]   ;
\filldraw [] (g30) circle [radius=1.5pt]   ;
\filldraw [] (g40) circle [radius=1.5pt]   ;

\draw [] (g01) circle [radius=1.5pt]   ;
\draw [] (g11) circle [radius=1.5pt]   ;
\draw [] (g21) circle [radius=1.5pt]   ;
\draw [] (g31) circle [radius=1.5pt]   ;
\draw [] (g41) circle [radius=1.5pt]   ;

\filldraw [] (g02) circle [radius=1.5pt]   ;
\filldraw [] (g12) circle [radius=1.5pt]   ;
\filldraw [] (g22) circle [radius=1.5pt]   ;
\filldraw [] (g32) circle [radius=1.5pt]   ;
\filldraw [] (g42) circle [radius=1.5pt]   ;

\draw [] (g03) circle [radius=1.5pt]   ;
\draw [] (g13) circle [radius=1.5pt]   ;
\draw [] (g23) circle [radius=1.5pt]   ;
\draw [] (g33) circle [radius=1.5pt]   ;
\draw [] (g43) circle [radius=1.5pt]   ;

\filldraw [] (g04) circle [radius=1.5pt]   ;
\filldraw [] (g14) circle [radius=1.5pt]   ;
\filldraw [] (g24) circle [radius=1.5pt]   ;
\filldraw [] (g34) circle [radius=1.5pt]   ;
\filldraw [] (g44) circle [radius=1.5pt]   ;


\drawtileflat{(g00)}{(g10)}{(g11)}{(g01)}{c1}{c2}{c3}{c4}{$t_4$};
\drawtileflat{(g10)}{(g20)}{(g21)}{(g11)}{c3}{c5}{c1}{c2}{$t_1$};
\drawtileflat{(g20)}{(g30)}{(g31)}{(g21)}{c1}{c5}{c3}{c2}{$t_0$};
\drawtileflat{(g30)}{(g40)}{(g41)}{(g31)}{c3}{c5}{c1}{c2}{$t_1$};

\drawtileflat{(g01)}{(g11)}{(g12)}{(g02)}{c2}{c4}{c1}{c5}{$t_2$};
\drawtileflat{(g11)}{(g21)}{(g22)}{(g12)}{c1}{c2}{c1}{c5}{$t_5$};
\drawtileflat{(g21)}{(g31)}{(g32)}{(g22)}{c1}{c2}{c3}{c5}{$t_6$};
\drawtileflat{(g31)}{(g41)}{(g42)}{(g32)}{c3}{c2}{c1}{c5}{$t_7$};

\drawtileflat{(g02)}{(g12)}{(g13)}{(g03)}{c1}{c5}{c3}{c2}{$t_0$};
\drawtileflat{(g22)}{(g32)}{(g33)}{(g23)}{c1}{c5}{c3}{c2}{$t_0$};
\drawtileflat{(g32)}{(g42)}{(g43)}{(g33)}{c3}{c5}{c1}{c2}{$t_1$};

\drawtileflat{(g03)}{(g13)}{(g14)}{(g04)}{c1}{c2}{c3}{c4}{$t_4$};
\drawtileflat{(g13)}{(g23)}{(g24)}{(g14)}{c3}{c2}{c1}{c2}{$t_8$};
\drawtileflat{(g23)}{(g33)}{(g34)}{(g24)}{c1}{c2}{c3}{c5}{$t_6$};
\drawtileflat{(g33)}{(g43)}{(g44)}{(g34)}{c3}{c2}{c1}{c5}{$t_7$};

\drawtileflat{(-1.2,2.6)}{(-0.2,2.6)}{(-0.2,3.6)}{(-1.2,3.6)}{c3}{c5}{c1}{c2}{$t_1$};

\draw [fill=white!10,opacity=0.45] (ttr)--(1.5,2.5)--(tbl)--(tbr)--cycle;
\draw [dashed, black!25, >=latex, ->, shorten >= 5pt] (1.5,2.5)--(g12);

\end{tikzpicture}

\caption{Сетка и укладка плиток домино}
\label{MRybakov:fig:Grid1}
\end{figure}

\subsubsection{Моделирование укладки домино}
\label{sec:tiling}

Опишем условие размещения плитки домино в сетке. Будем считать, что элемент сетки соответствует месту, куда помещается левый нижний угол плитки домино, см.~рис.~\ref{MRybakov:fig:Grid1}, справа. Пусть
$$
\begin{array}{rcl}
  F(x)
    & =
    & \neg\exists y\,P(x,y);
    \\
  \mathit{tile}_0(x)
    & =
    & \exists y\,(\neg G(y)\wedge x\prec y\wedge F(y));
    \\
  \mathit{tile}_1(x)
    & =
    & \exists z\,(\neg G(z)\wedge x\prec z\wedge \mathit{tile}_0(z));
    \\
  \mathit{tile}_{2k+2}(x)
    & =
    & \exists y\,(\neg G(y)\wedge x\prec y\wedge \mathit{tile}_{2k+1}(y));
    \\
  \mathit{tile}_{2k+3}(x)
    & =
    & \exists z\,(\neg G(z)\wedge x\prec z\wedge \mathit{tile}_{2k+2}(z)).
    \\
\end{array}
$$

Формула $\mathit{tile}_m(x)$ выражает следующее: $x$ видит непосредственного $P$-последователя вне сетки, который видит последний элемент за~$m$ $P$\nobreakdash-шагов, см.~рис.~\ref{MRybakov:fig:Grid}.

Подразумеваемый смысл введённых сокращений следующий:
$$
\begin{array}{rcl}
F(x)               & \mbox{---} & \mbox{$x$ является последним элементом;} \\
\mathit{tile}_m(x) & \mbox{---} & \mbox{$x$ содержит плитку типа $t_m$.} \\
\end{array}
$$

Для того, чтобы определить формулу $\mathit{tile}_m(y)$, надо одновременно заменить $x$ на $y$ и $y$ на $x$ в $\mathit{tile}_m(x)$: такой приём позволит не использовать новых переменных.

Теперь мы можем описать свойства укладки. Положим
$$
\begin{array}{rcl}
  T_0
    & =
    & \displaystyle
      \forall x\,(G(x)\to \bigvee\limits_{i=0}^n(\mathit{tile}_i(x)\wedge \bigwedge\limits_{j\ne i}\neg\mathit{tile}_j(x)));
    \smallskip\\
  T_1
    & =
    & \displaystyle
      \forall x\,(G(x)\to \bigwedge\limits_{i=0}^{n} (\mathit{tile}_i(x)\to \forall y\,(G(y)\wedge x\prec_H y\to \bigvee\limits_{\mathclap{\mathop{\scriptsize\rightsquare{0.21}} t_i\,{=}\,\mathop{\scriptsize\leftsquare{0.21}} t_j}}\mathit{tile}_j(y))));
    \smallskip\\
  T_2
    & =
    & \displaystyle
      \forall x\,(G(x)\to \bigwedge\limits_{i=0}^{n} (\mathit{tile}_i(x)\to \forall y\,(G(y)\wedge x\prec_V y\to \bigvee\limits_{\mathclap{\mathop{\scriptsize\upsquare{0.21}} t_i\,{=}\,\mathop{\scriptsize\downsquare{0.21}} t_j}}\mathit{tile}_j(y))));
    \smallskip\\
  \mathit{Tiling}_T
    & =
    & T_0\wedge T_1\wedge T_2.
\end{array}
$$

\begin{lemma}
\label{lem:main}
Формула $\mathit{Grid}\wedge\mathit{Tiling}_T$ выполнима тогда и только тогда, когда существует $T$-укладка плиток домино.
\end{lemma}

\begin{proof}
$(\Rightarrow)$
Пусть формула $\mathit{Grid}\wedge\mathit{Tiling}_T$ истинна в некоторой модели $\cModel{M}=\langle D,I\rangle$; покажем, как определить $T$-укладку $f\colon\numN\times\numN\to T$. Для этого введём вспомогательную функцию $g\colon\numN\times\numN\to \{\mathit{cl}_{\cModel{M}}(a) : a\in D\}$, задача которой~--- выбрать элементы из~$D$, из которых можно сформировать сетку~$\numN\times\numN$.

Согласно $G_6$, существует $a\in D$, такой, что $\cModel{M}\cmodels G(a)\wedge B(a)$; положим $g(0,0)=\mathit{cl}_{\cModel{M}}(a)$.

Пусть для всех $i,j\in \{0,\ldots,m\}$ уже определены значения $g(i,j)$; определим для каждых $i,j\in \{0,\ldots,m\}$ значение $g(m+1,j)$ и $g(i,m+1)$, а также определим значение~$g(m+1,m+1)$.

Пусть $b$~--- элемент из $D$, такой, что $\cModel{M}\cmodels g(m,j)\prec_H b$ и $\cModel{M}\cmodels G(b)$; такой элемент существует согласно~$G_1$. Положим $g(m+1,j)=\mathit{cl}_{\cModel{M}}(b)$. Формула $G_3$ гарантирует, что $c\in\mathit{cl}_{\cModel{M}}(b)$ для каждого $c\in D$, такого, что $\cModel{M}\cmodels g(m,j)\prec_H c$ и~$\cModel{M}\cmodels G(c)$.

Пусть $b$~--- элемент из $D$, такой, что $\cModel{M}\cmodels g(i,m)\prec_V b$ и $\cModel{M}\cmodels G(b)$; такой элемент существует согласно~$G_2$. Положим $g(i,m+1)=\mathit{cl}_{\cModel{M}}(b)$. Формула $G_4$ гарантирует, что $c\in\mathit{cl}_{\cModel{M}}(b)$ для каждого $c\in D$, такого, что $\cModel{M}\cmodels g(i,m)\prec_V c$ и~$\cModel{M}\cmodels G(c)$.

Пусть $b$~--- элемент из $D$, такой, что $\cModel{M}\cmodels g(m,m+1)\prec_H b$, $\cModel{M}\cmodels g(m+1,m)\prec_V b$ и $\cModel{M}\cmodels G(b)$; такой элемент существует согласно $G_1$, $G_2$ и~$G_5$. Положим $g(m+1,m+1) = \mathit{cl}_{\cModel{M}}(b)$. Любая из формул $G_3$ и $G_4$ гарантирует, что $c\in\mathit{cl}_{\cModel{M}}(b)$ для каждого $c\in D$, такого, что $\cModel{M}\cmodels g(m,m+1)\prec_H c$, $\cModel{M}\cmodels g(m+1,m)\prec_V c$ и~$\cModel{M}\cmodels G(c)$.

Для $i,j\in \numN$ выберем какой-нибудь индивид $b\in g(i,j)$.
Согласно $T_0$, существует единственное $k\in\{0,\ldots,n\}$, такое, что $\cModel{M}\cmodels \mathit{tile}_k(b)$. Положим $f(i,j)=t_k$.
Согласно $T_1$ и~$T_2$, функция $f\colon\numN\times\numN\to T$ является $T$-укладкой плиток домино.

$(\Leftarrow)$:
Пусть имеется $T$-укладка $f\colon\numN\times\numN\to T$. Чтобы получить модель, в которой истинна формула $\mathit{Grid}\wedge\mathit{Tiling}_T$, достаточно взять сетку $\numN\times\numN$, см.~рис.~\ref{MRybakov:fig:Grid1} (слева), и для каждой пары $(i,j)$ сетки сделать $P$-достижимой из $(i,j)$ $P$-цепь, число элементов в которой равно $k+1$, где $k$ определяется условием $f(i,j)=t_k$, см.~рис.~\ref{MRybakov:fig:Grid}.
\end{proof}


\begin{figure}
\centering
\begin{tikzpicture}[scale=1.5]

\coordinate (g00)      at ( 0.0, 0.0);
\coordinate (g00infr)  at ( 6.2, 2.9);
\coordinate (g00infl)  at (-3.2, 3.5);
\coordinate (gphantom) at ($(g00infr)+(g00infl)$);
\coordinate (gnninf)   at ($0.80*(gphantom)$);

\foreach \s in {0.956} 
{
  \coordinate (g10)      at ($0.2*(g00infr)-0.2*(g00)+(g00)$);
  \coordinate (g10inf)   at ($0.2*(gnninf)-0.2*(g00infl)+(g00infl)$);
  \coordinate (g20)      at ($(g10)   + \s*0.2*(g00infr)- \s*0.2*(g00)$);
  \coordinate (g20inf)   at ($(g10inf)+ \s*0.2*(gnninf) - \s*0.2*(g00infl)$);
  \coordinate (g30)      at ($(g20)   + \s*\s*0.2*(g00infr)- \s*\s*0.2*(g00)$);
  \coordinate (g30inf)   at ($(g20inf)+ \s*\s*0.2*(gnninf) - \s*\s*0.2*(g00infl)$);
  \coordinate (g40)      at ($(g30)   + \s*\s*\s*0.2*(g00infr)- \s*\s*\s*0.2*(g00)$);
  \coordinate (g40inf)   at ($(g30inf)+ \s*\s*\s*0.2*(gnninf) - \s*\s*\s*0.2*(g00infl)$);
  \coordinate (g50)      at ($(g40)   + \s*\s*\s*\s*0.2*(g00infr)- \s*\s*\s*\s*0.2*(g00)$);
  \coordinate (g50inf)   at ($(g40inf)+ \s*\s*\s*\s*0.2*(gnninf) - \s*\s*\s*\s*0.2*(g00infl)$);

  \coordinate (g01)      at ($0.25*(g00infl)-0.25*(g00)+(g00)$);
  \coordinate (g01inf)   at ($0.25*(gnninf)-0.25*(g00infr)+(g00infr)$);
  \coordinate (g02)      at ($(g01)+\s*0.25*(g00infl)-\s*0.25*(g00)$);
  \coordinate (g02inf)   at ($(g01inf)+\s*0.25*(gnninf)-\s*0.25*(g00infr)$);
  \coordinate (g03)      at ($(g02)+\s*\s*0.25*(g00infl)-\s*\s*0.25*(g00)$);
  \coordinate (g03inf)   at ($(g02inf)+\s*\s*0.25*(gnninf)-\s*\s*0.25*(g00infr)$);
  \coordinate (g04)      at ($(g03)+\s*\s*\s*0.25*(g00infl)-\s*\s*\s*0.25*(g00)$);
  \coordinate (g04inf)   at ($(g03inf)+\s*\s*\s*0.25*(gnninf)-\s*\s*\s*0.25*(g00infr)$);
}

\draw [name path = lineLR00] (g00)--(g00infr);
\draw [name path = lineRL00] (g00)--(g00infl);

\draw [name path = line 10] (g10)--(g10inf);
\draw [name path = line 20] (g20)--(g20inf);
\draw [name path = line 30] (g30)--(g30inf);
\draw [name path = line 40] (g40)--(g40inf);
\draw [name path = line 50] (g50)--($(g50)+0.8*(g50inf)-0.8*(g50)$);

\draw [name path = line 01] (g01)--(g01inf);
\draw [name path = line 02] (g02)--(g02inf);
\draw [name path = line 03] (g03)--(g03inf);
\draw [name path = line 04] (g04)--($(g04)+0.8*(g04inf)-0.8*(g04)$);

\draw [name intersections = {of = line 10 and line 01, by = {g11}}];
\draw [name intersections = {of = line 10 and line 02, by = {g12}}];
\draw [name intersections = {of = line 10 and line 03, by = {g13}}];
\draw [name intersections = {of = line 10 and line 04, by = {g14}}];
\draw [name intersections = {of = line 20 and line 01, by = {g21}}];
\draw [name intersections = {of = line 20 and line 02, by = {g22}}];
\draw [name intersections = {of = line 20 and line 03, by = {g23}}];
\draw [name intersections = {of = line 20 and line 04, by = {g24}}];
\draw [name intersections = {of = line 30 and line 01, by = {g31}}];
\draw [name intersections = {of = line 30 and line 02, by = {g32}}];
\draw [name intersections = {of = line 30 and line 03, by = {g33}}];
\draw [name intersections = {of = line 30 and line 04, by = {g34}}];
\draw [name intersections = {of = line 40 and line 01, by = {g41}}];
\draw [name intersections = {of = line 40 and line 02, by = {g42}}];
\draw [name intersections = {of = line 40 and line 03, by = {g43}}];
\draw [name intersections = {of = line 40 and line 04, by = {g44}}];
\draw [name intersections = {of = line 50 and line 01, by = {g51}}];
\draw [name intersections = {of = line 50 and line 02, by = {g52}}];
\draw [name intersections = {of = line 50 and line 03, by = {g53}}];
\draw [name intersections = {of = line 50 and line 04, by = {g54}}];

\begin{scope}[>=latex, ->, shorten >= 1.96pt, shorten <= 1.96pt]
\draw [] (g00) -- (g10);
\draw [] (g10) -- (g20);
\draw [] (g20) -- (g30);
\draw [] (g30) -- (g40);
\draw [] (g40) -- (g50);
\draw [] (g01) -- (g11);
\draw [] (g11) -- (g21);
\draw [] (g21) -- (g31);
\draw [] (g31) -- (g41);
\draw [] (g41) -- (g51);
\draw [] (g02) -- (g12);
\draw [] (g12) -- (g22);
\draw [] (g22) -- (g32);
\draw [] (g32) -- (g42);
\draw [] (g42) -- (g52);
\draw [] (g03) -- (g13);
\draw [] (g13) -- (g23);
\draw [] (g23) -- (g33);
\draw [] (g33) -- (g43);
\draw [] (g43) -- (g53);
\draw [] (g04) -- (g14);
\draw [] (g14) -- (g24);
\draw [] (g24) -- (g34);
\draw [] (g34) -- (g44);
\draw [] (g00) -- (g01);
\draw [] (g01) -- (g02);
\draw [] (g02) -- (g03);
\draw [] (g03) -- (g04);
\draw [] (g10) -- (g11);
\draw [] (g11) -- (g12);
\draw [] (g12) -- (g13);
\draw [] (g13) -- (g14);
\draw [] (g20) -- (g21);
\draw [] (g21) -- (g22);
\draw [] (g22) -- (g23);
\draw [] (g23) -- (g24);
\draw [] (g30) -- (g31);
\draw [] (g31) -- (g32);
\draw [] (g32) -- (g33);
\draw [] (g33) -- (g34);
\draw [] (g40) -- (g41);
\draw [] (g41) -- (g42);
\draw [] (g42) -- (g43);
\draw [] (g43) -- (g44);
\end{scope}

\foreach \d in {0.075pt}
{
\shade [ball color=black] (g00) circle [radius=2.0pt+5*\d]   ;
\shade [ball color=black] (g10) circle [radius=2.0pt+4*\d]   ;
\shade [ball color=black] (g20) circle [radius=2.0pt+3*\d]   ;
\shade [ball color=black] (g30) circle [radius=2.0pt+2*\d]   ;
\shade [ball color=black] (g40) circle [radius=2.0pt+1*\d]   ;
\shade [ball color=black] (g50) circle [radius=2.0pt+0*\d]   ;
\shade [ball color=white] (g01) circle [radius=2.0pt+4*\d]   ;
\shade [ball color=white] (g11) circle [radius=2.0pt+3*\d]   ;
\shade [ball color=white] (g21) circle [radius=2.0pt+2*\d]   ;
\shade [ball color=white] (g31) circle [radius=2.0pt+1*\d]   ;
\shade [ball color=white] (g41) circle [radius=2.0pt+0*\d]   ;
\shade [ball color=white] (g51) circle [radius=2.0pt-1*\d]   ;
\shade [ball color=black] (g02) circle [radius=2.0pt+3*\d]   ;
\shade [ball color=black] (g12) circle [radius=2.0pt+2*\d]   ;
\shade [ball color=black] (g22) circle [radius=2.0pt+1*\d]   ;
\shade [ball color=black] (g32) circle [radius=2.0pt+0*\d]   ;
\shade [ball color=black] (g42) circle [radius=2.0pt-1*\d]   ;
\shade [ball color=black] (g52) circle [radius=2.0pt-2*\d]   ;
\shade [ball color=white] (g03) circle [radius=2.0pt+2*\d]   ;
\shade [ball color=white] (g13) circle [radius=2.0pt+1*\d]   ;
\shade [ball color=white] (g23) circle [radius=2.0pt+0*\d]   ;
\shade [ball color=white] (g33) circle [radius=2.0pt-1*\d]   ;
\shade [ball color=white] (g43) circle [radius=2.0pt-2*\d]   ;
\shade [ball color=white] (g53) circle [radius=2.0pt-3*\d]   ;
\shade [ball color=black] (g04) circle [radius=2.0pt+1*\d]   ;
\shade [ball color=black] (g14) circle [radius=2.0pt+0*\d]   ;
\shade [ball color=black] (g24) circle [radius=2.0pt-1*\d]   ;
\shade [ball color=black] (g34) circle [radius=2.0pt-2*\d]   ;
\shade [ball color=black] (g44) circle [radius=2.0pt-3*\d]   ;
\shade [ball color=black] (g54) circle [radius=2.0pt-4*\d]   ;
}



\drawtile{(g30)}{(g40)}{(g41)}{(g31)}{c3}{c5}{c1}{c2}{0.069*0.75};

\drawtile{(g01)}{(g11)}{(g12)}{(g02)}{c2}{c4}{c1}{c5}{0.073*0.75};

\drawtile{(g22)}{(g32)}{(g33)}{(g23)}{c1}{c5}{c3}{c2}{0.067*0.75};


\drawtilemodela{(g22)}{0.5}{2.0pt+1*0.075pt}{0.35};
\drawtilemodelb{(g30)}{0.52}{2.0pt+2*0.075pt}{0.36};
\drawtilemodelc{(g01)}{0.56}{2.0pt+4*0.075pt}{0.40};

\node [below=2pt] at (g01) {$a$};
\node [below=2pt] at (g22) {$b$};
\node [below=2pt] at (g30) {$c$};
\node [right=2.5pt] at ($(g01)-(0,0.05)$) {$\mathit{tile}_2(a)$};
\node [right=2.5pt] at ($(g22)-(0,0.05)$) {$\mathit{tile}_0(b)$};
\node [right=2.5pt] at ($(g30)-(0,0.05)$) {$\mathit{tile}_1(c)$};

\fill [white,path fading=south] (-3.6,3.0) rectangle (6.2,4.8);
\fill [white,path fading=south] (-3.6,4.0) rectangle (6.2,4.8);

\end{tikzpicture}

\caption{Моделирование плиток домино}
\label{MRybakov:fig:Grid}
\end{figure}

\subsubsection{Уточнения теорем Чёрча и Трахтенброта}
\label{sec:theorems}

Мы получаем следующее уточнение теоремы Чёрча.

\begin{theorem}
\label{th:church}
Логика\/ $\logic{QCl}$ является\/ $\Sigma^0_1$\nobreakdash-полной в языке, содержащем бинарную предикатную букву и три предметные переменные.
\end{theorem}

\begin{proof}
Согласно лемме~\ref{lem:main}, формула $\mathit{Grid}\wedge\mathit{Tiling}_T$ выполнима тогда и только тогда, когда существует $T$-укладка домино. Формула $\mathit{Grid}\wedge\mathit{Tiling}_T$ не содержит предикатных букв, отличных от~$P$, и предметных переменных, отличных от $x$, $y$, и~$z$. Поэтому как проблема выполнимости, так и проблема общезначимости формул соответствующего фрагмента неразрешимы; $\Sigma^0_1$\nobreakdash-полнота мгновенно следует из $\Pi^0_1$\nobreakdash-полноты проблемы домино, которую мы рассматриваем, и из принадлежности логики $\logic{QCl}$ классу~$\Sigma^0_1$.
\end{proof}

Заметим, что в доказательстве леммы~\ref{lem:main} можно обойтись использованием только \defnotion{позитивных} формул, т.е. не содержащих вхождений константы~$\bot$. Для этого достаточно заменить вхождения константы $\bot$ вхождениями формулы $\forall x\forall y\,P(x,y)$, а также заменить выполнимость формулы $\mathit{Grid}\wedge\mathit{Tiling}_T$ на опровержимость формулы $\mathit{Grid}^\ast\wedge\mathit{Tiling}^\ast_T\to \forall x\forall y\,P(x,y)$, где $\mathit{Grid}^\ast$ и $\mathit{Tiling}^\ast_T$ получаются из $\mathit{Grid}$ и $\mathit{Tiling}_T$, соответственно, заменой каждого вхождения константы~$\bot$ вхождением формулы $\forall x\forall y\,P(x,y)$. Как следствие, получаем следующее уточнение теоремы~\ref{th:church}.

\begin{theorem}
\label{th:church:positive}
Позитивный фрагмент логики\/ $\logic{QCl}$ является\/ $\Sigma^0_1$\nobreakdash-полным в языке, содержащем бинарную предикатную букву и три предметные переменные.
\end{theorem}

Аналогичные уточнения можно получить и для теоремы Трахтенброта.\footnote{Здесь приводится слабая формулировка теоремы Трахтенброта; вариант теоремы Трахтенброта, когда утверждается рекурсивная неотделимость фрагментов $\logic{QCl}$ и $\logic{QCl}_{\mathit{fin}}$ тоже справедлив для рассматриваемого фрагмента, но мы не будем его использовать.}

\begin{theorem}
\label{th:trakhtenbrot}
Логика\/ $\logic{QCl}_{\mathit{fin}}$ является\/ $\Pi^0_1$\nobreakdash-полной в языке, содержащем бинарную предикатную букву и три предметные переменные.
\end{theorem}

\begin{proof}
Дадим детальный набросок доказательства.

Напомним, что для доказательства неразрешимости проблемы укладки домино, которую мы рассматриваем, можно свести к ней проблему неостановки машин Тьюринга: каждая строка в сетке $\numN\times\numN$ с укладкой соответствует некоторой конфигурации некоторой машины Тьюринга~$M$ (если $M$ не останавливается на входе, конечно).\footnote{Детали, связанные с машинами Тьюринга, можно посмотреть, например, в~\cite{Papadimitriou,Sipser12}.} Если мы зафиксируем первую строку так, чтобы она соответствовала начальной конфигурации с пустой лентой, то $k$-я строка будет соответствовать $k$-й конфигурации в вычислении $M$ на пустом входе, где $k\in\numN$. Тогда мы получаем ещё два условия, которые можно добавить к $(T_1)$ и~$(T_2)$, не меняя сложности проблемы укладки. Будем считать, что начальная конфигурация машины $M$ имеет вид $q_0\#\Box\Box\Box\ldots$, а заключительная~--- вид $q_e \#s_1s_2s_3\ldots$, где $q_0 $ и $q_e $~--- \defnotion{начальное состояние} и \defnotion{заключительное состояние} машины~$M$, $\#$ и $\Box $~--- \defnotion{граничный маркер} и \defnotion{символ пустой ячейки}, $s_1, s_2,s_3,\ldots$~--- некоторые символы на ленте; тип $t$ плитки домино соответствует либо символу, либо комбинации $qs$, где $q$ является состоянием, а $s$ символом, который видит машина~$M$ в состоянии~$q$. Пусть типы плиток $t_0$, $t_1$ и $t_2$ соответствуют пустой ячейке на ленте, граничному маркеру, просматриваемому $M$ в начальном состоянии, и граничному маркеру, просматриваемому $M$ в заключительном состоянии. Тогда мы можем добавить к $(T_1)$ и $(T_2)$ следующие условия:
\begin{itemize}
\item[]$(T_3)$\quad $f(0,0)=t_1$ и $f(i,0)=t_0$ для каждого $i\in\numN^+$;
\item[]$(T_4)$\quad $f(i,j)\ne t_2$ для любых $i,j\in\numN$.
\end{itemize}

Будем считать, что $T$-укладка удовлетворяет условиям $(T_1)$--$(T_4)$.
Переопределим формулы $G_5$ и~$G_6$:
$$
\begin{array}{lcl}
G'_5
  & =
  & G_5 \wedge
    \forall x\forall y\forall z\,(G(x)\wedge G(y)\wedge G(z) \wedge x\prec_H y \wedge x\prec_V z\to
    \exists z\, y\prec_V z);
  \\
G'_6
  & =
  & \exists x\,(B(x)\wedge G(x)\wedge {}
  \\
  &
  & \phantom{\exists x\,(}
    {} \wedge
    \forall y\,(G(y)\to P(x,y)\vee y\simeq x) \wedge {}
  \\
  &
  & \phantom{\exists x\,(}
    {} \wedge
    \mathit{tile}_1(x)\wedge {}
  \\
  &
  & \phantom{\exists x\,(}
    {} \wedge
    \forall y\,(G(y)\wedge\neg\exists z\,(W(z)\wedge P(x,z)\wedge P(z,y))\to \mathit{tile}_0(y))).
\end{array}
$$
Пусть также
$$
\begin{array}{lcl}
\mathit{Grid}'
  & =
  & G_0\wedge G_3\wedge G_4\wedge G'_5\wedge G'_6;
  \\
\mathit{Stop}
  & =
  & \exists x\,(G(x)\wedge\mathit{tile}_2(x)).
\end{array}
$$
Тогда формула $\mathit{Grid}'\wedge\mathit{Tiling}_T\to \neg\mathit{Stop}$ принадлежит логике $\logic{QCl}_{\mathit{fin}}$ тогда и только тогда, когда существует $T$-укладка плиток домино.
\end{proof}

Как и в случае с теоремой Чёрча, мы можем получить уточнение, связанное с позитивным фрагментом.

\begin{theorem}
\label{th:trakhtenbrot:positive}
Позитивный фрагмент логики\/ $\logic{QCl}_{\mathit{fin}}$ является\/ $\Pi^0_1$\nobreakdash-пол\-ным в языке, содержащем бинарную предикатную букву и три предметные переменные.
\end{theorem}

    \subsection{Теории бинарного отношения}

\subsubsection{Транзитивность и антисимметричность}
\label{sec:trans}

Фактически в доказательстве теорем~\ref{th:church}--\ref{th:trakhtenbrot:positive} (см.~лемму~\ref{lem:main}) было достаточно рассматривать модели с транзитивным и антисимметричным отношением, соответствующим букве~$P$. Поэтому можно убрать из формулы $\mathit{Grid}\wedge\mathit{Tiling}_T$ требования транзитивности и антисимметричности для~$P$, но наложить их на модели, которые мы рассматриваем. В результате мы получим, что теория транзитивного антисимметричного бинарного отношения неразрешима в языке с тремя предметными переменными, а соответствующая теория конечных моделей в том же языке не является рекурсивно перечислимой (с~уточнением, что в первом случае получаем $\Sigma^0_1$-полную теорию, а во втором~--- $\Pi^0_1$-полную). Конечно, эти результаты сохранятся для теории транзитивного бинарного отношения и теории антисимметричного бинарного отношения.

Тем не менее, описанные построения не дают ответ на вопрос о разрешимости, скажем, теории рефлексивного бинарного отношения в языке с тремя предметными переменными. То же можно сказать об иррефлексивности, симметричности и многих других свойствах, часто встречающихся в теориях, связанных с бинарными отношениями, но отсутствующих в моделях формул вида $\mathit{Grid}\wedge\mathit{Tiling}_T$.
Покажем, как можно модифицировать формулу $\mathit{Grid}\wedge\mathit{Tiling}_T$, чтобы получить аналогичные результаты для некоторых таких теорий.

\subsubsection{Рефлексивность и иррефлексивность}
\label{sec:ref:irref}

Заметим, что в доказательстве теоремы~\ref{th:church} и теоремы~\ref{th:trakhtenbrot} можно использовать только иррефлексивные элементы. Для этого достаточно переопределить свойство <<быть белым элементом>>. Пусть
$$
\begin{array}{lcl}
W'(x)
  & =
  & \mathit{tile}_{n+1}(x).
\end{array}
$$
Поскольку $T$ состоит из элементов $t_0,\ldots,t_n$, типа домино $t_{n+1}$ в~$T$ нет. Поэтому, использование формулы $\mathit{tile}_{n+1}(x)$ указанным способом не приводит к противоречию с другими формулами.

Приведём следствия сделанного наблюдения.

\begin{theorem}
\label{th:church:cor}
Следующие теории первого порядка, а также их позитивные фрагменты являются\/ $\Sigma^0_1$\nobreakdash-трудными в языке, содержащем одну бинарную предикатную букву и три предметные переменные:
\begin{itemize}
\item
теория бинарного предиката с любой комбинацией следующих свойств: связность, транзитивность, антисимметричность и иррефлексивность;
\item
теория бинарного предиката с любой комбинацией следующих свойств: связность, транзитивность, антисимметричность и рефлексивность.
\end{itemize}
\end{theorem}

\begin{proof}
Если формула $\mathit{Grid}\wedge \mathit{Tiling}_T$ выполнима, то она истинна в связной транзитивной антисимметричной модели; если мы заменим $W(x)$ на $W'(x)$, мы можем добавить к списку свойств и иррефлексивность.
Для рефлексивности достаточно заметить, что дополнение связного транзитивного антисимметричного иррефлексивного отношения является одновременно связным, транзитивным, антисимметричным и рефлексивным.
\end{proof}

Попутно заметим, что для определения антисимметричности бинарного отношения~$P$ мы используем равенство: $\forall x\forall y\,(P(x,y)\wedge P(y,x)\to x=y)$; тем не менее, теорема~\ref{th:church:cor} не требует наличия равенства в языке теории антисимметричного бинарного отношения. Кроме того, условие связности даже не является первопорядково определимым (что несложно показать, используя, например, игры Эренфойхта--Фрессе, см.~\cite[глава~3]{Libkin}).

\begin{theorem}
\label{th:trakhtenbrot:cor}
Следующие теории первого порядка, а также их позитивные фрагменты являются\/ $\Pi^0_1$\nobreakdash-трудными в языке, содержащем одну бинарную предикатную букву и три предметные переменные:
\begin{itemize}
\item
теория конечных моделей бинарного предиката с любой комбинацией следующих свойств: связность, транзитивность, антисимметричность и иррефлексивность;
\item
теория конечных бинарного предиката с любой комбинацией следующих свойств: связность, транзитивность, антисимметричность и рефлексивность.
\end{itemize}
\end{theorem}

\begin{proof}
Аналогично доказательству теоремы~\ref{th:church:cor}.
\end{proof}

\subsubsection{Симметричность}
\label{sec:sym}

Что можно сказать о разрешимости теории симметричного бинарного предиката при тех же ограничениях на средства языка? Известно (см.~\cite{NerodeShore80} и~\cite[приложение]{Kremer97}), что теория симметричного иррефлексивного бинарного отношения неразрешима. Покажем, что для доказательства неразрешимости классической логики первого порядка в языке, содержащем бинарную предикатную букву $P$ и три предметные переменные, достаточно использовать модели, интерпретирующие $P$ симметричным иррефлексивным бинарным отношением; для краткости такие отношения будем называть \defnotion{sib-отношениями},\index{отношение!sib@sib-отношение} а такие модели~--- \defnotion{sib-моделями}.\index{модель!sib@sib-модель} Теорию sib-отношений будем обозначать $\logic{SIB}$. Отметим, что как sib-модели, так и теория $\logic{SIB}$ представляют интерес не только для исследований в классической логике или теории графов (sib-модели~--- это, по сути, простые графы), но и для исследований в области неклассических логик~\cite{Kremer97,MR:2017:LI}.

Начнём с дополнительных определений и переопределений некоторых старых <<понятий>>, т.е. введённых сокращений. В определяемых ниже формулах можно понимать $P$ как sib-отношение (немного позже мы потребуем выполнение соответствующих свойств).

Пусть
$$
\begin{array}{lcl}
\mathit{tr}(x,y,z)
  & =
  & P(x,y)\wedge P(y,z)\wedge P(z,x);
  \smallskip\\
\mathit{Tr}(x)
  & =
  & \exists y\exists z\,\mathit{tr}(x,y,z);
  \smallskip\\
A_0(x)
  & =
  & \forall y\forall z\,(P(x,y)\wedge P(x,z)\to y\simeq z);
  \smallskip\\
A_1(x)
  & =
  & \neg\mathit{Tr}(x)\wedge \exists y\,(P(x,y)\wedge A_0(y));
  \smallskip\\
A_{k+2}(x)
  & =
  & \displaystyle
    \neg\mathit{Tr}(x)\wedge \exists y\,(P(x,y)\wedge A_{k+1}(y)\wedge \bigwedge\limits_{\mathclap{i=0}}^{k}\neg A_i(y));
  \smallskip\\
G'(x)
  & =
  & \displaystyle
    \neg\mathit{Tr}(x)\wedge \exists y\,(P(x,y)\wedge \mathit{Tr}(y))\wedge \neg\exists y\,(P(x,y)\wedge\bigvee\limits_{\mathclap{i=0}}^{n+2} A_i(y));
  \smallskip\\
\mathit{tile}'_k(x)
  & =
  & \exists y\,(P(x,y)\wedge\exists x\exists z\,(\mathit{tr}(x,y,z)\wedge {}
  \\
  &
  & \phantom{\exists y\,(}
    {} \wedge
    \exists y\,(P(x,y)\wedge A_k(y)) \wedge \neg\exists y\,(P(z,y)\wedge\neg\mathit{Tr}(y))));
  \smallskip\\
\mathit{num}_k(x)
  & =
  & \exists y\,(P(x,y)\wedge\exists x\exists z\,(\mathit{tr}(x,y,z)\wedge {}
  \\
  &
  & \phantom{\exists y\,(}
    {} \wedge
    \exists y\,(P(x,y)\wedge A_k(y)) \wedge \exists y\,(P(z,y)\wedge A_0(y))));
  \smallskip\\
x\prec'_H y
  & =
  & P(x,y) \wedge ((\mathit{num}_1(x)\wedge \mathit{num}_2(y))\vee (\mathit{num}_2(x)\wedge \mathit{num}_3(y))\vee{}
  \\
  &
  & \phantom{P(x,y) \wedge ((\mathit{num}_1(x)\wedge \mathit{num}_2(y))}\vee
    (\mathit{num}_3(x)\wedge \mathit{num}_1(y)));
  \smallskip\\
x\prec'_V y
  & =
  & P(x,y) \wedge ((\mathit{num}_4(x)\wedge \mathit{num}_5(y))\vee (\mathit{num}_5(x)\wedge \mathit{num}_6(y))\vee {}
  \\
  &
  & \phantom{P(x,y) \wedge ((\mathit{num}_4(x)\wedge \mathit{num}_5(y))} \vee
    (\mathit{num}_6(x)\wedge \mathit{num}_4(y))).
\end{array}
$$

Подразумеваемый смысл введённых сокращений следующий:
$$
\begin{array}{rcl}
\mathit{tr}(x,y,z) & \mbox{---} & \mbox{$x,y,z$ образуют треугольник;} \\
\mathit{Tr}(x)     & \mbox{---} & \mbox{$x$ входит в треугольник;} \\
A_k(x)             & \mbox{---} & \mbox{$x$ видит висячую вершину за $n$ шагов;} \\
G'(x)              & \mbox{---} & \mbox{$x$ является элементом сетки;} \\
x\prec'_H y & \mbox{---} & \mbox{$y$ находится непосредственно справа от $x$;} \\
x\prec'_V y & \mbox{---} & \mbox{$y$ находится непосредственно над $x$;} \\
\mathit{tile}'_k(x) & \mbox{---} & \mbox{$x$ содержит плитку типа $t_k$;} \\
\mathit{num}_k(x)   & \mbox{---} & \mbox{$x$ помечен числом $k$.} \\
\end{array}
$$

\begin{figure}
\centering
\begin{tikzpicture}[scale=1.5]

\coordinate (g00)   at (0,0);
\coordinate (g10)   at (1,0);
\coordinate (g20)   at (2,0);
\coordinate (g30)   at (3,0);
\coordinate (g40)   at (4,0);
\coordinate (g50)   at (4.75,0);
\coordinate (g60)   at (5,0);

\coordinate (g10f)   at ($(g10) - (0.5,0.5)$);
\coordinate (g10e)   at ($(g10) + (0.5,0.5)$);
\coordinate (g30f)   at ($(g30) - (0.5,0.5)$);
\coordinate (g30e)   at ($(g30) + (0.5,0.5)$);

\coordinate (t10)       at (1,1);
\coordinate (t10left)   at (0.5,1.866);
\coordinate (t10right)  at (1.5,1.866);
\coordinate (t10left0)  at (0.5,1.866+0.75*1);
\coordinate (t10left1)  at (0.5,1.866+0.75*2);
\coordinate (t10left2)  at (0.5,1.866+0.75*3);

\coordinate (t30)       at (3,1);
\coordinate (t30left)   at (2.5,1.866);
\coordinate (t30right)  at (3.5,1.866);
\coordinate (t30left0)  at (2.5,1.866+0.75*1);
\coordinate (t30left1)  at (2.5,1.866+0.75*2);
\coordinate (t30left2)  at (2.5,1.866+0.75*3);
\coordinate (t30right0) at (3.5,1.866+0.75*1);

\begin{scope}[>=latex, <->, shorten >= 1.96pt, shorten <= 1.96pt]
\draw [black, >=latex,  ->, shorten >= 1.96pt, shorten <= 9.96pt] (g00) -- (g10);
\draw [black, >=latex, <- , shorten >= 9.96pt, shorten <= 1.96pt] (g10) -- (g20);
\draw [black, >=latex,  ->, shorten >= 1.96pt, shorten <= 9.96pt] (g20) -- (g30);
\draw [black, >=latex, <- , shorten >= 9.96pt, shorten <= 1.96pt] (g30) -- (g40);
\draw [black, >=latex, <- , shorten >= 0.65*9.96pt, shorten <= 1.96pt] (g10) -- (g10f);
\draw [black, >=latex, <- , shorten >= 0.65*9.96pt, shorten <= 1.96pt] (g10) -- (g10e);
\draw [black, >=latex, <- , shorten >= 0.65*9.96pt, shorten <= 1.96pt] (g30) -- (g30f);
\draw [black, >=latex, <- , shorten >= 0.65*9.96pt, shorten <= 1.96pt] (g30) -- (g30e);

\draw [black] (g10)      -- (t10);
\draw [black] (t10)      -- (t10left);
\draw [black] (t10)      -- (t10right);
\draw [black] (t10left)  -- (t10right);
\draw [black] (t10left)  -- (t10left0);
\draw [black] (t10left0) -- (t10left1);
\draw [black] (t10left1) -- (t10left2);

\draw [black] (g30)      -- (t30);
\draw [black] (t30)      -- (t30left);
\draw [black] (t30)      -- (t30right);
\draw [black] (t30left)  -- (t30right);
\draw [black] (t30left)  -- (t30left0);
\draw [black] (t30left0) -- (t30left1);
\draw [black] (t30left1) -- (t30left2);
\draw [black] (t30right) -- (t30right0);

\end{scope}

\shade [ball color=black] (g10) circle [radius=1.75pt]   ;
\shade [ball color=black] (g30) circle [radius=1.75pt]   ;
\shade [ball color=black] (t10) circle [radius=1.75pt]   ;
\shade [ball color=black] (t30) circle [radius=1.75pt]   ;
\shade [ball color=black] (t10left)   circle [radius=1.75pt]   ;
\shade [ball color=black] (t10left0)  circle [radius=1.75pt]   ;
\shade [ball color=black] (t10left1)  circle [radius=1.75pt]   ;
\shade [ball color=black] (t10left2)  circle [radius=1.75pt]   ;
\shade [ball color=black] (t10right)  circle [radius=1.75pt]   ;
\shade [ball color=black] (t30left)   circle [radius=1.75pt]   ;
\shade [ball color=black] (t30left0)  circle [radius=1.75pt]   ;
\shade [ball color=black] (t30left1)  circle [radius=1.75pt]   ;
\shade [ball color=black] (t30left2)  circle [radius=1.75pt]   ;
\shade [ball color=black] (t30right)  circle [radius=1.75pt]   ;
\shade [ball color=black] (t30right0) circle [radius=1.75pt]   ;

\node [black] at (g00) {$\cdots$};
\node [black] at (g20) {$\cdots$};
\node [black] at (g40) {$\cdots$};
\node [black] at ($(g10f)+(0.012,0)$) {$\cdot$};
\node [black] at ($(g10f)+(0.05,0.05)+(0.012,0)$) {$\cdot$};
\node [black] at ($(g10f)-(0.05,0.05)+(0.012,0)$) {$\cdot$};
\node [black] at ($(g10e)+(0.012,0)$) {$\cdot$};
\node [black] at ($(g10e)+(0.05,0.05)+(0.012,0)$) {$\cdot$};
\node [black] at ($(g10e)-(0.05,0.05)+(0.012,0)$) {$\cdot$};
\node [black] at ($(g30f)+(0.012,0)$) {$\cdot$};
\node [black] at ($(g30f)+(0.05,0.05)+(0.012,0)$) {$\cdot$};
\node [black] at ($(g30f)-(0.05,0.05)+(0.012,0)$) {$\cdot$};
\node [black] at ($(g30e)+(0.012,0)$) {$\cdot$};
\node [black] at ($(g30e)+(0.05,0.05)+(0.012,0)$) {$\cdot$};
\node [black] at ($(g30e)-(0.05,0.05)+(0.012,0)$) {$\cdot$};

\node [above left ] at (g10) {$a$};
\node [above left ] at (g30) {$c$};
\node [below right] at (g10) {$\mathit{tile}'_2(a)$};
\node [below right] at (g30) {$\mathit{num}_2(c)$}  ;

\end{tikzpicture}

\caption{Моделирование плиток домино и чисел}
\label{MRybakov:fig:Grid2}
\end{figure}

Мы будем использовать треугольники, чтобы отделить элементы сетки от элементов, моделирующих типы плиток домино и числа, см.~рис.~\ref{MRybakov:fig:Grid2}.

Используя введённые <<понятия>>, можно описать сетку~$\mathds{N}\times\mathds{N}$:
\settowidth{\templength}{\mbox{$\forall x\,($}}
$$
\begin{array}{rcl}
  G''_0
    & =
    & \forall x\forall y\,(P(x,y)\to P(y,x))\wedge \forall x\,\neg P(x,x);
    \\
  G''_1
    & =
    & \forall x\,(G'(x)\to \exists y\,(G'(y)\wedge x\prec'_H y));
    \\
  G''_2
    & =
    & \forall x\,(G'(x)\to \exists y\,(G(y)\wedge x\prec'_V y));
    \\
  G''_3
    & =
    & \forall x\forall y\forall z\,(G'(x)\wedge G'(y)\wedge G'(z)\to (x\prec'_H y\wedge x\prec'_H z \to y\simeq z));
    \\
  G''_4
    & =
    & \forall x\forall y\forall z\,(G'(x)\wedge G'(y)\wedge G'(z)\to (x\prec'_V y\wedge x\prec'_V z \to y\simeq z));
    \\
  G''_5
    & =
    & \forall x\forall y\,(G'(x)\wedge G'(y)\to {}
    \\
    &
    & \phantom{\forall x\forall y\,(}
      {} \to
      (\exists z\,(G'(z)\wedge x\prec'_Hz\wedge z\prec'_Vy)\leftrightarrow {}
      \\
      && \hfill {} \leftrightarrow\exists z\,(G'(z)\wedge x\prec'_Vz\wedge z\prec'_Hy)));
    \\
  G''_6
    & =
    & \exists x\,(\mathit{num}_1(x)\wedge \mathit{num}_4(x)\wedge G'(x));
    \\
  G''_7
    & =
    & \forall x\,(
      (\mathit{num}_1(x)\to \neg \mathit{num}_2(x)\wedge \neg \mathit{num}_3(x)) \wedge {}
    \\
    &
    & 
      \parbox{\templength}{\hfill ${}\wedge{}$}
      (\mathit{num}_2(x)\to \neg \mathit{num}_1(x)\wedge \neg \mathit{num}_3(x)) \wedge {}
    \\
    &
    & 
      \parbox{\templength}{\hfill ${}\wedge{}$}
      (\mathit{num}_3(x)\to \neg \mathit{num}_1(x)\wedge \neg \mathit{num}_2(x)) \wedge {}
    \\
    &
    & 
      \parbox{\templength}{\hfill ${}\wedge{}$}
      (\mathit{num}_4(x)\to \neg \mathit{num}_5(x)\wedge \neg \mathit{num}_6(x)) \wedge {}
    \\
    &
    & 
      \parbox{\templength}{\hfill ${}\wedge{}$}
      (\mathit{num}_5(x)\to \neg \mathit{num}_4(x)\wedge \neg \mathit{num}_6(x)) \wedge {}
    \\
    &
    & 
      \parbox{\templength}{\hfill ${}\wedge{}$}
      (\mathit{num}_6(x)\to \neg \mathit{num}_4(x)\wedge \neg \mathit{num}_5(x)));
    \\
  \mathit{Grid}''
    & =
    & G''_0\wedge G''_1\wedge G''_2\wedge G''_3\wedge G''_4\wedge G''_5\wedge G''_6\wedge G''_7.
\end{array}
$$

\begin{figure}
\centering
\begin{tikzpicture}[scale=1.32]

\coordinate (g00)   at (0,0);
\coordinate (g10)   at (1,0);
\coordinate (g20)   at (2,0);
\coordinate (g30)   at (3,0);
\coordinate (g40)   at (4,0);
\coordinate (g50)   at (5,0);
\coordinate (g01)   at (0,1);
\coordinate (g11)   at (1,1);
\coordinate (g21)   at (2,1);
\coordinate (g31)   at (3,1);
\coordinate (g41)   at (4,1);
\coordinate (g51)   at (5,1);
\coordinate (g02)   at (0,2);
\coordinate (g12)   at (1,2);
\coordinate (g22)   at (2,2);
\coordinate (g32)   at (3,2);
\coordinate (g42)   at (4,2);
\coordinate (g52)   at (5,2);
\coordinate (g03)   at (0,3);
\coordinate (g13)   at (1,3);
\coordinate (g23)   at (2,3);
\coordinate (g33)   at (3,3);
\coordinate (g43)   at (4,3);
\coordinate (g53)   at (5,3);
\coordinate (g04)   at (0,4);
\coordinate (g14)   at (1,4);
\coordinate (g24)   at (2,4);
\coordinate (g34)   at (3,4);
\coordinate (g44)   at (4,4);
\coordinate (g54)   at (5,4);
\coordinate (g05)   at (0,5);
\coordinate (g15)   at (1,5);
\coordinate (g25)   at (2,5);
\coordinate (g35)   at (3,5);
\coordinate (g45)   at (4,5);
\coordinate (g55)   at (5,5);

\begin{scope}[>=latex, <->, shorten >= 1.5pt, shorten <= 1.5pt]
\draw [] (g00) -- (g10);
\draw [] (g10) -- (g20);
\draw [] (g20) -- (g30);
\draw [] (g30) -- (g40);
\draw [<-, shorten >= 14.5pt] (g40) -- (g50);
\draw [] (g01) -- (g11);
\draw [] (g11) -- (g21);
\draw [] (g21) -- (g31);
\draw [] (g31) -- (g41);
\draw [<-, shorten >= 14.5pt] (g41) -- (g51);
\draw [] (g02) -- (g12);
\draw [] (g12) -- (g22);
\draw [] (g22) -- (g32);
\draw [] (g32) -- (g42);
\draw [<-, shorten >= 14.5pt] (g42) -- (g52);
\draw [] (g03) -- (g13);
\draw [] (g13) -- (g23);
\draw [] (g23) -- (g33);
\draw [] (g33) -- (g43);
\draw [<-, shorten >= 14.5pt] (g43) -- (g53);
\draw [] (g04) -- (g14);
\draw [] (g14) -- (g24);
\draw [] (g24) -- (g34);
\draw [] (g34) -- (g44);
\draw [<-, shorten >= 14.5pt] (g44) -- (g54);
\draw [] (g00) -- (g01);
\draw [] (g01) -- (g02);
\draw [] (g02) -- (g03);
\draw [] (g03) -- (g04);
\draw [<-, shorten >= 14.5pt] (g04) -- (g05);
\draw [] (g10) -- (g11);
\draw [] (g11) -- (g12);
\draw [] (g12) -- (g13);
\draw [] (g13) -- (g14);
\draw [<-, shorten >= 14.5pt] (g14) -- (g15);
\draw [] (g20) -- (g21);
\draw [] (g21) -- (g22);
\draw [] (g22) -- (g23);
\draw [] (g23) -- (g24);
\draw [<-, shorten >= 14.5pt] (g24) -- (g25);
\draw [] (g30) -- (g31);
\draw [] (g31) -- (g32);
\draw [] (g32) -- (g33);
\draw [] (g33) -- (g34);
\draw [<-, shorten >= 14.5pt] (g34) -- (g35);
\draw [] (g40) -- (g41);
\draw [] (g41) -- (g42);
\draw [] (g42) -- (g43);
\draw [] (g43) -- (g44);
\draw [<-, shorten >= 14.5pt] (g44) -- (g45);
\end{scope}

\node [] at (g50)   {$\cdots$}     ;
\node [] at (g51)   {$\cdots$}     ;
\node [] at (g52)   {$\cdots$}     ;
\node [] at (g53)   {$\cdots$}     ;
\node [] at (g54)   {$\cdots$}     ;

\node [] at (g05)   {$\vdots$}     ;
\node [] at (g15)   {$\vdots$}     ;
\node [] at (g25)   {$\vdots$}     ;
\node [] at (g35)   {$\vdots$}     ;
\node [] at (g45)   {$\vdots$}     ;

\node [below right] at (g00)   {$1$}     ;
\node [below right] at (g10)   {$2$}     ;
\node [below right] at (g20)   {$3$}     ;
\node [below right] at (g30)   {$1$}     ;
\node [below right] at (g40)   {$2$}     ;
\node [below right] at (g01)   {$1$}     ;
\node [below right] at (g11)   {$2$}     ;
\node [below right] at (g21)   {$3$}     ;
\node [below right] at (g31)   {$1$}     ;
\node [below right] at (g41)   {$2$}     ;
\node [below right] at (g02)   {$1$}     ;
\node [below right] at (g12)   {$2$}     ;
\node [below right] at (g22)   {$3$}     ;
\node [below right] at (g32)   {$1$}     ;
\node [below right] at (g42)   {$2$}     ;
\node [below right] at (g03)   {$1$}     ;
\node [below right] at (g13)   {$2$}     ;
\node [below right] at (g23)   {$3$}     ;
\node [below right] at (g33)   {$1$}     ;
\node [below right] at (g43)   {$2$}     ;
\node [below right] at (g04)   {$1$}     ;
\node [below right] at (g14)   {$2$}     ;
\node [below right] at (g24)   {$3$}     ;
\node [below right] at (g34)   {$1$}     ;
\node [below right] at (g44)   {$2$}     ;

\node [above left ] at (g00)   {$4$}     ;
\node [above left ] at (g01)   {$5$}     ;
\node [above left ] at (g02)   {$6$}     ;
\node [above left ] at (g03)   {$4$}     ;
\node [above left ] at (g04)   {$5$}     ;
\node [above left ] at (g10)   {$4$}     ;
\node [above left ] at (g11)   {$5$}     ;
\node [above left ] at (g12)   {$6$}     ;
\node [above left ] at (g13)   {$4$}     ;
\node [above left ] at (g14)   {$5$}     ;
\node [above left ] at (g20)   {$4$}     ;
\node [above left ] at (g21)   {$5$}     ;
\node [above left ] at (g22)   {$6$}     ;
\node [above left ] at (g23)   {$4$}     ;
\node [above left ] at (g24)   {$5$}     ;
\node [above left ] at (g30)   {$4$}     ;
\node [above left ] at (g31)   {$5$}     ;
\node [above left ] at (g32)   {$6$}     ;
\node [above left ] at (g33)   {$4$}     ;
\node [above left ] at (g34)   {$5$}     ;
\node [above left ] at (g40)   {$4$}     ;
\node [above left ] at (g41)   {$5$}     ;
\node [above left ] at (g42)   {$6$}     ;
\node [above left ] at (g43)   {$4$}     ;
\node [above left ] at (g44)   {$5$}     ;


\filldraw [] (g00) circle [radius=1.5pt]   ;
\filldraw [] (g10) circle [radius=1.5pt]   ;
\filldraw [] (g20) circle [radius=1.5pt]   ;
\filldraw [] (g30) circle [radius=1.5pt]   ;
\filldraw [] (g40) circle [radius=1.5pt]   ;

\filldraw [] (g01) circle [radius=1.5pt]   ;
\filldraw [] (g11) circle [radius=1.5pt]   ;
\filldraw [] (g21) circle [radius=1.5pt]   ;
\filldraw [] (g31) circle [radius=1.5pt]   ;
\filldraw [] (g41) circle [radius=1.5pt]   ;

\filldraw [] (g02) circle [radius=1.5pt]   ;
\filldraw [] (g12) circle [radius=1.5pt]   ;
\filldraw [] (g22) circle [radius=1.5pt]   ;
\filldraw [] (g32) circle [radius=1.5pt]   ;
\filldraw [] (g42) circle [radius=1.5pt]   ;

\filldraw [] (g03) circle [radius=1.5pt]   ;
\filldraw [] (g13) circle [radius=1.5pt]   ;
\filldraw [] (g23) circle [radius=1.5pt]   ;
\filldraw [] (g33) circle [radius=1.5pt]   ;
\filldraw [] (g43) circle [radius=1.5pt]   ;

\filldraw [] (g04) circle [radius=1.5pt]   ;
\filldraw [] (g14) circle [radius=1.5pt]   ;
\filldraw [] (g24) circle [radius=1.5pt]   ;
\filldraw [] (g34) circle [radius=1.5pt]   ;
\filldraw [] (g44) circle [radius=1.5pt]   ;


\end{tikzpicture}
\caption{Сетка для sib-отношения}
\label{MRybakov:fig:Grid3}
\end{figure}

Такая сетка изображена на рис.~\ref{MRybakov:fig:Grid3}, где для каждого элемента $a$ сетки числа рядом с ним указывают на то, какие из утверждений $\mathit{num}_1(a),\ldots,\mathit{num}_6(a)$ истинны.

Чтобы описать свойства $T$-укладки, достаточно заменить $G$, $\prec_H$ и $\prec_V$ в формуле $\mathit{Tiling}_T$ на $G'$, $\prec'_H$ и $\prec'_V$, соответственно: пусть
$$
\begin{array}{rcl}
  T'_0
    & =
    & \displaystyle
      \forall x\,(G'(x)\to \bigvee\limits_{i=0}^n(\mathit{tile}'_i(x)\wedge \bigwedge\limits_{j\ne i}\neg\mathit{tile}'_j(x)));
    \smallskip\\
  T'_1
    & =
    & \displaystyle
      \forall x\,(G'(x)\to \bigwedge\limits_{i=0}^{n} (\mathit{tile}'_i(x)\to \forall y\,(G'(y)\wedge x\prec'_H y\to \bigvee\limits_{\mathclap{\mathop{\scriptsize\rightsquare{0.21}} t_i\,{=}\,\mathop{\scriptsize\leftsquare{0.21}} t_j}}\mathit{tile}'_j(y))));
    \smallskip\\
  T'_2
    & =
    & \displaystyle
      \forall x\,(G'(x)\to \bigwedge\limits_{i=0}^{n} (\mathit{tile}'_i(x)\to \forall y\,(G'(y)\wedge x\prec'_V y\to \bigvee\limits_{\mathclap{\mathop{\scriptsize\upsquare{0.21}} t_i\,{=}\,\mathop{\scriptsize\downsquare{0.21}} t_j}}\mathit{tile}'_j(y))));
    \smallskip\\
  \mathit{Tiling}'_T
    & =
    & T'_0\wedge T'_1\wedge T'_2.
\end{array}
$$

\begin{theorem}
\label{th:church:cor2}
Позитивный фрагмент теории класса всех моделей, определяемых тем, что бинарный предикат в них удовлетворяет любой комбинации таких свойств как связность, интранзитивность, симметричность, серийность и иррефлексивность, является\/ $\Sigma^0_1$\nobreakdash-трудным в языке с одной бинарной предикатной буквой и тремя предметными переменными.
\end{theorem}

\begin{proof}
Чтобы свести проблему укладки домино к проблеме выполнимости в такой теории, достаточно использовать формулу $\mathit{Grid}''\wedge \mathit{Tiling}'_T$, определяемую по задаче домино~$T$.
\end{proof}

\begin{corollary}
\label{cor:th:church:cor2}
Позитивный фрагмент теории\/ $\logic{SIB}$ является\/ $\Sigma^0_1$\nobreakdash-пол\-ным в языке с одной бинарной предикатной буквой и тремя предметными переменными.
\end{corollary}

    \subsection{Замечания}



Отметим, что теоремы~\ref{th:church}--\ref{th:church:cor2} могут быть расширены на некоторые другие теории первого порядка. Например, в теоремах~\ref{th:church:cor} и~\ref{th:trakhtenbrot:cor} мы можем добавить условие серийности: при моделировании типов плиток домино достаточно заменить все максимальные иррефлексивные элементы на рефлексивне и соответствующим образом переопределить формулы $\mathit{tile}_k(x)$ и~$W(x)$. Кроме того, мы можем добавить некоторые свойства для моделей <<бесплатно>>: так, в терминах теории графов, достаточно использовать модели, которые являются планарными, $2$-раскрашиваемыми ($3$-раскрашиваемыми в случае симметричного бинарного отношения) и~т.д. графами. Тем не менее, автор не знает, применим ли предлагаемый метод к разным видам деревьев или конечным sib-моделям; возможно, некоторые из таких вопросов являются темами для будущих исследований, но, возможно, ответы на некоторые из них получаются очень просто.

Метод, который был использован, фактически является модификацией метода моделирования модальных пропозициональных формул формулами от одной переменной. Те части модели, которые отвечали за моделирование типов плиток домино, очень похожи на модели Крипке, которые соответствовали формулам от одной переменной, использовавшимся для подстановки в модальную формулу вместо её переменных.
Также обратим внимание на то, что формулы типа $\mathit{tile}_k(x)$ соответствуют в моделях некоторым унарным предикатам, т.е. можно было бы проделать моделирование укладки плиток домино в более богатом языке (с~унарными предикатными буквами), а потом осуществить подстановку. Эту идею мы ещё применим в модальных предикатных логиках.

\setcounter{savefootnote}{\value{footnote}}
\chapter{Модальные логики}
\setcounter{footnote}{\value{savefootnote}}
      \label{ch:7}
  \section{Основные определения и факты}
    \subsection{Синтаксис}

Будем считать, что исходными символами модального предикатного языка $\lang{QML}$\index{уян@язык!qml@$\lang{QML}$} являются исходные символы классического предикатного языка $\lang{QL}$ и символ~$\Box$. \defnotion{Модальные предикатные формулы}\index{уяа@формула!предикатная!модальная} определяются так же, как и $\lang{QL}$-формулы, с добавлением условия, что если $\varphi$~--- формула, то $\Box\varphi$~--- тоже формула. Такие формулы будем называть также $\lang{QML}$-формулами, при их записи будем использовать введённые выше сокращения и договорённости. Формально под языком $\lang{QML}$ понимаем множество всех $\lang{QML}$-формул\index{уяа@формула!ql@$\lang{QML}$-формула}.

Под \defnotion{модальной предикатной логикой}\index{логика!предикатная!модальная} будем понимать любое множество $\lang{QML}$-формул, замкнутое относительно \defnotion{правила предикатной подстановки}\index{правило!предикатной подстановки}\footnote{Детальное обсуждение предикатной подстановки можно найти в~\cite[\S2.3, \S2.5]{GShS}. Отметим, что мы будем использовать предикатную подстановку в некоторых построениях, но в очень простых её вариантах; при этом все необходимые пояснения будут даны.} (по формуле $\varphi$ получается некоторый её подстановочный пример). Мы будем рассматривать только такие модальные предикатные логики, которые расширяют классическую логику предикатов~$\logic{QCl}$, замкнуты по \MP\ (по формулам $\varphi$ и $\varphi\to\psi$ получается формула~$\psi$) и \defnotion{правилу обобщения}\index{правило!обобщения} (по формуле $\varphi$ получается формула~$\forall x\,\varphi$, где $x$ может быть любой предметной переменной).

Пусть $L$~--- нормальная модальная пропозициональная логика. Определим логику $\logic{Q}L$ как наименьшее множество формул, содержащее $\logic{QCl}$, все подстановочные примеры формул из $L$ в языке~$\lang{QML}$, а также замкнутое по правилу предикатной подстановки, \MP, правилу Гёделя и правилу обобщения. Можно считать, что $\logic{Q}L$ содержит~$L$: мы получим такое включение, если в формулах языка логики $L$ каждое вхождение каждой пропозициональной переменной~$p$ заменим вхождением выражения~$p(\curlywedge)$; с учётом того, что мы договорились (раздел~\ref{sec:QCl:syntax:semantics}) вместо $p(\curlywedge)$ писать~$p$, в записи $\lang{QML}$-формул мы этих изменений не увидим. Логику $\logic{Q}L$ будем называть \defnotion{предикатным вариантом}, или \defnotion{предикатным напарником}\index{напарник!предикатный} логики~$L$.

Модальную предикатную логику называем \defnotion{нормальной},\index{логика!предикатная!нормальная модальная} если она содержит логику $\logic{QK}$ и замкнута по \MP, правилу обобщения и правилу Гёделя (по формуле $\varphi$ получается формула~$\Box\varphi$).

    \subsection{Семантика Крипке}

Для оценки истинности $\lang{QML}$-формул будем использовать семантику Крипке; дадим соответствующие определения.

Как и раньше, \defnotion{шкалой Крипке}\index{уяи@шкала!Крипке} называем пару $\kframe{F} = \langle W,R\rangle$, где $W$~--- непустое множество \defnotion{миров}, а $R$~--- бинарное \defnotion{отношение достижимости}\index{отношение!достижимости} на множестве~$W$.

Пусть $\kframe{F} = \otuple{W,R}$~--- шкала Крипке, $\mathcal{D}$~--- некоторое непустое множество, $\function{D}{W}{\Power{\mathcal{D}}}$. Пару $\kFrame{F} = \otuple{\kframe{F},D}$ называем \defnotion{шкалой Крипке с предметными областями},\index{уяи@шкала!с предметными областями} если $D(w)\ne\varnothing$ для любого $w\in W$, а также для любых $w,w'\in W$
$$
\begin{array}{lcl}
wRw' & \imply & D(w) \subseteq D(w').
\end{array}
\eqno{\mbox{$(\mathit{ED})$}}
$$
Множество $D(w)$ будем обозначать также $D_w$ и называть \defnotion{предметной областью мира\/~$w$},\index{область!предметная!мира} а про шкалу~$\kFrame{F}$ будем говорить, что она \defnotion{определена на шкале~$\kframe{F}$}, при этом наряду с записью $\kFrame{F} = \otuple{\kframe{F},D}$ будем использовать запись $\kFrame{F} = \otuple{W,R,D}$. Множество
$$
\begin{array}{lcl}
D^+ & = & \displaystyle\bigcup\limits_{\mathclap{w\in W}}D_w
\end{array}
$$
будем называть \defnotion{предметной областью}\index{область!предметная!уяи@шкалы} шкалы~$\kFrame{F}$; говорим также, что шкала $\kFrame{F}$ \defnotion{построена на шкале~$\kframe{F}$}.

Условие~\mbox{$(\mathit{ED})$} называется \defnotion{условием расширяющихся областей}.\index{условие!расширяющихся областей}
Мы будем рассматривать шкалы с предметной областью, удовлетворяющие более сильным условиям. Первое состоит в том, что для любых~$w,w'\in W$
$$
\begin{array}{lcl}
wRw' & \imply & D(w) = D(w'),
\end{array}
\eqno{\mbox{$(\mathit{LCD})$}}
$$
а второе~--- в том, что для любых~$w,w'\in W$
$$
\begin{array}{c}
D(w) = D(w').
\end{array}
\eqno{\mbox{$(\mathit{GCD})$}}
$$
Условие~\mbox{$(\mathit{LCD})$} называется \defnotion{условием локально постоянных областей},\index{условие!локально постоянных областей} а условие~\mbox{$(\mathit{GCD})$}~--- \defnotion{условием глобально постоянных областей}.\index{условие!глобально постоянных областей} Обсуждение этих условий пока отложим; отметим лишь, что из~\mbox{$(\mathit{GCD})$} следует~\mbox{$(\mathit{LCD})$}, а из~\mbox{$(\mathit{LCD})$} следует~\mbox{$(\mathit{ED})$}.

Если шкала $\kFrame{F}=\otuple{W,R,D}$ удовлетворяет условию~\mbox{$(\mathit{GCD})$}, $\mathcal{D}$~--- её предметная область (а~значит, $D(w)=\mathcal{D}$ для каждого $w\in W$) и $\kframe{F}=\otuple{W,R}$, то иногда будем использовать также записи $\kFrame{F}=\otuple{W,R,\mathcal{D}}$, $\kFrame{F}=\otuple{\kframe{F},\mathcal{D}}$ и $\kFrame{F}=\kframe{F}\odot\mathcal{D}$.

\defnotion{Моделью Крипке}\index{модель!Крипке} на шкале $\kFrame{F}=\otuple{W,R,D}$ называем пару $\kModel{M} = \otuple{\kFrame{F},I}$, где $I$~--- \defnotion{интерпретация предикатных букв}\index{интерпретация!предикатных букв} в предметных областях миров множества~$W$, т.е. функция, сопоставляющая каждому $w\in W$ и $n$-местной предикатной букве~$P$ некоторое $n$-местное отношение $I(w,P)$ на $D_w$, т.е. $I(w,P) \subseteq D_w^n$; в частности, если $p$~--- пропозициональная буква ($0$-местная предикатная буква), то $I(w,P) \subseteq D_w^0 = \set{\otuple{\curlywedge}}$. Интерпретацию предикатных букв называют также \defnotion{оценкой}.\index{оценка} В~дальнейшем вместо $I(w,P)$ будем иногда писать~$P^{I, w}$. Для обозначения модели $\kModel{M}$ на шкале $\kFrame{F}$ будем также использовать запись $\kModel{M} = \otuple{W,R,D,I}$. Говорим, что модель $\kModel{M} = \otuple{W,R,D,I}$ \defnotion{построена} (или \defnotion{определена}) \defnotion{на шкале} $\kFrame{F}=\otuple{W,R,D}$, а также на шкале $\kframe{F}=\otuple{W,R}$.

\defnotion{Приписыванием}\index{приписывание} в модели Крипке $\kModel{M} = \otuple{W,R,D,I}$ называем функцию~$g$, сопоставляющую каждой предметной переменной $x$ элемент $g(x)$ из предметной области шкалы, на которой определена модель $\kModel{M}$. Приписывания называют также \defnotion{интерпретациями предметных переменных}.\index{интерпретация!предметных переменных} Пишем $g' \stackrel{x}{=} g$, если приписывание $g'$ отличается от приписывания $g$ разве что значением на переменной~$x$.

Истинность $\lang{QML}$-формулы $\varphi$ в мире $w$ модели $\kModel{M} = \langle W,R,D,I\rangle$ при приписывании $g$ определяется рекурсивно:
\settowidth{\templength}{\mbox{$(\kModel{M},w)\models^g\varphi'$ или $(\kModel{M},w)\models^g\varphi''$;}}
\settowidth{\templengtha}{\mbox{$w$}}
\settowidth{\templengthb}{$(\kModel{M},w)\models^{\phantom{g}} \varphi$ для каждого мира $v$ модели $\kModel{M}$;}
\settowidth{\templengthc}{\mbox{$(\kModel{M},w)\models^g P(x_1,\ldots,x_n)$}}
$$
\begin{array}{lcl}
(\kModel{M},w)\models^g P(x_1,\ldots,x_n)
  & \leftrightharpoons
  & \parbox{\templengthb}{$\langle g(x_1),\ldots,g(x_n)\rangle \in P^{I, w}$,} \\
\end{array}
$$
\mbox{где $P$~--- $n$-местная предикатная буква;}
\settowidth{\templength}{\mbox{$(\kModel{M},w)\models^g\varphi'$ или $(\kModel{M},w)\models^g\varphi''$;}}
\settowidth{\templengtha}{\mbox{$w$}}
\settowidth{\templengthb}{\mbox{$(\kModel{M},w)\models^{g}\varphi'\to\varphi''$}}
\settowidth{\templengthc}{\mbox{$(\kModel{M},w)\models^g P(x_1,\ldots,x_n)$}}
\settowidth{\templengthd}{\mbox{$(\kModel{M},w)\models^{\phantom{g}} \varphi$ для каждого мира $w$ модели $\kModel{M}$;}}
$$
\begin{array}{lcl}
\parbox{\templengthc}{{}\hfill\parbox{\templengthb}{$(\kModel{M},w) \not\models^g \bot;$}}
  \\
\parbox{\templengthc}{{}\hfill\parbox{\templengthb}{$(\kModel{M},w)\models^g\varphi' \wedge \varphi''$}}
  & \leftrightharpoons
  & \parbox[t]{\templength}{$(\kModel{M},w)\models^g\varphi'$\hfill и\hfill $(\kModel{M},w)\models^g\varphi''$;}
  \\
\parbox{\templengthc}{{}\hfill\parbox{\templengthb}{$(\kModel{M},w)\models^g\varphi' \vee \varphi''$}}
  & \leftrightharpoons
  & \parbox[t]{\templength}{$(\kModel{M},w)\models^g\varphi'$ или $(\kModel{M},w)\models^g\varphi''$;}
  \\
\parbox{\templengthc}{{}\hfill\parbox{\templengthb}{$(\kModel{M},w)\models^g\varphi' \to \varphi''$}}
  & \leftrightharpoons
  & \parbox[t]{\templength}{$(\kModel{M},w)\not\models^g\varphi'$ или $(\kModel{M},w)\models^g\varphi''$;}
  \\
\parbox{\templengthc}{{}\hfill\parbox{\templengthb}{$(\kModel{M},w)\models^g\Box\varphi'$}}
  & \leftrightharpoons
  & \mbox{$(\kModel{M},\parbox{\templengtha}{$v$})\models^g\varphi'$ для каждого $v\in R(w)$;}
  \\
\parbox{\templengthc}{{}\hfill\parbox{\templengthb}{$(\kModel{M},w)\models^g\forall x\,\varphi'$}}
  & \leftrightharpoons
  & \parbox{\templengthd}{$(\kModel{M},w)\models^{h}\varphi'$ для каждого $h$, такого, что}
  \\
  &
  & \mbox{\phantom{$(\kModel{M},w)\models^{g'}\varphi'$ }$h \stackrel{x}{=} g$ и $h(x)\in D_w$;}
  \\
\parbox{\templengthc}{{}\hfill\parbox{\templengthb}{$(\kModel{M},w)\models^g\exists x\,\varphi'$}}
  & \leftrightharpoons
  & \parbox{\templengthd}{$(\kModel{M},w)\models^{h}\varphi'$ для некоторого $h$, такого, что}
  \\
  &
  & \mbox{\phantom{$(\kModel{M},w)\models^{g'}\varphi'$ }$h \stackrel{x}{=} g$ и $h(x)\in D_w$.}
\end{array}
$$

Пусть $\kModel{M}$, $\kFrame{F}$, $\kframe{F}$, $\scls{C}$ и $\Scls{C}$~--- это, соответственно, модель Крипке, шкала Крипке с предметными областями, шкала Крипке, класс шкал Крипке и класс шкал Крипке с предметными областями; пусть $w$~--- мир модели $\kModel{M}$; пусть $\varphi$~--- формула со свободными переменными $x_1,\ldots,x_n$. Положим
\settowidth{\templength}{\mbox{$(\kModel{M},w)\models^g P(x_1,\ldots,x_n)$}}
\settowidth{\templengthc}{\mbox{$\kModel{M}$}}
\settowidth{\templengthb}{\mbox{$w$}}
\settowidth{\templengtha}{$(\kModel{M},w)\models^{\phantom{g}} \varphi$ для каждого мира $w$ модели $\kModel{M}$;}
$$
\begin{array}{rcl}
\parbox{\templength}{{}\hfill$(\kModel{M},w)\models \varphi$}
  & \leftrightharpoons
  & \parbox[t]{\templengtha}{$(\kModel{M},w)\models^g \varphi$ для любого $g$, такого, что}
  \\
  &
  & \mbox{\phantom{$(\kModel{M},w)\models^g \varphi$ }$g(x_1),\ldots,g(x_n)\in D_w$;}
  \\
\parbox{\templength}{{}\hfill$\kModel{M}\models \varphi$}
  & \leftrightharpoons
  & \parbox[t]{\templengtha}{$(\kModel{M},\parbox{\templengthb}{$w$})\models^{\phantom{g}} \varphi$ для каждого мира $w$ модели $\kModel{M}$;}
  \\
\parbox{\templength}{{}\hfill$\kFrame{F}\models \varphi$}
  & \leftrightharpoons
  & \parbox[t]{\templengtha}{$\parbox{\templengthc}{$\kModel{M}$}\models \varphi$ для любой $\kModel{M}$, определённой на $\kFrame{F}$;}
  \\
\parbox{\templength}{{}\hfill$\kframe{F}\models \varphi$}
  & \leftrightharpoons
  & \parbox[t]{\templengtha}{$\parbox{\templengthc}{$\kModel{M}$}\models \varphi$ для любой $\kModel{M}$, определённой на $\kframe{F}$;}
  \\
\parbox{\templength}{{}\hfill$\Scls{C}\models \varphi$}
  & \leftrightharpoons
  & \parbox[t]{\templengtha}{$\parbox{\templengthc}{$\kFrame{F}$}\models \varphi$ для любой $\kFrame{F}\in\Scls{C}$;}
  \\
\parbox{\templength}{{}\hfill$\scls{C}\models \varphi$}
  & \leftrightharpoons
  & \parbox[t]{\templengtha}{$\parbox{\templengthc}{$\kframe{F}$}\models \varphi$ для любой $\kframe{F}\in\scls{C}$.}
  \\
\end{array}
$$
Если $\mathfrak{S}\models\varphi$, где $\mathfrak{S}$~--- мир модели, шкала или класс шкал, мы говорим, что формула $\varphi$ \defnotion{истинна} в~$\mathfrak{S}$; в противном случае говорим, что $\varphi$ \defnotion{опровергается} в~$\mathfrak{S}$.
Эти понятия и соответствующие обозначения распространяются на множества формул: если $X$~--- множество формул, то $\mathfrak{S}\models X$ означает, что $\mathfrak{S}\models\varphi$ для каждой~$\varphi\in X$.

Пусть $\kModel{M} = \otuple{W,R,D,I}$~--- модель Крипке и $w \in W$. Определим интерпретацию $I_w$ предикатных букв в $D_w$, положив $I_w(P) = I(w,P)$ для каждой предикатной буквы~$P$. Тогда пара $\kModel{M}_w = \langle D_w, I_w \rangle$ является моделью языка~$\lang{QL}$. Таким образом, мы можем смотреть на модель Крипке $\kModel{M}$ как на множество $\set{\kModel{M}_w : w\in W}$ моделей языка~$\lang{QL}$, структурированное в соответствии с отношением достижимости~$R$.

Пусть $\kModel{M} = \langle W,R,D,I\rangle$~--- модель Крипке, $w \in W$ и $a_1, \ldots, a_n \in D_w$; пусть также $\varphi(x_1, \ldots, x_n)$~--- формула, свободные переменные которой находятся в списке $x_1, \ldots, x_n$.  Будем писать $(\kModel{M},w) \models \varphi (a_1, \ldots, a_n)$, если $(\kModel{M},w) \models^g \varphi (x_1, \ldots, x_n)$, где $g(x_i) = a_i$ для каждого $i\in\set{1,\ldots,n}$.
Для формулы $\varphi(x_1, \ldots, x_n)$ и мира $w\in W$, определим $n$-местный предикат $\varphi^{I,w}$:
$$
\begin{array}{lcl}
\varphi^{I,w} & = & \{(a_1,\ldots,a_n)\in D_w^n : (\kModel{M},w)\models \varphi(a_1,\ldots,a_n)\}.
\end{array}
$$
Фактически это определение расширяет функцию $I$ на множество всех формул. Будем писать также $\varphi^{I,w}(a_1,\ldots,a_n)$, понимая эту запись как $(a_1,\ldots,a_n)\in\varphi^{I,w}$. Отметим, что такое определение и такое обозначение несут в себе неоднозначность: мы можем варьировать как множество переменных, так и их порядок. Эта неоднозначность снимается, если зафиксировать и то, и другое; нам будет важно лишь то, что ниже при использовании такого обозначения указанной неоднозначности возникать не будет.



Пусть $\Scls{C}$~--- класс шкал Крипке с предметными областями, $\kframe{F}$~--- шкала Крипке. Говорим, что $\Scls{C}$ \defnotion{содержит шкалу Крипке}~$\kframe{F}$, если $\kFrame{F}\in \Scls{C}$ для каждой предикатной шкалы~$\kFrame{F}$, построенной на~$\kframe{F}$. Говорим, что $\Scls{C}$~--- \defnotion{класс шкал Крипке},\index{класс!уяи@шкал Крипке} если для каждой шкалы Крипке~$\kframe{F}$ выполняется следующее: из того, что классу $\Scls{C}$ принадлежит некоторая шкала с предметными областями, построенная на~$\kframe{F}$, следует, что классу $\Scls{C}$ принадлежит любая шкала с предметными областями, построенная на~$\kframe{F}$. Говорим, что $\Scls{C}$~--- \defnotion{класс шкал Крипке с постоянными областями},\index{класс!уяи@шкал Крипке!с постоянными областями} если $\Scls{C}$ является пересечением некоторого класса шкал Крипке и класса всех шкал, удовлетворяющих условию~\mbox{$(\mathit{LCD})$}. Обратим внимание, что введённые понятия не отменяют возможности рассматривать классы, элементами которых являются обычные шкалы Крипке (что мы тоже будем делать).

Отметим, что в определении класса шкал Крипке с постоянными областями мы использовали условие локально постоянных областей. Используя определение истинности формул в шкалах с предметными областями, несложно понять, что если заменить это условие условием глобально постоянных областей, мы получим класс шкал, в котором истинны те же самые $\lang{QML}$-формулы. Нетрудно убедиться, что в любой шкале, удовлетворяющей условию~\mbox{$(\mathit{LCD})$}, истинна формула $\bm{bf}=\forall x\,\Box P(x)\to \Box\forall x\,P(x)$, где $P$~--- унарная предикатная буква; эта формула известна как \defnotion{формула Баркан}.\index{уяа@формула!Баркан} Если же шкала Крипке с предметными областями не удовлетворяет условию~\mbox{$(\mathit{LCD})$}, то формула Баркан в ней опровергается.

Шкалы Крипке с предметными областями обладают следующим свойством: множество $\lang{QML}$-формул, истинных в любом классе таких шкал, является нормальной модальной предикатной логикой. По этой причине, а также поскольку мы используем такие шкалы для оценки истинности именно предикатных формул, шкалы с предметными областями называют также \defnotion{предикатными шкалами Крипке}\index{уяи@шкала!Крипке!предикатная}. Отметим, что ситуация с обычными шкалами Крипке является сходной: множество $\lang{ML}$-формул, истинных в любом классе шкал Крипке, является нормальной модальной пропозициональной логикой. В~контексте исследований, затрагивающих как модальные предикатные, так и модальные пропозициональные логики, обычные шкалы Крипке называют также \defnotion{пропозициональными}.\index{уяи@шкала!Крипке!пропозициональная} Как естественное продолжение такой терминологии, модели Крипке также делят на \defnotion{предикатные}\index{модель!Крипке!предикатная} и \defnotion{пропозициональные}\index{модель!Крипке!пропозициональная}: первые строятся на предикатных шкалах за счёт добавления интерпретации предикатных букв в мирах, а вторые~--- на пропозициональных шкалах за счёт добавления оценки пропозициональных переменных.

Понятия \defnotion{подшкалы}, \defnotion{порождённой подшкалы}, \defnotion{корневой подшкалы}, \defnotion{корня} естественным образом переносятся с пропозициональных шкал Крипке на предикатные.

    \subsection{Логики}

Для множеств $\lang{QML}$-формул ${\Sigma}$ и ${\Sigma}'$ обозначим через ${\Sigma}+{\Sigma}'$ логику, получающуюся замыканием множества ${\Sigma}\cup{\Sigma}'$ по {\MP}, правилу обобщения и правилу подстановки, а через ${\Sigma}\oplus{\Sigma}'$~--- логику, получающуюся замыканием множества ${\Sigma}\cup{\Sigma}'$ по \MP, правилу обобщения, правилу Гёделя и правилу подстановки.

Мы будем рассматривать в основном нормальные модальные предикатные логики, причём в первую очередь являющиеся предикатными напарниками нормальных модальных пропозициональных логик. Примерами таких логик являются $\logic{QK}$, $\logic{QT}$, $\logic{QK4}$, $\logic{QS4}$, $\logic{QGL}$, $\logic{QGrz}$, $\logic{QwGrz}$, $\logic{QS5}$ и другие. При этом, как и в пропозициональном случае, нас будут интересовать не только отдельные системы, но и их классы, в том числе интервалы; интервал логик определим подобно тому, как раньше: $[L_1,L_2]=\set{L : L_1\subseteq L\subseteq L_2}$. Отметим, что результаты, полученные ниже для интервалов логик, будут справедливы не только для нормальных логик из таких интервалов, а вообще для любых множеств формул, находящихся между логиками, ограничивающими интервал; тем не менее, нам будет важно, что границами являются именно логики, поскольку мы будем пользоваться их замкнутостью по правилу подстановки.

Большое внимание мы будем уделять модальным предикатным логикам, которые определены не синтаксически (как, например, логики вида $\logic{Q}L$, где $L$~--- модальная пропозициональная логика), а семантически. При этом в качестве семантики мы будем использовать семантику Крипке. Необходимые определения будут даны ниже.

    \subsection{Необходимые факты}

            Для модальной предикатной логики $L$ определим $\kframes{L}$ как класс пропозициональных шкал Крипке, в которых истинны все формулы, принадлежащие логике~$L$, а также $\kFrames{L}$~--- как класс предикатных шкал Крипке, в которых истинны все формулы, принадлежащие логике~$L$. Если $\kframe{F}$~--- пропозициональная шкала Крипке и $\kframe{F}\models L$, то $\kframe{F}$ называем \defnotion{шкалой логики~$L$},\index{уяи@шкала!логики} или \defnotion{$L$-шкалой};\index{уяи@шкала!l@$L$-шкала} то же самое относится и к предикатным шкалам.

            Для класса шкал Крипке $\scls{C}$ определим $\mPlogic{\scls{C}}$ как множество модальных предикатных формул, истинных в~$\scls{C}$. Отметим, что $\mPlogic{\scls{C}}$ является нормальной модальной логикой. Логику $\mPlogic{\scls{C}}$ называем \defnotion{логикой класса~$\scls{C}$}.\index{логика!класса шкал} Аналогично для класса $\Scls{C}$ предикатных шкал.
            Для класса шкал Крипке $\scls{C}$ определим логику $\mPlogicC{\scls{C}}$ как множество формул, истинных в классе всех предикатных шкал с постоянными областями, построенных на шкалах из~$\scls{C}$. Будем писать $\mPlogic{\kframe{F}}$ и $\mPlogicC{\kframe{F}}$ вместо $\mPlogic{\set{\kframe{F}}}$ и $\mPlogicC{\set{\kframe{F}}}$ соответственно.

            Пусть $L$~--- нормальная модальная предикатная логика. Логика $L$ называется \defnotion{корректной}\index{логика!корректная} относительно класса шкал~$\scls{C}$, если $\scls{C}\subseteq\kframes{L}$, и относительно класса предикатных шкал~$\Scls{C}$, если $\Scls{C}\subseteq\kFrames{L}$. Логика $L$ называется \defnotion{полной} относительно класса~$\scls{C}$ шкал Крипке, если $L\subseteq\mPlogic{\scls{C}}$, и относительно класса~$\Scls{C}$ предикатных шкал Крипке, если $L\subseteq\mPlogic{\Scls{C}}$. Логика $L$ называется \defnotion{адекватной}\index{логика!адекватная} относительно класса шкал $\scls{C}$, если $L=\mPlogic{\scls{C}}$, и относительно класса предикатных шкал $\Scls{C}$, если $L=\mPlogic{\Scls{C}}$. Как и в случае пропозициональных логик, ниже под полнотой будем понимать адекватность. Будем говорить, что логика $L$ \defnotion{полна по Крипке},\index{логика!полная по Крипке} если существует такой класс $\Scls{C}$ предикатных шкал Крипке, что $L=\mPlogic{\Scls{C}}$. Логика $L$ называется \defnotion{финитно аппроксимируемой},\index{логика!уяа@финитно аппроксимируемая} если существует такой класс $\Scls{C}$ предикатных шкал Крипке, имеющих конечные множества миров и конечные предметные области, что~$L=\mPlogic{\Scls{C}}$.

Для класса предикатных шкал $\Scls{C}$ определим $\kframes{\Scls{C}}$ как класс шкал Крипке, на которых определены предикатные шкалы из~$\Scls{C}$.
Нетрудно понять, что $\kframes{\kFrames{\logic{Q}L}} = \kframes{L}$. При этом предикатные напарники некоторых полных по Крипке нормальных модальных пропозициональных логик тоже полны по Крипке (относительно же классов пропозициональных шкал или относительно некоторых классов предикатных шкал, определённых на пропозициональных шкалах исходной пропозициональной логики), но довольно часто это не так. Например~\cite{GShS},
$$
\begin{array}{lcl}
\logic{QK}  & = & \mPlogic{\kframes{\logic{K}}}; \\
\logic{QT}  & = & \mPlogic{\kframes{\logic{T}}}; \\
\logic{QK4} & = & \mPlogic{\kframes{\logic{K4}}}; \\
\logic{QS4} & = & \mPlogic{\kframes{\logic{S4}}}; \\
\logic{QS5} & = & \mPlogic{\kframes{\logic{S5}}}. \\
\end{array}
$$
В то же время~\cite{Montagna84,MR:2001:LI,MR:2002:LI:2}
$$
\begin{array}{lcl}
\logic{QGL}    & \ne & \mPlogic{\kframes{\logic{GL}}}; \\
\logic{QGLLin} & \ne & \mPlogic{\kframes{\logic{GLLin}}}; \\
\logic{QGrz}   & \ne & \mPlogic{\kframes{\logic{Grz}}}; \\
\logic{QGrz.3} & \ne & \mPlogic{\kframes{\logic{Grz.3}}}. \\
\end{array}
$$
При этом
для логик $\logic{QGL}$, $\logic{QGLLin}$, $\logic{QGrz}$, $\logic{QGrz.3}$, $\logic{QGL.bf}$, $\logic{QGLLin.bf}$, $\logic{QGrz.bf}$, $\logic{QGrz.3.bf}$ не существует таких классов предикатных шкал, относительно которых они полны.
Другие примеры неполных по Крипке нормальных модальных предикатных логик можно найти в~\cite{GShS}.

Для нормальной модальной пропозициональной логики $L$ определим логику $\kflogic{L}$, положив
$$
\begin{array}{lcl}
\kflogic{L} & = & \mPlogic{\kFrames{L}}.
\end{array}
$$
Нетрудно видеть, что $\kflogic{L}$ является минимальным нормальным модальным расширением логики $\logic{Q}L$, полным по Крипке. Отметим, что исходная логика $L$, вообще говоря, может при этом и не быть полной по Крипке. Логику $\kflogic{L}$ называем \defnotion{полным по Крипке предикатным напарником}\index{напарник!предикатный!полный по Крипке} логики~$L$. Как мы видели, иногда $\kflogic{L}=\logic{Q}L$, но в целом мы можем утверждать лишь, что $\logic{Q}L\subseteq\kflogic{L}$. Ввиду этого включения логику $\kflogic{L}$ называют также \defnotion{пополнением по Крипке}\index{пополнение по Крипке} логики~$\logic{Q}L$.

Для нормальной модальной предикатной логики $L$ определим логику $L\logic{.bf}$, положив
$$
\begin{array}{lcl}
L\logic{.bf} & = & L \oplus \bm{bf}.
\end{array}
$$
Шкалы Крипке таких логик удовлетворяют условию локально постоянных областей. Отметим, что некоторые логики уже содержат формулу Баркан, и её добавление не приводит к изменению логики; например, $\logic{QS5}=\kflogic{\logic{S5}}=\logic{QS5.bf}$, но, скажем, логики $\logic{QGL}$, $\kflogic{\logic{GL}}$, $\logic{QGL.bf}$ и $\kflogic{\logic{GL.bf}}$ попарно различны. Многие <<естественные>> логики формулу Баркан не содержат; например, это относится ко всем подлогикам логик $\logic{QGLLin}$ и $\logic{QGrz.3}$, а значит, к $\logic{QK}$, $\logic{QT}$, $\logic{QKB}$, $\logic{QKTB}$, $\logic{QK4}$, $\logic{QS4}$, $\logic{QGL}$, $\logic{QGrz}$ и многим другим.
Формула Баркан весьма полно описывает условие постоянства областей в предикатных шкалах Крипке: известно (см., например,~\cite[Proposition~3.4.2]{GShS}), что если $\Scls{C}$~--- класс предикатных шкал нормальной модальной предикатной логики~$L$, то подкласс класса $\Scls{C}$, состоящий из шкал с локально постоянными областями, совпадает с классом предикатных $L\logic{.bf}$-шкал.

    \subsection{Полимодальный случай}

Все приведённые выше определения переносятся естественным образом на случай, когда модальный предикатный язык содержит более одной модальности. В частности, условие $(\mathit{ED})$ требуется для каждого отношения достижимости в предикатной шкале; то же относится к условиям $(\mathit{LCD})$ и $(\mathit{GCD})$, если нужны шкалы с постоянными предметными областями; аналогично, формула Баркан выполняется для тех модальностей, которым в предикатных шкалах Крипке соответствуют отношения, удовлетворяющие условию~$(\mathit{LCD})$.

Детали соответствующих формальных определений опускаем; они легко восстанавливаются соединением определений для полимодального пропозиционального и модального предикатного случаев. Модальный предикатный язык с $n$ модальностями обозначим~$\lang{QML}_n$; в целом обозначения для полимодальных предикатных языков, как и в пропозициональном случае, будут варьироваться.

  \section{Неразрешимость логик унарного предиката}
	\subsection{Трюк Крипке}
    \label{sec:KripkeTrick:QModal}

\subsubsection{Предварительные сведения}

В 1962 году С.~Крипке сделал наблюдение, позволившее ему доказать неразрешимость очень большого класса модальных предикатных логик, язык которых содержит всего две одноместные предикатные буквы~\cite{Kripke62}. Формально, это наблюдение состоит в следующем.
\begin{itemize}
\item
Пусть $P$~--- бинарная, а $Q_1$ и $Q_2$~--- унарные предикатные буквы. Пусть $\varphi$~--- классическая предикатная формула, не содержащая предикатных букв, отличных от~$P$, а $\varphi^\ast$ получается из $\varphi$ подстановкой вместо $P(x,y)$ формулы $\Diamond(Q_1(x)\wedge Q_2(y))$. Тогда
$$
\begin{array}{lcl}
\varphi\in \logic{QCl} & \iff & \varphi^\ast\in \logic{QS5}.
\end{array}
\eqno{\mbox{$(\ast)$}}
$$
\end{itemize}
Фактически же, сам Крипке отмечает, что доказательство эквивалентности~\mbox{$(\ast)$} использует лишь тот факт, что логика $\logic{QS5}$ имеет такие реляционные модели\footnote{Называемые сейчас шкалами Крипке.}, в которых из некоторого мира достижимо бесконечно много миров. В~результате мы получаем, что в эквивалентности~\mbox{$(\ast)$} логику $\logic{QS5}$ можно заменить любой логикой, допускающей такие модели. Например, сюда попадают все модальные логики, содержащиеся в $\logic{QS5}$, $\logic{QGL}$ или~$\logic{QGrz}$ (и~не только они).

Было замечено, что трюк Крипке может быть применим и в некоторых других условиях. Ниже покажем, какие условия и как можно изменить, чтобы при этом трюк Крипке (или некоторая его модификация) продолжил работать, а также опишем ситуации, когда трюк Крипке применить невозможно.

Для этого мы сначала воспроизведём исходную идею, изложенную С.~Крипке~\cite{Kripke62}, затем перечислим важные условия, используемые в доказательстве, после чего обсудим каждое из них в контексте вопроса о возможности его ослабления.

\subsubsection{Исходная конструкция Крипке}

В исходной конструкции моделирования бинарной предикатной буквы с помощью модальности и двух унарных предикатных букв нам будет важна как формулировка, так и доказательство. Приведём краткое изложение оригинального доказательства~\cite{Kripke62}.
\begin{itemize}
\item 
{\bf Формулировка.} \\
Пусть $P$~--- бинарная, а $Q_1$ и $Q_2$~--- унарные предикатные буквы. Пусть $\varphi$~--- замкнутая классическая предикатная формула, не содержащая предикатных букв, отличных от~$P$, а $\varphi^\ast$ получается из $\varphi$ подстановкой вместо $P(x,y)$ формулы $\Diamond(Q_1(x)\wedge Q_2(y))$. Тогда
$$
\begin{array}{lcl}
\varphi\in \logic{QCl} & \iff & \varphi^\ast\in \logic{QS5}.
\end{array}
$$
\item 
{\bf Доказательство импликации $(\Rightarrow)$.} \\
Пусть $\varphi\in \logic{QCl}$. Заметим, что $\logic{QCl}\subseteq\logic{QS5}$, формула $\varphi^\ast$ получена из $\varphi$ с помощью правила подстановки, $\logic{QS5}$ замкнута по правилу подстановки. Но тогда $\varphi^\ast\in \logic{QS5}$.
\item 
{\bf Доказательство импликации $(\Leftarrow)$.} \\
Пусть $\varphi\not\in \logic{QCl}$. Тогда, по теореме Лёвенгейма--Сколема, существует такая классическая модель $\cModel{M}=\langle\numN,J\rangle$, что $\cModel{M}\not\cmodels \varphi$. Пусть
$$
\begin{array}{lcl}
W & = & \numN\times\numN; \\
R & = & W\times W; \\
\kframe{F} & = & \langle W,R\rangle. \\
\end{array}
$$
На предикатной шкале $\kframe{F}\odot\numN$ определим интерпретацию $I$, положив для каждого $a\in \numN$
$$
\begin{array}{lcl}
I(\otuple{m,n},Q_1) & = & \{\otuple{a} : \cModel{M}\cmodels P(a,n)\}; \\
I(\otuple{m,n},Q_2) & = & \{\otuple{b} : \cModel{M}\cmodels P(m,b)\}, \\
\end{array}
$$
и возьмём модель $\kmodel{M}=\langle \kframe{F}\odot\numN, I\rangle$. Из определения этой модели получаем, что для любых $a,b,m,n\in\numN$
$$
\begin{array}{lcl}
(\kmodel{M},\otuple{m,n})\models Q_1(a)\wedge Q_2(b)
  & \iff
  & \mbox{$m=a$, $n=b$ и $\cModel{M}\cmodels P(a,b)$,}
\end{array}
$$
откуда следует, что для любых $a,b\in\numN$
$$
\begin{array}{lcl}
\kmodel{M}\models \Diamond (Q_1(a)\wedge Q_2(b))
  & \iff
  & \cModel{M}\cmodels P(a,b).
\end{array}
\eqno{\mbox{$({\ast}{\ast})$}}
$$
Но тогда нетрудно видеть, что $\kmodel{M}\models\neg \varphi^\ast$, в частности, $\kmodel{M}\not\models \varphi^\ast$, а значит, $\varphi^\ast\not\in\logic{QS5}$.
\end{itemize}

В приведённом доказательстве мы опустили технические обоснования эквивалентностей, но они не представляют трудности. Кроме того, отметим, что хотя доказательство использует семантику, в силу теоремы о корректности можно говорить о проблеме выводимости в модальном исчислении для $\logic{QS5}$, что и происходит в исходной работе Крипке.

\subsubsection{Условия, используемые в трюке Крипке}
\label{sssec:KripkeTrick:conditions}

Перечислим условия, которые использовались в описанном выше трюке Крипке, а также некоторые наблюдения, которые, не являясь важными для осуществления трюка, тем не менее, важны для понимания границ его применимости.
\begin{enumerate}[nosep]
\item
\label{cond1}
В трюке Крипке используется подстановка формул.
\item
\label{cond2}
Множества индивидных переменных формул $\varphi$ и $\varphi^\ast$ совпадают.
\item
\label{cond3}
Доказательство использует теорему о полноте для логики $\logic{QCl}$ и лишь теорему о корректности для логики, в отношении которой использовался трюк Крипке.
\item
\label{cond6}
В случае, когда $\varphi\not\in \logic{QCl}$, для модели $\kmodel{M}$ доказывается, что $\kmodel{M}\models\neg \varphi^\ast$, хотя фактически используется лишь то, что $\kmodel{M}\not\models \varphi^\ast$.
\item
\label{cond4a}
Трюк Крипке проделан именно для логики $\logic{QS5}$.
\item
\label{cond4}
Логика, в отношении которой использовался трюк Крипке, является \defnotion{консервативным расширением}\index{расширение!консервативное}\footnote{В данном случае это означает, что безмодальный фрагмент рассматриваемой модальной логики (в трюке Крипке это $\logic{QS5}$) совпадает с логикой~$\logic{QCl}$.} логики $\logic{QCl}$.
\item
\label{cond5}
Доказательство опирается на теорему Лёвенгейма--Сколема.
\item
\label{cond7}
Формула $\varphi$ является классической.
\item
\label{cond8}
Буква $P$ является именно бинарной.
\item
\label{cond9}
Формула $\varphi$ не содержит предикатных букв, отличных от бинарной предикатной буквы~$P$.
\item
\label{cond10}
Число унарных букв, используемых для моделирования бинарной буквы, равно двум.
\item
\label{cond11}
Число миров модели $\kmodel{M}$ бесконечно, причём счётно.
\item
\label{cond12}
Предметные области миров модели $\kmodel{M}$ бесконечны, причём счётны.
\item
\label{cond13}
Предметные области миров модели $\kmodel{M}$ одинаковы.
\item
\label{cond14}
Логика, в отношении которой использовался трюк Крипке, является нормальной модальной.
\item
\label{cond15}
Исходная логика~--- это именно $\logic{QCl}$.
\item
\label{cond16}
Задействованные в трюке Крипке логики могут быть заданы в виде исчислений.
\end{enumerate}

\subsubsection{Несколько простых наблюдений}

Условие \ref{cond1} кажется важным: подстановка (с учётом условия~\ref{cond4}) обеспечивает простоту в доказательстве импликации $(\Rightarrow)$. Ниже мы покажем, что, тем не менее, иногда удобно делать подстановку не в исходную формулу, а в некоторую её модификацию.

Условие~\ref{cond2}, очевидно, не является существенным: добавление новых переменных к разрешимости не приведёт. Нам будет важно, что оно как раз не требуется.

Условие~\ref{cond3}, как и условие~\ref{cond2}, даёт удобство использования: хотя многие значимые для приложений модальные предикатные логики полны по Крипке, достаточно, что для них выполнялось лишь условие корректности, чем мы будем пользоваться.

Условие~\ref{cond6}, очевидно, можно нарушить: если формула опровергнется лишь в одном мире некоторой модели логики, то этого достаточно, чтобы она не принадлежала логике.

Ввиду сказанного мы не будем подробно обсуждать условия \ref{cond1}, \ref{cond2}, \ref{cond3} и~\ref{cond6}; а вот каждое из оставшихся условий будет подвергнуто сомнению, при этом мы приведём примеры, в которых при ослаблении указанных требований, но с сохранением условий \ref{cond1}, \ref{cond2}, \ref{cond3} и~\ref{cond6} получаются математические результаты, связанные с алгоритмическими свойствами неклассических\footnote{Не только модальных, но и, например, суперинтуиционистских логик, см.~раздел~\ref{sec:KripkeTrick:QInt}.} логик унарных предикатов.

\subsubsection{Использование теоремы Лёвенгейма--Сколема}

Условие~\ref{cond5} убрать из доказательства Крипке совсем просто. Теорема Лёвенгейма--Сколема использовалась для того, чтобы модель $\cModel{M}$, возникшая в доказательстве импликации $(\Leftarrow)$, содержала в качестве предметной области именно~$\numN$. Предположим, что это не так, т.е. $\cModel{M}=\langle M,J\rangle$ и при этом мы знаем лишь, что $M\ne \varnothing$ и $\cModel{M}\not\cmodels \varphi$. Тогда при определении шкалы $\kframe{F}$ можно взять $W=M\times M$ и $R=W\times W$, а модель $\kmodel{M}$ определить на шкале $\kframe{F}\odot M$ так, чтобы выполнялось условие~\mbox{$({\ast}{\ast})$}.

Попутно отметим, что условие~\mbox{$({\ast}{\ast})$} можно выполнить многими способами, а не только тем, который описан у Крипке. Так, например, определяя $\kmodel{M}$ на~$\kframe{F}\odot M$, можно задать интерпретацию~$I$, положив
$$
\begin{array}{lcl}
I(\langle s,t\rangle, Q_1)
  & =
  & \left\{
       \begin{array}{rl}
         \{\langle s\rangle\}, & \mbox{если $\cModel{M}\models P(s,t)$;} \\
         \varnothing,      & \mbox{если $\cModel{M}\not\models P(s,t)$;} \\
       \end{array}
    \right.
\medskip\\
I(\langle s,t\rangle, Q_2)
  & =
  & \left\{
       \begin{array}{rl}
         \{\langle t\rangle\}, & \mbox{если $\cModel{M}\models P(s,t)$;} \\
         \varnothing,      & \mbox{если $\cModel{M}\not\models P(s,t)$,} \\
       \end{array}
    \right.
\end{array}
\eqno{\mbox{$({\ast}{\ast}{\ast})$}}
$$
что, возможно, выглядит более естественно.

Итак, условие~\ref{cond5} не является принципиальным. Чуть позже мы покажем, как можно этим воспользоваться.

\subsubsection{Логики, отличные от $\logic{QS5}$}

О том, что условие~\ref{cond4a}, конечно же, легко снимается, говорил и сам Крипке, и многие другие авторы. Мы сформулируем здесь те хорошо известные синтаксические и семантические условия, которые обеспечивают возможность трюка Крипке для логик, отличных от~$\logic{QS5}$.

Пусть $W=\{w\}\cup (\numN\times\numN)$, $R=\{w\}\times (\numN\times\numN)$. Пусть $\bar{W}$~--- множество, расширяющее~$W$, $\bar{R}$~--- бинарное отношение на~$\bar{W}$, содержащее~$R$, и $\bar{\kframe{F}} = \langle\bar{W},\bar{R}\rangle$. Пусть $\bar{\kFrame{F}} = \langle\bar{\kframe{F}},\bar{D}\rangle$~--- предикатная шкала Крипке на $\bar{\kframe{F}}$, причём $\numN\subseteq D_w$ для каждого $w\in\bar{W}$. Тогда для любой формулы $\varphi$, не содержащей предикатных букв, отличных от бинарной буквы~$P$,
$$
\begin{array}{lcl}
\varphi\in \logic{QCl} & \iff & (\bar{\kFrame{F}},w)\models \varphi^\ast.
\end{array}
$$
Обоснование этой эквивалентности проводится так же, как и в исходном обосновании трюка Крипке для~$\logic{QS5}$, только теперь при доказательстве импликации $(\Leftarrow)$ формула $\varphi^\ast$ не обязательно будет опровергаться в каждом мире соответствующей модели, но нам это и не нужно: достаточно, что $\varphi^\ast$ опровергается в мире~$w$.

Обратим внимание, что такой формулировкой мы сняли очень много условий: это одновременно условия \ref{cond6} (нам важно, истинна ли формула $\varphi^\ast$ лишь в~$w$) и~\ref{cond13} (предметные области миров могут быть различными), а также частично условия \ref{cond11} и~\ref{cond12} (и число миров, и предметные области могут не быть счётными, но пока всё же остаются бесконечными).

Семантические условия описаны; обратимся к синтаксическим. Чтобы логика допускала подобные шкалы, она должны быть консервативна относительно $\logic{QCl}$ (т.е. нам пока важно условие~\ref{cond4}) и либо должна не содержать формул, ограничивающих \defnotion{ширину}\index{уяи@ширина шкалы}\footnote{Под шириной шкалы понимают наибольшую мощность антицепей этой шкалы.} шкал Крипке (при этом \defnotion{высоту}\index{бяа@высота!уяи@шкалы}\footnote{Под высотой (транзитивной) шкалы понимают наибольшую мощность цепей этой шкалы.} можно ограничить числом~$2$), либо должна допускать транзитивную шкалу, в которой из некоторого мира видна бесконечная цепь попарно различных миров. При этом полнота по Крипке, как и прежде, не требуется, т.е. условие~\ref{cond3} продолжает выполняться, а также требуется замкнутость по правилу подстановки, т.е. выполнение условия~\ref{cond1}. Такими логиками являются, например, все модальные предикатные логики, содержащие $\logic{QCl}$ и содержащиеся хотя бы в одной из следующих логик: $\logic{QS5}$, $\logic{QGL.3.bf}$, $\logic{QGrz.3.bf}$,
$\logic{QGL.bf}\oplus\bm{bd}_2$,
$\logic{QGrz.bf}\oplus\bm{bd}_2$,
$\logic{QK4.3.D.X.bf}$. Здесь $\bm{bd}_2$~--- формула, ограничивающая \defnotion{глубину}\index{бяб@глубина!уяи@шкалы}\footnote{Глубина шкалы~--- то же самое, что и её высота; буквы в формуле $\bm{bd}_2$ (и в общем случае в $\bm{bd}_n$ при $n\in\numNp$) взяты как первые буквы из <<bounded depth>>.} транзитивных шкал числом~$2$. Подобные формулы определяются следующим образом:
$$
\begin{array}{lcl}
\bm{bd}_1 & = & \Diamond\Box p_1 \to p_1; \\
\bm{bd}_{n+1} & = & \Diamond(\Box p_{n+1}\wedge \neg\bm{bd}_n) \to p_{n+1}, \\
\end{array}
$$
где $p_1,p_2,p_3,\ldots$~--- различные пропозициональные буквы.
Для определения логики $\logic{QK4.3.D.X.bf}$ достаточно сказать, что
$$
\begin{array}{lcl}
\logic{QK4.3.D.X} & = & \logic{QK4.3} \oplus \Diamond\top \oplus \Box\Box p\to p; \\
\logic{QK4.3}     & = & \logic{QK4}   \oplus \Box(\Box^+p\to q)\vee\Box(\Box^+q\to p). \\
\end{array}
$$

Обратим внимание на несколько моментов.
Прежде всего, описанные здесь условия, во-первых, извлекаются из обсуждения самого Крипке~\cite{Kripke62}, а во-вторых, являются лишь достаточными, но, как мы увидим, не необходимыми. Кроме того, класс логик, к которым применим трюк Крипке, не ограничен снизу, скажем, логикой~$\logic{QK}$; он применим и к квазинормальным, и даже к ненормальным модальным логикам (например, ко всем, содержащимся в ненормальной модальной предикатной логике\footnote{Чтобы определить $\logic{QK0}$, достаточно объединить $\logic{K0}$ и $\logic{QCl}$, после чего взять замыкание получившегося множества по \MP, правилу подстановки и правилу обобщения.} $\logic{QK0}+\Diamond\Diamond\bot$); это отчасти снимает ограничение в условии~\ref{cond14} (мы выходим за класс нормальных логик, но всё же остаёмся в классе модальных логик). Более того, трюк Крипке применим к, казалось бы, совсем простой логике~--- классической логике предикатов, язык которой всего лишь обогащён модальностью $\Box$ или~$\Diamond$ (она замкнута по правилу подстановки и содержится в~$\logic{QS5}$, а этого достаточно). Как следствие, все эти логики алгоритмически неразрешимы в языке с двумя одноместными предикатными буквами.

\subsubsection{Консервативность относительно $\logic{QCl}$}
\label{sec:conservativeness}

Фактически условие~\ref{cond4} состоит из двух частей: во-первых, логика, в отношении которой проводится трюк Крипке, содержит в себе $\logic{QCl}$, и во-вторых, безмодальный фрагмент этой логики совпадает с~$\logic{QCl}$. Конечно, из второй части условия~\ref{cond4} следует первая, тем не менее, первую часть можно рассмотреть без второй. В этом разделе мы ограничимся грубым подходом, рассмотрев лишь вторую часть условия; к первой же части в её самостоятельной формулировке вернёмся позже.

Для модальной предикатной логики $L$ определим логику $L_{\mathit{fin}}$ как логику класса предикатных шкал логики~$L$, предметная область каждого мира в которых является конечной. Тогда, например, логики $\logic{QK}_{\mathit{fin}}$, $\logic{QT}_{\mathit{fin}}$, $\logic{QS4}_{\mathit{fin}}$, $\logic{QGL}_{\mathit{fin}}$, $\logic{QGrz}_{\mathit{fin}}$, $\logic{QS5}_{\mathit{fin}}$ уже не будут консервативны относительно $\logic{QCl}$; зато они будут консервативны относительно $\logic{QCl}_{\mathit{fin}}$.

Пусть $\varphi$~--- классическая предикатная формула, не содержащая предикатных букв, отличных от бинарной буквы~$P$.

Если $\varphi\in\logic{QCl}_{\mathit{fin}}$, то $\varphi^\ast$ принадлежит любой из модальных предикатных логик, содержащих~$\logic{QCl}_{\mathit{fin}}$.

Пусть $\varphi\not\in\logic{QCl}_{\mathit{fin}}$. Тогда существует такая конечная классическая модель $\cModel{M}=\langle M,J\rangle$, что $\cModel{M}\not\cmodels \varphi$.

Пусть $W=\{w\}\cup (M\times M)$, $R=\{w\}\times (M\times M)$. Пусть $\bar{W}$~--- множество, расширяющее~$W$, $\bar{R}$~--- бинарное отношение на~$\bar{W}$, содержащее~$R$, $\bar{\kframe{F}} = \langle\bar{W},\bar{R}\rangle$. Определим интерпретацию $I$ в шкале $\bar{\kframe{F}}\odot M$ следующим образом: в мирах множества $M\times M$~--- в соответствии с~\mbox{$({\ast}{\ast}{\ast})$}, а для любого другого мира $u$ (в частности, для~$w$) положим $I(u,Q_1)=I(u,Q_2)=\varnothing$. Пусть $\kmodel{M}=\langle\bar{\kframe{F}}\odot M,I\rangle$. Тогда ясно, что для любых $a,b\in M$ выполняется эквивалентность, похожая на~\mbox{$({\ast}{\ast})$}:
$$
\begin{array}{lcl}
(\kmodel{M},w)\models \Diamond (Q_1(a)\wedge Q_2(b))
  & \iff
  & \cModel{M}\cmodels P(a,b),
\end{array}
$$
откуда получаем, что $(\kmodel{M},w)\not\models \varphi^\ast$.

Согласно теореме Трахтенброта~\cite{Trakhtenbrot50,Trakhtenbrot53} (см.~теоремы~\ref{th:trakhtenbrot} и~\ref{th:trakhtenbrot:positive}), логика $\logic{QCl}_{\mathit{fin}}$ не является рекурсивно перечислимой (причём уже в языке с одной бинарной буквой), поэтому как следствие описанной конструкции мы получаем, что любая логика, содержащая $\logic{QCl}_{\mathit{fin}}$ и содержащаяся в $\logic{QS5}_{\mathit{fin}}$, $\logic{QGL.3.bf}_{\mathit{fin}}$, $\logic{QGrz.3.bf}_{\mathit{fin}}$,
$\logic{QGL.bf}_{\mathit{fin}}\oplus\bm{bd}_2$,
$\logic{QGrz.bf}_{\mathit{fin}}\oplus\bm{bd}_2$
или $\logic{QK4.3.D.X.bf}_{\mathit{fin}}$ является $\Pi^0_1$-трудной в языке с двумя одноместными предикатными буквами (в частности, не является рекурсивно перечислимой).

Отметим, что здесь мы показали, что можно обойти не только условие~\ref{cond4}, но и условие~\ref{cond16}, поскольку ни $\logic{QCl}_{\mathit{fin}}$, ни модальные логики, консервативные относительно $\logic{QCl}_{\mathit{fin}}$, не могут быть заданы в виде исчислений, а также условия \ref{cond11} и~\ref{cond12}, поскольку нам оказалось достаточно шкал с конечными предметными областями и конечным множеством миров.

Кроме того, нам снова была важна замкнутость по подстановке (условие~\ref{cond1}) и корректность относительно определённых классов шкал (условие~\ref{cond3}), поэтому трюк Крипке оказывается применимым даже к логике $\logic{QCl}_{\mathit{fin}}$ в языке, обогащённом модальностью $\Box$ или~$\Diamond$.

\subsubsection{Логики шкал с конечным числом миров}
\label{sec:wfin3}

В предыдущем разделе нам не была важна конечность множества миров в шкалах Крипке (множество $\bar{W}$ могло быть как конечным, так и бесконечным), но зато была важна конечность предметной области мира~$w$. Покажем, что трюк Крипке работает, когда мы требуем, чтобы было конечным лишь множество миров шкалы, а ограничений на предметные области при этом нет (т.е. условие~\ref{cond11} нарушается, а условие~\ref{cond12} может при этом выполняться).

Для модальной предикатной логики $L$ определим логику $L_{\mathit{wfin}}$ как логику класса конечных шкал логики~$L$, т.е. таких шкал~$L$, множество миров в которых конечно. Если для $L$ имеется конечная шкала, то, в отличие от $L_{\mathit{fin}}$, логика $L_{\mathit{wfin}}$ будет консервативной относительно~$\logic{QCl}$. Тем не менее, в неё погружается логика~$\logic{QCl}_{\mathit{fin}}$. Покажем это, ограничившись случаем, когда в языке имеется лишь одна бинарная предикатная буква~$P$.
Далее пусть $L$~--- такая модальная предикатная логика, что для каждого $n\in\numN$ существует конечная $L$-шкала, в которой имеется мир, из которого достижимо не менее $n$~миров.

Пусть $T$~--- новая одноместная предикатная буква. С её помощью определим отношение эквивалентности, которое опишем формулой
$$
\begin{array}{lcl}
E(x,y)
  & =
  & \Box(T(x)\leftrightarrow T(y)).
\end{array}
$$
Теперь опишем условие, означающее, что это отношение эквивалентности является конгруэнтностью относительно~$P$:
$$
\begin{array}{lcl}
\mathit{congr}
  & =
  & \forall x\forall x'\forall y\forall y'\,(E(x,x')\wedge E(y,y')\wedge P(x,y)\to P(x',y')).
\end{array}
$$

Для $\lang{QL}$-формулы $\varphi$, не содержащей предикатных букв, отличных от~$P$, определим формулу $\varphi^\ast$ так же, как и раньше: $\varphi^\ast$ получается из $\varphi$ подстановкой вместо $P(x,y)$ формулы $\Diamond(Q_1(x)\wedge Q_2(y))$. Тогда
$$
\begin{array}{lcl}
\varphi \in \logic{QCl}_{\mathit{fin}}
  & \iff
  & \forall x\,\Diamond T(x)\wedge \mathit{congr}\to \varphi^\ast \in L_{\mathit{wfin}}.
\end{array}
$$
Покажем это.

Пусть $\varphi \not\in \logic{QCl}_{\mathit{fin}}$. Тогда существует такая конечная модель $\cModel{M}=\langle M,J\rangle$, что $\cModel{M}\not\cmodels \varphi$. Без ограничений общности можем (и будем) считать, что $M=\{1,\ldots,n\}$. Возьмём конечную $L$-шкалу $\kframe{F}=\langle W,R\rangle$, в которой имеется мир $w$, из которого достижимо не менее $n^2$ миров; пусть $\{w_{ij} : i,j\in\{1,\ldots,n\}\}$~--- множество попарно различных миры, достижимых из~$w$. На шкале $\kframe{F}\odot M$ определим интерпретацию~$I$, положив
$$
\begin{array}{lcl}
I(u,T)
  & =
  & \left\{
      \begin{array}{rl}
        \{\langle k\rangle\},  & \mbox{если $u=w_{kk}$;} \\
        \varnothing\phantom{,} & \mbox{в остальных случаях;}
      \end{array}
    \right.
\medskip \\
I(u,Q_1)
  & =
  & \left\{
      \begin{array}{rl}
        \{\langle k\rangle\},  & \mbox{если $u=w_{kj}$ и $\cModel{M}\cmodels P(k,j)$;} \\
        \varnothing\phantom{,} & \mbox{в остальных случаях;}
      \end{array}
    \right.
\medskip \\
I(u,Q_2)
  & =
  & \left\{
      \begin{array}{rl}
        \{\langle k\rangle\},  & \mbox{если $u=w_{ik}$ и $\cModel{M}\cmodels P(i,k)$;} \\
        \varnothing\phantom{,} & \mbox{в остальных случаях.}
      \end{array}
    \right.
\end{array}
$$
Тогда несложно убедиться, что $(\langle \kframe{F}\odot M\rangle,w)\not\models\forall x\,\Diamond T(x)\wedge \mathit{congr}\to \varphi^\ast$, а значит, $\forall x\,\Diamond T(x)\wedge \mathit{congr}\to \varphi^\ast \not\in L_{\mathit{wfin}}$.

Пусть $\forall x\,\Diamond T(x)\wedge \mathit{congr}\to \varphi^\ast \not\in L_{\mathit{wfin}}$. Тогда существуют такие модель $\kmodel{M}=\langle W,R,D,I\rangle$ на конечной $L$-шкале $\langle W,R\rangle$ и мир $w\in W$, что
$$
\begin{array}{lcl}
(\kmodel{M},w)\models \forall x\,\Diamond T(x)\wedge \mathit{congr}
  & \mbox{и}
  & \kmodel{M},w\not\models \varphi^\ast.
\end{array}
$$
Тогда $D_w$ разбивается отношением $E^{I,w}$ на классы конгруэнтности, причём число классов не может быть больше, чем~$2^{|R(w)|}$. Осталось заметить, что эти классы и дают конечную классическую модель, в которой опровергается формула~$\varphi$.

В итоге как следствие описанной конструкции мы получаем, что любая логика, содержащая $\logic{QCl}$ и содержащаяся в $\logic{QS5}_{\mathit{wfin}}$, $\logic{QGL.3.bf}$, $\logic{QGrz.3.bf}$,
$\logic{QGL.bf}_{\mathit{wfin}}\oplus\bm{bd}_2$,
$\logic{QGrz.bf}_{\mathit{wfin}}\oplus\bm{bd}_2$
или $\logic{QK4.3.D.X.bf}_{\mathit{wfin}}$ является одновременно $\Pi^0_1$-трудной и $\Sigma^0_1$-трудной в языке с тремя одноместными предикатными буквами (в частности, не является рекурсивно перечислимой).

\subsubsection{Валентность моделируемой буквы}

Теперь покажем, что условие~\ref{cond8} совершенно несущественно, и вместо бинарной буквы~$P$ можно взять $n$-арную предикатную букву для любого натурального $n\geqslant 2$. Достаточно заметить\footnote{На это обратил внимания автора А.\,В.~Чагров.}, что формулу $P(x_1,\ldots,x_n)$ можно промоделировать формулой $\Diamond (Q_1(x_1)\wedge\ldots\wedge Q_n(x_n))$. В этом случае в доказательстве импликации $(\Leftarrow)$ достаточно рассмотреть шкалы, в которых из мира $w$ достижимы миры множества $\numN^n$, предметные области которых равны или содержат~$\numN$.

Интерес к бинарной букве, тем не менее, имеется: для её моделирования с помощью трюка Крипке достаточно двух одноместных предикатных букв, в то время как при большей валентности используется большее число одноместных букв. Тем не менее, фактически достаточно одной одноместной буквы, и ниже мы это покажем.

\subsubsection{Количество одноместных предикатных букв}

Здесь мы покажем, как обойти условие~\ref{cond10}. Увеличить количество используемых одноместных букв несложно, поэтому вопрос в том, можно ли их уменьшить до одной? Если логика допускает иррефлексивное дерево бесконечного ветвления высоты $3$, то произвольный бинарный предикат можно промоделировать с помощью всего одной одноместной буквы: достаточно вместо $P(x,y)$ подставить формулу $\Diamond(Q(x)\wedge \Diamond Q(y))$. Если $\varphi$~--- классическая формула, не содержащая предикатных букв, отличных от бинарной буквы~$P$, и $\varphi^{\star}$ получается из $\varphi$ такой подстановкой, то
$$
\begin{array}{lcl}
\varphi\in \logic{QCl} & \iff & \varphi^\star \in \logic{QGL},
\end{array}
$$
и, конечно же, аналогичная эквивалентность справедлива для любой логики, содержащей $\logic{QCl}$ и содержащейся в~$\logic{QGL}$.

Такое моделирование не сработает для логик с рефлексивной семантикой (содержащих формулу $\Box p\to p$), поскольку в этом случае если в мире истинна формула $\Diamond(Q(a)\wedge \Diamond Q(b))$, то в нём же будет истинна и формула $\Diamond(Q(a)\wedge \Diamond Q(a))$, т.е. для исходного отношения, соответствующего букве~$P$ должно, как минимум, выполняться свойство $\forall x\,(\exists y\,P(x,y)\to P(x,x))$, а значит, такой подход позволяет моделировать не любое бинарное отношение.

Но для того, чтобы доказать неразрешимость модальных логик в языке с одноместными предикатными буквами, нет необходимости моделировать произвольный бинарный предикат. Дело в том, что теория $\logic{SIB}$ симметричного иррефлексивного бинарного отношения является неразрешимой (это следует из~\cite{NerodeShore80} и~\cite[приложение]{Kremer97}; см.~также следствие~\ref{cor:th:church:cor2}). В этом случае вместо $P(x,y)$ можно сделать подстановку $\Box(\neg Q(x)\vee \neg Q(y))$. И трюк Крипке снова работает. Покажем это.

Пусть $\bm{sib} = \forall x\forall y\,(P(x,y)\to P(y,x))\wedge \forall x\,\neg P(x,x)$.

Для формулы $\varphi$, не содержащей предикатных букв, отличных от $P$, обозначим через $\varphi^\circ$ формулу, получающуюся из $\varphi$ подстановкой вместо $P(x,y)$ формулы $\Box(\neg Q(x)\vee \neg Q(y))$.
$$
\begin{array}{lcl}
\varphi\in\logic{SIB}
  & \iff
  & \bm{sib}\to \varphi\in\logic{QCl}.
\end{array}
$$

Для формулы $B$, не содержащей предикатных букв, отличных от $P$, обозначим через $B^\circ$ формулу, получающуюся из $B$ подстановкой вместо $P(x,y)$ формулы $\Box(\neg Q(x)\vee \neg Q(y))$.

Если $\bm{sib}\to \varphi\in\logic{QCl}$, то $(\bm{sib}\to \varphi\in\logic{QCl})^\circ$ принадлежит любой из модальных предикатных логик, содержащих~$\logic{QCl}$ из-за замкнутости по правилу подстановки.

Пусть $\bm{sib}\to \varphi\not\in\logic{QCl}$. Тогда существует такая классическая модель $\cModel{M}=\langle M,J\rangle$, что $\cModel{M}\not\cmodels \bm{sib}\to \varphi\in\logic{QCl}$.

Пусть $W=\{w\}\cup (M\times M)$, $R=\{w\}\times (M\times M)$. Пусть $\bar{W}$~--- множество, расширяющее~$W$, $\bar{R}$~--- бинарное отношение на~$\bar{W}$, содержащее~$R$, $\bar{\kframe{F}} = \langle\bar{W},\bar{R}\rangle$. Определим интерпретацию~$I$ в шкале $\bar{\kframe{F}}\odot M$ следующим образом:
$$
\begin{array}{lcl}
I(\langle a,b\rangle, Q)
  & =
  & \left\{
      \begin{array}{rl}
        \{\langle a\rangle,\langle b\rangle\}, & \mbox{если $\cModel{M}\not\cmodels P(a,b)$;} \\
        \varnothing,                           & \mbox{если $\cModel{M}\cmodels P(a,b)$;}
      \end{array}
    \right.
\medskip \\
I(w, Q)
  & =
  & \varnothing.
\end{array}
$$
Заметим, что такое определение корректно в силу симметричности и иррефлексивности бинарного отношения, соответствующего букве $P$ в модели~$\cModel{M}$.
Пусть $\kmodel{M}=\langle\bar{\kframe{F}}\odot M,I\rangle$. Тогда нетрудно видеть, что для любых $a,b\in M$ выполняется эквивалентность, похожая на~\mbox{$({\ast}{\ast})$}:
$$
\begin{array}{lcl}
\kmodel{M},w\models \Box (\neg Q(a)\vee \neg Q(b))
  & \iff
  & \cModel{M}\cmodels P(a,b),
\end{array}
$$
откуда получаем, что $\kmodel{M},w\not\models \varphi^\circ$. Осталось заметить, что $\kmodel{M},w\models \bm{sib}^\circ$, а значит, $\kmodel{M},w\not\models (\bm{sib}\to \varphi)^\circ$.

В итоге получаем, что
$$
\begin{array}{lcl}
\varphi\in\logic{SIB}
  & \iff
  & \bar{\kframe{F}},w\models (\bm{sib}\to \varphi)^\circ.
\end{array}
$$

Как следствие, любая модальная предикатная логика, содержащая $\logic{QCl}$ и содержащаяся в
$\logic{QS5}$, $\logic{QGL.3.bf}$, $\logic{QGrz.3.bf}$,
$\logic{QGL.bf}\oplus\bm{bd}_2$,
$\logic{QGrz.bf}\oplus\bm{bd}_2$
или $\logic{QK4.3.D.X.bf}$, не является разрешимой в языке с одной одноместной предикатной буквой.
Фактически, справедлив более сильный результат, который мгновенно получается из следствия~\ref{cor:th:church:cor2} с помощью описанной модификации трюка Крипке.

\begin{theorem}
\label{th:KripkeTrick:3var:Qmodal}
Любая модальная предикатная логика, содержащая $\logic{QCl}$ и содержащаяся в
$\logic{QS5}$, $\logic{QGL.3.bf}$, $\logic{QGrz.3.bf}$,
$\logic{QGL.bf}\oplus\bm{bd}_2$,
$\logic{QGrz.bf}\oplus\bm{bd}_2$
или $\logic{QK4.3.D.X.bf}$, неразрешима в языке с одной одноместной предикатной буквой и тремя предметными переменными.
\end{theorem}

Обратим внимание, что этот результат справедлив не только для нормальных модальных предикатных формул, но и для квазинормальных, и для ненормальных. Более того, утверждение теоремы~\ref{th:KripkeTrick:3var:Qmodal} остаётся справедливым, если вместо логик с указанными свойствами рассматривать произвольные множества формул, для которых выполняются те же включения.

Отметим также, что в построениях мы оттолкнулись от теории симметричного иррефлексивного бинарного отношения. Вместо этого можно было бы сначала заметить, что она в определённом смысле эквивалентна теории симметричного рефлексивного бинарного отношения. Именно, пусть $\varphi'$ получается из $\varphi$ подстановкой формулы $\neg P(x,y)$ вместо $P(x,y)$; тогда несложно понять, что $\varphi$ принадлежит одной из этих теорий тогда и только тогда, когда $\varphi'$ принадлежит другой. Но, отталкиваясь от теории симметричного рефлексивного бинарного отношения, формулу $P(x,y)$ можно моделировать формулой $\Diamond(Q(x)\wedge Q(y))$, и отличия от исходного моделирования в трюке Крипке становятся минимальны.

Ещё одно замечание. Можно показать, что в случае логик конечных шкал тоже достаточно одной одноместной предикатной буквы. Но для этого требуется другое моделирование, позволяющее все имеющиеся одноместные предикатные буквы промоделировать формулами от одной одноместной предикатной буквы. Тем не менее, подготовка формул к такому моделированию связана именно с трюком Крипке, т.к. именно благодаря ему в формулах не остаётся предикатных букв, отличных от унарных. Ниже будут приведены соответствующие конструкции.

\subsubsection{Трюк Крипке внутри модальных формул}
\label{sssec:KripkeTrick:insidemodels}

Мы не затронули условия~\ref{cond7} и~\ref{cond9}, хотя как раз за счёт использования трюка Крипке внутри модальных формул, причём содержащих несколько предикатных букв, удалось получить ряд результатов, касающихся алгоритмических свойств логик одноместных предикатов.

При моделировании бинарной предикатной буквы~$P$ внутри классической формулы требовалось, чтобы в шкалах логики были миры, из которых достижимо достаточно много других миров. Эти другие миры были использованы в построениях только для целей моделирования, т.к. возникающая в формуле модальность была только в моделирующей части. Более точно, если для опровержения формулы $\varphi^\ast$ (или её модификаций) в мире $w$ некоторой шкалы требовались другие миры этой шкалы, то для опровержения формулы $\varphi$ (в случае её опровержимости) достаточно было бы обойтись лишь самим миром~$w$.

Сложность применения трюка Крипке внутри модальных формул состоит в том, что модальность в формуле могла быть изначально, и добавление новых миров в модель может нарушить истинность подформул исходной формулы. Так, например, формула $\varphi = \forall x\forall y\,P(x,y) \to \Diamond\top$ не принадлежит логике $\logic{QK}$ (как, впрочем, любой нормальной модальной предикатной логике, для которой имеется шкала с миром, из которого ничего не достижимо), а формула $\varphi^\ast$ логике~$\logic{QK}$ принадлежит. Чтобы всё-таки применить в таких случаях трюк Крипке, можно предварительно применить приём~\defnotion{релятивизации},\index{релятивизация} которым мы уже не раз пользовались в отношении пропозициональных логик.

Пусть $p$~--- пропозициональная буква (т.е. $0$-арная предикатная буква). Для формулы $\varphi$ определим рекурсивно $p$-релятивизацию $\rho_p\varphi$ формулы~$\varphi$:
$$
\begin{array}{lcll}
\rho_p\psi & = & \psi & \mbox{для атомарной формулы~$\psi$;} \\
\rho_p(\psi'\wedge \psi'') & = & \rho_p\psi'\wedge\rho_p\psi''; \\
\rho_p(\psi'\vee \psi'')   & = & \rho_p\psi'\vee\rho_p\psi''; \\
\rho_p(\psi'\to \psi'')    & = & \rho_p\psi'\to\rho_p\psi''; \\
\rho_p(\forall x\,\psi)    & = & \forall x\,\rho_p\psi; \\
\rho_p(\exists x\,\psi)    & = & \exists x\,\rho_p\psi; \\
\rho_p(\Box \psi)          & = & \Box(p\to\rho_p\psi).
\end{array}
$$

Если модальная предикатная логика $L$ полна по Крипке и является сабфреймовой, то для всякой формулы $\varphi$ и пропозициональной буквы~$p$, не входящей в~$\varphi$,
$$
\begin{array}{lcl}
\varphi\in L & \iff & p\to \rho_p \varphi\in L.
\end{array}
$$
Поясним, почему справедлива эта эквивалентность.

Если $p\to \rho_p \varphi\not\in L$, то имеется модель на $L$-шкале, в которой опровергается $p\to \rho_p \varphi\in L$. Если в этой модели оставить только те миры, в которых истинно~$p$, то получим подмодель, в которой будет опровергаться~$\varphi$. То, что получившаяся модель основана на $L$-шкале, следует из условия сабфреймовости логики~$L$.

Если $\varphi\not\not\in L$, то $\varphi$ опровергается в некотором мире $w$ некоторой модели на некоторой $L$-шкале; если переменную $p$ сделать истинной в этой $L$-модели, то в мире $w$ получившейся модели будет опровергаться формула $p\to \rho_p \varphi$.

Теперь обратимся к построению, о котором говорится в предыдущем абзаце. Если теперь к получившейся модели добавить новые миры и сделать в них переменную $p$ ложной, то их наличие никак не повлияет на то, что в мире $w$ опровергается формула $p\to \rho_p \varphi$. Этим можно воспользоваться для обоснования того, что трюк Крипке продолжает работать. Но так получится не всегда. Если мы можем добавлять независимо для каждого мира старой модели новые миры, делая их достижимыми только из этого мира, то действительно всё получится. А вот если из-за какого-то условия, выполняемого для всех $L$-шкал, придётся сделать миры, добавленные для одного мира, достижимыми и из другого мира, то может возникнуть техническая трудность.

Приведём пример. Пусть
$$
\begin{array}{lcl}
\varphi & = & \forall x\forall y\,(\neg P(x,y)\to \Box \neg P(x,y)).
\end{array}
$$
Если мы сделаем $p$-релятивизацию и затем подстановку, используемую в трюке Крипке, то получим формулу
$$
\begin{array}{lcl}
\varphi' & = & \forall x\forall y\,(\neg \Diamond (Q_1(x)\wedge Q_2(y))\to \Box(p\to \neg \Diamond (Q_1(x)\wedge Q_2(y)))).
\end{array}
$$
Формула $\varphi$ опровергается в двухэлементной шкале с двумя иррефлексивными мирами, где один достижим из другого, а обратной достижимости нет. Таким образом, в частности, $\varphi\not\in\logic{QK4}$. Но дело в том, что, как нетрудно убедиться, $p\to \varphi'\in\logic{QK4}$, поэтому трюк Крипке здесь не срабатывает. Причина кроется в транзитивности: если мы добавим миры, достижимые в модели из какого-то мира $u$, то в силу транзитивности мы должны сделать их достижимыми и из каждого мира, из которого достижим мир~$u$.

Тем не менее, в случае транзитивности выход есть. Трюк Крипке можно применять к формулам, для которых имеется модель $\kmodel{M}=\langle W,R,D,I\rangle$, удовлетворяющая \defnotion{условию наследственности вниз}:\index{условие!наследственности!вниз} для любых $w,u\in W$, таких, что $wRu$, и любых $a,b\in D_w$ из того, что $\kmodel{M},u\models P(a,b)$, следует, что $\kmodel{M},w\models P(a,b)$. Отметим, что в реальных ситуациях это условие довольно часто удаётся выполнить, а значит, трюк Крипке может быть применён.

Отметим, что условие наследственности вниз можно заменить на более привычное (знакомое по семантике для интуиционистских формул) \defnotion{условие наследственности вверх}:\index{условие!наследственности!бяа@вверх} для любых $w,u\in W$, таких что $wRu$, и любых $a,b\in D_w$ из того, что $\kmodel{M},w\models P(a,b)$, следует, что $\kmodel{M},u\models P(a,b)$. В случае модальных логик такое изменение несущественно, просто в формулах вместо $P(x,y)$ нужно подставлять не $\Diamond (Q_1(x)\wedge Q_2(y))$, а $\neg \Diamond (Q_1(x)\wedge Q_2(y))$, и всё.

Наконец, скажем о возможности для исходной формулы содержать предикатные буквы, отличные от единственной бинарной буквы~$P$. Если предварительно сделана $p$-релятивизация, то можно проделывать трюк Крипке выборочно для тех букв, для которых это требуется. Отметим, что если исходная формула не содержит модальностей, то $p$-релятивизация её не меняет, и трюк Крипке можно применять к различным буквам исходной формулы, причём независимо и выборочно. Конечно, для каждой предикатной буквы при этом нужно брать новые унарные предикатные буквы (одну, две или больше~--- зависит от ситуации). Что касается пропозициональной буквы~$p$, используемой в $p$-релятивизации, то её можно заменить, скажем, на формулу $\forall x\,Q(x)$ или~$\exists x\,Q(x)$.

Хороший пример, где всё это соединено вместе, можно найти в~\cite{KKZ05}. При этом в~\cite{KKZ05} используется важная особенность трюка Крипке: отмеченное выше свойство~\ref{cond2}. В результате, доказав неразрешимость модальных логик в языке с двумя предметными переменными, одноместными предикатными буквами и двумя бинарными предикатными буквами, с помощью трюка Крипке авторы получают неразрешимость модальных логик унарных предикатов в языке с двумя предметными переменными. Ещё один пример можно найти в~\cite{MR:2020:JLC:2}, где, используя теорему Трахтенброта и записав условие $\mathit{congr}$ более экономно, с тремя предметными переменными вместо четырёх, была доказана неразрешимость модальных предикатных логик различных классов конечных моделей в языке с унарным предикатом и тремя переменными.

\subsubsection{Полимодальные логики}

Здесь сделаем лишь короткое замечание: конечно же, трюк Крипке применим в полимодальных логиках. Более того, за счёт наличия нескольких модальностей становится возможным преодолеть некоторые технические трудности. Так, например, если мы имеем слияние двух логик с транзитивными модальностями $\Box_1$ и $\Box_2$, то можно рассмотреть модальность $\Box_1\Box_2$, которая уже не будет транзитивной, и позволит применить трюк Крипке. Другой пример: как мы отмечали выше, в мономодальных логиках, содержащих формулу $\mref$, не получается промоделировать произвольное бинарное отношение формулой $\Diamond(Q(x)\wedge \Diamond Q(y))$, в бимодальном же случае можно воспользоваться формулой $\Diamond_1(Q(x)\wedge \Diamond_2 Q(y))$, и всё получится.

\subsubsection{Трюк Крипке в интуиционистском случае}
\label{sec:int:KripkeTrick:short}

Обсуждение трюка Крипке для интуиционистского случая мы отложим до раздела~\ref{sec:KripkeTrick:QInt}: сейчас для его описания у нас, как минимум, нет необходимых синтаксических и семантических определений.

\subsubsection{О других ситуациях}
\label{sec:additions}

Бывают особые ситуации, в которых трюк Крипке применим, но чтобы это обосновать, требуются определённые технические построения. Автор не берётся предусмотреть (и тем более описать) все подобные ситуации, но пример такой ситуации ниже будет приведён (см.~раздел~\ref{sec:NatNumbersFrame:Qmodal}).

\subsubsection{Когда трюк Крипке заблокирован семантически}

О том, когда такой метод моделирования бинарной буквы унарными может не сработать, сказал сам Крипке~\cite{Kripke62}: необходимым условием должно быть отсутствие в шкалах логики миров, из которых достижимо бесконечно много миров. Мы видели, что это условие, вообще говоря, не является достаточным, и показали, что трюк Крипке можно применять для логик, определяемых классами конечных шкал. Тем не менее, нам при этом было важно, что имеются шкалы с мирами, множество достижимых миров из которых может содержать любое наперёд заданное конечное число элементов. Покажем, что если это условие не выполнено, то трюк Крипке может оказаться неприменимым.

Пусть для каждого $n\in \numN$
$$
\begin{array}{lcl}
\logic{QAlt}_n & = & \logic{QK} \oplus \bm{alt}_n,
\end{array}
$$
где
$$
\begin{array}{lcl}
\bm{alt}_n
  & =
  & \displaystyle
    \neg \bigwedge\limits_{i = 0}^n\Diamond(p_i\wedge \bigwedge_{j\ne i} \neg p_j).
\end{array}
$$

Формула $\bm{alt}_n$ требует для своей истинности в шкале Крипке, чтобы из каждого мира шкалы было достижимо не более чем $n$ миров. Таким образом, промоделировать в шкалах логики $\logic{QAlt}_n$ формулу вида $P(x,y)$ формулой вида $\Diamond(Q_1(x)\wedge Q_2(y))$ не получится ни для какого $n\in\numNp$. Более того, никакие модификации трюка Крипке здесь тоже не возможны; покажем это.

Поскольку логика $\logic{QAlt}_n$ содержит логику $\logic{QCl}$, фрагмент $\logic{QAlt}_n$ с одной бинарной предикатной буквой неразрешим. Покажем, что монадический фрагмент $\logic{QAlt}_n$, тем не менее, разрешим; это будет, в частности, означать, что моделирование бинарной буквы унарными в $\logic{QAlt}_n$ невозможно.

Для этого докажем техническую лемму. Для её формулировки введём несколько понятий.

Формулу $\varphi$ языка $\lang{QML}$ называем \defnotion{монадической},\index{уяа@формула!монадическая} если все входящие в неё предикатные буквы являются унарными. \defnotion{Монадическим фрагментом языка}\index{уяа@фрагмент!монадический}~$\lang{QML}$ называем множество всех монадических $\lang{QML}$-формул. \defnotion{Монадическим фрагментом логики}~$L$ называем множество всех монадических формул, принадлежащих~$L$.

\begin{lemma}
  \label{lem:main:unary:finiteframe}
  Пусть $\vp$~--- монадическая формула от $n$ предикатных букв, опровергающаяся в предикатной шкале
  $\kFrame{F} = \otuple{W, R, D}$ с конечным множеством~$W$. Тогда
  $\vp$ опровергается в некоторой шкале
  $\bar{\kFrame{F}} = \otuple{ W, R, \bar{D}}$, где
  $|\bar{D}^+| \leqslant 2^{|W|(n+1)}$; если при этом
  $\kFrame{F}$~--- шкала с постоянными областями, то $\bar{\kFrame{F}}$~--- тоже шкала с постоянными областями.
\end{lemma}

\begin{proof}
  Пусть $P_1, \ldots, P_n$~--- все унарные предикатные буквы, входящие в~$\varphi$. Пусть   $\kModel{M} = \otuple{W, R, D, I}$~--- модель на $\kFrame{F}$, такая, что $(\kModel{M}, w_0) \not\models \vp$ для некоторого $w_0 \in W$.

  Определим бинарное отношение $\sim$ на $D^+$, положив $a \sim b$, если для каждого $w \in W$ и каждого $i \in \{1, \ldots, n\}$
  $$
  \begin{array}{lllcl}
    (1) && a\in D_w & \iff & b\in D_w; \\
    (2) && (\kModel{M},w)\models P_i(a) & \iff & (\kModel{M},w)\models P_i(b).
  \end{array}
  $$
  Нетрудно видеть, что $\sim$ является эквивалентностью на~$D^+$. Для каждого $a \in D^+$ положим $\bar{a} = \{c\in D^+ : c\sim a\}$; пусть также $\mathcal{D} = \set{\bar{a} : a \in D^+}$.

  Определим функцию $\bar{D}$: для каждого $w \in W$ положим
  $$
  \begin{array}{lcl}
  \bar{D}_w & = & \{\bar{a} \in \mathcal{D} : a \in D_w\}.
  \end{array}
  $$
  Тогда $\bar{\kFrame{F}} = \langle W, R, \bar{D} \rangle$~--- предикатная шкала: для любых $w,v\in W$ условие $w R v$ влечёт, что $D_w\subseteq D_v$, откуда получаем, что
  $\bar{D}_w\subseteq \bar{D}_v$; кроме того, если $D_w = D_v$, то $\bar{D}_w = \bar{D}_v$
  (это даёт выполнение дополнительного условия в формулировке леммы). Осталось показать, что $\bar{\kFrame{F}}$~--- требуемая шкала.

  Пусть $\bar{I}$~--- интерпретация предикатных букв в $\bar{\kFrame{F}}$, такая, что для каждого $w \in W$ и каждого $i \in \{1, \ldots, n\}$
  $$
  \begin{array}{lcl}
  \bar{I}(w,P_i) & = & \{\langle\bar{a}\rangle : \langle {a}\rangle \in I(w,P_i)\};
  \end{array}
  $$
  пусть $\bar{\kModel{M}} = \langle \bar{\kFrame{F}}, \bar{I} \rangle$.
  Нетрудно видеть, что для каждой подформулы $\psi(x_1,\ldots,x_m)$ формулы $\varphi$, каждого $w \in W$ и любых $a_1,\ldots,a_m \in D_w$,
  $$
  \begin{array}{lcl}
    (\kModel{M},w) \models \psi(a_1,\ldots,a_m)
      & \iff
      & (\bar{\kModel{M}},w) \models \psi(\bar{a}_1,\ldots,\bar{a}_m).
  \end{array}
  $$
  В частности, получаем, что $(\bar{\kModel{M}}, w_0) \not\models \varphi$, а значит,
  $\bar{\kFrame{F}} \not\models \vp$.

  Остаётся показать, что $|\bar{D}^+| \leqslant 2^{|W|(n+1)}$, т.е. что $|\mathcal{D}| \leqslant 2^{|W|(n+1)}$.
  Условие~$(1)$ определяет отношение эквивалентности на $D^+$, разбивая $D^+$ на не более чем
  $2^{|W|}$ классов эквивалентности. Каждый из этих классов разбивается отношением эквивалентности, определяемым условием~$(2)$, на не более чем $2^{|W| \cdot n}$ классов эквивалентности; в итоге получаем разбиение на классы эквивалентности по отношению~$\sim$. Значит, $|\mathcal{D}| \leqslant 2^{|W|(n+1)}$. Поскольку $\bar{D}^+ = \mathcal{D}$, получаем, что $|\bar{D}^+| \leqslant 2^{|W|(n+1)}$.
\end{proof}

Вернёмся к логике $\logic{QAlt}_n$, где $n\in\numN^+$. Можно доказать~\cite{ShehtmanShkkatov20,ShSh23}, что логика $\logic{QAlt}_n$ полна по Крипке, и её шкалами являются в точности шкалы, где каждый мир видит не более $n$ миров; аналогичное утверждение справедливо и для логики $\logic{QAlt}_n\oplus \mref$, с~той разницей, что шкалы должны быть рефлексивными. Кроме того, если формула $\varphi$ не принадлежит логике $\logic{QAlt}_n\oplus \mref$, то она опровергается в некоторой конечной шкале Крипке этой логики, поскольку помимо ограничения на мощность ветвления в мире, модальная глубина формулы даёт нам дополнительное ограничение на глубину интересующей нас части шкалы, позволяя тем самым рассматривать лишь конечные шкалы, размер которых определяется числом $n$ и модальной глубиной формулы~$\varphi$. Именно, достаточно рассматривать шкалы, число миров в которых не превосходит числа
$$
  1+n+n^2+\ldots+n^{\mathop{\mathsf{md}}\varphi},
$$
т.е. числа
$$
\frac{n^{\mathop{\mathsf{md}\varphi+1}}-1}{n-1}.
$$

Используя сделанное наблюдение, а также лемму~\ref{lem:main:unary:finiteframe}, получаем следующую теорему.

\begin{theorem}
\label{th:QAlt_n:decidability}
Пусть $n\in\numN$ и $L\in \{\logic{QAlt}_n, \logic{QAlt}_n\oplus \mref\}$. Тогда монадические фрагменты логик $L$, $L\logic{.bf}$ разрешимы.
\end{theorem}

\begin{proof}
Следует из леммы~\ref{lem:main:unary:finiteframe} и полноты по Крипке указанных логик.
\end{proof}

Отметим, что
$$
\begin{array}{lcl}
\logic{QAlt}_0 & = & \logic{QVerum}; \\
\logic{QAlt}_0\oplus\mref & = & \lang{QML}; \\
\logic{QAlt}_1\oplus\mref & = & \logic{QTriv},
\end{array}
$$
и разрешимость монадических фрагментов этих логик тривиальна: $\lang{QML}$~--- множество всех формул, а $\logic{QVerum}$ и $\logic{QTriv}$~--- логики, соответственно, шкалы из одного иррефлексивного мира и шкалы из одного рефлексивного мира, поэтому разрешимость их монадических фрагментов мгновенно следует из разрешимости монадического фрагмента логики~$\logic{QCl}$~(см., например,~\cite[глава~21]{BBJ07} или~\cite[глава~25]{BJ-1994-1-rus}).

\subsubsection{Когда трюк Крипке заблокирован синтаксически}

На синтаксическом уровне необходимым условием для трюка Крипке является возможность написания формулы $\Diamond(Q_1(x)\wedge Q_2(y))$, $\Box(\neg Q_1(x)\vee \neg Q_2(y))$ или какой-то похожей. Во всех случаях в области действия модальности оказываются две свободные предметные переменные (в интуиционистском случае~--- в области действия импликации). Что будет, если мы рассмотрим фрагмент логики, в котором такие формулы запрещены?

При исследовании этого вопроса были получены результаты о разрешимости некоторых довольно богатых фрагментов. Именно, было введено понятие \defnotion{монодического}\index{уяа@фрагмент!монодический} фрагмента языка, определяемого следующим условием: модальность $\Box$ может использоваться для построения формул вида $\Box\varphi$, только если $\varphi$ содержит не более одной свободной переменной. При таких ограничениях трюк Крипке становится невозможным, поскольку в нём требуется использовать формулы, в которых в области действия модальности имеются две свободные переменные. Оказалось, что монодические фрагменты многих логик разрешимы~\cite{WZ01}, в частности, разрешимы монодические фрагменты монадических фрагментов модальных логик $\logic{QK}$, $\logic{QT}$, $\logic{QK4}$, $\logic{QS4}$~и~др. Отметим, что предложенная в~\cite{WZ01} техника была применена авторами~\cite{WZ01} для доказательства разрешимости монодических фрагментов не только модальных, но и полимодальных предикатных логик~\cite{WZ02}.

	\subsection{Логики с двумя переменными}
    \label{s7-2-2}
\subsubsection{Логики, содержащиеся в $\logic{QGL}$ и $\logic{QGrz}$}
\label{sec:undecidability:QGL:QGrz}

Выше мы показали (теорема~\ref{th:KripkeTrick:3var:Qmodal}), что многие модальные предикатные логики неразрешимы в языке с одной унарной предикатной буквой и тремя предметными переменными.
Покажем, что модальные предикатные логики, содержащиеся в $\kflogic{\logic{GL}}$ или в
$\kflogic{\logic{Grz}}$ неразрешимы в языке, содержащем одну унарную предикатную букву и лишь две предметные переменные.
В доказательстве будем пользоваться некоторой модификацией формул, описанных в работе Р.\,Кончакова, А.\,Куруш и М.\,Захарьящева~\cite{KKZ05}; эти формулы позволяют промоделировать в модальных логиках неразрешимую проблему укладки домино.

Напомним неразрешимую проблему укладки домино, которую мы рассматривали выше.
\defnotion{Плитки домино} имеют квадратную форму одного и того же фиксированного размера; для каждой такой плитки зафиксирована ориентация её сторон: указаны её \defnotion{верхняя}, \defnotion{нижняя}, \defnotion{правая} и \defnotion{левая} стороны. Каждая плитка имеет \defnotion{тип} $t$, определяемый \defnotion{цветами} $\leftsq t$, $\rightsq t$, $\upsq t$ и $\downsq t$ её сторон. \defnotion{Задача домино}\index{еяд@задача!домино} определяется набором $T = \{t_0, \ldots, t_{n}\}$ типов плиток домино и состоит в том, что нужно выяснить, существует ли \defnotion{$T$-укладка},\index{укладка плиток домино} т.е. функция $f\colon \mathds{N} \times \mathds{N} \to T$, такая, что для любых $i, j \in \mathds{N}$
\begin{itemize}
\item[]$(T_1)$\quad $\rightsq f(i,j) = \leftsq f(i+1,j)$;
\item[]$(T_2)$\quad $\upsq f(i,j) = \downsq f(i,j+1)$.
\end{itemize}

Пусть $\vp$~--- замкнутая монадическая формула,
$P_1, \ldots, P_n$~--- входящие в неё унарные предикатные буквы. Пусть $P_{n+1}$~--- унарная предикатная буква, отличная от $P_1, \ldots, P_n$, и пусть $B = \forall x\, P_{n+1}(x)$.
Определим перевод $\cdot'$:
$$
  \begin{array}{lcll}
    {P_i(x)}' & = & P_i(x) & \mbox{для $i \in \{1, \ldots, n \}$;} \\
    (\bot)' & = & \bot; \\
    (\phi \con \psi)' & = & \phi' \con \psi';  \\
    (\phi \dis \psi)' & = & \phi' \dis \psi';  \\
    (\phi \imp \psi)' & = & \phi' \imp \psi';  \\
    (\forall x\, \phi)' & = & \forall x\, \phi';  \\
    (\exists x\, \phi)' & = & \exists x\, \phi';  \\
    (\Box \phi)' & = & \Box (B \imp \phi').
  \end{array}
$$

\begin{lemma}
  \label{lem:vp-B}
  Пусть $L \in \{\logic{QK}, \kflogic{\logic{GL.bf}}, \kflogic{\logic{Grz}}\}$. Тогда
  $$
  \begin{array}{lcl}
  \mbox{$\vp$ $L$-выполнима} & \iff & \mbox{$B \con \vp'$ $L$-выполнима.}
  \end{array}
  $$
\end{lemma}

\begin{proof}
  Предположим, что $(\kModel{M},{w_0})\models{\vp}$ для некоторой модели Крипке $\kModel{M}=\otuple{W,R,D,I}$, определённой на $L$-шкале, и некоторого мира~$w_0\in W$. Пусть $\kModel{M}'=\otuple{W,R,D,I'}$~--- модель, отличающаяся от $\kModel{M}$ лишь тем, что $I'(w, P_{n+1}) = D(w)$ для каждого $w \in W$. Тогда нетрудно понять, что $(\kModel{M}',{w_0})\models{B \con \vp'}$.

  Теперь предположим, что $(\kModel{M},{w_0})\models{B \con \vp'}$ для некоторой модели Крипке $\kModel{M}=\otuple{W,R,D,I}$, определённой на $L$-шкале, и некоторого мира~$w_0\in W$.
  Пусть $\kModel{M}'=\otuple{W',R',D',I'}$~--- подмодель модели~$\kModel{M}$, где
  $W' = \set{w : (\kModel{M},{w})\models {B}} $.  Тогда $(\kModel{M}',{w_0})\models{\vp}$. Осталось заметить, что модель $\kModel{M}'$ определена на $L$-шкале.
\end{proof}


Пусть $P$~--- унарная предикатная буква; определим рекурсивно следующие формулы:
$$
\begin{array}{lcll}
    \delta_1(x)
      & =
      & P (x) \con \Diamond (\neg P(x) \con \Diamond \Box^+ P(x) );
      \\
    \delta_{m+1}(x)
      & =
      & P (x) \con \Diamond (\neg P(x) \con \Diamond \delta_{m}(x))
      & \mbox{для $m\in\numNp$}.
\end{array}
$$

Для каждого
$k \in \{1, \ldots, n+1 \}$ положим
$$
\begin{array}{lcl}
\alpha_k (x ) & = & \delta_{k}(x) \con \neg \delta_{k+1}(x) \con
\Diamond \Box^+ \neg P(x).
\end{array}
$$

Кроме того, для каждого $k \in \{1, \ldots, n+1 \}$ положим
$\kframe{F}_k = \langle W_k,R_k\rangle$, где
$W_k = \{w_k^0,\ldots,w_k^{2k}\}\cup\{w_k^\ast\}$ и $R_k$~--- транзитивное замыкание отношения
$\{\langle w_k^i,w_k^{i+1}\rangle : 0\leqslant i < 2k\}\cup\{\langle
w_k^0,w_k^\ast\rangle\}$; пусть также $\kframe{F}_k^r = \langle W_k,R_k^\ast\rangle$, где $R_k^\ast$~--- рефлексивное замыкание отношения~$R_k$.

Пусть $k\in \{1, \ldots, n+1 \}$, $\mathcal{D}$~--- непустое множество и $a\in \mathcal {D}$. Модель $\kModel{M}=\langle \kframe{F}_k\odot \mathcal{D},I\rangle$  называем \defnotion{$a$-подходящей}, если
$$
\begin{array}{lcl}
(\kModel{M},w)\models P(a) & \iff & \mbox{$w = w_k^{2i}$ для некоторого $i\in\{0,\ldots,k\}$};
\end{array}
$$
модель $\kModel{M}=\langle \kframe{F}_k^r\odot \mathcal{D},I\rangle$  называем \defnotion{$a$-подходящей}, если
$$
\begin{array}{lcl}
(\kModel{M},w)\models P(a) & \iff & \mbox{$w = w_k^{2i}$ для некоторого $i\in\{0,\ldots,k\}$}.
\end{array}
$$

\begin{lemma}
\label{lem:model-M_k}
Пусть $\kModel{M}_k=\langle \kframe{F}_k\odot \mathcal{D},I\rangle$ для $k\in \{1, \ldots, n+1 \}$ и непустого множества~$\mathcal{D}$.
Пусть $a\in\mathcal{D}$ и\/
$\kModel{M}_1,\ldots,\kModel{M}_{n+1}$~--- $a$-подходящие. Тогда
$$
\begin{array}{lcl}
(\kModel{M}_k,w)\models \alpha_m(a)
  & \iff
  & \mbox{$k=m$ и $w=w_k^0$}.
\end{array}
$$
\end{lemma}

\begin{proof}
Состоит в непосредственной проверке.
\end{proof}

\begin{lemma}
  \label{lem:ref-closures}
Пусть $\kModel{M}_k=\langle \kframe{F}_k^r\odot \mathcal{D},I\rangle$ для $k\in \{1, \ldots, n+1 \}$ и непустого множества~$\mathcal{D}$.
Пусть $a\in\mathcal{D}$ и\/
$\kModel{M}_1,\ldots,\kModel{M}_{n+1}$~--- $a$-подходящие. Тогда
$$
\begin{array}{lcl}
(\kModel{M}_k,w)\models \alpha_m(a)
  & \iff
  & \mbox{$k=m$ и $w=w_k^0$}.
\end{array}
$$
\end{lemma}

\begin{proof}
Состоит в непосредственной проверке.
\end{proof}

Для каждого $k \in \{1, \ldots, n+1 \}$ положим
$$
\begin{array}{lcl}
\beta_k(x) & = & \neg P(x)\con\Diamond \alpha_k(x).
\end{array}
$$

Пусть $\varphi^\ast$~--- формула, получающаяся из формулы $\varphi'$ подстановкой
формул $\beta_1(x),\ldots,\beta_{n+1}(x)$ вместо $P_1(x),\ldots,P_{n+1}(x)$ соответственно.

Модель $\kModel{M}$, определённую на $L$-шкале, называем \defnotion{$L$-подходящей}, если $\kModel{M}\models \Diamond Q(x)\to Q(x)$ для каждой одноместной предикатной буквы~$Q$.
Монадическую формулу $\psi$ называем \defnotion{$L$-подходящей}, если $\psi$ либо не является $L$\nobreakdash-выполнимой, либо $\psi$ истинна в некотором мире некоторой $L$-подходящей модели.

\begin{lemma}
  \label{lem:vp-ast}
  Пусть $L \in \{\logic{QK}, \kflogic{\logic{GL.bf}}, \kflogic{\logic{Grz.bf}}\}$ и пусть $\vp$~--- $L$-под\-хо\-дя\-щая формула. Тогда
  $$
  \begin{array}{lcl}
   \mbox{$B \con \vp'$ $L$-выполнима} & \iff & \mbox{$\forall x\, \beta_{n+1} (x) \con \vp^\ast$ $L$-выполнима.}
  \end{array}
  $$
\end{lemma}

\begin{proof}
  $(\Leftarrow)$
  Если формула $B \con \vp'$ не является $L$-выполнимой, то в силу замкнутости $L$ по правилу подстановки, получаем, что формула $\forall x\, \beta_{n+1} (x) \con \vp^\ast$ тоже не является $L$-выполнимой.

  $(\Rightarrow)$ Сначала рассмотрим случай, когда $L=\logic{QK}$.

  Пусть формула $B \con \vp'$ является $\logic{QK}$-выполнимой. Тогда существуют модель Крипке
  $\kModel{M}=\langle W,R,D,I\rangle$ и мир $w_0\in W$, такие, что ($\kModel{M},w_0)\models B \con \vp'$. Учитывая доказательство леммы~\ref{lem:vp-B}, можем (и будем) считать, что $\kModel{M}\models B$.

  Для каждого $w\in W$ и каждого $k\in\set{1, \ldots, n+1}$ определим шкалу
  $\kframe{F}_k^w = \langle \{w\}\times W_k,R_k^w\rangle$, являющуюся изоморфной копией шкалы~$\kframe{F}_k$. Для каждого $w\in W$ и каждого $k\in \set{1,\ldots,n+1}$ добавим множество $\{w\}\times W_k$ к $W$; получившееся множество обозначим~$W^\ast$. Определим отношение $R^\ast$ на $W^\ast$:
  $$
  \begin{array}{lcl}
  R^\ast
    & =
    & \displaystyle
      R\cup\bigcup 
  \big\{R_k^w \cup \{\otuple{w,\otuple{w,w_k^0}}\} : \mbox{$w\in W$, $1\leqslant k\leqslant n+1$}\big\}.
  \end{array}
  $$
  Таким образом, для каждого $w\in W$ мы делаем достижимыми из $w$ корни шкал
  $\kframe{F}_1^w,\ldots,\kframe{F}_{n+1}^w$. Пусть для каждого $u\in W^\ast$
$$
\begin{array}{lcl}
D^\ast(u)
  & =
  & \left\{
      \begin{array}{rl}
        D(u), & \mbox{если $u\in W$}, \\
        D(w), & \mbox{если $u\in \{w\}\times W_k$}.
      \end{array}
     \right.
\end{array}
$$
Положим для каждого $u\in W^\ast$ и каждого $a\in D^\ast(u)$
$$
\begin{array}{lcl}
\langle a \rangle \in I^\ast(u,P)
  & \bydef
  & \parbox[t]{320pt}{существуют такие $w\in W$, $k\in\{1,\ldots,n+1\}$ и $i\in\{0,\ldots,k\}$, что $u = \otuple{w, w_k^{2i}}$ и $(\kModel{M},w)\models P_k[a]$.}
\end{array}
$$
Пусть $\kModel{M}^\ast = \otuple{W^\ast,R^\ast,D^\ast,I^\ast}$. Из леммы~\ref{lem:model-M_k} получаем, что для каждого
$w\in W$, каждого $a\in D(w)$ и каждого $k\in\set{1,\ldots,n+1}$,
$$
\begin{array}{lcl}
(\kModel{M},w)\models P_k(a) & \iff & \kModel{M}^\ast,w\models \beta_k (a).
\end{array}
$$

Тогда для каждого $w\in W$, каждой подформулы $\psi(x_1,\ldots,x_m)$ формулы $\vp$ и любых
$a_1,\ldots,a_m\in D(w)$,
$$
\begin{array}{lcl}
(\kModel{M},w)\models \psi'(a_1, \ldots, a_m) & \iff &
(\kModel{M}^\ast,w)\models  \psi^\ast(a_1, \ldots, a_m),
\end{array}
$$
где формула $\psi^\ast(x_1,\ldots,x_m)$ получается из формулы $\psi'(x_1,\ldots,x_m)$ подстановкой формул $\beta_1(x),\ldots, \beta_{n+1}(x)$ вместо $P_1(x),\ldots,P_{n+1}(x)$ соответственно.

Доказательство этого утверждения проводится индукцией по построению $\psi(x_1,\ldots,x_m)$. Рассмотрим только случай, когда $\psi(x_1,\ldots,x_m) = \Box\chi(x_1,\ldots,x_m)$, а значит,
$$
\begin{array}{lcl}
\psi'(x_1,\ldots,x_m) & = & \Box(\forall x\, P_{n+1}(x) \to
\chi'(x_1,\ldots,x_m)); \\
\psi^\ast(x_1,\ldots,x_m) & = & \Box(\forall x\, \beta_{n+1}(x) \to
\chi^\ast(x_1,\ldots,x_m)).
\end{array}
$$
Импликация $(\Leftarrow)$ очевидна, и мы обоснуем только импликацию $(\Rightarrow)$. Если $(\kModel{M}^\ast,w)\not\models \psi^\ast(a_1,\ldots,a_m)$, то существует мир $w'\in W^\ast$, такой, что $wR^\ast w'$,
$(\kModel{M}^\ast,w')\models \forall x\, \beta_{n+1}(x)$ и
$(\kModel{M}^\ast,w')\not\models\chi^\ast(a_1,\ldots,a_m)$. Поскольку условие
$(\kModel{M}^\ast,w')\models \forall x\,\beta_{n+1}(x)$ гарантирует, что
$w'\in W$, можем применить индукционное предположение; получаем, что $(\kModel{M},w')\not\models\chi'(a_1,\ldots,a_m)$.

Таким образом,
$(\kModel{M}^\ast,w_0)\models \forall x\, \beta_{n+1} (x) \con \vp^\ast$,
а значит, формула $\forall x\, \beta_{n+1} (x) \con \vp^\ast$ является $\logic{QK}$-выполнимой.

Если $L=\kflogic{\logic{GL.bf}}$ или $L=\kflogic{\logic{Grz.bf}}$, то доказательство аналогично.
Разница в том, что когда строится модель
$\kModel{M}^{\ast}$, вместо $R^\ast$ нужно взять транзитивное замыкание получающегося отношения для $L=\kflogic{\logic{GL.bf}}$ и, соответственно, рефлексивно-транзитивное замыкание для $L=\kflogic{\logic{Grz.bf}}$. При этом важно, что в обоих случаях формулы $\vp$ и $B \wedge \vp'$ являются $L$-подходящими, поэтому модель $\kModel{M}$, в мире $w_0$ которой истинна формула $B \wedge \vp'$, можно (и~нужно) взять $L$-подходящей.
\end{proof}


\begin{theorem}
  \label{thr:Grz}
  \label{thrm:QGrz}
  Пусть $L$~--- модальная предикатная логика, такая, что\/
  $\logic{QK} \subseteq L\subseteq \kflogic{\logic{GL.bf}}$ или\/
  $\logic{QK} \subseteq L\subseteq \kflogic{\logic{Grz.bf}}$. Тогда $L$ неразрешима в языке с двумя предметными переменными и одной унарной предикатной буквой.
\end{theorem}

\begin{proof}
%

      Сведём к проблеме $L$-выполнимости формул указанного в формулировке фрагмента неразрешимую~\cite{Berger66} проблему укладки плиток домино в сетке~$\nat \times \nat$, описанную в начале данного раздела (см.~также раздел~\ref{sec:s6-2-1}).
      Для множества $T$ типов плиток домино возьмём формулу $\chi_T$, определённую в~\cite{KKZ05}; $\chi_T$ определяется как конъюнкция следующих формул:
      \[
      \begin{array}{ll}
      \arrayitem &
           \displaystyle \forall
           x\,\bigvee\limits_{t\in T}(P_t(x)\wedge\bigwedge\limits_{t'\ne
           t}P_{t'}(x));
           \arrayitemskip\\
      \arrayitem &
           \displaystyle \forall x\forall y\,(H(x,y)\to
           \bigwedge\limits_{\mathclap{\mathop{\scriptsize\rightsquare{0.21}} t\,{\ne}\,\mathop{\scriptsize\leftsquare{0.21}} t'}}\neg(P_t(x)\wedge
           P_{t'}(y)));
           \arrayitemskip\\
      \arrayitem &
           \displaystyle \forall x\forall y\,(V(x,y)\to
           \bigwedge\limits_{\mathclap{\mathop{\scriptsize\upsquare{0.21}} t\,{\ne}\,\mathop{\scriptsize\downsquare{0.21}} t'}}\neg(P_t(x)\wedge
           P_{t'}(y)));
           \arrayitemskip\\
      \arrayitem &
           \displaystyle \forall x\exists y\,H(x,y) \wedge \forall
           x\exists y\,V(x,y);
           \arrayitemskip\\
      \arrayitem &
           \displaystyle \forall x\forall y\,(H(x,y) \to \Box H(x,y));
           \arrayitemskip\\
      \arrayitem &
           \displaystyle \forall x\forall y\,(V(x,y) \to \Box V(x,y));
           \arrayitemskip\\
      \arrayitem &
           \displaystyle \forall x\forall y\,(\Diamond V(x,y) \to
           V(x,y));
           \arrayitemskip\\
      \arrayitem &
           \displaystyle \forall x\,\Diamond D(x);
           \arrayitemskip\\
      \arrayitem &
           \displaystyle \Box\forall x\forall y\,(V(x,y)\wedge \exists
           x\, (D(x)\wedge H(x,y)) \to {}
           \\
           & \hfill
             ~~~~~~~~~~~~~~~~~~~~~~~~~{} \to \forall y\,(H(x,y)\to \forall x\,(D(x)\to V(y,x))));
      \end{array}
      \]
         в~\cite{KKZ05} доказано, что
         $$
         \begin{array}{lcl}
           \mbox{$\chi_T$ $L$-выполнима} &
           \iff & \mbox{существует $T$-укладка $\nat\times \nat$.}
         \end{array}
           \eqno {({\ast})}
         $$

         По формуле $\chi_T$ построим $L$-подходящую формулу
         $\chi^\star_T$, для которой выполняется условие, аналогичное~($\ast$).

         Пусть $L = \logic{QK}$. Пусть $\chi^\circ_T$~--- формула, получающаяся из $\chi_T$ подстановкой $\neg D(x)$ вместо $D(x)$. Ясно, что $\chi^\circ_T$ $\logic{QK}$-выполнима тогда и только тогда, когда $\chi_T$ $\logic{QK}$-выполнима.
         Заменим в $\chi^\circ_T$ каждое вхождение формулы вида $\Box\psi$ на
         $\Box(\forall x\,Q(x)\to\psi)$, а также подставим
         $\Diamond(\neg Q_1^H(x)\wedge \neg Q_2^H(y))$ и
         $\Diamond(\neg Q_1^V(x)\wedge \neg Q_2^V(y))$ вместо, соответственно, $H(x,y)$ и $V(x,y)$.  Получившуюся формулу обозначим $\bar{\chi}^\circ_T$ и положим
         $\chi^\star_T=\forall x\,Q(x)\wedge\bar{\chi}^\circ_T$.
         Покажем, что
         $$
         \begin{array}{lcl}
           \mbox{$\chi^\star_T$ $L$-выполнима} &
           \iff & \mbox{$\chi_T^\circ$ $L$-выполнима.}
         \end{array}
         $$

         Пусть $\chi_T^\circ$ $L$-выполнима. Тогда, по ($\ast$), существует укладка $\tau\colon\nat\times\nat\to T$.  Определим по $\tau$ модель $\kModel{M}=\otuple{W,R,D,I}$ следующим образом:
         $$
         \begin{array}{lcl}
         W & =
           & \{w^\ast\} \cup \{w_{ij} : i,j\in \nat\} \cup \{w'_{ij} : i,j\in \nat\}; \\
         R & =
           & \big( \{w^\ast\} \times \{w_{ij} : i,j\in \nat\} \big) \cup {}
           \\
           &
           & \hfill{} \cup
             \big( \big( \{w^\ast\}\cup\{w_{ij} : i,j\in \nat\} \big)
             \times \{w'_{ij} : i,j\in \nat\} \big);
             \\
         D_w
           & =
           & \nat\times\nat \quad\mbox{для каждого $w\in W'$;}
           \\
         I(w,Q)
           & =
           & \left\{
               \begin{array}{rl}
                 \varnothing,    & \mbox{если $w=w'_{ij}$,}\\
                 \nat\times\nat, & \mbox{если $w\ne w'_{ij}$;}
               \end{array}
             \right.
             \\
         I(w,Q_1^H)
           & =
           & \left\{
             \begin{array}{rl}
               \nat\times\nat\setminus\{\langle i,j\rangle\}, & \mbox{если $w=w'_{ij}$,}\\
               \nat\times\nat, & \mbox{если $w\ne w'_{ij}$;}
             \end{array}
             \right.
             \\
         I(w,Q_2^H)
           & =
           & \left\{
             \begin{array}{rl}
               \nat\times\nat\setminus\{\langle i+1,j\rangle\}, & \mbox{если $w=w'_{ij}$,}\\
               \nat\times\nat, & \mbox{если $w\ne w'_{ij}$;}
             \end{array}
             \right.
             \\
         I(w,Q_1^V)
           & =
           & \left\{
             \begin{array}{rl}
               \nat\times\nat\setminus\{\langle i,j\rangle\}, & \mbox{если $w=w'_{ij}$,}\\
               \nat\times\nat, & \mbox{если $w\ne w'_{ij}$;}
             \end{array}
             \right.
             \\
         I(w,Q_2^V)
           & =
           & \left\{
             \begin{array}{rl}
               \nat\times\nat\setminus\{\langle i,j+1\rangle\}, & \mbox{если $w=w'_{ij}$,}\\
               \nat\times\nat & \mbox{если $w\ne w'_{ij}$;}
             \end{array}
             \right.
             \\
         I(w^\ast,D)  & = & \nat\times\nat; \\
         I(w_{ij},D)  & = & \nat\times\nat \setminus \{\langle i,j\rangle\}; \\
         I(w'_{ij},D) & = & \varnothing; \\
         I(w,P_t)     & = & \{\langle i,j\rangle\in \nat\times\nat : \tau(i,j)=t\}
                            \quad \mbox{для всех $w\in W$ и $t\in T$.} \\
         \end{array}
         $$
               Тогда $(\kModel{M}, w^\ast) \models \chi^\star_T$.

               Пусть теперь $\chi^\star_T$ $L$-выполнима, т.е.
               $(\kModel{M}_0, w_0) \models \chi^\star_T$ для некоторой модели Крипке
               $\kModel{M}_0$ и некоторого мира $w_0$ в~$\kModel{M}_0$. Удалим из $\kModel{M}_0$ все миры, в которых опровергается формула $\forall x\,Q(x)$ и определим интерпретацию для $H$ и $V$ в мире как множества пар элементов его предметной области, на которых истинны, соответственно, формулы
               $\Diamond(\neg Q_1^H(x)\wedge \neg Q_2^H(y))$ и
               $\Diamond(\neg Q_1^V(x)\wedge \neg Q_2^V(y))$. Тогда
               $\chi^\circ_T$ будет истинной в мире $w_0$ получившейся модели.

               Осталось заметить, что формула $\chi^\star_T$ является
               $\logic{QK}$-подходящей, поскольку модель $\kModel{M}$ является
               $\logic{QK}$-подходящей.

               Пусть теперь
               $L \in \set{\kflogic{\logic{GL.bf}}, \kflogic{\logic{Grz.bf}}}$.
               Для $L=\kflogic{\logic{GL}}$ доказательство проводится так же, как и для $\logic{QK}$, поскольку описанные построения позволяют получить из $\kflogic{\logic{GL.bf}}$-шкалы $\kflogic{\logic{GL.bf}}$-шкалу. Для $L=\kflogic{\logic{Grz}}$ построения отличаются лишь тем, что нужно брать рефлексивное замыкание возникающего отношения достижимости.


             Пусть $F = \set{\neg \chi^\star_T : \mbox{$T$~--- набор типов плиток домино}}$. Тогда $F$ содержит только $L$-подходящие монадические формулы в языке с двумя предметными переменными, причём $\logic{QK}\cap F = \kflogic{\logic{GL.bf}}\cap F =
             \kflogic{\logic{Grz.bf}}\cap F$ и $\logic{QK}\cap F$~--- неразрешимое множество.


Тогда, согласно леммам~\ref{lem:vp-B} и~\ref{lem:vp-ast}, для
$L\in\set{\logic{QK},\kflogic{\logic{GL.bf}},\kflogic{\logic{Grz.bf}}}$ справедлива эквивалентность
$$
\begin{array}{lcl}
  \neg \chi^\star_T \in L\cap F & \iff & \forall x\, \beta_{n+1} (x) \imp
                                                        \neg(\neg \chi^\star_T)^\ast \in L,
  \end{array}
  $$
откуда следует справедливость доказываемой теоремы.
\end{proof}

\begin{corollary}
  \label{cor:QK-QGrz}
  Логики\/ $\logic{QK}$, $\logic{QT}$, $\logic{QD}$, $\logic{QK4}$, $\logic{QS4}$, $\logic{QwGrz}$, $\logic{QGL}$, $\logic{QGrz}$, а также $\logic{QK.bf}$, $\logic{QT.bf}$, $\logic{QD.bf}$, $\logic{QK4.bf}$, $\logic{QS4.bf}$, $\logic{QwGrz.bf}$, $\logic{QGL.bf}$, $\logic{QGrz.bf}$ неразрешимы в языке с двумя предметными переменными и одной унарной предикатной буквой.
\end{corollary}

Отметим, что полученные результаты в определённом смысле оптимальны: дальнейшее уменьшение числа предметных переменных часто приводит к разрешимым фрагментам логик, причём даже без ограничений на количество и валентность имеющихся в языке предикатных букв. Кроме того, из~\cite[теорема~5.1]{WZ01} следует, что монодические фрагменты монадических фрагментов таких логик как $\logic{QK}$, $\logic{QT}$, $\logic{QK4}$, $\logic{QS4}$ и др., в языке с двумя предметными переменными разрешимы.

\subsubsection{Логики, содержащиеся в $\logic{QKTB}$}

Покажем, что аналогичный результат справедлив и для логик из интервала $[\logic{QK},\logic{QKTB}]$; отметим, что $\logic{QKTB.bf}=\logic{QKTB}$.

Для каждого $k \in \{1, \ldots, n+1 \}$ определим следующие формулы, используя рекурсию:
$$
\begin{array}{lcll}
    \delta_{k+1}^k(x) & = & \Box^+ P(x); \\
    \delta_k^k(x)     & = & \Box^{\leqslant k} \neg P(x)\con
                            \Diamond^{k+1} P(x)\con \Diamond^{k+2}\delta_{k+1}^k(x); \\
    \delta_{i}^k(x)   & = & \Box^{\leqslant i}\neg P(x)\con
                            \Box^+\Diamond^{i+1}P(x)\con
                            \Diamond^{2i+3}\delta^k_{i+1}(x),
                          & \mbox{где $1\leqslant i < k$.}
\end{array}
$$
%
Для каждого $k \in \{1, \ldots, n+1 \}$ переопределим формулы $\alpha_k(x)$ и~$\beta_k(x)$ следующим образом:
$$
\begin{array}{lcll}
\alpha_k (x) & = & P(x) \con \Diamond^2 \delta^k_1(x)\con \neg\Diamond^3 \delta^k_2(x); \\
\beta_k (x)  & = & \neg P(x) \con \Box\Diamond P(x)\con \Diamond \alpha^k(x).
\end{array}
$$

Пусть имеется непустое множество $\mathcal{D}$ и элемент $a\in \mathcal{D}$. Для каждого $k \in \{1, \ldots, n+1 \}$ определим модель $\kModel{M}_k=\otuple{\otuple{W_k,R_k}\odot\mathcal{D},I_k}$.
Для удобства описания некоторые миры определяемой модели будем называть \defnotion{$a$-мирами}, остальные миры будем называть \defnotion{$\bar{a}$-мирами}. Модель $\kModel{M}_k$ содержит корень $r_k$, являющийся $a$\nobreakdash-миром. Остальные миры выстраиваются в $R_k$-последовательность без повторяющихся элементов, подчинённую следующей схеме: корень ($a$\nobreakdash-мир), потом три $\bar{a}$\nobreakdash-мира, потом $a$-мир, потом пять $\bar{a}$\nobreakdash-миров, потом снова $a$-мир, и так далее; то есть после каждого $i$-го $a$\nobreakdash-мира идёт $R_k$\nobreakdash-список, в котором $2i+1$ $\bar{a}$\nobreakdash-миров; заканчивается последовательность двумя $a$-мирами. Отношение $R_k$~--- минимальное рефлексивно-симметричное отношение, содержащее описанную конструкцию.

Говорим, что модель $\kModel{M}_k$ является \defnotion{$a$-подходящей}, если
$$
\begin{array}{lcl}
(\kModel{M}_k,w)\models P(a) & \iff & \mbox{$w$ является $a$-миром}.
\end{array}
$$

Тогда справедлив следующий аналог леммы~\ref{lem:model-M_k}.
\begin{lemma}
  \label{lem:model-M_k-QKTB}
Пусть $a$~--- элемент предметных областей $a$-подходящих моделей
$\kModel{M}_1,\ldots,\kModel{M}_{n+1}$. Тогда
$$
\begin{array}{lcl}
(\kModel{M}_k,w)\models \alpha_m(a) & \iff & \mbox{$k=m$ и $w=r_k$}.
\end{array}
$$
\end{lemma}
\begin{proof}
  Состоит в проверке.
\end{proof}

Для каждого $k \in \{1, \ldots, n+1 \}$ переопределим формулу $\beta_k$:
$$
\begin{array}{lcl}
\beta_k(x) & = & \neg P(x)\con\Diamond \alpha_k(x).
\end{array}
$$
Пусть $\varphi^\ast$~--- результат подстановки в $\varphi'$ формул $\beta_1(x),\ldots,\beta_{n+1}(x)$ вместо, соответственно, $P_1(x),\ldots,P_{n+1}(x)$.

Получаем следующий аналог леммы~\ref{lem:vp-ast}.

\begin{lemma}
  \label{lem:vp-ast-QKTB}
  Пусть $L \in \set{\logic{QK}, \logic{QKTB}}$. Тогда
  $$
  \begin{array}{lcl}
  \mbox{$B \con \vp'$ $L$-выполнима}
  & \iff
  & \mbox{$\forall x\, \beta_{n+1} (x) \con \vp^\ast$  $L$-выполнима.}
  \end{array}
  $$
\end{lemma}

\begin{proof}
Ход доказательства аналогичен доказательству леммы~\ref{lem:vp-ast}; достаточно отметить, что истинность формул $\alpha_1,\ldots,\alpha_{n+1}$ не меняется в мирах моделей $\kModel{M}_1,\ldots,\kModel{M}_{n+1}$ при их <<присоединении>> к модели $\kModel{M}$, в некотором мире которой истинна формула $B \con \vp'$, с целью получения модели $\kModel{M}^\ast$, где в этом же мире будет истинна формула $\forall x\, \beta_{n+1} (x) \con\vp^\ast$.
\end{proof}

\begin{theorem}
  \label{thrm:QBT}
  Пусть $L$~--- модальная предикатная логика, такая, что
  $\logic{QK} \subseteq L\subseteq \logic{QKTB}$. Тогда $L$ неразрешима в языке с двумя предметными переменными и одной унарной предикатной буквой.
\end{theorem}

\begin{proof}
Аналогично доказательству теоремы~\ref{thr:Grz}.
\end{proof}

\begin{corollary}
  \label{cor:KTB}
  Логики\/ $\logic{QKB}$ и\/ $\logic{QKTB}$ неразрешимы в языке с двумя предметными переменными и одной унарной предикатной буквой.
\end{corollary}

	\subsection{Замечания}

Мы рассмотрели модальные предикатные логики, являющиеся нормальными, но нетрудно видеть, что теоремы~\ref{thrm:QGrz} и~\ref{thrm:QBT} останутся справедливыми и для квазинормальных модальных предикатных логик (т.е. не замкнутых по правилу Гёделя расширений логики~$\logic{QK}$), и даже для произвольных множеств формул, находящихся между $\logic{QK}$ и одной из логик $\logic{QGL}$, $\logic{QGrz}$ или $\logic{QKTB}$. Кроме того, нижнюю границу можно <<отодвинуть>>, заменив $\logic{QK}$ на ненормальную модальную предикатную логику $\logic{QK0}$, определяемую соединением логик $\logic{QCl}$ и~$\logic{K0}$. Верхние границы, как мы увидим ниже (разделы~\ref{sec:NatNumbersFrame:Qmodal} и~\ref{sec:NoetherianFramea:Qmodal}), тоже можно <<отодвинуть>>, но за счёт применения несколько иной техники.

В то же время используемая техника такова, что неясно, как её перенести, например, на такие логики как $\logic{QS5}$, $\logic{QK45}$, $\logic{QKD45}$, $\logic{QKB4}$. Теорема~\ref{th:KripkeTrick:3var:Qmodal} гарантирует, что фрагменты этих логик с одной унарной предикатной буквой неразрешимы в языке с тремя предметными переменными, а из~\cite{KKZ05} следует, что монадические фрагменты этих логик неразрешимы в языке с двумя предметными переменными (но число используемых в доказательстве унарных предикатных букв при этом бесконечно).

\begin{problem}
Разрешимы ли логики\/ $\logic{QS5}$, $\logic{QK45}$, $\logic{QKD45}$, $\logic{QKB4}$ в языке с одной унарной предикатной буквой и двумя предметными переменными?
\end{problem}

Сформулированный вопрос актуален, если в языке имеется даже лишь фиксированное конечное множество унарных предикатных букв, а не только одна.

  \section{Разрешимые логики унарных предикатов}
  \label{s7-3}
	\subsection{Нормальные логики}

Несмотря на то, что теоремы~\ref{thrm:QGrz} и~\ref{thrm:QBT} дают нам большие классы логик, имеющих неразрешимые монадические фрагменты с двумя предметными переменными, за пределами этих классов имеются логики, монадические фрагменты которых разрешимы даже с произвольным количеством предметных переменных в языке. Примеры таких логик мы видели выше (теорема~\ref{th:QAlt_n:decidability}), и здесь приведём некоторые другие примеры.

\begin{proposition}
  \label{prop:finite-frame}
  Пусть\/ $\kframe{F}$~--- конечная шкала Крипке. Тогда монадические фрагменты логик\/ $\mPlogic{\kframe{F}}$ и\/ $\mPlogicC{\kframe{F}}$ разрешимы.
\end{proposition}

\begin{proof}
  Пусть $\kframe{F} = \langle W, R \rangle$.  По лемме~\ref{lem:main:unary:finiteframe}, чтобы для монадической формулы $\vp$, содержащей $n$ предикатных букв, выяснить, верно ли, что $\vp \in \mPlogic{\kframe{F}}$, достаточно проверить,
  истинна ли $\vp$ в каждой предикатной шкале вида
  $\langle \kframe{F}, D \rangle$, для которой $|D^+| \leqslant 2^{|W|(n+1)}$.
  Аналогично для $\mPlogicC{\kframe{F}}$, только нужно рассматривать шкалы с постоянными областями.
\end{proof}

\begin{proposition}
  \label{prop:finite-class}
  Пусть\/ $\scls{C}$~--- конечный класс конечных шкал Крипке. Тогда монадические фрагменты логик\/ $\mPlogic{\scls{C}}$ и\/ $\mPlogicC{\scls{C}}$ разрешимы.
\end{proposition}

\begin{proof}
  Заметим, что раздельное объединение конечного множества конечных шкал Крипке является конечной шкалой Крипке, и применим предложение~\ref{prop:finite-frame}.
\end{proof}

Заметим, что аналогичные утверждения справедливы и для логик предикатных шкал Крипке. Напомним, что предикатную шкалу называем конечной, если конечны как множество миров шкалы, так и предметная область каждого её мира.

\begin{proposition}
  \label{prop:finite-frame-pred}
  Пусть\/ $\kFrame{F}$~--- конечная предикатная шкала Крипке. Тогда монадический фрагмент логики\/ $\mPlogic{\kFrame{F}}$ разрешим.
\end{proposition}

\begin{proposition}
  \label{prop:finite-class-pred}
  Пусть\/ $\Scls{C}$~--- конечный класс конечных предикатных шкал Крипке. Тогда монадический фрагмент логики\/ $\mPlogic{\Scls{C}}$ разрешим.
\end{proposition}

	\subsection{Ненормальные и квазинормальные логики}

Заметим, что все примеры нормальных модальных предикатных логик с разрешимыми монадическими фрагментами, которые были приведены выше, находятся в классе расширений логики $\logic{QAlt}_n$ для некоторого $n\in\numN$. Можно ли снять это ограничение? Да, можно; но для этого придётся наложить ограничение на предметные области, поскольку в противном случае становится возможным трюк Крипке, и мы получим неразрешимость. Если же наложить ограничение на предметные области, скажем, потребовав от них наличия не более чем $k$ элементов для заранее зафиксированного $k\in\numNp$, то получаем, что истинность формул вида $\forall x\,\varphi(x)$ в мирах моделей, определённых на таких шкалах, фактически равносильна истинности конечной конъюнкции $\varphi(a_1)\wedge\ldots\wedge\varphi(a_k)$, где $a_1,\ldots,a_k$~--- все элементы предметной области мира. В итоге при проверке истинности формулы в шкале становится возможным элиминировать кванторы. Как следствие получаем, что, например, логика класса всех предикатных шкал, предметные области которых содержат не более $k$ элементов, разрешима; можно добавить условия типа рефлексивности, симметричности, транзитивности и др., и результат будет тот же. 

Покажем, что в случае квазинормальных и ненормальных логик ситуация иная. Построим примеры квазинормальных и ненормальных модальных предикатных логик, обладающих следующими свойствами:
\begin{itemize}
\item 
    их безмодальные фрагменты совпадают с~$\logic{QCl}$;
\item 
    их монадические фрагменты разрешимы;
\item 
    для каждого $n\in\numN$ имеется шкала логики и мир в ней, видящий не менее $n$ миров.
\end{itemize}
В частности, это будет означать, что такие логики не содержатся в логике $\logic{QAlt}_n$ ни для какого $n\in\numN$.

Положим
$$
\begin{array}{lcll}
W_0 & = & \{0\}; 
\smallskip\\
W_{n+1} & = & W_n\times \{0,\ldots,n\} ~\hfill \mbox{для $n\in\numN$;} 
\smallskip\\
W & = & \displaystyle \bigcup\limits_{n=0}^\infty W_n; 
\smallskip\\
R & = & \set{\otuple{x,y} : \mbox{$x\in W$ и $y=\otuple{x,k}$ для некоторого $k\in\numN$}}; 
\smallskip\\
\kframe{F}    & = & \otuple{W,R}; 
\smallskip\\
\kframe{F}_n  & = & \otuple{W_n,R\upharpoonright W_n} ~\hfill \mbox{для $n\in\numN$;} 
\smallskip\\
\kframe{F}'   & = & \otuple{W\cup\set{0^\ast},R\cup\set{\otuple{0,0^\ast}}}; 
\smallskip\\
\kframe{F}'_n & = & \otuple{W_n\cup\set{0^\ast},R\upharpoonright W_n\cup\set{\otuple{0,0^\ast}}} ~\hfill \mbox{для $n\in\numN$.}
\smallskip\\
\end{array}
$$
Проще говоря, шкала Крипке $\kframe{F}$ устроена так: корень $0$ видит один мир, который видит два мира, каждый из которых видит по три мира, каждый из которых видит по четыре мира, и так далее. Шкала $\kframe{F}_n$ является подшкалой шкалы $\kframe{F}$. Шкала $\kframe{F}'$ является расширением шкалы $\kframe{F}$ взрывающимся миром~$0^\ast$, достижимым из корня; $\kframe{F}'_n$ получается из $\kframe{F}_n$ аналогично. 

Пусть теперь
$$
\begin{array}{lcll}
\logic{L}    & = & \{\varphi\in\lang{QML} : (\kframe{F},0)\models\varphi\}; \\
\logic{L}_n  & = & \{\varphi\in\lang{QML} : (\kframe{F}_n,0)\models\varphi\} 
                 & \mbox{для $n\in\numN$;} \\
\logic{L}'   & = & \{\varphi\in\lang{QML} : (\kframe{F}',0)\models\varphi\}; \\
\logic{L}'_n & = & \{\varphi\in\lang{QML} : (\kframe{F}'_n,0)\models\varphi\}
                 & \mbox{для $n\in\numN$;} \\
\end{array}
$$
Нетрудно понять, что логики $\logic{L},\logic{L}_0,\logic{L}_1,\logic{L}_2,\ldots$ являются квазинормальными, а логики $\logic{L}',\logic{L}'_0,\logic{L}'_1,\logic{L}'_2,\ldots$ являются ненормальными.

Теперь заметим, что для всякой формулы $\varphi$ имеют место следующие эквивалентности:
$$
\begin{array}{lcll}
\varphi\in\logic{L}  & \iff & \varphi\in\logic{L}_{\md\varphi}; \\
\varphi\in\logic{L}' & \iff & \varphi\in\logic{L}'_{\md\varphi}. \\
\end{array}
$$

Теперь, учитывая лемму~\ref{lem:main:unary:finiteframe}, получаем, что для каждого $n\in\numN$ монадический фрагмент логики $\logic{L}_n$ разрешим; но тогда из первой эквивалентности получаем разрешимость монадического фрагмента логики~$\logic{L}$. Заметим, что доказательство леммы~\ref{lem:main:unary:finiteframe} существенно не изменится, если в конечной шкале Крипке будут взрывающиеся миры, поэтому, рассуждая аналогично, получаем, что монадический фрагмент логики~$\logic{L}'$ тоже разрешим.

Отметим, что $\Box^n\bm{alt}_n\not\in\logic{L}$ и $\Box(\Box\top\to\Box^n\bm{alt}_n)\not\in\logic{L}'$ для любого $n\in\numN$, поэтому логики $\logic{L}$ и $\logic{L}'$ не содержатся в логике $\logic{QAlt}_n$ ни для какого $n\in\numN$.

Нетрудно понять, что меняя шкалы $\kframe{F}$ и $\kframe{F}'$, можно получить бесконечно много квазинормальных и ненормальных модальных предикатных логик с описанными свойствами. Зафиксируем сделанное наблюдение в виде следующего предложения.

\begin{proposition}
Существует бесконечно много ненормальных и квазинормальных модальных предикатных логик с разрешимым монадическим фрагментом, безмодальные фрагменты которых совпадают с\/ $\logic{QCl}$, не содержащихся в логике\/ $\logic{QAlt}_n$ ни для какого $n\in\numN$.
\end{proposition}

	\subsection{Логики с равенством}

Известно, что в логике $\logic{QCl}^=$ разрешим также и монадический фрагмент с равенством (см., например, \cite[глава~21]{BBJ07} или \cite[глава~25]{BJ-1994-1-rus}), поэтому возникает естественный вопрос о том, нельзя ли распространить лемму~\ref{lem:main:unary:finiteframe} на модальные предикатные логики с равенством.

Прежде чем говорить о распространении леммы~\ref{lem:main:unary:finiteframe} на логики с равенством, нужно уточнить, что означает наличие равенства в модальной предикатной логике. В классической логике равенство можно интерпретировать в моделях как предикат совпадения, и такие модели называются \defnotion{нормальными},\index{модель!нормальная} а можно интерпретировать как \defnotion{конгруэнтность},\index{конгруэнтность} расширяя тем самым класс моделей. Известно~\cite{Mendelson-1976-1-rus}, что оба подхода оказываются эквивалентными в том смысле, что в обоих случаях в качестве множества формул, истинных во всех моделях, мы получаем~$\logic{QCl}^=$. В~частности, это означает, что достаточно ограничиться рассмотрением нормальных моделей.

Аналог требования к равенству, как в нормальных моделях классического языка первого порядка, приводит к следующему определению равенства в мире $w$ модели Крипке $\kmodel{M}=\otuple{W,R,D,I}$ при приписывании~$g$:
\[
\begin{array}{llcl}
\arrayitem
  & (\kmodel{M},w) \models^g x = y
  & \bydef
  & \otuple{g(x),g(y)}\in \mathit{Id}_{D_w},
\end{array}
\]
где $\mathit{Id}_{D_w} = \set{\otuple{a,a} : a\in D_w}$~--- диагональ множества~$D_w$.

Другой подход состоит в том, что мы просто добавляем к логике \defnotion{аксиомы равенства},\index{аксиомы равенства} утверждающие лишь то, что равенство является эквивалентностью и допускает \defnotion{замену равных} (\defnotion{замену эквивалентных}), т.е. является \defnotion{конгруэнтностью}. Это приводит к тому, что в модели Крипке $\kmodel{M}=\otuple{W,R,D,I}$ требуется, чтобы для каждого $w\in W$ в $w$ выполнялись аксиомы равенства, в частности, чтобы отношение $=^{I,w}$ было конгруэнтностью на~$D_w$. Одна из аксиом замены эквивалентных выглядит как
$$
\forall x\forall y\,(x=y\to \Box(x=x\to x=y)),
$$
откуда несложно вывести формулу
$$
\forall x\forall y\,(x=y\to \Box(x=y)).
$$
Таким образом, при втором подходе равенство (конгруэнтность) объектов в некотором мире сохраняется при переходе к достижимым из него мирам. В то же время нетрудно понять, что формула
$$
\forall x\forall y\,(x\ne y\to \Box(x\ne y)),
$$
где $x\ne y$~--- сокращение для формулы $\neg(x=y)$, при таком подходе может быть опровергнута в двухэлементной шкале Крипке, где один из миров виден из другого, но не наоборот. Действительно, достаточно взять предметную область каждого из миров двухэлементной, положив эти элементы не равными в том мире, из которого виден второй, и равными во втором мире (при этом все отличные от равенства предикатные буквы можно интерпретировать, например, пустым множеством). Отметим, что при первом подходе эту формулу опровергнуть невозможно.

Формально при втором подходе рассматривают не предикатные шкалы Крипке, а \defnotion{предикатные шкалы Крипке с равенством};\index{уяи@шкала!Крипке!предикатная с равенством} такая шкала представляет собой набор $\kFrame{F} = \otuple{W,R,D,\simeq}$, где $\otuple{W,R,D}$~--- предикатная шкала Крипке, а~$\simeq$~--- это совокупность $(\simeq_w)_{w\in W}$, состоящая из отношений эквивалентности в предметных областях миров, причём для любых $w,w'\in W$, таких, что $wRw'$, и любых $a,b\in D_w$
$$
\begin{array}{lcl}
a\simeq_w b
  & \imply
  & a\simeq_{w'} b.
\end{array}
$$
Говорим, что модель Крипке $\kModel{M}=\otuple{W,R,D,I}$ является моделью Крипке на предикатной шкале Крипке с равенством $\kFrame{F} = \otuple{W,R,D,\simeq}$, если для каждого $w\in W$ выполнено следующее условие: $I(w,=)$~--- это отношение $\simeq_w$, причём $\simeq_w$~--- конгруэнтность на~$D_w$. При определении истинности формул в шкалах с равенством учитывается истинность формул только в таких моделях.

В итоге мы приходим к тому, что в модальных предикатных логиках разные понимания равенства приводят, в отличие от классической логики, к разным логикам с равенством (см.~\cite{GShS}).
Введём следующие обозначения:
\begin{itemize}
\item
пусть $\lang{QML}^=$\index{уян@язык!qml@$\lang{QML}^=$}~--- множество всех модальных предикатных формул в языке с равенством;
\item
для полной по Крипке модальной предикатной логики $L$ посредством $L^=$ обозначим множество всех $\lang{QML}^=$-формул\index{уяа@формула!ql@$\lang{QML}^=$-формула}, истинных в шкалах Крипке логики $L$, где равенство в мире интерпретируется диагональю предметной области мира;
\item
для полной по Крипке модальной предикатной логики $L$ посредством $L^\simeq$ обозначим множество всех $\lang{QML}^=$-формул, истинных в шкалах Крипке логики $L$, где равенство в мире интерпретируется конгруэнтностью;
\item
для модальной предикатной логики $L$ посредством $L\logic{.eq}$ обозначим наименьшую\footnote{Здесь требуется уточнение, в каком классе, что будет восстанавливаться по контексту: это может быть нормальная, ненормальная или квазинормальная логика.} логику, содержащую $L$ и аксиомы равенства.
\end{itemize}

Будем также использовать обозначения $\mPlogicx{\mathfrak{S}}{=}$, $\mPlogicCx{\mathfrak{S}}{=}$, $\mPlogicx{\mathfrak{S}}{\simeq}$ и\/ $\mPlogicCx{\mathfrak{S}}{\simeq}$, где $\mathfrak{S}$~--- шкала Крипке, предикатная шкала Крипке или класс шкал, а верхние индексы указывают на наличие в языке равенства и способ его интерпретации.

Отметим, что модальные предикатные логики с равенством, получающиеся описанными способами, могут как совпадать, так и отличаться. Например, несложно понять, что
$$
\begin{array}{lclclcl}
\kflogic{\logic{Triv}}^= & = & \kflogic{\logic{Triv}}^\simeq & = & \logic{QTriv.eq};
\\
\kflogic{\logic{Verum}}^= & = & \kflogic{\logic{Verum}}^\simeq & = & \logic{QVerum.eq};
\\
\kflogic{\logic{GL}}^= & \ne & \kflogic{\logic{GL}}^\simeq & \ne & \logic{QGL.eq} & \ne & \kflogic{\logic{GL}}^=;
\\
\kflogic{\logic{GL.bf}}^= & \ne & \kflogic{\logic{GL.bf}}^\simeq & \ne & \logic{QGL.bf.eq} & \ne & \kflogic{\logic{GL.bf}}^=.
\end{array}
$$

Теперь вернёмся к лемме~\ref{lem:main:unary:finiteframe}. Поскольку она утверждает что-то о логиках, определяемых семантически, мы можем ставить вопрос о переносе этой леммы на логики с равенством, определяемые семантически. И такой перенос возможен.

Формулу языка $\lang{QML}^=$ будем называть \defnotion{монадической формулой с равенством},\index{уяа@формула!монадическая!с равенством} если $\varphi$~--- формула в монадическом фрагменте языка $\lang{QML}$, обогащённом символом равенства; множество таких формул будем называть \defnotion{монадическим фрагментом с равенством}\index{уяа@фрагмент!монадический!с равенством} языка~$\lang{QML^=}$.

\begin{lemma}
  \label{lem:main:unary:finiteframe:eq}
  Пусть $\vp$~--- монадическая формула с равенством от $n$ унарных предикатных букв, опровергающаяся в предикатной шкале $\kFrame{F} = \otuple{W, R, D}$ с конечным множеством~$W$ при условии, что равенство понимается как диагональ предметной области.
  Тогда существует вычисляемое по $\varphi$ число $c_\varphi$, такое, что $\vp$ опровергается в некоторой шкале $\bar{\kFrame{F}} = \otuple{ W, R, \bar{D}}$ при том же понимании равенства, где $|\bar{D}^+| \leqslant c_\varphi$; если при этом $\kFrame{F}$~--- шкала с постоянными областями, то $\bar{\kFrame{F}}$~--- тоже шкала с постоянными областями.
\end{lemma}

\begin{proof}
Достаточно заметить, что в каждом классе эквивалентности, определённом в доказательстве леммы~\ref{lem:main:unary:finiteframe}, с помощью равенства можно различить не более чем $r$~элементов, где $r$~--- число предметных переменных, входящих в~$\varphi$.
\end{proof}

\begin{lemma}
  \label{lem:main:unary:finiteframe:congr}
  Пусть $\vp$~--- монадическая формула с равенством от $n$ унарных предикатных букв, опровергающаяся в предикатной шкале $\kFrame{F} = \otuple{W, R, D, \simeq}$ с конечным множеством~$W$ при условии, что равенство понимается как конгруэнтность.
  Тогда существует вычисляемое по $\varphi$ число $c_\varphi$, такое, что $\vp$ опровергается в некоторой шкале $\bar{\kFrame{F}} = \otuple{ W, R, \bar{D}}$ при том же понимании равенства, где $|\bar{D}^+| \leqslant c_\varphi$; если при этом $\kFrame{F}$~--- шкала с постоянными областями, то $\bar{\kFrame{F}}$~--- тоже шкала с постоянными областями.
\end{lemma}

\begin{proof}
Аналогично доказательству леммы~\ref{lem:main:unary:finiteframe:eq}. Отметим, что классов конгруэнтности может оказаться больше, чем классов эквивалентности, которые позволяет выделить равенство, интерпретируемое диагональю предметной области; нам важно лишь, что из-за конечности множества миров число таких классов будет конечным и вычислимым по~$\varphi$. Более точно, можно взять $c_\varphi = |W|\cdot r\cdot 2^{|W|\cdot(n+1)}$, где $r$~--- число предметных переменных, входящих в~$\varphi$, а $n$~--- число унарных предикатных букв, входящих в~$\varphi$.
\end{proof}

Извлечём из этих лемм очевидные следствия.

\begin{proposition}
  \label{prop:finite-frame:eq}
  Пусть\/ $\kframe{F}$~--- конечная шкала Крипке. Тогда монадические фрагменты с равенством логик\/ $\mPlogicx{\kframe{F}}{=}$, $\mPlogicCx{\kframe{F}}{=}$, $\mPlogicx{\kframe{F}}{\simeq}$ и\/ $\mPlogicCx{\kframe{F}}{\simeq}$ разрешимы.
\end{proposition}

\begin{proof}
  Аналогично доказательству предложения~\ref{prop:finite-frame} с применением лемм~\ref{lem:main:unary:finiteframe:eq} и~\ref{lem:main:unary:finiteframe:congr}.
\end{proof}

\begin{proposition}
  \label{prop:finite-class:eq}
  Пусть\/ $\scls{C}$~--- конечный класс конечных шкал Крипке. Тогда монадические фрагменты с равенством логик\/ $\mPlogicx{\scls{C}}{=}$, $\mPlogicCx{\scls{C}}{=}$, $\mPlogicx{\scls{C}}{\simeq}$ и\/ $\mPlogicCx{\scls{C}}{\simeq}$ разрешимы.
\end{proposition}

\begin{proof}
  Раздельное объединение конечного множества конечных шкал Крипке является конечной шкалой Крипке; остаётся применить предложение~\ref{prop:finite-frame:eq}.
\end{proof}

И конечно, верны следующие наблюдения, не требующие для их обоснования полученных выше лемм.

\begin{proposition}
  \label{prop:finite-frame-pred:eq}
  Пусть\/ $\kFrame{F}$~--- конечная предикатная шкала Крипке. Тогда монадические фрагменты с равенством логик\/ $\mPlogicx{\kFrame{F}}{=}$ и\/ $\mPlogicx{\kFrame{F}}{\simeq}$ разрешимы.
\end{proposition}

\begin{proposition}
  \label{prop:finite-class-pred:eq}
  Пусть\/ $\Scls{C}$~--- конечный класс конечных предикатных шкал Крипке. Тогда монадические фрагменты с равенством логик\/ $\mPlogicx{\Scls{C}}{=}$ и\/ $\mPlogicx{\Scls{C}}{\simeq}$ разрешимы.
\end{proposition}

Отметим, что аналоги предложений~\ref{prop:finite-frame:eq}--\ref{prop:finite-class-pred:eq} справедливы и для ненормальных, и для квазинормальных модальных предикатных логик, определяемых конечными шкалами Крипке, конечными предикатными шкалами Крипке, конечными классами конечных шкал Крипке или конечными классами конечных предикатных шкал Крипке. Соответствующие доказательства не имеют существенных отличий от приведённых выше.

  \section{Логики элементарно определимых классов шкал} \label{sec:7-4}
	\subsection{Элементарная определимость классов шкал}
    \label{s7-4-1}

\providecommand{\uW}{\dismath{\ulined{W}}}
\providecommand{\uR}{\dismath{\ulined{R}}}
\providecommand{\uD}{\dismath{\ulined{D}}}
\providecommand{\smodel}[1]{\dismath{{\argument{#1}^{}}^{\starletfill[scale=0.45]}}}
\providecommand{\slang}[2]{\dismath{{\lang{\argument{#1}}^{}}^{\starletfill[scale=0.45]}_{\argument{#2}}}}
\providecommand{\mstared}[2]{\dismath{{\argument{#1}^{}}^{\starletfill[scale=0.45]}_{\argument{#2}}}}
\providecommand{\mst}[1]{\mstared{#1}{}}

Пусть $\uR_1,\ldots,\uR_n$~--- бинарные предикатные буквы, $\lang{QL}^=\upharpoonright\set{=,\uR_1,\ldots,\uR_n}$~--- фрагмент языка $\lang{QL}^=$, формулы которого содержат только предикатные буквы $=$ и~$\ulined{R}_1,\ldots,\ulined{R}_n$. Заметим, что шкалу Крипке $\kframe{F}=\otuple{W,R_1,\ldots,R_n}$ можно рассматривать как нормальную модель языка $\lang{QL}^=\upharpoonright\set{=,\uR_1,\ldots,\uR_n}$, где буквы $\uR_1,\ldots,\uR_n$ интерпретируются отношениями~$R_1,\ldots,R_n$, соответственно; в этом случае класс $\scls{C}$ шкал Крипке с $n$ отношениями достижимости называется \defnotion{первопорядково определимым},\index{класс!уяи@шкал Крипке!первопорядково определимый} или \defnotion{элементарно определимым}\index{класс!уяи@шкал Крипке!уял@элементарно определимый}, если существует такая замкнутая формула $\Phi$ языка $\lang{QL}^=\upharpoonright\set{=,\uR_1,\ldots,\uR_n}$, что для любой шкалы $\kframe{F}=\otuple{W,R_1,\ldots,R_n}$
$$
\begin{array}{lcl}
\kframe{F}\in\scls{C} & \iff & \kframe{F}\cmodels \Phi,
\end{array}
$$
при этом говорим также, что формула $\Phi$ \defnotion{определяет} класс~$\scls{C}$, или что класс~$\scls{C}$ \defnotion{определяется} формулой~$\Phi$.

Так, например,
\begin{itemize}
\item формула $\FOrefp(\uR)$ определяет класс всех рефлексивных шкал с одним отношением достижимости;
\item формула $\FOsymp(\uR)$ определяет класс всех симметричных шкал с одним отношением достижимости;
\item формула $\FOtrap(\uR)$ определяет класс всех транзитивных шкал с одним отношением достижимости;
\item класс всех конечных шкал Крипке с $n$ отношениями достижимости не является первопорядково определимым;
\item класс всех шкал Крипке с одним отношением достижимости, в которых истинна формула $\mla$, не является первопорядково определимым;
\item класс всех шкал Крипке с одним отношением достижимости, в которых истинна формула $\mgrz$, не является первопорядково определимым.
\end{itemize}

Если мы хотим говорить об элементарной определимости предикатных шкал Крипке, то помимо описания свойств отношений достижимости, нужно ещё описать и свойства предметных областей миров. Кроме того, предикатная шкала уже не является моделью языка первого порядка, поскольку её носитель содержит лишь миры, а мы хотим говорить также об индивидах в предметных областях миров шкалы. Как следствие, нам потребуется не только расширить язык средствами для описания предметных областей миров, но и сопоставить предикатным шкалам модели получившегося языка. Сделаем это.

Определим язык $\slang{QL}{}$ как фрагмент языка $\lang{QL}^=$, содержащий унарную предикатную букву $\uW$, бинарные предикатные буквы $\uD,=,\uR_1,\uR_2,\uR_3,\ldots{}$, а также для каждой $m$\nobreakdash-арной предикатной буквы~$P$ инъективно сопоставленную ей $(m+1)$\nobreakdash-арную предикатную букву~$\mst{P}$.
Букву $\uR_1$ иногда будем также обозначать~$\uR$.
%
Отметим, что пока нам понадобятся только буквы $\uW,\uD,=,\uR_1,\uR_2,\uR_3,\ldots{}$, а роль букв вида~$\mst{P}$ прояснится чуть позже (раздел~\ref{s7-4-2}).
Для каждого $n\in\numNp$ определим язык $\slang{QL}{n}$ как фрагмент языка $\slang{QL}{}$, в котором из предикатных букв имеются только буквы $\uW,\uD,=,\uR_1,\ldots,\uR_n$, а остальные отсутствуют; определим язык $\slang{QL}{\infty}$ как фрагмент языка $\slang{QL}{}$, содержащий лишь буквы $\uW,\uD,=,\uR_1,\uR_2,\uR_3,\ldots{}$, т.е. в котором отсутствуют все предикатные буквы вида~$\mst{P}$.

Теперь определим теорию $\logic{TKF}$, которую будем называть \defnotion{теорией предикатных шкал Крипке}.\index{теория!предикатных шкал Крипке} Пусть $\logic{TKF}$~--- множество, содержащее следующие $\slang{QL}{\infty}$\nobreakdash-формулы:
\[
\begin{array}{llcl}
\arrayitem &
  \exists x\, \uW(x) \wedge \forall x\,(\uW(x)\to \exists y\,\uD(x,y));
  \arrayitemskip\\
\arrayitem &
  \forall x\forall y\,(\uR_i(x,y)\to \uW(x)\wedge \uW(y))
  && \mbox{для каждого $i\in\numNp$;}
  \arrayitemskip\\
\arrayitem &
  \forall x\forall y\,(\uD(x,y)\to \uW(x));
  \arrayitemskip\\
\arrayitem &
  \forall x\forall y\forall z\,(\uR_i(x,y)\wedge \uD(x,z)\to \uD(y,z))
  && \mbox{для каждого $i\in\numNp$;}
  \arrayitemskip\\
\arrayitem &
  \forall x\,(\uW(x)\vee \exists y\,\uD(y,x)).
\end{array}
\]
Пусть также $\logic{TKF}_n$~--- фрагмент теории $\logic{TKF}$ в языке~$\slang{QL}{n}$. Теория $\logic{TKF}_n$, в отличие от $\logic{TKF}$, содержит лишь конечное множество формул, что нам будет важно; пусть $\bm{tkf}_n$~--- их конъюнкция.

Нам будут интересны модели теории $\logic{TKF}$, поэтому уделим внимание описанию их устройства.

Пусть $\kFrame{F}=\otuple{\kframe{F},D}$~--- предикатная шкала Крипке, определённая на шкале Крипке $\kframe{F}=\otuple{W,R_1,\ldots,R_n}$. Для шкалы $\kFrame{F}$ определим классическую модель $\mst{\kFrame{F}}=\otuple{U,J}$ языка $\slang{QL}{\infty}$, положив
\[
\begin{array}{llcll}
\arrayitem &
  U & = & W\cup D^+;
  \arrayitemskip\\
\arrayitem &
  J(\uW) & = & \set{\otuple{w} : w\in W};
  \arrayitemskip\\
\arrayitem &
  J(\hfill\uD\hfill) & = & \set{\otuple{w,a} : a\in D_w};
  \arrayitemskip\\
\arrayitem &
  J(\hfill\uR_i\hfill) & = & \set{\otuple{w,w'} : w R_i w'} & \mbox{для $i\in\set{1,\ldots,n}$};
  \arrayitemskip\\
\arrayitem &
  J(\hfill\uR_i\hfill) & = & \varnothing & \mbox{для $i\in\numNp\setminus\set{1,\ldots,n}$}.
  \arrayitemskip\\
\end{array}
\]
Нетрудно видеть, что модель $\mst{\kFrame{F}}$ является и моделью языка $\slang{QL}{n}$ для каждого $n\in\numNp$. Более того, имеет место следующая лемма.

\begin{lemma}
\label{lem:tkf-model-1}
Пусть\/ $\kFrame{F}=\otuple{\kframe{F},D}$~--- предикатная шкала Крипке, определённая на шкале Крипке\/ $\kframe{F}=\otuple{W,R_1,\ldots,R_n}$. Тогда\/ $\mst{\kFrame{F}}\cmodels \logic{TKF}$.
\end{lemma}

\begin{proof}
Следует из определения модели~$\mst{\kFrame{F}}$.
\end{proof}

Из леммы~\ref{lem:tkf-model-1}, в частности, получаем, что $\mst{\kFrame{F}}\cmodels \logic{TKF}_n$ для каждого $n\in\numNp$, т.е. $\mst{\kFrame{F}}\cmodels \bm{tkf}_n$.

Пусть теперь $\cModel{M}=\otuple{U,J}$~--- модель языка $\slang{QL}{n}$, такая, что $\cModel{M}\cmodels\logic{TKF}_n$. Положим
\[
\begin{array}{llcll}
\arrayitem &
  W & = & \set{a\in U : \cModel{M}\cmodels \uW(a)};
  \arrayitemskip\\
\arrayitem &
  R_i & = & \set{\otuple{a,b}\in U\times U : \cModel{M}\cmodels \uR_i(a,b)} & \mbox{для $i\in\set{1,\ldots,n}$};
  \arrayitemskip\\
\arrayitem &
  D_a & = & \set{b\in U : \cModel{M}\cmodels \uD(a,b)}.
  \arrayitemskip\\
\end{array}
\]
Положим $D(a) = D_a$ для всякого $a\in W$. Нетрудно видеть, что в этом случае $\kframe{F}=\otuple{W,R_1,\ldots,R_n}$~--- шкала Крипке, а $\kFrame{F}=\otuple{\kframe{F},D}$~--- предикатная шкала Крипке, построенная на шкале~$\kframe{F}$.

Таким образом, мы установили соответствие между предикатными шкалами Крипке и моделями теории~$\logic{TKF}_n$, где~$n\in\numNp$.

Класс $\Scls{C}$ предикатных шкал Крипке с $n$ отношениями достижимости называем \defnotion{первопорядково определимым}, или \defnotion{элементарно определимым}, если существует замкнутая $\slang{QL}{n}$\nobreakdash-формула $\Phi$, такая, что для каждой предикатной шкалы~$\kFrame{F}$
$$
\begin{array}{lcl}
\kFrame{F}\in\Scls{C} & \iff & \mst{\kFrame{F}}\cmodels \Phi,
\end{array}
\eqno{\mbox{$({\ast})$}}
$$
при этом говорим также, что формула $\Phi$ \defnotion{определяет} класс~$\Scls{C}$, или что класс~$\Scls{C}$ \defnotion{определяется} формулой~$\Phi$.

Нетрудно видеть, что условие \mbox{$({\ast})$} эквивалентно условию\footnote{Для множества формул $X$ и формулы $\varphi$ понимаем $X\cmodels \varphi$ как \defnotion{семантическое следование}\index{следование!семантическое} формулы~$\varphi$ из~$X$, т.е. когда в каждой модели множества $X$ истинна формула~$\varphi$.}
$$
\begin{array}{lcl}
\kFrame{F}\in\Scls{C} & \iff & \logic{TKF}_n\cmodels\Phi,
\end{array}
\eqno{\mbox{$({\ast}{\ast})$}}
$$
а также условию
$$
\begin{array}{lcl}
\kFrame{F}\in\Scls{C} & \iff & \bm{tkf}_n\to\Phi\in\logic{QCl}^=.
\end{array}
\eqno{\mbox{$({\ast}{\ast}{\ast})$}}
$$

Отметим, что для шкал с бесконечным набором отношений достижимости аналога условия \mbox{$({\ast}{\ast}{\ast})$} нет; при этом аналог условия \mbox{$({\ast}{\ast})$} возможен с заменой $\logic{TKF}_n$ на $\logic{TKF}$, если, например, требуется описать свойства лишь конечного множества отношений достижимости (а остальные могут быть произвольными); кроме того, в случае условий \mbox{$({\ast})$} и \mbox{$({\ast}{\ast})$} вместо формулы $\Phi$ можно было бы рассматривать бесконечное множество $\slang{QL}{\infty}$\nobreakdash-формул. Для дальнейших построений подобные обобщения нам не потребуются, поэтому не будем уделять им внимания.

	\subsection{Стандартный перевод}
    \label{s7-4-2}

Определим \defnotion{стандартный перевод}\index{перевод!стандартный}\footnote{См., например,~\cite[раздел~3.12]{GShS}.} $\lang{QML}_n$-формул в $\slang{QL}{}$\nobreakdash-формулы. Пусть в следующих ниже формулах $z$~--- новая предметная переменная по отношению к формулам в левой части равенств; положим
\[
\begin{array}{llcll}
\arrayitem &
  \mstared{P(x_1,\ldots,x_m)}{z}
  & =
  & \mst{P}(z,x_1,\ldots,x_m);
  \arrayitemskip\\
\arrayitem &
  \mstared{\bot}{z}
  & =
  & \bot;
  \arrayitemskip\\
\arrayitem &
  \mstared{(\varphi\wedge\psi)}{z}
  & =
  & \mstared{\varphi}{z}\wedge\mstared{\psi}{z};
  \arrayitemskip\\
\arrayitem &
  \mstared{(\varphi\vee\psi)}{z}
  & =
  & \mstared{\varphi}{z}\vee\mstared{\psi}{z};
  \arrayitemskip\\
\arrayitem &
  \mstared{(\varphi\to\psi)}{z}
  & =
  & \mstared{\varphi}{z}\to\mstared{\psi}{z};
  \arrayitemskip\\
\arrayitem &
  \mstared{(\forall x\,\varphi)}{z}
  & =
  & \forall x\,(\uD(z,x)\to \mstared{\varphi}{z});
  \arrayitemskip\\
\arrayitem &
  \mstared{(\exists x\,\varphi)}{z}
  & =
  & \exists x\,(\uD(z,x)\hfill \wedge\hfill \mstared{\varphi}{z});
  \arrayitemskip\\
\arrayitem &
  \mstared{(\Box_i\varphi)}{z}
  & =
  & \forall y\,(\uR_i(z,y)\to \mstared{\varphi}{y})
  & \mbox{для $i\in\set{1,\ldots,n}$,}
  \arrayitemskip\\
\end{array}
\]
где $P$~--- $m$-арная предикатная буква для некоторого $m\in\numN$, а предметная переменная $y$ не входит в~$\varphi$.

Пусть $\kframe{F}=\otuple{W,R_1,\ldots,R_n}$~--- шкала Крипке, $\kFrame{F}=\otuple{\kframe{F},D}$~--- построенная на ней предикатная шкала и $\kModel{M}=\otuple{\kFrame{F},I}$~--- модель Крипке на~$\kFrame{F}$.
Для модели $\kModel{M}$ определим классическую модель $\mst{\kModel{M}}=\otuple{U,J}$ языка $\slang{QL}{}$, положив
\[
\begin{array}{llcll}
\arrayitem &
  U & = & W\cup D^+;
  \arrayitemskip\\
\arrayitem &
  J(\hfill\uW\hfill) & = & \set{\otuple{w} : w\in W};
  \arrayitemskip\\
\arrayitem &
  J(\hfill\uD\hfill) & = & \set{\otuple{w,a} : a\in D_w};
  \arrayitemskip\\
\arrayitem &
  J(\hfill\uR_i\hfill) & = & \set{\otuple{w,w'} : w R_i w'}
  \phantom{\varnothing}~ \mbox{для $i\in\set{1,\ldots,n}$};
  \arrayitemskip\\
\arrayitem &
  J(\hfill\uR_i\hfill) & = & \varnothing
  \phantom{\set{\otuple{w,w'} : w R_i w'}}~ \mbox{для $i\in\numNp\setminus\set{1,\ldots,n}$};
  \arrayitemskip\\
\arrayitem &
  J(\hfill\mst{P}\hfill) & = & \set{\otuple{w,a_1,\ldots,a_m} : (\kModel{M},w)\models P(a_1,\ldots,a_m)},
  \arrayitemskip\\
\end{array}
\]
где $P$~--- $m$-арная предикатная буква.

Имеет место следующая лемма.

\begin{lemma}
\label{lem:tkf-model-2}
Пусть $\varphi$~--- $\lang{QML}_n$-формула, $\kModel{M}$~--- модель Крипке, определённая на шкале с $n$ отношениями достижимости, $w$~--- мир в~$\kModel{M}$. Тогда $\mst{\kModel{M}}\cmodels\logic{TKF}$ и
$$
\begin{array}{lcl}
(\kModel{M},w)\models\varphi
  & \iff
  & \mst{\kModel{M}}\cmodels \mstared{\varphi}{w}.
\end{array}
$$
\end{lemma}

\begin{proof}
Следует из определения модели~$\mst{\kModel{M}}$.
\end{proof}

Пусть теперь $\cModel{M}=\otuple{U,J}$~--- модель языка $\slang{QL}{}$, такая, что $\cModel{M}\cmodels\logic{TKF}_n$. Положим
\[
\begin{array}{llcll}
\arrayitem &
  W & = & \set{a\in U : \cModel{M}\cmodels \uW(a)};
  \arrayitemskip\\
\arrayitem &
  R_i & = & \set{\otuple{a,b}\in U\times U : \cModel{M}\cmodels \uR_i(a,b)} & \mbox{для $i\in\set{1,\ldots,n}$};
  \arrayitemskip\\
\arrayitem &
  D_a & = & \set{b\in U : \cModel{M}\cmodels \uD(a,b)}.
  \arrayitemskip\\
\end{array}
\]
Положим $D(a) = D_a$ для всякого $a\in W$. Как было отмечено выше, в этом случае $\kframe{F}=\otuple{W,R_1,\ldots,R_n}$~--- шкала Крипке, а $\kFrame{F}=\otuple{\kframe{F},D}$~--- предикатная шкала Крипке, построенная на шкале~$\kframe{F}$. На шкале $\kFrame{F}$ построим модель $\kModel{M}=\otuple{\kFrame{F},I}$, где для каждой $m$-арной предикатной буквы~$P$ и каждого $w\in W$ полагаем
\[
\begin{array}{llcll}
\arrayitem &
  I(w,P) & = & \set{\otuple{a_1,\ldots,a_m} : \cModel{M}\cmodels \mst{P}(w,a_1,\ldots,a_m)}. \\
\end{array}
\]
Тогда нетрудно понять\footnote{Потребуется индукция по построению формулы, задействующая истинность формул в мирах при интерпретациях предметных переменных.}, что для всякой замкнутой $\lang{QML}_n$-формулы $\varphi$ и всякого $w\in W$
$$
\begin{array}{lcl}
  (\kModel{M},w)\models\varphi
    & \iff
    & \cModel{M}\cmodels \mstared{\varphi}{w}.
\end{array}
$$

	\subsection{Погружения в $\logic{QCL}^=$}
    \label{s7-4-3}

Используя стандартный перевод модальных формул, можно построить погружения большого класса модальных предикатных логик в~$\logic{QCl}^=$. Следующее утверждение хорошо известно~\cite{MR:2000:IFRAS,GShS}.

\begin{proposition}
\label{prop:Mpred:std-translation}
Пусть $L$~--- нормальная модальная предикатная логика в языке\/ $\lang{QML}_n$, полная относительно некоторого элементарно определимого класса предикатных шкал Крипке, и пусть\/ $\Phi$~--- $\slang{QL}{n}$-формула, определяющая этот класс. Тогда для всякой $\lang{QML}_n$-формулы $\varphi$
$$
\begin{array}{lcl}
\varphi\in L 
  & \iff 
  & \bm{tkf}_n\wedge \Phi \to \forall z\,(\uW(z)\to \mstared{\varphi}{z}) \in \logic{QCl}^=.
\end{array}
$$
\end{proposition}

\begin{proof}
Следует из леммы~\ref{lem:tkf-model-2}.
\end{proof}

\begin{corollary}
\label{cor:prop:Mpred:std-translation}
Пусть $L$~--- нормальная модальная предикатная логика в языке\/ $\lang{QML}_n$, полная относительно некоторого элементарно определимого класса предикатных шкал Крипке. Тогда $L$ является рекурсивно перечислимой.
\end{corollary}

\begin{proof}
Рекурсивная перечислимость логики $L$ следует из предложения~\ref{prop:Mpred:std-translation} с учётом рекурсивной перечислимости логики~$\logic{QCl}^=$.
\end{proof}

Следствие~\ref{cor:prop:Mpred:std-translation} гарантирует наличие рекурсивной перечислимости (а значит, и наличие рекурсивной аксиоматики) для логик вида $\mPlogic{\kframes{L_1\fusion\ldots\fusion L_n}}$ и $\mPlogicC{\kframes{L_1\fusion\ldots\fusion L_n}}$, где в качестве каждой из логик $L_1,\ldots, L_n$ можно взять такие логики как $\logic{K}$, $\logic{T}$, $\logic{D}$, $\logic{K4}$, $\logic{S4}$, $\logic{KB}$, $\logic{KTB}$, $\logic{S5}$, $\logic{Triv}$, $\logic{Verum}$ и~др. При этом следствие~\ref{cor:prop:Mpred:std-translation} ничего не может сказать о логиках $\mPlogic{\kframes{L}}$ и $\mPlogicC{\kframes{L}}$, где $L$~--- одна из таких логик как $\logic{GL}$, $\logic{GLLin}$, $\logic{Grz}$, $\logic{Grz.3}$, $\logic{wGrz}$, $\logic{wGrz.3}$, а также о других модальных предикатных логиках, определяемых классами шкал Крипке, но не полных относительно элементарно определимых классов шкал Крипке. Ниже мы покажем, что ситуация с рекурсивной перечислимостью таких логик может быть разной: во многих естественных случаях такие логики не являются рекурсивно перечислимыми, при этом могут быть даже $\Pi^1_1$\nobreakdash-трудными, но имеются и примеры, когда логики с такой семантикой рекурсивно перечислимы.

  \section{Логики классов шкал, не определимых элементарно}
    \subsection{Логики классов конечных шкал}
\providecommand{\arrayitembackspace}{\!\!\!\!\!\!\!\!}

\subsubsection{Определения и обозначения}

Пусть $L$~--- нормальная модальная предикатная логика, для которой существуют $L$-шкалы с конечным множеством миров. Определим логику $\wfin{L}$ как множество формул, истинных в классе всех конечных шкал Крипке логики~$L$. Обратим внимание на то, что логика $L$ может как быть полной по Крипке, так и не быть; логика $\wfin{L}$ полна по Крипке по её определению.

Как мы видели (теоремы~\ref{th:trakhtenbrot} и~\ref{th:trakhtenbrot:positive}), логика $\logic{QCl}_{\mathit{fin}}$ является $\Pi^0_1$-полной в языке с одной бинарной предикатной буквой и тремя предметными переменными; пусть $\QCLFs$~--- соответствующий её фрагмент, а $\langQCLFs$~--- соответствующий фрагмент языка~$\lang{QL}$. В построениях ниже будем считать, что $\langQCLFs$ содержит бинарную букву $S$ и предметные переменные~$x,y,z$.

\subsubsection{Погружение логики ${\bf QCl}_{\mathit{fin}}$ в логики конечных шкал}
\label{wfin:sec:modal-reduction}


Построим погружение $\Pi^0_1$-трудного множества
$\QCLFs$ в логики интервалов $[\logic{QK}_{\mathit{wfin}}, \logic{ QGL.3.bf}_{\mathit{wfin}}]$,
$[\logic{QK}_{\mathit{wfin}}, \logic{QGrz.3.bf}_{\mathit{wfin}}]$ и
$[\logic{QK}_{\mathit{wfin}}, \logic{QS5}_{\mathit{wfin}}]$.


Пусть $\vp$~--- замкнутая $\langQCLFs$-формула.

Пусть $T$ и $\approx$~--- унарная и бинарная предикатные буквы, причём $\approx$ не входит в формулу~$\vp$.

Положим
$$
\begin{array}{lcl}
  A_1 & = & \forall x\, \Diamond T( x ); \smallskip\\
  A_2 & = &  \forall x \forall y\, ( x \approx
            y \equivalence \Box (T(x ) \equivalence T(y ))). \\
\end{array}
$$

Заметим, что истинность формулы $A_2$ в мире модели Крипке влечёт, что букве $\approx$ в этом мире соответствует отношение эквивалентности.

Пусть $A = A_1 \con A_2$ и пусть
$$
\begin{array}{lcl}
\mathit{Congr}
  & =
  & \forall x\forall y\forall z\, (x\approx y\imp((S(z,x)\imp S(z,y))
\wedge(S(x,z) \imp S(y,z))).
\end{array}
$$

Положим
$$
\begin{array}{lcl}
\bar{\vp} & = & A \con \mathit{Congr} \imp \vp.
\end{array}
$$
Заметим, что $\bar{\vp}$ содержит лишь три предметные переменные.

\begin{lemma}
  \label{wfin:lem:QCl-to-QK}
  Пусть $L \in \{ \logic{QK}, \logic{QS5}, \logic{QGL.3}, \logic{QGrz.3} \}$.
  Тогда следующие условия эквивалентны друг другу:
  \[
  \begin{array}{ll}
  \arrayitembackspace(1) & \vp \in \logic{QCl}_{\mathit{fin}};
    \arrayitemskip\\
  \arrayitembackspace(2) & \bar{\vp} \in L_{\mathit{wfin}};
    \arrayitemskip\\
  \arrayitembackspace(3) & \bar{\vp} \in L\logic{.bf}_{\mathit{wfin}}.
  \end{array}
  \]
\end{lemma}

\begin{proof}
  Докажем импликацию \mbox{$(1) \Rightarrow (2)$}.

  Пусть
  $(\kModel{M},w) \not\models \bar{\vp}$ для некоторой модели
  $\kModel{M} = \langle W, R, D, I \rangle$, определённой на $L$-шкале,
  и некоторого мира $w \in W$. Определим конечную классическую модель, в которой опровергается~$\vp$.

  Поскольку $(\kModel{M},w) \models A_2 \con \mathit{Congr}$, заключаем, что множество $D_w$ разбивается отношением ${\approx}^{I, w}$ на классы конгруэнтности (по отношению к~$S^{I,w}$).  Пусть
  $\bar{a} = \{ b \in D_{w} : b \approx^{I, w} a \}$, где $a\in D_w$, и
  $\bar{D}_{w} = \{ \bar{a} : a \in D_{w} \}$.

  Покажем, что множество $\bar{D}_{w}$ является конечным.
  Для каждого $a \in D_w$ положим
  $$
  \begin{array}{lcl}
  V(a) & = & \{ v \in R(w) : (\kModel{M},v) \models T(a) \}.
  \end{array}
  $$
  Поскольку $(\kModel{M},w) \models A_2$, элементы $a, b \in D_w$, для которых
  $V(a) = V(b)$, принадлежат одному и тому же классу конгруэнтности по отношению~${\approx}^{I, w}$.

  Пусть $\Sigma = \{ V(a) : a \in D_w \}$. Нетрудно понять, что
  $|\Sigma| \leqslant 2^{|R(w)|} \leqslant 2^{|W|}$.
  Поскольку $(\kModel{M},w) \models A_1$, получаем, что
  $|\bar{D}_{w}| \leqslant |\Sigma|$, и в силу конечности множества $W$ получаем, что $\bar{D}_{w}$ тоже конечно.

  Теперь определим конечную классическую модель $\cModel{M} = \langle \bar{D}_{w}, J \rangle$ языка $\langQCLFs$, положив
  $$
  \begin{array}{lcl}
  J(S)
    & =
    & \set{\otuple{\bar{a},\bar{c}} : (\kModel{M},w)\models S(a,c)}.
  \end{array}
  $$

  Приписывания $\bar{g}$ в $\cModel{M}$ и $g$ в
  $\kModel{M}$ называем \defnotion{согласованными}, если для каждой переменной $x$ и каждого $a\in D_w$
  $$
  \begin{array}{rcl}
  \bar{g}(x) = \bar{a} & \Longleftrightarrow & g(x) = a.
  \end{array}
  $$

  Тогда для каждой подформулы $\theta$ формулы $\vp$ и любых двух согласованных приписываний $\bar{g}$~и~$g$
  $$
  \begin{array}{rcl}
  \cModel{M} \cmodels^{\bar{g}} \theta
    & \iff
    & (\kModel{M},w) \models^g \theta,
  \end{array}
  $$
  что обосновывается индукцией по построению~$\theta$.

  Как следствие, $\cModel{M} \not\cmodels \vp$, а значит,
  $\vp \not\in \logic{QCl}_{\mathit{fin}}$.

  Импликация \mbox{$(2) \Rightarrow (3)$} следует из того, что
  $L_{\mathit{wfin}} \subseteq L\logic{.bf}_{\mathit{wfin}}$.

  Докажем импликацию \mbox{$(3) \Rightarrow (1)$}, для чего рассмотрим каждый из четырёх возможных случаев для~$L$.

  Пусть $L=\logic{QK.bf}_{\mathit{wfin}}$. Пусть $\cModel{M} \not \cmodels \vp$ для некоторой модели $\cModel{M} = \langle \mathcal{D}, J \rangle$ с конечным множеством $\mathcal{D}$; будем считать, что $\mathcal{D} = \{ a_0, a_1, \ldots, a_n \}$.
  Положим
  \[
  \begin{array}{llcl}
  \arrayitem
    & W & = & \{ w^\ast, w_0, w_1, \ldots, w_n \};
    \arrayitemskip\\
  \arrayitem
    & R & = & \{\langle w^\ast,w_i\rangle : 0\leqslant i\leqslant n\};
    \arrayitemskip\\
  \arrayitem
    & D(w) & = & \mathcal{D}~ \qquad \mbox{для каждого $w \in W$}.
  \end{array}
  \]
  Определим интерпретацию $I$ в предикатной шкале $\otuple{W,R,D}$, положив
  \[
  \begin{array}{llcll}
  \arrayitem
    & I (w^\ast,\hfill T) & = & \mathcal{D};
    \arrayitemskip\\
  \arrayitem
    & I (w_i,\hfill T) & = & \{\langle a_i \rangle \} & \mbox{для $i \in \{0, \ldots, n\}$;}
    \arrayitemskip\\
  \arrayitem
    & I (w^\ast,\hfill \approx) & = & \{ \langle a, a \rangle : a\in \mathcal{D} \};
    \arrayitemskip\\
  \arrayitem
    & I (w_i,\hfill \approx) & = & \varnothing & \mbox{для $i \in \{0, \ldots, n\}$;}
    \arrayitemskip\\
  \arrayitem
    & I(w^\ast,\hfill S) & = & J(S);
    \arrayitemskip\\
  \arrayitem
    & I(w_i,\hfill S) & = & \varnothing & \mbox{для $i \in \{0, \ldots, n\}$.}
  \end{array}
  \]
  Пусть $\kModel{M} = \langle W, R, D, I \rangle$; модель
  $\kModel{M}$ изображена на рис.~\ref{wfin:fig0}.

    \begin{figure}
  \centering
  \begin{tikzpicture}[scale=1.5]

\coordinate (wn)     at (10.0, 2.0);
\coordinate (wnm1)   at (8.0 , 2.0);
\coordinate (w2)     at (5.0 , 2.0);
\coordinate (w1)     at (3.0 , 2.0);
\coordinate (w0)     at (1.0 , 2.0);
\coordinate (wast)   at (5.5 , 0.5);

\coordinate (dots)   at (6.5 , 2.0);

\draw [fill] (wn)    circle [radius=2.0pt] ;
\draw [fill] (wnm1)  circle [radius=2.0pt] ;
\draw [fill] (w2)    circle [radius=2.0pt] ;
\draw [fill] (w1)    circle [radius=2.0pt] ;
\draw [fill] (w0)    circle [radius=2.0pt] ;
\draw [fill] (wast)  circle [radius=2.0pt] ;

\draw [] (dots) node {$\ldots$};

\begin{scope}[>=latex]
\draw [->, shorten >= 2.75pt, shorten <= 2.75pt] (wast)    -- (w0);
\draw [->, shorten >= 2.75pt, shorten <= 2.75pt] (wast)    -- (w1);
\draw [->, shorten >= 2.75pt, shorten <= 2.75pt] (wast)    -- (w2);
\draw [->, shorten >= 2.75pt, shorten <= 2.75pt] (wast)    -- (wnm1);
\draw [->, shorten >= 2.75pt, shorten <= 2.75pt] (wast)    -- (wn);

\end{scope}

\node [above right] at (wn)    {$w_n$}   ;
\node [above right] at (wnm1)  {$w_{n-1}$}   ;
\node [above right] at (w2)    {$w_2$}   ;
\node [above right] at (w1)    {$w_1$}   ;
\node [above right] at (w0)    {$w_0$}   ;
\node [below right] at (wast)  {$w^{\ast}$};

\node [above left ] at (wn)    {$T(a_n)$}   ;
\node [above left ] at (wnm1)  {$T(a_{n-1})$}   ;
\node [above left ] at (w2)    {$T(a_2)$}   ;
\node [above left ] at (w1)    {$T(a_1)$}   ;
\node [above left ] at (w0)    {$T(a_0)$}   ;
\node [below left ] at (wast)  {$T(a_0),\,T(a_1),\,\ldots,\,T(a_n)$};

\end{tikzpicture}
\caption{Модель $\kModel{M}$}
  \label{wfin:fig0}
\end{figure}

Нетрудно убедиться, что $(\kModel{M},w^\ast) \models A \con \mathit{Congr}$.

Теперь заметим, что для каждой подформулы $\theta$ формулы $\varphi$ и каждого приписывания $g$ в модели $\cModel{M}$ (а тем самым, и в модели~$\kModel{M}$) справедлива эквивалентность
$$
\begin{array}{lcl}
(\kModel{M},w^\ast) \models^g \theta & \iff & \cModel{M} \cmodels^g \theta,
\end{array}
$$
которая обосновывается индукцией по построению~$\theta$.

Как следствие, $(\kModel{M},w^\ast) \not\models \bar{\vp}$, а значит,
$\bar{\vp} \not\in \logic{QK}_{\mathit{wfin}}$.

Чтобы получить доказательства для $\logic{QS5}_{\mathit{wfin}}$, $\logic{QGL.3}_{\mathit{wfin}}$ и $\logic{QGrz.3}_{\mathit{wfin}}$, изменим определение модели $\kModel{M}$ подходящим образом; фактически достаточно изменить лишь отношение достижимости.

Пусть $L = \logic{QS5}_{\mathit{wfin}}$.
Положим $\kModel{M}_1 = \langle W, W \times W, D, I \rangle$.
Тогда $(\kModel{M}_1, w^\ast) \not\models \bar{\vp}$.

Пусть $L = \logic{QGL.3}_{\mathit{wfin}}$.
Пусть шкала Крипке $\langle W, R_2 \rangle$ представляет собой иррефлексивную цепь $w^\ast R_2 w_0 R_2 w_2 R_2 \ldots R_2 w_n$ и пусть $\kModel{M}_2 = \langle W, R_2, D, I \rangle$. Тогда $(\kModel{M}_2, w^\ast) \not\models \bar{\vp}$.

Пусть $L = \logic{QGrz.3}_{\mathit{wfin}}$.
Пусть $R_3$~--- рефлексивное замыкание отношения $R_2$ и пусть $\kModel{M}_3 = \langle W, R_3, D, I \rangle$. Тогда $\kModel{M}_3, w^\ast \not\models \bar{\vp}$.
\end{proof}

Сделаем замечание, связанное со свойствами моделей, определённых в доказательстве импликации \mbox{$(3) \Rightarrow (1)$} леммы~\ref{wfin:lem:QCl-to-QK}.

\begin{remark}
  \label{wfin:rem:Grz-GL-KTB}
  Заметим, что
  \[
  \begin{array}{ll}
  \arrayitembackspace(1) &
    \parbox[t]{445pt}{модель $\kModel{M} = \langle W, R, D, I \rangle$, определённая в ходе обоснования импликации \mbox{$(3) \Rightarrow (1)$}, построена на $\logic{GL}$-шкале с постоянной областью~$\mathcal{D}$;}
    \arrayitemskip\\
  \arrayitembackspace(2) &
    \parbox[t]{445pt}{модель $\kModel{M}$ обладает свойством наследственности <<вниз>> для~$S$ и~$\approx$: $(\kModel{M},w) \models \Diamond S(a, b) \imp S(a, b)$ и $(\kModel{M},w) \models \Diamond (a \approx b) \imp a\approx b$ для любых $w \in W$ и $a, b \in \mathcal{D}$;}
    \arrayitemskip\\
  \arrayitembackspace(3) &
    \parbox[t]{445pt}{без ограничений общности можно считать, что $\mathcal{D}$, т.е. предметная область моделей $\cModel{M}$, $\kModel{M}$, $\kModel{M}_1$, $\kModel{M}_2$ и $\kModel{M}_3$, содержит не менее двух элементов.}
  \end{array}
  \]
\end{remark}

Из леммы~\ref{wfin:lem:QCl-to-QK} получаем следующую теорему.

\begin{theorem}
  \label{wfin:thr:three-variables}
  Каждая логика из интервалов\/
  $[\logic{QK}_{\mathit{wfin}}, \logic{QGL.3.bf}_{\mathit{wfin}}]$,
  $[\logic{QK}_{\mathit{wfin}}, \logic{QGrz.3.bf}_{\mathit{wfin}}]$\/ и
  $[\logic{QK}_{\mathit{wfin}}, \logic{QS5}_{\mathit{wfin}}]$ является\/
  $\Pi^0_1$-трудной в языке с тремя предметными переменными, одной унарной предикатной буквой и двумя бинарными предикатными буквами.
\end{theorem}

\begin{proof}
  По лемме~\ref{wfin:lem:QCl-to-QK}, функция
  $f\colon \vp \mapsto \bar{\vp}$ погружает $\Pi^0_1$-трудное множество
  $\QCLFs$ в каждую логику из указанных интервалов.
\end{proof}

\begin{remark}
  \label{wfin:rem:three-variables}
  Каждая логика в интервалах
  $[\logic{QK}_{\mathit{wfin}}, \logic{QGL.3.bf}_{\mathit{wfin}}]$,
  $[\logic{QK}_{\mathit{wfin}}, \logic{QGrz.3.bf}_{\mathit{wfin}}]$ и
  $[\logic{QK}_{\mathit{wfin}}, \logic{QS5}_{\mathit{wfin}}]$ является консервативным расширением логики $\logic{QCl}$, в частности, консервативным расширением логики $\QCLFs$, и по теореме~$\ref{th:trakhtenbrot}$,  $\Sigma^0_1$-трудна в языке с тремя предметными переменными и одной бинарной предикатной буквой.
\end{remark}

\begin{corollary}
  \label{wfin:cor:three-variables}
  Пусть $L$~--- одна из логик $\logic{QK}$, $\logic{QT}$, $\logic{QD}$,
  $\logic{QK4}$, $\logic{QS4}$, $\logic{QS4.3}$, $\logic{QGL}$, $\logic{QGL.3}$,
  $\logic{QGrz}$, $\logic{QGrz.3}$, $\logic{QwGrz}$, $\logic{QwGrz.3}$, $\logic{QKB}$, $\logic{QKTB}$,
  $\logic{QS5}$.  Тогда логики $L_{\mathit{wfin}}$ и
  $L\logic{.bf}_{\mathit{wfin}}$ являются $\Pi^0_1$-трудными и
  $\Sigma^0_1$-трудными в языке с тремя предметными переменными, одной унарной предикатной буквой и двумя бинарными предикатными буквами.
\end{corollary}

\subsubsection{Элиминация бинарных предикатных букв}
\label{wfin:sec:elim-binary-pred-letters}




Чтобы элиминировать бинарные предикатные буквы, мы будем использовать незначительную модификацию трюка Крипке~\cite{Kripke62} и конструкцию, возникшую в доказательстве леммы~\ref{wfin:lem:QCl-to-QK}.

Заметим, что формула $\bar{\vp}$ содержит лишь две бинарные предикатные буквы~--- $\approx$ и $S$; будем обозначать их $P_1$ и $P_2$.

Пусть $F_1$, $F_2$, $G_1$, $G_2$~--- различные унарные предикатные буквы, отличные от буквы~$T$.
Пусть $\cdot^{\sigma}$~--- функция, осуществляющая подстановку формулы
$\Diamond (F_j(x) \con G_j (y) \con \neg \forall x\, F_j(x))$ вместо
$P_j(x, y)$, где $j \in \{1,2\}$.

\begin{lemma}
  \label{wfin:lem:Kripke}
  Для логики $L \in \{ \logic{QK}, \logic{QS5}, \logic{QGL.3}, \logic{QGrz.3} \}$ следующие условия эквивалентны друг другу:
  \[
  \begin{array}{ll}
  \arrayitembackspace(1) &
    \vp  \in \QCLFs;
    \arrayitemskip\\
  \arrayitembackspace(2) &
    \bar{\vp}^\sigma \in L_{\mathit{wfin}};
    \arrayitemskip\\
  \arrayitembackspace(3) &
    \bar{\vp}^\sigma \in L\logic{.bf}_{\mathit{wfin}}.
  \end{array}
  \]
\end{lemma}

\begin{proof}
  Импликация \mbox{$(1) \Rightarrow (2)$} следует из леммы~\ref{wfin:lem:QCl-to-QK} и замкнутости логик по правилу предикатной подстановки.

  Импликация \mbox{$(2) \Rightarrow (3)$} следует из того, что
  $L_{\mathit{wfin}} \subseteq L\logic{.bf}_{\mathit{wfin}}$.

  Для доказательства импликации \mbox{$(3) \Rightarrow (1)$} рассмотрим четыре возможных случая для~$L$.

  Пусть $L=\logic{QK.bf}_{\mathit{wfin}}$. Пусть
  $\vp \not\in \QCLFs$ и пусть
  $\kModel{M} = \langle W, R, D, I \rangle$~--- модель, определённая в доказательстве импликации \mbox{$(3) \Rightarrow (1)$} леммы~\ref{wfin:lem:QCl-to-QK}; как мы видели, в этом случае
  $\vp \not\in \QCLFs$ влечёт, что
  $(\kModel{M}, w^\ast) \not\models \bar{\vp}$. Напомним, что
  $\kModel{M}$ определена на конечной шкале с конечной предметной областью~$\mathcal{D}$.

  Будем использовать $\kModel{M}$, чтобы определить некоторую конечную модель
  $\kModel{M}' = \langle W', R', D', I' \rangle$, опровергающую формулу~$\bar{\vp}^\sigma$.

  Положим
  $$
  \begin{array}{lcl}
  W' & = & W \cup (\{ w^\ast \} \times (\mathcal{D} \times \mathcal{D}));
  \end{array}
  $$
  мир
  $\otuple{w^\ast, \otuple{a, b}} \in W' \setminus W$ будем обозначать $w^\ast_{ab}$.

  Пусть
  $$
  \begin{array}{lcl}
  R' & = & R \cup (\{ w^\ast \} \times (W' \setminus W)).
  \end{array}
  $$

  Положим $D'(w) = \mathcal{D}$ для каждого~$w \in W'$.

  Пусть, наконец, для каждого $j \in \{1,2\}$,
  \settowidth{\templength}{\mbox{$^\ast$}}
  $$
  \begin{array}{lcl}
    (\kModel{M}', w^\ast_{ab}) \models \hfill F_j (c)
      & \bydef
      & c = a  \mbox{ и }  (\kModel{M}, w^\ast) \models P_j(a,b);
      \smallskip\\
    (\kModel{M}', w^\ast_{ab}) \models \hfill G_j (c)
      & \bydef
      & c = b  \mbox{ и }  (\kModel{M}, w^\ast) \models P_j(a,b);
      \smallskip\\
    (\kModel{M}', w^\ast)_{\phantom{ab}}\hspace{-\templength} \models \hfill F_j (c)
      & \bydef
      & c \in \mathcal{D};
      \smallskip\\
    (\kModel{M}', w^\ast)_{\phantom{ab}}\hspace{-\templength} \models \hfill G_j (c)
      & \bydef
      & c \in \mathcal{D};
      \smallskip\\
    (\kModel{M}', w)_{\phantom{ab}} \not\models \hfill F_j (c)
      & \bydef
      & c \in \mathcal{D},
      ~\qquad\hfill \mbox{где $w \in W \setminus \{w^\ast\}$};
      \smallskip\\
    (\kModel{M}', w)_{\phantom{ab}} \not\models \hfill G_j (c)
      & \bydef
      & c \in \mathcal{D},
      ~\qquad\hfill \mbox{где $w \in W \setminus \{w^\ast\}$};
      \smallskip\\
    (\kModel{M}', w)_{\phantom{ab}} \models \hfill T (c)
      & \bydef
      & \mbox{$w\in W$, $c\in \mathcal{D}$ и $(\kModel{M}', w) \models T(c)$.}
  \end{array}
  $$
  Поскольку как $W$, так и $\mathcal{D}$ конечны, $W'$ тоже конечно.

  Как было отмечено (пункт~(3) замечания~\ref{wfin:rem:Grz-GL-KTB}), можем считать, что множество $\mathcal{D}$ содержит не менее двух элементов, поэтому для каждого~$j \in \{1,2\}$
  $$
  \begin{array}{lcl}
   (\kModel{M}', w) \models \neg \forall x\, F_j(x)
     & \iff
     & w \ne w^\ast.
  \end{array}
   \eqno {(\ast)}
  $$

  Учитывая $(\ast)$, нетрудно видеть, что
  $(\kModel{M}', w^\ast) \models A^\sigma \con \mathit{Congr}^\sigma$.

  Покажем, что $(\kModel{M}', w^\ast) \not\models \vp^\sigma$.

  Для этого достаточно доказать, что для каждой подформулы~$\theta$
  формулы $\vp$ и каждого приписывания~$g$
  $$
  \begin{array}{lcl}
  (\kModel{M}', w^\ast) \models^g \theta^\sigma
    & \iff
    & (\kModel{M}, w^\ast) \models^g \theta.
  \end{array}
   \eqno {({\ast}{\ast})}
  $$
  Доказательство $({\ast}{\ast})$ проводится индукцией по построению формулы~$\theta$ и не содержит каких-либо технических трудностей.

  Как следствие, $(\kModel{M}', w^\ast) \not\models \bar{\vp}^{\sigma}$,
  а значит,
  $\bar{\vp}^{\sigma} \not\in \logic{QK.bf}_{\mathit{wfin}}$.


  Пусть $L=\logic{QS5}_{\mathit{wfin}}$. Положим
  $\kModel{M}'_1 = \langle W', W' \times W', D', I' \rangle$. Тогда
  $(\kModel{M}'_1, w^\ast) \not\models \bar{\vp}^\sigma$.

  Пусть $\logic{QGL.3.bf}_{\mathit{wfin}}$. Пусть
  $\kModel{M}_2 = \langle W, R_2, I, D \rangle$~--- модель, определённая в доказательстве леммы~\ref{wfin:lem:QCl-to-QK}.
  Напомним, что $\langle W, R_2 \rangle$~--- иррефлексивная цепь и
  $D(w^\ast) = D(w_0) = \ldots = D(w_n) = \mathcal{D}$. Пусть
  $k = |\mathcal{D}|^2$.

  Пусть $\langle W'_2, R'_2 \rangle$~--- иррефлексивная цепь, полученная добавлением к $\langle W, R_2 \rangle$ миров $w_{n+1}, \ldots, w_{n+k}$ так, что $w^\ast R'_2 w_0 R'_2 \ldots R'_2 w_n R'_2 w_{n+1} R'_2 \ldots R'_2 w_{n+k}$. Пусть
  $D'_2(w) = \mathcal{D}$ для каждого $w \in W'_2$. Пусть интерпретация $I'_2$ определена аналогично $I'$, где
  $w_{n+1}, \ldots, w_{n+k}$ используются вместо миров вида
  $w^\ast_{ab}$. Пусть
  $\kModel{M}'_2 = \langle W'_2, R'_2, D'_2, I'_2 \rangle$. Тогда
  $(\kModel{M}'_2, w^\ast) \not\models \bar{\vp}^\sigma$.

  Пусть $L=\logic{QGrz.3.bf}_{\mathit{wfin}}$. Определим $R'_3$ как рефлексивное замыкание отношения $R'_2$ и положим
  $\kModel{M}'_3 = \langle W'_2, R'_3, D'_2, I'_2 \rangle$.
  Тогда $(\kModel{M}'_3, w^\ast) \not\models \bar{\vp}^\sigma$.
\end{proof}

\begin{theorem}
  \label{wfin:thr:Kripke}
  Каждая логика в интервалах\/
  $[\logic{QK}_{\mathit{wfin}}, \logic{QGL.3.bf}_{\mathit{wfin}}]$,
  $[\logic{QK}_{\mathit{wfin}}, \logic{QGrz.3.bf}_{\mathit{wfin}}]$ и\/
  $[\logic{QK}_{\mathit{wfin}}, \logic{QS5}_{\mathit{wfin}}]$ является\/
  $\Pi^0_1$-трудной в языке с тремя предметными переменными и пятью унарными предикатными буквами.
\end{theorem}

\begin{proof}
  Функция
  $f\colon \vp \mapsto \bar{\vp}^\sigma$ погружает $\QCLFs$ в соответствующий фрагмент такой логики.
\end{proof}

\begin{corollary}
  \label{wfin:cor:Kripke-2}
  Пусть $L$~--- одна из логик\/ $\logic{QK}$, $\logic{QT}$, $\logic{QD}$,
  $\logic{QK4}$, $\logic{QS4}$, $\logic{QS4.3}$, $\logic{QGL}$, $\logic{QGL.3}$,
  $\logic{QGrz}$, $\logic{QGrz.3}$, $\logic{QwGrz}$, $\logic{QwGrz.3}$, $\logic{QKB}$, $\logic{QKTB}$,
  $\logic{QS5}$. Тогда логики $L_{\mathit{wfin}}$ и
  $L\logic{.bf}_{\mathit{wfin}}$ являются $\Pi^0_1$-трудными в языке с тремя предметными переменными и пятью унарными предикатными буквами.
\end{corollary}


\subsubsection{Элиминация унарных предикатных букв}
\label{wfin:sec:elim-monadic-pred-letters}



Построим одно погружение для логик из интервалов
$[\logic{QK}_{\mathit{wfin}}, \logic{QGL.bf}_{\mathit{wfin}}]$ и
$[\logic{QK}_{\mathit{wfin}}, \logic{QGrz.bf}_{\mathit{wfin}}]$, а другое~--- для логик из интервала $[\logic{QK}_{\mathit{wfin}}, \logic{QKTB}_{\mathit{wfin}}]$.

Пусть $P_1, \ldots, P_n$~--- унарные предикатные буквы, входящие в $\bar{\vp}^\sigma$ (мы знаем, что $n\leqslant 5$, но нам это неважно). Пусть $P_{n+1}$~--- унарная буква, отличная от $P_1, \ldots, P_n$, и пусть $B = \forall x\, P_{n+1}(x)$.

Определим функцию $\cdot'$ рекурсией по сложности формулы:
$$
  \begin{array}{llll}
    (P_i(x))'
      & =
      & P_i(x) & \mbox{для $i \in \{1, \ldots, n \}$;}
      \\
    \bot'
      & =
      & \bot;
      \\
    (\psi \con \chi)'
      & =
      & \psi' \con \chi';
      \\
    (\psi \dis \chi)'
      & =
      & \psi' \dis \chi';
      \\
    (\psi \imp \chi)'
      & =
      & \psi' \imp \chi';
      \\
    (\forall x\, \psi)'
      & =
      & \forall x\, \psi';
      \\
    (\exists x\, \psi)'
      & =
      & \exists x\, \psi';
      \\
    (\Box \psi)'
      & =
      & \Box (B \imp \psi').
  \end{array}
$$

\begin{lemma}
  \label{wfin:lem:vp-B}
  Для логики $L \in \{ \logic{QK}, \logic{QGL}, \logic{QGrz}, \logic{QKTB} \}$ следующие условия эквивалентны друг другу:
  \[
  \begin{array}{ll}
  \arrayitembackspace (1) & \vp \in \QCLFs;
    \arrayitemskip\\
  \arrayitembackspace (2) & B \imp (\bar{\vp}^\sigma)' \in L_{\mathit{wfin}};
    \arrayitemskip\\
  \arrayitembackspace (3) & B \imp (\bar{\vp}^\sigma)' \in L\logic{.bf}_{\mathit{wfin}}.
  \end{array}
  \]
\end{lemma}

\begin{proof}
  Докажем импликацию \mbox{$(1) \Rightarrow (2)$}.

  Пусть
  $\kModel{M}, w_0 \not\models B \imp (\bar{\vp}^\sigma)'$ для некоторой модели $\kModel{M} = \otuple{W,R,D,I}$, определённой на конечной $L$-шкале, и некоторого $w_0 \in W$.

  Пусть $W' = \{ w \in W : \kModel{M},w \models B \}$. Поскольку
  $(\kModel{M}, w_0) \models B$, получаем, что $w_0 \in W'$, а значит, $W' \ne \varnothing$. Пусть
  $\kModel{M}' = \otuple{W',R',D',I'}$~--- модель, где
  $$
  \begin{array}{lcll}
  R' & = & R \upharpoonright W'; \\
  D'(w) & = & D(w) & \mbox{для кадого $w \in W'$;} \\
  I'(w, P) & = & I(w, P) & \mbox{для всех $w \in W'$ и $P \in \{ P_1, \ldots, P_n, P_{n+1} \}$.}
  \end{array}
  $$
  Заметим, что $\kModel{M}' \models B$.

  Покажем, что $(\kModel{M}', w_0) \not\models \bar{\vp}^\sigma$.
  Для этого покажем, что для любой подформулы $\theta$ формулы $\bar{\vp}^\sigma$, любого мира $w \in W'$ и любого приписывания~$g$
  $$
  \begin{array}{lcl}
  (\kModel{M}'_{\phantom{0}}, w) \models^g \theta
   & \iff
   & (\kModel{M}, w) \models^g \theta'.
   \end{array}
  $$

  Если формула $\theta$ является атомарной, то $\theta' = \theta$, и справедливость эквивалентности очевидна.
  Случаи, когда $\theta = \psi \con \chi$, $\theta = \psi \dis \chi$, $\theta = \psi \imp \chi$, $\theta = \forall x\, \psi$ и $\theta = \exists x\, \psi$, не содержат трудностей в обосновании.

  Пусть $\theta = \Box \psi$. Тогда $\theta' = \Box (B \imp \psi')$.

  Пусть $(\kModel{M}', w) \not\models^g \Box \psi$. Тогда
  $(\kModel{M}', v) \not\models^g \psi$ для некоторого $v \in W'$, такого, что
  $w R' v$. Но тогда $w R v$, и по индукционному предположению,
  $(\kModel{M}, v) \not\models^g \psi'$. Поскольку $v \in W'$, получаем, что
  $(\kModel{M}, v) \models B$. Следовательно,
  $(\kModel{M}, w) \not\models^g \Box (B \imp \psi')$.

  Пусть теперь $(\kModel{M}, w) \not\models^g \Box (B \imp \psi')$.
  Тогда $(\kModel{M}, v) \models B$ и $(\kModel{M}, v) \not\models^g \psi'$ для некоторого $v \in W$, такого, что $w R v$. Поскольку
  $(\kModel{M}, v) \models B$, получаем, что $v \in W'$. Но тогда $w R' v$, и по индукционному предположению, $(\kModel{M}', v) \not\models^g \psi$. Значит, $(\kModel{M}', w) \not\models^g \Box \psi$.

  Заметим, что если модель $\kModel{M}$ определена на конечной $L$-шкале, где
  $L \in \{ \logic{QGL}, \logic{QGrz}, \logic{QKTB} \}$, то модель
  $\kModel{M}'$ тоже определена на конечной $L$\nobreakdash-шкале.  Следовательно,
  $\bar{\vp}^\sigma \notin L_{\mathit{wfin}}$, и по лемме~\ref{wfin:lem:Kripke}, $\vp \notin \QCLFs$.

  Для обоснования импликации \mbox{$(2) \Rightarrow (3)$} достаточно заметить, что
  $L_{\mathit{wfin}} \subseteq L\logic{.bf}_{\mathit{wfin}}$.

  Чтобы обосновать импликацию \mbox{$(3) \Rightarrow (1)$}, рассмотрим четыре возможных случая для логики~$L$.

  Пусть $L = \logic{QK.bf}_{\mathit{wfin}}$.
  Пусть $\vp \notin \QCLFs$. Рассмотрим модель Крипке
  $\kModel{M}' = \otuple{W',R',D',I'}$, определённую в доказательстве импликации \mbox{$(3) \Rightarrow (1)$} леммы~\ref{wfin:lem:Kripke}. Как было показано, из того, что
  $\vp \notin \QCLFs$, следует, что
  $(\kModel{M}', w^\ast) \not\models \bar{\vp}^\sigma$. Напомним, что модель
  $\kModel{M}'$ конечна.

  Используя $\kModel{M}'$, определим модель
  $\kModel{M}'' = \langle W', R', D', I'' \rangle$, где $I''$ определяется согласно следующим условиям:
  $$
  \begin{array}{lcll}
    I''(w, P_{n+1}) & = & \mathcal{D}        & \mbox{для каждого $w \in W'$;} \\
    I''(w, P)       & = & I'(w, P) & \mbox{для любых $w \in W'$ и $P \in \{P_1, \ldots, P_n\}$.}
  \end{array}
  $$

  Поскольку $I''(w^\ast, P_{n+1}) = \mathcal{D}$,
  получаем, что $(\kModel{M}'', w^\ast) \models B$.

  Покажем, что
  $(\kModel{M}'', w^\ast) \not\models (\bar{\vp}^\sigma)'$.
  Для этого покажем, что для каждой подформулы $\theta$ формулы
  $\bar{\vp}^\sigma$, каждого мира $w \in W'$ и каждого приписывания~$g$
  $$
  \begin{array}{lcl}
  (\kModel{M}'', w) \models^g \theta'
    & \iff
    & (\kModel{M}', w) \models^g \theta.
  \end{array}
  $$
  Доказательство этой эквивалентности проводится индукцией по построению формулы~$\theta$ и не содержит каких-либо трудностей; мы разберём только случай, когда $\theta = \Box \psi$, и значит, $\theta' = \Box (B \imp \psi')$.

  Пусть $(\kModel{M}', w) \not\models^g \Box \psi$, т.е.
  $(\kModel{M}', v) \not\models^g \psi$ для некоторого $v \in R'(w)$. По индукционному предположению, $(\kModel{M}'', v) \not\models^g \psi'$. Поскольку $I''(v, P_{n+1}) = \mathcal{D}$, получаем, что $(\kModel{M}'', v) \models B$. Значит,
  $(\kModel{M}'', w) \not\models^g \Box (B \imp \psi')$.

  Пусть теперь
  $(\kModel{M}'', w) \not\models^g \Box (B \imp \psi')$, т.е.
  $(\kModel{M}'', v) \not\models^g \psi'$ для некоторого $v \in R'(w)$. По индукционному предположению,
  $(\kModel{M}', v) \not\models^g \psi$; следовательно,
  $(\kModel{M}', w) \not\models^g \Box \psi$.

  Как следствие доказанного получаем, что
  $(\kModel{M}'', w^\ast) \not\models B \imp (\bar{\vp}^\sigma)'$.
  Значит,
  $B \imp (\bar{\vp}^\sigma)' \notin \logic{QK.bf}_{\mathit{wfin}}$.

  Пусть $L = \logic{QGL.bf}_{\mathit{wfin}}$.
  Заметим, что шкала $\otuple{W', R'}$ является $\logic{GL}$-шкалой (пункт~$(1)$ замечания~\ref{wfin:rem:Grz-GL-KTB}), поэтому приведённое выше доказательство для $\logic{QK.bf}_{\mathit{wfin}}$ полностью переносится и на
  $\logic{QGL.bf}_{\mathit{wfin}}$.

  Пусть $L = \logic{QGrz.bf}_{\mathit{wfin}}$. Пусть ${R}^r$~--- рефлексивное замыкание отношения $R'$ и пусть
  ${\kModel{M}^{r}} = \otuple{W',R^r,D',I''}$. Тогда модель
  ${\kModel{M}^{r}}$ определена на $\logic{Grz}$-шкале с постоянной предметной областью и
  $({\kModel{M}^{r}}, w^\ast) \not\models B \imp (\bar{\vp}^\sigma)'$.

  Пусть $L = \logic{QKTB}_{\mathit{wfin}}$. В этом случае достаточно повторить описанное выше построение в случае логики $\logic{QK.bf}_{\mathit{wfin}}$ к модели
  $\kModel{M}'_1$, определённой в доказательстве импликации
  \mbox{$(3) \Rightarrow (1)$} леммы~\ref{wfin:lem:Kripke} для случая логики
  $\logic{QS5}_{\mathit{wfin}}$; мы получим некоторую модель
  $\kModel{M}^{rs} = \langle W', W' \times W', D', I'' \rangle$. Тогда
  $\kModel{M}^{rs}$ определена на конечной $\logic{KTB}$-шкале\footnote{Фактически $\kModel{M}^{rs}$ определена на конечной $\logic{S5}$-шкале.} и
  $(\kModel{M}^{rs}, w^\ast) \not\models B \imp (\bar{\vp}^\sigma)'$.
\end{proof}

\begin{remark}
  \label{wfin:rem:B} Заметим, что
  \[
  \begin{array}{ll}
  \arrayitembackspace(1) &
    \parbox[t]{445pt}{модели ${\kModel{M}}''$ и ${\kModel{M}^{r}}$, определённые в доказательстве импликации \mbox{$(3) \Rightarrow (1)$} леммы~$\ref{wfin:lem:vp-B}$ обладают свойством наследственности <<вниз>> для предикатных букв формулы
    $B \imp (\bar{\vp}^\sigma)'$: ${\kModel{M}}'' \models \Diamond P(a) \imp
    P(a)$ и ${\kModel{M}}^r \models \Diamond P(a) \imp
    P(a)$ для любых
    $a \in \mathcal{D}$ и $P \in \{P_1, \ldots, P_{n+1} \}$;}
    \arrayitemskip\\
  \arrayitembackspace(2) &
    \parbox[t]{445pt}{учитывая пункт~$(3)$ замечания~$\ref{wfin:rem:Grz-GL-KTB}$, можем считать что предметная область $\mathcal{D}$ модели $\kModel{M}^{rs}$, определённой в доказательстве импликации \mbox{$(3) \Rightarrow (1)$} леммы~$\ref{wfin:lem:vp-B}$, содержит не менее двух элементов.}
\end{array}
\]
\end{remark}

Дальнейшие построения для логик из интервалов $[\logic{QK}_{\mathit{wfin}}, \logic{QGL.bf}_{\mathit{wfin}}]$ и
$[\logic{QK}_{\mathit{wfin}}, \logic{QGrz.bf}_{\mathit{wfin}}]$ будут отличаться от построений для логик из интервала
$[\logic{QK}_{\mathit{wfin}}, \logic{QKTB}_{\mathit{wfin}}]$; опишем их по-отдельности.

\subsubsection{Подлогики логик $\logic{QGL.bf}_{\mathit{wfin}}$ и
  $\logic{QGrz.bf}_{\mathit{wfin}}$}
\label{wfin:sec:GL-Grz}

Пусть $P$~--- унарная предикатная буква, отличная от
$P_1, \ldots, P_{n+1}$.
Определим рекурсивно следующие формулы:
$$
\begin{array}{lcll}
    \delta_0(x)
      & =
      & \Box^+ P(x);
      \\
    \delta_{m+1}(x)
      & =
      & P (x) \con \Diamond (\neg P(x) \con \Diamond \delta_{m}(x)).
\end{array}
$$
Для каждого $k \in \{1, \ldots, n+1 \}$ положим
$$
\begin{array}{lcll}
\alpha_k (x )
  & =
  & \delta_{k}(x) \con \neg \delta_{k+1}(x) \con \Diamond \Box^+ \neg P(x);
  \\
\beta_k(x)
  & =
  & \neg P(x)\con\Diamond \alpha_k(x).
\end{array}
$$
Пусть $(\bar{\vp}^\sigma)^\diamond$~--- формула, получающаяся из $(\bar{\vp}^\sigma)'$ подстановкой $\beta_k(x)$ вместо $P_k(x)$ для каждого
$k\in\{1,\ldots,n+1\}$.

\begin{figure}
  \centering
  \begin{tikzpicture}[scale=1.5]

\coordinate (w2kk)    at (1, 6.5);
\coordinate (w2km1k)  at (1, 5.5);
\coordinate (w3k)     at (1, 4.0);
\coordinate (w2k)     at (1, 3.0);
\coordinate (w1k)     at (1, 2.0);
\coordinate (w0k)     at (1, 1.0);
\coordinate (wastk)   at (2, 2.0);

\coordinate (dots)    at (1, 4.75);

\draw [fill] (w2kk)   circle [radius=2.0pt] ;
\draw [fill] (w2km1k) circle [radius=2.0pt] ;
\draw [fill] (w3k)    circle [radius=2.0pt] ;
\draw [fill] (w2k)    circle [radius=2.0pt] ;
\draw [fill] (w1k)    circle [radius=2.0pt] ;
\draw [fill] (w0k)    circle [radius=2.0pt] ;
\draw [fill] (wastk)  circle [radius=2.0pt] ;

\draw [] (dots) node {$\vdots$};

\begin{scope}[>=latex]
\draw [->, shorten >= 2.75pt, shorten <= 2.75pt] (w0k)    -- (wastk);
\draw [->, shorten >= 2.75pt, shorten <= 2.75pt] (w0k)    -- (w1k);
\draw [->, shorten >= 2.75pt, shorten <= 2.75pt] (w1k)    -- (w2k);
\draw [->, shorten >= 2.75pt, shorten <= 2.75pt] (w2k)    -- (w3k);
\draw [->, shorten >= 2.75pt, shorten <= 2.75pt] (w2km1k) -- (w2kk);

\end{scope}

\node [right] at (w2kk)   {$w_{2k}^k$}  ;
\node [right] at (w2km1k) {$w_{2k-1}^k$};
\node [right] at (w3k)    {$w_{3}^k$}   ;
\node [right] at (w2k)    {$w_{2}^k$}   ;
\node [right] at (w1k)    {$w_{1}^k$}   ;
\node [right] at (w0k)    {$w_{0}^k$}   ;
\node [below right] at (wastk)  {$v^{k}$};
\node [left ] at (w2kk)   {$P(a)~$}  ;
\node [left ] at (w2km1k) {$\neg P(a)~$};
\node [left ] at (w3k)    {$\neg P(a)~$}   ;
\node [left ] at (w2k)    {$P(a)~$}   ;
\node [left ] at (w1k)    {$\neg P(a)~$}   ;
\node [left ] at (w0k)    {$P(a)~$}   ;
\node [above right] at (wastk)  {$\neg P(a)$};

\end{tikzpicture}

\caption{Шкала $\kframe{F}_k$}
  \label{wfin:fig_Mm}
\end{figure}

\begin{lemma}
  \label{wfin:lem:vp-ast}
  Пусть $L \in \{ \logic{QK}, \logic{QGL}, \logic{QGrz} \}$. Следующие условия эквивалентны другу другу:
  \[
  \begin{array}{ll}
  \arrayitembackspace (1) & \vp \in \QCLFs;
    \arrayitemskip\\
  \arrayitembackspace (2) & \forall x\, \beta_{n+1} (x) \imp
    (\bar{\vp}^\sigma)^\diamond \in L_{\mathit{wfin}};
    \arrayitemskip\\
  \arrayitembackspace (3) & \forall x\, \beta_{n+1} (x) \imp (\bar{\vp}^\sigma)^\diamond
    \in L{\bf.bf}_{\mathit{wfin}}.
  \end{array}
  \]
\end{lemma}

\begin{proof}
  Импликация \mbox{$(1) \Rightarrow (2)$} следует из леммы~\ref{wfin:lem:vp-B} и замкнутости логик по предикатной подстановке.

  Импликация \mbox{$(2) \Rightarrow (3)$} следует из того, что
  $L_{\mathit{wfin}} \subseteq L\logic{.bf}_{\mathit{wfin}}$.

  Импликация \mbox{$(3) \Rightarrow (1)$} обосновывается с помощью конструкции, описанной в доказательстве леммы~\ref{lem:vp-B}.

  Пусть $L=\logic{QK.bf}_{\mathit{wfin}}$ или $L=\logic{QGL.bf}_{\mathit{wfin}}$.
  Пусть
  $\vp \notin \QCLFs$.
  Пусть
  $\kModel{M}'' = \langle W', R', D', I'' \rangle$~--- модель Крипке, определённая в доказательстве импликации \mbox{$(3) \Rightarrow (1)$} of
  леммы~\ref{wfin:lem:vp-B}. Как мы видели, из того, что
  $\vp \notin \QCLFs$, следует, что
  $(\kModel{M}'', w^\ast) \not\models (\bar{\vp}^\sigma)'$. Напомним, что модель $\kModel{M}''$ определена на конечной $\logic{GL}$-шкале с конечной постоянной областью~$\mathcal{D}$ и
  $\kModel{M}'' \models \forall x\, P_{n+1} (x)$.

  По модели $\kModel{M}''$ построим модель $\kModel{M}^\diamond$, определённую на $\logic{GL}$-шкале с постоянной областью, опровергающую формулу
  $\forall x\, \beta_{n+1} (x) \imp (\bar{\vp}^\sigma)^\diamond$.

  Для каждого~$k \in \{1, \ldots, n+1 \}$ определим шкалу Крипке
  $\kframe{F}_k = \otuple{W_k,R_k}$, положив
  $W_k = \{w^k_0,\ldots,w^k_{2k}\}\cup\{v^k\}$ а в качестве $R_k$ взяв транзитивное замыкание отношения $\{\langle w^k_i,w^k_{i+1}\rangle : 0\leqslant i < 2k\}\cup\{\langle w^k_0, v^k \rangle\}$,
  см.~рис.~\ref{wfin:fig_Mm}.

  Для $w\in W'$, $a \in \mathcal{D}$ и
  $k \in \{1, \ldots, n +1\}$ определим шкалу Крипке
  $\kframe{F}_k^{wa} = \langle \{\langle w, a \rangle \}\times W_k^{\phantom{0}},R_k^{wa}\rangle$
  как изоморфную копию шкалы $\kframe{F}_k$ при изоморфизме
  $f\colon v \mapsto \langle w, a, v \rangle$.

  Пусть
  $$
  \begin{array}{lcl}
  W^\diamond
    & =
    & \displaystyle
      W' \cup \bigcup\limits_{k=1}^{n+1} (W' \times \mathcal{D} \times W_k).
  \end{array}
  $$
  Поскольку множества $W'$, $\mathcal{D}$ и $W_k$ (где $k \in \{1,\ldots,n+1\}$) конечны, получаем, что $W^\diamond$ тоже конечно.
  Пусть $R^\diamond$~--- транзитивное замыкание отношения
  $$
  \begin{array}{lcl}
  S
    & =
    & \displaystyle
      R' \cup\bigcup_{k=1}^{n+1}
  \big\{R_k^{wa} \cup \{\langle w, \langle w, a, w^k_0 \rangle
  \rangle\} : w\in W',\ a \in \mathcal{D} \big\},
  \end{array}
  $$
  т.е. для любых $w\in W'$ и $a \in \mathcal{D}$ корни шкал
  $\frak{F}_1^{wa},\ldots,\frak{F}_{n+1}^{wa}$ $R^\diamond$\nobreakdash-достижимы из~$w$.
  Для каждого $u\in W^\diamond$ положим $D^\diamond(u) = \mathcal{D}$.
  Для $u\in W^\diamond$ и $a\in \mathcal{D}$ положим
  $$
  \begin{array}{lcl}
  \langle a \rangle \in I^\diamond(u,P)
    & \bydef
    & \parbox[t]{335pt}{существуют такие $w\in
    W'$, $k\in\{1,\ldots,n+1\}$ и $i\in\{0,\ldots,k\}$, что $u
    = \langle w, a, w^k_{2i} \rangle$ и $(\kModel{M}'',w)\models
    P_k(a)$,}
  \end{array}
  $$
  см.~рис.~\ref{wfin:fig_Mm}.
  Пусть
  $\kModel{M}^\diamond = \langle
  W^\diamond,R^\diamond,D^\diamond,I^\diamond\rangle$.
  Заметим, что модель $\kModel{M}^\diamond$ определена на $\logic{GL}$\nobreakdash-шкале.

  \begin{sublemma}
    \label{wfin:lem:model-M_k}
    Для любых $u \in W^\diamond$, $a \in \mathcal{D}$ и $k\in\{1,\ldots,n+1\}$
    $$
    \begin{array}{lcl}
    (\kModel{M}^\diamond, u) \models \alpha_k(a)
      & \iff
      & \parbox[t]{250pt}{$u = \langle w, a, w^k_0 \rangle$ для некоторого $w \in W'$, такого, что $(\kModel{M}'',w)\models P_k(a)$.}
    \end{array}
    $$
  \end{sublemma}

\begin{proof}
  Заметим, что
  $$
  \begin{array}{lll}
    \phantom{{}\con \neg \delta_{j+1} (a)}
    (\kModel{M}^\diamond, u) \models \delta_0 (a)
    & \iff &
    \parbox[t]{225pt}{$u = \langle w, a, w^m_{2m} \rangle$ для некоторых $m \in \{1, \ldots, n+1\}$ и $w \in W'$, таких, что $(\kModel{M}'',w)\models P_m(a)$.}
  \end{array}
  $$
  Кроме того, если $(\kModel{M}^\diamond, u) \models \delta_j (a)$,
  то $(\kModel{M}^\diamond, u) \models P (a)$, и значит,
  $u \notin W'$.
  Используя это наблюдение, индукцией по $j \in \{ 1, \ldots, k \}$ несложно показать, что
  $$
  \begin{array}{lcl}
     (\kModel{M}^\diamond, u) \models \delta_j (a) \con \neg \delta_{j+1} (a)
             & \iff &
             \parbox[t]{225pt}{$u = \langle w, a, w^m_{2m - 2j} \rangle$ для некоторых $m \in \{1, \ldots, n+1\}$ и $w \in W'$, таких, что $(\kModel{M}'',w)\models P_m(a)$.}
  \end{array}
  $$

  Из того, что $(\kModel{M}'', w) \models P_m(a)$ и
  $(\kModel{M}^\diamond, \langle w, a, w^m_{i} \rangle) \models \Diamond
  \Box^+ \neg P(a)$, следует, что $i = 0$.
  Поэтому $(\kModel{M}^\diamond, u) \models \alpha_k (a)$ тогда и только тогда, когда $u = \langle w, a, w^m_{2m - 2k} \rangle$ для некоторого
  $m \in \{1, \ldots, n+1\}$ и некоторого $w \in W'$, такого, что
  $(\kModel{M}'',w)\models P_m(a)$ и $2m - 2k = 0$, т.е. $k = m$.
\end{proof}

  Покажем, что
  $(\kModel{M}^\diamond, w^\ast) \not\models \forall x\, \beta_{n+1} (x) \imp
  (\bar{\vp}^\sigma)^\diamond$.

  Для этого сначала покажем, что для любых $w\in W'$,
  $a\in \mathcal{D}$ и $k\in\{1,\ldots,n+1\}$
  $$
  \begin{array}{lcl}
  \kModel{M}'',w\models P_k(a)
    & \iff
    & (\kModel{M}^\diamond,w)\models \beta_k (a).
  \end{array}
  \eqno{(\ast)}
  $$
  Пусть $(\kModel{M}'',w)\models P_k(a)$. По определению,
  $(\kModel{M}^\diamond, w) \not\models P(a)$ и
  $w R^\diamond \langle w, a, w^k_0 \rangle$.
  Значит, по подлемме~\ref{wfin:lem:model-M_k},
  $(\kModel{M}^\diamond, w) \models \beta_k (a)$.

  Пусть теперь $(\kModel{M}^\diamond, w) \models \beta_k (a)$. Согласно определению модели $\kModel{M}^\diamond$ и подлемме~\ref{wfin:lem:model-M_k},
  либо $w R^\diamond \langle w, a, w^k_0 \rangle$, либо
  $w R^\diamond v R^\diamond \langle v, a, v^k_0 \rangle$ для некоторого
  $v \in W'$.  Значит, по определению модели $\kModel{M}^\diamond$, либо
  $(\kModel{M}'',w)\models P_k(a)$, что и требовалось, либо
  $(\kModel{M}'',v) \models P_k (a)$, и значит, согласно пункту~$(1)$
  замечания~\ref{wfin:rem:B}, тоже получаем, что $(\kModel{M}'', w) \models P_k (a)$.

  Теперь покажем, что
  $$
  \begin{array}{lcl}
  (\kModel{M}^\diamond, w) \models \forall x\, \beta_{n+1} (x)
    & \iff
    & w \in W'.
  \end{array}
  \eqno{({\ast}{\ast})}
  $$

  Импликация $(\Leftarrow)$ следует из того, что
  $\kModel{M}'' \models \forall x\, P_{n+1} (x)$ и условия~$(\ast)$. Для обоснования импликации $(\Rightarrow)$ заметим, что согласно определению $\kModel{M}^\diamond$, только миры множества $W'$ видят миры вида $\langle w, a, w_0^{n+1} \rangle$.
  Значит, из того, что $(\kModel{M}^\diamond, w) \models \beta_{n+1} (a)$ следует, что $w \in W'$.

  Поскольку $w^\ast \in W'$, by $(\ast\ast)$, получаем, что
  $(\kModel{M}^\diamond, w^\ast) \models \forall x\, \beta_{n+1} (x)$.

  Осталось показать, что $(\kModel{M}^\diamond, w^\ast) \not\models
  (\bar{\vp}^\sigma)^\diamond$.

  Покажем, что для каждой подформулы
  $\theta$ формулы $\bar{\vp}^\sigma$, каждого мира $w\in W'$ и каждого приписывания~$g$,
  $$
  \begin{array}{lcl}
  (\kModel{M}'',w)\models^g \theta'
    & \iff
    & (\kModel{M}^\diamond,w)\models^g \theta^\diamond,
  \end{array}
  $$
  где $\theta^\diamond$~--- результат подстановки в $\theta'$ формул
  $\beta_1(x),\ldots, \beta_{n+1}(x)$ вместо, соответственно,
  $P_1(x),\ldots,P_{n+1}(x)$.

  Обоснование проведём индукцией по построению~$\theta$.

  Если $\theta = P_k (x)$, то $\theta' = P_k (x)$ и
  $\theta^\diamond = \beta_k (x)$; в этом случае получаем требуемое по~$(\ast)$.

  Если $\theta = \bot$, $\theta = \psi \con \chi$, $\theta = \psi \dis \chi$, $\theta = \psi \imp \chi$, $\theta = \forall x\, \psi$, $\theta = \exists x\, \psi$, то обоснование не вызывает трудности.

  Пусть $\theta = \Box \psi$. Тогда
  $$
  \begin{array}{lcr}
    \theta' = \Box(\forall x\, P_{n+1}(x) \imp
    \psi') & \mbox{и} & \theta^\diamond = \Box(\forall x\, \beta_{n+1}(x)
                            \imp \psi^\diamond).
  \end{array}
  $$

  Пусть $(\kModel{M}^\diamond,w)\not\models^g \theta^\diamond$. Тогда
  существует мир $v \in W^\diamond$, такой, что $wR^\diamond v$,
  $(\kModel{M}^\diamond, v) \models \forall x\, \beta_{n+1}(x)$ и
  $(\kModel{M}^\diamond, v) \not\models^g \psi^\diamond$. По
  $({\ast}{\ast})$ получаем, что $v \in W'$. Значит, по индукционному предположению,
  $(\kModel{M}'', v) \not\models^g \psi'$.  Поскольку $w R^\diamond v$ и
  $w, v \in W'$, получаем, что $w R' v$.  Поскольку
  $\kModel{M}'' \models \forall x\, P_{n+1} (x)$, получаем, что
  $(\kModel{M}'', w) \not\models^g \theta'$.

  Пусть теперь $(\kModel{M}'', w) \not\models^g \theta'$.  Тогда существует мир
  $v \in W'$, такой, что $w R' v$,
  $(\kModel{M}'', v) \models \forall x\, P_{n+1}(x)$ и
  $(\kModel{M}'', v) \not\models^g \psi'$.  По $(\ast)$ получаем, что
  $(\kModel{M}^\diamond, v) \models \forall x\, \beta_{n+1}(x)$. По индукционному предположению,
  $(\kModel{M}^\diamond, v) \not\models^g \psi^\diamond$. Значит,
  $(\kModel{M}^\diamond,w)\not\models^g \theta^\diamond$.

  Следовательно,
  $(\kModel{M}^\diamond, w^\ast) \not\models \forall x\, \beta_{n+1} (x)
  \imp (\bar{\vp}^\sigma)^\diamond$.

  Значит,
  $\forall x\, \beta_{n+1} (x) \imp (\bar{\vp}^\sigma)^\diamond
  \notin \logic{QGL.bf}_{\mathit{wfin}}$,
  откуда получаем также, что
  $\forall x\, \beta_{n+1} (x) \imp (\bar{\vp}^\sigma)^\diamond
  \notin \logic{QK.bf}_{\mathit{wfin}}$.

  Пусть $L = \logic{QGrz.bf}_{\mathit{wfin}}$.
  Пусть $S^{\ast}$~--- рефлексивно-транзитивное замыкание отношения $S$, определённого выше, и пусть $\kModel{M}^{\ast} = \langle W^\diamond,
  S^{\ast},D^\diamond,I^\diamond\rangle$.
  Тогда $\langle W^\diamond, S^{\ast} \rangle$~--- конечная
  $\logic{Grz}$-шкала с постоянной областью, причём
  $(\kModel{M}^{\ast}, w^\ast) \not\models \forall x\, \beta_{n+1} (x)
  \imp (\bar{\vp}^\sigma)^\diamond$.
  Доказательство проводится так же; отметим только, что при обосновании $({\ast}{\ast})$ надо пользоваться тем, что $\forall x\, \beta_{n+1}(x)$ опровергается в мире
  $\langle w, a, w_0^{n+1} \rangle$ для любых $w \in W'$ и $a \in \mathcal{D}$, что следует из того, что условие
  $(\kModel{M}^\ast, v) \models \beta_{n+1}(a)$ влечёт, что
  $(\kModel{M}^\ast, v) \models \neg P(a)$.
\end{proof}

В результате получаем следующую теорему.

\begin{theorem}
  \label{wfin:thr:GL-Grz}
  Каждая логика из интервалов\/
  $[\logic{QK}_{\mathit{wfin}}, \logic{QGL.bf}_{\mathit{wfin}}]$ и\/
  $[\logic{QK}_{\mathit{wfin}}, \logic{QGrz.bf}_{\mathit{wfin}}]$ является\/
  $\Pi^0_1$-трудной в языке с тремя предметными переменными и одной унарной предикатной буквой.
\end{theorem}

\begin{proof}
  По лемме~\ref{wfin:lem:vp-ast}, функция
  $f\colon \vp \mapsto \forall x\, \beta_{n+1} (x) \imp
  (\bar{\vp}^\sigma)^\diamond$
  погружает $\Pi^0_1$-трудное множество $\QCLFs$
  в соответствующие фрагменты логик из интервалов
  $[\logic{QK}_{\mathit{wfin}}, \logic{QGL.bf}_{\mathit{wfin}}]$ и
  $[\logic{QK}_{\mathit{wfin}}, \logic{QGrz.bf}_{\mathit{wfin}}]$.
\end{proof}


\begin{corollary}
  \label{wfin:cor:GL-Grz}
  Пусть $L$~--- одна из логик\/ $\logic{QK}$, $\logic{QT}$, $\logic{QD}$,
  $\logic{QK4}$, $\logic{QS4}$, $\logic{QGL}$, $\logic{QGrz}$, $\logic{QwGrz}$.  Тогда логики
  $L_{\mathit{wfin}}$ и $L\logic{.bf}_{\mathit{wfin}}$ являются\/
  $\Pi^0_1$-трудными в языке с тремя предметными переменными и одной унарной предикатной буквой.
\end{corollary}

\subsubsection{Подлогики логики $\logic{QKTB}_{\mathit{wfin}}$}
\label{wfin:sec:KTB}


Пусть, как и выше, $P$~--- унарная предикатная буква, отличная от букв
$P_1, \ldots, P_{n+1}$, и пусть $k \in \{1, \ldots, n+1 \}$.

Определим рекурсивно следующие формулы:
$$
\begin{array}{lcll}
    \delta_{k+1}^k(x) & = & \Box^+ P(x); \\
    \delta_k^k(x)     & = & \Box^{\leqslant k} \neg P(x)\con
                            \Diamond^{k+1} P(x)\con \Diamond^{k+2}\delta_{k+1}^k(x); \\
    \delta_{i}^k(x)   & = & \Box^{\leqslant i}\neg P(x)\con
                            \Box^+\Diamond^{i+1}P(x)\con
                            \Diamond^{2i+3}\delta^k_{i+1}(x),
                          & \mbox{где $1\leqslant i < k$.}
\end{array}
$$
Для каждого $k \in \{1, \ldots, n+1 \}$ переопределим формулы $\alpha_k(x)$ и~$\beta_k(x)$ следующим образом:
$$
\begin{array}{lcll}
\alpha_k (x) & = & P(x) \con \Diamond^2 \delta^k_1(x)\con \neg\Diamond^3 \delta^k_2(x); \\
\beta_k (x)  & = & \neg P(x) \con \Box\Diamond P(x)\con \Diamond \alpha^k(x).
\end{array}
$$
Пусть $(\varphi^\sigma)^\diamond$~--- формула, получающаяся из $(\varphi^\sigma)'$ подстановкой
$\beta_k(x)$ вместо $P_k(x)$ для каждого $k\in\{1,\ldots,n+1\}$.

\begin{figure}
  \centering
  \begin{tikzpicture}[scale=1.50]

\coordinate (wk0)     at ( 1, 1);   \node [below=2pt] at (wk0)     {$w^k_0$}         ;
\coordinate (vk11)    at ( 2, 1);   \node [below=2pt] at (vk11)    {$v^{k1}_1$}      ;
\coordinate (vk12)    at ( 3, 1);   \node [below=2pt] at (vk12)    {$v^{k1}_2$}      ;
\coordinate (vk13)    at ( 4, 1);   \node [below=2pt] at (vk13)    {$v^{k1}_3$}      ;
\coordinate (wk1)     at ( 5, 1);   \node [below=2pt] at (wk1)     {$w^k_1$}         ;

\coordinate (vki1)    at ( 1,-1);   \node [below=2pt] at (vki1)    {$v^{kj}_1$}      ;
\coordinate (vkiim1)  at ( 3,-1);   \node [below=2pt] at (vkiim1)  {$v^{kj}_{j}$}    ;
\coordinate (vkii)    at ( 4,-1);   \node [below=2pt] at (vkii)    {$v^{kj}_{j+1}$}  ;
\coordinate (vkiip1)  at ( 5,-1);   \node [below=2pt] at (vkiip1)  {$v^{kj}_{j+2}$}  ;
\coordinate (vki2ip1) at ( 7,-1);   \node [below=2pt] at (vki2ip1) {$v^{kj}_{2j+1}$} ;
\coordinate (wki)     at ( 8,-1);   \node [below=2pt] at (wki)     {$w^k_j$}         ;

\coordinate (vkk1)    at ( 4,-3);   \node [below=2pt] at (vkk1)    {$v^{kk}_1$}      ;
\coordinate (vkkk)    at ( 5,-3);
\coordinate (vkk2kp1) at ( 6,-3);   \node [below=2pt] at (vkk2kp1) {$v^{kk}_{2k+1}$} ;
\coordinate (wkk)     at ( 7,-3);   \node [below=2pt] at (wkk)     {$w^k_k$}         ;
\coordinate (ek)      at ( 8,-3);   \node [below=2pt] at (ek)      {$e^k$}           ;

\draw [] (wk0)     circle [radius=2.0pt] ;
\draw [] (vk11)    circle [radius=2.0pt] ;
\draw [] (vk12)    circle [radius=2.0pt] ;
\draw [] (vk13)    circle [radius=2.0pt] ;
\draw [] (wk1)     circle [radius=2.0pt] ;

\draw [] (vki1)    circle [radius=2.0pt] ;
\draw [] (vkiim1)  circle [radius=2.0pt] ;
\draw [] (vkii)    circle [radius=2.0pt] ;
\draw [] (vkiip1)  circle [radius=2.0pt] ;
\draw [] (vki2ip1) circle [radius=2.0pt] ;
\draw [] (wki)     circle [radius=2.0pt] ;

\draw [] (vkk1)    circle [radius=2.0pt] ;
\draw [] (vkk2kp1) circle [radius=2.0pt] ;
\draw [] (wkk)     circle [radius=2.0pt] ;
\draw [] (ek)      circle [radius=2.0pt] ;


\begin{scope}[>=latex]

\draw [<->, shorten >= 2.75pt, shorten <= 2.75pt, dashed] (vki1)   -- (vkiim1) ;
\draw [<->, shorten >= 2.75pt, shorten <= 2.75pt, dashed] (vkiip1) -- (vki2ip1);
\draw [<->, shorten >= 2.75pt, shorten <= 2.75pt, dashed] (vkk1)   -- (vkk2kp1);

\draw [<->, shorten >= 2.75pt, shorten <= 2.75pt] (wk0)     -- (vk11)  ;
\draw [<->, shorten >= 2.75pt, shorten <= 2.75pt] (vk11)    -- (vk12)  ;
\draw [<->, shorten >= 2.75pt, shorten <= 2.75pt] (vk12)    -- (vk13)  ;
\draw [<->, shorten >= 2.75pt, shorten <= 2.75pt] (vk13)    -- (wk1)   ;

\draw [<->, shorten >= 2.75pt, shorten <= 2.75pt] (vkiim1)  -- (vkii)  ;
\draw [<->, shorten >= 2.75pt, shorten <= 2.75pt] (vkii)    -- (vkiip1);
\draw [<->, shorten >= 2.75pt, shorten <= 2.75pt] (vki2ip1) -- (wki)   ;

\draw [<->, shorten >= 2.75pt, shorten <= 2.75pt] (vkk2kp1) -- (wkk)   ;
\draw [<->, shorten >= 2.75pt, shorten <= 2.75pt] (wkk)     -- (ek)    ;

\draw [<->, shorten >= 2.75pt, shorten <= 2.75pt, dashed]
      (wk1)  .. controls (5.75,1)   and (6.4,-0.45) ..
      (3,0)  .. controls (-0.1,0.5) and (0.25,-1)   ..
      (vki1);

\draw [<->, shorten >= 2.75pt, shorten <= 2.75pt, dashed]
      (wki)  .. controls (8.75,-1)  and (9.4,-2.45) ..
      (6,-2) .. controls (2.7,-1.5) and (3.25,-3)   ..
      (vkk1);

\end{scope}

\draw (1.75,1.15)       -- (1.75, 1.25)       -- (4.25, 1.25)       -- (4.25, 1.15)      ;
\draw (1.75-1,1.15-2)   -- (1.75-1, 1.25-2)   -- (4.25+4-1, 1.25-2) -- (4.25+4-1, 1.15-2);
\draw (1.75+3-1,1.15-4) -- (1.75+3-1, 1.25-4) -- (4.25+3-1, 1.25-4) -- (4.25+3-1, 1.15-4);

\node [above=10pt] at (vk12)  {$\neg P(a)$};
\node [above=10pt] at (vkii)  {$\neg P(a)$};
\node [above=10pt] at (vkkk)  {$\neg P(a)$};

\node [above=10pt] at (wk0)   {$P(a)$};
\node [above=10pt] at (wk1)   {$P(a)$};
\node [above=10pt] at (wki)   {$P(a)$};
\node [above=10pt] at (wkk)   {$P(a)$};
\node [above=10pt] at (ek)    {$P(a)$};

\end{tikzpicture}

\caption{Шкала $\kframe{G}_k$}
  \label{wfin:fig2}
\end{figure}

\begin{lemma}
  \label{wfin:lem:vp-ast-QKTB}
  Для логики $L \in \{ \logic{QK}, \logic{QKTB} \}$ следующие условия эквивалентны друг другу:
  \[
  \begin{array}{ll}
  \arrayitembackspace(1)
    & \vp \in \QCLFs;
    \arrayitemskip \\
  \arrayitembackspace(2)
    & \forall x\, \beta_{n+1} (x) \imp (\bar{\vp}^\sigma)^\diamond
    \in L_{\mathit{wfin}}.
  \end{array}
  \]
\end{lemma}

\begin{proof}
  Импликация \mbox{$(1) \Rightarrow (2)$} следует из леммы~\ref{wfin:lem:vp-B}
  и замкнутости логик по предикатной подстановке.

  Докажем импликацию \mbox{$(2) \Rightarrow (1)$}.

  Пусть $\vp \notin \QCLFs$. Пусть
  $\kModel{M}^{rs} = \langle W', W' \times W', D', I'' \rangle$~--- модель, определённая в доказательстве импликации
  \mbox{$(3) \Rightarrow (1)$} леммы~\ref{wfin:lem:vp-B} для
  $\logic{QKTB}_{\mathit{wfin}}$. Модель $\kModel{M}^{rs}$ построена на
  конечной шкале Крипке с универсальным отношением достижимости и с конечной постоянной областью~$\mathcal{D}$. Как было показано, из того, что
  $\vp \notin \QCLFs$, следует, что
  $(\kModel{M}^{rs}, w^\ast) \not\models (\bar{\vp}^\sigma)'$. Напомним, что $\kModel{M}^{rs} \models \forall x\, P_{n+1} (x)$.

  Мы воспользуемся моделью $\kModel{M}^{rs}$, чтобы построить модель $\kModel{M}^\diamond$, определённую на $\logic{KTB}$-шкале и опровергающую формулу
  $\forall x\, \beta_{n+1} (x) \imp (\bar{\vp}^\sigma)^\diamond$.

  Для каждого~$k \in \{1, \ldots, n+1 \}$ определим шкалу Крипке
  $\frak{G}_k = \langle W_k,R_k\rangle$, где
  $$
  \begin{array}{rcl}
  W_k & = & \{ w_s^k : 0 \leqslant s \leqslant k\} \cup \{ v_s^{ki} : 1
  \leqslant i \leqslant k, 1 \leqslant s \leqslant 2i + 1\} \cup \{
  e^k \}
  \end{array}
  $$
  а $R_k$~--- рефлексивно-симметричное замыкание отношения
  $$
  \begin{array}{l}
    \{ \langle \mathstrut w^{\mathstrut k}_i, v_1^{\mathstrut k(i+1)} \rangle : 0 \leqslant i < k \}  \cup \{ \langle
  v^{ki}_{s}, v_{s+1}^{ki} \rangle : 1 \leqslant i \leqslant k, 1
  \leqslant s \leqslant 2i\} \cup {}
  \smallskip\\
  {} \cup \{ \langle v^{ki}_{2i + 1}, w_{i}^{k} \rangle : 1 \leqslant i
  \leqslant k\} \cup \{ \langle w^k_{k}, e^k \rangle \},
  \end{array}
  $$
  см.~рис.~\ref{wfin:fig2}.

  Для $w\in W'$, $a \in \mathcal{D}$ и
  $k \in \{1, \ldots, n +1\}$ определим шкалу Крипке
  $\kframe{G}_k^{wa} = \langle \{\langle w, a \rangle \}\times
  W_k^{\phantom{0}},R_k^{wa}\rangle$
  как изоморфную копию шкалы $\kframe{G}_k$ при изоморфизме
  $f\colon v \mapsto \langle w, a, v \rangle$.

  Пусть
  $$
  \begin{array}{rcl}
  W^\diamond
    & =
    & \displaystyle
      W' \cup \bigcup\limits_{k=1}^{n+1} (W' \times \mathcal{D} \times W_k).
  \end{array}
  $$
  Поскольку множества $W'$, $\mathcal{D}$ и $W_k$ (для любого $k \in \{1,\ldots,n+1\}$) конечны, множество~$W^\diamond$ тоже конечно.
  Пусть $R^\diamond$~--- рефлексивно-симметричное замыкание отношения
  $$
  (W' \times W')
  \cup\bigcup\limits_{k=1}^{n+1} 
  \big\{R_k^{wa} \cup \{\langle w, \langle w, a, w_0^k \rangle
  \rangle\} : w\in W',\ a \in \mathcal{D} \big\}.
  $$
  Для каждого $u\in W^\diamond$ положим $D^\diamond(u) = \mathcal{D}$.
  Для каждого $u\in W^\diamond$ и $a\in \mathcal{D}$ положим
  $$
  \begin{array}{lcl}
  \langle a \rangle \in I^\diamond(u,P)
  & \leftrightharpoons &
  \parbox[t]{250pt}{существуют $w\in
    W'$ и $k\in\{1,\ldots,n+1\}$, такие, что $(\kModel{M}^{rs},w)\models
    P_k(a)$ и либо $u = \langle w, a, e^k \rangle$, либо $u = \langle
    w, a, w_i^{k} \rangle$ для некоторого $i\in\{0,\ldots,k\}$,}
  \end{array}
  $$
  см.~рис.~\ref{wfin:fig2}.
  Пусть $\kModel{M}^\diamond = \langle
  W^\diamond,R^\diamond,D^\diamond,I^\diamond\rangle$.

  \begin{sublemma}
    \label{wfin:lem:model-M_k-QKTB}
    Для любых $u \in W^\diamond$ и $a \in \mathcal{D}$
    $$
    \begin{array}{lcl}
    (\kModel{M}^\diamond, u) \models \alpha_k(a)
    & \iff
    & \mbox{$u = \langle w, a, w^k_0 \rangle$ для некоторого $w \in W'$,}
    \\
    &
    & \mbox{такого, что $(\kModel{M}^{rs},w)\models P_k(a)$}.
    \end{array}
    $$
  \end{sublemma}

  \begin{proof}
    Заметим, что
    $$
  \begin{array}{lll}
    (\kModel{M}^\diamond, u) \models \Box^+ P(a)
      & \iff
      & \parbox[t]{225pt}{$u = \langle w, a, e^k \rangle$ для некоторых $k \in \{1, \ldots, n+1\}$ и $w \in W'$, таких, что $(\kModel{M}^{rs},w) \models P_k(a)$.}
  \end{array}
  $$
  Используя это наблюдение, индукцией по $j \in \{ 1, \ldots, k \}$ несложно показать, что
  $$
  \begin{array}{lll}
  (\kModel{M}^\diamond, u) \models \delta^k_{j}(a)
    & \iff
    & \mbox{$u = \langle w, a, v^{kj}_{j+1} \rangle$ и $(\kModel{M}^{rs},w)\models P_k(a)$.}
  \end{array}
  $$
  В частности,
  $$
    \begin{array}{lcl}
      (\kModel{M}^\diamond, u) \models \delta^k_1(a)
      & \iff &
      \mbox{$u = \langle w, a, v^{k1}_{2} \rangle$ и $(\kModel{M}^{rs},w)\models P_k(a)$.}
    \end{array}
    $$
    Значит,
    $(\kModel{M}^\diamond, u) \models P(a) \con \Diamond^2 \delta^k_1(a)$
    тогда и только тогда, когда $(\kModel{M}^{rs},w)\models P_k(a)$ и при этом либо
    $u = \langle w, a, w^k_{0} \rangle$, либо $u = \langle w, a, w^k_{1} \rangle$.
    Осталось заметить, что если $(\kModel{M}^{rs},w)\models P_k(a)$, то
      $(\kModel{M}^\diamond, \langle w, a, w^k_{0} \rangle) \not\models
      \Diamond^3 \delta^k_2(a)$ и $(\kModel{M}^\diamond, \langle w, a, w^k_{1} \rangle) \models
      \Diamond^3 \delta^k_2(a)$.
  \end{proof}

  Покажем, что
  $(\kModel{M}^\diamond, w^\ast) \not\models \forall x\, \beta_{n+1} (x) \imp
  (\bar{\vp}^\sigma)^\diamond$.

  Для этого сначала покажем, что для любых $w\in W'$,
  $a\in \mathcal{D}$ и $k\in\{1,\ldots,n+1\}$
  $$
  \begin{array}{lcl}
  (\kModel{M}^{rs},w)\models P_k(a)
    & \iff
    & (\kModel{M}^\diamond,w)\models \beta_k (a).
  \end{array}
  \eqno{(\ast)}
  $$

  Пусть $(\kModel{M}^{rs},w) \models P_k(a)$. По определению,
  $(\kModel{M}^\diamond, w) \not\models P(a)$ и
  $w R^\diamond \langle w, a, w^k_0 \rangle$. Значит, по подлемме~\ref{wfin:lem:model-M_k-QKTB},
  $(\kModel{M}^\diamond,w)\models \beta_k (a)$.

  Пусть теперь $(\kModel{M}^\diamond, w) \models \beta_k (a)$, т.е.
  $(\kModel{M}^\diamond, w) \models \neg P(a)$ и
  $(\kModel{M}^\diamond, w) \models \Diamond \alpha_k (a)$. По подлемме~\ref{wfin:lem:model-M_k-QKTB},
  $w R^\diamond \langle w, a, w^k_0 \rangle$. Значит,
  $\kModel{M}^{rs},w \models P_k(a)$ по определению модели~$\kModel{M}^\diamond$.

  Теперь покажем, что
  $$
  \begin{array}{clcl}
  (\kModel{M}^\diamond, u) \models \forall x\, \beta_{n+1} (x)
    & \iff
    & u \in W'.
  \end{array}
  \eqno{({\ast}{\ast})}
  $$

  Импликация $(\Leftarrow)$ следует из того, что
  $\kModel{M}^{rs} \models \forall x\, P_{n+1} (x)$, и условия~$(\ast)$.

  Докажем импликацию $(\Rightarrow)$. Пусть
  $(\kModel{M}^\diamond, u) \models \forall x\, \beta_{n+1} (x)$. Значит,
  $(\kModel{M}^\diamond, u) \not\models P(a)$ и
  $(\kModel{M}^\diamond, u) \models \Diamond \alpha_{n+1} (a)$ для каждого
  $a \in \mathcal{D}$.

  По подлемме~\ref{wfin:lem:model-M_k-QKTB},
  $(\kModel{M}^\diamond, u) \models \Diamond \alpha_{n+1} (a)$ тогда и только тогда, когда $u R^\diamond \langle w, a, w^{n+1}_0 \rangle$ и
  $(\kModel{M}^{rs}, w) \models P_{n+1}(a)$ для некоторого $w \in W'$. По определению, $u R^\diamond \langle w, a, w^{n+1}_0 \rangle$ тогда и только тогда, когда
  либо $u = w$ и $w \in W'$, либо
  $u = \langle w, a, w^{n+1}_0 \rangle$, либо
  $u = \langle w, a, v^{(n+1)1}_1 \rangle$. Покажем, что последние два случая невозможны.

  Согласно пункту~$(2)$ замечания~\ref{wfin:rem:B}, можно считать, что предметная область модели $\kModel{M}^{rs}$ содержит не менее двух элементов; пусть $b \in \mathcal{D} \setminus \set{a}$.  По определению,
  $(\kModel{M}^\diamond, \langle w, a, v^{(n+1)1}_1 \rangle) \not\models
  \Diamond \alpha_{n+1} (b)$,
  а значит,
  $(\kModel{M}^\diamond, \langle w, a, v^{({n+1})1}_1 \rangle) \not\models
  \forall x\, \beta_{n+1} (x)$.
  Следовательно, $u \ne \langle w, a, v^{({n+1})1}_1 \rangle$. Кроме того, поскольку
  $(\kModel{M}^\diamond, \langle w, a, w^{n+1}_0 \rangle) \models P(a)$,
  получаем, что $u \ne \langle w, a, w^{n+1}_0 \rangle$.

  Значит, $u \in W'$, и условие $({\ast}{\ast})$ доказано.

  Поскольку $w^\ast \in W'$, по $({\ast}{\ast})$ получаем, что
  $(\kModel{M}^\diamond, w^\ast) \models \forall x\, \beta_{n+1} (x)$.

  Остаётся показать, что
  $(\kModel{M}^\diamond, w^\ast) \not\models (\bar{\vp}^\sigma)^\diamond$.

  Для этого покажем, что для каждой подформулы
  $\theta$ формулы $\bar{\vp}^\sigma$, каждого мира $w\in W'$ и каждого приписывания~$g$
  $$
  \begin{array}{lcl}
  (\kModel{M}^{rs},w)\models^g \theta'
    & \iff
    & (\kModel{M}^\diamond,w)\models^g \theta^\diamond,
  \end{array}
  $$
  где $\theta^\diamond$~--- результат подстановки в формулу $\theta'$ формул
  $\beta_1(x),\ldots, \beta_{n+1}(x)$ вместо, соответственно,
  $P_1(x),\ldots,P_{n+1}(x)$.

  Доказательство проведём индукцией по построению~$\theta$.

  Если $\theta = P_k (x)$, то $\theta' = P_k (x)$ и
  $\theta^\diamond = \beta_k (x)$; тогда требуемое утверждение получается по~$(\ast)$.

  Если $\theta = \bot$, $\theta = \psi \con \chi$, $\theta = \psi \dis \chi$, $\theta = \psi \imp \chi$, $\theta = \forall x\, \psi$ или $\theta = \exists x\, \psi$, то обоснование шага индукции не содержит трудностей.

  Пусть $\theta = \Box \psi$. Тогда
  $$
  \begin{array}{lcr}
    \theta' = \Box(\forall x\, P_{n+1}(x) \imp
    \psi') & \mbox{и} & \theta^\diamond = \Box(\forall x\, \beta_{n+1}(x)
                            \imp \psi^\diamond).
  \end{array}
  $$

  Пусть $(\kModel{M}^\diamond,w)\not\models^g \theta^\diamond$. Тогда существует $v \in W^\diamond$, такой, что $wR^\diamond v$,
  $(\kModel{M}^\diamond, v) \models \forall x\, \beta_{n+1}(x)$ и
  $(\kModel{M}^\diamond, v) \not\models^g \psi^\diamond$. По
  $({\ast}{\ast})$ получаем, что $v \in W'$. Значит, по индукционному предположению,
  $(\kModel{M}^{rs}, v) \not\models^g \psi'$. Поскольку модель~$\kModel{M}^{rs}$ определена на шкале Крипке с универсальным отношением достижимости и
  $\kModel{M}^{rs} \models \forall x\, P_{n+1} (x)$, получаем, что
  $(\kModel{M}^{rs}, w) \not\models^g \theta'$.

  Пусть теперь $(\kModel{M}^{rs}, w) \not\models^g \theta'$, т.е.
  $(\kModel{M}^{rs}, v) \models \forall x\, P_{n+1}(x)$ и
  $(\kModel{M}^{rs}, v) \not\models^g \psi'$ для некоторого $v \in W'$. По
  $(\ast)$ получаем, что $(\kModel{M}^\diamond, v) \models \forall x\,
  \beta_{n+1}(x)$.
  По индукционному предположению,
  $(\kModel{M}^\diamond, v) \not\models^g \psi^\diamond$. Значит,
  $(\kModel{M}^\diamond,w)\not\models^g \theta^\diamond$.

  Таким образом,
  $(\kModel{M}^\diamond, w^\ast) \not\models \forall x\, \beta_{n+1} (x)
  \imp (\bar{\vp}^\sigma)^\diamond$.

  Значит,
  $\forall x\, \beta_{n+1} (x) \imp (\bar{\vp}^\sigma)^\diamond
  \notin \logic{QKTB}_{\mathit{wfin}}$,
  откуда, в частности, следует, что
  $\forall x\, \beta_{n+1} (x) \imp (\bar{\vp}^\sigma)^\diamond
  \notin \logic{QK}_{\mathit{wfin}}$.
\end{proof}

\begin{theorem}
  \label{wfin:thr:QKTB}
  Каждая логика из интервала\/
  $[\logic{QK}_{\mathit{wfin}}, \logic{QKTB}_{\mathit{wfin}}]$ является\/
  $\Pi^0_1$-трудной в языке с тремя предметными переменными и одной унарной предикатной буквой.
\end{theorem}

\begin{proof}
  По лемме~\ref{wfin:lem:vp-ast-QKTB}, функция
  $f\colon \vp \mapsto \forall x\, \beta_{n+1} (x) \imp
  (\bar{\vp}^\sigma)^\diamond$
  погружает $\Pi^0_1$-трудное множество $\QCLFs$
  в соответствующий фрагмент каждой логики из интервала
  $[\logic{QK}_{\mathit{wfin}}, \logic{QKTB}_{\mathit{wfin}}]$.
\end{proof}

В частности, получаем следующее утверждение.

\begin{corollary}
  \label{wfin:cor:QKTB}
  Логики\/ $\logic{QKB}_{\mathit{wfin}}$, $\logic{QKB.bf}_{\mathit{wfin}}$,
  $\logic{QKTB}_{\mathit{wfin}}$ являются\/ $\Pi^0_1$-трудными в языке с тремя предметными переменными и одной унарной предикатной буквой.
\end{corollary}

\subsubsection{Полнота в классе $\Pi^0_1$}

Используя предложение~\ref{prop:finite-class:eq}, утверждения теорем~\ref{wfin:thr:GL-Grz} и~\ref{wfin:thr:QKTB} можно усилить.
Прежде всего заметим, что из предложения~\ref{prop:finite-class:eq} вытекает следующая теорема.

\begin{theorem}
  \label{wfin:thr:Pi01}
  Пусть\/ $\scls{C}$~--- рекурсивно перечислимый класс конечных шкал Крипке. Тогда монадические фрагменты с равенством логик\/ $\mPlogicx{\scls{C}}{=}$, $\mPlogicCx{\scls{C}}{=}$, $\mPlogicx{\scls{C}}{\simeq}$ и\/ $\mPlogicCx{\scls{C}}{\simeq}$ находятся в классе\/ $\Pi^0_1$.
\end{theorem}

\begin{proof}
Покажем, что дополнения монадических фрагментов с равенством логик\/ $\mPlogicx{\scls{C}}{=}$, $\mPlogicCx{\scls{C}}{=}$, $\mPlogicx{\scls{C}}{\simeq}$ и\/ $\mPlogicCx{\scls{C}}{\simeq}$ рекурсивно перечислимы, откуда и будет следовать, что сами эти фрагменты находятся в классе~$\Pi^0_1$. Имеются два рекурсивных перечисления:
\begin{itemize}
\item перечисление $\varphi_1,\varphi_2,\varphi_3,\ldots$ всех $\lang{QML}^=$-формул;
\item перечисление $\kFrame{F}_1,\kFrame{F}_2,\kFrame{F}_3,\ldots$ всех шкал из класса~$\scls{C}$.
\end{itemize}
Чтобы построить рекурсивное перечисление дополнения к $\mPlogicx{\scls{C}}{=}$, достаточно последовательно проверять на $k$\nobreakdash-м шаге (где $k\in\numNp$), принадлежат ли формулы $\varphi_1,\ldots,\varphi_k$ логике $\mPlogicx{\set{\kFrame{F}_1,\ldots,\kFrame{F}_k}}{=}$, занося в последовательность те формулы, для которых проверка показала, что они не принадлежат этой логике. Такая проверка возможна, поскольку, согласно предложению~\ref{prop:finite-class:eq}, логика $\mPlogicx{\set{\kFrame{F}_1,\ldots,\kFrame{F}_k}}{=}$ разрешима. Аналогично поступаем с логиками $\mPlogicCx{\scls{C}}{=}$, $\mPlogicx{\scls{C}}{\simeq}$ и\/ $\mPlogicCx{\scls{C}}{\simeq}$.
\end{proof}

\begin{corollary}
\label{wfin:cor:Pi-0-1-eq-3}
Пусть $L$~--- одна из логик\/ $\logic{K}$, $\logic{T}$, $\logic{D}$, $\logic{KB}$, $\logic{K4}$, $\logic{S4}$, $\logic{KTB}$, $\logic{GL}$, $\logic{Grz}$, $\logic{wGrz}$. Тогда монадические фрагменты и монадические фрагменты с равенством логик\/ $\logic{Q}L^=_{\mathit{wfin}}$, $\logic{Q}L\logic{.bf}^=_{\mathit{wfin}}$, $\logic{Q}L^\simeq_{\mathit{wfin}}$, $\logic{Q}L\logic{.bf}^\simeq_{\mathit{wfin}}$, $\logic{Q}L_{\mathit{wfin}}\logic{.eq}$, $\logic{Q}L\logic{.bf}_{\mathit{wfin}}\logic{.eq}$ являются\/ $\Pi^0_1$\nobreakdash-полными, если их язык содержит хотя бы три предметные переменные.
\end{corollary}

\begin{proof}
Следует из предложения~\ref{prop:finite-class:eq} и, в зависимости от $L$, из теоремы~\ref{wfin:thr:GL-Grz} или теоремы~\ref{wfin:thr:QKTB}.
\end{proof}

Следствие~\ref{wfin:cor:Pi-0-1-eq-3}, в частности, означает, что ни одна из логик $\logic{Q}L^=_{\mathit{wfin}}$, $\logic{Q}L\logic{.bf}^=_{\mathit{wfin}}$, $\logic{Q}L^\simeq_{\mathit{wfin}}$, $\logic{Q}L\logic{.bf}^\simeq_{\mathit{wfin}}$, $\logic{Q}L_{\mathit{wfin}}\logic{.eq}$, $\logic{Q}L\logic{.bf}_{\mathit{wfin}}\logic{.eq}$ в полном языке не погружается рекурсивно ни в свой монадический фрагмент, ни в свой монадический фрагмент с равенством. Это следует из $\Sigma^0_1$\nobreakdash-трудности каждой из указанных логик, поскольку их безмодальные фрагменты совпадают с классической логикой предикатов, которая $\Sigma^0_1$\nobreakdash-полна.

\subsubsection{Замечания}
\label{wfin:sec:discussion}

Представленные в этом разделе результаты могут быть распространены на некоторые другие
логики, не рассмотренные здесь, включая ненормальные и квазинормальные модальные предикатные логики.

Однако имеются важные логики, на которые представленные здесь результаты не распространяются.
Так, теоремы~\ref{wfin:thr:GL-Grz} и~\ref{wfin:thr:QKTB} не затрагивают такие логики как $\logic{QGL.3}$, $\logic{QGrz.3}$, $\logic{QKD45}$, $\logic{QS5}$. Ниже мы покажем, как можно получить аналогичные результаты для некоторых других логик, применяя несколько иную технику.

    \subsection{Логики натурального ряда}
    \label{sec:NatNumbersFrame:Qmodal}

\subsubsection{Предварительные сведения}

В этом разделе под логикой натурального ряда будем понимать модальную предикатную логику шкалы Крипке $\otuple{\numN,\leqslant}$. Мы покажем, что монадический фрагмент этой логики в языке с двумя предметными переменными является $\Pi^1_1$\nobreakdash-трудным; более того, этот результат сохраняется, если язык содержит две переменные, одну унарную предикатную букву и одну пропозициональную букву.

Используя возникающие при этом конструкции, мы сможем получить сходные результаты для некоторых других логик.

Порядок построений такой: сначала мы сведём к проблеме выполнимости для рассматриваемой логики некоторую $\Sigma^1_1$-трудную проблему укладки домино~\cite[теорема~6.4]{Harel86}, потом, используя идею трюка Крипке~\cite{Kripke62}, промоделируем возникающие бинарные буквы унарными, а затем, используя идею, близкую к предложенной в~\cite{BS93} для моделирования пропозициональных переменных в пропозициональных логиках, промоделируем все унарные буквы с помощью одной унарной и одной пропозициональной.

\subsubsection{Моделирование проблемы укладки домино}
\label{nat:sec:reduction}

Будем использовать проблему укладки домино, сходную с той, которую мы рассматривали выше, но с одним дополнительным условием. Пусть, как и раньше,
\defnotion{плитки домино} имеют квадратную форму одинакового размера; для каждой плитки зафиксирована ориентация её сторон: указаны её \defnotion{верхняя}, \defnotion{нижняя}, \defnotion{правая} и \defnotion{левая} стороны. Каждая плитка имеет \defnotion{тип} $t$, определяемый \defnotion{цветами} $\leftsq t$, $\rightsq t$, $\upsq t$ и $\downsq t$ её сторон. \defnotion{Задача домино}\index{еяд@задача!домино} определяется набором $T = \{t_0, \ldots, t_{n}\}$ типов плиток домино, но теперь состоит в том, что нужно выяснить, существует ли \defnotion{$T$-укладка},\index{укладка плиток домино} т.е. функция $f\colon \mathds{N} \times \mathds{N} \to T$, такая, что для любых $i, j \in \mathds{N}$
\begin{itemize}
\item[]$(T_1)$\quad $\rightsq f(i,j) = \leftsq f(i+1,j)$;
\item[]$(T_2)$\quad $\upsq f(i,j) = \downsq f(i,j+1)$,
\end{itemize}
при этом
\begin{itemize}
\item[]$(T_3)$\quad $\{ j \in \mathds{N} : f(0, j) = t_0 \}$~--- бесконечное множество,
\end{itemize}
т.е. плитка типа $t_0$ встречается в первом (имеющем номер~$0$) вертикальном ряду укладки бесконечно много раз. Известно, что эта проблема укладки домино $\Sigma^1_1$\nobreakdash-полна~\cite[теорема~6.4]{Harel86}.

Идея кодирования, которое будет использовано, основана на работе~\cite[теорема~2]{HWZ00} (см.~также~\cite[теорема~11.1]{GKWZ}; похожие конструкции использовались и в других работах~\cite{Spaan-1993-1,Marx99,WZ01,KKZ05}), но приводимая ниже конструкция не использует модальностей типа~$\Next$ (см.~раздел~\ref{s4-4-2}), позволяя тем самым получить более сильные результаты.

Пусть $\triangleleft$~--- бинарная предикатная буква, $M$ и $P_t$ (для каждого $t \in T$)~--- унарные предикатные буквы, $p$~--- пропозициональная буква.

Для формулы $\vp$ в языке с перечисленными буквами положим
$$
\begin{array}{lcll}
  \pDiamond\varphi & = & \Diamond(p \con\Diamond( \neg p \con\varphi));
  \\
  \pDiamond^0 \varphi & = & \varphi;
  \\
  \pDiamond^{n+1} \varphi & = & \pDiamond^{\phantom{n}}\pDiamond^n\varphi
  & \mbox{для каждого $n \in \nat$;}
  \\
  \pDiamond^{=n} \varphi & = & \pDiamond^n\varphi\wedge\neg\pDiamond^{n+1}\varphi
  & \mbox{для каждого $n \in \nat$.}
\end{array}
$$
Модальность $\pDiamond$, в отличие от $\Diamond$, в моделях Крипке вынуждает делать переход в другой мир, чтобы в текущем мире формула $\pDiamond \vp$ была истинна. Для более точного описания для модели $\kModel{M} = \otuple {\nat,\leqslant,D,I}$ определим бинарное отношение
$R_{\scriptsize\pDiamond}$ на $\nat$, положив для всяких $w,v\in \nat$
$$
\begin{array}{lcl}
w R_{\scriptsize\pDiamond} v
  & \leftrightharpoons
  & \mbox{$v \not\models p$ и существует $u \in \nat$, такой, что $w \leqslant u \leqslant v$ и $(\kModel{M},u) \models p$.}
\end{array}
$$
Получаем, что $R_{\scriptsize\pDiamond}$~--- иррефлексивное и транзитивное отношение. Отметим, что $R_{\scriptsize\pDiamond}$ зависит от $\kModel{M}$, но ниже всегда будет ясно, для какой именно модели определяется такое отношение.

Для удобства введём следующее сокращение:
$$
\begin{array}{lcl}
  U (x) & = & \displaystyle\bigwedge\limits_{t \in T} \neg P_t(x).
\end{array}
$$

Определим формулы, описывающие укладку домино:
$$
\begin{array}{rcl}
  A_0 & = & \exists x\,  \Box\, U (x); \\
  A_1 & = &  \exists x\, (\neg U (x) \con M(x)); \medskip\\
  A_2 & = & \forall x \exists y\,  (x \triangleleft y); \medskip\\
  A_3 & = & \forall x \forall y\,  (x \triangleleft y \imp \Box (
            \exists x\, M(x) \imp x
            \triangleleft y)); \medskip\\
  A_4 & = & \forall x \forall y\, (  x \triangleleft y \imp \Box (
            M(x) \equivalence \neg p \con\pDiamond^{=1} M(y)
            ); \medskip\\
  A_5 & = & \displaystyle
            \forall x \forall y\, \Box \bigwedge\limits_{t \in T}
            \big( M(x) \con P_t (y) \imp \Box (M(x) \imp P_{t} (y)) \big) ; \medskip\\
  A_6 & = & \displaystyle
            \forall x\, \Box \bigwedge\limits_{t \in T}    (P_t(x) \imp
            \bigwedge\limits_{\mathclap{t' \ne t}} \neg P_{t'}(x)); \medskip\\
  A_7 & = & \displaystyle
            \forall x \forall y\, \Box  \bigwedge\limits_{t \in T}( x \triangleleft y
            \con P_t(x) \imp  \bigvee\limits_{\mathclap{\mathop{\scriptsize\rightsquare{0.21}} t\,{=}\,\mathop{\scriptsize\leftsquare{0.21}} t'}} P_{t'} (y)) ; \medskip\\
  A_8 & = & \displaystyle
            \forall x \forall y\, \Box \bigwedge\limits_{t \in T} \big(
            M(x) \con P_t (y) \imp \Box (\exists y\, (x
            \triangleleft y \con M(y)) \imp \bigvee\limits_{\mathclap{\mathop{\scriptsize\upsquare{0.21}} t\,{=}\,\mathop{\scriptsize\downsquare{0.21}} t'}} P_{t'} (y)) \big); \medskip\\
  A_9 & = & \forall x\, (M(x) \imp \Box \pDiamond P_{t_0} (x)).
\end{array}
$$

Пусть $A=A_0\wedge\ldots\wedge A_{9}$. Заметим, что формула~$A$
содержит только две предметные переменные.

Содержательно можно понимать $\triangleleft$ как \defnotion{отношение следования} в области ${D}_0$ мира $0$, где формула $A$ оказалась истинной. Элемент $a \in {D}_0$, для которого в мире $w$ истинно $M(a)$, можно понимать как \defnotion{метку} мира~$w$.

\begin{lemma}
  \label{nat:lem:tiling-reduction}
  Для множества $T$ типов плиток домино верно следующее:
  $$
  \begin{array}{lcl}
  \langle \nat, \leqslant \rangle \not\models \neg A
    & \iff
    & \mbox{существует $T$-укладка с условиями $(T_1)$--$(T_3)$.}
  \end{array}
  $$
\end{lemma}

\begin{proof}
  $(\Rightarrow)$
  Пусть $(\kModel{M}, w_0) \models A$ для некоторой модели Крипке
  $\kModel{M} = \otuple{\nat, \leqslant, D, I}$ и некоторого мира
  $w_0 \in \nat$. Без ограничений общности можем (и будем) считать, что~$w_0=0$.

  Поскольку $(\kModel{M},0) \models A_1$, существует $a_0 \in D_0$, такой, что
  $(\kModel{M},0) \not\models U (a_0)$ и $(\kModel{M},0) \models M(a_0)$.  Поскольку
  $(\kModel{M},0) \models A_2$, получаем бесконечную последовательность
  $a_0, a_1, a_2, \ldots$ элементов области $D_0$, такую, что
  $a_k \triangleleft^{I,0} a_{k+1}$ для каждого $k\in\numN$. Поскольку $(\kModel{M},0) \models A_3$, получаем, что
  $a_0 \triangleleft^{I,w} a_1 \triangleleft^{I,w} a_2
  \triangleleft^{I,w} \ldots$ для каждого $w \in \nat$, такого, что
  $(\kModel{M},w) \models \exists x\, M(x)$.

  Поскольку $(\kModel{M},0) \models A_4$, получаем, что для любых $w,n \in \nat$,
  $$
  \begin{array}{lcl}
    (\kModel{M},w) \models M(a_n) & \iff  &
    \left\{
         \begin{array}{ll}
             \bullet &
             (\kModel{M},w\phantom{'}) \not\models p;
               \smallskip\\
             \bullet &
             \mbox{$(\kModel{M},w') \models M(a_{n+1})$}
             \\
             & \mbox{для некоторого $w' \in R_{\scriptsize\pDiamond}(w)$;}
               \smallskip\\
             \bullet &
             \mbox{$w'' \in R_{\scriptsize\pDiamond}^2 (w)$ $\imply$ $w'' \not\models M(a_{n+1})$.}
         \end{array}
    \right.
  \end{array}
  \eqno (1)
  $$
  Значит, метка меняется с $a_n$ на $a_{n+1}$, если мы проходим через непустой блок миров, в каждом из которых верно $p$, попадая в мир, где $p$ опровергается.

  Покажем, что метка не меняется в мирах, пока мы не достигнем мир, в котором верно $p$, т.е. что для любых $u, u', n \in \nat$,
  $$
  \begin{array}{l}
    \mbox{если $u\models M(a_n)$, $u \not\models p$, $u' \not\models p$,
    и нет такого $v$, что $u \lessgtr v \lessgtr u'$} \\
    \mbox{и $(\kModel{M},v) \models p$, то $(\kModel{M},u') \models M(a_n)$.}
  \end{array}
  \eqno (2)
  $$
  Пусть $(\kModel{M},u) \models M(a_n)$, $(\kModel{M},u) \not\models p$,
  $(\kModel{M},u') \not\models p$ и не существует такого $v$, что
  $u \lessgtr v \lessgtr u'$ и $(\kModel{M},v) \models p$. Тогда, по $(1)$, существует мир
  $w' \in R_{\scriptsize\pDiamond} (u)$, такой, что
  $(\kModel{M},w') \models M(a_{n+1})$, а также $(\kModel{M},w'') \not\models M(a_{n+1})$ для некоторого
  $w'' \in R^2_{\scriptsize\pDiamond} (u)$.
  Из сделанного допущения следует, что для любых
  $w, n \in \nat$,
  $$
  \begin{array}{lcl}
  w \in R^n_{\scriptsize\pDiamond} (u)
    & \iff
    & w \in R^n_{\scriptsize\pDiamond} (u').
  \end{array}
  $$
  Поэтому из того, что $w' \in R_{\scriptsize\pDiamond} (u')$ и
  $w'' \in R^2_{\scriptsize\pDiamond} (u')$ можно сделать вывод, что
  $(\kModel{M},w'') \not\models M(a_{n+1})$. Поскольку
  $u' \not\models p$, по $(1)$ получаем, что $(\kModel{M},u') \models M(a_n)$.

  Теперь покажем, что метка в каждом мире единственна, т.е. что для каждых
  $w,n \in \nat$ и каждого $j \in \nat^+$
  $$
  \begin{array}{lcl}
  (\kModel{M},w) \models M(a_n)
    & \imply
    & (\kModel{M},w) \not\models M(a_{n+j}).
  \end{array}
  \eqno (3)
  $$
  Пусть $(\kModel{M},w) \models M(a_n)$. Согласно $(1)$, существует
  $w' \in R^{j+1}_{\scriptsize\pDiamond} (w)$, такой, что
  $(\kModel{M},w') \models M(a_{n+j+1})$. Поскольку $j \geqslant 1$ и
  отношение $R_{\scriptsize\pDiamond}$ транзитивно, получаем, что
  $w' \in R_{\scriptsize\pDiamond}^{2} (w)$. Значит, по $(1)$,
  $(\kModel{M},w) \not\models M(a_{n+j})$.

  Теперь покажем, что для любых $w, m, n \in \nat$,
  $$
  \begin{array}{lcl}
  (\kModel{M},w) \models M(a_m)
    & \imply
    & \mbox{$(\kModel{M},w) \models P_t (a_n)$ для некоторого типа $t \in T$.}
  \end{array}
  \eqno (4)
  $$

  Используем индукцию по~$m$.

  Как мы уже видели, $(\kModel{M},0) \not\models U(a_0)$, т.е.
  $(\kModel{M},0) \models P_t (a_0)$ для некоторого $t \in T$. Поскольку $(\kModel{M},0) \models A_7$, получаем, что
  для каждого $n \in \nat$ существует $t \in T$, такой, что
  $(\kModel{M},0) \models P_t (a_n)$. Поскольку $(\kModel{M},0) \models A_5$, получаем, что для любых
  $w, v, m, n \in \nat$ и $t \in T$
  $$
  \begin{array}{lcl}
  \left.
  \begin{array}{l}
  (\kModel{M},w) \models M(a_m) \\
  (\kModel{M},v\hfill) \models M(a_m) \\
  (\kModel{M},w) \models P_t(a_n)
  \end{array}
  \right\}
     & \imply
     & \mbox{$(\kModel{M},v) \models P_t(a_n)$.}
  \end{array}
  \eqno (5)
  $$
  Значит, $(4)$ выполняется при $m = 0$.

  Пусть теперь условие $(4)$ выполняется для некоторого $m \geqslant 0$ и пусть
  $(\kModel{M},w) \models M(a_{m+1})$. Покажем, что существует мир $w'$, такой, что $w' < w$ и $(\kModel{M},w') \models M(a_m)$. Для этого сначала заметим, что существует $w'$, такой, что $(\kModel{M},w') \not\models p$
  и $w \in R_{\scriptsize\pDiamond} (w')$: в противном случае, согласно~$(2)$,
  $(\kModel{M},w) \models M(a_0)$, что противоречит условию~$(3)$.
  Зафиксируем такой мир~$w'$.
  Теперь покажем, что $w'' \in R^2_{\scriptsize\pDiamond} (w')$ влечёт, что
  $(\kModel{M},w'') \not\models M(a_{m+1})$. Предположим, что
  $w'' \in R^2_{\scriptsize\pDiamond} (w')$. Тогда
  $w'' \in R_{\scriptsize\pDiamond} (w)$. Поскольку
  $(\kModel{M},w) \models M(a_{m+1})$, по $(1)$ и $(2)$ получаем, что $(\kModel{M},w'') \models M(a_{m+2})$, а значит, по $(3)$, $(\kModel{M},w'') \not\models M(a_{m+1})$. Наконец, поскольку
  $(\kModel{M},w') \not\models p$, по $(1)$ получаем, что $(\kModel{M},w') \models M(a_m)$, и тем самым требуемое обосновано.

  Пусть теперь имеется $n \in \nat$. По индукционному предположению, существует тип $t$, такой, что $(\kModel{M},w') \models P_t(a_n)$. Поскольку $(\kModel{M},0) \models A_8$,
  получаем, что $(\kModel{M},w) \models P_{t'}(a_n)$ для некоторого $t' \in T$. Значит, условие $(4)$ доказано.

  Учитывая $(5)$, для каждого $m \in \nat$ можно выбрать произвольный мир, имеющий метку $a_m \in D_0$, сделав его частью укладки домино. Для определённости пусть для каждого~$m \in \nat$,
  $$
  \begin{array}{lcl}
  w_m & = & \min \set{ w \in \nat : (\kModel{M},w) \models M(a_m) }.
  \end{array}
  $$

  Согласно $(4)$, для любых $n, m \in \nat$ существует тип $t \in T$, такой, что $(\kModel{M},w_m) \models P_t (a_n)$; из того, что $(\kModel{M},0) \models A_6$, получаем, что такой тип $t$ является единственным. Значит, мы можем определить функцию
  $\function{f}{\nat \times \nat}{T}$, положив
  $$
  \begin{array}{lcl}
    f(n, m) = t
    & \bydef & (\kModel{M},w_m) \models P_t(a_n).
  \end{array}
  $$

  Покажем, что $f$ удовлетворяет условиям $(T_1)$--$(T_3)$.

  Поскольку $(\kModel{M},0) \models A_7$, условие $(T_1)$ выполнено.

  Чтобы показать, что выполняется $(T_2)$, предположим, что $f(n, m) = t$. Тогда
  $(\kModel{M},w_m) \models P_t (a_n)$ согласно определению~$f$. Из определения $w_m$ следует, что $(\kModel{M},w_m) \models M(a_m)$. Поскольку $(\kModel{M},0) \models A_8$,
  из того, что $v \geqslant w_m$ и $(\kModel{M},v) \models M(a_{m+1})$, следует, что $(\kModel{M},v) \models P_{t'} (a_n)$ для некоторого типа $t'$, удовлетворяющего условию $\upsq t = \downsq t'$. Покажем, что $w_{m+1} \geqslant w_m$. Предположим, что это не так, т.е.
  $w_{m+1} < w_m$. Из определения $w_{m+1}$ получаем, что
  $(\kModel{M},w_{m+1}) \models M(a_{m+1})$. Тогда, по $(2)$ и $(3)$,
  $(\kModel{M},w_m) \in R_{\scriptsize\pDiamond} (w_{m+1})$. Поскольку
  $(\kModel{M},w_m) \models M(a_m)$, существует, согласно $(1)$,
  $w' \in R_{\scriptsize\pDiamond}^2 (w_m)$, такой, что
  $(\kModel{M},w') \models M(a_{m+2})$. Поскольку отношение $R_{\scriptsize\pDiamond}$ транзитивно, получаем, что $w' \in R_{\scriptsize\pDiamond}^2 (w_{m+1})$, а это противоречит третьему пункту условия~$(1)$. Значит, $w_{m+1} \geqslant w_m$. Следовательно,
  $(\kModel{M},w_{m+1}) \models P_{t'} (a_n)$ для некоторого $t'$, удовлетворяющего условию $\upsq t = \downsq t'$. Таким образом, $(T_2)$ выполняется.

  Осталось показать, что выполняется~$(T_3)$. Поскольку
  $(\kModel{M},0) \models M(a_0)$ и $(\kModel{M},0) \models A_9$, множество
  $\set{ w \in \nat : \mbox{$(\kModel{M},w) \not\models p$ и $(\kModel{M},w) \models P_{t_0}(a_0)$}}$ является бесконечным. Учитывая, что $(\kModel{M},0) \models M(a_0)$, а также $(1)$ и $(2)$, получаем, что для каждого $w \in \nat$, такого, что $(\kModel{M},w) \not\models p$, существует $m \in \nat$, для которого $(\kModel{M},w) \models M(a_m)$. Поэтому, согласно $(5)$, множество $\set{ w_m : \mbox{$m \in \nat$ и $(\kModel{M},w_m) \models P_{t_0} (a_0)$}}$ бесконечно. Значит, $(T_3)$ выполняется.

  Таким образом, $f$~--- требуемая функция.

\begin{figure}
  \centering
  \begin{tikzpicture}[scale=1.25]

\coordinate (w0)   at (1, 1.0);
\coordinate (w1)   at (1, 2.50);
\coordinate (w2)   at (1, 4.00);
\coordinate (w3)   at (1, 5.50);
\coordinate (w0')  at (1, 1.75);
\coordinate (w1')  at (1, 3.25);
\coordinate (w2')  at (1, 4.75);
\coordinate (w3')  at (1, 6.25);

\coordinate (dots) at (1, 7.00);

\coordinate (a)    at (1, 0.25);

\draw [] (w0)  circle [radius=2.0pt] ;
\draw [] (w1)  circle [radius=2.0pt] ;
\draw [] (w2)  circle [radius=2.0pt] ;
\draw [] (w3)  circle [radius=2.0pt] ;
\draw [] (w0') circle [radius=2.0pt] ;
\draw [] (w1') circle [radius=2.0pt] ;
\draw [] (w2') circle [radius=2.0pt] ;
\draw [] (w3') circle [radius=2.0pt] ;

\draw [] (dots) node {$\vdots$};

\begin{scope}[>=latex]
\draw [->, shorten >= 2.750pt, shorten <= 2.750pt] (w0)  -- (w0');
\draw [->, shorten >= 2.750pt, shorten <= 2.750pt] (w0') -- (w1) ;
\draw [->, shorten >= 2.750pt, shorten <= 2.750pt] (w1)  -- (w1');
\draw [->, shorten >= 2.750pt, shorten <= 2.750pt] (w1') -- (w2) ;
\draw [->, shorten >= 2.750pt, shorten <= 2.750pt] (w2)  -- (w2');
\draw [->, shorten >= 2.750pt, shorten <= 2.750pt] (w2') -- (w3) ;
\draw [->, shorten >= 2.750pt, shorten <= 2.750pt] (w3)  -- (w3');
\end{scope}

\node [left=5pt] at (w0)   {$0$}     ;
\node [left=5pt] at (w0')  {$1$}     ;
\node [left=5pt] at (w1)   {$2$}     ;
\node [left=5pt] at (w1')  {$3$}     ;
\node [left=5pt] at (w2)   {$4$}     ;
\node [left=5pt] at (w2')  {$5$}     ;
\node [left=5pt] at (w3)   {$6$}     ;
\node [left=5pt] at (w3')  {$7$}     ;

\node [right=5pt] at (w0)  {$M(0)$}  ;
\node [right=5pt] at (w1)  {$M(1)$}  ;
\node [right=5pt] at (w2)  {$M(2)$}  ;
\node [right=5pt] at (w3)  {$M(3)$}  ;
\node [right=5pt] at (w0') {$p$}     ;
\node [right=5pt] at (w1') {$p$}     ;
\node [right=5pt] at (w2') {$p$}     ;
\node [right=5pt] at (w3') {$p$}     ;

\node [right = 40pt +10pt] at (w0) {$P_{f(0,0)}(0)$};
\node [right = 95pt +20pt] at (w0) {$P_{f(1,0)}(1)$};
\node [right = 150pt+30pt] at (w0) {$P_{f(2,0)}(2)$};
\node [right = 205pt+40pt] at (w0) {$P_{f(3,0)}(3)$};
\node [right = 265pt+50pt] at (w0) {$\ldots$}       ;

\node [right = 40pt +10pt] at (w1) {$P_{f(0,1)}(0)$};
\node [right = 95pt +20pt] at (w1) {$P_{f(1,1)}(1)$};
\node [right = 150pt+30pt] at (w1) {$P_{f(2,1)}(2)$};
\node [right = 205pt+40pt] at (w1) {$P_{f(3,1)}(3)$};
\node [right = 265pt+50pt] at (w1) {$\ldots$}       ;

\node [right = 40pt +10pt] at (w2) {$P_{f(0,2)}(0)$};
\node [right = 95pt +20pt] at (w2) {$P_{f(1,2)}(1)$};
\node [right = 150pt+30pt] at (w2) {$P_{f(2,2)}(2)$};
\node [right = 205pt+40pt] at (w2) {$P_{f(3,2)}(3)$};
\node [right = 265pt+50pt] at (w2) {$\ldots$}       ;

\node [right = 40pt +10pt] at (w3) {$P_{f(0,3)}(0)$};
\node [right = 95pt +20pt] at (w3) {$P_{f(1,3)}(1)$};
\node [right = 150pt+30pt] at (w3) {$P_{f(2,3)}(2)$};
\node [right = 205pt+40pt] at (w3) {$P_{f(3,3)}(3)$};
\node [right = 265pt+50pt] at (w3) {$\ldots$}       ;

\node [right = 60pt +10pt] at (dots) {$\vdots$};
\node [right = 115pt+20pt] at (dots) {$\vdots$};
\node [right = 170pt+30pt] at (dots) {$\vdots$};
\node [right = 225pt+40pt] at (dots) {$\vdots$};

\node [right =  40pt + 20pt +10pt] at (a) {$0$};
\node [right =  65pt + 20pt +15pt] at (a) {$\lhd$};
\node [right =  95pt + 20pt +20pt] at (a) {$1$};
\node [right = 120pt + 20pt +25pt] at (a) {$\lhd$};
\node [right = 150pt + 20pt +30pt] at (a) {$2$};
\node [right = 175pt + 20pt +35pt] at (a) {$\lhd$};
\node [right = 205pt + 20pt +40pt] at (a) {$3$};
\node [right = 245pt + 20pt +50pt] at (a) {$\ldots$}       ;

\end{tikzpicture}

\caption{Модель $\kModel{M}_0$}
  \label{nat:fig1}
\end{figure}

$(\Leftarrow)$
Пусть $f$~--- $T$-укладка, удовлетворяющая условиям $(T_1)$--$(T_3)$. Построим модель на шкале $\otuple{\nat,\leqslant}$, в корне которой истинна формула~$A$.

  Пусть $\mathcal{D} = \nat \cup \{-1\}$ и $D(w) = \mathcal{D}$ для каждого $w \in \nat$.

  Пусть $\kModel{M}_0 = \otuple{\nat,\leqslant,D,I}$~--- модель, в которой для каждого $w \in \nat$ и любых $a,b\in \mathcal{D}$
  $$
  \begin{array}{lcl}
    (\kModel{M}_0,w) \models a\triangleleft b
      & \bydef
      & \mbox{$w$ чётно и $b=a+1$;}
      \smallskip \\
    (\kModel{M}_0,w) \models p
      & \bydef
      & \mbox{$w$ нечётно;}
      \smallskip \\
    (\kModel{M}_0,w) \models M(a)
      & \bydef
      & \mbox{$w=2a$;}
      \smallskip \\
    (\kModel{M}_0,w) \models P_t(a)
      & \bydef
      & \mbox{существует $m\in\nat$, для которого}
      \\
      &&\mbox{$w=2m$ и $f(a,m) = t$.}
  \end{array}
  $$

  Тогда несложная проверка показывает, что $(\kModel{M}_0, 0) \models A$.
\end{proof}


\subsubsection{Элиминация бинарной предикатной буквы}
\label{nat:sec:elim-binary}

Для элиминации бинарной предикатной буквы $\triangleleft$ в формуле $A$ будем использовать незначительную модификацию трюка Крипке~\cite{Kripke62}; как мы видели (раздел~\ref{sec:KripkeTrick:QModal}), количество предметных переменных при этом не увеличивается.

Ниже для упрощения обозначений будем считать, что $A$ содержит унарные предикатные буквы $P_0, \ldots, P_s$ вместо $P_{t_0}, \ldots, P_{t_s}$ соответственно.





Пусть $P_{s+1}$ и $P_{s+2}$~--- новые унарные предикатные буквы; определим функцию~$\cdot'$ как подстановку формулы $\pDiamond (P_{s+1} (x) \con P_{s+2} (y))$ вместо~$x \triangleleft y$.

\begin{lemma}
  \label{nat:lem:Kripke}
  Для множества $T$ типов плиток домино верно следующее:
  $$
  \begin{array}{lcl}
  \langle \nat, \leqslant \rangle \not\models \neg A'
    & \iff
    & \mbox{существует $T$-укладка с условиями $(T_1)$--$(T_3)$.}
  \end{array}
  $$
\end{lemma}

\begin{proof}
  $(\Rightarrow)$
  Пусть $(\kModel{M}, w_0) \models A'$ для некоторой модели Крипке $\kModel{M} = \otuple{\nat,\leqslant,D,I}$ и некоторого мира $w_0$; можем (и~будем) считать, что~$w_0=0$.

  По существу дальше доказательство такое же, как в части $(\Rightarrow)$ доказательства леммы~\ref{nat:lem:tiling-reduction}. Единственное несущественное отличие состоит в том, что вместо $x \triangleleft y$ надо теперь рассматривать формулу
  $\pDiamond (P_{s+1} (x) \con P_{s+2} (y))$: для каждого $w \in \nat$ отношение
  $I(w, \triangleleft)$ заменяется отношением
  $$
  \set{ \otuple{a,b} \in D_w \times D_w : (\kModel{M}, w) \models
  \pDiamond (P_{s+1} (a) \con P_{s+2} (b)) }.
  $$

  $(\Leftarrow)$
  Пусть $f$~--- $T$-укладка, удовлетворяющая условиям $(T_1)$--$(T_3)$. Пусть $\kModel{M}_0 = \otuple{\nat, \leqslant, D, I}$~--- модель, определённая в части $(\Leftarrow)$ доказательства леммы~\ref{nat:lem:tiling-reduction}. Как мы видели,
  $(\kModel{M}_0, 0) \models A$. Используя модель $\kModel{M}_0$, построим модель, в которой опровергается формула~$A'$.

  Пусть $\alpha$~--- бесконечная последовательность
  $$
  0, ~ 0, 1, ~ 0, 1, 2, ~ 0, 1, 2, 3,  ~ 0, 1, 2, 3, 4, ~ 0, 1, 2, 3, 4, 5,~  \ldots
  $$
  и пусть $\alpha_k$~--- $k$-й элемент последовательности~$\alpha$, где $k \in \nat$.

  Пусть $\kModel{M}'_0 = \otuple{\nat, \leqslant, D, I'}$~--- модель Крипке, определённая следующим образом: для любых~$w,c\in\nat$
  $$
  \begin{array}{lcl}
    (\kModel{M}'_0, w) \models P_{s+1} (c)
    & \bydef &
    \mbox{существует $m\in\nat$, такое, что $w=2m$ и $c=\alpha_m$;}
    \smallskip \\
    (\kModel{M}'_0, w) \models P_{s+2} (c)
    & \bydef &
    \mbox{существует $m\in\nat$, такое, что $w=2m$ и $c=\alpha_m+1$,}
  \end{array}
  $$
  а также для любого $w \in \nat$ и любого $S \in \{P_0, \ldots, P_s, M, p\}$,
  $$
  \begin{array}{lcl}
  I'(w,S) & = & I(w,S).
  \end{array}
  $$

  Покажем, что $(\kModel{M}'_0, 0) \models A'$.

  Поскольку $(\kModel{M}_0, 0) \models A$, достаточно показать, что для любых $m \in \nat$ и $a, b \in \mathcal{D}$
  $$
  \begin{array}{lcl}
  (\kModel{M}_0, 2m) \models a \triangleleft b
    & \iff
    & (\kModel{M}'_0, 2m) \models \pDiamond (P_{s+1}(a) \con P_{s+2}(b)).
  \end{array}
  $$

  Пусть $(\kModel{M}_0, 2m) \models a \triangleleft b$. Тогда
  $b = a + 1$, согласно определению модели~$\kModel{M}_0$. Выберем $k \in \nat$ так, что
  $k > m$ и $\alpha_k = a$; согласно определению $\alpha$, такое число~$k$ существует. Из определения модели $\kModel{M}'_0$ следует, что
  $(\kModel{M}'_0, 2k) \not\models p$ и
  $(\kModel{M}'_0, 2k) \models P_{s+1} (a) \con P_{s+2} (b)$. Кроме того, $(\kModel{M}'_0, 2k-1) \models p$. Значит,
  $(\kModel{M}'_0, 2m) \models \pDiamond (P_{s+1} (a) \con P_{s+2} (b))$.

  Пусть теперь
  $(\kModel{M}'_0, 2m) \models \pDiamond (P_{s+1} (a) \con P_{s+2} (b))$.
  Тогда существует $v > 2m$, для которого $(\kModel{M}'_0, v) \not\models p$ и
  $(\kModel{M}'_0, v) \models P_{s+1} (a) \con P_{s+2} (b)$. По определению модели $\kModel{M}'_0$ получаем, что $(\kModel{M}_0, v) \not\models p$; значит $v = 2k$ для некоторого $k > m$.
  Из определения модели $\kModel{M}'_0$ также следует, что $a = \alpha_k$ и
  $b = \alpha_k + 1$; значит, $b = a + 1$. Но тогда
  $(\kModel{M}_0, 2m) \models a \triangleleft b$ согласно определению модели~$\kModel{M}_0$.
\end{proof}

\subsubsection{Элиминация унарных предикатных букв}
\label{nat:sec:elim-monadic}

Промоделируем вхождения букв $p, M, P_0, \ldots, P_{s+2}$ в формуле~$A'$ с помощью одной унарной предикатной и одной пропозициональной буквы, не увеличивая при этом число предметных переменных.

Пусть $P$~--- новая унарная предикатная буква, а $q$~--- новая пропозициональная буква.

Для произвольной формулы $\vp$ положим
$$
\begin{array}{lcll}
  \PDiamond\varphi
    & =
    & \Diamond(\forall x\, P(x) \con \Diamond( \neg \forall x\, P(x) \con\varphi));
      \medskip\\
  \PDiamond^0 \varphi
    & =
    & \varphi;
      \medskip\\
  \PDiamond^{n+1} \varphi
    & =
    & \PDiamond^{\phantom{n}}\PDiamond^n\varphi & \mbox{для каждого $n \in \nat$;}
      \medskip\\
  \PDiamond^{=n} \varphi
    & =
    & \PDiamond^{n}\varphi\wedge\neg\PDiamond^{n+1}\varphi & \mbox{для каждого $n \in \nat$,}
\end{array}
$$
а также для каждого $n \in \{0, \ldots, s + 2\}$ положим
$$
\begin{array}{rcl}
  \beta_n (x)
    & =
    & \exists y\,\big(\PDiamond^{=s+4}(q\con P(y))\con 
      \PDiamond^{\phantom{n}}(\PDiamond^{=n+1}(q\con P(y))\con P(x))\big);
      \medskip\\
  \beta_n (y)
    & =
    & \exists x\,\big(\PDiamond^{=s+4}(q \con P(x))\con 
      \PDiamond^{\phantom{n}}(\PDiamond^{=n+1}(q\con P(x))\con P(y))\big).
\end{array}
$$

Для каждого $i\in\set{1,2,3,5,6,7,8}$ определим формулу $A_i^\ast$ как результат одновременной замены в формуле~$A'_i$
\begin{itemize}
\item $P_n(x)$ на $\beta_n (x)$ для каждого $n \in \{0, \ldots, s + 2\}$;
\item $P_n(y)$ на $\beta_n (y)$ для каждого $n \in \{0, \ldots, s + 2\}$;
\item $M(x)$ на $q \con P(x)$;
\item $M(y)$ на $q \con P(y)$;
\end{itemize}
пусть также
$$
\begin{array}{lcl}
  A_4^{\ast}
   & =
   & \forall x \forall y\, \big(\PDiamond^{\phantom{n}} (\beta_{s+1} (x) \con \beta_{s+2} (y) ) \imp {}
     \smallskip\\
   &
   & {}~~\,\imp
     \Box \big(q \con P(x) \equivalence 
     \neg \forall x\,P(x)\con
     \PDiamond^{=s+4} ( q \con P(y)) \big) \big);
     \medskip\\
  A^\ast_{9}
    & =
    & \forall x\, (q \con P(x) \imp \Box \PDiamond \beta_{0} (x)).
\end{array}
$$

Положим $A^\ast = A_0^\ast\wedge\ldots\wedge A_9^\ast$. Заметим, что $A^\ast$ содержит только две предметные переменные, унарную предикатную букву~$P$ и пропозициональную букву~$q$.


\settowidth{\templength}{\mbox{$v^{s+}$}}
\begin{figure}
  \centering
  \begin{tikzpicture}[scale=1.5]

\coordinate (w0)    at (1, -0.5);
\coordinate (ws)    at (1, 1.5);
\coordinate (ws`)   at (1, 2.75);
\coordinate (vsn2)  at (1, 3.50);
\coordinate (vsn2`) at (1, 4.25);
\coordinate (vsn1)  at (1, 5.0);
\coordinate (vsn1`) at (1, 5.75);
\coordinate (vsn)   at (1, 6.50);
\coordinate (vs1)   at (1, 7.50);
\coordinate (vs1`)  at (1, 8.25);
\coordinate (vs0)   at (1, 9.0);
\coordinate (vs0`)  at (1, 9.75);
\coordinate (ws')   at (1, 11.00);
\coordinate (ws'')  at (1, 11.75);

\coordinate (dots)  at (1, 11.8+0.35);
\coordinate (dots0) at (1, 0.6);
\coordinate (dotsv) at (1, 7.1);

\draw [] (w0)    circle [radius=2.0pt] ;
\draw [] (ws)    circle [radius=2.0pt] ;
\draw [] (ws')   circle [radius=2.0pt] ;
\draw [] (vs0)   circle [radius=2.0pt] ;
\draw [] (vs1)   circle [radius=2.0pt] ;
\draw [] (vsn)   circle [radius=2.0pt] ;
\draw [] (vsn1)  circle [radius=2.0pt] ;
\draw [] (vsn2)  circle [radius=2.0pt] ;

\draw [] (ws`)   circle [radius=2.0pt] ;
\draw [] (vs0`)  circle [radius=2.0pt] ;
\draw [] (vs1`)  circle [radius=2.0pt] ;
\draw [] (vsn1`) circle [radius=2.0pt] ;
\draw [] (vsn2`) circle [radius=2.0pt] ;

\draw (dots)  node {$\vdots$};

\begin{scope}[>=latex]
\draw [->, shorten >= 2.75pt, shorten <= 2.75pt] (ws)    -- (ws`)  ;
\draw [->, shorten >= 2.75pt, shorten <= 2.75pt] (ws`)   -- (vsn2) ;
\draw [->, shorten >= 2.75pt, shorten <= 2.75pt] (vsn2)  -- (vsn2`);
\draw [->, shorten >= 2.75pt, shorten <= 2.75pt] (vsn2`) -- (vsn1) ;
\draw [->, shorten >= 2.75pt, shorten <= 2.75pt] (vsn1)  -- (vsn1`);
\draw [->, shorten >= 2.75pt, shorten <= 2.75pt] (vsn1`) -- (vsn)  ;
\draw [->, shorten >= 2.75pt, shorten <= 2.75pt] (vs1)   -- (vs1`) ;
\draw [->, shorten >= 2.75pt, shorten <= 2.75pt] (vs1`)  -- (vs0)  ;
\draw [->, shorten >= 2.75pt, shorten <= 2.75pt] (vs0)   -- (vs0`) ;
\draw [->, shorten >= 2.75pt, shorten <= 2.75pt] (vs0`)  -- (ws')  ;

\draw [->, shorten >= 2.75pt, shorten <= 2.75pt, dashed] (w0)  -- (ws)   ;
\draw [->, shorten >= 2.75pt, shorten <= 2.75pt, dashed] (vsn) -- (vs1)  ;
\draw [->, shorten >= 2.75pt, shorten <= 2.75pt, dashed] (ws') -- (ws'') ;
\end{scope}

\node [left = 5pt] at (w0)    {{$w_0$}}              ;
\node [left = 5pt] at (ws)    {\parbox{\templength}{$w_m$}}              ;
\node [left = 2pt] at (ws')   {{$w_{m+1}$}}          ;
\node [left = 5pt] at (vs0)   {\parbox{\templength}{$v_m^0$}}            ;
\node [left = 5pt] at (vs1)   {\parbox{\templength}{$v_m^1$}}            ;
\node [left = 5pt] at (vsn)   {\parbox{\templength}{$v_m^s$}}            ;
\node [left = 5pt] at (vsn1)  {\parbox{\templength}{$v_m^{s+1}$}}        ;
\node [left = 5pt] at (vsn2)  {\parbox{\templength}{$v_m^{s+2}$}}        ;
\node [left = 5pt] at (ws`)   {\parbox{\templength}{$\bar w_m$}}         ;
\node [left = 5pt] at (vs0`)  {\parbox{\templength}{$\bar v_m^0$}}       ;
\node [left = 5pt] at (vs1`)  {\parbox{\templength}{$\bar v_m^1$}}       ;
\node [left = 5pt] at (vsn1`) {\parbox{\templength}{$\bar v_m^{s+1}$}}   ;
\node [left = 5pt] at (vsn2`) {\parbox{\templength}{$\bar v_m^{s+2}$}}   ;

\node [right=5pt]  at (w0)    {$q$}                ;
\node [right=5pt]  at (ws)    {$q$}                ;
\node [right=5pt]  at (ws')   {$q$}                ;
\node [right=5pt]  at (w0)    {$\phantom{\forall x\,}P(0)$}             ;
\node [right=5pt]  at (ws)    {$\phantom{\forall x\,}P(m)$}             ;
\node [right=5pt]  at (ws')   {$\phantom{\forall x\,}P(m+1)$}           ;
\node [right=5pt]  at (ws`)   {$\forall x\, P(x)$} ;
\node [right=5pt]  at (vs0`)  {$\forall x\, P(x)$} ;
\node [right=5pt]  at (vs1`)  {$\forall x\, P(x)$} ;
\node [right=5pt]  at (vsn1`) {$\forall x\, P(x)$} ;
\node [right=5pt]  at (vsn2`) {$\forall x\, P(x)$} ;

\node [right = 40pt  + 15pt + 25pt +00pt] at (w0) {$\beta_{f(0,\,\,0)}(0)$};
\node [right =  95pt + 15pt + 35pt +10pt] at (w0) {$\beta_{f(1,\,\,0)}(1)$};
\node [right = 150pt + 15pt + 45pt +20pt] at (w0) {$\beta_{f(2,\,\,0)}(2)$};
\node [right = 205pt + 15pt + 55pt +30pt] at (w0) {$\beta_{f(3,\,\,0)}(3)$};
\node [right = 275pt + 15pt + 65pt +40pt] at (w0) {$\ldots$}               ;

\node [right =  40pt + 15pt + 25pt +00pt] at (ws) {$\beta_{f(0,m)}(0)$}    ;
\node [right =  95pt + 15pt + 35pt +10pt] at (ws) {$\beta_{f(1,m)}(1)$}    ;
\node [right = 150pt + 15pt + 45pt +20pt] at (ws) {$\beta_{f(2,m)}(2)$}    ;
\node [right = 205pt + 15pt + 55pt +30pt] at (ws) {$\beta_{f(3,m)}(3)$}    ;
\node [right = 275pt + 15pt + 65pt +40pt] at (ws) {$\ldots$}               ;

\node [right =  40pt + 25pt + 25pt +00pt] at (ws') {$\beta_{f(0,m+1)}(0)$} ;
\node [right =  95pt + 25pt + 35pt +10pt] at (ws') {$\beta_{f(1,m+1)}(1)$} ;
\node [right = 150pt + 25pt + 45pt +20pt] at (ws') {$\beta_{f(2,m+1)}(2)$} ;
\node [right = 205pt + 25pt + 55pt +30pt] at (ws') {$\beta_{f(3,m+1)}(3)$} ;
\node [right = 275pt + 15pt + 65pt +40pt] at (ws') {$\ldots$}              ;

\node [right = 5pt] at (vs0)  {$\phantom{\forall x\,}P(n) \iff f(n,m) = t_0$};
\node [right = 5pt] at (vs1)  {$\phantom{\forall x\,}P(n) \iff f(n,m) = t_1$};
\node [right = 5pt] at (vsn)  {$\phantom{\forall x\,}P(n) \iff f(n,m) = t_s$};
\node [right = 5pt] at (vsn1) {$\phantom{\forall x\,}P(\alpha_m)$};
\node [right = 5pt] at (vsn2) {$\phantom{\forall x\,}P(\alpha_m + 1)$};

\end{tikzpicture}

\caption{Модель $\kModel{M}_0^\ast$}
  \label{nat:fig2}
\end{figure}


\begin{lemma}
  \label{nat:lem:monadic}
  Для множества $T$ типов плиток домино верно следующее:
  $$
  \begin{array}{lcl}
  \langle \nat, \leqslant \rangle \not\models \neg A^\ast
    & \iff
    & \mbox{существует $T$-укладка с условиями $(T_1)$--$(T_3)$.}
  \end{array}
  $$
\end{lemma}

\begin{proof}
  $(\Rightarrow)$
  Пусть $(\kModel{M}, w_0) \models A^\ast$ для некоторой модели Крипке $\kModel{M} = \otuple{\nat, \leqslant, D, I}$ и некоторого мира $w_0$; как и раньше, считаем, что~$w_0=0$.

  Дальше остаётся повторить рассуждение из части $(\Rightarrow)$ доказательства леммы~\ref{nat:lem:Kripke}, используя в нём
  \[
  \begin{array}{llcl}
  \arrayitem & \beta_n(x)  & \mbox{вместо} & P_n(x); \\
  \arrayitem & \beta_n(y)  & \mbox{вместо} & P_n(y); \\
  \arrayitem & q \con P(x) & \mbox{вместо} & M(x); \\
  \arrayitem & q \con P(y) & \mbox{вместо} & M(y).
  \end{array}
  \]

  $(\Leftarrow)$
  Пусть $f$~--- $T$-укладка, удовлетворяющая условиям $(T_1)$--$(T_3)$. Пусть $\kModel{M}'_0 = \otuple{\nat, \leqslant, D, I'}$~--- модель, определённая в части $(\Leftarrow)$ доказательства леммы~\ref{nat:lem:Kripke}. Как мы видели,
  $(\kModel{M}'_0, 0) \models A'$. Используя модель $\kModel{M}'_0$, построим модель, в которой опровергается формула~$A^{\ast}$.

  Для удобства обозначим элементы множества $\nat$ следующим образом, располагая их по возрастанию:
  $$
  \begin{array}{l}
  w_0^{\phantom{i}}, \bar{w}_0^{\phantom{i}}, v^{s+2}_0,
  \bar{v}^{s+2}_0, \ldots, v^0_0, \bar{v}^0_0, \\
  w_1^{\phantom{i}}, \bar{w}_1^{\phantom{i}}, v^{s+2}_1,
  \bar{v}^{s+2}_1, \ldots, v^0_1, \bar{v}^0_1, \\
    w_2^{\phantom{i}}, \bar{w}_2^{\phantom{i}}, \ldots\, ,
  \end{array}
  $$
  т.е. считаем, что $w_0 = 0$, $\bar{w}_0^{\phantom{i}} = 1$, $v^{s+2}_0 = 2$, и так далее.

  Пусть $\kModel{M}_0^\ast = \otuple{\nat, \leqslant, D, I^\ast}$~--- модель, определяемая тем, что для каждого~$u \in \nat$
  $$
  \begin{array}{lcll}
    (\kModel{M}_0^\ast, u) \models q
      & \bydef
      & \mbox{$u = w_m$ для некоторого $m \in \nat$,}
  \end{array}
  $$
  а также для каждого $u \in \nat$ и каждого
  $a \in \mathcal{D}$ отношение $(\kModel{M}_0^\ast, u) \models P(a)$
  выполняется тогда и только тогда, когда имеет место одно из следующих условий:
  \begin{itemize}
  \item $u = w_m$ и $(\kModel{M}'_0, 2m) \models M(a)$ для некоторого $m \in \nat$;
  \item $u = v^n_m$ и $(\kModel{M}'_0, 2m) \models P_n(a)$ для некоторых $m \in \nat$ и $n \in \set{0, \ldots, s+2}$;
  \item $u = \bar{w}_m$ для некоторого $m \in \nat$;
  \item $u = \bar{v}^n_m$ для некоторых $m \in \nat$ и $n \in \set{0, \ldots, s+2}$.
  \end{itemize}

  Тогда, согласно определению модели $\kModel{M}^\ast_0$,
  $$
  \begin{array}{lcl}
    (\kModel{M}^\ast_0, w_m) \models q \con P(a) & \iff & a = m.
  \end{array}
  \eqno (6)
  $$

  Покажем, что $(\kModel{M}_0^\ast, w_0) \models A^{\ast}$.

  Для этого сначала покажем, что
  $$
  \begin{array}{lcl}
  (\kModel{M}^\ast_0, u) \models \forall x\, P(x)
  & \iff &
  \left[
  \begin{array}{l}
    u \in \set{ \bar{w}_m : m \in \nat };
    \\
    u \in \set{\bar{v}^n_m : m \in \nat, 0 \leqslant n \leqslant s + 2}.
  \end{array}
  \right.
  \end{array}
  \eqno (7)
  $$

  Импликация $(\Leftarrow)$ мгновенно следует из определения модели~$\kModel{M}^\ast_0$.

  Обоснуем импликацию $(\Rightarrow)$. Пусть мир $u\in\nat$ таков, что
  $u \notin \set{ \bar{w}_m : m \in \nat }$ и
  $u \notin \set{ \bar{v}^n_m : m \in \nat, 0 \leqslant n \leqslant s + 2 }$.

  Рассмотрим возможные случаи для~$u$.

  Пусть $u = w_m$. Из определения модели $\kModel{M}^\ast_0$ следует, что
  $(\kModel{M}^\ast_0, w_m) \not\models P(c)$ для каждого
  $c \in \mathcal{D} \setminus \set{m}$. Поскольку
  $\mathcal{D} \setminus \set{m} \ne \varnothing$, получаем, что
  $(\kModel{M}^\ast_0, w_m) \not\models \forall x\, P(x)$.

  Пусть $u = v^{s+1}_m$. Из определения модели $\kModel{M}^\ast_0$ следует, что
  $(\kModel{M}^\ast_0, v^{s+1}_m) \models P(a)$ тогда и только тогда, когда
  $(\kModel{M}'_0, 2m) \models P_{s+1} (a)$, что, согласно определению модели
  $\kModel{M}'_0$, выполняется тогда и только тогда, когда
  $(\kModel{M}_0, 0) \models a \triangleleft b$ и $\alpha_m = a$.
  Значит, $(\kModel{M}^\ast_0, v^{s+1}_m) \not\models P(c)$ для каждого
  $c \in \mathcal{D} \setminus \set{\alpha_m}$. Поскольку
  $\mathcal{D} \setminus \set{\alpha_m} \ne \varnothing$, получаем, что
  $(\kModel{M}^\ast_0, v^{s+1}_m) \not\models \forall x\, P(x)$.

  Пусть $u = v^{s+2}_m$. Из определения модели $\kModel{M}^\ast_0$ следует, что
  $(\kModel{M}^\ast_0, v^{s+2}_m) \models P(a)$ тогда и только тогда, когда
  $\kModel{M}'_0, 2m \models P_{s+2} (a)$, что, согласно определению модели
  $\kModel{M}'_0$, выполняется тогда и только тогда, когда $a = \alpha_m +1$. Значит,
  $(\kModel{M}^\ast_0, v^{s+2}_m) \not\models P(c)$ для каждого
  $c \in \mathcal{D} \setminus \set{\alpha_m + 1}$. Поскольку
  $\mathcal{D} \setminus \set{\alpha_m + 1} \ne \varnothing$, получаем, что
  $(\kModel{M}^\ast_0, v^{s+2}_m) \not\models \forall x\, P(x)$.

  Пусть $u = v^{n}_m$, где $n \in \set{0, \ldots, s}$. Согласно тому, как определены модели
  $\kModel{M}^\ast_0$ и~$\kModel{M}'_0$, получаем, что
  $$
  \begin{array}{lclcl}
  (\kModel{M}^\ast_0, v^{n}_m) \models P(a)
  & \iff &
  (\kModel{M}'_0, 2m) \models P_{n} (a)
  & \iff &
  (\kModel{M}_0, 2m) \models P_{n} (a).
  \end{array}
  $$
  Из определения модели $\kModel{M}_0$ получаем, что
  $(\kModel{M}_0, 2m) \not\models P_n(-1)$, а значит,
  $(\kModel{M}^\ast_0, v^{n}_m) \not\models P(-1)$; следовательно,
  $(\kModel{M}^\ast_0, v^{n}_m) \not\models \forall x\, P(x)$.

  Таким образом, $(\kModel{M}^\ast_0, u) \not\models \forall x\, P(x)$, и эквивалентность
  $(7)$ доказана.

  Теперь покажем, что для любых $m \in \nat$,
  $n \in \{0, \ldots, s+2\}$ и $a \in \mathcal{D}$
  $$
  \begin{array}{lcl}
    (\kModel{M}_0^\ast, w_m)\models\beta_n(a) & \iff & (\kModel{M}'_0, 2m) \models P_n(a).
  \end{array}
  \eqno (8)
  $$

  Определим отношение ${R}_{\scriptsize\PDiamond}$ на $\nat$, положив
  $$
  \begin{array}{lcl}
  w R_{\scriptsize\PDiamond} v
    & \bydef
    & \parbox[t]{320pt}{$(\kModel{M}_0^\ast,v) \not\models
  \forall x\, P(x)$ и существует мир $u \in \nat$, такой, что $w
  \leqslant u \leqslant v$ и $(\kModel{M}_0^\ast,u) \models \forall x\, P(x)$.}
  \end{array}
  $$

  Пусть $(\kModel{M}'_0, 2m) \models P_n(a)$.
  По~$(6)$ получаем, что $(\kModel{M}_0^\ast, w_{m+1}) \models q \con P(m+1)$.
  По~$(7)$ и определению модели~$\kModel{M}_0^\ast$ получаем, что
  \begin{itemize}
  \item $w_{m+1} \in R^{s+4}_{\scriptsize\PDiamond}(w_m) \setminus R^{s+5}_{\scriptsize\PDiamond}(w_m)$;
  \item $w_m < v^n_m$;
  \item $w_{m+1} \in R^{n+1}_{\scriptsize\PDiamond}(v^n_m) \setminus R^{n+2}_{\scriptsize\PDiamond}(v^n_m)$;
  \item $(\kModel{M}_0^\ast, v^n_m) \models P(a)$.
  \end{itemize}
  Значит, $(\kModel{M}_0^\ast, w_m) \models \beta_n (a)$.

  Пусть теперь $(\kModel{M}_0^\ast, w_m) \models \beta_n(a)$.
  Тогда $(\kModel{M}_0^\ast, w_m) \models \exists y\, \PDiamond^{=s+4} (q \con P(y))$.
  Значит, существуют $u \in \nat$ и $b \in \mathcal{D}$, такие, что
  $$
  \begin{array}{lcl}
  (\kModel{M}_0^\ast, u) \models q \con P(b)
    & \mbox{и}
    & u \in R^{s+4}_{\scriptsize\PDiamond}(w_m) \setminus R^{s+5}_{\scriptsize\PDiamond}(w_m).
  \end{array}
  $$
  Согласно определению модели $\kModel{M}^\ast_0$ и условию $(6)$, это возможно только при $u=w_{m+1}$ и $b=m+1$. Значит,
  $(\kModel{M}_0^\ast, w_m) \models \PDiamond^{\phantom{1}} (\PDiamond^{=n+1} (q \con
  P(m+1)) \con P(a))$.
  Тогда по определению модели $\kModel{M}_0^\ast$ получаем, что
  $(\kModel{M}_0^\ast, v^n_m) \models P(a)$, а значит,
  $(\kModel{M}'_0, 2m) \models P_n (a)$. Следовательно, условие $(8)$ доказано.

  Из $(6)$, $(7)$ и $(8)$ получаем, что
  $(\kModel{M}_0^\ast, w_0) \models A_i^{\ast}$ для каждого $i\in\set{0, \ldots, 9}$.
  Значит, $(\kModel{M}_0^\ast, w_0) \models A^{\ast}$.
\end{proof}

\begin{theorem}
  \label{nat:thr:nat-ref-sat}
  Проблема выполнимости для логики\/ $\mPlogic{\otuple{\nat,\leqslant}}$ является\/
  $\Sigma^1_1$\nobreakdash-трудной в языке с двумя предметными переменными, одной унарной предикатной буквой и одной пропозициональной буквой.
\end{theorem}

\begin{proof}
Следует из леммы~\ref{nat:lem:monadic}.
\end{proof}

    \subsection{Логики бесконечных ординалов и плотных порядков}

\providecommand{\mynext}{\ocircle}

\subsubsection{Логики дискретных линейных порядков}
\label{nat:sec:discrete}

Обобщим теорему~\ref{nat:thr:nat-ref-sat} на логики дискретных линейных порядков, отличных от~$\otuple{\nat,\leqslant}$.

Сначала рассмотрим логики шкал, основанных на множестве~$\nat$. Пусть $R$~--- бинарное отношение на $\nat$, находящееся между отношениями $<$~и~$\leqslant$.
Тогда отношение~$\leqslant$ является рефлексивным замыканием отношения~$R$, и следовательно, отношением, соответствующим модальности~$\Box^{+}$: для каждой модели
$\kModel{M} = \otuple{\nat,R,D,I}$, каждого мира $w \in \nat$, каждого приписывания~$g$ и каждой формулы~$\varphi$
$$
\begin{array}{lcl}
  (\kModel{M},w)\models^g \Box^+\varphi
    & \iff
    & \mbox{$(\kModel{M},w')\models^g \varphi$ для каждого $w'\in \nat$,}
      \\
    &
    & \mbox{такого, что $w\leqslant w'$.}
\end{array}
$$

Пусть $A^+$~--- формула, получающаяся из $A^\ast$ заменой каждого вхождения $\Box$ на~$\Box^+$.
Тогда получаем следующий аналог леммы~\ref{nat:lem:monadic}.

\begin{lemma}
  \label{nat:lem:monadic-gen}
  Пусть $R$~--- бинарное отношение на множестве\/~$\nat$, находящееся между отношениями\/ $<$~и\/~$\leqslant$.
  Тогда для множества~$T$ типов плиток домино верно следующее:
  $$
  \begin{array}{lcl}
  \langle \nat, R \rangle \not\models \neg A^+
    & \iff
    & \mbox{существует $T$-укладка с условиями $(T_1)$--$(T_3)$.}
  \end{array}
  $$
\end{lemma}

\begin{proof}
Аналогично доказательству леммы~\ref{nat:lem:monadic}.
\end{proof}

Из леммы~\ref{nat:lem:monadic-gen} получаем аналог теоремы~\ref{nat:thr:nat-ref-sat} для $\mPlogic{\otuple{\nat,R}}$. Более того,
поскольку доказательство нигде не использует никаких предположений о расширении или постоянстве предметных областей, мы получаем следующее обобщение теоремы~\ref{nat:thr:nat-ref-sat}.

\begin{theorem}
  \label{nat:thr:nat}
  Пусть $R$~--- бинарное отношение на множестве\/~$\nat$, находящееся между отношениями\/ $<$~и\/~$\leqslant$,
  и пусть $L$~--- логика, такая, что
  $\mPlogic{\otuple{\nat,R}} \subseteq L \subseteq \mPlogicC{\otuple{\nat,R}}$. Тогда проблема $L$\nobreakdash-выполнимости является\/ $\Sigma^1_1$\nobreakdash-трудной в языке с двумя предметными переменными, одной унарной предикатной буквой и одной пропозициональной буквой.
\end{theorem}

\begin{figure}
  \centering
  \begin{tikzpicture}[scale=1.5]

\coordinate (w0)     at (1, 1.00);
\coordinate (wn`)    at (1, 2.00);
\coordinate (wn)     at (1, 2.75);
\coordinate (wn')    at (1, 3.50);
\coordinate (wn'')   at (1, 4.25);
\coordinate (wdots)  at (1, 1.60);
\coordinate (wdots') at (1, 4.85);

\coordinate (v0)     at (5, 1.00);
\coordinate (vn`)    at (5, 2.00);
\coordinate (vn)     at (5, 2.75);
\coordinate (vn')    at (5, 3.50);
\coordinate (vn'')   at (5, 4.25);
\coordinate (vdots)  at (5, 1.60);
\coordinate (vdots') at (5, 4.85);

\draw [fill] (w0)     circle [radius=2.0pt] ;
\draw [fill] (wn`)    circle [radius=2.0pt] ;
\draw []     (wn)     circle [radius=2.0pt] ;
\draw []     (wn')    circle [radius=2.0pt] ;
\draw []     (wn'')   circle [radius=2.0pt] ;
\draw []     (wdots)  node   {$\vdots$}     ;
\draw []     (wdots') node   {$\vdots$}     ;

\draw []     (v0)     circle [radius=2.0pt] ;
\draw []     (vn`)    circle [radius=2.0pt] ;
\draw [fill] (vn)     circle [radius=2.0pt] ;
\draw [fill] (vn')    circle [radius=2.0pt] ;
\draw [fill] (vn'')   circle [radius=2.0pt] ;
\draw []     (vdots)  node   {$\vdots$}     ;
\draw []     (vdots') node   {$\vdots$}     ;

\begin{scope}[>=latex]
\draw [->, shorten >= 2.750pt, shorten <= 2.750pt] (wn`)  -- (wn)  ;
\draw [->, shorten >= 2.750pt, shorten <= 2.750pt] (wn)   -- (wn') ;
\draw [->, shorten >= 2.750pt, shorten <= 2.750pt] (wn')  -- (wn'');
\draw [->, shorten >= 2.750pt, shorten <= 2.750pt] (vn`)  -- (vn)  ;
\draw [->, shorten >= 2.750pt, shorten <= 2.750pt] (vn)   -- (vn') ;
\draw [->, shorten >= 2.750pt, shorten <= 2.750pt] (vn')  -- (vn'');
\end{scope}

\node [right=2pt] at (w0)   {$0$}  ;
\node [right=2pt] at (wn`)  {$n-1$};
\node [right=2pt] at (wn)   {$n$}  ;
\node [right=2pt] at (wn')  {$n+1$};
\node [right=2pt] at (wn'') {$n+2$};

\node [right=2pt] at (v0)   {$0$}  ;
\node [right=2pt] at (vn`)  {$n-1$};
\node [right=2pt] at (vn)   {$n$}  ;
\node [right=2pt] at (vn')  {$n+1$};
\node [right=2pt] at (vn'') {$n+2$};

\node [below = 10pt ] at (w0) {$\frak{G}_n$};
\node [below = 10pt ] at (v0) {$\frak{H}_n$};

\end{tikzpicture}

\caption{Шкалы $\frak{G}_n$ и $\frak{H}_n$}
  \label{nat:fig3}
\end{figure}

Отметим, что теорема~\ref{nat:thr:nat} затрагивает бесконечно много логик, причём среди них бесконечно много пар, состоящих из логик, несравнимых между собой. Для этого сначала заметим, что логики $\mPlogic{\otuple{\nat, <}}$ и $\mPlogic{\otuple{\nat, \leqslant}}$ являются несравнимыми. Пусть
$$
\begin{array}{lcl}
  \formula{Z}   & = & \Box (\Box p \imp p) \imp (\Diamond \Box p \imp \Box p);\\
  \formula{ref} & = & \Box p \imp p.
\end{array}
$$
Известно~\cite{Goldblatt92}, что
$\otuple{\nat, <} \models \formula{Z}$, но при этом
$\otuple{\nat, \leqslant} \not\models \formula{Z}$; следовательно,
$\mPlogic{\otuple{\nat, <}} \not\subseteq \mPlogic{\otuple{\nat, \leqslant}}$. Также несложно понять, что
$\otuple{\nat, \leqslant} \models \formula{ref}$, но при этом
$\otuple{\nat, <} \not\models \formula{ref}$; значит,
$\mPlogic{\otuple{\nat, \leqslant}} \not\subseteq \mPlogic{\otuple{\nat, <}}$.

Обобщим это наблюдение, чтобы получить бесконечно много логик, о которых говорилось выше.
Пусть
$$
\begin{array}{lcl}
\XBox \vp
  & =
  & (q \con \Box (\neg q \imp \vp)) \dis (\neg q \con \Box (q \imp \vp)).
\end{array}
$$

Пусть $\kFrame{G}_n$~--- цепь, состоящая из иррефлексивной подцепи $0, \ldots, n - 1$ и идущей за ней рефлексивной подцепи $n, n+1, n+2 \ldots{}$, см.~рис.~\ref{nat:fig3}, слева.  Пусть $\kFrame{H}_n$~--- цепь, состоящая из рефлексивной подцепи $0, \ldots, n - 1$ и идущей за ней иррефлексивной подцепи
$n, n+1, n+2, \ldots\,$, см.~рис.~\ref{nat:fig3}, справа.

Покажем, что $\mPlogic{\kFrame{G}_k} \ne \mPlogic{\kFrame{G}_m}$ и
$\mPlogic{\kFrame{H}_k} \ne \mPlogic{\kFrame{H}_m}$ при $k \ne m$.
Действительно, условие $k > m$ влечёт, что
$\mPlogic{\kFrame{G}_k} \subseteq \mPlogic{\kFrame{G}_m}$: если
$\kFrame{G}_m \not\models \vp$, то $\kFrame{G}_k \not\models \vp$, поскольку
$\kFrame{G}_m$~--- порождённая подшкала шкалы~$\kFrame{G}_k$. Кроме того,
$\kFrame{G}_n \models \Box^n \formula{ref}$, но
$(\kFrame{G}_{n+1}, 0) \not\models \Box^n \formula{ref}$. Значит,
$\mPlogic{\kFrame{G}_k} \ne \mPlogic{\kFrame{G}_m}$ при $k \ne m$. Аналогично, используя формулу $\XBox^n \formula{Z}$, нетрудно показать, что
$\mPlogic{\kFrame{H}_k} \ne \mPlogic{\kFrame{H}_m}$ при $k \ne m$.

Теперь заметим, что для любых
$k, m \in \nat$ логики $\mPlogic{\kFrame{G}_k}$ и
$\mPlogic{\kFrame{H}_m}$ несравнимы:
$\Box^{k+m} \formula{ref} \in \mPlogic{\kFrame{G}_k} \setminus \mPlogic{\kFrame{H}_m}$ и
$\XBox^{k+m} \formula{Z} \in \mPlogic{\kFrame{H}_m} \setminus \mPlogic{\kFrame{G}_k}$.

Из теоремы~\ref{nat:thr:nat} получаем очевидное следствие.

\begin{corollary}
  \label{nat:cor:nat}
  Проблема выполнимости для логик\/ $\mPlogicC{\otuple{\nat, \leqslant}}$,
  $\mPlogic{\otuple{\nat, <}}$ и\/ $\mPlogicC{\otuple{\nat, <}}$ является\/
  $\Sigma^1_1$\nobreakdash-трудной в языке с двумя предметными переменными, одной унарной предикатной буквой и одной пропозициональной буквой.
\end{corollary}

Шкала $\otuple{\nat,<}$ изоморфна\footnote{Если считать, что $\nat=\omega$, то даже равна.} шкале
$\otuple{\omega,<}$, где $\omega$~--- наименьший бесконечный ординал, а отношение
$<$ совпадает с отношением $\in$ на~$\omega$. Обобщим теорему~\ref{nat:thr:nat} на логики шкал, представляющих собой ординалы особого вида, содержащие ординал~$\omega$.

\begin{theorem}
  \label{nat:thr:ordinals}
  Пусть $\alpha = \omega \cdot m + k$, где $1 \leqslant m < \omega$ и $k < \omega$. Пусть $R$~--- бинарное отношение на $\alpha$, находящееся между\/ $<$ и\/ $\leqslant$. Тогда проблема выполнимости для логик\/ $\mPlogic{\otuple{\alpha,R}}$ и\/ $\mPlogicC{\otuple{\alpha,R}}$ является\/ $\Sigma^1_1$\nobreakdash-трудной в языке с двумя предметными переменными, одной унарной предикатной буквой и одной пропозициональной буквой.
\end{theorem}

\begin{proof}
  Доказательство аналогично доказательству теоремы~\ref{nat:thr:nat}. Покажем лишь, как получить аналог леммы~\ref{nat:lem:tiling-reduction} для $\mPlogic{\otuple{\alpha,\leqslant}}$ (а~тем самым и для $\mPlogicC{\otuple{\alpha,\leqslant}}$). Затем для отношения, находящего между $<$ и $\leqslant$, достаточно использовать в формулах $\Box^+$ вместо~$\Box$.

  Поскольку шкала $\otuple{\alpha,\leqslant}$ может содержать мир, который не видит ни одного другого мира, нам требуется переопределить формулу~$A_9$ (см.~раздел~\ref{nat:sec:reduction}) подходящим образом:
  $$
  \begin{array}{lcl}
    A^{\bullet}_9 & = & \forall x\, \big( M(x) \imp \Box (\exists y\,
                        M(y) \imp \pDiamond (\exists y\, M(y) \imp
                        P_{t_0} (x))) \big).
  \end{array}
  $$

  Пусть $A^{\bullet}=A_0\con\ldots\con A_8\con A_9^\bullet$.
  Покажем, что
  $$
  \begin{array}{lcl}
  \langle \alpha, \leqslant \rangle \not\models \neg A^{\bullet}
    & \iff
    & \mbox{существует $T$-укладка с условиями $(T_1)$--$(T_3)$.}
  \end{array}
  $$

  Докажем импликацию~$(\Rightarrow)$. Пусть $(\kModel{M}, u_0) \models A^{\bullet}$ для некоторой модели $\kModel{M} = \otuple{\alpha, \leqslant, D, I}$ и некоторого мира
  $u_0 \in \alpha$. Тогда в $\alpha$ существует последняя копия ординала
  $\omega$, обладающая следующим свойством: она содержит мир $w$
  помеченный некоторым элементом $a_k$ из цепи
  $a_0 \triangleleft^{I,u_0} a_1 \triangleleft^{I,u_0} a_2
  \triangleleft^{I,u_0} \ldots{}$ элементов из $D(u_0)$, которые существуют, поскольку $(\kModel{M}, u_0) \models A_2$. Тогда укладка может быть получена
  из этой копии ординала $\omega$ аналогично тому, как описано в части $(\Rightarrow)$ доказательства леммы~\ref{nat:lem:tiling-reduction}: вертикальные ряды помечаются элементами
  $a_0,a_1,a_2,\ldots$ of $D(u_0)$, а горизонтальные~--- мирами
  $w_k,w_{k+1},w_{k+2},\ldots$, такими, что
  \begin{itemize}
  \item $w_k = w$;
  \item $w_k < w_{k+1} < w_{k+2} < \ldots{}$;
  \item $(\kModel{M},w_{k+n})\models M(a_{k+n})$ для каждого $n \in \nat$.
  \end{itemize}
  Сама $T$-укладка $f\colon \nat\times\nat \to T$ может быть определена следующим образом:
  $$
  \begin{array}{lcl}
    f(n,m) = t & \bydef & (\kModel{M},w_{k+m})\models P_t(a_n).
  \end{array}
  $$
  Нетрудно понять, что $f$ удовлетворяет условиям $(T_1)$ и~$(T_2)$. А поскольку
  $(\kModel{M}, u_0) \models A^{\bullet}_9$, множество
  $\set{w_{k+m} : \mbox{$m \in \nat$ и $(\kModel{M},w_{k+m}) \models P_{t_0} (a_0)$}}$
  является бесконечным, и значит, $f$ удовлетворяет также условию~$(T_3)$.

  Для обоснования импликации $(\Leftarrow)$ будем использовать первую копию ординала $\omega$, содержащуюся в~$\alpha$ для того, чтобы обеспечить истинность формулы~$A^{\bullet}$ в корне. Сначала определим интерпретацию всех предикатных букв, входящих в $A^{\bullet}$, в мирах первой копии ординала $\omega$ так же, как в части $(\Leftarrow)$ доказательства леммы~\ref{nat:lem:tiling-reduction}. Затем определим интерпретацию букв $p$, $M$, $P_t$ (для каждого $t \in T$) и~$\triangleleft$ в мирах, не принадлежащих первой копии ординала~$\omega$, сопоставив каждой из них в каждом из таких миров пустое множество. Тогда $A^{\bullet}$ будет истинна в корне получившейся модели.
\end{proof}

\subsubsection{Логики плотных и непрерывных линейных порядков}
\label{nat:sec:dense}

Пока неясно, можно ли получить аналог теоремы~\ref{nat:thr:ordinals} для линейных порядков, отличных от рассмотренных выше.
Довольно интересными из не подпадающих под условие теоремы~\ref{nat:thr:ordinals}, представляются логики шкал $\otuple{\numQ,\leqslant}$, $\otuple{\numQ,<}$, $\otuple{\numR,\leqslant}$ и $\otuple{\numR,<}$.

Доказательство леммы~\ref{nat:lem:tiling-reduction} не переносится ни на логику шкалы $\otuple{\numQ,\leqslant}$, ни на логику шкалы $\otuple{\numR,\leqslant}$, поскольку в этом случае истинность формулы $A$ в мире модели, определённой на какой-либо из этих шкал, не даёт возможности построить укладку домино, удовлетворяющую условию~$(T_3)$. Для логики $\mPlogic{\otuple{\numQ, \leqslant}}$ аналог леммы~\ref{nat:lem:tiling-reduction} не доказать в принципе: известно~\cite{Corsi93}, что $\mPlogic{\otuple{\numQ, \leqslant}}$ рекурсивно перечислима, а следовательно, $\Sigma^0_1$-полна. Ситуация с логикой $\mPlogic{\otuple{\numR, \leqslant}}$ менее ясна, но может оказаться сходной: неизвестно\footnote{При этом известно, что суперинтуиционистские логики шкал $\otuple{\numQ, \leqslant}$
и $\otuple{\numR, \leqslant}$ совпадают~\cite[стр.~701]{Skvortsov2009} и
$\Sigma^0_1$-полны~\cite[теорема~1]{Takano87}; с другой стороны,
суперинтуиционистские, а значит, и модальные, логики шкал с постоянными областями, определённых на  $\otuple{\numQ, \leqslant}$
и $\otuple{\numR, \leqslant}$, различаются~\cite[теорема~2]{Takano87}.}, различаются ли логики $\mPlogic{\otuple{\numR, \leqslant}}$ и $\mPlogic{\otuple{\numQ, \leqslant}}$.

Однако незначительная модификация доказательства леммы~\ref{nat:lem:tiling-reduction} позволяет обосновать $\Sigma^0_1$-трудность логик $\mPlogic{\otuple{\numR, \leqslant}}$ и $\mPlogic{\otuple{\numQ, \leqslant}}$, откуда уже несложно следует их $\Sigma^0_1$-трудность в языке с двумя предметными переменными, унарной предикатной буквой и пропозициональной буквой. Для этого достаточно отказаться от условия~$(T_3)$ и использовать $\Pi^0_1$-полную проблему домино~\cite{Berger66} (см.~также~\cite[Приложение~A.4]{BGG97}). Мы получим более общий результат, касающийся и других логик.

Пусть $B=A_0\wedge\ldots\wedge A_8$.

\begin{lemma}
  \label{nat:lem:tiling-dense}
  Пусть $\otuple{W,\leqslant}$~--- нестрогий линейный порядок, содержащий бесконечно возрастающую цепь.
  Тогда для множества $T$ типов плиток домино верно следующее:
  $$
  \begin{array}{lcl}
  \otuple{W,\leqslant} \not\models \neg B
    & \iff
    & \mbox{существует $T$-укладка с условиями $(T_1)$ и $(T_2)$.}
  \end{array}
  $$
\end{lemma}

\begin{proof}
  Импликация $(\Rightarrow)$ обосновывается
  аналогично части $(\Rightarrow)$ доказательства леммы~\ref{nat:lem:tiling-reduction}, за исключением, что теперь не требуется проверка условия~$(T_3)$.

  Докажем импликацию $(\Leftarrow)$.
  Пусть $f$~--- $T$-укладка, удовлетворяющая условиям $(T_1)$ и~$(T_2)$. Определим модель на шкале $\otuple{W,\leqslant}$, в некотором мире которой истинна формула~$B$.

  Пусть $\mathcal{D} = \nat \cup \{-1\}$. Чтобы определить интерпретацию $I$ предикатных букв на шкале $\otuple{W,\leqslant}\odot \mathcal{D}$, будем использовать бесконечно возрастающую цепь
  $w_0 \leqslant w_1 \leqslant w_2 \leqslant \ldots{}$ миров из~$W$, существующую по условию: определим~$I$ так, что для любых
  $k \in \nat$ и $a, b \in \mathcal{D}$
  $$
  \begin{array}{lcl}
  (\kModel{M},w_k) \models a\triangleleft b
  & \bydef &
  \mbox{$k$ чётно и $b=a+1$;}
  \smallskip\\
  (\kModel{M},w_k) \models p
  & \bydef &
  \mbox{$k$ нечётно;}
  \smallskip\\
  (\kModel{M},w_k) \models M(a)
  & \bydef &
  \mbox{$k=2a$;}
  \smallskip\\
  (\kModel{M},w_k) \models P_t(a)
  & \bydef &
  \mbox{$k=2m$ и $f(a,m) = t$ для некоторого $m\in\nat$,}
  \end{array}
  $$
  а также для каждого $v \not\in \{w_i : i \in \nat \}$ и каждой предикатной буквы $S$, входящей в формулу~$B$,
  $$
  \begin{array}{lcll}
  I(v, S) & = & I(w_m, S), & \mbox{где $m = \min \set{k\in\nat : v \leqslant w_k}$.}
  \end{array}
  $$

  Тогда нетрудно убедиться, что $(\kModel{M}, w_0) \models B$.
\end{proof}

Используя модификацию формулы $B$, полученную заменой вхождений $\Box$ на~$\Box^+$, получаем следующий аналог теоремы~\ref{nat:thr:nat}.

\begin{theorem}
  \label{nat:thr:ascending-chain}
  Пусть $\langle W, < \rangle$~--- строгий линейных порядок, содержащий бесконечно возрастающую цепь. Пусть
  $\leqslant$~--- рефлексивное замыкание отношения~$<$, а $R$~--- бинарное отношение, находящееся между $<$ и~$\leqslant$. Пусть $L$~--- логика, такая, что
  $\mPlogic{\otuple{W, R}} \subseteq L \subseteq \mPlogicC{\otuple{W, R}}$.
  Тогда проблема $L$\nobreakdash-выполнимости является\/ $\Pi^0_1$\nobreakdash-трудной в языке с двумя предметными переменными, одной унарной предикатной буквой и одной пропозициональной буквой.
\end{theorem}

\begin{proof}
Аналогично доказательству теоремы~\ref{nat:thr:nat}: нужно использовать лемму~\ref{nat:lem:tiling-dense} так же, как в доказательстве теоремы~\ref{nat:thr:nat}
используется лемма~\ref{nat:lem:tiling-reduction}.
\end{proof}

\begin{corollary}
  \label{nat:cor:q-and-r}
  Логики\/
  $\mPlogic{\otuple{\numQ, \leqslant}}$,
  $\mPlogicC{\otuple{\numQ, \leqslant}}$,
  $\mPlogic{\otuple{\numQ, <}}$,
  $\mPlogicC{\otuple{\numQ, <}}$,
  $\mPlogic{\otuple{\numR, \leqslant}}$,
  $\mPlogicC{\otuple{\numR, \leqslant}}$,
  $\mPlogic{\otuple{\numR, <}}$ и\/
  $\mPlogicC{\otuple{\numR, <}}$
  являются\/ $\Sigma^0_1$\nobreakdash-трудными в языке с двумя предметными переменными, одной унарной предикатной буквой и одной пропозициональный буквой.
\end{corollary}

Некоторые логики, упомянутые в следствии~\ref{nat:cor:q-and-r}, совпадают с хорошо известными аксиоматически определёнными системами. Напомним, что
$$
\begin{array}{lcl}
\logic{K4.3}
  & =
  & \logic{K4} \oplus \Box (\Box^+ p \imp q) \dis \Box (\Box^+ q \imp p);
  \smallskip\\
\logic{K4.3.D.X}
  & =
  & \logic{K4.3}
    \oplus \Diamond \top
    \oplus \Box \Box p \imp \Box p;
  \smallskip\\
\logic{S4.3}
  & =
  & \logic{S4}
    \oplus \Box (\Box p \imp q) \dis \Box (\Box q \imp p).
\end{array}
$$

Известно~\cite{Corsi93}, что
$\logic{QS4.3} = \mPlogic{\otuple{\numQ, \leqslant}}$ и
$\logic{QK4.3.D.X} = \mPlogic{\otuple{\numQ, <}}$. Учитывая рекурсивную перечислимость этих логик, получаем справедливость следующего утверждения.

\begin{corollary}
  \label{nat:cor:QS4.3}
  Логики\/ $\logic{QS4.3}$ и\/ $\logic{QK4.3.D.X}$ являются\/
  $\Sigma^0_1$\nobreakdash-полными в языке с двумя предметными переменными, одной унарной предикатной буквой и одной пропозициональной переменной.
\end{corollary}


\subsubsection{Некоторые другие логики}
\label{nat:sec:rest}

Извлечём несколько следствий из теоремы~\ref{nat:thr:ascending-chain}, касающихся логик, не попавших в следствие~\ref{nat:cor:q-and-r}.

Первое следствие касается логик бесконечных ординалов с естественно определёнными на них отношениями $<$ или $\leqslant$.

\begin{corollary}
  \label{nat:cor:inf-ordinals}
  Пусть $\alpha$~--- бесконечный ординал. Тогда\/
  $\mPlogic{\otuple{\alpha,<}}$, $\mPlogicC{\otuple{\alpha,<}}$, $\mPlogic{\otuple{\alpha,\leqslant}}$ и\/
  $\mPlogic{\otuple{\alpha,\leqslant}}$,
  являются\/ $\Sigma^0_1$\nobreakdash-трудными в языке с двумя предметными переменными, одной бинарной предикатной буквой и одной пропозициональной буквой.
\end{corollary}

Второе следствие касается логик \defnotion{нестандартных моделей}\index{модель!нестандартная} элементарных теорий моделей $\otuple{\numN,<}$, $\otuple{\numN,\leqslant}$, $\otuple{\numQ,<}$, $\otuple{\numQ,\leqslant}$, $\otuple{\numR,<}$ и $\otuple{\numR,\leqslant}$.

\begin{corollary}
  \label{nat:cor:q-r-n-ast}
  Пусть $\mathfrak{A}$~--- одна из структур\/
$\otuple{\numN,<}$, $\otuple{\numN,\leqslant}$, $\otuple{\numQ,<}$, $\otuple{\numQ,\leqslant}$, $\otuple{\numR,<}$ и\/ $\otuple{\numR,\leqslant}$, а $\kFrame{F}$~--- нестандартная модель классической теории $\mathit{Th}(\mathfrak{A})$.
Тогда\/ $\mPlogic{\kFrame{F}}$ и\/ $\mPlogicC{\kFrame{F}}$ являются\/ $\Sigma^0_1$\nobreakdash-трудными в языке с двумя предметными переменными, одной унарной предикатной буквой и одной пропозициональной буквой.
\end{corollary}

Третье следствие касается логики~$\logic{QK4.3}$. Известно~\cite{Corsi93}, что $\logic{QK4.3}$ является логикой строгих линейных порядков. Рассуждая аналогично тому, как описано в разделе~\ref{nat:sec:dense}, нетрудно обосновать справедливость следующего утверждения.

\begin{corollary}
  \label{nat:cor:QK4.3}
  Логика\/ $\logic{QK4.3}$ является\/ $\Sigma^0_1$\nobreakdash-полной в языке с двумя предметными переменными, одной унарной предикатной буквой и одной пропозициональной буквой.
\end{corollary}

\begin{proof}
  Несложно убедиться, что формула $B^+$, получающаяся из~$B$ заменой вхождений $\Box$ на $\Box^+$, истинна в некотором мире модели, определённой на
  $\logic{K4.3}$-шкале, тогда и только тогда, когда существует $T$-укладка, удовлетворяющая условиям $(T_1)$ и~$(T_2)$.
\end{proof}

    \subsection{Логики деревьев}
    \label{sec:TreesFrame:Qmodal}

Покажем, как полученные выше результаты можно перенести на логики различных классов деревьев; при этом не потребуются какие-либо существенно новые построения, и мы ограничимся пояснениями, связанными с применимостью описанных выше конструкций.

В теории графов под \defnotion{деревом}\index{дерево} принято понимать \defnotion{простой связный ациклический граф}.\index{бяб@граф!простой}\index{бяб@граф!связный}\index{бяб@граф!ациклический} В~соответствии с таким подходом шкала Крипке $\kframe{T}=\otuple{W,R}$ называется \defnotion{деревом}, если она \defnotion{$R$-связна} (т.е. если существует \defnotion{$R$-путь} между любыми её элементами), не содержит \defnotion{простых $R$\nobreakdash-циклов} и при этом $R$~--- симметричное иррефлексивное отношение. Ниже мы дадим другие определения для понятия <<дерево>>, которыми при изучении модальных логик пользуются довольно часто, и чтобы не возникало путаницы, будем называть деревья, являющиеся таковыми в смысле общепринятого определения, \defnotion{иррефлексивными простыми деревьями}.\index{дерево!иррефлексивное простое}

Для бинарного отношения $R$ на множестве $W$ определим его \defnotion{иррефлексивизацию}\index{иррефлексивизация} $R^-$ и \defnotion{обратное отношение}\index{отношение!обратное}~$R^{-1}$:
\[
\begin{array}{llcl}
\arrayitem
  & R^-
  & =
  & R \setminus \mathit{id}_W;
  \arrayitemskip\\
\arrayitem
  & R^{-1}
  & =
  & \set{\otuple{w,w'} : \otuple{w',w} \in R},
\end{array}
\]
где, как и раньше, $\mathit{id}_W = \set{\otuple{w,w} : w\in W}$~--- \defnotion{диагональ}\index{диагональ множества} множества~$W$. Кроме того, элементы $w,w'\in W$ называем \defnotion{$R$-сравнимыми}, если $w=w'$, $wRw'$ или $w'Rw$.

Шкалу Крипке $\kframe{T}=\otuple{W,R}$ называем:
\begin{itemize}
\item
\defnotion{простым деревом},\index{дерево!простое} если шкала Крипке $\otuple{W,R^-}$~--- иррефлексивное простое дерево;
\item
\defnotion{интранзитивным деревом},\index{дерево!интранзитивное} если $\kframe{T}$~--- корневая шкала с иррефлексивным антисимметричным отношением $R$ и множество $W$ с симметричным замыканием отношения $R$ образует простое дерево;
\item
\defnotion{квази-интранзитивным деревом},\index{дерево!квази-интранзитивное} если шкала Крипке $\otuple{W,R^-}$~--- интранзитивное дерево;
\item
\defnotion{транзитивным деревом},\index{дерево!транзитивное} если $\kframe{T}$~--- корневая шкала, $R$~--- транзитивное отношение и для любого $w\in W$ любые два элемента множества $R^{-1}(w)$ являются $R$-сравнимыми.
\end{itemize}
Нетрудно понять, что интранзитивные деревья являются квази-ин\-тран\-зи\-тив\-ны\-ми. Отметим, что простые, интранзитивные, транзитивные деревья, вообще говоря, могут не быть деревьями в обычном понимании; тем не менее, шкалу $\kframe{T}=\otuple{W,R}$ называем
\begin{itemize}
\item \defnotion{деревом},\index{дерево} если $\kframe{T}$ является простым, квази-ин\-тран\-зи\-тив\-ным или транзитивным деревом.
\end{itemize}
Говорим, что дерево $\kframe{T}$ рефлексивно, иррефлексивно, серийно, и т.д., если соответствующим свойством обладает отношение~$R$.

Перейдём к формулировкам алгоритмических свойств модальных предикатных логик классов деревьев и фрагментов этих логик.

\begin{proposition}
Пусть\/ $\kframe{T}$~--- конечное дерево. Тогда монадические фрагменты логик\/ $\mPlogic{\kframe{T}}$ и\/ $\mPlogicC{\kframe{T}}$, а также монадические фрагменты с равенством логик\/ $\mPlogicx{\kframe{T}}{=}$, $\mPlogicCx{\kframe{T}}{=}$, $\mPlogicx{\kframe{T}}{\simeq}$ и\/ $\mPlogicCx{\kframe{T}}{\simeq}$ разрешимы.
\end{proposition}

\begin{proof}
Следует из предложений~\ref{prop:finite-frame} и~\ref{prop:finite-frame:eq}.
\end{proof}

\begin{proposition}
Пусть\/ $\scls{C}$~--- рекурсивно перечислимый класс конечных деревьев. Тогда монадические фрагменты логик\/ $\mPlogic{\scls{C}}$ и\/ $\mPlogicC{\scls{C}}$, а также монадические фрагменты с равенством логик\/ $\mPlogicx{\scls{C}}{=}$, $\mPlogicCx{\scls{C}}{=}$, $\mPlogicx{\scls{C}}{\simeq}$ и\/ $\mPlogicCx{\scls{C}}{\simeq}$ находятся в классе\/~$\Pi^0_1$.
\end{proposition}

\begin{proof}
Следует из теоремы~\ref{wfin:thr:Pi01}.
\end{proof}

Из этих предложений получаем следствие для логик естественных классов конечных деревьев.
Для краткости формулировки введём обозначения, которые будем использовать до конца раздела. Пусть\/ $\scls{T}$~--- один из следующих классов деревьев:
\begin{itemize}
\item класс всех простых деревьев;
\item класс всех рефлексивных простых деревьев;
\item класс всех иррефлексивных простых деревьев;
\item класс всех интранзитивных деревьев;
\item класс всех квази-интранзитивных деревьев;
\item класс всех рефлексивных квази-интранзитивных деревьев;
\item класс всех транзитивных деревьев;
\item класс всех рефлексивных транзитивных деревьев;
\item класс всех иррефлексивных транзитивных деревьев.
\end{itemize}
Пусть также $\scls{T}_{\!\!\mathit{fin}}$~--- класс всех конечных деревьев из~$\scls{T}$.

\begin{corollary}
\label{cor:fin:trees:QModal}
Монадические фрагменты логик\/ $\mPlogic{\scls{T}_{\!\!\mathit{fin}}}$ и\/ $\mPlogicC{\scls{T}_{\!\!\mathit{fin}}}$, а также монадические фрагменты с равенством логик\/ $\mPlogicx{\scls{T}_{\!\!\mathit{fin}}}{=}$, $\mPlogicCx{\scls{T}_{\!\!\mathit{fin}}}{=}$, $\mPlogicx{\scls{T}_{\!\!\mathit{fin}}}{\simeq}$ и\/ $\mPlogicCx{\scls{T}_{\!\!\mathit{fin}}}{\simeq}$ являются\/ $\Pi^0_1$\nobreakdash-полными. \end{corollary}

\begin{proof}
Достаточно заметить, что указанные классы конечных деревьев разрешимы.
\end{proof}

Результат, приведённый в следствии~\ref{cor:fin:trees:QModal}, останется верным, если в языке из унарных букв имеется лишь одна, а количество предметных переменных равно трём.

\begin{proposition}
Пусть\/ $\kframe{T}$~--- дерево, в котором имеется мир, видящий бесконечно много миров. Тогда фрагменты логик\/ $\mPlogic{\kframe{T}}$ и\/ $\mPlogicC{\kframe{T}}$ с одной унарной предикатной буквой и тремя предметными переменными\/ $\Sigma^0_1$\nobreakdash-трудны.
\end{proposition}

\begin{proof}
Получается с помощью трюка Крипке (раздел~\ref{sec:KripkeTrick:QModal}); следует из теоремы~\ref{th:KripkeTrick:3var:Qmodal}.
\end{proof}

\begin{corollary}
\label{cor:all:trees:QModal}
Логики\/ $\mPlogic{\scls{T}}$ и\/ $\mPlogicC{\scls{T}}$ являются\/ $\Sigma^0_1$\nobreakdash-трудными в языке с одной унарной предикатной буквой и двумя предметными переменными.
\end{corollary}

\begin{proof}
Достаточно применить технику, описанную в разделе~\ref{s7-2-2}.
\end{proof}

Пусть $\kframe{T}=\otuple{W,R}$~--- дерево; определим $\kframe{T}^+=\otuple{W,R^+}$ как дерево, получающееся из $\kframe{T}$ заменой отношения $R$ на его рефлексивное замыкание~$R^+$.

\begin{proposition}
\label{prop:nat:trees:QModal}
Пусть\/ $\kframe{T}$~--- дерево, такое, что дерево\/ $\kframe{T}^+$ содержит бесконечные цепи, причём каждая бесконечная цепь в\/ $\kframe{T}^+$ изоморфна цепи\/ $\otuple{\numN,\leqslant}$. Тогда логики\/ $\mPlogic{\kframe{T}}$ и\/ $\mPlogicC{\kframe{T}}$ являются $\Pi^1_1$\nobreakdash-трудными в языке с двумя предметными переменными, одной унарной предикатной буквой и одной пропозициональной буквой.
\end{proposition}

\begin{proof}
Достаточно заметить, что в этом случае работают все шаги моделирования проблемы домино, описанные в разделе~\ref{sec:NatNumbersFrame:Qmodal}.
\end{proof}

Нетрудно понять, что этот результат останется справедливым, если вместо одного дерева рассматривать класс деревьев с указанным свойством.

    \subsection{Логики шкал логик доказуемости}
    \label{sec:NoetherianFramea:Qmodal}

\newlength{\tmplength}

\newcommand{\doublesymbol}[2]
{
\settowidth{\tmplength}{\mbox{#1}}
\mbox{#1}\hspace{-\tmplength}\hspace{#2}\mbox{#1}
}

\newcommand{\doublesymboltwo}[2]{%
\begin{picture}(#2,#2)
\put(0,0){#1}
\put(0.71,0.71){#1}
\end{picture}%
}

\newcommand{\doubleDiamondOne}{\doublesymbol{$\Diamond$}{0.10em}}
\newcommand{\dDiamond}{\doubleDiamondOne}
\newcommand{\doubleBoxOne}{\doublesymboltwo{$\Box$}{9.9}}
\newcommand{\dBox}{\doubleBoxOne}

\subsubsection{Предварительные сведения}
\label{Noetherian:sec:pre}

Под логиками доказуемости имеются в виду в первую очередь логики $\logic{GL}$ и $\logic{Grz}$, а также их расширения. Шкалы этих логик являются нётеровыми порядками~--- строгими в случае нормальных расширений~$\logic{GL}$ и нестрогими в случае нормальных расширений~$\logic{Grz}$. При этом в таких шкалах допускаются бесконечные возрастающие цепи, но, будучи бесконечными, они не могут быть бесконечно возрастающими. Примерами таких цепей являются шкалы $\otuple{\omega^\ast,>}$ и $\otuple{\omega^\ast,\geqslant}$, где первую можно понимать как $\ldots >3>2>1>0$, а вторую~--- как <<рефлексивное замыкание>> первой.

Мы видели, что монадические фрагменты логик шкал $\otuple{\omega,<}$ и $\otuple{\omega,\leqslant}$ являются $\Pi^1_1$-трудными. Используя результаты выше, несложно понять, что логики шкал $\otuple{\omega^\ast,>}$ и $\otuple{\omega^\ast,\geqslant}$ являются $\Pi^0_1$-трудными и $\Sigma^0_1$-трудными в полном языке, а их монадические фрагменты~$\Pi^0_1$-полными. Что будет, если к этим шкалам добавить ещё один мир в качестве корня? То есть, какова трудность логик и монадических фрагментов логик шкал $\otuple{1+\omega^\ast,>}$ и $\otuple{1+\omega^\ast,\geqslant}$? Эти шкалы являются транзитивными деревьями, содержат бесконечную цепь, но они не содержат цепей, изоморфных множеству натуральных чисел с отношением < или~$\leqslant$, а потому не удовлетворяют условию предложения~\ref{prop:nat:trees:QModal}. То же самое происходит с любой шкалой логик $\logic{GL}$ и $\logic{Grz}$, имеющей бесконечные цепи.

Ниже мы покажем, что, тем не менее, в этом случае можно получить результаты, аналогичные представленным в разделе~\ref{sec:TreesFrame:Qmodal}. Мы снова будем моделировать $\Sigma^1_1$\nobreakdash-полную проблему укладки домино, описанную в разделе~\ref{nat:sec:reduction}.

\subsubsection{Моделирование натуральных чисел}
\label{Noetherian:sec:tech}

Для моделирования проблемы домино сделаем предварительную подготовку: промоделируем натуральные числа с отношением~$<$, что позволит нам получить решётку~$\numN\times\numN$.


Зафиксируем унарную предикатную букву~$Q$. Пусть $\varphi$~--- формула в языке, содержащем букву~$Q$; положим
$$
\begin{array}{lcll}
  \hfill\sigma
    & =
    & \forall x\,Q(x); \\
  \dDiamond\varphi
    & =
    & \Diamond(\sigma \wedge\Diamond(\neg \sigma \wedge\varphi));
    \\
  \dDiamond^0 \varphi
    & =
    & \varphi;
    \\
  \dDiamond^{n+1} \varphi
    & =
    & \dDiamond\dDiamond^n\varphi
    & \mbox{для каждого $n \in \mathds{N}$;}
    \\
  \dDiamond^{=n}\varphi
    & =
    & \dDiamond^{n}\varphi \wedge \neg\dDiamond^{=n+1}\varphi
    & \mbox{для каждого $n \in \mathds{N}$;}
    \\
  \dBox\varphi
    & =
    & \neg\dDiamond\neg\varphi.
\end{array}
$$
Используя предикатную букву $Q$ и модальность $\dDiamond$, введём следующие сокращения:
$$
\begin{array}{rcll}
  x\approx y
    & =
    & \dBox(Q(x)\leftrightarrow Q(y));
    \\
  x \prec y
    & =
    & \dBox(Q(y) \to \dDiamond Q(x));
    \\
  x \lhd y
    & =
    & \dBox(Q(y) \leftrightarrow \dDiamond^{=1} Q(x));
    \\
  M(x)
    & =
    & \neg\exists y\, y\prec x.
\end{array}
$$

Интуитивно, $x \approx y$, $x \prec y$ и $x \lhd y$ можно понимать, соответственно, как $x=y$, $x<y$ и $x+1 = y$, где $x$ и $y$~--- натуральные числа. Тогда $M(x)$ означает, что $x$~--- наименьший элемент в смысле отношения~$\prec$.
Определим формулы, которые требуют выполнения соответствующих условий для $\approx$, $\prec$, $\lhd$ и~$M$:
$$
\begin{array}{lcll}
  A_1 & = &
      \forall x\,\dDiamond Q(x);
      \\
  A_2 & = &
      \forall x\exists y\, x\lhd y;
      \\
  A_3 & = &
      \forall x\,(\neg M(x)\to \exists y\, y\lhd x);
      \\
  A_4 & = &
      \forall x\,\neg(x\prec x);
      \\
  A_5 & = &
      \forall x\forall y\,(x\prec y\wedge y\prec x\to x\approx y);
      \\
  A_6 & = &
      \forall x\forall y\,(x\prec y\vee x\approx y\vee y\prec x);
      \\
  A   & = &
      A_1\wedge A_2\wedge A_3\wedge A_4\wedge A_5\wedge A_6.
\end{array}
$$

Пусть до конца этого раздела
\begin{itemize}
\item
$L$~--- логика из множества $[\kflogic{\logic{wGrz}},\kflogic{\logic{GL.3}\logic{.bf}}] \cup [\kflogic{\logic{wGrz}},\kflogic{\logic{Grz.3}\logic{.bf}}]$,
\item
$\kframe{F}=\langle W,R\rangle$~--- $L$-шкала,
\item
$\kModel{M}=\langle W,R,D,I\rangle$~--- модель, определённая на~$\kframe{F}$,
\item
$w$~--- мир шкалы $\kframe{F}$, такой, что $(\kModel{M},w)\models A$.
\end{itemize}
Пусть также $R_{\scriptsize\dDiamond}$~--- бинарное отношение на~$W$, зависящее от модели~$\kModel{M}$, определённое следующим образом:
$$
\begin{array}{lcl}
w R_{\scriptsize\dDiamond} w'
  & \leftrightharpoons
  & \mbox{$(\kModel{M},w') \not\models \sigma$ и существует мир $u \in W$, такой, что}
    \\
  &
  & \mbox{$w R u R w'$ и $(\kModel{M},u) \models \sigma$.}
\end{array}
$$
Нетрудно понять, что $R_{\scriptsize\dDiamond}$ транзитивно и иррефлексивно.

\begin{proposition}
Формула\/ $A$ является\/ $L$-выполнимой.
\end{proposition}

\begin{proof}
Рассмотрим два возможных случая для~$L$.

Пусть $L\in [\kflogic{\logic{wGrz}},\kflogic{\logic{GL.3}\logic{.bf}}]$.
Пусть $W_0=\numN \cup\set{w^\ast}$, где $w^\ast\not\in\numN$. Определим бинарное отношение $R_0$ на $W_0$, положив
$$
\begin{array}{lcl}
s R_0 t
  & \leftrightharpoons
  & \mbox{либо $s=w^\ast$ и $t\ne w^\ast$,}
  \\
  &
  & \mbox{либо $s,t\in\numN$ и $s > t$.}
\end{array}
$$
Определим модель $\kModel{M}_0$ на шкале $\langle W_0,R_0\rangle\odot\numN$ так, чтобы для любых $s\in W_0$ и $k\in \numN$ выполнялось следующее условие:
$$
\begin{array}{lcl}
(\kModel{M}_0,s)\models Q(k)
  & \iff
  & \mbox{либо $s\in 2\numN +1$, либо $s = 2k$.}
\end{array}
$$
Тогда нетрудно видеть, что $\langle W_0,R_0\rangle\odot\numN$~--- $L$-шкала и $(\kModel{M}_0,w^\ast)\models A$.

Пусть $L\in [\kflogic{\logic{wGrz}},\kflogic{\logic{Grz.3}\logic{.bf}}]$. Определим модель $\kModel{M}_1$ заменой отношения $R_0$ в $\kModel{M}_0$ на его рефлексивное замыкание~$R_0^\ast$. Тогда $\langle W_0^{\mathstrut{}},R_0^\ast\rangle\odot\numN$~--- $L$-шкала и $(\kModel{M}_1,w^\ast)\models A$.
\end{proof}

\begin{lemma}
\label{Noetherian:lem:cong1}
Отношение\/ $\approx^{I,w}$ является конгруэнтностью в\/ $\langle D_w,\prec^{I,w}\rangle$.
\end{lemma}

\begin{proof}
Ясно, что $\approx^{I,w}$~--- эквивалентность, поскольку для любых $a,b\in D_w$
$$
\begin{array}{lcl}
(\kModel{M},w)\models a\approx b & \iff & (\kModel{M},w)\models \dBox(Q(a)\leftrightarrow Q(b)).
\end{array}
$$

Для доказательства того, что $\approx^{I,w}$ является конгруэнтностью, предположим, что
$a\approx^{I,w}a'$, $b\approx^{I,w}b'$ и $a\prec^{I,w}b$ для некоторых $a,b,a',b'\in D_w$, и покажем, что $a'\prec^{I,w}b'$, т.е. $(\kModel{M},w)\models \dBox(Q(b')\to \dDiamond Q(a'))$.
Пусть $wR_{\scriptsize\dDiamond}u$ и $(\kModel{M},u)\models Q(b')$. Тогда $(\kModel{M},u)\models Q(b)$, поскольку $b\approx^{I,w}b'$. Из того, что $a\prec^{I,w}b$, следует, что существует мир $u'$ такой, что $u R_{\scriptsize\dDiamond} u'$ и $(\kModel{M},u')\models Q(a)$. Тогда $(\kModel{M},u)\models Q(a')$, поскольку $a\approx^{I,w}a'$. Значит, $a'\prec^{I,w}b'$.
\end{proof}

\begin{lemma}
\label{Noetherian:lem:cong2}
Отношение\/ $\approx^{I,w}$ является конгруэнтностью в\/ $\langle D_w,\lhd^{I,w}\rangle$.
\end{lemma}

\begin{proof}
Аналогично доказательству леммы~\ref{Noetherian:lem:cong1}.
\end{proof}

Для элемента $a\in D_w$ положим $\bar{a}=\set{b\in D_w : b\approx^{I,w} a}$.
Пусть
$$
\begin{array}{lclcl}
\bar{D}_w & = & D_w/{\approx^{I,w}} & = & \{\bar{a} : a\in D_w\}.
\end{array}
$$

Благодаря леммам~\ref{Noetherian:lem:cong1} и~\ref{Noetherian:lem:cong2} мы можем определить отношения $\prec_w$ и $\lhd_w$ на $\bar{D}_w$, положив для любых $a,c\in \bar{D}_w$
$$
\begin{array}{lcl}
  \bar{a} \hfill \prec_w \bar{c}
    & \leftrightharpoons
    & a \hfill \prec^{I,w} c;
    \\
  \bar{a} \hfill \lhd_w \bar{c}
    & \leftrightharpoons
    & a \hfill \lhd^{I,w} c.
\end{array}
$$

\begin{lemma}
\label{Noetherian:lem:rel3}
Отношение $\prec_w$ является расширением отношения~$\lhd_w$.
\end{lemma}

\begin{proof}
Следует из определений этих отношений.
\end{proof}

\begin{lemma}
\label{Noetherian:lem:rel1}
Отношение\/ $\prec_w$ является строгим линейным порядком на~$\bar{D}_w$.
\end{lemma}

\begin{proof}
Поскольку $(\kModel{M},w)\models A_4\wedge A_5\wedge A_6$, достаточно доказать только транзитивность отношения~$\prec_w$. Пусть $\bar{a}\prec_w \bar{c}$ и $\bar{c}\prec_w \bar{e}$ для некоторых $a,c,e\in D_w$; нам нужно показать, что $\bar{a}\prec_w \bar{e}$. Предположим, что $wR_{\scriptsize\dDiamond}u$ и $(\kModel{M},u)\models Q(e)$ для некоторого $u\in W$. Тогда, поскольку $\bar{c}\prec_w \bar{e}$, существует мир $u'\in W$, такой, что $uR_{\scriptsize\dDiamond}u'$ и $(\kModel{M},u')\models Q(c)$. Значит, поскольку $\bar{a}\prec_w \bar{c}$, существует мир $u''\in W$, такой, что $u'R_{\scriptsize\dDiamond}u''$ и $(\kModel{M},u')\models Q(a)$. Поскольку $R_{\scriptsize\dDiamond}$ транзитивно, получаем, что $uR_{\scriptsize\dDiamond}u''$; значит, $(\kModel{M},u)\models Q(e)\to \dDiamond Q(a)$. В силу произвольности выбора $u$ получаем, что $(\kModel{M},w)\models \dBox(Q(e)\to \dDiamond Q(a))$, т.е. $\bar{a}\prec_w \bar{e}$.
\end{proof}

\begin{lemma}
\label{Noetherian:lem:rel2}
Пусть $\bar{a}\lhd_w \bar{c}$ для некоторых $a,c\in D_w$. Тогда не существует такого $e\in D_w$, что $\bar{a}\prec_w \bar{e}\prec_w \bar{c}$.
\end{lemma}

\begin{proof}
Предположим, что $\bar{a}\prec_w \bar{e}\prec_w \bar{c}$ для некоторого $e\in D_w$. Поскольку $(\kModel{M},w)\models A_1$, существует мир $u\in W$, такой, что $wR_{\scriptsize\dDiamond}u$ и $(\kModel{M},u)\models Q(c)$. Из того, что $\bar{a}\prec_w \bar{e}\prec_w \bar{c}$, следует, что существуют $u',u''\in W$, такие, что $uR_{\scriptsize\dDiamond}u'R_{\scriptsize\dDiamond}u''$, $(\kModel{M},u')\models Q(e)$ и $(\kModel{M},u'')\models Q(a)$. Тогда $(\kModel{M},u)\models \dDiamond^2 Q(a)$. Но из того, что $\bar{a}\lhd_w \bar{c}$ и $\kModel{M},u\models Q(c)$, получаем, что $(\kModel{M},u)\models \dDiamond^{=1} Q(a)$, а следовательно, $(\kModel{M},u)\not\models \dDiamond^2 Q(a)$, что даёт противоречие.
\end{proof}

\begin{lemma}
\label{Noetherian:lem:deschain}
Не существует бесконечно убывающей $\prec_w$-цепи в~$\bar{D}_w$.
\end{lemma}

\begin{proof}
Предположим, что существует бесконечно убывающая цепь $\ldots\prec_w \bar{a}_2\prec_w \bar{a}_1\prec_w \bar{a}_0$. Значит, существуют миры $u_0,u_1,u_2,\ldots\in W$, такие, что $wR_{\scriptsize\dDiamond}u_0 R_{\scriptsize\dDiamond}u_1 R_{\scriptsize\dDiamond}u_2 R_{\scriptsize\dDiamond}\ldots{}$, причём $(\kModel{M},u_k)\models Q(a_k)$ для каждого $k\in\numN$. Нетрудно понять, что $u_0,u_1,u_2\ldots$ попарно различны; но рефлексивное замыкание отношения $R$ является нестрогим нётеровым порядком, следовательно, такая $R$-цепь невозможна, что даёт противоречие.
\end{proof}

\begin{lemma}
\label{Noetherian:lem:min}
Существует элемент $a\in D_w$, такой, что $M^{I,w}(a)$.
\end{lemma}

\begin{proof}
Предположим, что для каждого $a\in D_w$ существует $a'\in D_w$, такой, что $\bar{a}'\prec_w \bar{a}$. Тогда мы получаем бесконечно убывающую $\prec_w$\nobreakdash-цепь, что противоречит лемме~\ref{Noetherian:lem:deschain}.
\end{proof}

\begin{lemma}
\label{Noetherian:lem:nat}
$\langle \bar{D}_w,\prec_w\rangle \cong \langle\numN ,<\rangle$.
\end{lemma}

\begin{proof}
Согласно лемме~\ref{Noetherian:lem:rel1}, $\prec_w$~--- строгий линейный порядок $\bar{D}_w$. По лемме~\ref{Noetherian:lem:min}, существует $a_0\in D_w$, для которого верно $M^{I,w}(a_0)$; ясно, что $\bar{a}_0$~--- наименьший в смысле отношения $\prec_w$ элемент в~$\bar{D}_w$. Поскольку $(\kModel{M},w)\models A_2$, существуют элементы $a_0,a_1,a_2,\ldots\in D_w$, такие, что $\bar{a}_k\lhd_w\bar{a}_{k+1}$ для каждого $k\in\numN$. По лемме~\ref{Noetherian:lem:rel3}, $\prec_w$ является расширением отношения $\lhd_w$, следовательно $\bar{a}_k\prec_w\bar{a}_{k+1}$ для каждого $k\in\numN$; более того, лемма~\ref{Noetherian:lem:rel2} гарантирует, что не существует элементов между $\bar{a}_k$ и~$\bar{a}_{k+1}$.

Покажем, что для каждого $b\in D_w$ существует $k\in\numN$, такое, что $b\in \bar{a}_k$.

Предположим, что это не так, т.е. существует элемент $b_0\in D_w$, такой, что $b_0\not\in \bar{a}_k$ для каждого $k\in\numN$. Тогда, в частности, $\neg M^{I,w}(b_0)$, поскольку иначе $b_0\in\bar{a}_0$. Поскольку $(\kModel{M},w)\models A_3$, существует $b_1\in D_w$, такой, что $b_1\lhd_w b_0$. Снова $\neg M^{I,w}(b_1)$, поскольку иначе $b_1\in\bar{a}_0$, и значит, $b_0\in\bar{a}_1$. Рассуждая сходным образом, получаем, что существуют $b_0,b_1,b_2,\ldots \in D_w$, такие, что $\ldots \lhd_w b_2\lhd_w b_1\lhd_w b_0$ (и~$b_j\not\in \bar{a}_k$ для любых $j,k\in\numN$). По лемме~\ref{Noetherian:lem:rel3}, $\ldots \prec_w b_2\prec_w b_1\prec_w b_0$, а значит, мы получаем бесконечно убывающую $\prec_w$\nobreakdash-цепь, что противоречит лемме~\ref{Noetherian:lem:deschain}.

Это означает, что
$$
\begin{array}{lcl}
D_w & = & \displaystyle\bigcup\limits_{k=0}^{\infty}\bar{a}_k,
\end{array}
$$
и значит, $\langle \bar{D}_w,\prec_w\rangle \cong \langle\numN ,<\rangle$ при изоморфизме $f\colon \bar{a}_k\mapsto k$.
\end{proof}

Таким образом, формула $A$ фактически даёт нам натуральные числа. Используя это, можно построить погружение истинностной арифметики, т.е. первопорядковой теории $\logic{TA}$ алгебры $\langle \numN ,+,\cdot,0,1,= \rangle$, в логику~$L$. Действительно, положим
$$
\begin{array}{lcl}
\mathit{Zero}(x)
  & =
  & M(x); \smallskip\\
\mathit{One}(x)
  & =
  & \forall y\,(\mathit{Zero}(y)\leftrightarrow y\lhd x); \smallskip\\
\mathit{Sum}(x,y,z)
  & =
  & (\phantom{\neg}\mathit{Zero}(y)\to z\approx x) \wedge {}
  \\ &&
    (\neg\mathit{Zero}(y)\to \forall y'\forall z'\,(y'\lhd y\wedge \mathit{Sum}(x,y',z')\to z'\lhd z)); \smallskip\\
\mathit{Prod}(x,y,z)
  & =
  & (\phantom{\neg}\mathit{Zero}(y)\to \mathit{Zero}(z)) \wedge {}
  \\ &&
    (\neg\mathit{Zero}(y)\to \forall y'\forall z'\,(y'\lhd y\wedge \mathit{Prod}(x,y',z')\to \mathit{Sum}(z',y,z))). \\
\end{array}
$$
Нетрудно понять, что $\approx^{I,w}$ является конгруэнтностью по отношению к $\mathit{Zero}^{I,w}$, $\mathit{One}^{I,w}$, $\mathit{Sum}^{I,w}$ и~$\mathit{Prod}^{I,w}$. Пусть $a_0,a_1,a_2,\ldots$~--- элементы, определённые так же, как в доказательстве леммы~\ref{Noetherian:lem:nat}; тогда для любых $k,l,m\in\numN$
$$
\begin{array}{lcl}
(\kModel{M},w)\models a_k \approx a_l
  & \iff
  & k = l; \\
(\kModel{M},w)\models \mathit{Zero}(a_k)
  & \iff
  & k = 0; \\
(\kModel{M},w)\models \mathit{One}(a_k)
  & \iff
  & k = 1; \\
(\kModel{M},w)\models \mathit{Sum}(a_k,a_l,a_m)
  & \iff
  & k+l=m; \\
(\kModel{M},w)\models \mathit{Prod}(a_k,a_l,a_m)
  & \iff
  & k\hfill \cdot\hfill l = m. \\
\end{array}
$$
С этим наблюдением уже несложно построить погружение $\logic{TA}$ в $L$, т.е. вычислимую функцию $f$, такую, что для каждой замкнутой арифметической формулы~$\varPhi$
$$
\begin{array}{lcl}
\varPhi\in \logic{TA} & \iff & f(\varPhi)\in L.
\end{array}
$$


\begin{theorem}
\label{Noetherian:theorem:TA}
Если $L\in [\kflogic{\logic{wGrz}},\kflogic{\logic{GL.3}\logic{.bf}}] \cup [\kflogic{\logic{wGrz}},\kflogic{\logic{Grz.3}\logic{.bf}}]$, то фрагмент $L$ в языке с одной унарной предикатной буквой и бесконечным множеством предметных переменных не является арифметическим.
\end{theorem}

\begin{proof}
Следует из существования описанного погружения.
\end{proof}

Используя проблему укладки домино со свойствами $(T_1)$--$(T_3)$, получим утверждения, близкие теореме~\ref{Noetherian:theorem:TA}, для языков с тремя или даже двумя предметными переменными и одной дополнительной пропозициональной буквой.

\subsubsection{Моделирование $\Sigma^1_1$-полной проблемы укладки домино}
\label{Noetherian:sec:tiling}

По техническим причинам ниже нам будет удобно, чтобы нулю соответствовал не наименьший элемент, а следующий за наименьшим. Пусть
$$
\begin{array}{lcl}
  Z(x)
    & =
    & \forall y\, (M(y)\leftrightarrow y\lhd x).
\end{array}
$$
Неформально, $Z(x)$ означает что $x=0$. Чтобы избежать недоразумений из-за используемой терминологии (поскольку ноль~--- это наименьший элемент в~$\numN$), будем называть элемент $x$, для которого выполнено $M(x)$, \defnotion{магическим}, или \defnotion{волшебным элементом}. Такое название выбрано намеренно. Введём особый тип $t^\ast$ плиток домино, соответствующий \defnotion{волшебным плиткам}, рёбра которых окрашены в \defnotion{волшебный цвет}: формально это означает, что для любого типа $t$ плиток домино (включая~$t^\ast$) выполнены условия
$$
\begin{array}{llll}
\rightsq t^\ast = \leftsq  t, &
\leftsq  t^\ast = \rightsq t, &
\upsq    t^\ast = \downsq  t, &
\downsq  t^\ast = \upsq    t.
\end{array}
$$
Ниже мы будем использовать индивиды в предметных областях миров как места для укладки плиток домино, и волшебные индивиды будут использоваться для укладки волшебных плиток.

Конечно, можно избежать этой <<волшебной>> терминологии, понимая $M(x)$ как число~$x=-1$, а место, где лежит плитка типа~$t^\ast$,~--- как пустое место.

Для набора $T=\{t_0,\ldots,t_n\}$ типов плиток домино определим набор $T^\ast = T\cup\{t^\ast\}$. Пусть $P_{t_0},\ldots,P_{t_n},P_{t^\ast}$~--- новые попарно различные унарные предикатные буквы.

\bigskip

\textbf{Первое кодирование.}
\label{Noetherian:subsec:first}
Начнём с простого кодирования, которое затем будем постепенно усложнять. Пусть
$$
\begin{array}{lcl}
B_1 & = & \displaystyle
    \forall x\forall y\,(x\approx y\to \bigwedge\limits_{t\in T^\ast}\dBox(P_t(x)\to P_t(y));
    \smallskip\\
B_2 & = & \displaystyle
    \forall x\forall y\,\bigwedge\limits_{t\in T^\ast}(\dDiamond (Q(x)\wedge P_t(y)) \to \dBox(Q(x)\to P_t(y)));
    \smallskip\\
B_3 & = & \displaystyle
    \forall x\forall y\,(\neg M(x)\to \dBox(Q(x)\to \bigvee_{t\in T^\ast}(P_t(y)\wedge \bigwedge\limits_{t'\ne t}\neg P_{t'}(y))));
    \smallskip\\
B_4 & = & \displaystyle
    \forall x\forall y\,(M(x) \to
    \dBox (Q(x)\to \bigwedge\limits_{t\in T^\ast} \neg P_{t}(y)));
    \smallskip\\
B_5 & = & \displaystyle
    \forall x\forall y\,(M(x)\wedge \neg M(y) \to
    \dBox (\neg Q(x)\to P_{t^\ast}(x)\wedge \neg P_{t^\ast}(y)));
    \smallskip\\
B_6 & = & \displaystyle
    \forall x\forall y\,(x\lhd y\to \dBox\bigwedge\limits_{t\in T^\ast}(P_{t}(x)\to
    \bigvee\limits_{\mathclap{\mathop{\scriptsize\upsquare{0.21}} t=\mathop{\scriptsize\downsquare{0.21}} t'}}P_{t'}(y)));
    \smallskip\\
B_7 & = & \displaystyle
    \forall x\forall y\,(x\lhd y\to \forall z\,\bigwedge\limits_{t\in T^\ast}(\dDiamond (Q(y)\wedge P_{t}(z))\to
    \dBox\bigvee\limits_{\mathclap{\mathop{\scriptsize\rightsquare{0.21}} t'=\mathop{\scriptsize\leftsquare{0.21}} t}}(Q(x)\wedge P_{t'}(z))));
    \smallskip\\
B_8 & = & \displaystyle
    \forall x\exists y\,(x\prec y\wedge \forall x\,(Z(x)\to \dBox(Q(x)\to P_{t_0}(y))));
    \smallskip\\
B_T & = & B_1\wedge\ldots\wedge B_8. 
\end{array}
$$

\begin{lemma}
\label{Noetherian:lem:tiling1}
Если $L\in[\kflogic{\logic{wGrz}},\kflogic{\logic{GL.3}\logic{.bf}}] \cup [\kflogic{\logic{wGrz}},\kflogic{\logic{Grz.3}\logic{.bf}}]$, то
$$
\begin{array}{lcl}
\mbox{$A\wedge B_T$ $L$-выполнима}
  & \iff
  & \mbox{существует $T$-укладка}
  \\
  &
  & \mbox{с условиями $(T_1)$--$(T_3)$.}
\end{array}
$$
\end{lemma}

\begin{proof}
$(\Rightarrow)$
Пусть формула $A\wedge B_T$ является $L$-выполнимой. Тогда существуют модель $\kModel{M}=\otuple{W,R,D,I}$, определённая на $L$-шкале $\otuple{W,R}$, и мир $w\in W$, такие, что $(\kModel{M},w)\models A\wedge B_T$. Поскольку $(\kModel{M},w)\models A$,
\begin{itemize}
\item
по леммам~\ref{Noetherian:lem:cong1} и~\ref{Noetherian:lem:cong2}, $\approx^{I,w}$ является конгруэнтностью относительно $\prec^{I,w}$ и $\lhd^{I,w}$; пусть $\bar{a}=\{b\in D_w : b\approx^{I,w}a\}$ для каждого $a\in D_w$;
\item
по лемме~\ref{Noetherian:lem:nat}, $\langle D_w/{\approx^{I,w}},\prec_w\rangle \cong \langle \numN ,<\rangle$; пусть $\tau\colon \numN \to D_w/{\approx^{I,w}}$~--- соответствующий изоморфизм;
\item
по леммам~\ref{Noetherian:lem:rel3} и~\ref{Noetherian:lem:rel2}, для любых $k,m\in\numN$
$$
\begin{array}{lcl}
k+1 = m & \iff & \tau(k)\lhd_w \tau(m);
\end{array}
$$
\item
по лемме~\ref{Noetherian:lem:min}, $\tau(0) = \{a\in D_w : (\kModel{M},w)\models M(a)\}$~--- наименьший элемент по отношению~$\prec_w$;
\item
значит, $\tau(1) = \{a\in D_w : (\kModel{M},w)\models Z(a)\}$.
\end{itemize}

Из того, что $(\kModel{M},w)\models B_1$, получаем, что для любого $u\in R(w)$, для любых $a,a'\in D_w$, таких, что $a\approx^{I,w}a'$, и для любого $t\in T^\ast$
$$
\begin{array}{lcl}
(\kModel{M},u)\models P_t(a) & \iff & (\kModel{M},u)\models P_t(a').
\end{array}
\eqno{\mbox{$(\ast)$}}
$$

Пусть $W_k=\{u\in R(w) : \kModel{M},u\models Q(a_k)\}$ для каждого $k\in \numN$. Поскольку $(\kModel{M},w)\models B_2$, получаем, что для любых $k\in\numN$, $u,u'\in W_k$, $a\in D_w$ и $t\in T^\ast$
$$
\begin{array}{lcl}
(\kModel{M},u)\models P_t(a) & \iff & (\kModel{M},u')\models P_t(a).
\end{array}
\eqno{\mbox{$({\ast}{\ast})$}}
$$

Из \mbox{$({\ast})$}, \mbox{$({\ast}{\ast})$} и того, что $(\kModel{M},w)\models B_3\wedge B_4$, получаем, что функция $f\colon \numN \times\numN \to T^\ast$, заданная условием
$$
\begin{array}{lcl}
f(k,m) = t
  & \bydef
  & \mbox{либо $(\kModel{M},u)\models P_t(a)$, где $a\in \tau(m)$, $u\in W_k$ и $k\ne 0$,}
  \\
  &
  & \mbox{либо $k = 0$ и $t=t^\ast$,}
\end{array}
$$
определена корректно.

Поскольку $(\kModel{M},w)\models B_6\wedge B_7$, получаем, что $f$ является $T^\ast$-укладкой, удовлетворяющей условиям $(T_1)$ и~$(T_2)$. Чтобы получить $T$-укладку с теми же условиями, заметим, что поскольку $(\kModel{M},w)\models B_5$,
$$
\begin{array}{lcl}
f(k,m) = t^\ast
  & \iff
  & \mbox{$k=0$ or $m=0$.}
\end{array}
$$
Следовательно, функция $g\colon \numN \times\numN \to T^\ast$, определённая как
$$
\begin{array}{lcl}
g(k,m)
  & =
  & f(k+1,m+1),
\end{array}
$$
является $T$-укладкой, удовлетворяющей условиям $(T_1)$ и~$(T_2)$. Осталось заметить, что поскольку $(\kModel{M},w)\models B_8$, функция $g$ удовлетворяет также и условию~$(T_3)$.

$(\Leftarrow)$
Пусть имеется $T$-укладка $f\colon \numN \times\numN \to T$, удовлетворяющая условиям $(T_1)$--$(T_3)$.

Пусть $L\in [\kflogic{\logic{wGrz}},\kflogic{\logic{GL.3}\logic{.bf}}]$.
Пусть $W_0=\numN \cup\{w^\ast\}$, где $w^\ast\not\in\numN$. Определим бинарное отношение $R_0$ на $W_0$, положив
$$
\begin{array}{lcl}
s R_0 t & \bydef & \mbox{либо $s=w^\ast$ и $t\ne w^\ast$,}
        \\
        &
        & \mbox{либо $s,t\in\numN$ и $s > t$.}
\end{array}
$$
Определим модель $\kModel{M}_0$ на шкале $\langle W_0,R_0\rangle\odot W_0$ так, что для любых $s,k\in W_0$ и $t\in T^\ast$,
$$
\begin{array}{lcl}
(\kModel{M}_0,s)\models Q(k)
  & \iff
  & \mbox{либо $s\in 2\numN + 1$,}
  \\
  &
  & \mbox{либо $k\in \numN$ и $s = 2(k+1)$,}
  \\
  &
  & \mbox{либо $k = w^\ast$ и $s = 0$;}
  \smallskip\\
(\kModel{M}_0,s)\models P_t(k)
  & \iff
  & \mbox{либо $f(m,k)=t$, где $s=2m+2$ и $m,k\in\numN$,}
  \\
  &
  & \mbox{либо $k=w^\ast$, $s \in 2\numN +2$ и $t=t^\ast$.}
\end{array}
$$
Тогда нетрудно понять, что $\langle W_0,R_0\rangle\odot W_0$~--- $L$\nobreakdash-шкала и $(\kModel{M}_0,w^\ast)\models A\wedge B_T$.

Пусть теперь $L\in [\kflogic{\logic{wGrz}},\kflogic{\logic{Grz.3}\logic{.bf}}]$. Определим модель $\kModel{M}_1$ путём замены отношения $R_0$ в $\kModel{M}_0$ на его рефлексивное замыкание~$R_0^\ast$. Тогда $\langle W_0^{\mathstrut{}},R_0^\ast\rangle\odot W_0^{\mathstrut{}}$~--- $L$\nobreakdash-шкала и $(\kModel{M}_1,w^\ast)\models A\wedge B_T$.
\end{proof}


\bigskip

\textbf{Второе кодирование.}
\label{Noetherian:subsec:second}
Заметим, что количество унарных предикатных букв в формуле~$B_T$ зависит от~$T$. Поэтому лемма~\ref{Noetherian:lem:tiling1} гарантирует, что рассматриваемые логики $\Pi^1_1$-трудны в языке с тремя переменными и бесконечным множеством унарных предикатных букв.
Тем не менее, несложно промоделировать все унарные предикатные буквы в формуле $A\wedge B_T$ с помощью буквы~$Q$ и дополнительной пропозициональной переменной~$p$.

Для этого сначала переопределим формулу~$A$.
Для удобства введём следующие модальности как сокращения:
$$
\begin{array}{lcl}
\dBox_p\varphi
  & =
  & \dBox(p\to\varphi);
  \\
\dDiamond_p\varphi
  & =
  & \dDiamond(p\wedge\varphi);
  \\
\dBox_p^+\varphi
  & =
  & \varphi\wedge \dBox_p\varphi;
\end{array}
$$
где $n\in\numN$. Положим
$$
\begin{array}{rcll}
  x \prec' y
    & =
    & \dBox_p(Q(y) \to \dDiamond_pQ(x));
    \\
  x \lhd' y
    & =
    & \dBox_p(Q(y) \leftrightarrow \dDiamond^{=n+3}(p\wedge Q(x));
    \\
  M'(x)
    & =
    & \neg\exists y\, y\prec' x;
    \\
  Z'(x)
    & =
    & \forall y\, (M(y)\leftrightarrow y\lhd' x).
\end{array}
$$
Пусть $A'_1,\ldots,A'_6$~--- формулы, получающиеся из $A_1,\ldots,A_6$ заменой $\prec$, $\lhd$, $M$ и~$\dDiamond$ на $\prec'$, $\lhd'$, $M'$ и~$\dDiamond_p$ соответственно, и пусть $A'$~--- их конъюнкция:
$$
\begin{array}{lcll}
  A'_1 & = &
      \forall x\,\dDiamond_pQ(x);
      \\
  A'_2 & = &
      \forall x\exists y\, x\lhd' y;
      \\
  A'_3 & = &
      \forall x\,(\neg M'(x)\to \exists y\, y\lhd' x);
      \\
  A'_4 & = &
      \forall x\,\neg(x\prec' x);
      \\
  A'_5 & = &
      \forall x\forall y\,(x\prec' y\wedge y\prec' x\to x\approx y);
      \\
  A'_6 & = &
      \forall x\forall y\,(x\prec' y\vee x\approx y\vee y\prec' x);
      \\
  A'   & = &
      A'_1\wedge A'_2\wedge A'_3\wedge A'_4\wedge A'_5\wedge A'_6.
\end{array}
$$

Нетрудно понять, что мы можем повторить утверждения раздела~\ref{Noetherian:sec:tech}, заменяя $A$ на~$A'$ в каждом из них.

Вернёмся к формуле~$B$. Напомним, что $T=\set{t_0,\ldots,t_n}$ и $T^\ast=T\cup\set{t^\ast}$, где $t^\ast$~--- это тип, соответствующий волшебным плиткам домино. Для типа $t\in T^\ast$ определим его \defnotion{индекс} $\#t$, положив
$$
\begin{array}{lcll}
\#t_k
  & =
  & k,
  & \mbox{где $0\leqslant k\leqslant n$;}
  \\
\#t^\ast
  & =
  & n+1.
\end{array}
$$
Заметим, что индекс типа $t$ плиток домино (в частности, индекс типа~$t^\ast$) зависит от~$T$, поскольку один и тот же тип плиток домино может входит в разные задачи домино с разными индексами.

Пусть для каждого $t\in T^\ast$
$$
\begin{array}{lcl}
S_t^{z'}(x_1)
  & =
  & \exists z'\,(\dDiamond^{=n+3}(p\wedge Q(z')) \wedge\dDiamond(Q(x_1)\wedge\dDiamond^{=\#t+1}(p\wedge Q(z')))).
\end{array}
$$

Идея дальнейших построений состоит в том, что мы хотели бы выполнить подстановку формулы $S_{t}^{z'}(x_1)$ вместо $P_{t}(x_1)$ (для каждого $t\in T^\ast$) в формулу~$B$ (точнее, в некоторую её модификацию).
Но, как минимум, для буквы $P_{t_0}$ в формуле~$B$ имеются подформулы $P_{t_0}(x)$, $P_{t_0}(y)$ и~$P_{t_0}(z)$, поэтому, чтобы реализовать такую подстановку, нужно, чтобы переменная~$z'$ была отлична и~от~$x$, и~от~$y$, и~от~$z$.

Чтобы сохранить число предметных переменных в формулах, заметим, что для каждого~$t\in T^\ast$
$$
\begin{array}{lcl}
S_t^{z'}(x)\leftrightarrow S_t^y(x) & \in & \logic{QK}; \\
S_t^{z'}(y)\leftrightarrow S_t^x(y) & \in & \logic{QK}; \\
S_t^{z'}(z)\leftrightarrow S_t^x(z) & \in & \logic{QK}.
\end{array}
$$

Из этого наблюдения следует, что вместо указанной подстановки можно для каждого $t\in T^\ast$ заменить вхождения подформул $P_{t}(x)$, $P_{t}(y)$, $P_{t}(z)$ на $S_t^y(x)$, $S_t^x(y)$, $S_t^x(z)$ соответственно.\footnote{Фактически это тоже подстановка формул, известная как \defnotion{непрямая},\index{подстановка!непрямая} в то время как замена $P_{t}(x)$, $P_{t}(y)$, $P_{t}(z)$ на $S_t^{z'}(x)$, $S_t^{z'}(y)$, $S_t^{z'}(z)$ является \defnotion{прямой}\index{подстановка!прямая} подстановкой формул.}

Теперь переопределим формулу~$B$ аналогично тому, как выше переопределили формулу~$A$:
$$
\begin{array}{lcl}
B'_1 & = & \displaystyle
    \forall x\forall y\,(x\approx y\to \bigwedge\limits_{t\in T^\ast}\dBox_p(S_t^y(x)\to S_t^x(y));
    \smallskip\\
B'_2 & = & \displaystyle
    \forall x\forall y\,\bigwedge\limits_{t\in T^\ast}(\dDiamond_p (Q(x)\wedge S_t^x(y)) \to \dBox_p(Q(x)\to S_t^x(y)));
    \smallskip\\
B'_3 & = & \displaystyle
    \forall x\forall y\,(\neg M'(x)\to \dBox_p(Q(x)\to \bigvee_{t\in T^\ast}(S_t^x(y)\wedge \bigwedge\limits_{t'\ne t}\neg S_{t'}^x(y))));
    \smallskip\\
B'_4 & = & \displaystyle
    \forall x\forall y\,(M'(x) \to
    \dBox_p (Q(x)\to \bigwedge\limits_{t\in T^\ast} \neg S_{t}^x(y)));
    \smallskip\\
B'_5 & = & \displaystyle
    \forall x\forall y\,(M'(x)\wedge \neg M'(y) \to
    \dBox_p (\neg Q(x)\to S_{t^\ast}^y(x)\wedge \neg S_{t^\ast}^x(y)));
    \smallskip\\
B'_6 & = & \displaystyle
    \forall x\forall y\,(x\lhd' y\to \dBox_p\bigwedge\limits_{t\in T^\ast}(S_{t}^y(x)\to
    \bigvee\limits_{\mathclap{\mathop{\scriptsize\upsquare{0.21}} t=\mathop{\scriptsize\downsquare{0.21}} t'}}S_{t'}^x(y)));
    \smallskip\\
B'_7 & = & \displaystyle
    \forall x\forall y\,(x\lhd' y\to \forall z\,\bigwedge\limits_{t\in T^\ast}(\dDiamond_p (Q(y)\wedge S_{t}^x(z))\to
    \dBox_p\bigvee\limits_{\mathclap{\mathop{\scriptsize\rightsquare{0.21}} t'=\mathop{\scriptsize\leftsquare{0.21}} t}}(Q(x)\wedge S_{t'}^x(z))));
    \smallskip\\
B'_8 & = & \displaystyle
    \forall x\exists y\,(x\prec' y\wedge \forall x\,(Z'(x)\to \dBox_p(Q(x)\to S_{t_0}^x(y))));
    \smallskip\\
B'_T & = & B'_1\wedge\ldots\wedge B'_8.
\end{array}
$$

\begin{lemma}
\label{Noetherian:lem:tiling2}
Если $L\in[\kflogic{\logic{wGrz}},\kflogic{\logic{GL.3}\logic{.bf}}] \cup [\kflogic{\logic{wGrz}},\kflogic{\logic{Grz.3}\logic{.bf}}]$, то
$$
\begin{array}{lcl}
\mbox{$A'\wedge B'_T$ $L$-выполнима}
  & \iff
  & \mbox{существует $T$-укладка}
  \\
  &
  & \mbox{с условиями $(T_1)$--$(T_3)$.}
\end{array}
$$
\end{lemma}

\begin{proof}
%
Для доказательства импликации $(\Rightarrow)$ достаточно повторить аргументацию части~$(\Rightarrow)$ доказательства леммы~\ref{Noetherian:lem:tiling1}.

Докажем импликацию~$(\Leftarrow)$.
Пусть имеется $T$-укладка $f\colon \numN \times\numN \to T$, удовлетворяющая условиям $(T_1)$--$(T_3)$.

Пусть $L\in [\kflogic{\logic{wGrz}},\kflogic{\logic{GL.3}\logic{.bf}}]$.
Пусть снова $W_0=\numN \cup\{w^\ast\}$, где $w^\ast\not\in\numN$, и пусть отношение $R_0$ на $W_0$ определено следующим образом:
$$
\begin{array}{lcl}
s R_0 t & \bydef & \mbox{либо $s=w^\ast$ и $t\ne w^\ast$,}
        \\
        &   & \mbox{либо $s,t\in\numN$ и $s > t$.}
\end{array}
$$
Определим модель $\kModel{M}_0$ на $L$-шкале $\langle W_0,R_0\rangle\odot W_0$ так, чтобы для любых $s,k\in W_0$ и $t\in T^\ast$ выполнялись следующие условия:
$$
\begin{array}{lcl}
(\kModel{M}_0,s)\models p
  & \iff
  & \mbox{$s \in 2(n+3)\numN$;}
  \smallskip\\
(\kModel{M}_0,s)\models Q(k)
  & \iff
  & \mbox{либо $s\in 2\numN +1$,}
  \\
  &
  & \mbox{либо $k\in \numN$ и $s = 2(n+3)(k+1)$,}
  \\
  &
  & \mbox{либо $k = w^\ast$ и $s = 0$,}
  \\
  &
  & \mbox{либо $f(m,k)=t$, где $s = 2(n+3)m+2(\#t+1)$}
  \\
  &
  & \hfill\mbox{и $m,k\in\numN$,}
  \\
  &
  & \mbox{либо $k=w^\ast$ и $s\in 2(n+3)\numN +2(\#t^\ast+1)$.}
\end{array}
$$

Покажем, что $(\kModel{M}_0,w^\ast)\models A'\wedge B'_T$.

Сначала заметим, что, благодаря использованию волшебных плиток домино,
$$
\begin{array}{lcl}
(\kModel{M}_0, s)\models \sigma
  & \iff
  & s\in 2\numN +1.
\end{array}
\eqno{\mbox{$({\ast}{\ast}{\ast})$}}
$$
Действительно, импликация $(\Leftarrow)$ выполняется по определению модели~$\kModel{M}_0$. Чтобы доказать импликацию~$(\Rightarrow)$, предположим, что $s\not\in 2\numN +1$. Тогда либо $s = 2(n+3)(k+1)$ для некоторого $k\in\numN$, либо $s = 0$, либо $s = 2(n+3)m+2(\#t+1)$ для некоторых $m\in\numN$ и~$t\in T^\ast$. Если $s = 2(n+3)(k+1)$, то $(\kModel{M}_0,s)\not\models Q(l)$ при $l\ne k$. Если $s = 0$, то $(\kModel{M}_0,s)\not\models Q(l)$ при $l\ne w^\ast$. Если $s = 2(n+3)m+2(\#t+1)$ для $t\ne t^\ast$, то $(\kModel{M}_0,s)\not\models Q(w^\ast)$; если $s = 2(n+3)m+2(\#t^\ast+1)$, то $(\kModel{M}_0,s)\not\models Q(l)$ при $l\ne w^\ast$. В любом случае, $(\kModel{M}_0, s)\not\models \sigma$, что обосновывает импликацию~$(\Rightarrow)$.

Теперь чтобы обосновать, что $(\kModel{M}_0,w^\ast)\models A'$, обратим внимание на определение $\lhd'$ и определение~$\kModel{M}_0$: для $m\in\numNp$ и $k\in W_0$,
$$
\begin{array}{lcl}
(\kModel{M}_0, 2(n+3)(m+1))\models Q(k)
  & \iff
  & k=m;
  \\
(\kModel{M}_0, 2(n+3)(m+1))\models \dDiamond^{=n+3}(p\wedge Q(k))
  & \iff
  & k=m-1,
\end{array}
$$
поскольку, согласно~\mbox{$({\ast}{\ast}{\ast})$}, мир $2(n+3)m$ является $R_{\scriptsize\dDiamond}$-достижимым из мира $2(n+3)(m+1)$ за $n+3$ шага, но не достижим за $n+4$ шага; аналогично, для $k\in W_0$,
$$
\begin{array}{lcl}
(\kModel{M}_0, 2(n+3))\models Q(k)
  & \iff
  & k=0;
  \\
(\kModel{M}_0, 2(n+3))\models \dDiamond^{=n+3}(p\wedge Q(k))
  & \iff
  & k=w^\ast.
\end{array}
$$
Можем заключить, что $(\kModel{M}_0,w^\ast)\models A'_1\wedge A'_2\wedge A'_3$; оставшаяся часть доказательства не содержит трудностей.

Ключевое наблюдение, важное для проверки того, что $(\kModel{M}_0,w^\ast)\models B'_T$, состоит в том, что для любых $m\in\numN$ и $k\in W_0$
$$
\begin{array}{lcl}
(\kModel{M}_0,2(n+3)(m+1))\models S_t^{z'}(k)
  & \iff
  & \mbox{либо $f(m,k) = t$ и $k\in\numN$,}
  \\
  &
  & \mbox{либо $t=t^\ast$ и $k=w^\ast$,}
\end{array}
$$
где $z'$ может быть переменной $x$, $y$ или~$z$. Оставшаяся часть снова не содержит трудностей.

Пусть $L\in [\kflogic{\logic{wGrz}},\kflogic{\logic{Grz.3}\logic{.bf}}]$. Определим модель $\kModel{M}_1$ путём замены отношения $R_0$ в $\kModel{M}_0$ на его рефлексивное замыкание~$R_0^\ast$. Тогда $\langle W_0^{\mathstrut{}},R_0^\ast\rangle\odot W_0^{\mathstrut{}}$~--- $L$-шкала и $(\kModel{M}_1,w^\ast)\models A\wedge B_T$.
\end{proof}

\bigskip

\textbf{Третье кодирование.}
\label{Noetherian:subsec:third}
Основная цель следующей модификации моделирования проблемы укладки домино состоит в том, чтобы использовать только две предметные переменные. Заметим, что формулы $B_7$ и $B'_7$ содержат три предметные переменные, взаимодействующие друг с другом. Уменьшение числа переменных до двух будет основано на том, чтобы использовать $\lhd$ не только в корне модели, где истинна формула~$A\wedge B_T$ (точнее, некоторая её модификация). Заметим, что если мы оставим определение для $\lhd$ тем же, то получим бинарные отношения, соответствующие $\lhd$, меняющимися при движении от одного мира к другому, что не подходит для наших целей. Чтобы избежать этого, будем использовать бинарную предикатную букву~$P_\lhd$.

Следующая формула описывает требуемое свойство для~$P_\lhd$:
$$
\begin{array}{lcl}
C & = & \forall x\forall y\, (x\lhd' y \to \dBox_p^+ P_\lhd (x,y)).
\end{array}
$$

Переопределим $B'_T$:
$$
\begin{array}{lcl}
B''_7 & = & \displaystyle
    \forall x\forall y\,\dBox_p\bigwedge\limits_{\mathclap{t\in T^\ast}}(Q(y)\wedge P_t(x)\to \dBox_p(\exists x\,(P_\lhd(x,y)\wedge Q(x))\to \bigvee\limits_{\mathclap{\mathop{\scriptsize\rightsquare{0.21}} t'=\mathop{\scriptsize\leftsquare{0.21}} t}} S_{t'}^y(x)));
    \smallskip\\
B''_T & = & B'_1\wedge \ldots\wedge B'_6\wedge B''_7\wedge B'_8.
\end{array}
$$

\begin{lemma}
\label{Noetherian:lem:tiling3}
Если $L\in[\kflogic{\logic{wGrz}},\kflogic{\logic{GL.3}\logic{.bf}}] \cup [\kflogic{\logic{wGrz}},\kflogic{\logic{Grz.3}\logic{.bf}}]$, то
$$
\begin{array}{lcl}
\mbox{$A'\wedge B''_T\wedge C$ $L$-выполнима}
  & \iff
  & \mbox{существует $T$-укладка}
  \\
  &
  & \mbox{с условиями $(T_1)$--$(T_3)$.}
\end{array}
$$
\end{lemma}

\begin{proof}
Для доказательства импликации $(\Rightarrow)$
снова достаточно следовать части $(\Rightarrow)$ доказательства леммы~\ref{Noetherian:lem:tiling1}.

Докажем импликацию $(\Leftarrow)$.
Пусть $f\colon \numN \times\numN \to T$~--- $T$-укладка, удовлетворяющая условиям $(T_1)$--$(T_3)$.
Достаточно следовать части $(\Leftarrow)$ доказательства леммы~\ref{Noetherian:lem:tiling2}, добавив к определению~$\kModel{M}_0$ следующие условия:
$$
\begin{array}{lcl}
(\kModel{M}_0,s)\models P_\lhd (k,m)
  & \iff
  & \mbox{либо $k=w^\ast$ и $m=0$,}
  \\
  &
  & \mbox{либо $k,m\in \numN$ и $k+1=m$.}
  \\
\end{array}
$$
Тогда $(\kModel{M}'_0,w^\ast)\models A'\wedge B''_T\wedge C$ и $(\kModel{M}'_1,w^\ast)\models A'\wedge B''_T\wedge C$.
\end{proof}


\bigskip

\textbf{Четвёртое кодирование: утрата линейности.}
\label{Noetherian:subsec:forth}
Элиминируем бинарную букву~$P_\lhd$, для чего применим трюк Крипке~\cite{Kripke62}, состоящей в замене подформул вида $P(x,y)$, где $P$~--- бинарная предикатная буква, на $\Diamond(P_1(x)\wedge P_2(y))$, где $P_1$ и $P_2$~--- унарные предикатные буквы.


Для $m$-арной предикатной буквы $P$ и формулы $\psi$ со свободными переменными $x_1,\ldots,x_m$ обозначим через $[\psi/P]\varphi$ результат подстановки формулы $\psi$ вместо $P(x_1,\ldots,x_m)$ в~$\varphi$; при этом мы предполагаем, что переменные $x_1,\ldots,x_m$ не входят в~$\varphi$; например, если $P$~--- бинарная предикатная буква и $\psi=\Diamond(P_1(x_1)\wedge P_2(x_2))$, то $[\psi/P]\forall x\forall y\,P(x,y) = \forall x\forall y\,\Diamond(P_1(x)\wedge P_2(y))$.

Пусть $\psi=\Diamond(P_1(x_1)\wedge P_2(x_2))$ и $E=A'\wedge B''_T\wedge C$.

\begin{lemma}
\label{Noetherian:lem:tiling4}
Если $L\in[\kflogic{\logic{wGrz}},\kflogic{\logic{GL}\logic{.bf}}] \cup [\kflogic{\logic{wGrz}},\kflogic{\logic{Grz}\logic{.bf}}]$, то
$$
\begin{array}{lcl}
\mbox{$[\psi/P_\lhd]E$ $L$-выполнима} & \iff & \mbox{$E$ $L$-выполнима.}
\end{array}
$$
\end{lemma}

\begin{proof}
$(\Rightarrow)$
Если формула $E$ не является $L$-выполнимой, то любой подстановочный пример $E$ тоже не будет $L$-выполнимым (поскольку $L$ замкнута по правилу предикатной подстановки).

$(\Leftarrow)$
Пусть формула $E$ является $L$-выполнимой. Тогда, согласно лемме~\ref{Noetherian:lem:tiling3}, существует $T$-укладка с условиями $(T_1)$--$(T_3)$, и, согласно доказательству леммы~\ref{Noetherian:lem:tiling3}, формула $E$ истинна в мире $w^\ast$ модели $\kModel{M}_0$, определённой на шкале $\langle W_0,R_0\rangle\odot W_0$.
Пусть $W'=\{w_k : k\in W_0^{\mathstrut{}}\}$, $W'_0=W_0^{\mathstrut{}}\cup W'$, $R'_0=R_0^{\mathstrut{}}\cup W_0^{\mathstrut{}} \times W'$; определим модель $\kModel{M}'_0$ на шкале $\langle W'_0,R'_0\rangle\odot W_0^{\mathstrut{}}$ так, что для любых $s\in W'_0$, $k,l\in W_0^{\mathstrut{}}$ и $t\in T^\ast$
$$
\begin{array}{lcl}
(\kModel{M}'_0,s)\models p
  & \iff
  & \mbox{$s \in 2(n+3)\numN$;}
  \smallskip\\
(\kModel{M}'_0,s)\models Q(k)
  & \iff
  & \mbox{либо $s\in 2\numN +1$,}
  \\
  &
  & \mbox{либо $k\in \numN$ и $s = 2(n+3)(k+1)$,}
  \\
  &
  & \mbox{либо $k = w^\ast$ и $s = 0$,}
  \\
  &
  & \mbox{либо $f(m,k)=t$, где $s = 2(n+3)m+2(\#t+1)$}
  \\
  &
  & \hfill \mbox{и $m,k\in\numN$,}
  \\
  &
  & \mbox{либо $k=w^\ast$ и $s\in 2(n+3)\numN +2(\#t^\ast+1)$;}
  \smallskip\\
(\kModel{M}'_0,s)\models P_1(k)
  & \iff
  & \mbox{$s=w_k$,}
  \smallskip\\
(\kModel{M}'_0,s)\models P_2(k)
  & \iff
  & \mbox{либо $k\in \numNp$ и $s=w_{k-1}$,}
  \\
  &
  & \mbox{либо $k=0$ и $s=w_{w^\ast}$.}
\end{array}
$$
Тогда нетрудно понять, что для любых $s,k,m\in W_0$
$$
\begin{array}{lcl}
(\kModel{M}'_0,s)\models \Diamond (P_1(k)\wedge P_2(m))
  & \iff
  & (\kModel{M}_0,s)\models P_\lhd(k,m).
\end{array}
$$
Поскольку $(\kModel{M}'_0,w_k)\not\models p$ для каждого $w_k\in W'$, благодаря использованию модальностей $\dDiamond_p$ и $\dBox_p$, получаем, что $(\kModel{M}'_0,w^\ast)\models [\psi/P_\lhd]E$. Если $\kModel{M}'_1$~--- модель, получающаяся из $\kModel{M}'_0$ заменой отношения $R'_0$ на его рефлексивное замыкание, то также $(\kModel{M}'_1,w^\ast)\models [\psi/P_\lhd]E$. В любом случае, формула $[\psi/P_\lhd]E$ является $L$-выполнимой.
\end{proof}

Отметим, что доказательство леммы~\ref{Noetherian:lem:tiling4} таково, что мы <<теряем>> все логики, содержащие формулы, ограничивающие ширину шкал; мы <<вернём>> их позже, в разделе~\ref{Noetherian:subsec:sixth}. Пока обратимся к трюку Крипке с целью усилить возникшую конструкцию.

\bigskip

\textbf{Пятое кодирование.}
\label{Noetherian:subsec:fifth}
Теперь покажем, как промоделировать предикатные буквы $P_1$ и $P_2$ с помощью букв $Q$ и~$p$. Пусть
$$
\begin{array}{lcl}
\psi_1 & = & \neg\sigma\wedge \Box^+\neg p\wedge Q(x_1); \\
\psi_2 & = & \dDiamond Q(x_1).
\end{array}
$$
Заметим, что $[\psi_1/P_1][\psi_2/P_2]\Diamond(P_1(x)\wedge P_2(y)) = \Diamond(\neg\sigma\wedge \Box^+\neg p\wedge Q(x)\wedge \dDiamond Q(y))$.

\begin{lemma}
\label{Noetherian:lem:tiling5}
Если $L\in[\kflogic{\logic{wGrz}},\kflogic{\logic{GL}\logic{.bf}}] \cup [\kflogic{\logic{wGrz}},\kflogic{\logic{Grz}\logic{.bf}}]$, то
$$
\begin{array}{lcl}
\mbox{$[\psi_1/P_1][\psi_2/P_2][\psi/P_\lhd]E$ $L$-выполнима} & \iff & \mbox{$[\psi/P_\lhd]E$ $L$-выполнима.}
\end{array}
$$
\end{lemma}

\begin{proof}
$(\Rightarrow)$
Если формула $[\psi/P_\lhd]E$ не является $L$-выполнимой, то любой её подстановочный пример тоже не будет $L$-выполнимым; значит, формула $[\psi_1/P_1][\psi_2/P_2][\psi/P_\lhd]E$ не является $L$-выполнимой.

$(\Leftarrow)$
Пусть формула $[\psi/P_\lhd]E$ является $L$-выполнимой. Тогда, как мы видели, $(\kModel{M}'_0,w^\ast)\models [\psi/P_\lhd]E$, где $\kModel{M}'_0$~--- модель, определённая в доказательстве леммы~\ref{Noetherian:lem:tiling4}. Для каждого мира $w_k\in W'_0$ добавим новые миры $w'_k$ и $w''_k$ так, что $w_k$ видит $w'_k$, а $w'_k$ видит $w''_k$, после чего возьмём транзитивное замыкание получившегося отношения достижимости. Пусть предметная область каждого из добавленных миров равна~$W_0$. Положим $p$ ложной в каждом из добавленных миров. Положим $Q(m)$ истинным в $w_k$, если $(\kModel{M}'_0,w_k)\models P_1(m)$; положим $Q(m)$ истинным в $w'_k$ для каждого $m\in W_0$; положим $Q(m)$ истинным в $w''_k$, если $(\kModel{M}'_0,w_k)\models P_2(m)$. Тогда формула $[\psi_1/P_1][\psi_2/P_2][\psi/P_\lhd]E$ истинна в мире $w^\ast$ получившейся модели; аналогично для $\kModel{M}'_1$. Следовательно, формула $[\psi_1/P_1][\psi_2/P_2][\psi/P_\lhd]E$ является $L$-выполнимой.
\end{proof}

\bigskip

\textbf{Шестое кодирование: возвращение линейности.}
\label{Noetherian:subsec:sixth}
Нетрудно видеть, что лемма~\ref{Noetherian:lem:tiling4} останется справедливой, если мы заменим в её формулировке $\kflogic{\logic{GL}\logic{.bf}}$ на $\kflogic{\logic{GL.3}\logic{.bf}}$ и $\kflogic{\logic{Grz}\logic{.bf}}$ на $\kflogic{\logic{Grz.3}\logic{.bf}}$: можно незначительно изменить доказательство, превратив~$W'$ в цепь. Покажем, что вместо двух дополнительных унарных предикатных букв достаточно использовать лишь одну.

Пусть
$$
\begin{array}{lclclcl}
\psi'
  & =
  & [Q/P_2]\psi
  & =
  & \Diamond(P_1(x_1)\wedge Q(x_2)).
\end{array}
$$

\begin{lemma}
\label{Noetherian:lem:tiling6}
Если  $L\in[\kflogic{\logic{wGrz}},\kflogic{\logic{GL.3}\logic{.bf}}] \cup [\kflogic{\logic{wGrz}},\kflogic{\logic{Grz.3}\logic{.bf}}]$, то
$$
\begin{array}{lcl}
\mbox{$[\psi'/P_\lhd]E$ $L$-выполнима} & \iff & \mbox{$E$ $L$-выполнима.}
\end{array}
$$
\end{lemma}

\begin{proof}
$(\Rightarrow)$
Если формула $E$ не является $L$-выполнимой, то формула $[\psi'/P_\lhd]E$ тоже не является $L$-выполнимой как подстановочный пример формулы~$E$.

$(\Leftarrow)$
Пусть формула $E$ является $L$-выполнимой. Тогда рассмотрим модель $\kModel{M}'_0$, определённую в доказательстве леммы~\ref{Noetherian:lem:tiling4}, и расширим отношение достижимости $R'_0$ до $R''_0$ так, что последовательность
$$
w^\ast,\ldots,2,1,0,\ldots,w_2,w_1,w_0,w_{w^\ast}
$$
образует $R''_0$-цепь, а также положим $Q(m)$ истинным в $w_k$, если $(\kModel{M}'_0,w_k)\models P_2(m)$. Тогда формула $[\psi'/P_\lhd]E$ будет истинна в мире $w^\ast$ получившейся модели; аналогично для~$\kModel{M}'_1$. Следовательно, формула $[\psi'/P_\lhd]E$ является $L$-выполнимой.
\end{proof}

\subsubsection{Теоремы}
\label{Noetherian:sec:theorems}

Полученные выше построения и леммы дают нам теоремы о высокой алгоритмической сложности модальных предикатных логик, определяемых нётеровыми порядками.

\begin{theorem}
\label{Noetherian:theorem:1}
Каждая логика из интервалов\/ $[\kflogic{\logic{wGrz}},\kflogic{\logic{GL.3}\logic{.bf}}]$ и\/ $[\kflogic{\logic{wGrz}},\kflogic{\logic{Grz.3}\logic{.bf}}]$ является\/ $\Pi^1_1$-трудной в языке, где имеется хотя бы один из следующих наборов предикатных букв и предметных переменных:
\begin{itemize}
\item
$1$ пропозициональная буква, $1$ унарная предикатная буква, $3$ предметные переменные;
\item
$1$ пропозициональная буква, $2$ унарные предикатные буквы, $2$ предметные переменные.
\end{itemize}
\end{theorem}

\begin{proof}
Следует из лемм~\ref{Noetherian:lem:tiling2} и~\ref{Noetherian:lem:tiling6}.
\end{proof}

\begin{theorem}
\label{Noetherian:theorem:2}
Каждая логика из интервалов\/ $[\kflogic{\logic{wGrz}},\kflogic{\logic{GL}\logic{.bf}}]$ и\/ $[\kflogic{\logic{wGrz}},\kflogic{\logic{Grz}\logic{.bf}}]$ является\/ $\Pi^1_1$-трудной, если в её языке имеются
\begin{itemize}
\item
$1$ пропозициональная буква, $1$ унарная предикатная буква, $2$ предметные переменные.
\end{itemize}
\end{theorem}

\begin{proof}
Следует из леммы~\ref{Noetherian:lem:tiling5}.
\end{proof}

\subsubsection{Замечания}
\label{Noetherian:sec:discussion}

Полученные выше результаты несложно расширить и на другие логики.
Пусть $<$ и $\leqslant$~--- естественные отношения на ординалах, а $>$ и $\geqslant$~--- обратные к ним отношения. Нетрудно видеть, что
$$
\begin{array}{lclclclclcl}
\kflogic{\logic{GL.3}\logic{.bf}}
  & \subseteq
  & \mPlogicC{\langle 1+\omega^\ast\cdot 2,>\rangle}
  & \subseteq
  \\
  & \subseteq
  & \mPlogicC{\langle 1+\omega^\ast,>\rangle}
  & \subseteq
  & \mPlogicC{\langle \omega^\ast,>\rangle};
  \medskip \\
\kflogic{\logic{Grz.3}\logic{.bf}}
  & \subseteq
  & \mPlogicC{\langle 1+\omega^\ast\cdot 2,\geqslant\rangle}
  & \subseteq
  \\
  & \subseteq
  & \mPlogicC{\langle 1+\omega^\ast,\geqslant\rangle}
  & \subseteq
  & \mPlogicC{\langle \omega^\ast,\geqslant\rangle},
\end{array}
$$
при этом все указанные включения являются строгими. Для мира $w$ транзитивной шкалы Крипке $\mathfrak{F}=\langle W,R\rangle$ положим
$$
\begin{array}{lclclclclcl}
\mathop{\mathit{cl}}(w) & = & \{w\}\cup\{u\in W : wRuRw\},
\end{array}
$$
и множество $\mathop{\mathit{cl}}(w)$ будем называть \defnotion{кластером}, или \defnotion{сгустком}, порождённым миром~$w$.
Заметим, что если логика $L$ является расширением логики $\logic{QwGrz}$, то для каждого мира $w$ каждой $L$-шкалы кластер, порождённый миром~$w$, содержит только~$w$. Если $R$~--- транзитивное отношение достижимости в шкале~$\mathfrak{F}$, то положим для каждых $w,w'\in W$
$$
\begin{array}{lclclclclcl}
\mathop{\mathit{cl}}(w)\,\bar{R}\,{\mathop{\mathit{cl}}(w')} & \leftrightharpoons & wRw'.
\end{array}
$$
Бинарное отношение $R$ будем называть \defnotion{слабым нётеровым порядком}, если рефлексивное замыкание отношения $\bar{R}$ является нестрогим нётеровым порядком. Пусть $\mathscr{C}_{\mathit{wN}}$~--- класс всех шкал Крипке, отношения достижимости в которых являются слабыми нётеровыми порядками.
Тогда нетрудно видеть, что
$$
\begin{array}{lcl}
\mPlogic{\mathscr{C}_{\mathit{wN}}}
  & \subseteq
  & \kflogic{\logic{wGrz}}
\end{array}
$$
и что это включение является строгим.

\begin{corollary}
\label{Noetherian:cor:theorem:1a}
Каждая логика, принадлежащая интервалу $[\mPlogic{\mathscr{C}_{\mathit{wN}}},\mPlogicC{\langle 1+\omega^\ast,>\rangle}]$ или интервалу\/ $[\mPlogic{\mathscr{C}_{\mathit{wN}}},\mPlogicC{\langle 1+\omega^\ast,\geqslant\rangle}]$, является\/ $\Pi^1_1$-трудной в языке, содержащем\/
$1$ пропозициональную букву, $1$ унарную предикатную букву и\/ $3$ предметные переменные.
\end{corollary}

\begin{proof}
Следует из доказательства леммы~\ref{Noetherian:lem:tiling2}.
\end{proof}

\begin{corollary}
\label{Noetherian:cor:theorem:1b}
Каждая логика, принадлежащая интервалу\/ $[\mPlogic{\mathscr{C}_{\mathit{wN}}},\mPlogicC{\langle 1+\omega^\ast\cdot 2,>\rangle}]$ или интервалу\/ $[\mPlogic{\mathscr{C}_{\mathit{wN}}},\mPlogicC{\langle 1+\omega^\ast\cdot 2,\geqslant\rangle}]$, является\/ $\Pi^1_1$-трудной в языке, содержащем\/
$1$ пропозициональную букву, $2$ унарные предикатные буквы и\/ $2$ индивидные переменные.
\end{corollary}

\begin{proof}
Следует из доказательства леммы~\ref{Noetherian:lem:tiling6}.
\end{proof}

\begin{corollary}
\label{Noetherian:cor:theorem:2a}
Каждая логика, принадлежащая интервалу\/ $[\mPlogic{\mathscr{C}_{\mathit{wN}}},\kflogic{\logic{GL}\logic{.bf}}]$ или интервалу\/ $[\mPlogic{\mathscr{C}_{\mathit{wN}}},\kflogic{\logic{Grz}\logic{.bf}}]$, является\/ $\Pi^1_1$-трудной в языке, содержащем\/
$1$ пропозициональную букву, $1$ унарную предикатную букву, $2$ предметные переменные.
\end{corollary}

\begin{proof}
Следует из доказательства леммы~\ref{Noetherian:lem:tiling5}.
\end{proof}



\begin{hypothesis}
Каждая логика, принадлежащая интервалу $[\kflogic{\logic{wGrz}\logic{.bf}},\kflogic{\logic{GL.3}\logic{.bf}}]$ или интервалу\/ $[\kflogic{\logic{Grz}\logic{.bf}},\kflogic{\logic{Grz.3}\logic{.bf}}]$, является\/ $\Pi^1_1$-трудной в языке, содержащем одну унарную предикатную букву и две предметные переменные.
\end{hypothesis}

	\subsection{Контрпримеры}

\providecommand{\md}{\mathop{\mathit{md}}}

\subsubsection{Предварительные сведения}
\label{c-ex:sec:pre}

Мы видели, что, во-первых, логики элементарно определимых классов шкал Крипке являются рекурсивно перечислимыми, а во-вторых, логики многих элементарно не определимых классов шкал рекурсивно перечислимыми не являются. Возникает естественный вопрос о том, нет ли здесь общей закономерности: существуют ли логики, которые полны по Крипке, при этом не полны относительно элементарно определимых классов шкал Крипке и являются рекурсивно перечислимыми? Оказывается, такие логики существуют. Первый пример такой логики был найден автором в классе квазинормальных модальных предикатных логик~\cite{MR:2019:AiML}, но при этом рассматривались только корневые шкалы. Позже были найдены другие примеры, в том числе для нормальных логик и для классов произвольных шкал (не обязательно корневых), и ниже будут приведены некоторые из таких примеров.

\subsubsection{Пример для класса корневых шкал}
\label{c-ex:sec:example-1}

Построим нормальную модальную предикатную логику~$L_0$, которая рекурсивно перечислима, полна по Крипке, но не полна относительно элементарно определимых классов корневых шкал. Ограничение, связанное с рассмотрением именно корневых шкал существенно: мы покажем, что $L_0$ полна относительно элементарно определимого класса шкал, если не требуется, чтобы шкалы были корневыми.

Для каждого $n\in\numNp$ определим шкалу Крипке~$\kframe{F}_n=\otuple{W_n,R_n}$, положив
$W_n = \set{ w_1, \ldots, w_n, w^\ast }$ и
$R_n = \set{ \otuple{ w_i, w_{i+1}} : 1 \leqslant i < n } \cup \set{
\otuple{ w_1, w^\ast } }$, см.~рис.~\ref{c-ex:fig}. Пусть $\scls{C}^\ast$~--- класс всех таких шкал. Пусть также
$\scls{C}_0 = \{ \kframe{F}_{2n} \in \scls{C}^\ast : n \geqslant 1 \}$.
Положим $L_0 = \mPlogic{\scls{C}_0}$.

\begin{figure}
\centering
\begin{tikzpicture}[scale=1.5]

\coordinate (w1)   at (+0.0000,+0.0000);
\coordinate (w2)   at (+1.0000,+0.0000);
\coordinate (w3)   at (+2.0000,+0.0000);
\coordinate (w4ph) at (+3.0000,+0.0000);
\coordinate (dts)  at (+3.5000,+0.0000);
\coordinate (w5ph) at (+4.0000,+0.0000);
\coordinate (wn-1) at (+5.0000,+0.0000);
\coordinate (wn)   at (+6.0000,+0.0000);
\coordinate (w*)   at (+0.0000,+1.0000);

\begin{scope}[>=latex]
\draw [->,  shorten >= 2.5pt, shorten <= 2.5pt]
(w1) -- (w2);
\draw [->,  shorten >= 2.5pt, shorten <= 2.5pt]
(w2) -- (w3);
\draw [->,  shorten >= 2.5pt, shorten <= 2.5pt]
(w3) -- (w4ph);
\draw [->,  shorten >= 2.5pt, shorten <= 2.5pt]
(w5ph) -- (wn-1);
\draw [->,  shorten >= 2.5pt, shorten <= 2.5pt]
(wn-1) -- (wn);
\draw [->,  shorten >= 2.5pt, shorten <= 2.5pt]
(w1) -- (w*);
\end{scope}

\node [below = 2pt      ] at (w1)   {${w_1}$}     ;
\node [below = 2pt      ] at (w2)   {${w_2}$}     ;
\node [below = 2pt      ] at (w3)   {${w_3}$}     ;
\node [                 ] at (dts)  {${\cdots}$}  ;
\node [above = 2pt      ] at (w*)   {${w^\ast}$}  ;
\node [below = 2pt      ] at (wn-1) {${w_{n-1}}$} ;
\node [below = 2pt      ] at (wn)   {${w_n}$}     ;


\filldraw [] (w1)   circle [radius=2pt]   ;
\filldraw [] (w2)   circle [radius=2pt]   ;
\filldraw [] (w3)   circle [radius=2pt]   ;
\filldraw [] (w*)   circle [radius=2pt]   ;
\filldraw [] (wn-1) circle [radius=2pt]   ;
\filldraw [] (wn)   circle [radius=2pt]   ;

\end{tikzpicture}

\caption{Шкала $\kframe{F}_n$}
\label{c-ex:fig}
\end{figure}

Логика $L_0$ полна по Крипке согласно её определению. Покажем, что
$L_0$ рекурсивно перечислима и не полна относительно элементарно определимых классов корневых шкал Крипке.

Для того, чтобы доказать рекурсивную перечислимость логики $L_0$, построим её погружение в логику $\logic{QCl}^=$, для чего будем использовать стандартный перевод (см.~разделы~\ref{s7-4-2} и~\ref{s7-4-3}).

Ниже мы будем придерживаться обозначений, введённых в разделах~\ref{s7-4-1} и~\ref{s7-4-2}; будем также использовать обозначение~$\bm{tkf}$ вместо~$\bm{tkf}_1$.

Для каждого $n\leqslant 2$ положим $\kframe{F}^{\uplus}_n = \bigsqcup\set{\kframe{F}_k\in \scls{C}_0 : 1\leqslant k \leqslant n}$. Пусть $F_n$~--- формула в языке с буквами $\uR$ и $=$, описывающая шкалу~$\kframe{F}^{\uplus}_n$; нетрудно понять, что $F_n$ строится эффективно.
%
%

Для замкнутой модальной предикатной формулы $\vp$ положим
$$
\begin{array}{rcl}
  \widehat{\vp} & = & \bm{tkf}\wedge F_{\mathop{\mathit{md}}\vp + 3} \imp \forall x\, (\uW(x) \imp \mstared{\varphi}{x}).
\end{array}
$$

\begin{lemma}
\label{c-ex:lem:Lw0intoQClE}
Для каждой замкнутой модальной предикатной формулы~$\vp$ имеет место следующая эквивалентность:
$$
\begin{array}{lcl}
\vp \in L_0 & \iff & \widehat{\vp} \in \logic{QCl}^=.
\end{array}
$$
\end{lemma}

\begin{proof}
  Достаточно заметить, что для опровержения формулы $\varphi$ в некотором мире некоторой модели нужно учитывать лишь миры, достижимые из этого мира не более чем за $\mathop{\mathit{md}}\vp$ шагов и использовать конструкцию, описанную в разделе~\ref{s7-4-3}.
\end{proof}

\begin{lemma}
  \label{c-ex:lem:L0re}
  Логика $L_0$ рекурсивно перечислима.
\end{lemma}

\begin{proof}
  Следует из леммы~\ref{c-ex:lem:Lw0intoQClE}.
\end{proof}

Покажем, что $L_0$ не полна относительно элементарно определимых классов корневых шкал Крипке.

Для каждого $n\in\numNp$ положим
$$
\begin{array}{lcl}
\alpha_n & = & \Diamond \Box {\bot} \con \Diamond^n \Box {\bot}.
\end{array}
$$

\begin{lemma}
  \label{c-ex:lem:alphas}
  Для каждого $n\in\numNp$
  $$
  \begin{array}{lcl}
    \neg \alpha_n \in L_0
      & \iff
      & \mbox{$n$ чётно.}
  \end{array}
$$
\end{lemma}

\begin{proof}
  Следует из определения формулы $\alpha_n$, шкал вида~$\kframe{F}_m$ и логики~$L_0$.
\end{proof}


\begin{lemma}
  \label{c-ex:lem:L0notFOdef}
  Логика $L_0$ не полна относительно элементарно определимых классов корневых шкал Крипке.
\end{lemma}

\begin{proof}
  Предположим, что $L_0$ полна относительно элементарно определимого класса $\scls{C}$ корневых шкал Крипке.

  Покажем, что для каждого $n\geqslant 3$
  $$
  \begin{array}{lcl}
  \kframe{F}_n\in \scls{C}
    & \iff
    & \mbox{$n$ чётно.}
  \end{array}
  $$

  Пусть $n$ нечётно. Тогда, согласно лемме~\ref{c-ex:lem:alphas},
  $\neg\alpha_{n-1}\in L_0$. Поскольку
  $(\kframe{F}_n, w_1) \not\models\neg\alpha_{n-1}$, получаем, что
  $\kframe{F}_n\not\in \scls{C}$.

  Пусть $n$ чётно. Тогда, согласно лемме~\ref{c-ex:lem:alphas},
  $\neg\alpha_{n-1}\not\in L_0$. Значит, существуют корневая шкала
  $\kframe{F}' \in\scls{C}$ и мир $w$, такие, что
  $(\kframe{F}',w) \models\alpha_{n-1}$. Покажем, что, с точностью до изоморфизма,
  $\kframe{F}' = \kframe{F}_n$.

  Пусть $\zeta = \Diamond p \imp \Box p$. Несложно видеть, что
  $\zeta$ истинна в мире $w$ шкалы тогда и только тогда, когда из $w$ достижимо не более одного мира. Поскольку только корни шкал класса $\scls{C}_0$ видят более одного мира,
  $\Box \zeta, \neg \zeta \imp \Diamond \Box {\bot} \in L_0$.
  Поскольку
  $\bm{alt}_2, \Box \zeta, \neg \zeta \imp \Diamond \Box {\bot} \in
  L_0$,
  из того, что $\kframe{F} \in \scls{C}$, следует, что $\kframe{F}$ имеет ветвление не более двух, никакой мир из $\kframe{F}$, достижимый из другого мира, не видит более одного мира (т.е. только корень шкалы $\kframe{F}$ может видеть два мира), а также если корень шкалы $\kframe{F}$ видит два мира, то один из них слепой, т.е. не видит ничего.
  Поскольку $(\kframe{F}', w) \models\alpha_{n-1}$, получаем, что мир $w$~--- корень шкалы $\kframe{F}'$ и $w$ видит слепые миры за один шаг и за $n-1$ шагов. Следовательно, шкала~$\kframe{F}'$ изоморфна шкале~$\kframe{F}_n$.

  Чтобы получить противоречие, осталось заметить, что не существует классической формулы первого порядка, которая различала бы конечные последовательности с чётным и нечётным числом элементов.
  (см,~например,~\cite[следствие~3.12]{Libkin}).
\end{proof}


\begin{theorem}
  Существует полная по Крипке нормальная модальная предикатная логика, которая рекурсивно перечислима, но не полна относительно элементарных классов корневых шкал.
\end{theorem}

\begin{proof}
  Можно взять $L_0$ в качестве такой логики.
\end{proof}

Покажем, что условие наличия корня в шкалах является существенным.

\begin{figure}
\centering
\begin{tikzpicture}[scale=1.6]

\coordinate (w1)   at (+0.0000,+0.0000);
\coordinate (w2)   at (+1.0000,+0.0000);
\coordinate (w3)   at (+2.0000,+0.0000);
\coordinate (w4)   at (+3.0000,+0.0000);
\coordinate (w5ph) at (+4.0000,+0.0000);
\coordinate (dts)  at (+4.5000,+0.0000);
\coordinate (w6ph) at (+5.0000,+0.0000);
\coordinate (w2k-1)at (+6.0000,+0.0000);
\coordinate (w2k)  at (+7.0000,+0.0000);
\coordinate (y)    at (+8.0000,+0.0000);
\coordinate (x)    at (+9.0000,+0.0000);
\coordinate (z)    at (+9.0000,-1.0000);
\coordinate (w*)   at (+0.0000,+1.0000);
\coordinate (w2*)  at (+1.0000,-1.0000);
\coordinate (w4*)  at (+3.0000,-1.0000);
\coordinate (w2k*) at (+7.0000,-1.0000);

\coordinate (p1)   at ( -0.3,-0.40);
\coordinate (p2)   at (1-0.3,-0.40);
\coordinate (p3)   at (1-0.3,-1.45);
\coordinate (p4)   at (1+0.3,-1.45);
\coordinate (p5)   at (1+0.3,+0.40);
\coordinate (p6)   at ( +0.3,+0.40);
\coordinate (p7)   at ( +0.3,+1.40);
\coordinate (p8)   at ( -0.3,+1.40);

\coordinate (p11)  at (2  -0.3,-0.40);
\coordinate (p21)  at (2+1-0.3,-0.40);
\coordinate (p31)  at (2+1-0.3,-1.45);
\coordinate (p41)  at (2+1+0.3,-1.45);
\coordinate (p51)  at (2+1+0.3,+0.40);
\coordinate (p61)  at (2  -0.3,+0.40);

\coordinate (p12)  at (6  -0.4,-0.40);
\coordinate (p22)  at (6+1-0.3,-0.40);
\coordinate (p32)  at (6+1-0.3,-1.45);
\coordinate (p42)  at (6+1+0.3,-1.45);
\coordinate (p52)  at (6+1+0.3,+0.40);
\coordinate (p62)  at (6  -0.4,+0.40);

\coordinate (p13)  at (8  -0.3,-0.40);
\coordinate (p23)  at (8+1-0.3,-0.40);
\coordinate (p33)  at (8+1-0.3,-1.45);
\coordinate (p43)  at (8+1+0.3,-1.45);
\coordinate (p53)  at (8+1+0.3,+0.40);
\coordinate (p63)  at (8  -0.3,+0.40);

\coordinate (text0) at (0.3,1.00);
\coordinate (text1) at (1.5,0.45);
\coordinate (text2) at (5.5,0.45);
\coordinate (text3) at (7.5,0.45);
\node [above right] at (text0)   {{<<нижняя часть>> шкалы}} ;
\node [above right] at (text1)   {{<<регулярно повторяющиеся>> фрагменты шкалы}} ;

\begin{scope}[>=latex]
\draw [->,  shorten >= 2.5pt, shorten <= 2.5pt]
(w1) -- (w2);
\draw [->,  shorten >= 2.5pt, shorten <= 2.5pt]
(w2) -- (w3);
\draw [->,  shorten >= 2.5pt, shorten <= 2.5pt]
(w3) -- (w4);
\draw [->,  shorten >= 2.5pt, shorten <= 2.5pt]
(w4) -- (w5ph);
\draw [->,  shorten >= 2.5pt, shorten <= 2.5pt]
(w6ph) -- (w2k-1);
\draw [->,  shorten >= 2.5pt, shorten <= 2.5pt]
(w2k-1) -- (w2k);
\draw [->,  shorten >= 2.5pt, shorten <= 2.5pt]
(w2k) -- (y);
\draw [->,  shorten >= 2.5pt, shorten <= 2.5pt]
(y) -- (x);
\draw [->,  shorten >= 2.5pt, shorten <= 2.5pt]
(z) -- (x);
\draw [->,  shorten >= 2.5pt, shorten <= 2.5pt]
(w2*) -- (w2);
\draw [->,  shorten >= 2.5pt, shorten <= 2.5pt]
(w4*) -- (w4);
\draw [->,  shorten >= 2.5pt, shorten <= 2.5pt]
(w2k*) -- (w2k);
\draw [->,  shorten >= 2.5pt, shorten <= 2.5pt]
(w1) -- (w*);
\end{scope}

\draw[dashed, rounded corners=1.5]
(p1) -- (p2) -- (p3) -- (p4) -- (p5) -- (p6) -- (p7) -- (p8) -- cycle;
\draw[dashed, rounded corners=1.5]
(p11) -- (p21) -- (p31) -- (p41) -- (p51) -- (p61) -- cycle;
\draw[dashed, rounded corners=1.5]
(p12) -- (p22) -- (p32) -- (p42) -- (p52) -- (p62) -- cycle;
\draw[dashed, rounded corners=1.5]
(p13) -- (p23) -- (p33) -- (p43) -- (p53) -- (p63) -- cycle;

\node [below      ] at (w1)   {${w_1}$}         ;
\node [above      ] at (w2)   {${w_2}$}         ;
\node [above      ] at (w3)   {${w_3}$}         ;
\node [above      ] at (w4)   {${w_4}$}         ;
\node [above      ] at (w*)   {${w^\ast}$}      ;
\node [           ] at (dts)  {${\cdots}$}      ;
\node [above      ] at (w2k-1){${w_{2k-1}}$}    ;
\node [above      ] at (w2k)  {${w_{2k}}$}      ;
\node [above      ] at (y)    {${y}$}           ;
\node [above      ] at (x)    {${x}$}           ;
\node [below      ] at (z)    {${z}$}           ;
\node [below      ] at (w2*)  {${w_2^\ast}$}    ;
\node [below      ] at (w4*)  {${w_4^\ast}$}    ;
\node [below      ] at (w2k*) {${w_{2k}^\ast}$} ;

\filldraw [] (w1)   circle [radius=2pt];
\filldraw [] (w2)   circle [radius=2pt];
\filldraw [] (w3)   circle [radius=2pt];
\filldraw [] (w4)   circle [radius=2pt];
\filldraw [] (w*)   circle [radius=2pt];
\filldraw [] (w2k-1)circle [radius=2pt];
\filldraw [] (w2k)  circle [radius=2pt];
\filldraw [] (x)    circle [radius=2pt];
\filldraw [] (y)    circle [radius=2pt];
\filldraw [] (z)    circle [radius=2pt];
\filldraw [] (w2*)  circle [radius=2pt];
\filldraw [] (w4*)  circle [radius=2pt];
\filldraw [] (w2k*) circle [radius=2pt];

\end{tikzpicture}

\caption{Регулярный вид шкал класса $\scls{C}_0^\ast$}
\label{c-ex:C0ast}
\end{figure}

\begin{proposition}
  Существует элементарно определимый класс $\scls{C}^\ast_0$ шкал Крипке, такой, что $L_0 = \mPlogic{\scls{C}^\ast_0}$.
\end{proposition}

\begin{proof}
  Построим $\scls{C}^\ast_0$ так, что он будет содержать шкалы, сходные со шкалами из~$\scls{C}_0$, но каждый мир вида $w_{2k}$ будет <<помечен>> миром $w^\ast_{2k}$ видящим мир~$w_{2k}$.
  Отметим, что $\scls{C}^\ast_0$ может содержать также и шкалы, устроенные иначе, нежели было сказано; в частности, он может содержать бесконечные шкалы; для нас это будет неважно.

  Определим $\scls{C}^\ast_0$ как класс шкал, удовлетворяющих первопорядковому условию (см.~рис.~\ref{c-ex:C0ast}).

  Скажем, что шкалы класса $\scls{C}^\ast_0$ иррефлексивна и не содержат транзитивных цепей из более чем двух элементов:
  $$
  \begin{array}{lcl}
  \Phi_1
    & =
    & \forall x\,\neg \uR(x,x) \wedge \forall x\forall y\forall z\,(\uR(x,y)\wedge \uR(y,z) \imp \neg \uR(x,z)).
  \end{array}
  $$
  Теперь опишем устройство <<нижней части>> шкалы из $\scls{C}^\ast_0$:
  $$
  \begin{array}{rcl}
    \Phi_2 (w_1) & = &
                       \exists w^\ast \exists w_2\exists w^\ast_2\,
                       \big(
                       \forall x\,(\uR(w_1,x)\leftrightarrow x=w^\ast\vee x=w_2)
                       \con {}
    \\&&
         {}\con \uR(w^\ast_2,w_2)
         \con \neg \exists x\, \uR(x, w_1)
         \con \neg \exists x\, \uR(x, w^\ast_2)
         \con {}
    \\&&
         {}\con \forall x\,(\uR(x,w^\ast)\imp x = w_1)
         \con \neg \exists x\, \uR(w^\ast,x)
         \con w_1 \ne w^\ast_2
         \big).
  \end{array}
  $$
  Скажем, что <<нижний>> мир $w_1$~--- это единственный мир, который может видеть более одного мира, и что нет миров, которые видимы из более чем двух миров:
  $$
  \begin{array}{rcl}
    \Phi_3 (w_1) & = &
                       \forall x\,(x\ne w_1 \imp \neg\exists y\exists z\,(y\ne z\wedge \uR(x,y)\con \uR(x,z))) \con {}
    \\&&
         {}\con \forall w\neg\exists x\exists y\exists z\,(x\ne y \con y\ne z\con x\ne z \con {}
    \\&&
         \phantom {{}\con \forall w\neg\exists x\exists y\exists z\,(}
         {}\con \uR(x,w)\con \uR(y,w)\con \uR(z,w)).
  \end{array}
  $$
  Наконец, опишем процедуру расширения шкал класса $\scls{C}^\ast_0$ путём возможной <<приклейки>> к <<верхнему>> миру $w$ двух миров $y$ и $z$, отличных от $w$ и друг от друга, таких, что
  $wRyRx$ и $z$ не лежит в <<цепи>>, но видит мир~$x$:
  $$
  \begin{array}{rcl}
    \Phi_4 & = &
                 \forall w\,\big(\exists y\exists z\,(y\ne z \con \uR(y,w)\con \uR(z,w)) \imp
    \\&&
         \neg\exists x\,\uR(w,x) \vee {}
    \\&&
         {} \vee
         \exists x\exists y\exists z\,(\uR(w,y)\con \uR(y,x)\con \uR(z,x) \con
         y\ne z\con x\ne w \con{}
    \\&&
         {}\con
         \forall u\,(\uR(u,y)\imp u = w) \con \neg\exists u\,\uR(u,z))\big).
  \end{array}
  $$

  Определим $\scls{C}^\ast_0$ как класс шкал Крипке, удовлетворяющих следующей первопорядковой формуле:
  $$
  \begin{array}{rcl}
  \Phi
    & =
    & \exists w_1\, ( \Phi_1 \con \Phi_2 (w_1) \con \Phi_3 (w_1) \con \Phi_4).
  \end{array}
  $$
  Покажем, что $\mPlogic{\scls{C}^\ast_0} = \mPlogic{\scls{C}_0}$.

  Чтобы показать, что $\mPlogic{\scls{C}_0} \subseteq \mPlogic{\scls{C}^\ast_0}$, предположим, что
  $(\kframe{F}, w) \not\models \vp$ для некоторых $\kframe{F}\in\scls{C}^\ast_0$
  и $w$ в~$\kframe{F}$. Пусть $\kframe{F}_w$~--- подшкала шкалы
  $\kframe{F}$, порождённая миром~$w$.

  Пусть $w \ne w_1$, где $w_1$~--- мир, существование которого утверждает формула~$\Phi$. Поскольку только нижний мир $w_1$ шкалы $\kframe{F}$ может видеть два мира, получаем, что
  $\kframe{F}_w$~--- цепочка последовательно достижимых друг из друга миров. Если взять начальный сегмент
  $\kframe{F}_w'$ этой цепочки, длина которой не превосходит $\md \vp + 1$, то $\kframe{F}_w'$ будет изоморфна некоторой порождённой подшкале некоторой шкалы из~$\scls{C}_0$ и при этом
  $(\kframe{F}_w', w) \not\models \vp$.  Значит, в этом случае
  $\vp \notin \mPlogic{\scls{C}_0}$.

  Пусть $w= w_1$. Тогда $w$ видит слепой мир
  $w^\ast$ и последовательность $w_1Rw_2R\ldots$ миров, построенную добавление по два мира, начиная с $w_1$ и $w_2$; следовательно, цепочка
  $w_1Rw_2R\ldots$ либо бесконечна, либо содержит чётное число миров. В любом из этих случаем существует шкала
  $\kframe{F}' \in \scls{C}_0$, являющаяся подшкалой шкалы $\kframe{F}$, такая, что $(\kframe{F}', w_1) \not\models \vp$;
  значит, снова $\vp \notin \mPlogic{\scls{C}_0}$.

  Чтобы показать, что $\mPlogic{\scls{C}^\ast_0} \subseteq \mPlogic{\scls{C}_0}$, заметим, что если $\kframe{F}_n \in\scls{C}_0$, где
  $\kframe{F}_n = \langle W_n, R_n \rangle$, то мы можем сделать каждый мир вида
  $w_{2k} \in {W}_n$ достижимым из нового мира
  $w^\ast_{2k}$, и тогда получившаяся шкала $\kframe{F}^\ast_n$ окажется в~$\scls{C}^\ast_0$.  Поскольку для каждого $w \in W_n$ подшкала шкалы
  $\kframe{F}_n$, порождённая миром $w$, совпадает с подшкалой шкалы
  $\kframe{F}^\ast_n$, порождённой этим же миром $w$, получаем, что
  из того, что $\vp \notin \mPlogic{\scls{C}_0}$, следует, что $\vp \notin \mPlogic{\scls{C}^\ast_0}$.

  Поскольку $L_0 = \mPlogic{\scls{C}_0}$, заключаем, что
  $L_0 = \mPlogic{\scls{C}^\ast_0}$.
\end{proof}


\subsubsection{Пример для класса всех шкал}
\label{c-ex:sec:example-2}

Построим нормальную модальную предикатную логику $L_1$, которая рекурсивно перечислима, полна по Крипке, но при этом не полна относительно элементарно определимых классов шкал.

Для каждого $n \geqslant 2$ определим шкалу Крипке $\kframe{G}_n=\otuple{W_n, R_n}$, положив
$W_n = \set{ w_1, w_2, \ldots, w_n, w^\ast }$ и
$R_n = \set{ \otuple{ w_i, w_{i+1} } : 1 \leqslant i < n } \cup \set{
\otuple{ w_n, w_1 }, \otuple{ w_1, w^\ast } }$, см.~рис.~\ref{c-ex:fig-2}. Пусть
$\scls{C}'$~--- множество всех таких шкал Крипке и пусть
$\scls{C}_1 = \set{ \kframe{G}_n \in \scls{C}' : \mbox{$n$ чётно} }$.
Положим $L_1 = \mPlogic{\scls{C}_1}$.

\begin{figure}
\centering
\begin{tikzpicture}[scale=1.5]

\coordinate (w1) at (+0.0000,-2.0000);
\coordinate (w2) at (+1.4142,-1.4142);
\coordinate (w3) at (+2.0000,+0.0000);
\coordinate (w4) at (+1.4142,+1.4142);
\coordinate (w5) at (+0.0000,+2.0000);
\coordinate (w6) at (-1.4142,+1.4142);
\coordinate (w*) at (+0.0000,-1.0000);
\coordinate (wn) at (-1.4142,-1.4142);
\coordinate (w7) at (-2.0000,+0.0000);

\begin{scope}[>=latex]
\draw [->,  shorten >= 2.5pt, shorten <= 2.5pt]
(wn) -- (w1);
\draw [->,  shorten >= 2.5pt, shorten <= 2.5pt]
(w1) -- (w2);
\draw [->,  shorten >= 2.5pt, shorten <= 2.5pt]
(w2) -- (w3);
\draw [->,  shorten >= 2.5pt, shorten <= 2.5pt]
(w3) -- (w4);
\draw [->,  shorten >= 2.5pt, shorten <= 2.5pt]
(w4) -- (w5);
\draw [->,  shorten >= 2.5pt, shorten <= 2.5pt]
(w5) -- (w6);
\draw [->,  shorten >= 2.5pt, shorten <= 2.5pt]
(w1) -- (w*);
\end{scope}

\draw[dashed, rounded corners=1.5] (w6) -- (w7) -- (wn);

\node [below      ] at (w1) {${w_1}$}           ;
\node [below right] at (w2) {${w_2}$}           ;
\node [      right] at (w3) {${w_3}$}           ;
\node [above right] at (w4) {${w_4}$}           ;
\node [above      ] at (w5) {${w_5}$}           ;
\node [above left ] at (w6) {${w_6}$}           ;
\node [above      ] at (w*) {${w^\ast}$}        ;
\node [below left ] at (wn) {${w_n}$}           ;
\node [      left ] at (w7) {${\phantom{w_7}}$} ;


\filldraw [] (w1) circle [radius=2pt];
\filldraw [] (w2) circle [radius=2pt];
\filldraw [] (w3) circle [radius=2pt];
\filldraw [] (w4) circle [radius=2pt];
\filldraw [] (w5) circle [radius=2pt];
\filldraw [] (w6) circle [radius=2pt];
\filldraw [] (w*) circle [radius=2pt];
\filldraw [] (wn) circle [radius=2pt];

\end{tikzpicture}

\caption{Шкала $\kframe{G}_n$}
\label{c-ex:fig-2}
\end{figure}

Чтобы показать, что $L_1$ рекурсивно перечислима, построим её погружение в логику~$\logic{QCl}^=$.
Пусть $G_n$~--- формула, описывающая устройство шкалы $\kframe{G}^{\uplus}_n=\bigsqcup\set{\kframe{G}_m\in \scls{C}_1 : m \leqslant n}$; нетрудно понять, что $G_n$ строится эффективно.

Для модальной предикатной формулы~$\vp$ положим
$$
\begin{array}{rcl}
  \bar{\vp} & = & \bm{tkf}\wedge G_{\md \vp + 3} \imp \forall x\, (\uW(x) \imp \mstared{\varphi}{x}).
\end{array}
$$

\begin{lemma}
\label{c-ex:lem:L1intoQClE}
Для каждой замкнутой модальной предикатной формулы $\vp$ имеет место следующая эквивалентность:
$$
\begin{array}{lcl}
\vp \in L_1
 & \iff
 & \bar{\vp} \in \logic{QCl}^=.
\end{array}
$$
\end{lemma}

\begin{proof}
  Импликация $(\Rightarrow)$ обосновывается аналогично тому, как это сделано в доказательстве леммы~\ref{c-ex:lem:Lw0intoQClE}.

  Для обоснования импликации $(\Leftarrow)$ предположим, что $\vp \notin L_1$; это означает, что
  $(\kModel{M}, \bar{w}) \not\models \vp$ для некоторой модели $\kModel{M}$, определённой на шкале
  $\kframe{G}_n\in\scls{C}_1$, и некоторого мира $\bar{w}$ в~$\kModel{M}$. Покажем, что $\vp$ опровергается в некоторой шкале $\kframe{G}_m\in\scls{C}_1$, где $m \leqslant \md \vp + 3$.

  Если $\bar{w} = w^\ast$, тогда нетрудно видеть, что
  $(\kModel{M}', w^\ast) \not\models \vp$ для некоторой модели
  $\kModel{M}'$, определённой на шкале~$\kframe{G}_2$.

  Пусть $\bar{w} = w_k$ для некоторого $k \in \set{1, \ldots, n}$.
  Если $n \leqslant \md\vp + 3$, то доказывать нечего.
  В противном случае пусть $\kframe{G}'$~--- либо шкала $\kframe{G}_{\md \vp + 2}$, либо шкала
 $\kframe{G}_{\md \vp + 3}$, в зависимости от того, которая из них принадлежит классу
  $\scls{C}_1$. Нетрудно видеть, что
  $\kframe{G}'$ содержит мир $w'$, такой, что подшкала шкалы
  $\kframe{G}'$, образованная мирами, достижимыми из $w'$ не более чем за
  $\md \vp$ шагов, изоморфна подшкале шкалы $\kframe{G}_n$, образованной мирами, достижимыми из $\bar{w}$ не более чем за $\md \vp$ шагов.
  Значит, $(\kModel{M}', w') \not\models \vp$ для некоторой модели $\kModel{M}'$, определённой на шкале~$\kframe{G}'$.

  Следовательно, согласно свойствам стандартного перевода модальных формул в классические, существует классическая модель, в которой опровергается формула $\bm{tkf}\wedge G_{\md\vp + 3} \imp \forall x\,
  (\uW(x) \imp \mstared{\varphi}{x})$.
  Значит, $\bar{\vp} \notin \logic{QCl}^=$.
\end{proof}

\begin{lemma}
  \label{c-ex:lem:L1re}
  Логика $L_1$ рекурсивно перечислима.
\end{lemma}

\begin{proof}
  Следует из леммы~\ref{c-ex:lem:L1intoQClE}.
\end{proof}

Осталось показать, что $L_1$ не полна относительно элементарно определимых классов шкал. Чтобы сделать это, определим следующие формулы:
$$
\begin{array}{lclr}
  \beta_n & = & \displaystyle
                \Diamond \Box {\bot} \con \Diamond^n \Diamond \Box
                {\bot} \con \bigwedge\limits_{k=2}^{n-1} \neg \Diamond^k \Diamond \Box
                {\bot};
  \\
  \gamma & = & \Diamond \Box {\bot} \dis (\Diamond p \imp \Box p);
         & \smallskip\\
  \delta^k_{n} & = & \Diamond^k \beta_n \con p \imp \Diamond^n p;
         & \smallskip\\
  \varepsilon_n & = & \beta_n \con p \imp \Box^n (\beta_n \con p),
\end{array}
$$
где $p$~--- пропозициональная буква.

\begin{lemma}
  \label{c-ex:lem:formulas}
  Пусть $\kframe{F}$~--- шкала Крипке и $w$~--- мир в~$\kframe{F}$. Тогда имеют место следующие факты:
  \begin{itemize}
  \item[$(1)$]
  $(\kframe{F},w)\models\beta_n$ тогда и только тогда, когда $w$
  видит слепой мир, видит за $n$ шагов мир, видящий слепой мир, но не видит за $k$ шагов мир, видящий слепой мир, где $k \in \{2, \ldots, n-1\}$;
  \item[$(2)$]
  $\kframe{F}\models\gamma$ тогда и только тогда, когда в $\kframe{F}$ только миры, видящие слепой мир, могут видеть более одного мира;
  \item[$(3)$]
  $\kframe{F}\models\delta^k_{n}$ тогда и только тогда, когда каждый мир в $\kframe{F}$, который может видеть за $k$ шагов мир, где истинна формула $\beta_n$, может также видеть себя за $n$ шагов;
  \item[$(4)$]
  $\kframe{F}\models\varepsilon_n$ тогда и только тогда, когда каждый мир шкалы $\kframe{F}$, в котором истинна формула $\beta_n$, не может видеть за $n$ шагов мир, отличный от себя.
  \end{itemize}
\end{lemma}

\begin{proof}
  Фактически описанные условия являются прочтениями соответствующих формул. Дадим пояснение только для~$\varepsilon_n$.

  Если $\kframe{F}$ содержит мир $w'$, такой, что $(\kframe{F},w) \models \beta_n$ и из $w'$ можно за $n$ попасть в мир $w''$, отличный от~$w'$, то
  $(\kModel{M}, w') \not\models \varepsilon_n$, если $\kModel{M}$~--- модель, определённая на шкале~$\kframe{F}$, в которой $(\kModel{M}, v) \models p$ тогда и только тогда, когда~$v = w$.
\end{proof}


\begin{lemma}
  \label{c-ex:lem:valid_in_L1}
  Логика $L_1$ содержит следующие формулы: $\bm{alt}_2$, $\gamma$,
  $\delta^k_{n}$ для любых $k$ и~$n$, а также $\varepsilon_n$ для любого~$n$.
\end{lemma}

\begin{proof}
  Нетрудно видеть, что каждая шкала из класса $\scls{C}_1$
  удовлетворяет тем свойствам, о которых говорится в лемме~\ref{c-ex:lem:formulas}.
\end{proof}

\begin{lemma}
  \label{c-ex:lem:betas}
  Для каждого $n\in\numNp$
  $$
  \begin{array}{lcl}
  \neg \beta_n \in L_1
   & \iff
   & \mbox{$n$ нечётно.}
  \end{array}
  $$
\end{lemma}

\begin{proof}
  Пусть $n$ нечётно.  Пусть $\kframe{G}_m \in \scls{C}_1$ и $w$~--- мир в~$\kframe{G}_m$. Предположим, что
  $(\kframe{G}_m, w) \models \Diamond \Box {\bot}$. По лемме~\ref{c-ex:lem:formulas}, это возможно только если $w = w_1$. Поскольку ни в одной шкале из класса $\scls{C}_1$ мир $w_1$ не достижим из себя за нечётное число шагов, получаем, что $(\kframe{G}_m, w) \models \neg \beta_n$ для любого мира $w$ шкалы~$\kframe{G}_m$; значит, $\neg \beta_n \in L_1$.

  Пусть $n$ чётно. Тогда, согласно лемме~\ref{c-ex:lem:formulas},
  $(\kframe{G}_{n}, w_1) \models \beta_n$, а значит,
  $(\kframe{G}_{n}, w_1) \not\models \neg \beta_n$. Поскольку
  $\kframe{G}_{n} \in \scls{C}_1$, получаем, что
  $\neg \beta_n \notin L_1$.
\end{proof}

\begin{lemma}
  \label{c-ex:lem:L1notFOdef}
  Логика $L_1$ не полна относительно элементарно определимых классов шкал Крипке.
\end{lemma}

\begin{proof}
  Предположим, что $L_1$ полна относительно некоторого элементарно определимого класса $\scls{C}$ шкал Крипке. Пусть
  $\scls{C}$ определяется первопорядковой формулой $\Phi$ и пусть $n$~--- \defnotion{кванторный ранг}\index{ранг!кванторный}\footnote{Наибольшая глубина вложенных кванторов; определяется аналогично модальной глубине формулы.} формулы~$\Phi$.

  Поскольку ${2^n}$ чётно, по лемме~\ref{c-ex:lem:betas} получаем, что
  $\neg \beta_{2^n} \notin L_1$. Следовательно, $\scls{C}$ содержит такую шкалу $\kframe{F} = \langle W, R \rangle$, что
  $(\kframe{F}, w_1) \models \beta_{2^n}$ для некоторого $w_1 \in W$. Значит, в $\kframe{F}$ имеются миры $w_1, \ldots, w_{2^n + 1}, w_1^\ast, w_{2^n + 1}^\ast$, такие, что $w_1 R w_2 \ldots w_{2^n} R w_{2^n + 1}$, а также
  $w_1 R w_1^\ast$ и $w_{2^n + 1} R w_{2^n + 1}^\ast$. Поскольку
  $\scls{C} \models \gamma$, каждый мир $w_i \in \{ w_2, \ldots, w_{2^n}\}$
  может видеть только один мир, т.е. мир~$w_{i+1}$. Поскольку
  $\scls{C} \models \varepsilon_{2^n}$, по лемме~\ref{c-ex:lem:formulas} получаем, что $w_1 = w_{2^n+1}$. Поскольку
  $\scls{C} \models alt_2$, получаем, что $w^\ast_{2^n+1} = w^\ast_1$ или
  $w^\ast_{2^n+1} = w_2$.  Поскольку $\scls{C} \models \delta^k_{2^n}$ для каждого $k$, можно утверждать, что никакой мир из $X = \{w_1, \ldots, w_{2^n} \}$ не достижим из миров множества $W \setminus X$. Действительно, предположим, что это не так, т.е.
  $v R w$ для некоторых $v \in W \setminus X$ и $w \in X$.
  Тогда $v R^k w_1$ для некоторого $k \in \numN$. Поскольку
  $(\kframe{F},v) \models \delta^k_{2^n}$, существует путь длины $2^n$ из $v$
  в~$v$. Заметим, что этот путь не может содержать миры из~$X$;
  значит, $w$ не входит в этот путь, а значит, $v$ видит не менее двух миров, не являющихся слепыми. Но поскольку
  $(\kframe{F},v) \models \gamma$, мир $v$ должен видеть слепой мир, что противоречит тому, что $(\kframe{F},v) \models \bm{alt}_2$.

  Значит, $\kframe{F}$ выглядит как <<кольцо>> из $2^n$ миров, один элемент которого видит слепой мир, и при этом не существует миров вне этого <<кольца>>, которые видят миры <<кольца>>. Рассмотрим шкалу $\kframe{F}'$, которая выглядит как $\kframe{F}$, но является <<кольцом>>, состоящим из $2^{n}+1$ миров вместо $2^{n}$ миров. Поскольку, согласно лемме~\ref{c-ex:lem:betas},
  $\neg \beta_{2^{n}+1} \in L_1$, получаем, что $\kframe{F}' \notin \scls{C}$.
  Однако, используя игры Эренфойхта--Фрессе (см., например,
  \cite[глава~3]{Libkin}), несложно показать, что $\Phi$ не может отличить
  $\kframe{F}$ от $\kframe{F}'$, что даёт противоречие.
\end{proof}

\begin{theorem}
  Существует рекурсивно перечислимая полная по Крипке нормальная модальная предикатная логика, которая не полна относительно элементарно определимых классов шкал Крипке.
\end{theorem}

\begin{proof}
  Можно взять $L_1$ в качестве такой логики.
\end{proof}

\subsubsection{Замечания}
\label{c-ex:sec:discussion}

Несложно понять, что примеров таких логик можно построить бесконечно много: например, вместо кратности индекса шкал числу два можно использовать кратность трём, четырём, и т.д. Кроме того, справедливость всех утверждений сохранится, если в построениях ограничиться рассмотрением шкал с постоянным областями. Приведённые примеры несложно перестраиваются на случай рефлексивных шкал Крипке. При этом автор не знает, как получить аналогичные примеры логик, полных в классах транзитивных шкал.

\setcounter{savefootnote}{\value{footnote}}
\chapter{Суперинтуиционистские логики}
\setcounter{footnote}{\value{savefootnote}}
      \label{ch:8}
\label{ch:QInt}
  \section{Основные определения и факты}
	\subsection{Синтаксис и семантика Крипке}

Считаем, что интуиционистский предикатный язык~--- это тот же язык $\lang{QL}$, который был описан как классический предикатный язык. Соответственно, \defnotion{интуиционистские предикатные формулы}\index{уяа@формула!предикатная!интуиционистская}~--- это $\lang{QL}$-формулы.

Для оценки истинности интуиционистских предикатных формул будем использовать семантику Крипке.
\defnotion{Интуиционистской шкалой Крипке}\index{уяи@шкала!Крипке!интуиционистская} называем шкалу $\kframe{F} = \langle W,R\rangle$, в которой $R$ рефлексивно, транзитивно и антисимметрично.
\defnotion{Интуиционистской моделью Крипке}\index{модель!Крипке!интуиционистская} на интуиционистской шкале $\kFrame{F}=\otuple{W,R,D}$ называем модель $\kModel{M} = \otuple{\kFrame{F},I}$, в которой интерпретация $I$ удовлетворяет \defnotion{условию наследственности}: для всякой предикатной буквы $P$ и всяких миров $w,w'\in W$
$$
\begin{array}{lcl}
\mbox{$wRw'$} & \imply & I(w,P)\subseteq I(w',P).
\end{array}
$$

Истинность $\lang{QL}$-формулы $\varphi$ в мире $w$ интуиционистской модели $\kModel{M} = \langle W,R,D,I\rangle$ при приписывании $g$ определяется рекурсивно:
\settowidth{\templength}{\mbox{$(\kModel{M},w)\imodels^g\varphi'$ или $(\kModel{M},w)\imodels^g\varphi''$;}}
\settowidth{\templengtha}{\mbox{$w$}}
\settowidth{\templengthb}{$(\kModel{M},w)\imodels^{\phantom{g}} \varphi$ для каждого мира $v$ модели $\kModel{M}$;}
\settowidth{\templengthc}{\mbox{$(\kModel{M},w)\imodels^g P(x_1,\ldots,x_n)$}}
$$
\begin{array}{lcl}
(\kModel{M},w)\imodels^g P(x_1,\ldots,x_n)
  & \leftrightharpoons
  & \parbox{\templengthb}{$\langle g(x_1),\ldots,g(x_n)\rangle \in P^{I, w}$,} \\
\end{array}
$$
\mbox{где $P$~--- $n$-местная предикатная буква;}
\settowidth{\templength}{\mbox{$(\kModel{M},w)\imodels^g\varphi'$ или $(\kModel{M},w)\imodels^g\varphi''$;}}
\settowidth{\templengtha}{\mbox{$w$}}
\settowidth{\templengthb}{\mbox{$(\kModel{M},w)\imodels^{g}\varphi'\to\varphi''$}}
\settowidth{\templengthc}{\mbox{$(\kModel{M},w)\imodels^g P(x_1,\ldots,x_n)$}}
\settowidth{\templengthd}{\mbox{$(\kModel{M},w)\imodels^{\phantom{g}} \varphi$ для каждого мира $w$ модели $\kModel{M}$;}}
$$
\begin{array}{lcl}
\parbox{\templengthc}{{}\hfill\parbox{\templengthb}{$(\kModel{M},w) \not\imodels^g \bot;$}}
  \\
\parbox{\templengthc}{{}\hfill\parbox{\templengthb}{$(\kModel{M},w)\imodels^g\varphi' \wedge \varphi''$}}
  & \leftrightharpoons
  & \parbox[t]{\templength}{$(\kModel{M},w)\imodels^g\varphi'$\hfill и\hfill $(\kModel{M},w)\imodels^g\varphi''$;}
  \\
\parbox{\templengthc}{{}\hfill\parbox{\templengthb}{$(\kModel{M},w)\imodels^g\varphi' \vee \varphi''$}}
  & \leftrightharpoons
  & \parbox[t]{\templength}{$(\kModel{M},w)\imodels^g\varphi'$ или $(\kModel{M},w)\imodels^g\varphi''$;}
  \\
\parbox{\templengthc}{{}\hfill\parbox{\templengthb}{$(\kModel{M},w)\imodels^g\varphi' \to \varphi''$}}
  & \leftrightharpoons
  & \parbox[t]{\templengthd}{для каждого мира $w'\in R(w)$ имеет место $(\kModel{M},w')\not\imodels^g\varphi'$ или $(\kModel{M},w')\imodels^g\varphi''$;}
  \\
\parbox{\templengthc}{{}\hfill\parbox{\templengthb}{$(\kModel{M},w)\imodels^g\forall x\,\varphi'$}}
  & \leftrightharpoons
  & \parbox[t]{\templengthd}{$(\kModel{M},w')\imodels^{h}\varphi'$ для каждого $w'\in R(w)$ и
  каждого $h$, такого, что $h \stackrel{x}{=} g$ и $h(x)\in D_{w'}$;}
  \\
\parbox{\templengthc}{{}\hfill\parbox{\templengthb}{$(\kModel{M},w)\imodels^g\exists x\,\varphi'$}}
  & \leftrightharpoons
  & \parbox[t]{\templengthd}{$(\kModel{M},w)\imodels^{h}\varphi'$ для некоторого $h$, такого, что $h \stackrel{x}{=} g$ и $h(x)\in D_w$.}
\end{array}
$$

Пусть $\kModel{M}$, $\kFrame{F}$, $\kframe{F}$, $\scls{C}$ и $\Scls{C}$~--- это, соответственно, интуиционистская модель Крипке, интуиционистская шкала Крипке с предметными областями, интуиционистская шкала Крипке, класс интуиционистских шкал Крипке и класс интуиционистских шкал Крипке с предметными областями; пусть $w$~--- мир модели $\kModel{M}$; пусть $\varphi$~--- формула со свободными переменными $x_1,\ldots,x_n$. Положим
\settowidth{\templength}{\mbox{$(\kModel{M},w)\imodels^g P(x_1,\ldots,x_n)$}}
\settowidth{\templengthc}{\mbox{$\kModel{M}$}}
\settowidth{\templengthb}{\mbox{$w$}}
\settowidth{\templengtha}{$(\kModel{M},w)\imodels^{\phantom{g}} \varphi$ для каждого мира $w$ модели $\kModel{M}$;}
$$
\begin{array}{rcl}
\parbox{\templength}{{}\hfill$\kModel{M}\imodels \varphi$}
  & \leftrightharpoons
  & \parbox[t]{\templengtha}{$(\kModel{M},\parbox{\templengthb}{$w$})\imodels^{g} \varphi$ для каждого мира $w$ модели $\kModel{M}$}
  \\
  &
  & \mbox{и любого $g$, такого, что $g(x_1),\ldots,g(x_n)\in D_w$;}
  \\
\parbox{\templength}{{}\hfill$\kFrame{F}\imodels \varphi$}
  & \leftrightharpoons
  & \parbox[t]{\templengtha}{$\parbox{\templengthc}{$\kModel{M}$}\imodels \varphi$ для любой $\kModel{M}$, определённой на $\kFrame{F}$;}
  \\
\parbox{\templength}{{}\hfill$\kframe{F}\imodels \varphi$}
  & \leftrightharpoons
  & \parbox[t]{\templengtha}{$\parbox{\templengthc}{$\kModel{M}$}\imodels \varphi$ для любой $\kModel{M}$, определённой на $\kframe{F}$;}
  \\
\parbox{\templength}{{}\hfill$\Scls{C}\imodels \varphi$}
  & \leftrightharpoons
  & \parbox[t]{\templengtha}{$\parbox{\templengthc}{$\kFrame{F}$}\imodels \varphi$ для любой $\kFrame{F}\in\Scls{C}$;}
  \\
\parbox{\templength}{{}\hfill$\scls{C}\imodels \varphi$}
  & \leftrightharpoons
  & \parbox[t]{\templengtha}{$\parbox{\templengthc}{$\kframe{F}$}\imodels \varphi$ для любой $\kframe{F}\in\scls{C}$.}
  \\
\end{array}
$$
Если $\mathfrak{S}\imodels\varphi$, где $\mathfrak{S}$~--- мир модели, шкала или класс шкал, мы говорим, что формула $\varphi$ \defnotion{истинна} в~$\mathfrak{S}$; в противном случае говорим, что $\varphi$ \defnotion{опровергается} в~$\mathfrak{S}$.
Эти понятия и соответствующие обозначения распространяются на множества формул: если $X$~--- множество формул, то $\mathfrak{S}\imodels X$ означает, что $\mathfrak{S}\imodels\varphi$ для каждой~$\varphi\in X$.

Отметим, что в интуиционистских шкалах с предметными областями, удовлетворяющих~\mbox{$(\mathit{LCD})$}, истинна формула $\bm{cd}=\forall x\,(P(x)\vee q) \to \forall x\,P(x) \vee q$, где $P$~--- унарная предикатная буква, а $q$~--- пропозициональная буква; если же шкала Крипке с предметными областями не удовлетворяет условию~\mbox{$(\mathit{LCD})$}, то формула $\bm{cd}$ в ней опровергается.

	\subsection{Логики}

Определим \defnotion{интуиционистскую предикатную логику} $\logic{QInt}$ как множество $\lang{QL}$-формул, истинных в классе всех интуиционистских шкал Крипке; отметим, что логику $\logic{QInt}$ можно задать с помощью исчисления (см., например,~\cite{GShS}). 

Под \defnotion{суперинтуиционистской предикатной логикой}\index{логика!предикатная!суперинтуиционистская} будем понимать любое множество $\lang{QL}$-формул, замкнутое относительно правила предикатной подстановки. 

Для множества $\lang{QL}$-формул ${\Sigma}$ обозначим через $\logic{QInt}+{\Sigma}$ суперинтуиционистскую логику, получающуюся замыканием множества $\logic{QInt}\cup{\Sigma}$ по {\MP}, правилу обобщения и правилу подстановки. Для суперинтуиционистской пропозициональной логики $L$ определим логику $\logic{Q}L$, положив $\logic{Q}L = \logic{QInt}+L$. Так, например, $\logic{QKC} = \logic{QInt}+\logic{KC}$.

  \section{Неразрешимость логик унарного предиката}
	\subsection{Трюк Крипке}
    \label{sec:KripkeTrick:QInt}
\subsubsection{Интуиционистский случай: сложности переноса}

Выше (раздел~\ref{sec:KripkeTrick:QModal}) было показано, как устроен трюк Крипке и как можно его модифицировать; речь шла о модальных предикатных логиках, и теперь обратимся к суперинтуиционистским.

В своей работе~\cite{Kripke62} Крипке отмечает, что можно доказать неразрешимость интуиционистской логики в языке с только лишь унарными предикатными буквами. Но автору неизвестно о работах Крипке, где бы это было сделано.\footnote{Судя по всему, таких работ у Крипке и нет. Более того, аналог трюка Крипке в интуиционистском случае, видимо, не был известен до недавнего времени: если бы он был известен, то он был бы применён в~\cite{KKZ05}, где авторы как раз указывают на неясность того, как это сделать.} Неразрешимость интуиционистской логики унарного предиката доказана~\cite{MMO-1965-1,Gabbay81}, но наша цель~--- показать именно модификацию трюка Крипке для интуиционистского случая.

В интуиционистском языке можно определить модальность $\Box$ следующим образом: $\Box B = B$. Действительно, для интуиционистской модели $\kmodel{M}=\langle W,R,D,I\rangle$ и мира $w\in W$ мы получаем, что
$$
\begin{array}{lclclc}
\kmodel{M},w\imodels B & \iff & \mbox{$\kmodel{M},u\imodels B$ для любого $u\in R(w)$.}
\end{array}
$$
Тогда, действуя по аналогии с модальным случаем, получаем, что $\Diamond B = \neg\Box\neg B = \neg\neg B$. Формула~$\neg\neg B$ обладает нужным свойством:
$$
\begin{array}{lclclc}
\kmodel{M},w\imodels \neg\neg B & \iff & \mbox{$\kmodel{M},u\imodels B$ для некоторого $u\in R(w)$.}
\end{array}
$$
Казалось бы, всё замечательно, мы получили модальности $\Box$ и $\Diamond$, и трюк Крипке можно применять. Но всё не так просто.

Пусть $\varphi = \forall x\forall y\,(\neg\neg P(x,y)\to P(x,y))$. Подставляя вместо $P(x,y)$ формулу $\Diamond (Q_1(x)\wedge Q_2(y))$, т.е. формулу $\neg\neg (Q_1(x)\wedge Q_2(y))$, получаем формулу
$\varphi' = \forall x\forall y\,(\neg\neg \neg\neg (Q_1(x)\wedge Q_2(y))\to \neg\neg (Q_1(x)\wedge Q_2(y)))$. Теперь заметим, что $\varphi\not\in \logic{QInt}$, но $\varphi'\in \logic{QInt}$, т.е. применить трюк Крипке в таком виде не получилось.

Почему так? Давайте вспомним, что в случае транзитивных шкал в модальном случае для применения трюка Крипке в моделях мы использовали условие наследственности вниз (раздел~\ref{sssec:KripkeTrick:insidemodels}). А~в интуиционистских моделях это условие не выполнено. Зато выполнено условие наследственности вверх. Но тогда, казалось бы, вот и решение: поскольку в интуиционистских моделях выполнено условие наследственности вверх, надо вместо $\Diamond (Q_1(x)\wedge Q_2(y))$ использовать формулу $\Box(\neg Q_1(x)\vee \neg Q_2(y))$.

Что ж, давайте попробуем? Заметим, что тогда из формулы $\varphi$ выше получим формулу $\varphi'' = \forall x\forall y\,(\neg\neg (\neg Q_1(x)\vee \neg Q_2(y))\to (\neg Q_1(x)\vee \neg Q_2(y)))$. И теперь получилось хорошо, т.к. $\varphi''\not\in\logic{QInt}$.

Но это не всё, т.к. нужно ещё убедиться, что мы сможем обосновать трюк Крипке для такой подстановки в произвольные формулы. Приведём другой пример. Пусть $\psi = \forall x\forall y\forall z\,(R(x,y)\to R(x,z)\vee R(z,y))$. Эта формула опровергается в модели из одного мира, предметная область которого содержит всего два элемента. Сделав предложенную подстановку, получим формулу $\psi' = \forall x\forall y\forall z\,((\neg Q_1(x)\vee \neg Q_2(y))\to (\neg Q_1(x)\vee \neg Q_2(z))\vee (\neg Q_1(z)\vee \neg Q_2(y)))$, которая, очевидно, принадлежит~$\logic{QInt}$.

Заметим, что эта же формула $\psi$ будет контрпримером и для подстановки формулы $Q_1(x)\vee Q_2(y)$ вместо $P(x,y)$.

Мы не проверяли вариант с импликацией; попробуем: будем использовать подстановку формулы $Q_1(x)\to Q_2(y)$ вместо $P(x,y)$. Выполнив эту подстановку в формулу
$\chi=\forall x\forall y\forall z\forall u\,\neg(\neg P(x,y)\wedge\neg P(z,u)\wedge P(x,u))$,
получим формулу
$\chi'=\forall x\forall y\forall z\forall u\,\neg(\neg (Q_1(x)\to Q_2(y))\wedge\neg (Q_1(z)\to Q_2(u))\wedge (Q_1(x)\to Q_2(u)))$. Осталось заметить, что $\chi$ опровергается в одноэлементной шкале, а $\chi'$ принадлежит~$\logic{QInt}$.

Остановимся с тем, чтобы приводить примеры, ведущие к неудаче. Наша цель была лишь в том, чтобы показать, что предложенные пути не приводят к решению, в результате чего может сложиться впечатление, что с трюком Крипке в интуиционистском случае всё не так просто, как в модальном. Но аналог трюка Крипке для интуиционистского случая, как оказалось, есть. Найденное моделирование выглядит несложным; просто покажем его.

\subsubsection{Интуиционистский случай: трюк Крипке}

Как и раньше, формулу интуиционистского языка будем называть \defnotion{позитивной}, если она не содержит~$\bot$ (в частности, не содержит отрицания). Сначала покажем аналог трюка Крипке для позитивных интуиционистских формул, а затем сделаем замечание о произвольных формулах.

Пусть $\varphi$~--- позитивная интуиционистская формула, содержащая бинарную предикатную букву~$P$ (и, быть может, ещё какие-то предикатные буквы). Пусть $\bar{\varphi}$ получается из $\varphi$ подстановкой формулы $(Q_1(x)\wedge Q_2(y)\to q)\vee p$ вместо $P(x,y)$, где $p$ и $q$~--- пропозициональные буквы.

\begin{lemma}
\label{lem:s-8-2-1:Kripke:trick:int}
Для любой позитивной $\lang{QL}$-формулы~$\varphi$ справедлива следующая эквивалентность:
$$
\begin{array}{lcl}
\varphi\in\logic{QInt} & \iff & \bar{\varphi}\in\logic{QInt}.
\end{array}
$$
\end{lemma}

\begin{proof}
Если $\varphi\in\logic{QInt}$, то $\bar{\varphi}\in\logic{QInt}$ в силу замкнутости интуиционистской логики по правилу подстановки.

Пусть $\varphi\not\in\logic{QInt}$. Тогда
существует модель $\kModel{M} = \otuple{W, R, D, I}$ и мир $w_0 \in W$, такие, что $(\kModel{M},{w_0})\not\imodels{\varphi}$. Модифицируем $\kModel{M}$, расширив её до модели $\kModel{M}' = \otuple{W', R', D', I'}$, опровергающей формулу~$\bar\varphi$. Для каждого $w \in W$ и всех $a, b \in D(w)$, таких, что $(\kModel{M},{w})\not\imodels{P(a, b)}$, добавим к $W$ новый мир $w_{a, b}$, положив $w R' w_{a, b}$, $D'(w_{a,b}=D(w)$, а также
  $$
  \begin{array}{lcl}
    (\kModel{M}', w_{a, b}) \not\imodels q; &  &   \\
    (\kModel{M}', w_{a, b}) \imodels p; & &   \\
    (\kModel{M}', w_{a, b}) \imodels Q_1 (d) & \leftrightharpoons & d = a; \\
    (\kModel{M}', w_{a, b}) \imodels Q_2 (d) & \leftrightharpoons & d = b;
  \end{array}
  $$
пусть при этом всем предикатным буквам, отличным от $Q_1$ и $Q_2$ в новых мирах сопоставлены отношения, истинные на любых наборах элементов предметной области, а самим $Q_1$ и $Q_2$ в мирах из $W$ сопоставлены пустые отношения. Пусть, наконец, $(\kModel{M}', w) \not\models q$.

Тогда для каждой $\psi\in\sub\varphi$, для каждого $w\in W$ и любых $a_1, \ldots, a_m \in D(w)$
$$
\begin{array}{lcl}
(\kModel{M},{w})\imodels{\psi}(a_1, \ldots, a_m)
  & \iff
  & (\kModel{M}',{w})\imodels{\bar\psi}(a_1, \ldots, a_m).
\end{array}
$$
Это утверждение доказывается индукцией по построению формулы~$\psi$.

Если $\psi$~--- атомарная формула, то нетривиален лишь случай, когда $\psi=P(x,y)$.
Если $(\kModel{M},{w})\not\imodels{P(a, b)}$, то в модели $\kModel{M}'$ существует мир $w_{a,b}$, обеспечивающий выполение условия $(\kModel{M}',{w})\not\imodels{(Q_1(a) \con Q_2(b) \imp q) \dis p}$; если же $(\kModel{M},{w})\imodels{P(a, b)}$, то $(\kModel{M}',{w})\imodels{(Q_1(a) \con Q_2(b) \imp q) \dis p}$, поскольку $(\kModel{M}',{u})\not\imodels{Q_1(a)}$ или $(\kModel{M}',{u})\not\imodels{Q_2(b)}$ для каждого $u\in R'(w)$.

Если $\psi = \psi_1 \dis \psi_2$, $\psi = \psi_1 \con \psi_2$ или $\psi = \exists x\, \psi_1$, то доказательство тривиально.

Пусть $\psi = \psi_1 \imp \psi_2$. Предположим, что $(\kModel{M}',{w})\not\imodels{\bar\psi(a_1, \ldots, a_m)}$. Тогда $(\kModel{M}',{u})\imodels{\bar\psi_1(a_1, \ldots, a_m)}$ и $(\kModel{M}',{u})\not\imodels{\bar\psi_2(a_1, \ldots, a_m)}$ для некоторого $u \in R'(w)$. Заметим, что $\bar\psi_2$ позитивна, все её атомарные подформулы, не содержащие $Q_1$ и $Q_2$ истинны в мирах из $W'\setminus W$, формулы вида $(Q_1(x) \con Q_2(y) \imp q) \dis p$ тоже истинны в мирах из $W'\setminus W$. Следовательно, $\bar\psi_2$ истинна в каждом мире из $W'\setminus W$ при любом приписывании. Значит,   $u \in W$, и мы можем применить индукционное предположение. Но тогда $(\kModel{M},{w})\not\imodels{\psi(a_1, \ldots, a_m)}$. Обоснование обратной импликации тривиально.

Случай, когда $\psi = \forall x\, \psi_1$ обосновывается аналогично.
\end{proof}

Как видно из доказательства, переменная $q$ ведёт себя, как константа~$\bot$, но, не используя~$\bot$, мы остались в позитивном фрагменте $\logic{QInt}$. Переменная $p$ используется для тех же целей, для каких использовалась переменная $p$ в модальном случае, когда мы использовали $p$-релятивизацию (раздел~\ref{sssec:KripkeTrick:insidemodels}). Конечно же, эти переменные могут быть заменены формулами с одной одноместной предикатной буквой, например, $\forall x\,Q(x)$ и $\forall x\,Q'(x)$. Таким образом, мы соединили четыре условия: моделирование бинарной буквы унарными (или унарными и пропозициональными) буквами, $p$-релятивизацию, позитивные формулы, сохранение исходного множества предметных переменных в формуле.

Насколько существенным является ограничение на формулы, состоящее в их позитивности? Несложно понять, что это условие не создаёт непреодолимого препятствия, о чём уже говорилось выше (см. раздел~\ref{sec:notes:int} и, в частности, предложение~\ref{prop:Int:positivization}). Действительно, во-первых, довольно часто при описании даже очень сложных условий можно обойтись без константы~$\bot$, во-вторых, даже если этого не удалось избежать, можно сопоставить интуиционистской формуле дедуктивно эквивалентную ей в $\logic{QInt}$ позитивную формулу; покажем, как это можно сделать в предикатном случае. Пусть $\varphi$~--- некоторая интуиционистская формула. Зафиксируем новую пропозициональную букву~$f$. Заменим каждое вхождение константы $\bot$ в $\varphi$ вхождением буквы~$f$, и получившуюся формулу обозначим~$\varphi^f$. Для каждой $n$-арной предикатной буквы $S$, входящей в $\varphi$, определим формулу $f\to \forall x_1\ldots\forall x_n\, S(x_1,\ldots,x_n)$, и конъюнкцию всех таких формул обозначим~$F^f$.

\begin{lemma}
\label{lem:s-8-2-1:pos:int}
Для всякой $\lang{QL}$-формулы $\varphi$ справедлива следующая эквивалентность:
$$
\begin{array}{lcl}
\varphi\in\logic{QInt} & \iff & F^f\to \varphi^f\in\logic{QInt}.
\end{array}
$$
\end{lemma}

\begin{proof}
Пусть $F^\bot$ и $\varphi^\bot$ получаются, соответственно, из $F^f$ и $\varphi^f$ подстановкой $\bot$ вместо~$f$.

Если $F^f\to \varphi^f\in\logic{QInt}$, то $F^\bot\to \varphi^\bot\in\logic{QInt}$. Заметим, что $F^\bot\in\logic{QInt}$, поэтому $\varphi^\bot\in\logic{QInt}$. Осталось заметить, что $\varphi^\bot = \varphi$, т.е. мы показали, что в этом случае $\varphi\in\logic{QInt}$.

Пусть теперь $F^f\to \varphi^f\not\in\logic{QInt}$. Тогда существуют такие интуиционистская модель $\kmodel{M}=\langle W,R,D,I\rangle$ и мир $w\in W$, что $(\kmodel{M},w)\imodels F^f$ и $(\kmodel{M},w)\not\imodels \varphi^f$. Пусть $\kmodel{M}'=\langle W',R',D',I'\rangle$~--- подмодель модели $\kmodel{M}$, в которой $W'=\{u\in W : (\kmodel{M},u)\not\imodels f\}$. Заметим, что $w\in W'$; несложно понять, что $\kmodel{M}'\imodels f\leftrightarrow\bot$ и, как следствие, $(\kmodel{M}',w)\not\imodels \varphi^\bot$. Поскольку $\varphi^\bot=\varphi$, получаем, что $\varphi\not\in\logic{QInt}$.
\end{proof}

Заметим, что при таком моделировании, вообще говоря, множество предметных переменных формулы $F^f\to \varphi^f$ может оказаться шире, чем множество предметных переменных исходной формулы~$\varphi$; чтобы этого не произошло, в формуле~$F^f$ нужно использовать только те переменные, которые имеются в~$\varphi$.

Приведём примеры применения трюка Крипке в интуиционистском случае. В~\cite{KKZ05} доказано, что $\logic{QInt}$ неразрешима в языке с двумя предметными переменными, при этом в доказательстве использовались две бинарные предикатные буквы и бесконечно много унарных. Превращая исходные формулы в позитивные и применяя к получившимся формулам трюк Крипке, получаем, что позитивный монадический фрагмент $\logic{QInt}$ неразрешим, причём уже в языке лишь с двумя предметными переменными. Поскольку позитивный фрагмент $\logic{QInt}$ совпадает с позитивным фрагментом логики слабого закона исключённого третьего $\logic{QKC}=\logic{QInt}+\neg p\vee\neg\neg p$, получаем, что аналогичный результат справедлив для любой логики $L$, содержащейся между $\logic{QInt}$ и~$\logic{QKC}$. Похожий трюк использовался в~\cite{MR:2021:JLC:2}, чтобы доказать отсутствие рекурсивной перечислимости у логики конечных моделей в языке с тремя предметными переменными и унарными предикатными буквами.\footnote{Там показано, что при этом достаточно одной унарной буквы и трёх переменных, но для этого после трюка Крипке использовалось уже другое моделирование.}

Конечно, в интуиционистском случае все условия \ref{cond4}--\ref{cond16} (см. раздел~\ref{sssec:KripkeTrick:conditions}), как и в модальном случае, могут быть ослаблены. Сам трюк Крипке может быть использован и для других логик, например, для предикатных вариантов базисной и формальной логик Виссера. Детальное обсуждение этого мы опускаем, поскольку существенных изменений по отношению к модальному случаю здесь нет.

	\subsection{Позитивные фрагменты с двумя переменными}

Как следует из предыдущего раздела, позитивный монадический фрагмент логики $\logic{QInt}$ неразрешим. Покажем, что это утверждение можно усилить.

Действительно, в~\cite{KKZ05} доказана неразрешимость логики $\logic{QInt}$ в языке с двумя предметными переменными, двумя бинарными предикатными буквами и бесконечным множеством унарных предикатных букв. Тогда, используя лемму~\ref{lem:s-8-2-1:pos:int}, можем элиминировать константу~$\bot$ в формулах, погрузив указанный фрагмент $\logic{QInt}$ в его позитивную часть. Затем, используя лемму~\ref{lem:s-8-2-1:Kripke:trick:int} (т.е. трюк Крипке для интуиционистского случая), моделируем обе бинарные предикатные буквы позитивными формулами от унарных предикатных букв. Как следствие, получаем следующее утверждение.

\begin{proposition}
\label{prop:s8-2-2:pos:unary:int;two:var}
Позитивный монадический фрагмент логики\/ $\logic{QInt}$ неразрешим в языке с двумя предметными переменными.
\end{proposition}

\begin{proof}
Следует из~\cite{KKZ05} и лемм~\ref{lem:s-8-2-1:pos:int} и~\ref{lem:s-8-2-1:Kripke:trick:int}.
\end{proof}

Теперь заметим, что, как и в пропозициональном случае~\cite{Jankov-1968-1-rus}, позитивный фрагмент логики $\logic{QInt}$ совпадает с позитивным фрагментом логики $\logic{QKC}$~\cite{MR:2024:AiML} (но, в отличие от пропозиционального случая, логика $\logic{QKC}$ уже не является наибольшей логикой с таким же позитивным фрагментом, как у логики $\logic{QInt}$). Это даёт нам следующую теорему.

\begin{theorem}
\label{th:prop:s8-2-2:pos:unary:int:two:var}
Позитивный монадический фрагмент любой суперинтуиционистской логики\/ $L\in[\logic{QInt},\logic{QKC}]$ неразрешим в языке с двумя предметными переменными.
\end{theorem}

\begin{proof}
Следует из предложения~\ref{prop:s8-2-2:pos:unary:int;two:var} с учётом того, что позитивные фрагменты логик $\logic{QInt}$ и $\logic{QKC}$ совпадают.
\end{proof}

Ниже мы покажем, что утверждение этой теоремы можно усилить, сократив число унарных букв языка до одной.

Прежде сделаем ещё одно наблюдение. Заметим, что в~\cite{KKZ05} при доказательстве неразрешимости $\logic{QInt}$ в языке с двумя переменными, двумя бинарными предикатными буквами и бесконечным множеством унарных предикатных букв для опровержения возникающих формул строятся модели с постоянными областями, т.е. фактически там получен следующий результат: любая логика из интервала $[\logic{QInt},\logic{QInt.cd}]$ неразрешима в этом языке, причём для всех логик из указанного интервала имеется одно и то же подмножество формул, неразрешимость которого фактически и доказана в~\cite{KKZ05}. Это даёт возможность усилить утверждение теоремы~\ref{th:prop:s8-2-2:pos:unary:int:two:var}.

\begin{theorem}
\label{th:2:prop:s8-2-2:pos:unary:int:two:var}
Позитивный монадический фрагмент любой суперинтуиционистской логики\/ $L\in[\logic{QInt},\logic{QKC.cd}]$ неразрешим в языке с двумя предметными переменными.
\end{theorem}

\begin{proof}
Следует из предложения~\ref{prop:s8-2-2:pos:unary:int;two:var} с учётом того, что логики $\logic{QInt}$ и $\logic{QKC.cd}$ содержат один и тот же неразрешимый фрагмент внутри позитивного монадического фрагмента с двумя предметными переменными.
\end{proof}

	\subsection{Позитивные фрагменты с двумя переменными и одним унарным предикатом}
	   \label{ss:8-2-3}

Покажем, как промоделировать все унарные предикатные буквы языка с помощью формул, содержащих одну унарную предикатную букву, причём без увеличения числа используемых предметных переменных.

Пусть $\varphi$~--- позитивная формула монадического фрагмента $\lang{QL}$ и пусть $P_1, \ldots, P_s$~--- все входящие в неё унарные предикатные буквы. Считаем, что $s \geqslant 2$, поскольку иначе $\varphi$ уже является формулой с одной унарной предикатной буквой. Пусть $P$~--- унарная предикатная буква, отличная от $P_1, \ldots, P_s$.

Определим шкалу
$\kframe{F}_0 = \langle W_0, R_0 \rangle$, которая будет играть важную роль в дальнейших построениях.

Шкала $\kframe{F}_0$ изображена на рис.~\ref{s8-2-3:fig:F-2}; она состоит из миров, образующих \defnotion{уровни}. Имеется четыре верхних мира
$\delta_1$, $\delta_2$, $\delta'_2$ и $\delta_3$; уровень $0$
образуют миры $\alpha^0_1$, $\alpha^0_2$, $\beta^0_1$ и $\beta^0_2$;
уровень $1$ содержит миры $\alpha^1_1$, $\alpha^1_2$, $\alpha^1_3$,
$\beta^1_1$, $\beta^1_2$ и $\beta^1_3$; отношение достижимости между указанными мирами соответствует стрелкам. Остальные миры определяются рекурсивно.

Уровень $k$, где $k \in \{2, \ldots, s + 1\}$, содержит миры
$\alpha^k_l$ и $\beta^k_l$ для каждого $l \in \{1, \ldots, n_k \}$,
где $n_k$ определяется рекурсивно: $n_1 = 3$;
$n_{k+1} = (n_k - 1)^2$. Чтобы определить отношение достижимости между мирами уровня $k+1$ и мирами уровня $k$, возьмём лексико-графическое упорядочение пар элементов множества $\{2, \ldots, n_k\}$ и, считая, что $\langle i, j \rangle$~--- $m$-я пара в этом упорядочении, положим
$$
\begin{array}{lll}
  \alpha^{k+1}_{m} R_0\, \beta^k_1,
    & \alpha^{k+1}_{m} R_0\, \alpha^k_i,
    & \alpha^{k+1}_{m} R_0\, \beta^k_j,
    \smallskip\\
  \beta^{k+1}_{m} R_0\, \alpha^k_1,
    & \beta^{k+1}_{m} R_0\, \alpha^k_i,
    & \beta^{k+1}_{m} R_0\, \beta^k_j.
\end{array}
$$
Заметим, что $n_k \geqslant k$ для каждого $k \geqslant 1$; в частности,
$n_{s+1} \geqslant s+1$.  Следовательно, уровень $s+1$ содержит не менее $s+1$
миров вида $\alpha^{s+1}_i$ и столько же миров вида~$\beta^{s+1}_i$.

Пусть $\kModel{N}_a = \langle \kframe{F}_0\odot\cal{A}, I \rangle$~--- интуиционистская модель Крипке с постоянной областью $\mathcal{A}$, содержащей~$a$; считаем также, что $\mathcal{A}$ содержит не менее трёх элементов. Говорим, что $\kModel{N}_a$ является \defnotion{$a$-подходящей}, если для некоторого $a' \in \mathcal{A}\setminus \{a\}$,
\begin{itemize}
\item $I(\delta_2, P) = \mathcal{A} \setminus \{a\}$;
\item $I(\delta'_2, P) = \{a'\}$;
\item $I(\delta_3, P) = \{a, a'\}$;
\item $I(\beta_1^0, P) = \{a'\}$;
\item $I(w, P) = \varnothing$ для каждого
  $w \in W_0 \setminus \{ \delta_2^{\phantom{1}}, \delta'_2,
  \delta_3^{\phantom{1}}, \beta_1^0\}$.
\end{itemize}

\begin{figure}
  \centering
  \begin{tikzpicture}[scale=2.25]

\coordinate (d1)  at (2, 7);
\coordinate (d3)  at (3, 7);
\coordinate (d2)  at (4, 7);
\coordinate (d2a) at (5, 7);
\coordinate (a01) at (2, 6);
\coordinate (a02) at (3, 6);
\coordinate (b01) at (4, 6);
\coordinate (b02) at (5, 6);
\coordinate (b13) at (1, 5);
\coordinate (b12) at (2, 5);
\coordinate (b11) at (3, 5);
\coordinate (a13) at (4, 5);
\coordinate (a12) at (5, 5);
\coordinate (a11) at (6, 5);

\coordinate (dots1) at (3.5, 4.25);

\coordinate (ak1) at (1, 3.5);
\coordinate (aki) at (2, 3.5);
\coordinate (bkj) at (5, 3.5);
\coordinate (bk1) at (6, 3.5);

\coordinate (dots1ak) at (1.5, 3.5);
\coordinate (dots2ak) at (2.5, 3.5);
\coordinate (dots1bk) at (5.5, 3.5);
\coordinate (dots2bk) at (4.5, 3.5);

\coordinate (akp1m) at (5, 2.5);
\coordinate (bkp1m) at (2, 2.5);

\coordinate (dots1akp1m) at (4.5, 2.5);
\coordinate (dots2akp1m) at (5.5, 2.5);
\coordinate (dots1bkp1m) at (1.5, 2.5);
\coordinate (dots2bkp1m) at (2.5, 2.5);

\coordinate (dots2) at (3.5, 1.75);

\coordinate (as1)  at (1, 1);
\coordinate (as2)  at (2, 1);
\coordinate (asns) at (3, 1);
\coordinate (bs1)  at (6, 1);
\coordinate (bs2)  at (5, 1);
\coordinate (bsns) at (4, 1);

\coordinate (dots1as) at (2.5, 1);
\coordinate (dots1bs) at (4.5, 1);

\draw [] (d1)  circle [radius=2.0pt] ;
\draw [] (d3)  circle [radius=2.0pt] ;
\draw [] (d2)  circle [radius=2.0pt] ;
\draw [] (d2a) circle [radius=2.0pt] ;
\draw [] (a01) circle [radius=2.0pt] ;
\draw [] (a02) circle [radius=2.0pt] ;
\draw [] (b01) circle [radius=2.0pt] ;
\draw [] (b02) circle [radius=2.0pt] ;
\draw [] (b13) circle [radius=2.0pt] ;
\draw [] (b12) circle [radius=2.0pt] ;
\draw [] (b11) circle [radius=2.0pt] ;
\draw [] (a13) circle [radius=2.0pt] ;
\draw [] (a12) circle [radius=2.0pt] ;
\draw [] (a11) circle [radius=2.0pt] ;

\draw [] (dots1) node {$\ldots$};

\draw [] (ak1) circle [radius=2.0pt] ;
\draw [] (aki) circle [radius=2.0pt] ;
\draw [] (bkj) circle [radius=2.0pt] ;
\draw [] (bk1) circle [radius=2.0pt] ;

\draw [] (dots1ak) node {$\ldots$};
\draw [] (dots2ak) node {$\ldots$};
\draw [] (dots1bk) node {$\ldots$};
\draw [] (dots2bk) node {$\ldots$};

\draw [] (akp1m) circle [radius=2.0pt] ;
\draw [] (bkp1m) circle [radius=2.0pt] ;

\draw [] (dots1akp1m) node {$\ldots$};
\draw [] (dots2akp1m) node {$\ldots$};
\draw [] (dots1bkp1m) node {$\ldots$};
\draw [] (dots2bkp1m) node {$\ldots$};

\draw [] (dots2) node {$\ldots$};

\draw [] (as1)  circle [radius=2.0pt] ;
\draw [] (as2)  circle [radius=2.0pt] ;
\draw [] (asns) circle [radius=2.0pt] ;
\draw [] (bs1)  circle [radius=2.0pt] ;
\draw [] (bs2)  circle [radius=2.0pt] ;
\draw [] (bsns) circle [radius=2.0pt] ;

\draw [] (dots1as) node {$\ldots$};
\draw [] (dots1bs) node {$\ldots$};

\begin{scope}[>=latex]
\draw [->, shorten >= 4.00pt, shorten <= 4.00pt] (a01) -- (d1);
\draw [->, shorten >= 4.00pt, shorten <= 4.00pt] (a01) -- (d3);
\draw [->, shorten >= 4.00pt, shorten <= 4.00pt] (a02) -- (d1);
\draw [->, shorten >= 4.00pt, shorten <= 4.00pt] (a02) -- (d2);
\draw [->, shorten >= 4.00pt, shorten <= 4.00pt] (b01) -- (d3);
\draw [->, shorten >= 4.00pt, shorten <= 4.00pt] (b01) -- (d2);
\draw [->, shorten >= 4.00pt, shorten <= 4.00pt] (b02) -- (d1);
\draw [->, shorten >= 4.00pt, shorten <= 4.00pt] (b02) -- (d2);
\draw [->, shorten >= 4.00pt, shorten <= 4.00pt] (b02) -- (d3);
\draw [->, shorten >= 4.00pt, shorten <= 4.00pt] (a02) -- (d2a);
\draw [->, shorten >= 4.00pt, shorten <= 4.00pt] (b01) -- (d2a);
\draw [->, shorten >= 4.00pt, shorten <= 4.00pt] (b02) -- (d2a);

\draw [->, shorten >= 4.00pt, shorten <= 4.00pt] (b13) -- (a01);
\draw [->, shorten >= 4.00pt, shorten <= 4.00pt] (b13) -- (a02);
\draw [->, shorten >= 4.00pt, shorten <= 4.00pt] (b12) -- (a01);
\draw [->, shorten >= 4.00pt, shorten <= 4.00pt] (b12) -- (b01);
\draw [->, shorten >= 4.00pt, shorten <= 4.00pt] (b11) -- (a01);
\draw [->, shorten >= 4.00pt, shorten <= 4.00pt] (b11) -- (b02);
\draw [->, shorten >= 4.00pt, shorten <= 4.00pt] (a13) -- (a02);
\draw [->, shorten >= 4.00pt, shorten <= 4.00pt] (a13) -- (b01);
\draw [->, shorten >= 4.00pt, shorten <= 4.00pt] (a12) -- (a02);
\draw [->, shorten >= 4.00pt, shorten <= 4.00pt] (a12) -- (b02);
\draw [->, shorten >= 4.00pt, shorten <= 4.00pt] (a11) -- (b01);
\draw [->, shorten >= 4.00pt, shorten <= 4.00pt] (a11) -- (b02);

\draw [->, shorten >= 4.00pt, shorten <= 4.00pt] (bkp1m) -- (ak1);
\draw [->, shorten >= 4.00pt, shorten <= 4.00pt] (bkp1m) -- (aki);
\draw [->, shorten >= 4.00pt, shorten <= 4.00pt] (bkp1m) -- (bkj);
\draw [->, shorten >= 4.00pt, shorten <= 4.00pt] (akp1m) -- (bk1);
\draw [->, shorten >= 4.00pt, shorten <= 4.00pt] (akp1m) -- (aki);
\draw [->, shorten >= 4.00pt, shorten <= 4.00pt] (akp1m) -- (bkj);

\end{scope}

\node [      left ] at (d1)  {$\delta_1\mbox{~}$}    ;
\node [      right] at (d2)  {$\mbox{~}\delta_2$}    ;
\node [      right] at (d2a) {$\mbox{~}\delta'_2$}   ;
\node [      left ] at (d3)  {$\delta_3\mbox{~}$}    ;
\node [      left ] at (a01) {$\alpha^0_1\mbox{~}$}  ;
\node [      left ] at (a02) {$\alpha^0_2\mbox{~~}$} ;
\node [      left ] at (b01) {$\beta^0_1\mbox{~~}$}  ;
\node [      right] at (b02) {$\mbox{~}\beta^0_2$}   ;

\node [below right] at (a11) {$\alpha^1_1$} ;
\node [below right] at (a12) {$\alpha^1_2$} ;
\node [below right] at (a13) {$\alpha^1_3$} ;
\node [below right] at (b11) {$\beta^1_1$}  ;
\node [below right] at (b12) {$\beta^1_2$}  ;
\node [below right] at (b13) {$\beta^1_3$}  ;

\node [above right] at (ak1) {$\alpha^k_1$} ;
\node [above right] at (aki) {$\alpha^k_i$} ;
\node [above right] at (bk1) {$\beta^k_1$}  ;
\node [above right] at (bkj) {$\beta^k_j$}  ;

\node [below right] at (akp1m) {$\alpha^{k+1}_m$} ;
\node [below right] at (bkp1m) {$\beta^{k+1}_m$}  ;

\node [above right] at (as1)  {$\alpha^{s+1}_1$}         ;
\node [above right] at (as2)  {$\alpha^{s+1}_2$}         ;
\node [above right] at (asns) {$\alpha^{s+1}_{n_{s+1}}$} ;
\node [above right] at (bs1)  {$\beta^{s+1}_1$}          ;
\node [above right] at (bs2)  {$\beta^{s+1}_2$}          ;
\node [above right] at (bsns) {$\beta^{s+1}_{n_{s+1}}$}  ;

\node [above      ] at (d3)   {$\mbox{~}\begin{array}{l}P(a)\\P(a')\end{array}$}   ;
\node [above      ] at (d2)   {$\mbox{~}\begin{array}{l}P(b),\\b\ne a\end{array}$} ;
\node [      right] at (b01)  {$\mbox{~}P(a')$}         ;
\node [above right] at (d2a)  {$\mbox{~}P(a')$}         ;

\end{tikzpicture}

\caption{Общий вид $a$-подходящей модели на шкале $\kframe{F}_0$}
  \label{s8-2-3:fig:F-2}
\end{figure}

Определим формулы в языке с одной предметной переменной и единственной предикатной буквой~$P$, соответствующие мирам $a$-подходящей модели.
Пусть
$$
\begin{array}{lcl}
  D_1      & = & \exists x\, P(x); \\
  D_2(x)   & = & \exists x\, P(x) \imp P(x); \\
  D_3(x)   & = & P(x) \imp \forall x\, P(x); \\
  A_1^0(x) & = & D_2(x) \imp D_1 \dis D_3(x); \\
  A_2^0(x) & = & D_3(x) \imp D_1 \dis D_2(x); \\
  B_1^0(x) & = & D_1 \imp D_2(x) \dis D_3(x); \\
  B_2^0(x) & = & A_1^0(x) \con A_2^0(x) \con B_1^0(x) \imp D_1 \dis D_2(x) \dis D_3(x); \\
  A_1^1(x) & = & A_1^0(x) \con A_2^0(x) \imp B_1^0(x) \dis B_2^0(x);   \\
  A_2^1(x) & = & A_1^0(x) \con B_1^0(x) \imp A_2^0(x) \dis B_2^0(x); \\
  A_3^1(x) & = & A_1^0(x) \con B_2^0(x) \imp A_2^0(x) \dis B_1^0(x);  \\
  B_1^1(x) & = & A_2^0(x) \con B_1^0(x) \imp A_1^0(x) \dis B_2^0(x); \\
  B_2^1(x) & = & A_2^0(x) \con B_2^0(x) \imp A_1^0(x) \dis B_1^0(x); \\
  B_3^1(x) & = & B_1^0(x) \con B_2^0(x) \imp A_1^0(x) \dis A_2^0(x). \\
\end{array}
$$
Если $\otuple{i,j}$~--- $m$-я пара в лексико-графическом перечислении пар элементов множества $\set{2,\ldots,n_k}$, то положим
$$
\begin{array}{lcl}
  A_m^{k+1}(x) & = & A_1^k(x) \imp B_1^k(x) \dis A_i^k(x) \dis B_j^k(x); \\
  B_m^{k+1}(x) & = & B_1^k(x) \imp A_1^k(x) \dis A_i^k(x) \dis B_j^k(x).
\end{array}
$$

\begin{lemma}
  \label{s8-2-3:lem:frame-F}
  Пусть $\kModel{N}_a = \langle W_0, R_0, D, I \rangle$~--- некоторая $a$-подходящая модель, $w \in W_0$, $k \in \{1, \ldots, s+1 \}$ и $m \in \{ 1, \ldots, n_{s+1} \}$.
  Тогда
  $$
  \begin{array}{lcl}
  \kModel{N}_a, {w} \not\imodels {A_m^k}(a) & \Longleftrightarrow & w R_0 \alpha_m^k; \\
  \kModel{N}_a, {w} \not\imodels {B_m^k}(a) & \Longleftrightarrow & w R_0 \beta_m^k.
  \end{array}
  $$
\end{lemma}

\begin{proof}
Сначала заметим, что
  $$
  \begin{array}{lcl}
    \{w\in W_0: (\kModel{N}_a,w) \not\imodels D_1\}\cap\{\delta_1,\delta_2,\delta'_2,\delta_3\}
    & =
    & \{\delta_1\};
	\\
    \{w\in W_0: (\kModel{N}_a,w) \not\imodels D_2(a)\}\cap\{\delta_1,\delta_2,\delta'_2,\delta_3\}
    & =
    & \{\delta_2,\delta'_2\};
	\\
    \{w\in W_0: (\kModel{N}_a,w) \not\imodels D_3(a)\}\cap\{\delta_1,\delta_2,\delta'_2,\delta_3\}
    & =
    & \{\delta_3\}.
	\\
  \end{array}
  $$

  Покажем, что для всякого $w\in W_0$ и всякого $i \in \{1, 2, 3\}$
  $$
  \begin{array}{lcl}
    (\kModel{N}_a,w) \not\imodels D_i
    & \Longleftrightarrow
    & \mbox{$w R_0 \delta_i$ или $w R_0 \delta'_i$.}
  \end{array}
  $$
  Замечание: конечно, здесь имеется в виду, что $\delta'_i$ может быть достижимым только при $i=2$, поскольку $\delta'_1$ и $\delta'_3$ не определялось.

  Пусть $i=1$. Заметим, что
  $$
  \begin{array}{lcl}
  \{ w\in W_0: (\kModel{N}_a,w) \imodels \exists x\, P(x) \}
    & =
    & \{ \delta_2, \delta'_2, \delta_3, \beta^0_1 \}.
  \end{array}
  $$

  Заметим, что верхними мирами, начиная с уровня $0$ и выше, которые не видят $\delta_1$, являются $\delta_2$, $\delta'_2$, $\delta_3$ и $\beta^0_1$. Кроме того,
  каждый мир уровня $1$ видит~$\delta_1$, а значит, для каждого $k \geqslant 1$ каждый мир уровня $k+1$ тоже видит~$\delta_1$. Поэтому
  $$
  \begin{array}{lclcl}
    \{ w\in W_0 : w R_0 \delta_1\} & = & W_0 \setminus \{ \delta_2, \delta'_2, \delta_3,
                                  \beta^0_1 \} & = & \{ w\in W_0: (\kModel{N}_a,w) \not\imodels D_1 \}.
  \end{array}
  $$

  Пусть $i=2$. Заметим, что
  $$
  \begin{array}{lcl}
  \{ w\in W_0: (\kModel{N}_a,w) \imodels \exists x\, P(x) \mbox{ и } (\kModel{N}_a,w) \not\imodels P(a) \}
    & =
    & \{ \delta_2, \delta'_2, \beta^0_1 \}.
  \end{array}
  $$
  Поэтому
  $$
  \begin{array}{lcl}
  \{ w\in W_0: (\kModel{N}_a,w) \not\imodels D_2(a) \}
      & =
      & \{ w\in W_0: w R_0 \delta_2 \mbox{ или } w R_0 \delta'_2 \mbox{ или } w R_0 \beta^0_1 \}.
  \end{array}
  $$
  Поскольку $\beta^0_1 R_0 \delta_2$, $\beta^0_1 R_0 \delta'_2$ и $R$ транзитивно, получаем, что
  $$
  \begin{array}{lcl}
  \{ w\in W_0: (\kModel{N}_a,w) \not\imodels D_2(a) \}
      & =
      & \{ w\in W_0: w R_0 \delta_2 \} \cup \{ w\in W_0 : w R_0 \delta'_2 \}.
  \end{array}
  $$

  Пусть $i=3$. Заметим, что
  $$
  \begin{array}{lcl}
  \{ w\in W_0: (\kModel{N}_a,w) \imodels P(a) \mbox{ и } (\kModel{N}_a,w) \not\imodels \forall x\, P(x) \}
    & =
    & \{ \delta_3 \}.
  \end{array}
  $$
  Поэтому
  $$
  \begin{array}{lcl}
    \{ w\in W_0 : (\kModel{N}_a,w) \not\imodels D_3(a) \}
       & =
       & \{ w\in W_0: w R_0 \delta_3 \}.
  \end{array}
  $$

  Теперь докажем утверждение леммы индукцией по~$k$.

  Пусть $k = 0$. Заметим, что $(\kModel{N}_a,w) \not\imodels A^0_1(a)$ выполнено тогда и только тогда, когда существует такой мир $w' \in R_0(w)$, что
  $$
  \begin{array}{lll}
  (\kModel{N}_a,w') \imodels D_2(a);
    & (\kModel{N}_a,w') \not\imodels D_1;
    & (\kModel{N}_a,w') \not\imodels D_3(a),
  \end{array}
  $$
  т.е.
  $$
  \begin{array}{llll}
  \delta_2 \notin R_0(w');
    & \delta'_2 \notin R_0(w');
    & w' R_0 \delta_1;
    & w' R_0 \delta_3.
  \end{array}
  $$
  Эти условия дают, что $w' = \alpha^0_1$. Следовательно,
  $(\kModel{N}_a,w) \not\imodels A^0_1(a)$ тогда и только тогда, когда $w R_0 \alpha^0_1$. Обоснование для $A^0_2$, $B^0_1$ и $B^0_2$ проводится аналогично.

  Пусть $k = 1$. Тогда $(\kModel{N}_a,w) \not\imodels A^1_1(a)$ означает, что существует такой мир $w' \in R_0(w)$, что
  $$
  \begin{array}{llll}
     (\kModel{N}_a,w') \imodels A^0_1(a);
    & (\kModel{N}_a,w') \imodels A^0_2(a);
    & (\kModel{N}_a,w') \not\imodels B^0_1(a);
    & (\kModel{N}_a,w') \not\imodels B^0_2(a),
  \end{array}
  $$
  т.е.
  $$
  \begin{array}{llll}
    \alpha^0_1 \notin R_0(w');
    & \alpha^0_2 \notin R_0(w');
    & w' R_0 \beta^0_1;
    & w' R_0 \beta^0_2.
  \end{array}
  $$
  Тогда $w' = \alpha^1_1$. Следовательно,
  $(\kModel{N}_a,w) \not\imodels A^1_1(a)$ тогда и только тогда, когда $w R_0 \alpha^1_1$. Обоснование для $A^1_2$, $A^1_3$, $B^1_1$, $B^1_2$ и $B^1_3$ проводится аналогично.

  Пусть утверждение леммы верно для $k$, где $1\leqslant k\leqslant s$; покажем, что оно будет справедливым и для $k + 1$.

  Действительно, $(\kModel{N}_a,w) \not\imodels A^{k+1}_m(a)$ тогда и только тогда, когда существует такой мир $w' \in R_0(w)$, что
  $$
  \begin{array}{llll}
     (\kModel{N}_a,w') \imodels A^k_1(a);
    & (\kModel{N}_a,w') \not\imodels B^k_1(a);
    & (\kModel{N}_a,w') \not\imodels A^k_i(a);
    & (\kModel{N}_a,w') \not\imodels B^k_j(a),
  \end{array}
  $$
  т.е., согласно индукционному предположению,
  $$
  \begin{array}{llll}
  \alpha^k_1 \notin R_0(w');
    & w' R_0 \beta^k_1;
    & w' R_0 \alpha^k_i;
    & w' R_0 \beta^k_j.
  \end{array}
  $$

  Покажем, что $w'$ является миром уровня $k + 1$.

  Каждый мир $\beta_r^{k+1}$, видит мир
  $\alpha^k_1$; каждый мир уровня $k + 2$ видит мир
  $\beta_l^{k+1}$ для некоторого $l$; каждый мир уровня $t \geqslant k + 2$ видит как минимум один мир уровня $k + 2$. Поэтому каждый мир уровня $t \geqslant k + 2$ видит
  $\alpha^k_1$. Следовательно, $w'$ является миром некоторого уровня $t \leqslant k + 1$.
  Никакой мир уровня $t \leqslant k$ не видит одновременно миры
  $\beta^k_1$, $\alpha^k_i$ и $\beta^k_j$. Значит, $w'$ является миром уровня
  $k+1$.

  Единственным миром уровня $k+1$, который видит $\beta^k_1$,
  $\alpha^k_i$ и $\beta^k_j$, является $\alpha^{k+1}_m$, а значит,
  $w' = \alpha^{k+1}_m$.

  Обоснование для $B^{k+1}_m$ проводится аналогично.
\end{proof}

\begin{lemma}
  \label{s8-2-3:lem:frame-F-b}
  Пусть $\kModel{N}_a = \langle W_0, R_0, D, I \rangle $~--- некоторая $a$-подходящая модель и пусть
  $b$~--- элемент её предметной области, отличный от~$a$.  Тогда для каждого $w \in W_0$, каждого
  $k \in \{2, \ldots, s+1 \}$ и каждого
  $m \in \{ 1, \ldots, n_{s+1} \}$,
  $$
  \begin{array}{lcl}
  (\kModel{N}_a, {w}) \imodels {A_m^k}(b)
    & \mbox{и}
    & (\kModel{N}_a, {w}) \imodels {B_m^k}(b).
  \end{array}
  $$
\end{lemma}

\begin{proof}
  Рассмотрим два случая: $b=a'$ и $b\ne a'$.

  Пусть $b = a'$. Заметим, что для каждого $w \in W_0$ условие
  $P^{I, w} \ne \varnothing$ влечёт, что $(\kModel{N}_a,w) \imodels P(a')$.
  Поэтому
  $$
  \begin{array}{lcl}
    \{ w\in W_0 : (\kModel{N}_a,w) \imodels D_2(a') \} & = & W_0.
  \end{array}
  $$
  Следовательно,
  $$
  \begin{array}{ccccc}
  \{ w\in W_0 : (\kModel{N}_a,w) \imodels A^0_2(a') \}
     & =
     & \{ w\in W_0 : (\kModel{N}_a,w) \imodels B^0_1(a') \}
     & = \\
     & =
     & \{ w\in W_0 : (\kModel{N}_a,w) \imodels B^0_2(a') \}
     & =
     & W_0,
  \end{array}
  $$
  а значит, $(\kModel{N}_a,w) \imodels A^1_i (a')$ и $(\kModel{N}_a,w) \imodels B^1_i (a')$ для каждого
  $w \in W_0$ и каждого $i \in \{1, 2, 3\}$.

  Пусть $b \ne a'$. Заметим, что тогда
  $$
  \begin{array}{lcl}
    \{ w\in W_0 : (\kModel{N}_a,w) \imodels D_1 \}     & = & \{ \delta_2, \delta'_2, \delta_3, \beta^0_1 \}; \\
    \{ w\in W_0 : (\kModel{N}_a,w) \imodels D_2 (b) \} & = & \{ \delta_1, \delta_2 \}; \\
    \{ w\in W_0 : (\kModel{N}_a,w) \imodels D_3 (b) \} & = & \{ \delta_1, \delta'_2, \delta_3, \alpha^0_1 \},
  \end{array}
  $$
  а значит,
  $$
  \begin{array}{lcl}
    \{ w\in W_0 : (\kModel{N}_a,w) \imodels A^0_1(b) \} & = & W_0.
  \end{array}
  $$
  Поэтому $(\kModel{N}_a,w) \imodels B^1_i (b)$ для каждого $w \in W_0$ и каждого $i \in \{1, 2, 3\}$.
\end{proof}

Пусть ${\vp}^\hash$~--- формула, получающаяся из зафиксированной выше позитивной формулы $\varphi$ монадического фрагмента языка $\lang{QL}$ подстановкой
$$
\begin{array}{lcl}
A_r^{s+1}(x) \dis B_r^{s+1}(x) & \mbox{вместо} & P_r(x)
\end{array}
$$
для каждого $r \in \{1, \ldots, s\}$.

\begin{lemma}
  \label{s8-2-3:lem:QInt-main-lemma}
Пусть $L \in \{ \logic{QInt}, \logic{QKC} \}$.
Тогда имеют место следующие эквивалентности:
$$
\begin{array}{lcl}
\varphi\in L & \iff & \varphi^\hash\in L; \\
\varphi\in L\logic{.cd} & \iff & \varphi^\hash\in L\logic{.cd}. \\
\end{array}
$$
\end{lemma}

\begin{proof}
  Импликация \mbox{$(\Rightarrow)$} выполняется во всех случаях, т.к. логики замкнуты по правилу подстановки.

  Докажем импликацию \mbox{$(\Leftarrow)$} в первой эквивалентности для случая $L=\logic{QInt}$.

  Пусть
  $\vp \notin \logic{QInt}$.  Пусть
  $\kModel{M} = \langle W, R, D, I \rangle$~--- интуиционистская модель, в мире $w_0$ которой опровергается формула~$\varphi$. Без ограничений общности можем считать, что предметная область каждого мира этой модели содержит не менее трёх элементов.

  Для каждой пары $w\in W$ и $a \in {D}_w$ определим
  $\kframe{F}^{a}_w = \langle \{ w \}\times\{ a \}\times W_0^{\phantom{0}}, R^{a}_w
  \rangle$
  как изоморфную копию шкалы $\kframe{F}_0$ при изоморфизме $f\colon v \mapsto \langle w, a, v \rangle$.

  Пусть
  $$
  \begin{array}{lcl}
  W' & = & \displaystyle W \cup \bigcup\set{\{ w \}\times\{ a \}\times W_0^{\phantom{0}} : \mbox{$w\in W$, $a\in D_w$}}.
  \end{array}
  $$

  Пусть $S$~--- наименьшее отношение на $W'$, удовлетворяющее следующим условиям:
  \begin{itemize}
  \item $R \subseteq S$;
  \item $R^{a}_w \subseteq S$ для каждого $w\in W$ и каждого $a\in D_w$;
  \item для каждого $w \in W$, каждого $v \in W' \setminus W$, каждого
    $a \in {D}_w$ и каждого $r \in \{ 1, \ldots, s \}$,
    $$
    \begin{array}{lcll}
      w S v
        & \leftrightharpoons
        & \mbox{либо}
        & \mbox{$v\in \{\langle w, a, \alpha^{s+1}_r \rangle, \langle w, a, \beta^{s+1}_r \rangle \}$
          и $(\kModel{M}, w) \not\imodels P_r(a)$,}
        \\
        &
        & \mbox{либо}
        & \mbox{$v\in\{\langle w, a,\alpha^{s+1}_{s+1}\rangle,
          \langle w, a,\beta^{s+1}_{s+1}\rangle\}$.}
    \end{array}
    $$
  \end{itemize}
  Определим $R'$ как рефлексивно-транзитивное замыкание отношения~$S$.

  Пусть $D'_w = {D}_w$ и $D'_{u} = {D}_w$ для каждого $w \in W$ и каждого $u\in R'(w)\setminus W$.

  Определим $I'$ на шкале
  $\langle W', R',D' \rangle$ так, чтобы для любых $w\in W$ и $a\in D_w$ были выполнены следующие условия:
  \begin{itemize}
  \item $I'(\langle w, a, \delta_2 \rangle, P) = {D}_w \setminus
    \{a\}$;
  \item $I'(\langle w, a, \delta'_2 \rangle, P) = \{a'\}$, где $a'$~--- некоторый фиксированный элемент из $D_w$, отличный от $a$ и определяемый однозначно по $a$ и~$w$;
  \item $I'(\langle w, a, \delta_3 \rangle, P) = \{a, a'\}$;
  \item $I'(\langle w, a, \beta_1^0 \rangle, P) = \{a'\}$;
  \item $I'(u, P) = \varnothing$ в остальных случаях.
  \end{itemize}
  Положим $\kModel{M}' = \langle W', R',D',I' \rangle$.

\pagebreak[3]

  Заметим, что $I'$ удовлетворяет условию наследственности; следовательно,
  $\kModel{M}'$ является интуиционистской моделью.

  Нетрудно видеть, что для каждого $w\in W$ и каждого $a \in {D}_w$ подмодель модели
  $\kModel{M}'$, порождённая множеством миров
  $$
  \{\langle w, a, \alpha^{s+1}_1 \rangle, \ldots, \langle w, a, \alpha^{s+1}_{n_{s+1}}
  \rangle, \langle w, a, \beta^{s+1}_1 \rangle, \ldots, \langle w, a,
  \beta^{s+1}_{n_{s+1}} \rangle \},
  $$
  является $a$-подходящей моделью на шкале, изоморфной шкале $\kframe{F}_0$ при изоморфизме $f\colon v \mapsto \langle w, a, v \rangle$. Тогда по~\ref{s8-2-3:lem:frame-F}, для любых $w\in W$, $a \in {D}_w$, $v \in \{ w \} \times \{ a \} \times W_0$, $k \in \{1, \ldots, s+1 \}$ и $m \in \{ 1, \ldots, n_{s+1} \}$
  $$
  \begin{array}{lcl}
           (\kModel{M}', {v}) \not\imodels {A_m^k}(a)
             & \Longleftrightarrow
             & v R' \langle w, a, \alpha_m^k \rangle;
    \smallskip \\
           (\kModel{M}', {v}) \not\imodels {B_m^k}(a)
             & \Longleftrightarrow
             & v R'' \langle w, a, \beta_m^k \rangle.
  \end{array}
  \eqno{(\ast)}
  $$
  Кроме того, по лемме~\ref{s8-2-3:lem:frame-F-b}, если $b \in {D}_w \setminus \{a\}$, то
  $$
  \begin{array}{lcl}
    (\kModel{M}', {v}) \imodels {A_m^k}(b)
      & \mbox{и}
      & (\kModel{M}', {v}) \imodels {B_m^k}(b).
  \end{array}
  \eqno{({\ast}{\ast})}
  $$

  \begin{sublemma}
    \label{s8-2-3:lem:sublemma-2}
    Для любых $w \in W$ и $a \in {D}_w$
    $$
    \begin{array}{lcl}
      (\kModel{M}', {w}) \not\imodels {A_1^s}(a)
        & \mbox{и}
        & (\kModel{M}', {w}) \not\imodels {B_1^s}(a).
    \end{array}
    $$
  \end{sublemma}

  \begin{proof}
    Согласно определению $R'$, получаем, что
    $w R' \langle w, a, \alpha^{s+1}_{s+1} \rangle$ и
    $w R' \langle w, a, \beta^{s+1}_{s+1} \rangle$.

    Поскольку
    $\langle w, a, \alpha^{s+1}_{s+1} \rangle R' \langle w, a,
    \beta^{s}_{1} \rangle$,
     получаем
    $(\kModel{M}',\langle w, a, \alpha^{s+1}_{s+1} \rangle) \not\imodels B^s_1(a)$ по ($\ast$), и тогда $(\kModel{M}',w) \not\imodels B^s_1(a)$. Аналогично, $(\kModel{M}',w) \not\imodels A^s_1(a)$.
  \end{proof}

  Для формулы $\psi \in \sub \vp$ обозначим через $\psi'$ формулу, получающуюся из $\psi$ подстановкой
  $$
  \begin{array}{lcl}
    A_r^{s+1}(x) \dis B_r^{s+1}(x) & \mbox{вместо} & P_r(x)
  \end{array}
  $$
  для каждого $r \in \{1, \ldots, s\}$. В частности, $\varphi'=\varphi^\hash$.

  \begin{sublemma}
    \label{s8-2-3:lem:sublemma}
    Для каждой $\psi \in \sub \vp$, каждого $v \in W' \setminus W$ и любого приписывания~$g$ выполнено условие
    $(\kModel{M}', v) \imodels^g \psi'$.
  \end{sublemma}

  \begin{proof}
  Поскольку $\vp$ позитивна, формула $\psi'$ строится из формул
    $A_i^{s+1}(x) \dis B_i^{s+1}(x)$, где
    $i \in \{1, \ldots, n_{s+1}\}$, с помощью $\con$, $\dis$, $\imp$, а также кванторов.
    Следовательно, достаточно показать, что для каждого $v \in W' \setminus W$,
каждого приписывания $g$ и каждого $i \in \{1, \ldots, n_{s+1}\}$
$$
(\kModel{M}',v) \imodels^g A_i^{s+1}(x) \dis B_i^{s+1}(x).
$$

    Из того, что $v \in W' \setminus W$, следует, что
    $v = \langle u, a, w \rangle$ для некоторых $u\in W$, $a \in {D}_u$ и $w \in W_0$.
    Рассмотреть два случая: $g(x) = a$ и $g(x) \ne a$.

    Пусть $g(x) = a$. По определению шкалы $\kframe{F}^a_u$, для каждого
    $i \in \{ 1, \ldots, n_{s+1} \}$ мир $\langle u, a, w \rangle$
    не видит $\langle u, a, \alpha_i^{s+1} \rangle$ или не видит
    $\langle u, a, \beta_i^{s+1} \rangle$.  Тогда по ($\ast$) получаем, что
    $(\kModel{M}',v) \imodels^g A_i^{s+1}(x)$ или $(\kModel{M}',v) \imodels^g B_i^{s+1}(x)$,
    откуда следует, что $(\kModel{M}',v) \imodels^g A_i^{s+1}(x) \dis B_i^{s+1}(x)$.

    Пусть $g(x) \ne a$. В этом случае
    $(\kModel{M}',v) \imodels^g A_i^{s+1}(x) \dis B_i^{s+1}(x)$ по~(${\ast}{\ast}$).
  \end{proof}

  Теперь покажем, что $(\kModel{M}', w_0) \not\imodels {\vp}^\hash$.

  Для этого мы докажем, что каждой формулы $\theta \in \sub \vp$, для каждого мира $w \in W$ и каждого приписывания~$g$
  $$
  \begin{array}{lcl}
  (\kModel{M}, w) \imodels^g \theta
    & \iff
    & (\kModel{M}', w) \imodels^g \theta'.
  \end{array}
  $$

  Обоснование проведём индукцией по~$\theta$.

  Пусть $\theta = P_r (x)$ для некоторого $r \in \{1,\ldots, s\}$. Тогда
  $\theta' = A_r^{s+1}(x) \dis B_r^{s+1}(x)$.

  Пусть $(\kModel{M}, w) \not\imodels P_r (a)$. По определению
  $\kModel{M}'$ получаем, что $w R' \langle w, a, \alpha^{s+1}_r \rangle$ и
  $w R' \langle w, a, \beta^{s+1}_r \rangle$.  По условию~($\ast$) получаем, что
  $(\kModel{M}', \langle w, a, \alpha^{s+1}_r \rangle) \not\imodels A_r^{s+1}(a)$
  и
  $(\kModel{M}', \langle w, a, \beta^{s+1}_r \rangle) \not\imodels B_r^{s+1}(a)$.
  Но тогда $(\kModel{M}', w) \not\imodels A_r^{s+1}(a) \dis B_r^{s+1}(a)$.

  Пусть
  $(\kModel{M}', w) \not\imodels A_r^{s+1}(a) \dis B_r^{s+1}(a)$. В этом случае
  $(\kModel{M}', w) \not\imodels A_r^{s+1}(a)$ и
  $(\kModel{M}', w) \not\imodels B_r^{s+1}(a)$. Значит, существуют такие
  $u', u'' \in W'$ и $i, j \in \{2, \ldots, n_s\}$, что
  $u', u'' \in R'(w)$ и
  $$
  \begin{array}{lcl}
  \begin{array}{l}
    (\kModel{M}',u') \imodels\hfill A^s_1 (a); \\
    (\kModel{M}',u') \not\imodels\hfill B^s_1 (a); \\
    (\kModel{M}',u') \not\imodels\hfill A^s_i (a); \\
    (\kModel{M}',u') \not\imodels\hfill B^s_j (a); \\
  \end{array}
  & &
  \begin{array}{l}
    (\kModel{M}',u'') \imodels\hfill B^s_1 (a); \\
    (\kModel{M}',u'') \not\imodels\hfill A^s_1 (a);\\
    (\kModel{M}',u'') \not\imodels\hfill A^s_i (a);\\
    (\kModel{M}',u'') \not\imodels\hfill B^s_j (a).\\
  \end{array}
  \end{array}
  \eqno{({\ast}{\ast}{\ast})}
  $$

  Покажем, что $u' = \langle w, a, \alpha^{s+1}_r \rangle$ и
  $u'' = \langle w, a, \beta^{s+1}_r \rangle$.

  Поскольку $(\kModel{M}',u') \imodels A^s_1 (a)$ и $(\kModel{M}',u'') \imodels B^s_1 (a)$, по подлемме~\ref{s8-2-3:lem:sublemma-2} получаем, что $u', u'' \in W' \setminus W$. Тогда из того, что $(\kModel{M}',u') \not\imodels B^s_1 (a)$ и $(\kModel{M}',u'') \not\imodels A^s_1 (a)$ по (${\ast}{\ast}$) получаем, что $u', u'' \in \{w\}\times \{a\} \times W_0$.
  Тогда из (${\ast}{\ast}{\ast}$) и ($\ast$) следует, что для некоторых $i, j \in \{ 2, \ldots, n_s \}$,
  $$
  \begin{array}{lcl}
  \begin{array}{l}
    u' \notin\hfill R'(\langle w, a, \alpha^s_1 \rangle);\\
    u' \in\hfill R'(\langle w, a, \beta^s_1 \rangle);\\
    u' \in\hfill R'(\langle w, a, \alpha^s_i \rangle);\\
    u' \in\hfill R'(\langle w, a, \beta^s_j \rangle); \\
  \end{array}
    & &
  \begin{array}{l}
    u'' \notin\hfill R'(\langle w, a, \beta^s_1 \rangle);\\
    u'' \in\hfill R'(\langle w, a, \alpha^s_1 \rangle);\\
    u'' \in\hfill R'(\langle w, a, \alpha^s_i \rangle);\\
    u'' \in\hfill R'(\langle w, a, \beta^s_j \rangle).\\
  \end{array}
  \end{array}
  $$
  В шкале $\kframe{F}_0$, а значит, и в $\kframe{F}^a_w$, только миры уровня $s+1$ видят более одного мира уровня~$s$. Значит, $u'$ и $u''$ являются мирами уровня $s+1$. Тогда, с учётом (${\ast}{\ast}{\ast}$), получаем, что
  $(\kModel{M}', u') \not\imodels A_r^{s+1}(a)$ и
  $(\kModel{M}', u'') \not\imodels B_r^{s+1}(a)$. Тогда, с учётом ($\ast$), получаем, что
  $u' R' \langle w, a, \alpha_r^{s+1} \rangle$ и
  $u'' R' \langle w, a, \beta_r^{s+1} \rangle$.

  Значит, $u' = \langle w, a, \alpha^{s+1}_r \rangle$ и $u'' = \langle w, a, \beta^{s+1}_r \rangle$, и следовательно,
  $w R' \langle w, a, \alpha^{s+1}_r \rangle$ и
  $w R' \langle w, a, \beta^{s+1}_r \rangle$.

  Тогда $(\kModel{M}', w) \not\imodels P_r(a)$.

  Если $\theta = \psi \dis \chi$, $\theta = \psi \con \chi$ или
  $\theta = \exists x\, \psi$, то доказательство тривиально.

  Рассмотрим случай, когда $\theta = \psi \imp \chi$.

  Пусть $(\kModel{M}, w) \not\imodels^g \psi \imp \chi$. Тогда
  $(\kModel{M}, v) \imodels^g \psi$ и
  $(\kModel{M}, v) \not\imodels^g \chi$ для некоторого $v\in R(w)$. Тогда, согласно индукционному предположению,
  $(\kModel{M}', v) \imodels^g \psi'$ и
  $(\kModel{M}', v) \not\imodels^g \chi'$, а следовательно,
  $(\kModel{M}', w) \not\imodels^g \psi' \imp \chi'$.

  Пусть $(\kModel{M}', w) \not\imodels^g \psi' \imp \chi'$.
  Тогда $(\kModel{M}', v) \imodels^g \psi'$ и
  $(\kModel{M}', v) \not\imodels^g \chi'$ для некоторого $v\in R'(w)$. По подлемме~\ref{s8-2-3:lem:sublemma}, $v \in W$, и значит, $w R v$. Тогда, согласно индукционному предположению,
  $(\kModel{M}, v) \imodels^g \psi$ и
  $(\kModel{M}, v) \not\imodels^g \chi$, а значит,
  $(\kModel{M}, w) \not\imodels^g \psi \imp \chi$.

  Рассмотрим случай, когда $\theta = \forall x\,\psi$.

  Пусть $(\kModel{M}, w) \not\imodels^g \forall x\, \psi$. Тогда
  $(\kModel{M}, v) \not\imodels^{g'} \psi$ для некоторого $v\in R(w)$ и некоторого приписывания $g'$, такого, что $g' \stackrel{x}{=} g$. Согласно индукционному предположению,
  $(\kModel{M}', v) \imodels^{g'} \psi'$, а значит,
  $(\kModel{M}', w) \not\imodels^g \forall x\, \psi'$.

  Пусть $(\kModel{M}', w) \not\imodels^g \forall x\, \psi'$.
  Тогда $(\kModel{M}', v) \not\imodels^{g'} \psi'$ для некоторого мира $v\in R'(w)$ и некоторого приписывания $g'$, такого, что $g' \stackrel{x}{=} g$.
  По подлемме~\ref{s8-2-3:lem:sublemma}, $v \in W$, и значит, $w R v$.  Тогда, согласно индукционному предположению, $(\kModel{M}, v) \not\imodels^{g'} \psi$, и значит, $(\kModel{M}, w) \not\imodels^g \forall x\, \psi$.

  Теперь, поскольку $w_0 \in W$ и $\varphi'=\varphi^\hash$, получаем, что $(\kModel{M}', w_0) \not\imodels {\vp}^\hash$, а значит,
  ${\vp}^\hash \notin \logic{QInt}$.

Рассуждения в для $\logic{QInt.cd}$ аналогичны: достаточно заметить, что если в построениях выше модель $\kModel{M}$ удовлетворяла условию $(\mathit{LCD})$, то и модель $\kModel{M}'$ тоже ему удовлетворяет.

  Теперь докажем импликацию \mbox{$(\Leftarrow)$} в первой эквивалентности для случая $L=\logic{QKC}$.

  Изменим определение модели
  $\kModel{M}'$ в конструкции выше, чтобы получить модель на шкале логики $\logic{QKC}$ (при этом теперь можно считать, что исходная модель $\kModel{M}$ конвергентна). Пусть
  \begin{itemize}
  \item $W'' = W' \cup \{ w^\ast \}$;
  \item $R'' = R' \cup \{ \langle w, w^\ast \rangle : w \in W'' \}$;
  \item $D''_w = {D}'(w)$ для каждого $w \in W'$;
  \item $D''_{w^\ast} = D^+$.
  \end{itemize}
  Пусть также $I''$~--- интерпретация на шкале
  $\langle W'', R'', D'' \rangle$, удовлетворяющая следующим условиям:
  \begin{itemize}
  \item $I'' (w, P) = I'(w, P)$ для каждого $w \in W'$;
  \item $I'' (w^\ast, P) = {D}^+$.
  \end{itemize}
  Наконец, пусть $\kModel{M}'' = \langle W'', R'', D'', I'' \rangle$.

  Поскольку каждый мир множества $W''$ видит $w^\ast$, шкала
  $\langle W'', R'' \rangle$ является конвергентной, т.е. это
  $\logic{QKC}$-шкала. Поскольку $I''$ удовлетворяет условию наследственности, $\kModel{M}''$~--- интуиционистская модель. Заметим, что добавление к $W'$ мира, в котором каждая элементарная формула истинна, не меняет истинность позитивных формул в других мирах множества~$W$,\footnote{Это наблюдение получило ряд обобщений, см.~\cite{MR:2024:AiML}.} поэтому
  $\kModel{M}'', w_0 \not\imodels {\vp}^\hash$, а значит,
  ${\vp}^\hash \notin \logic{QKC}$.

Рассуждения в для $\logic{QKC.cd}$ аналогичны: достаточно заметить, что если в построениях выше исходная $\logic{QKC}$-модель $\kModel{M}$ удовлетворяла условию $(\mathit{LCD})$, то и модель $\kModel{M}''$ тоже ему удовлетворяет.
\end{proof}

Теперь, используя тот факт, что $\logic{QInt}$ неразрешима в языке с двумя переменными, двумя бинарными предикатными буквами и бесконечным множеством унарных предикатных букв~\cite{KKZ05}, получаем следующее утверждение.

\begin{theorem}
  \label{s8-2-3:thr:int-main}
  Пусть $L\in [\logic{QInt}, \logic{QKC.cd}]$. Тогда позитивный фрагмент $L$ в языке с двумя предметными переменными и одной унарной предикатной буквой является\/ $\Sigma^0_1$-трудным.
\end{theorem}

\begin{proof}
Следует из теоремы~\ref{th:2:prop:s8-2-2:pos:unary:int:two:var} и леммы~\ref{s8-2-3:lem:QInt-main-lemma}.
\end{proof}

Получим следствия, касающиеся некоторых конкретных логик. Первое из них касается интуиционистской предикатной логики, а также предикатных вариантов логики Крайзеля--Патнэма, логики Медведева и логики Янкова.

\begin{corollary}
  \label{s8-2-3:cor:int-main}
  Пусть $L \in \{\logic{QInt}, \logic{QKP}, \logic{QLM}, \logic{QKC} \}$.  Тогда $L$ и $L\logic{.cd}$ являются\/ $\Sigma^0_1$-полными в языке с двумя предметными переменными и одной унарной предикатной буквой.
\end{corollary}

Поскольку каждая непротиворечивая суперинтуиционистская пропозициональная логика, отличная от классической пропозициональной логики $\logic{Cl}$ и
аксиоматизируемая формулой с одной пропозициональной переменной, является
подлогикой логики $\logic{KC}$~\cite{Nishimura60}, из теоремы~\ref{s8-2-3:thr:int-main} получаем ещё одно следствие.

\begin{corollary}
  \label{s8-2-3:cor:int-main-2}
  Пусть $L = \logic{Int} + \vp$, где $\vp$~--- с одной пропозициональной переменной, и пусть $L \subset \logic{Cl}$. Тогда логики\/
  $\logic{Q}L$ и\/ $\logic{Q}L\logic{.cd}$ являются\/
  $\Sigma^0_1$-полными в языке с двумя предметными переменными и одной унарной предикатной буквой.
\end{corollary}

  \section{Разрешимые логики унарных предикатов}

Покажем, что результаты, представленные в разделе~\ref{s7-3} для модальных предикатных логик, переносятся и на суперинтуиционистские. Действительно, воспользуемся \defnotion{переводом Гёделя}\index{перевод!Гёделя}~$\mathsf{T}$ интуиционистских формул в модальные:
$$
\begin{array}{rcl}
\mathsf{T}(\bot) & = & \Box\bot;
\\
\mathsf{T}(P(x_1,\ldots,x_n)) & = & \Box P(x_1,\ldots,x_n);
\\
\mathsf{T}(\varphi\wedge\psi) & = & \mathsf{T}(\varphi)\wedge\mathsf{T}(\psi);
\\
\mathsf{T}(\varphi\vee\psi) & = & \mathsf{T}(\varphi)\vee\mathsf{T}(\psi);
\\
\mathsf{T}(\varphi\to\psi) & = & \Box(\mathsf{T}(\varphi)\to\mathsf{T}(\psi));
\\
\mathsf{T}(\exists x\,\varphi) & = & \exists x\, \mathsf{T}(\varphi);
\\
\mathsf{T}(\forall x\,\varphi) & = & \Box\forall x\, \mathsf{T}(\varphi),
\end{array}
$$
где $P(x_1,\ldots,x_n)$~--- атомарная формула. Если $\kFrame{F}$~--- интуиционистская предикатная шкала, то для всякой $\lang{QL}$-формулы~$\varphi$
$$
\begin{array}{rcl}
\kFrame{F}\imodels\varphi & \iff & \kFrame{F}\models\mathsf{T}(\varphi).
\end{array}
$$

Для класса интуиционистских шкал Крипке $\scls{C}$ определим $\iPlogic{\scls{C}}$ как множество интуиционистских предикатных формул, истинных в~$\scls{C}$. Отметим, что $\iPlogic{\scls{C}}$ является суперинтуиционистской предикатной логикой. Логику $\iPlogic{\scls{C}}$ называем \defnotion{логикой класса~$\scls{C}$}.\index{логика!класса шкал} Аналогично для класса $\Scls{C}$ предикатных шкал.
Для класса шкал Крипке $\scls{C}$ определим логику $\iPlogicC{\scls{C}}$ как множество формул, истинных в классе всех предикатных шкал с постоянными областями, построенных на шкалах из~$\scls{C}$. Будем писать $\iPlogic{\kframe{F}}$ и $\iPlogicC{\kframe{F}}$ вместо $\iPlogic{\set{\kframe{F}}}$ и $\iPlogicC{\set{\kframe{F}}}$ соответственно, где $\kframe{F}$~--- интуиционистская шкала Крипке.

\begin{proposition}
  \label{prop:finite-frame:int}
  Пусть\/ $\kframe{F}$~--- конечная интуиционистская шкала Крипке. Тогда монадические фрагменты логик\/ $\iPlogic{\kframe{F}}$ и\/ $\iPlogicC{\kframe{F}}$ разрешимы.
\end{proposition}

\begin{proof}
Следует из предложения~\ref{prop:finite-frame} с учётом перевода Гёделя.
\end{proof}

\begin{proposition}
  \label{prop:finite-frame-pred:int}
  Пусть\/ $\kFrame{F}$~--- конечная интуиционистская предикатная шкала Крипке. Тогда монадический фрагмент логики\/ $\iPlogic{\kFrame{F}}$ разрешим.
\end{proposition}

\begin{proof}
Следует из предложения~\ref{prop:finite-frame-pred} с учётом перевода Гёделя.
\end{proof}

То же касается и логик с равенством. В этом случае перевод Гёделя содержит следующий пункт для равенства:
$$
\begin{array}{rcl}
\mathsf{T}(x=y) & = & \Box (x=y).
\end{array}
$$
Интуиционистские предикатные шкалы с равенством определяются так же, как и для модального языка с равенством (см.~\cite{GShS}). Аналогом $\lang{QML}^=$\nobreakdash-формулы $\forall x\forall y\,(x\ne y\to \Box(x\ne y))$ служит $\lang{QL}^=$\nobreakdash-формула $\forall x\forall y\,(x=y\vee x\ne y)$, а равенство, удовлетворяющее этой формуле, называется \defnotion{разрешимым}. Естественным образом перенося определения и обозначения, введённые для модальных логик с равенством, на суперинтуиционистские, получаем следующее утверждение.

\begin{proposition}
  \label{prop:finite-frame:eq:int}
  Пусть\/ $\kframe{F}$~--- конечная интуиционистская шкала Крипке. Тогда монадические фрагменты с равенством логик\/ $\iPlogicx{\kframe{F}}{=}$, $\iPlogicCx{\kframe{F}}{=}$, $\iPlogicx{\kframe{F}}{\simeq}$ и\/ $\iPlogicCx{\kframe{F}}{\simeq}$ разрешимы.
\end{proposition}

\begin{proof}
Следует из предложения~\ref{prop:finite-frame:eq} с учётом перевода Гёделя.
\end{proof}

  \section{Логики первопорядково определимых классов шкал}

В этом разделе сделаем лишь короткое замечание. Поскольку, по модулю перевода Гёделя, суперинтуиционистские предикатные логики могут рассматриваться как фрагменты модальных предикатных логик, из конструкций, описанных в разделе~\ref{sec:7-4}, получаем следующее утверждение.

\begin{proposition}
\label{prop:cor:prop:Mpred:std-translation:int}
Пусть $L$~--- суперинтуиционистская предикатная логика, полная относительно некоторого элементарно определимого класса интуиционистских предикатных шкал Крипке. Тогда $L$ является рекурсивно перечислимой.
\end{proposition}

\begin{proof}
Аналогично следствию~\ref{cor:prop:Mpred:std-translation}.
\end{proof}

Это известный факт~\cite{MR:2000:IFRAS,GShS}, и мы не будем останавливаться подробно на его обсуждении.

  \section{Логики конечных моделей}
	\subsection{Неперечислимость}
	   \label{ss:8-5-1}

Отсутствие рекурсивной перечислимости предикатной суперинтуиционистской логики конечных интуиционистских шкал Крипке следует из работ Д.П.\,Скворцова~\cite{Skv88,Skv95,Skvortsov05JSL}. Покажем, как можно получить подобные результаты из теоремы Трахтенброта, причём используя всего три предметные переменные и одну унарную предикатную букву.

Пусть $L$~--- суперинтуиционистская предикатная логика, для которой существуют $L$-шкалы с конечным множеством миров. Определим логику $\wfin{L}$ как множество формул, истинных в классе всех конечных шкал Крипке логики~$L$.

Логика $\logic{QCl}_{\mathit{fin}}$ является $\Pi^0_1$-полной в языке с одной бинарной предикатной буквой и тремя предметными переменными, причём таковым является уже позитивный фрагмент этого языка (см.~теорему~\ref{th:trakhtenbrot:positive}); пусть $\pQCLFs$~--- соответствующий её фрагмент, а $\plangQCLFs$~--- соответствующий фрагмент языка~$\lang{QL}$. В построениях ниже будем считать, что $\plangQCLFs$ содержит бинарную букву $S$ и предметные переменные~$x,y,z$.

Покажем, как свести $\Pi^0_1$-полное множество $\pQCLFs$ к позитивному фрагменту каждой
логики интервала
$[\logic{QInt}_{\mathit{wfin}}, \logic{QLC}_{\mathit{wfin}}.\logic{cd}]$,
содержащему только формулы от не более чем трёх предметных
переменных и предикатных букв арности не более двух.

Пусть $\vp$~--- замкнутая $\plangQCLFs$-формула, содержащая не более трёх предметных
переменных и одну бинарную предикатную букву~$S$.

Пусть $q$~--- пропозициональная буква, $T$~--- унарная предикатная буква, а $\prec$ и $\approx$~--- бинарные предикатные буквы, отличные от~$S$.
Пусть
$$
\begin{array}{rcl}
  \mathit{Min}( x ) & = &  \forall y\, (  x \prec y \dis x \approx y ); \\
  \mathit{Max}( x ) & = & \forall y\, (y \prec x); \\
  x \triangleleft y & = &  x \prec y \con  \forall z\, ( x \prec z  \con
                           z \prec y \imp z \approx y). \\
\end{array}
$$
Определим следующие формулы:
$$
\begin{array}{rcl}
  A_1 & = & \forall x \exists y\,  (x \triangleleft y); \\
  A_2 & = & \exists x\, \mathit{Min}( x ); \\
  A_3 & = & \exists x\, \mathit{Max}( x ); \\
  A_4 & = & \forall x \forall y \forall z\, (  x \prec y \con y \prec z \imp
            x \prec z ); \\
  A_5 & = & \forall x \forall y\,  ( x \prec y \dis  y \prec x \dis x
            \approx y ); \\
  A_6 & = & \forall x \forall y\, ( x \prec y \con  y \prec x \imp x
            \approx y ); \\
  A_7 & = & \forall x \forall y\, ( x \prec y \con x \approx y \imp
   \mathit{Max}( x )); \\
  A_8 & = & \forall x \forall y\, ( ( T( x ) \imp T( y )) \imp x \prec
            y \dis x \approx y ); \\
  A_9 & = & \forall x \forall y\, (  x \prec y \imp ( T( x ) \imp T( y )) ).
\end{array}
$$
Пусть $A$~--- конъюнкция формул $A_1,\ldots,A_9$. Пусть
$$
\begin{array}{rcl}
  B_1 & = & \displaystyle
            \forall x \forall y \forall z\, \bigcon\limits_{\mathclap{\psi \in \sub\vp}} (q
            \imp \psi); \medskip\\
  B_2 & = & \displaystyle
            \forall x \forall y \forall z\, \bigcon\limits_{\mathclap{\psi \in \sub\vp}} (\psi \dis (\psi
            \imp q)),
\end{array}
$$
а также $B = B_1 \con B_2$.
Пусть $C$~--- формула, утверждающая, что $\approx$ является эквивалентностью и конгруэнцией по отношению к~$S$, т.е. является конъюнкцией формул
$$
\forall x\, (x \approx x) \con \forall x\forall y\, (x\approx y \imp y
\approx x ) \con \forall x\forall y\forall z\, (x\approx y \con y
\approx z \imp x \approx z)
$$
и
$$
\forall x\forall y\forall z\, \big( x\approx y\imp((S(z,x)\imp S(z,y))
\wedge(S(x,z) \imp S(y,z))\big).
$$
Наконец, пусть
$$
\begin{array}{lcl}
\bar{\vp} & = & A \con B \con C \imp \vp.
\end{array}
$$

\begin{lemma}
  \label{lem:QCl-to-QInt}
  Пусть $L \in \{ \logic{QInt}, \logic{QLC} \}$. Тогда следующие условия эквивалентны друг другу:
  \begin{itemize}
  \item[$(1)$] $\vp \in \logic{QCl}_{\mathit{fin}}$;
  \item[$(2)$] $\bar{\vp} \in L_{\mathit{wfin}}$;
  \item[$(3)$] $\bar{\vp} \in L_{\mathit{wfin}}\logic{.cd}$.
  \end{itemize}
\end{lemma}

\begin{proof}
  \mbox{$(1) \Rightarrow (2)$}: Пусть
  $\bar{\vp} \notin L_{\mathit{wfin}}$. Тогда
  $(\kModel{M}, w) \imodels A \con B \con C$ и
  $(\kModel{M}, w) \not\imodels \vp$ для некоторой модели
  $\kModel{M} = \langle W, R, D, I \rangle$, определённой на конечной $L$-шкале,
  и некоторого $w \in W$. Чтобы доказать, что
  $\vp \notin \logic{QCl}_{\mathit{fin}}$, построим конечную модель, опровергающую формулу~$\vp$.

  Из того, что $(\kModel{M},w) \imodels C$, следует, что множество $D_w$ разбивается на классы эквивалентности по отношению ${\approx}^{I, w}$. Пусть
  $\bar{a} = \{ b \in D_{w} : \langle a, b \rangle \in
  {\approx}^{I, w} \}$
  и $\bar{D}_{w} = \{ \bar{a} : a \in D_{w} \}$.

  Покажем, что множество $\bar{D}_{w}$ конечно.

  Поскольку $(\kModel{M},w) \imodels A_2$, получаем, что существует элемент $a_0 \in D_w$, для которого
  $(\kModel{M},w) \imodels \mathit{Min}(a_0)$. Поскольку $(\kModel{M},w) \imodels A_1$, получаем бесконечную последовательность
  $a_0 \triangleleft^{I,w} a_1 \triangleleft^{I,w} a_2
  \triangleleft^{I,w} \ldots{}$
  элементов множества $D_w$.

  Покажем, что последовательность $a_0, a_1, a_2, \ldots{}$ содержит лишь конечное множество попарно неэквивалентных по отношению ${\approx}^{I, w}$ элементов.

  Предположим, что это не так. Покажем, что в этом случае
  $a_k \not\approx^{I,w} a_{k+1}$ для каждого $k \geqslant 0$.

  Для этого предположим, что $a_k \approx^{I,w} a_{k+1}$ для некоторого $k \geqslant 0$. Поскольку $a_k \prec^{I,w} a_{k+1}$ и
  $(\kModel{M},w) \imodels A_7$, получаем, что $(\kModel{M},w) \imodels \mathit{Max}(a_k)$, и значит,
  $(\kModel{M},a_{k+s}) \prec^{I,w} a_k$ для каждого $s \geqslant 0$. Поскольку $(\kModel{M},w) \imodels A_4$, получаем $a_k \prec^{I,w} a_{k+s}$ для каждого
  $s \geqslant 1$. Но тогда $a_k \approx^{I,w} a_{k+s}$ для каждого
  $s \geqslant 1$, т.к. $(\kModel{M},w) \imodels A_6$. Это противоречит предположению о том, что последовательность $a_0, a_1, a_2, \ldots{}$ содержит бесконечно много неэквивалентных элементов.

  Следовательно, действительно $a_k \not\approx^{I,w} a_{k+1}$ для каждого $k \geqslant 0$.

  Поскольку $a_k \prec^{I,w} a_{k+1}$ для каждого $k \geqslant 0$ и
  $(\kModel{M},w) \imodels A_6$, получаем, что $a_{k+1} \not\prec^{I,w} a_{k}$ для каждого $k \geqslant 0$.  Поскольку $(\kModel{M},w) \imodels A_8$, получаем, что
  $(\kModel{M},w) \not\imodels T(a_{k+1}) \imp T(a_{k})$ для каждого $k \geqslant 0$.
  Значит, для каждого $k \geqslant 0$ существует мир $w_k \in R(w)$, такой, что
  $(\kModel{M},w_k) \imodels T(a_{k+1})$ и $(\kModel{M},w_k) \not\imodels T(a_{k})$.
  Поскольку $(\kModel{M},w) \imodels A_9$, получаем, что $(\kModel{M},w_k) \imodels T(a_{k+s})$ для каждого $s \geqslant 1$, а также $(\kModel{M},w_k) \not\imodels T(a_{k-s})$ для каждого
  $s \in \{ 0, \ldots, k\}$. Поскольку существует бесконечно много пар
  $\langle a_{k}, a_{k+1} \rangle$, таких, что
  $a_k \prec^{I,w} a_{k+1}$, $a_{k+1} \not\prec^{I,w} a_k$ и
  $a_{k} \not\approx^{I,w} a_{k+1}$, множество $\{ w_k : k \in \nat \}$
  должно быть бесконечным, что невозможно в силу конечности множества~$W$.

  Значит, последовательность $a_0, a_1, a_2, \ldots{}$ действительно содержит лишь конечное множество попарно неэквивалентных элементов.

  Следовательно, существуют такие $m \geqslant 0$ и $s \geqslant 1$, что
  $a_{m} \approx^{I,w} a_{m+s}$. Поскольку $(\kModel{M},w) \imodels A_4 \con A_7$, отсюда следует, что $(\kModel{M},w) \imodels \mathit{Max}(a_{m})$.

  Чтобы доказать, что множество $\bar{D}_{w}$ конечно, теперь достаточно показать, что каждый элемент
  $b \in D_w$ эквивалентен по отношению
  ${\approx}^{I, w}$ некоторому элементу $a_i$, где $i \geqslant 0$.

  Предположим, что это не так, т.е. существует такой элемент $b \in D_w$, что
  $b \not\approx^{I,w} a_i$ для каждого $i \geqslant 0$.

  Индукцией по $i$ покажем, что тогда $a_i \prec^{I,w} b$ для каждого $i \geqslant 0$. Базис индукции, т.е. когда $i=0$, мгновенно следует из того, что
  $b \not\approx^{I,w} a_0$ и $(\kModel{M},w) \imodels \mathit{Min}(a_0)$. Для обоснования индукционного шага предположим, что $a_j \prec^{I,w} b$. Поскольку $(\kModel{M},w) \imodels A_5$
  и $b \not\approx^{I,w} a_{j+1}$, получаем, что $b \prec^{I,w} a_{j+1}$ или $a_{j+1} \prec^{I,w} b$. Но
  $b \prec^{I,w} a_{j+1}$ невозможно:
  из того, что $a_j \triangleleft^{I,w} a_{j+1}$, $a_j \prec^{I,w} b$ и
  $b \prec^{I,w} a_{j+1}$, следует, что $b \approx^{I,w} a_{j+1}$, что противоречит сделанному для $b$ предположению. Значит, $a_{j+1} \prec^{I,w} b$.
  Итак, действительно $a_i \prec^{I,w} b$ для каждого $i \geqslant 0$.

  Как мы видели, существует $m \geqslant 0$, также, что
  $(\kModel{M},w) \imodels \mathit{Max}(a_m)$. Значит, $b \prec^{I,w} a_{m}$. Но тогда, как только что было показано,
  $a_{m} \prec^{I,w} b$. Поскольку $(\kModel{M},w) \imodels A_6$, получаем, что
  $b \approx^{I, w} a_m$, а это противоречит предположению, сделанному для~$b$.

  Как итог, множество $\bar{D}_{w}$ действительно является конечным.

  Теперь определим конечную классическую модель $\cModel{M}$, положив $\cModel{M}=\otuple{\mathcal{D},\mathcal{I}}$, где $\mathcal{D} = \bar{D}_w$ и $\mathcal{I}(S) = \set{\otuple{\bar{a},\bar{c}} : (\kModel{M},w)\imodels S(a,c)}$.

  Покажем, что $\cModel{M}\not\cmodels\varphi$.

  Говорим, что приписывания $\bar{g}$ в $\cModel{M}$ и $g$ в
  $\kModel{M}$ \defnotion{согласованы}, если для любой переменной $x$ и любого
  $a$ из предметной области шкалы $\langle W, R, D \rangle$
  $$
  \begin{array}{lcl}
  \bar{g}(x) = \bar{a} & \Longleftrightarrow & g(x) = a.
  \end{array}
  $$

  Чтобы доказать, что $\cModel{M} \not\cmodels \vp$, покажем, что
  $\cModel{M} \cmodels^{\bar{g}} \theta$ тогда и только тогда, когда
  $(\kModel{M},w) \imodels^g \theta$ для любой $\theta \in \sub \vp$ и любой пары согласованных приписываний
  $\bar{g}$ и~$g$. Доказательство последнего утверждения проведём индукцией по~$\theta$.

  Если $\theta$ является атомарной, получаем требуемое по определению модели~$\cModel{M}$.

  Если $\theta = \psi \con \chi$, $\theta = \psi \dis \chi$ или
  $\theta = \exists x\, \psi(x)$, то доказательство тривиально.

  Рассмотрим случай, когда $\theta = \psi\to\chi$.

  Пусть $\cModel{M} \not\cmodels^{\bar{g}} \psi \imp \chi$, т.е.
  $\cModel{M} \cmodels^{\bar{g}} \psi$ и
  $\cModel{M} \not\cmodels^{\bar{g}} \chi$. По индукционному предположению,
  $(\kModel{M},w) \imodels^g \psi$ и $(\kModel{M},w) \not\imodels^g \chi$, а значит,
  $(\kModel{M},w) \not\imodels^g \psi \imp \chi$.

  Пусть $(\kModel{M},w) \not\imodels^g \psi \imp \chi$, т.е.
  $(\kModel{M},w') \imodels^g \psi$ и $(\kModel{M},w') \not\imodels^g \chi$ для некоторого
  $w' \in R(w)$. По наследственности, $(\kModel{M},w) \not\imodels^g \chi$.  Покажем, что
  $(\kModel{M},w) \imodels^g \psi$. Поскольку $(\kModel{M},w) \imodels B_1$ и
  $\chi \in \sub\vp$, выполнено условие $(\kModel{M},w) \imodels^g q \imp \chi$; значит,
  $(\kModel{M},w') \not\imodels^g q$, и следовательно, $(\kModel{M},w) \not\imodels^g \psi \imp q$. Поскольку
  $(\kModel{M},w) \imodels B_2$ и $\psi \in \sub\vp$, выполнено условие
  $(\kModel{M},w) \imodels^g \psi \dis (\psi \imp q)$; значит, $(\kModel{M},w) \imodels^g \psi$.
  Тогда, по индукционному предположению, $\cModel{M} \cmodels^{\bar{g}} \psi$ и
  $\cModel{M} \not\cmodels^{\bar{g}} \chi$, а значит,
  $\cModel{M} \not\cmodels^{\bar{g}} \psi \imp \chi$.

  Рассмотрим случай, когда $\theta = \forall x\,\psi$.

  Пусть $\cModel{M} \not\cmodels^{\bar{g}} \forall x\, \psi$, т.е.
  $\cModel{M} \not\cmodels^{\bar{g}'} \psi$ для некоторого
  $\bar{g}' \stackrel{x}{=} \bar{g}$. Пусть
  $a \in \bar{g}'(x)$ и
  $$
  \begin{array}{lcl}
  g'(y) & = &
  \left\{
  \begin{array}{rl}
    g(y), & \mbox{если $y \ne x$;}\\
    a,    & \mbox{если $y = x$.}
  \end{array}
  \right.
  \end{array}
  $$
  Тогда $g'$ и $\bar{g}'$ согласованы. Значит, по индукционному предположению,
  $(\kModel{M},w) \not\imodels^{g'} \psi$. Поскольку
  $g'\stackrel{x}{=} g$, получаем, что
  $(\kModel{M},w) \not\imodels^{g} \forall x\, \psi$.

  Пусть $(\kModel{M},w) \not\imodels^g \forall x\, \psi$ и пусть
  $\bar{g}$~--- приписывание в $\cModel{M}$, согласованное с~$g$.
  Поскольку $(\kModel{M},w) \not\imodels^g \forall x\, \psi$, получаем, что
  $(\kModel{M},w') \not\imodels^{g'} \psi$ для некоторого мира $w' \in R(w)$ и некоторого приписывания
  $g'$, такого, что $g' \stackrel{x}{=} g$ и $g'(x) \in D(w')$.

  Покажем, что $g'(x) \approx^{I, w'} a$ для некоторого $a \in D(w)$.

  Как было показано, существует бесконечная последовательность
  $a_0 \triangleleft^{I,w} a_1 \triangleleft^{I,w} a_2
  \triangleleft^{I,w} \ldots{}$ элементов множества
  $D(w)$, такая, что $(\kModel{M},w) \imodels \mathit{Min}(a_0)$ и
  $(\kModel{M},w) \imodels \mathit{Max}(a_{m})$ для некоторого $m \geqslant 0$. По наследственности,
  $a_0 \triangleleft^{I,w'} a_1 \triangleleft^{I,w'} a_2
  \triangleleft^{I,w'} \ldots{}$,
  а также $(\kModel{M},w') \imodels \mathit{Min}(a_0)$ и $(\kModel{M},w') \imodels \mathit{Max}(a_{m})$; значит, что $g'(x) \prec^{I,w'} a_{m}$, откуда нетрудно заключить, что $g'(x) \approx^{I, w'} a_i$ для некоторого $i \in \set{0,\ldots,m}$.

  Пусть
  $$
  \begin{array}{lcl}
  g''(y) & = &
  \left\{
  \begin{array}{rl}
    g'(y), & \mbox{если $y \ne x$;} \\
    a,     & \mbox{если $y = x$.}\\
  \end{array}
  \right.
  \end{array}
  $$
  Поскольку $g'(x) \approx^{I, w'} g''(x)$ и $(\kModel{M},w) \imodels C$, получаем, что
  $(\kModel{M},w') \not\imodels^{g''} \psi$. Заметим, что $g''(y) \in D(w)$ для каждой переменной~$y$; следовательно, по наследственности,
  $(\kModel{M},w) \not\imodels^{g''} \psi$.

  Пусть $\bar{g}''$~--- приписывание в $\cModel{M}$, согласованное с~$g''$.
  Тогда, по индукционному предположению,
  $\cModel{M} \not\cmodels^{\bar{g}''} \psi$. Поскольку $\bar{g}''$ согласовано с~$g''$, $g'' \stackrel{x}{=} g$ и $\bar{g}$ согласовано с~$g$, получаем, что $\bar{g}'' \stackrel{x}{=} \bar{g}$.
  Следовательно, $\cModel{M} \not\cmodels^{\bar{g}} \forall x\, \psi(x)$.

  На этом обоснование индукционного шага завершено.

  Как следствие получаем, что $\cModel{M} \not\cmodels \vp$, и значит, $\vp \not\in \logic{QCl}_{\mathit{fin}}$.

  \mbox{$(2) \Rightarrow (3)$}: Очевидно, поскольку
  $L_{\mathit{wfin}} \subseteq L_{\mathit{wfin}}\logic{.cd}$.

  \begin{figure}
  \centering
  \begin{tikzpicture}[scale=1.5,
   node distance=5mm,
   terminal/.style={
   rectangle,minimum size=6mm,rounded corners=3mm,
   draw=black!50,
   }
   ]

\coordinate (wn)   at (1, 1.0);
\coordinate (wnm1) at (1, 2.0);
\coordinate (w2)   at (1, 3.5);
\coordinate (w1)   at (1, 4.5);
\coordinate (w0)   at (1, 5.5);

\coordinate (dots) at (1, 2.75);

\draw [] (w0)  circle [radius=2.0pt] ;
\draw [] (w1)  circle [radius=2.0pt] ;
\draw [] (w2)  circle [radius=2.0pt] ;
\draw [] (wnm1) circle [radius=2.0pt] ;
\draw [] (wn) circle [radius=2.0pt] ;

\draw [] (dots) node {$\vdots$};

\begin{scope}[>=latex]
\draw [->, shorten >= 2.75pt, shorten <= 2.75pt] (wn) -- (wnm1);
\draw [->, shorten >= 2.75pt, shorten <= 2.75pt] (w2) -- (w1)  ;
\draw [->, shorten >= 2.75pt, shorten <= 2.75pt] (w1) -- (w0)  ;
\end{scope}

\node [below right] at (wn)   {$w_n$}     ;
\node [below right] at (wnm1) {$w_{n-1}$} ;
\node [below right] at (w2)   {$w_2$}     ;
\node [below right] at (w1)   {$w_1$}     ;
\node [below right] at (w0)   {$w_0$}     ;

\node (wna0)     [right = 30pt] at (wn)     {$a_0^{\phantom{.}}$};
  \node (wna1)   [right = 10pt] at (wna0)   {$a_1^{\phantom{.}}$};
  \node (wna2)   [right = 10pt] at (wna1)   {$a_2^{\phantom{.}}$};
  \node (wndots) [right = 10pt] at (wna2)   {$\ldots$};
  \node (wnanm1) [right = 12pt] at (wndots) {$a_{n-1}^{\phantom{.}}$};
  \node (wnan)   [terminal, right = 14pt] at (wnanm1) {$a_{n}^{\phantom{.}}$};

\node (wnm1a0)     [right = 30pt] at (wnm1)     {$a_0^{\phantom{.}}$};
  \node (wnm1a1)   [right = 10pt] at (wnm1a0)   {$a_1^{\phantom{.}}$};
  \node (wnm1a2)   [right = 10pt] at (wnm1a1)   {$a_2^{\phantom{.}}$};
  \node (wnm1dots) [right = 10pt] at (wnm1a2)   {$\ldots$};
  \node (wnm1anm1) [terminal, right = 12pt] at (wnm1dots) {$a_{n-1}^{\phantom{.}}\hspace{8pt} a_n^{\phantom{.}}$};

\node (w2a0)     [right = 30pt] at (w2)     {$a_0^{\phantom{.}}$};
  \node (w2a1)   [right = 10pt] at (w2a0)   {$a_1^{\phantom{.}}$};
  \node (w2a2)   [terminal, right = 10pt] at (w2a1)   {$a_2^{\phantom{.}}\hspace{8pt}\ldots\hspace{8pt}a_{n-1}^{\phantom{.}}\hspace{8pt} a_n^{\phantom{.}}$};

\node (w1a0)     [right = 30pt] at (w1)     {$a_0^{\phantom{.}}$};
  \node (w1a1)   [terminal, right = 10pt] at (w1a0)   {$a_1^{\phantom{.}}\hspace{8pt}a_2^{\phantom{.}}\hspace{8pt}\ldots\hspace{8pt}a_{n-1}^{\phantom{.}}\hspace{8pt} a_n^{\phantom{.}}$};

\node (w0a0)   [terminal, right = 30pt] at (w0)   {$a_0^{\phantom{.}}\hspace{8pt}a_1^{\phantom{.}}\hspace{8pt}a_2^{\phantom{.}}\hspace{8pt}\ldots\hspace{8pt}a_{n-1}^{\phantom{.}}\hspace{8pt} a_n^{\phantom{.}}$};

\end{tikzpicture}

\caption{Model $\kModel{M}$}
  \label{s8-5-1:fig31}
\end{figure}

\mbox{$(3) \Rightarrow (1)$}: Пусть
$\vp \not\in \logic{QCl}_{\mathit{fin}}$. Тогда $\cModel{M} \not \cmodels \vp$ для некоторой конечной классической
$\cModel{M} = \langle \mathcal{D}, \mathcal{I} \rangle$; будем считать, что
$\mathcal{D} = \{ a_0, a_1, \ldots, a_n \}$ для некоторого $n \in \nat$.
Пусть $\kModel{M} = \langle W, R, D, I \rangle$~--- интуиционистская модель Крипке, см.\ рис.~\ref{s8-5-1:fig31}, где
  \begin{itemize}
  \item $W = \{ w_0, w_1, \ldots, w_n \}$;

  \pagebreak[3]

  \item $R$~--- это рефлексивно-транзитивное замыкание бинарного отношения
    $\{ \langle w_k, w_{k-1} \rangle: 1 \leqslant k \leqslant n \}$;

  \pagebreak[2]

    \item $D_w = \mathcal{D}$ для каждого $w \in W$,
  \end{itemize}
  а интерпретация $I$ удовлетворяет следующим условиям:
  \begin{itemize}
  \item $(\kModel{M},w_k) \imodels T(a_s)$ $~\leftrightharpoons~$ $k \leqslant s$;
  \item $(\kModel{M},w_k)
    \imodels a_s \prec a_t$ $~\leftrightharpoons~$ либо $s <
    t$, либо $s \geqslant k$ и $t \geqslant k$;
  \item $(\kModel{M},w_k) \imodels a_s \approx a_t$
    $~\leftrightharpoons~$ либо $s = t$, либо $s \geqslant k$ и $t
    \geqslant k$;
  \item $(\kModel{M},w_k) \imodels q$ $~\leftrightharpoons~$ $k \ne n$;
  \item $I(w_n, S) = \mathcal{I} (S)$.
  \end{itemize}
  Заметим, что модель $\kModel{M}$ определена на конечной $\logic{QLC}$-шкале с постоянными областями.

  Покажем, что $(\kModel{M},w_n) \not\imodels \bar{\vp}$.

  Поскольку формула $\vp$ является позитивной, получаем, что для любой формулы $\psi \in \sub(\vp)$, любого числа $k \in \{1, \ldots, n-1 \}$ и любого приписывания~$g$
  $$
  \mbox{$(\kModel{M}, w_k) \imodels^g \psi$. }
  \eqno{(\ast)}
  $$

  Учитывая $(\ast)$ и интерпретацию буквы $q$, получаем, что
  $(\kModel{M},w_n) \imodels^g q \imp \psi$ и
  $(\kModel{M},w_n) \imodels^g \psi \dis (\psi \imp q)$ для любой формулы
  $\psi \in \sub(\vp)$ и любого приписывания~$g$. Следовательно,
  $(\kModel{M},w_n) \imodels B$.

  Нетрудно видеть, что $(\kModel{M},w_n) \imodels A \con C$; покажем лишь, что $(\kModel{M},w_n) \imodels A_9$. Пусть $(\kModel{M},w_k) \imodels a_s \prec a_t$ для некоторого $w_k \in W$.
  Тогда, согласно определению $\kModel{M}$, либо $s < t$, либо
  $s \geqslant k$ и $t \geqslant k$. Покажем, что если $m \leqslant k$, то
  $(\kModel{M},w_m) \imodels T(a_s)$ влечёт, что $(\kModel{M},w_m) \imodels T(a_t)$. Пусть $m \leqslant k$. Сначала предположим, что
  $s < t$. Тогда из того, что $(\kModel{M},w_m) \not\imodels T(a_t)$ следует, согласно определению модели
  $\kModel{M}$, что $t < m$; значит, $s < m$ и
  $(\kModel{M},w_m) \not\imodels T(a_s)$. Теперь предположим, что $s \geqslant k$ и
  $t \geqslant k$.  Тогда $s \geqslant m$ и $t \geqslant m$, а следовательно,
  по определению модели $\kModel{M}$, получаем, что $(\kModel{M},w_m) \imodels T(a_s)$ и
  $(\kModel{M},w_m) \imodels T(a_t)$.

  Итак, $(\kModel{M},w_n) \imodels A \con B \con C$.

  Чтобы показать, что $(\kModel{M},w_n) \not\imodels \vp$, докажем, что для любой формулы $\theta \in \sub(\vp)$ и любого приписывания~$g$ имеет место следующая эквивалентность:
  $\cModel{M} \cmodels^g \theta$ тогда и только тогда, когда $(\kModel{M},w_n) \imodels^g \theta$. Обоснование  проведём индукцией по~$\theta$.

  Если формула $\theta$ является атомарной, то требуемое мгновенно следует из определения модели~$\kModel{M}$.

  В случаях, когда $\theta = \psi \con \chi$, $\theta = \psi \dis \chi$ или
  $\theta = \exists x\, \psi(x)$, доказательство является тривиальным.

  Рассмотрим случай, когда $\theta = \psi \to \chi$.

  Пусть $\cModel{M} \not\cmodels^g \psi \imp \chi$, т.е. $\cModel{M} \cmodels^g \psi$
  и $\cModel{M} \not\cmodels^g \chi$. По индукционному предположению,
  $(\kModel{M},w_n) \imodels^g \psi$ и $(\kModel{M},w_n) \not\imodels^g \chi$, а значит,
  $(\kModel{M},w_n) \not\imodels^g \psi \imp \chi$.

  Пусть $(\kModel{M},w_n) \not\imodels^g \psi \imp \chi$, т.е.
  $(\kModel{M},w_k) \imodels^g \psi$ и $(\kModel{M},w_k) \not\imodels^g \chi$ для некоторого
  $k \leqslant n$. Поскольку $(\kModel{M},w_k) \not\imodels^g \chi$ и
  $\chi \in \sub\vp$, по ($\ast$) получаем, что $k = n$. Тогда,
  по индукционному предположению, $\cModel{M} \cmodels^g \psi$ и
  $\cModel{M} \not\cmodels^g \chi$, а значит,
  $\cModel{M} \not\cmodels^g \psi \imp \chi$.

  Рассмотрим случай, когда $\theta = \forall x\,\psi$.

  Пусть $\cModel{M} \not\cmodels^g \forall x\, \psi$, т.е.
  $\cModel{M} \not\cmodels^{g'} \psi$ для некоторого приписывания $g'$, такого, что
  $g' \stackrel{x}{=} g$. По индукционному предположению,
  $(\kModel{M},w_n) \not\imodels^{g'} \psi$, а значит,
  $(\kModel{M},w_n) \not\imodels^{g} \forall x\, \psi$.

  Пусть $(\kModel{M},w_n) \not\imodels^{g} \forall x\, \psi$, т.е.
  $(\kModel{M},w_k) \not\imodels^{g'} \psi$ для некоторого числа $k \leqslant n$ и некоторого приписывания $g'$, такого, что $g' \stackrel{x}{=} g$ и $g'(x) \in D(w_k)$. Поскольку
  $(\kModel{M},w_k) \not\imodels^{g'} \psi$, по ($\ast$) получаем, что $k = n$.
  Следовательно, по индукционному предположению, $\cModel{M} \not\cmodels^{g'} \psi$, а значит, $\cModel{M} \not\cmodels^{g} \forall x\, \psi$.

  Поскольку $\kModel{M}$ определена на конечной $\logic{QLC}$-шкале с постоянными областями, получаем, что $\bar{\vp} \notin \logic{QLC}_{\mathit{wfin}}\logic{.cd}$.

  Поскольку
  $\logic{QInt}_{\mathit{wfin}}\logic{.cd} \subseteq \logic{QLC}_{\mathit{wfin}}\logic{.cd}$, получаем также, что $\bar{\vp} \notin \logic{QInt}_{\mathit{wfin}}\logic{.cd}$.
\end{proof}

Из леммы~\ref{lem:QCl-to-QInt} получаем следующую теорему.

\begin{theorem}
  \label{thr:three-variables-int}
  Пусть\/
  $L\in [\logic{QInt}_{\mathit{wfin}}, \logic{QLC}_{\mathit{wfin}}\logic{.cd}]$. Тогда позитивный фрагмент~$L$ в языке с тремя предметными переменными, является\/ $\Pi^0_1$-трудным.
\end{theorem}

Эта теорема даёт очевидное следствие.

\begin{corollary}
  \label{cor:three-variables-int}
  Пусть
  $L \in \{\logic{QInt}, \logic{QKP}, \logic{QLM}, \logic{QKC}, \logic{QLC}\}$.
  Тогда $L_{\mathit{wfin}}$ и $L_{\mathit{wfin}}\logic{.cd}$ являются как\/
  $\Pi^0_1$-трудными, так и\/ $\Sigma^0_1$-трудными в языке с тремя предметными переменными.
\end{corollary}

Отметим, что в обоих случаях требуется всего лишь несколько предикатных букв, арность которых не превосходит двух.

	\subsection{Фрагменты с одноместным предикатом}

Покажем, что результаты предыдущего раздела можно усилить, если вместо интервала $[\logic{QInt}_{\mathit{wfin}}, \logic{QLC}_{\mathit{wfin}}\logic{.cd}]$ рассмотреть его подынтервал $[\logic{QInt}_{\mathit{wfin}}, \logic{QKC}_{\mathit{wfin}}\logic{.cd}]$.

\begin{theorem}
  \label{s8-5-2:thr:int-main}
  Пусть\/
  $L\in [\logic{QInt}_{\mathit{wfin}}, \logic{QKC}_{\mathit{wfin}}\logic{.cd}]$. Тогда позитивный фрагмент $L$ в языке с тремя предметными переменными и одной унарной предикатной буквой является\/
  $\Pi^0_1$-трудным.
\end{theorem}

\begin{proof}
Достаточно сначала применить к формуле $\bar{\varphi}$, определённой в разделе~\ref{ss:8-5-1}, трюк Крипке (см.\ лемму~\ref{lem:s-8-2-1:Kripke:trick:int}), а затем использовать моделирование унарных предикатных букв формулами от одной унарной предикатной буквы, описанное в разделе~\ref{ss:8-2-3} (см.\ лемму~\ref{s8-2-3:lem:QInt-main-lemma}). Заметим, что на каждом из этих этапов возникают конечные модели (с~конечными множествами миров и с конечными предметными областями), а итоговая формула будет позитивной, при этом будет содержать три предметные переменные и одну унарную предикатную букву.
\end{proof}


\begin{corollary}
  \label{s8-3-1:cor:int-main}
  Пусть $L \in \{\logic{QInt}, \logic{QKP}, \logic{QLM}, \logic{QKC} \}$. Тогда
  $L_{\mathit{wfin}}$ и $L_{\mathit{wfin}}\logic{.cd}$ являются\/
  $\Pi^0_1$-трудными в языке с тремя предметными переменными и одной унарной предикатной буквой.
\end{corollary}


\begin{corollary}
  \label{s8-3-1:cor:int-main-2}
  Пусть $L = \logic{Int} + \vp$, где $\vp$~--- формула от одной пропозициональной переменной, и пусть $L \subset \logic{Cl}$. Тогда\/
  $\logic{Q}L_{\mathit{wfin}}$ и $\logic{Q}L_{\mathit{wfin}}\logic{.cd}$
  являются\/ $\Pi^0_1$-трудными в языке с тремя предметными переменными и одной унарной предикатной буквой.
\end{corollary} 

Возможны дальнейшие усиления и уточнения полученных результатов; некоторых из них приводятся в работе~\cite{MR:2025:RecInsep}. Мы вернёмся к обсуждению дальнейших продвижений полученных здесь теорем в главе~\ref{ch:10}.

\setcounter{savefootnote}{\value{footnote}}
\chapter{Логика квазиарных предикатов}
\setcounter{footnote}{\value{savefootnote}}
      \label{ch:9}
  \section{Основные определения и факты}
   \subsection{Предварительные сведения}

Рассмотрение логики частичных квазиарных предикатов~\cite{Shkilniak10,NT12} мотивировано её возможными приложениями в анализе, синтезе и верификации программ; исследования в этой области получили дальнейшее развитие у ряда
авторов~\cite{NS15,NS17,NS18,KIN18}.

Основная особенность логики частичных квазиарных
предикатов, отличающая её от классической логики предикатов и давшая ей название, заключается в том, что предикаты не имеют арности (или, в определённом смысле, имеют счётную арность): они применяются не
к заранее определённому конечному числу параметров, а ко всем параметрам языка сразу. Такой подход отчасти мотивирован необходимостью, возникающей при рассуждении о состояниях программы, ссылаться на
свойства наборов данных неопределенного размера, например, утверждая, что числовой массив отсортирован в порядке возрастания: верно это утверждение или нет, не зависит от
размера массива.\footnote{Более детально см.~\cite{Shkilniak10,NT12}.} Другой отличительной особенностью логики частичных квазиарных предикатов является то, что параметры могут не иметь значений, точно так же, как переменные, введённые в языке программирования могут быть неинициализированы при выполнении программы, и следовательно,
значения предикатов могут быть не определены для некоторых последовательностей параметров, т.е.
быть <<частичными>>.

Погружение логики частичных квазиарных предикатов в
классическую логику предикатов $\logic{QCl}$ было дано в~\cite{NT12}. Это погружение можно рассматривать как классическую интерпретацию логики частичных квазиарных предикатов, близкую стандартному переводу неклассических формул в $\logic{QCl}$, см., например,~\cite[раздел~3.12]{GShS}.

Здесь будет приведён пример
интерпретации классических формул формулами языка логики квазиарных предикатов, за счёт чего будет построено погружение логики $\logic{QCl}$ в логику
частичных квазиарных предикатов. В качестве следствий получим, что логика частичных квазиарных предикатов неразрешима ($\Sigma^0_1$-полна), а её расширение, определяемое классом конечных моделей, неперечислимо ($\Pi^0_1$-полно), причём в обоих случаях достаточно одного квазиарного предиката и пяти переменных.

	\subsection{Синтаксис и семантика}

\providecommand{\qModel}[1]{\cModel{#1}}

Опишем кратко логику частичных квазиарных предикатов;
хотя приводимые здесь определения отличаются от определений, представленных в~\cite{Shkilniak10,NT12}, различия несущественны.

Язык содержит следующие символы:
\begin{itemize}
\item множество $\{x_i : i\in\numN^+\}$ \defnotion{существенных предметных переменных};\index{переменная!предметная!существенная}
\item множество $\{y_i : i\in\numN^+\}$
\defnotion{несущественных предметных переменных};\index{переменная!предметная!несущественная}
\item множество $\{p_i : i\in\numN^+\}$
  \defnotion{квазиарных предикатных букв};\index{буква!предикатная!квазиарная}
\item логические символы $\bot$, $\wedge$, $\vee$, $\to$, $\exists$, $\forall$ и~$\mathcal{R}$;
\item скобки и запятую.
\end{itemize}
Символ $\mathcal{R}$ служит для обозначения \defnotion{оператора переименования переменных}\index{оператор!переименования переменных} (называемым также \defnotion{реноминация}\index{реноминация}), суть которого поясним ниже.


Пусть $\mathit{EVar}$, $\mathit{IVar}$ и $\mathit{QPred}$~--- множества существенных переменных, несущественных переменных и квазиарных предикатных букв, соответственно; по определению, все они счётны.
Предполагаем, что $\mathit{EVar}$, $\mathit{IVar}$ и $\mathit{QPred}$ попарно не пересекаются.  Обозначим через $V$ множество
$\mathit{EVar} \cup \mathit{IVar}$, состоящее из всех предметных
переменных.
Заметим, что, в отличие от предикатных букв языка классической логики
предикатов, квазиарным предикатным буквам не приписана никакая арность.

Формулы языка логики квазиарных предикатов, или $\lang{QL}\mathit{q}$-формулы\index{уяа@формула!ql@$\lang{QL}\mathit{q}$-формула}, определяются с помощью BNF-выражения следующим образом:
$$
\begin{array}{lcl}
\vp & := & \bot \mid p_i \mid (\vp \wedge \vp) \mid (\vp \vee \vp) \mid (\vp \to \vp) \mid \exists u\,\varphi \mid \forall u\,\varphi \mid
\mathcal{R}^{w}_{u}\varphi,
\end{array}
$$
где $w, u \in V$.
Множество всех $\lang{QL}\mathit{q}$-формул обозначаем через~$\lang{QL}\mathit{q}$.

Интуитивный смысл оператора $\mathcal{R}^{w}_{u}$ заключается в том, что
переменная $w$ получает значение, приписанное переменной~$u$ (как бы <<переименовывается>> в~$u$).
Применение оператора $\mathcal{R}^{w}_{u}$ напоминает присваивание в
процедурных языках: $\mathcal{R}^{w}_{u} \vp$ может быть понято как
утверждение, что $\vp$ истинно, как только процедурное присваивание $w:=u$ было осуществлено.

Несущественные переменные~--- это инструмент для преобразования формул без изменения их значения. В классической логике переменные формулы могут
быть переименованы в те, которые не встречаются в формуле, так, что,
например, $\forall x\, P(x) \vee \forall x\, Q(x)$ можно привести к
префиксной нормальной форме $\forall x \forall y\, (P(x) \vee Q(y))$, которая эквивалентна в $\logic{QCl}$ исходной формуле. Поскольку, как мы увидим, $\lang{QL}\mathit{q}$-формулы неявно ссылаются на значения сразу всех переменных языка, таких новых переменных ни для одной формулы просто нет. Вместо этого используются несущественные переменные~--- их значения, как мы увидим, не влияют на истинностные значения $\lang{QL}\mathit{q}$-формул.

Приоритет логических операторов, определённый для классической логики, расширяем таким образом, что оператор переименования переменных имеет тот же приоритет, что и кванторы.

Чтобы определить понятие $\lang{QL}\mathit{q}$-модели, сначала определим понятие частичного квазиарного предиката.

Пусть $D \ne \varnothing$.  \defnotion{Частичное приписывание}\index{приписывание!частичное} в $D$~--- это частичная функция $\alpha\colon V \to D$. Пишем
${\downarrow}\alpha(u)$ в случае, когда значение функции $\alpha$ определено на $u \in V$; если оно не определено, то пишем ${\uparrow}\alpha(u)$. Множество всех частичных приписываний в~$D$ Обозначим через~${{}^{V}}\!\!D$.

Частичные приписывания $\alpha$ и $\beta$ называем \defnotion{существенно неразличимыми}\index{приписывания!существенно неразличимые}, если для каждого $x_i\in \mathit{EVar}$
$$
\begin{array}{lrl}
{\downarrow}\alpha(x_i)
  & \Longleftrightarrow
  & {\downarrow}\beta(x_i); \smallskip\\
{\downarrow}\alpha(x_i)
  & \Longrightarrow
  & \alpha(x_i) = \beta(x_i).
\end{array}
\eqno{(\mathit{ES})}
$$
Если $\alpha$ и $\beta$ существенно неразличимы, то пишем $\alpha\simeq\beta$.

Нетрудно видеть, что отношение $\simeq$ является эквивалентностью на~${{}^V}\!\!D$.

Пусть $\bm{p} \colon {}^V\!\!D \to \{0,1\}$~--- частичная функция. Пишем ${\downarrow}\bm{p}(\alpha)$, если значение функции
$\bm{p}$ определено на $\alpha\in{}^V\!\!D$; если оно не определено, то пишем ${\uparrow}\bm{p}(\alpha)$.

Частичная функция $\bm{p} \colon {}^V\!\!D \to \{1,0\}$ называется \defnotion{частичным квазиарным предикатом}\index{предикат!частичный квазиарный} на~$D$, если для любых
$\alpha,\beta\in{{}^V}\!\!D$, таких, что $\alpha\simeq\beta$,
$$
\begin{array}{lrl}
{\downarrow}\bm{p}(\alpha)
  & \Longleftrightarrow
  & {\downarrow}\bm{p}(\beta); \smallskip\\
{\downarrow}\bm{p}(\alpha)
  & \Longrightarrow
  & \bm{p}(\alpha) = \bm{p}(\beta).
\end{array}
\eqno{(\mathit{QP})}
$$

\defnotion{Моделью} языка $\lang{QL}\mathit{q}$, или \defnotion{$\lang{QL}\mathit{q}$-моделью}, называется пара $\qModel{M}=\langle D,I\rangle$, где
$D\ne\varnothing$ и $I$~--- \defnotion{интерпретация}, приписывающая
каждой букве $p_i \in \mathit{QPred}$ некоторый частичный квазиарный предикат $p_i^I$ на~$D$. Множество $D$ называется \defnotion{предметной областью} модели~$\qModel{M}$. Модель называется \defnotion{конечной}, если её предметная область конечна.

Если $\qModel{M}=\langle D,I\rangle$ и $\alpha\in{}^V\!\!D$, то говорим также, что $\alpha$~--- \defnotion{частичное приписывание} в~$\qModel{M}$.

Пусть $U\subseteq V$. Частичные приписывания $\alpha$ и $\beta$ называем \defnotion{неразличимыми вне $U$}\index{приписывания!неразличимые вне $U$} (пишем $\alpha\stackrel{U}{=}\beta$),
если для каждой переменной $u \in V\setminus U$ либо ${\uparrow}\alpha(u)$ и
${\uparrow}\beta(u)$, либо $\alpha(u)=\beta(u)$; если
$U = \{u_1,\ldots, u_n\}$, то пишем
$\alpha\stackrel{u_1,\ldots, u_n}{=}\beta$ вместо
$\alpha\stackrel{\{u_1,\ldots,u_n\}}{=}\beta$.
Заметим, что
$$
\begin{array}{ccc}
\alpha\hfill\stackrel{\varnothing}{=}\hfill\beta & \Longleftrightarrow & \alpha = \beta;
\smallskip\\
\alpha\stackrel{\mathit{IVar}}{=}\beta & \Longleftrightarrow & \alpha \simeq \beta.
\end{array}
$$

Для $\lang{QL}\mathit{q}$-модели $\qModel{M}$, частичного приписывания $\alpha$ и формулы $\vp$ определим рекурсивно отношения
${\downarrow}\qModel{M}\models^\alpha\varphi$ (формула $\varphi$ истинна в модели $\qModel{M}$ при частичном приписывании~$\alpha$),
${\downarrow}\qModel{M}\not\models^\alpha\varphi$ (формула $\varphi$~ложна в модели $\qModel{M}$ при частичном приписывании~$\alpha$) и
${\uparrow}\qModel{M}\models^\alpha\varphi$ (значение формулы~$\varphi$ в
модели $\qModel{M}$ при частичном приписывании~$\alpha$ не определено):
\begin{itemize}
\item
${\downarrow}\qModel{M}\models^\alpha\bot$ ~~никогда;
\nopagebreak\\[2pt]
${\downarrow}\qModel{M}\not\models^\alpha\bot$ ~~всегда;
\nopagebreak\\[2pt]
${\uparrow}\qModel{M}\models^\alpha\bot$ ~~никогда;
\smallskip

\pagebreak[3]

\item
  ${\downarrow}\qModel{M}\models^\alpha p_i
  ~ \leftrightharpoons ~
  p_i^I(\alpha) = 1$;
  \nopagebreak\\[2pt]
  ${\downarrow}\qModel{M}\not\models^\alpha p_i
  ~ \leftrightharpoons ~
  p_i^I(\alpha) = 0$;
  \nopagebreak\\[2pt]
  ${\uparrow}\qModel{M}\models^\alpha p_i
  ~ \leftrightharpoons ~
  {\uparrow}p_i^I(\alpha)$;
\smallskip

\pagebreak[3]

\item
${\downarrow}\qModel{M}\models^\alpha \varphi_1 \wedge\varphi_2
~ \leftrightharpoons ~
{\downarrow}\qModel{M}\models^\alpha \varphi_1 ~~\mbox{ и }\,~ {\downarrow}\qModel{M}\models^\alpha \varphi_2$;
\nopagebreak\\[2pt]
${\downarrow}\qModel{M}\not\models^\alpha \varphi_1 \wedge\varphi_2
~ \leftrightharpoons ~
{\downarrow}\qModel{M}\not\models^\alpha \varphi_1 \mbox{ или } {\downarrow}\qModel{M}\not\models^\alpha \varphi_2$;
\nopagebreak\\[2pt]
${\uparrow}\qModel{M}\models^\alpha \varphi_1 \wedge \varphi_2$
~в остальных случаях;
\smallskip

\pagebreak[3]

\item
${\downarrow}\qModel{M}\models^\alpha \varphi_1 \vee\varphi_2
~ \leftrightharpoons ~
{\downarrow}\qModel{M}\models^\alpha \varphi_1 \mbox{ или } {\downarrow}\qModel{M}\models^\alpha \varphi_2$;
\nopagebreak\\[2pt]
${\downarrow}\qModel{M}\not\models^\alpha \varphi_1 \vee\varphi_2
~ \leftrightharpoons ~
{\downarrow}\qModel{M}\not\models^\alpha \varphi_1 ~~\mbox{ и }\,~ {\downarrow}\qModel{M}\not\models^\alpha \varphi_2$;
\nopagebreak\\[2pt]
${\uparrow}\qModel{M}\models^\alpha \varphi_1 \vee \varphi_2$
~в остальных случаях;
\smallskip

\pagebreak[3]

\item
${\downarrow}\qModel{M}\models^\alpha \varphi_1 \to\varphi_2
~ \leftrightharpoons ~
{\downarrow}\qModel{M}\not\models^\alpha \varphi_1 \mbox{ или } {\downarrow}\qModel{M}\models^\alpha \varphi_2$;
\nopagebreak\\[2pt]
${\downarrow}\qModel{M}\not\models^\alpha \varphi_1 \to\varphi_2
~ \leftrightharpoons ~
{\downarrow}\qModel{M}\models^\alpha \varphi_1 ~~\mbox{ и }\,~ {\downarrow}\qModel{M}\not\models^\alpha \varphi_2$;
\nopagebreak\\[2pt]
${\uparrow}\qModel{M}\models^\alpha \varphi_1 \imp \varphi_2$
~в остальных случаях;
\smallskip

\pagebreak[3]

\item
  ${\downarrow}\qModel{M}\models^\alpha \exists u\,\varphi_1 ~
  \leftrightharpoons ~ {\downarrow}\qModel{M}\models^\beta \varphi_1$
  для некоторого $\beta$, такого, что ${\downarrow}\beta(u)$ и
  $\beta\stackrel{u}{=}\alpha$;\nopagebreak\\[2pt]
  ${\downarrow}\qModel{M}\not\models^\alpha \exists u\,\varphi_1 ~
  \leftrightharpoons ~ {\downarrow}\qModel{M}\not\models^\beta
  \varphi_1$
  для всех таких $\beta$, что ${\downarrow}\beta(u)$ и
  $\beta\stackrel{u}{=}\alpha$;\nopagebreak\\[2pt]
  ${\uparrow}\qModel{M}\models^\alpha \exists u\,\varphi_1$ ~в остальных случаях;
\smallskip

\pagebreak[3]

\item
  ${\downarrow}\qModel{M}\models^\alpha \forall u\,\varphi_1 ~
  \leftrightharpoons ~ {\downarrow}\qModel{M}\models^\beta \varphi_1$
  для всех таких $\beta$, что ${\downarrow}\beta(u)$ и
  $\beta\stackrel{u}{=}\alpha$;\nopagebreak\\[2pt]
  ${\downarrow}\qModel{M}\not\models^\alpha \forall u\,\varphi_1 ~
  \leftrightharpoons ~ {\downarrow}\qModel{M}\not\models^\beta
  \varphi_1$
  для некоторого $\beta$, такого, что ${\downarrow}\beta(u)$ и
  $\beta\stackrel{u}{=}\alpha$;\nopagebreak\\[2pt]
  ${\uparrow}\qModel{M}\models^\alpha \forall u\,\varphi_1$ ~в остальных случаях;
\smallskip

\pagebreak[3]

\item
  ${\downarrow}\qModel{M}\models^\alpha \mathcal{R}^{w}_{u}\varphi_1 ~
  \leftrightharpoons ~ {\downarrow}\qModel{M}\models^\beta \varphi_1$,
  где $\beta$ таково, что $\beta\stackrel{w}{=}\alpha$ и
  $\beta(w) = \alpha (u)$;\nopagebreak\\[2pt]
  ${\downarrow}\qModel{M}\not\models^\alpha \mathcal{R}^{w}_{u}\varphi_1
  ~ \leftrightharpoons ~ {\downarrow}\qModel{M}\not\models^\beta
  \varphi_1$,
  где $\beta$ таково, что $\beta\stackrel{w}{=}\alpha$ и
  $\beta(w) = \alpha (u)$;\nopagebreak\\[2pt]
  ${\uparrow}\qModel{M}\models^\alpha \mathcal{R}^{w}_{u}\varphi_1$
  ~в остальных случаях.
\end{itemize}

Формула $\varphi$ \defnotion{истинна} в\/ $\qModel{M}$ (пишем
$\qModel{M}\models\varphi$), если
${\downarrow}\qModel{M}\not\models^\alpha\varphi$ не выполняется ни для какого~$\alpha\in{}^V\!\!D$; иначе $\varphi$ \defnotion{опровергается}
  в\/~$\qModel{M}$.  Формула \defnotion{$\lang{QL}\mathit{q}$-общезначима}, если она истинна в каждой
$\lang{QL}\mathit{q}$-модели. Формула \defnotion{конечно\/ $\lang{QL}\mathit{q}$-общезначима}, если она истинна
в каждой конечной $\lang{QL}\mathit{q}$-модели. Формула $\vp$  \defnotion{$\lang{QL}\mathit{q}$-выполнима}, если
${\downarrow} \qModel{M}\models^\alpha \varphi$ для некоторой $\lang{QL}\mathit{q}$-модели
$\qModel{M}$ и некоторого~$\alpha$. Формула $\vp$ \defnotion{конечно\/
  $\lang{QL}\mathit{q}$-выполнима}, если
${\downarrow} \qModel{M}\models^\alpha \varphi$ для некоторой конечной
$\lang{QL}\mathit{q}$-модели $\qModel{M}$ и некоторого~$\alpha$.

Определим две логики квазиарных предикатов:
$$
\begin{array}{lcl}
\logic{QPL}
  & =
  & \{\varphi \in \lang{QL}\mathit{q} :
  \mbox{$\varphi$ является $\lang{QL}\mathit{q}$-общезначимой}\};
  \smallskip\\
\logic{QPL}_{\it fin}
  & =
  & \{\varphi \in \lang{QL}\mathit{q} :
  \mbox{$\vp$ является конечно $\lang{QL}\mathit{q}$-общезначимой}\}.
\end{array}
$$

  \section{Алгоритмические свойства}
   \subsection{Погружение классической логики предикатов в логику квазиарных предикатов}   

Опишем погружение фрагмента логики \mbox{$\logic{QCl}$} в языке с бинарной предикатной буквой в логику $\logic{QPL}$; как мы увидим, полученная функция будет также и погружением соответствующего фрагмента логики $\mbox{$\logic{QCl}$}_{\mathit{fin}}$ в логику $\mbox{$\logic{QPL}$}_{\mathit{fin}}$.


Для дальнейших построений зафиксируем бинарную предикатную букву~$P$ и квазиарную предикатную букву~$p$. Кроме того, для удобства будем считать, что $\set{v_i : i\in\numNp}$~--- это пересчёт всех предметных переменных языка~$\lang{QL}$. Пусть также $\lang{QL}^{\mathit{bin}}$~--- фрагмент языка $\lang{QL}$, содержащий бинарную предикатную букву~$P$ и не содержащий других предикатных букв.

Определим рекурсивно перевод ${\mathsf{Tr}}\colon \lang{QL}^{\mathit{bin}} \to \lang{QL}\mathit{q}$:
$$
\begin{array}{lcl} {\mathsf{Tr}}(\bot) & = & \bot; \smallskip\\
  {\mathsf{Tr}}(P(v_i,v_j)) & = & \mathcal{R}^{x_1}_{y_i}\mathcal{R}^{x_2}_{y_j}\,p; \smallskip\\
  {\mathsf{Tr}}(\psi\wedge\chi) & = & {\mathsf{Tr}}(\psi)\wedge {\mathsf{Tr}}(\chi); \smallskip\\
  {\mathsf{Tr}}(\psi\vee\chi) & = & {\mathsf{Tr}}(\psi)\vee {\mathsf{Tr}}(\chi); \smallskip\\
  {\mathsf{Tr}}(\psi\to\chi) & = & {\mathsf{Tr}}(\psi)\to {\mathsf{Tr}}(\chi); \smallskip\\
  {\mathsf{Tr}}(\exists v_i\,\psi) & = & \forall y_i\,{\mathsf{Tr}}(\psi); \smallskip\\
  {\mathsf{Tr}}(\forall v_i\,\psi) & = & \forall y_i\,{\mathsf{Tr}}(\psi).
\end{array}
$$

Для каждой формулы $\vp \in \lang{QL}^{\mathit{bin}}$ положим
$$
\begin{array}{lcl}
  \varphi^\ast & = & \forall y_1\forall
                     y_2\,(\mathcal{R}^{x_1}_{y_1} \mathcal{R}^{x_2}_{y_2} p\vee\neg
                     \mathcal{R}^{x_1}_{y_1} \mathcal{R}^{x_2}_{y_2} p) \to \mathsf{Tr}(\varphi).
\end{array}
$$

\begin{lemma}
\label{lem_Tr}
Пусть $\varphi$~--- замкнутая\/ $\lang{QL}^{\mathit{bin}}$-формула. Тогда
$$
\begin{array}{lcl}
\varphi\in \mbox{$\logic{QCl}$}
  & \iff
  & \varphi^\ast\in \mbox{$\logic{QPL}$}.
\end{array}
$$
\end{lemma}

\begin{proof}
  Докажем импликацию $({\Leftarrow})$.
  Пусть $\varphi \not\in \mbox{$\logic{QCl}$}$.  Тогда
  $\cModel{M} \not\models^{\alpha_0} \vp$ для некоторой $\lang{QL}$-модели
  $\cModel{M} =\langle D,I\rangle$ и некоторого приписывания~$\alpha_0$.
  Пусть $P^I = I(P)$.

  Определим $\lang{QL}\mathit{q}$-модель $\cModel{M}'$, опровергающую формулу~$\vp^\ast$.
  Для этого сначала определим частичный квазиарный предикат $\bm{p}\colon {}^V\!\!D \to \{1, 0\}$, положив для всякого частичного приписывания~$\gamma$
  $$
  \begin{array}{lcl}
      \bm{p} (\gamma) & = & \left\{
      \begin{array}{ll}
        1, & \mbox{если $\otuple{\gamma(x_1),\gamma(x_2)}\in P^I$;}
        \\
        0 & \mbox{в остальных случаях.}
      \end{array}
      \right.
  \end{array}
  $$
  Затем определим интерпретации $I'$, положив для каждого $k\in\mathds{N}^+$
  $$
  \begin{array}{lcl}
  I'(p_k) & = & \bm{p}.
  \end{array}
  $$
  Наконец, положим $\cModel{M}' = \langle D, I' \rangle$.

  Зафиксируем элемент $a_0 \in D$. Для каждого приписывания $\alpha$ в
  $\cModel{M}$ определим частичное приписывание $\bar\alpha$ в~$\cModel{M}'$
  так, чтобы для каждого $k\in\numNp$ выполнялись условия
  $$
  \begin{array}{lcl}
  \bar{\alpha}(x_k) = a_0 & \mbox{ и } & \bar{\alpha}(y_k) = \alpha(v_k).
  \end{array}
  $$
  Покажем, что для любой формулы $\psi\in\sub\varphi$ и любого приписывания $\alpha$ в~$\cModel{M}$
  $$
  \begin{array}{lcl}
    \cModel{M}\cmodels^\alpha\psi & \Longleftrightarrow & {\downarrow}\cModel{M}'\models^{\bar{\alpha}}{\mathsf{Tr}}(\psi);
    \\
    \cModel{M}\not\cmodels^\alpha\psi & \Longleftrightarrow & {\downarrow}\cModel{M}'\not\models^{\bar{\alpha}}{\mathsf{Tr}}(\psi).
  \end{array}
  \eqno{(\ast)}
  $$

  Обоснование для $(\ast)$ проведём индукцией по~$\psi$.

  Если $\psi=\bot$, то ${\mathsf{Tr}}(\psi)=\bot$, и $(\ast)$ выполняется очевидным образом.

  Пусть $\psi = P(v_i, v_j)$; тогда ${\mathsf{Tr}}(\psi) =
  \mathcal{R}^{x_1}_{y_i} \mathcal{R}^{x_2}_{y_j}\,p$.

  Пусть $\cModel{M}\cmodels^\alpha P(v_i, v_j)$. Тогда
  $\otuple{\alpha(v_i),\alpha(v_j)}\in P^I$, и по определению
  для $\bar{\alpha}$ получаем, что
  $\otuple{\bar{\alpha}(y_i),\bar{\alpha}(y_j)}\in P^I$. Пусть $\bar{\beta}$~--- частичное приписывание, определённое следующими условиями:
  $$
  \begin{array}{lll}
  \bar{\beta} \stackrel{x_1, x_2}{=} \bar{\alpha}, & \bar{\beta} (x_1) =
  \bar{\alpha} (y_i), & \bar{\beta} (x_2) = \bar{\alpha}(y_j).
  \end{array}
  $$
  Тогда
  $$
  \begin{array}{lcl}
    \otuple{\bar{\alpha}(y_i),\bar{\alpha}(y_j)}\in P^I
    & \Longleftrightarrow & \langle{\bar{\beta}(x_1),\bar{\beta}(x_2)}\rangle\in P^I
                            \smallskip \\
    & \Longleftrightarrow &  \bm{p} (\bar{\beta}) = 1 \smallskip \\
    & \Longleftrightarrow &
                            {\downarrow}\cModel{M}'\models^{\bar{\alpha}}
                            \mathcal{R}^{x_1}_{y_i} \mathcal{R}^{x_2}_{y_j}\,p, \medskip \\
  \end{array}
  $$
  и тем самым первая эквивалентность в $(\ast)$ обоснована. Обоснование второй эквивалентности в $(\ast)$ проводится аналогично.

  Пусть $\psi
  = \psi_1 \to \psi_2$; тогда ${\mathsf{Tr}}(\psi)={\mathsf{Tr}}(\psi_1) \to
  {\mathsf{Tr}}(\psi_2)$.

  С учётом индукционного предположения получаем, что
  $$
  \begin{array}{lcl}
    \cModel{M}\cmodels^\alpha \psi_1 \to \psi_2
    & \Longleftrightarrow &
                            \cModel{M}\not\cmodels^\alpha \psi_1 \mbox{ или } \cModel{M}\cmodels^\alpha \psi_2
                            \smallskip \\
    & \Longleftrightarrow &
                            {\downarrow}\cModel{M}'\not\models^{\bar{\alpha}}
                            {\mathsf{Tr}}(\psi_1) \mbox{ или }  {\downarrow}\cModel{M}'\models^{\bar{\alpha}} {\mathsf{Tr}}(\psi_2)
                            \smallskip \\
    & \Longleftrightarrow &
                            {\downarrow}\cModel{M}'\models^{\bar{\alpha}} {\mathsf{Tr}}(\psi_1) \to {\mathsf{Tr}}(\psi_2);
                            \medskip \\
    \cModel{M}\not\cmodels^\alpha \psi_1 \to \psi_2
    & \Longleftrightarrow &
                            \cModel{M}\cmodels^\alpha \psi_1 \mbox{ и } \cModel{M}\not\cmodels^\alpha \psi_2
                            \smallskip \\
    & \Longleftrightarrow &
                            {\downarrow}\cModel{M}'\models^{\bar{\alpha}}
                            {\mathsf{Tr}}(\psi_1) \mbox{ и } {\downarrow}\cModel{M}'\not\models^{\bar{\alpha}} {\mathsf{Tr}}(\psi_2)
                            \smallskip \\
    & \Longleftrightarrow &
                            {\downarrow}\cModel{M}'\not\models^{\bar{\alpha}} {\mathsf{Tr}}(\psi_1) \to {\mathsf{Tr}}(\psi_2).
\end{array}
$$

Если $\psi = \psi_1 \wedge \psi_2$ или $\psi = \psi_1 \vee \psi_2$, то доказательство аналогично.

Пусть $\psi = \forall v_i\,\psi_1$; тогда ${\mathsf{Tr}}(\psi) = \forall
y_i\,{\mathsf{Tr}}(\psi_1)$.

Пусть $\cModel{M}\not\cmodels^\alpha\forall v_i\,\psi_1$. Тогда
$\cModel{M}\not\cmodels^\beta \psi_1$ для некоторого $\beta$, такого, что
$\beta\stackrel{v_i}{=}\alpha$. По индукционному предположению,
${\downarrow}\cModel{M}'\not\models^{\bar{\beta}} {\mathsf{Tr}}(\psi_1)$.
Поскольку $\beta\stackrel{v_i}{=}\alpha$, по определению $\bar{\alpha}$
и $\bar{\beta}$ получаем, что
$\bar{\beta}\stackrel{y_i}{=}\bar{\alpha}$. Значит,
${\downarrow}\cModel{M}'\not\models^{\bar{\alpha}} \forall y_i\,{\mathsf{Tr}}(\psi_1)$.

Пусть
${\downarrow}\cModel{M}'\not\models^{\bar{\alpha}} \forall y_i\,{\mathsf{Tr}}(\psi_1)$.
Тогда ${\downarrow}\cModel{M}'\not\models^{\gamma} {\mathsf{Tr}}(\psi_1)$ для некоторого частичного приписывания $\gamma$, такого, что
${\downarrow}\gamma(y_i)$ и $\gamma\stackrel{y_i}{=}\bar{\alpha}$.
Пусть $\beta$~--- приписывание, такое, что $\beta(v_k) = \gamma(y_k)$ для каждого $k\in\numNp$. Тогда $\bar{\beta}=\gamma$, и значит,
${\downarrow}\cModel{M}'\not\models^{\bar{\beta}} {\mathsf{Tr}}(\psi_1)$. По индукционному предположению, $\cModel{M}\not\cmodels^\beta \psi_1$. Поскольку
$\bar{\beta}\stackrel{y_i}{=}\bar{\alpha}$, по определению
$\bar{\alpha}$ и $\bar{\beta}$ получаем, что
$\beta\stackrel{v_i}{=}\alpha$. Значит,
$\cModel{M}\not\cmodels^\alpha\forall v_i\,\psi_1$.

Доказательство второй эквивалентности в $({\ast})$ проводится аналогично.

Если $\psi = \exists v_i\,\psi_1$, то доказательство $({\ast})$ тоже проводится аналогично.

Из (${\ast}$) получаем, что
${\downarrow}\cModel{M}'\not\models^{\bar{\alpha}_0} {\mathsf{Tr}}(\varphi)$.

Теперь заметим, что $\bm{p}$~--- тотальная функция. Но тогда получаем, что
${\downarrow} \cModel{M}'\models \forall y_1\forall
y_2\,(\mathcal{R}^{x_1}_{y_1} \mathcal{R}^{x_2}_{y_2} p \vee \neg
\mathcal{R}^{x_1}_{y_1} \mathcal{R}^{x_2}_{y_2} p)$.
Значит,
${\downarrow}\cModel{M}'\not\models^{\bar{\alpha}_0} \varphi^\ast$.

Следовательно, $\cModel{M}' \not\models \vp^\ast$, и значит,
$\vp^\ast \notin \mbox{$\logic{QPL}$}$.


Докажем импликацию $({\Rightarrow})$.
Пусть $\vp^\ast \not\in \mbox{$\logic{QPL}$}$. Тогда для некоторой
$\lang{QL}\mathit{q}$-модели $\cModel{M} = \langle D, I \rangle$ и некоторого частичного приписывания~$\alpha_0$
$$
\begin{array}{lcl}
{\downarrow}\cModel{M}\models^{\alpha_0}\forall y_1\forall
y_2\,(\mathcal{R}^{x_1}_{y_1} \mathcal{R}^{x_2}_{y_2}p\vee\neg
\mathcal{R}^{x_1}_{y_1} \mathcal{R}^{x_2}_{y_2} p)
  & \mbox{ и }
  & {\downarrow}\cModel{M}\not\models^{\alpha_0} {\mathsf{Tr}}(\varphi).
\end{array}
$$

Пусть $\psi \in \lang{QL}^{\mathit{bin}}$; частичные приписывания $\alpha,\beta\in{^V\!\!D}$ называем \defnotion{$\psi$-согла\-со\-ван\-ными}, если $\alpha\simeq\beta$ и для каждого
$k \in \nat^+$ выполняется следующее условие: если $v_k$ входит свободно в~$\psi$, то $\alpha(y_k)=\beta(y_k)$.

\begin{sublemma}
  \label{lem_agree}
  Пусть $\psi \in \lang{QL}^{\mathit{bin}}$, а $\alpha$ и $\beta$~--- $\psi$-согласованные частичные приписывания. Тогда
  $$
  \begin{array}{lcl}
    {\downarrow}\cModel{M}\models^\alpha {\mathsf{Tr}}(\psi) & \Longleftrightarrow & {\downarrow}\cModel{M}\models^\beta {\mathsf{Tr}}(\psi);
    \\
    {\downarrow}\cModel{M}\not\models^\alpha {\mathsf{Tr}}(\psi) & \Longleftrightarrow & {\downarrow}\cModel{M}\not\models^\beta {\mathsf{Tr}}(\psi);
    \\
    {\uparrow}\cModel{M}\models^\alpha {\mathsf{Tr}}(\psi) & \Longleftrightarrow & {\uparrow}\cModel{M}\models^\beta {\mathsf{Tr}}(\psi).
  \end{array}
  $$
\end{sublemma}

\begin{proof}
  Индукция по~$\psi$.

  Если $\psi = \bot$, то ${\mathsf{Tr}}(\psi) = \bot$, и справедливость утверждения очевидна.

  Пусть $\psi = P(v_i, v_j)$; тогда
  ${\mathsf{Tr}}(\psi) = \mathcal{R}^{x_1}_{y_i}\mathcal{R}^{x_2}_{y_j}\,p$.

  Пусть
  ${\downarrow}\cModel{M}\models^\alpha \mathcal{R}^{x_1}_{y_i}
  \mathcal{R}^{x_2}_{y_j}\,p$.
  Тогда ${\downarrow}\cModel{M}\models^{\alpha'} p$, где $\alpha'$ определено условиями $\alpha' \stackrel{x_1, x_2}{=} \alpha$,
  $\alpha'(x_1) = \alpha(y_i)$ и $\alpha'(x_2) = \alpha(y_j)$.
  Поскольку $v_i, v_j$ свободны в~$\psi$, получаем, что
  $\beta(y_i) = \alpha(y_i)$ и $\beta(y_j) = \alpha(y_j)$. Пусть
  $\beta'$~--- частичное приписывание, определённое условиями
  $\beta' \stackrel{x_1, x_2}{=} \beta$, $\beta'(x_1) = \beta(y_i)$
  и $\beta'(x_2) = \beta(y_j)$. Тогда $\beta' = \alpha'$, и значит,
  ${\downarrow}\cModel{M}\models^{\beta'} p$. Следовательно,
  ${\downarrow}\cModel{M}\models^\beta \mathcal{R}^{x_1}_{y_i}
  \mathcal{R}^{x_2}_{y_j}\,p$. Обратная импликация обосновывается так~же.

  Доказательство эквивалентностей
  $$
  \begin{array}{lcl}
    {\downarrow}\cModel{M}\not\models^\alpha \mathcal{R}^{x_1}_{y_i}
    \mathcal{R}^{x_2}_{y_j}\,p & \Longleftrightarrow & {\downarrow}\cModel{M}\not\models^\beta \mathcal{R}^{x_1}_{y_i}
                                                       \mathcal{R}^{x_2}_{y_j}\,p
  \end{array}
  $$
и
  $$
  \begin{array}{lcl}
    {\uparrow}\cModel{M}\models^\alpha \mathcal{R}^{x_1}_{y_i}
    \mathcal{R}^{x_2}_{y_j}\,p & \Longleftrightarrow & {\uparrow}\cModel{M}\models^\beta \mathcal{R}^{x_1}_{y_i}
                                                       \mathcal{R}^{x_2}_{y_j}\,p
  \end{array}
  $$
  проводится аналогично.

  Если $\psi = \psi_1 \to \psi_2$, $\psi = \psi_1 \wedge \psi_2$ или $\psi = \psi_1 \vee \psi_2$, то доказательство тривиально.

  Пусть $\psi = \forall v_i\,\psi_1$; тогда
  ${\mathsf{Tr}}(\psi) = \forall y_i\,{\mathsf{Tr}}(\psi_1)$. Значит,
  множество свободных переменных формулы $\psi_1$ содержится во множестве свободных переменных формулы $\psi$, расширенном переменной~$v_i$.

  Пусть
  ${\downarrow}\cModel{M}\models^\alpha \forall y_i\,{\mathsf{Tr}}(\psi_1)$.
  Пусть $\beta'$~--- произвольное частичное приписывание, такое, что
  ${\downarrow}\beta'(y_i)$ и $\beta'\stackrel{y_i}{=}\beta$.
  Определим частичное приписывание $\alpha'$, положив
  $\alpha'\stackrel{y_i}{=}\alpha$ и $\alpha'(y_i)=\beta'(y_i)$.
  Поскольку $\alpha'\stackrel{y_i}{=}\alpha$, получаем, что
  ${\downarrow}\cModel{M} \models^{\alpha'} {\mathsf{Tr}}(\psi_1)$. Нетрудно видеть, что $\alpha'$ и $\beta'$ являются $\psi_1$-согласованными. Значит, по индукционному предположению,
  ${\downarrow}\cModel{M} \models^{\beta'} {\mathsf{Tr}}(\psi_1)$. Поскольку
  $\beta'$ было выбрано произвольно, получаем, что
  ${\downarrow}\cModel{M}\models^\beta \forall y_i\,{\mathsf{Tr}}(\psi_1)$.
  Обратная импликация доказывается аналогично.

  Пусть
  ${\downarrow}\cModel{M}\not\models^\alpha \forall y_i\,{\mathsf{Tr}}(\psi_1)$.
  Тогда ${\downarrow}\cModel{M}\not\models^{\alpha'} {\mathsf{Tr}}(\psi_1)$ для некоторого $\alpha'$, такого, что ${\downarrow}\alpha'(y_i)$ и
  $\alpha'\stackrel{y_i}{=}\alpha$. Определим частичное приписывание
  $\beta'$, положив $\beta'\stackrel{y_i}{=}\beta$ и
  $\beta'(y_i)=\alpha'(y_i)$. Тогда $\alpha'$ и $\beta'$ являются
  $\psi_1$-согласованными. Значит, по индукционному предположению,
  ${\downarrow}\cModel{M}\not\models^{\beta'} {\mathsf{Tr}}(\psi_1)$.
  Следовательно,
  ${\downarrow}\cModel{M}\not\models^\beta \forall y_i\,{\mathsf{Tr}}(\psi_1)$. Обратная импликация доказывается аналогично.

  Пусть
  ${\uparrow}\cModel{M}\models^\alpha \forall y_i\,{\mathsf{Tr}}(\psi_1)$.
  Предположим, что условие
  ${\uparrow}\cModel{M}\models^\beta \forall y_i\,{\mathsf{Tr}}(\psi_1)$ не выполняется. Тогда либо
  ${\downarrow}\cModel{M}\models^\beta \forall y_i\,{\mathsf{Tr}}(\psi_1)$, либо
  ${\downarrow}\cModel{M}\not\models^\beta \forall y_i\,{\mathsf{Tr}}(\psi_1)$,
  и значит, как мы видели выше, либо
  ${\downarrow}\cModel{M}\models^\alpha \forall y_i\,{\mathsf{Tr}}(\psi_1)$,
  либо
  ${\downarrow}\cModel{M}\not\models^\alpha \forall y_i\,{\mathsf{Tr}}(\psi_1)$, что даёт противоречие.
\end{proof}

Теперь определим $\lang{QL}$-модель $\cModel{M}'$, опровергающую формулу~$\vp$.
Поскольку $\varphi$ замкнута, можем считать, согласно подлемме~\ref{lem_agree}, что ${\downarrow}\alpha_0(y_k)$ для каждой переменной $y_k \in \mathit{IVar}$.
Для каждой пары элементов $a, c \in D$ обозначим через $\alpha_{ac}$ частичное приписывание, определённое следующим образом:
$$
\begin{array}{lll}
\alpha_{ac} (x_1) = a,
  & \alpha_{ac} (x_2) = c,
  & \alpha_{ac}\stackrel{x_1,x_2}{=}\alpha_0.
\end{array}
$$
Пусть $I'$~--- такая интерпретация, что
$$
\begin{array}{lcl}
  I'(P)
  & =
  & \set{\otuple{a,c}\in D^2 : p^I(\alpha_{ac}) = 1}.
\end{array}
$$
Положим $\cModel{M}' = \langle D, I' \rangle$.

Для каждого $\alpha \in {{}^{V}}\!\!D$, удовлетворяющего $y_k \in \mathit{IVar}$ условиям
$$
\begin{array}{lcl}
\alpha\simeq\alpha_0 & \mbox{и} & {\downarrow}\alpha(y_k)
\end{array}
\eqno{({\ast}{\ast})}
$$
определим приписывание $\bar{\alpha}$, положив
$\bar{\alpha}(v_k) = \alpha(y_k)$ для каждого $k \in \numNp$.

Покажем, что для каждой $\psi\in\sub\varphi$ и каждой
$\alpha \in {{}^{V}}\!\!D$, удовлетворяющей условию (${\ast}{\ast}$), \nopagebreak
$$
\begin{array}{lcl}
  {\downarrow}\cModel{M}\models^\alpha {\mathsf{Tr}}(\psi)
  & \Longleftrightarrow
  & \cModel{M}'\cmodels^{\bar{\alpha}}\psi;
  \\
  {\downarrow}\cModel{M}\not\models^\alpha {\mathsf{Tr}}(\psi)
  & \Longleftrightarrow
  & \cModel{M}'\not\cmodels^{\bar{\alpha}}\psi.
\end{array}
\eqno{({\ast}{\ast}{\ast})}
\nopagebreak
$$
Поскольку
${\downarrow}\cModel{M}\models^{\alpha_0}\forall y_1\forall
y_2\,(\mathcal{R}^{x_1}_{y_1} \mathcal{R}^{x_2}_{y_2} p \vee \neg
\mathcal{R}^{x_1}_{y_1} \mathcal{R}^{x_2}_{y_2} p)$,
получаем, что частичное приписывание $\alpha_0$ является тотальной функцией; а значит, и $\alpha$ тоже. Но тогда эквивалентности в (${\ast}{\ast}{\ast}$) следуют друг из друга, и достаточно доказать лишь одну из них; докажем первую.

Обоснование проведём индукцией по~$\psi$.

Если $\psi = \bot$, то ${\mathsf{Tr}}(\psi) = \bot$, и доказывать нечего.

Пусть $\psi = P(v_i,v_j)$; тогда
${\mathsf{Tr}}(\psi) = \mathcal{R}^{x_1}_{y_i} \mathcal{R}^{x_2}_{y_j}\,p$.

Пусть $\alpha$~--- частичное приписывание, такое, что
$\alpha\simeq\alpha_0$. Тогда для $\beta$, определённого условиями
$\beta(x_1) = \alpha(y_i)$, $\beta(x_2) = \alpha(y_j)$ и
$\beta\stackrel{x_1,x_2}{=}\alpha$, получаем, что
$$
\begin{array}{lcl}
{\downarrow}\cModel{M}\models^\alpha \mathcal{R}^{x_1}_{y_i}
  \mathcal{R}^{x_2}_{y_j}\,p
  & \Longleftrightarrow & {\downarrow}\cModel{M}\models^\beta p.
\end{array}
$$

Пусть $\beta'$~--- частичное приписывание, определённое следующим образом: $\beta'(x_1)=\beta(x_1)$, $\beta'(x_2)=\beta(x_2)$ и
$\beta'\stackrel{x_1,x_2}{=}\alpha_0$. Покажем, что
$\beta'\simeq\beta$. По определению, $\beta'(x_1)=\beta(x_1)$ и
$\beta'(x_2)=\beta(x_2)$; кроме того, поскольку
$\beta'\stackrel{x_1,x_2}{=}\alpha_0$, $\alpha\simeq\alpha_0$ и
$\beta\stackrel{x_1,x_2}{=}\alpha$, получаем, что для каждого
$k\in\numNp\setminus\{1,2\}$
$$
\beta'(x_k) ~=~ \alpha_0(x_k) ~=~ \alpha(x_k) ~=~ \beta(x_k).
$$
Значит, $\beta'\simeq\beta$. Тогда, по $(\mathit{QP})$,
$$
\begin{array}{lclcl}
{\downarrow}\cModel{M}\models^\beta p & \Longleftrightarrow &
{\downarrow}\cModel{M}\models^{\beta'} p & \Longleftrightarrow &
p^I(\beta') = 1.
\end{array}
$$
Поскольку $\beta'\stackrel{x_1,x_2}{=}\alpha_0$,
$\beta'(x_1) = \alpha(y_i)$ и $\beta'(x_2) = \alpha(y_j)$, получаем, что
$$
\begin{array}{lclcl}
p^I(\beta') = 1 & \Longleftrightarrow &
P^{I'}(\beta'(x_1),\beta'(x_2)) = 1 & \Longleftrightarrow &
P^{I'}(\alpha(y_i),\alpha(y_j)) = 1.
\end{array}
$$
Поскольку $\bar{\alpha}(x_i)=\alpha(y_i)$ и
$\bar{\alpha}(x_j)=\alpha(y_j)$, получаем, что
$$
\begin{array}{lclcl}
\langle\alpha(y_i),\alpha(y_j)\rangle\in I'(P) & \Longleftrightarrow &
\langle\bar{\alpha}(x_i),\bar{\alpha}(x_j)\rangle\in I'(P) &
\Longleftrightarrow & \cModel{M}'\cmodels^{\bar{\alpha}} P(x_i,x_j).
\end{array}
$$

Таким образом, в итоге получаем, что
$$
\begin{array}{lcl}
{\downarrow}\cModel{M}\models^\alpha \mathcal{R}^{x_1}_{y_i}\mathcal{R}^{x_2}_{y_j}\,p
& \Longleftrightarrow &
\cModel{M}'\cmodels^{\bar{\alpha}} P(x_i,x_j).
\end{array}
$$

Пусть $\psi = \psi_1\to\psi_2$; тогда
${\mathsf{Tr}}(\psi)={\mathsf{Tr}}(\psi_1)\to {\mathsf{Tr}}(\psi_2)$.

С учётом индукционного предположения,
$$
\begin{array}{lcl}
{\downarrow}\cModel{M}\models^\alpha {\mathsf{Tr}}(\psi_1\to\psi_2)
  & \Longleftrightarrow &
                          {\downarrow}\cModel{M}\not\models^\alpha
                          {\mathsf{Tr}}(\psi_1) \mbox{ или } {\downarrow}\cModel{M}\models^\alpha {\mathsf{Tr}}(\psi_2)
                          \smallskip \\
  & \Longleftrightarrow &
                          \cModel{M}'\not\cmodels^{\bar{\alpha}} \psi_1 \mbox{ или }\cModel{M}'\cmodels^{\bar{\alpha}} \psi_2
                          \smallskip \\
  & \Longleftrightarrow &
                          \cModel{M}'\cmodels^{\bar{\alpha}} \psi_1 \to \psi_2.
\end{array}
$$

Аналогично, если $\psi = \psi_1\wedge\psi_2$ или $\psi = \psi_1\vee\psi_2$.

Пусть $\psi = \forall v_i\,\psi_1$; тогда
${\mathsf{Tr}}(\psi) = \forall y_i\,{\mathsf{Tr}}(\psi_1)$.

Пусь
${\downarrow}\cModel{M}\models^\alpha\forall y_i\,{\mathsf{Tr}}(\psi_1)$.
Пусть $\gamma$~--- произвольное приписывание, такое, что
$\gamma\stackrel{v_i}{=} \bar\alpha$. Определим частичное приписывание
$\beta$, положив $\beta(y_i)=\gamma(v_i)$ и
$\beta\stackrel{y_i}{=}\alpha$. Поскольку ${\downarrow}\beta(y_i)$ и
$\beta\stackrel{y_i}{=}\alpha$, по индукционному предположению получаем, что
${\downarrow}\cModel{M}\models^\beta {\mathsf{Tr}}(\psi_1)$. Очевидно, что
$\beta \simeq \alpha_0$ и ${\downarrow}\beta(y_k)$ для каждого
$y_k \in \mathit{IVar}$. Значит, по индукционному предположению,
$\cModel{M}'\cmodels^{\bar \beta} \psi_1$. Но $\gamma = \bar\beta$;
значит, $\cModel{M}'\cmodels^{\gamma} \psi_1$. В силу произвольности выбора $\gamma$ получаем, что $\cModel{M}'\cmodels^{\bar{\alpha}} \forall v_i\,\psi_1$.

Пусть
$\cModel{M}'\cmodels^{\bar{\alpha}} \forall v_i\,\psi_1$. Пусть $\beta$~--- произвольное частичное приписывание, такое, что ${\downarrow}\beta(y_i)$ и
$\beta\stackrel{y_i}{=}\alpha$. Определим приписывание $\gamma$ условиями
$\gamma(v_i)=\beta(y_i)$ и $\gamma\stackrel{v_i}{=}\bar\alpha$.
Поскольку $\gamma\stackrel{v_i}{=}\bar\alpha$, по индукционному предположению получаем, что
$\cModel{M}'\cmodels^{\gamma} \psi_1$. Но $\gamma = \bar\beta$; значит,
$\cModel{M}'\cmodels^{\bar \beta} \psi_1$. Очевидно, что
$\beta \simeq \alpha_0$ и ${\downarrow}\beta(y_k)$ для каждого
$y_k \in IVar$. Значит, по индукционному предположению,
${\downarrow}\cModel{M}\models^\beta {\mathsf{Tr}}(\psi_1)$. В силу произвольности выбора $\beta$ получаем, что
${\downarrow}\cModel{M}\models^\alpha\forall y_i\,{\mathsf{Tr}}(\psi_1)$.

Если $\psi = \exists v_i\,\psi_1$, то обоснование проводится аналогично.

Таким образом, условие (${\ast}{\ast}{\ast}$) доказано.

Поскольку частичное приписывание $\alpha_0$ удовлетворяет условию (${\ast}{\ast}$), получаем по
(${\ast}{\ast}{\ast}$), что
$\cModel{M}'\not\models^{\bar{\alpha}_0}\varphi$. Значит,
$\vp \notin \mbox{$\logic{QCl}$}$.
\end{proof}

Заметим, что в доказательстве леммы~\ref{lem_Tr} строились модели, предметная область которых совпадала с предметной областью исходных моделей, где опровергалась формула. Значит, имеет место следующее утверждение.

\begin{proposition}
  \label{cor:Tr}
  Пусть $\varphi$~--- замкнутая\/ $\lang{QL}^{\mathit{bin}}$-формула.  Тогда
$$
\begin{array}{lcl}
\varphi\in \mbox{$\logic{QCl}_{\mathit{fin}}$}
  & \iff
  & \varphi^\ast\in \mbox{$\logic{QPL}_{\mathit{fin}}$}.
\end{array}
$$
\end{proposition}

   \subsection{Аналоги теоремы Чёрча и теоремы Трахтенброта}   

Используя лемму~\ref{lem_Tr} и предложение~\ref{cor:Tr}, можно получить результаты об алгоритмических свойствах логик $\mbox{$\logic{QPL}$}$ и $\mbox{$\logic{QPL}$}_{\mathit{fin}}$, а также их фрагментов.

Из леммы~\ref{lem_Tr} и теоремы Чёрча~\cite{Church36} получаем следующий факт.

\begin{theorem}
  \label{thr:QPL}
  Логика\/ $\mbox{$\logic{QPL}$}$ алгоритмически неразрешима.
\end{theorem}

При этом проблема общезначимости $\lang{QL}\mathit{q}$-формул $\Sigma^0_1$-трудна, а проблема выполнимости $\lang{QL}\mathit{q}$-формул $\Pi^0_1$-трудна.

Из предложения~\ref{cor:Tr} и аналогичных результатов для
$\mbox{$\logic{QCl}$}_{\mathit{fin}}$~\cite{Trakhtenbrot50,Trakhtenbrot53} (см.~также~\cite{Libkin}), получаем ещё один факт.

\begin{theorem}
  \label{thr:QPL-fin}
  Логика\/ $\mbox{$\logic{QPL}$}_{\mathit{fin}}$ не является рекурсивно перечислимой.
\end{theorem}

Здесь уточнение следующее: проблема конечной общезначимости $\lang{QL}\mathit{q}$-формул $\Pi^0_1$-трудна, а проблема выполнимости в классе конечных $\lang{QL}\mathit{q}$-моделей является $\Sigma^0_1$-трудной.

Конечно, аналог теоремы Трахтенброта сохраняется и в более сильной формулировке. Благодаря~\cite{NT12}, известно, что $\mbox{$\logic{QPL}$}$ и $\mbox{$\logic{QPL}$}_{\mathit{fin}}$ рекурсивно погружаются в, соответственно, $\mbox{$\logic{QCl}$}$ и $\mbox{$\logic{QCl}$}_{\mathit{fin}}$. Поэтому получаем следующее утверждение.

\begin{theorem}
  \label{thr:inseperability}
  Множества\/ $\mbox{$\logic{QPL}$}$ и\/ $\lang{QL}\mathit{q}\setminus \mbox{$\logic{QPL}$}_{\mathit{fin}}$
  образуют неотделимую пару рекурсивно перечислимых множеств.
\end{theorem}

Формулу $\varphi$ языка $\lang{QL}\mathit{q}$ будем называть \defnotion{позитивной}, если она не содержит вхождений константы~$\bot$.

Заметим, что перевод $\mathsf{Tr}$ сохраняет позитивность формул, а переменных в формуле $\mathsf{Tr}(\varphi)$ не более чем на две больше, чем в исходной формуле $\varphi$ от бинарной буквы~$P$. Тогда, с учётом теорем~\ref{th:church:positive} и~\ref{th:trakhtenbrot:positive}, получаем ещё два уточнения.

\begin{corollary}
  \label{cor:QPL}
  Позитивный фрагмент логики\/ $\mbox{$\logic{QPL}$}$ с одной квазиарной буквой, двумя существенными переменными и тремя несущественными переменными является\/ $\Sigma^0_1$-полным.
\end{corollary}

\begin{corollary}
  \label{cor:QPL-fin}
  Позитивный фрагмент логики\/ $\mbox{$\logic{QPL}_{\mathit{fin}}$}$ с одной квазиарной буквой, двумя существенными переменными и тремя несущественными переменными является\/ $\Pi^0_1$-полным.
\end{corollary}

По всей видимости, число используемых переменных невозможно уменьшить с сохранением неразрешимости; но это лишь предположение. Тем не менее, при некоторых дополнительных ограничениях использование лишь одной существенной переменной приводит к разрешимому фрагменту; мы покажем это в следующем разделе.

   \subsection{Разрешимые фрагменты}   

Заметим, что в доказательстве неразрешимости в предыдущем разделе мы
применили оператор переименования $\mathcal{R}$ только к двум существенным
переменным, $x_1$ и $x_2$.  Более того, кванторы применялись только к несущественным переменным, значения которых были немедленно
<<заменены>> с помощью $\mathcal{R}$ на $x_1$ и $x_2$. Это наводит на мысль, что одним из способов алгоритмически упростить $\logic{QPL}$ может быть разрешение использовать $\mathcal{R}$ только с одной существенной переменной $x$, и
ограничить применение кванторов, разрешая только кванторы по несущественным переменным, значения которых заменяются с помощью $\mathcal{R}$ на~$x$.  
Результирующий фрагмент обозначим через $\lang{QL}\mathit{q}_1$; он может быть определён следующим BNF-выражением:
$$
\begin{array}{lcl}
\vp & := & \bot \mid p_i \mid (\vp \wedge \vp) \mid (\vp \vee \vp) \mid (\vp \to \vp) \mid
\mathcal{R}^{x}_{y_i}\varphi \mid \exists y_i\,\varphi' \mid \forall y_i\,\varphi',
\end{array}
$$
где переменная $x \in \mathit{EVar}$ зафиксирована и $\vp'$ удовлетворяет дополнительному условию, состоящему в том, что каждая входящая в $\vp'$ квазиарная предикатная буква находится в области действия оператора~$\mathcal{R}^{x}_{y_i}$ для некоторой несущественной переменной~$y_i$.

Поскольку формула $\mathcal{R}^{x}_{y_i} \forall y_j \mathcal{R}^{x}_{y_j} \vp$ эквивалентна в $\logic{QPL}$ формуле
$\forall y_j \mathcal{R}^{x}_{y_i} \mathcal{R}^{x}_{y_j} \vp$ и формула
$\mathcal{R}^{x}_{y_i} \mathcal{R}^{x}_{y_j} \vp$ эквивалентна в $\logic{QPL}$ формуле
$\mathcal{R}^{x}_{y_j} \vp$, получаем, что каждая формула из $\lang{QL}\mathit{q}_1$ эквивалентна в $\logic{QPL}$ некоторой формуле, построенной из $\bot$ и формул вида $\mathcal{R}^{x}_{y_i} p_k$ с использованием конъюнкции, дизъюнкции, импликации и кванторов по несущественным переменным. Обозначим фрагмент языка $\lang{QL}\mathit{q}$, состоящий из таких формул, через $\lang{QL}\mathit{q}_1'$. Нетрудно понять, что $\lang{QL}\mathit{q}_1'$ является образом монадического фрагмента языка $\lang{QL}$ при следующем переводе~$\mathsf{Tr}'$:
$$
\begin{array}{lcl} \mathsf{Tr}'(\bot) & = & \bot; \smallskip\\
  \mathsf{Tr}'(P_k(v_i)) & = & \mathcal{R}^{x}_{y_i} p_k; \smallskip\\
  \mathsf{Tr}'(\psi\wedge\chi) & = & \mathsf{Tr}'(\psi)\wedge \mathsf{Tr}'(\chi); \smallskip\\
  \mathsf{Tr}'(\psi\vee\chi) & = & \mathsf{Tr}'(\psi)\vee \mathsf{Tr}'(\chi); \smallskip\\
  \mathsf{Tr}'(\psi\to\chi) & = & \mathsf{Tr}'(\psi)\to \mathsf{Tr}'(\chi); \smallskip\\
  \mathsf{Tr}'(\exists v_i\,\psi) & = & \exists y_i\,\mathsf{Tr}'(\psi); \smallskip\\
  \mathsf{Tr}'(\forall v_i\,\psi) & = & \forall y_i\,\mathsf{Tr}'(\psi).
\end{array}
$$

Нетрудно видеть, что $\mathsf{Tr}'$ является биекцией между монадическим фрагментом языка $\lang{QL}$ и фрагментом $\lang{QL}\mathit{q}_1'$. Поскольку монадические фрагменты логик $\logic{QCl}$ и $\logic{QCl}_{\mathit{fin}}$ совпадают и разрешимы~\cite{Lowenheim15,Behmann22}, получаем, что $\lang{QL}\mathit{q}_1'$-фрагменты и $\lang{QL}\mathit{q}_1$-фрагменты логик $\logic{QPL}$ и $\logic{QPL}_{\mathit{fin}}$ тоже разрешимы.

В итоге получаем следующие утверждения.

\begin{proposition}
Множество\/ $\lang{QL}\mathit{q}_1\cap\logic{QPL}$ разрешимо.
\end{proposition}

\begin{proposition}
Множество\/ $\lang{QL}\mathit{q}_1\cap\logic{QPL}_{\mathit{fin}}$ разрешимо.
\end{proposition}

Конечно, разрешимыми будут и все фрагменты логики $\logic{QPL}$, являющиеся образами разрешимых фрагментов логики $\logic{QCl}$ при погружениях, построенных с помощью переводов, сходных с описанными выше $\mathsf{Tr}$ и~$\mathsf{Tr}'$; то же самое касается разрешимых фрагментов логик $\logic{QCl}_{\mathit{fin}}$ и $\logic{QPL}_{\mathit{fin}}$; детали соответствующих построений опускаем.


\chapter*{Заключение}
\addcontentsline{toc}{chapter}{Заключение}
\setcounter{footnote}{\value{savefootnote}}
      \label{ch:10}

Результаты, представленные в главах~\ref{chapter:ML}--\ref{chapter:PML}, можно коротко описать следующим образом:
\begin{itemize}
\item очень многие естественные мономодальные, полимодальные, суперинтуиционистские и виссеровские логики полиномиально погружаются в свои собственные фрагменты от небольшого числа (обычно от нуля до двух) переменных;
\item интервалы, ограниченные логиками, полными в некотором классе сложности и погружающимися в свой фрагмент от небольшого числа переменных, состоят из логик, фрагменты которых от такого же числа переменных трудны в этом классе сложности;
\item сложность контрмоделей многих естественных логик, трудных в классе $\cclass{PSPACE}$ или выше, как правило, высока для их фрагментов от небольшого числа переменных;
\item тем не менее, в классах расширений многих естественных логик существуют логики, которые линейно аппроксимируемы шкалами Крипке, но при этом могут иметь сколь угодно высокую сложность проблемы разрешения, включая степени неразрешимости.
\end{itemize}
При этом, когда речь идёт о погружениях логик в свои фрагменты от малого числа переменных, в очень многих случаях эти погружения построены явно; в остальных случаях существование таких погружений следует из сложности логики в полном языке и сложности соответствующего её фрагмента (и~логика, и её фрагмент оказываются полны в некотором, причём одном и том же, классе сложности, а потому полиномиально эквивалентны).

Отметим, что предложенная здесь техника имеет ограничения. Так, она неприменима к логикам, семантика которых запрещает ветвление, таким как $\logic{LTL}$ или темпорально-эпистемические логики линейного времени~\cite{HV89,GorSh09b}; тем не менее, как было отмечено выше, в подобных случаях возможны другие подходы для получения результатов о том, что фрагмент от конечного числа переменных логики и сама логика полиномиально эквивалентны. Кроме того, описанный выше метод погружения логической системы в её фрагмент от конечного числа переменных не может быть применён в случае, когда система не замкнута по правилу подстановки. Примером такой системы является \defnotion{логика публичных объявлений}\index{логика!публичных объявлений}\footnote{Английский термин~--- public announcement logic.} $\logic{PAL}$~\cite{Plaza89,DHK08} и близкие к ней логики. Тем не менее, известно, что~$\logic{PAL}$ содержит фрагмент, замкнутый по правилу подстановки, который имеет ту же вычислительную сложность, что и вся система~$\logic{PAL}$~\cite{Lutz06}, и этот фрагмент можно полиномиально свести к его фрагменту от конечного числа переменных с помощью описанной здесь техники.

Ещё один момент, на который обратим внимание, связан с тем, что фактически мы рассматривали проблему \defnotion{локальной выполнимости}\index{бяа@выполнимость!локальная} формул: вопрос о непринадлежности отрицания формулы полной по Крипке логике эквивалентен вопросу об истинности этой формулы в некотором мире некоторой модели этой логики. Но иногда интересен вопрос о \defnotion{глобальной выполнимости}\index{бяа@выполнимость!глобальная} формул, т.е. об их истинности в каждом мире некоторой модели логики. Например, такой вопрос возникает, когда модальные логики используются для моделирования \defnotion{дескрипционных логик}\index{логика!дескрипционная}\footnote{Английский термин~--- description logic.}~\cite{BCGNP:2003}. Имеются некоторые основания предполагать, что возможна модификация описанного здесь метода, чтобы промоделировать проблему глобальной выполнимости модальных формул в полном языке формулами в языке с фиксированным конечным числом переменных; это одна из возможных тем для дальнейших исследований.

Наконец, стоит сказать о новых результатах, которые были получены, но не были отражены в предыдущих главах диссертации. 
Во-первых, это уже упоминавшаяся логика $\logic{HC}$ и близкая к ней логика $\logic{H4}$~\cite{Mel:I,Mel:II}, а также интуиционистские эпистемические логики $\logic{IEL}^-$, $\logic{IEL}$, $\logic{IEL}^+$~\cite{ArtemovProtopopescu}. Из представленных здесь результатов следует $\ccls{PSPACE}$-трудность позитивного фрагмента $\logic{HC}$ в языке с двумя переменными; оказалось, что $\ccls{PSPACE}$-полными являются уже фрагменты $\logic{HC}$ с одной переменной (при этом переменная может иметь любой из двух возможных типов~--- <<задача>> или <<высказывание>>). Соответствующие результаты были представлены в~\cite{MR:2024:Nsk:1}. Во-вторых, это результаты, касающиеся погружения интуиционистских модальных логик в их позитивные фрагменты от одной переменной~\cite{MR:2023:SCAN:3,MR:2025:JLC}, которые тоже используют развитие идей, представленных в диссертации.
В-третьих, это результаты, касающиеся сложности константных фрагментов и фрагментов от одной переменной различных ненормальных модальных логик~\cite{MR:2025:Kudinov:SR, MR:2025:Kudinov:arXive, MR:2025:MR:1}, а также нормальных модальных логик, содержащих аксиому конвергентности или близкую к ней~\cite{MR:2025:Shcherbakov:SR, MR:2025:Shcherbakov:arXiv}. Некоторые результаты вошли в обзор~\cite{MR:2026:Z}.

Результаты, представленные в главах~\ref{ch:6}--\ref{ch:9}, можно коротко описать следующим образом:
\begin{itemize}
\item очень многие естественные первопорядковые классические теории неразрешимы в языке с одной бинарной предикатной буквой и тремя предметными переменными, при этом для доказательства требуются лишь позитивные формулы;
\item очень многие естественные модальные и суперинтуиционистские предикатные логики неразрешимы в языке с одной унарной предикатной буквой и двумя предметными переменными;
\item монадические фрагменты модальных и суперинтуиционистских предикатных логик, обогащённые равенством, разрешимы для логик, полных относительно одной конечной шкалы Крипке;
\item во многих естественных случаях логики неэлементарных классов шкал Крипке не являются рекурсивно перечислимыми или даже являются неарифметическими в языке с одной-двумя унарными предикатными буквами и двумя-тремя предметными переменными;
\item при этом существуют полные по Крипке модальные предикатные логики, не определяемые элементарными классами шкал, которые, тем не менее, погружаются в классическую логику предикатов и являются рекурсивно перечислимыми.
\end{itemize}

Результаты и методы, представленные в диссертации, переносятся и на другие классы логик. Перечислим некоторые результаты, полученные с помощью описанных здесь техник или их модификаций.

Результаты о разрешимых фрагментах модальных логик (см.~раздел~\ref{s7-3}) переносятся на полимодальные предикатные логики, в том числе логики с равенством~\cite{MR:2023:SCAN:1}. То же относится к полимодальным предикатным логикам, семантика модальных операторов которых неэлементарна~\cite{MR:2016:Tver,MR:2013:LI}, включая неполноту по Крипке различных исчислений~\cite{MR:2015:LI}.

Классическая логика с оператором композиции бинарных отношений неразрешима в языке с одной бинарной буквой и двумя предметными переменными, а если её язык обогатить некоторыми неэлементарными средствами (например, оператором транзитивного замыкания или оператором проверки транзитивности бинарного отношения), то получим $\Sigma^1_1$-трудное расширение классической логики предикатов~\cite{MR:2022:DAN:rus,MR:2023:LI} (см.~также~\cite{GOR-1999}); отметим, что алгоритмическая сложность логики конечных моделей при этом не увеличивается.

Заметное продвижение получается, если обратиться к понятию рекурсивной неотделимости множеств. Так, удалось доказать, что рекурсивно неотделимыми являются $\logic{QCl}$ и дополнение к теории $\logic{SIB}_{\mathit{fin}}$ конечных моделей симметричного иррефлексивного бинарного отношения в языке с одной бинарной предикатной буквой и тремя предметными переменными~\cite{MR:2024:MR,MR:2025:RecInsep}. Использование этого результата позволяет усилить и результаты соответствующих разделов в главах~\ref{ch:7} и~\ref{ch:8}, касающиеся поиска <<минимальных>> неразрешимых фрагментов~\cite{MR:2024:MR,MR:2025:RecInsep}. В~частности, удалось доказать рекурсивную неотделимость логики $\logic{QK}$ и дополнения к логике $\logic{QS5}_{\mathit{wfin}}$ в языке с одной унарной предикатной буквой и двумя предметными переменными. Из рекурсивной неотделимости $\logic{QS5}$ и $\logic{QS5}_{\mathit{wfin}}$ (а она следует из полученного результата) мгновенно получаем неожиданное усиление теоремы Трахтенброта: логика $\logic{QCl}$ и дополнение к логике $\logic{QCl}_{\mathit{fin}}$ имеют рекурсивно неотделимые фрагменты, состоящие из формул, построенных лишь из двух атомарных формул $P(x,z)$ и $P(y,z)$, где $P$~--- бинарная предикатная буква. В случае суперинтуиционистских логик удалось доказать неразрешимость позитивных фрагментов с одной унарной предикатной буквой и двумя предметными переменными для всех логик, содержащихся в логике конечных шкал Крипке высоты семь или в логике конечных конвергентных шкал высоты восемь.

Ещё одним результатом, о котором стоит упомянуть, является решение старой проблемы о разрешимости фрагмента предикатного варианта логики Гёделя--Дамметта $\logic{QLC}$ в языке с двумя предметными переменными (см., например,~\cite{CMRT:2022}). Оказалось, что неразрешимая проблема домино может быть промоделирована в логике $\logic{QLC}$ (и некоторых её расширениях) с помощью позитивных формул, содержащих одну бинарную предикатную букву, бесконечно много унарных предикатных букв и две предметные переменные~\cite{MR:2024:CMCR,MR:2025:QLC-2var}; бесконечное множество унарных предикатных букв можно заменить одной бинарной~\cite{MR:2024:MR:2}. При этом неясно, возможно ли дальнейшее упрощение языка с сохранением полученных результатов.

\setcounter{savefootnote}{\value{footnote}}



\renewcommand{\bibname}{Библиография}
 \newcommand{\noop}[1]{}
  \newcommand{\nosort}[1]{}\newcommand{\titlefont}{\rm}\providecommand{\nosort}[1]{}

\clearpage
\phantomsection
\addcontentsline{toc}{chapter}{Предметный указатель}
\printindex

\end{document}